\documentclass[sn-mathphys,Authoryear]{sn-jnl}
\usepackage[utf8]{inputenc}
\usepackage{graphicx}%
\usepackage{multirow}%
\usepackage{amsmath,amssymb,amsfonts}%
\usepackage{amsthm}%
\usepackage{mathrsfs}%
\usepackage[title]{appendix}%
\usepackage{xcolor}%
\usepackage{textcomp}%
\usepackage{manyfoot}%
\usepackage{booktabs}%
\usepackage{algorithm}%
\usepackage{algorithmicx}%
\usepackage{algpseudocode}%
\usepackage{listings}%
\usepackage{comment}
\usepackage{multicol}
\usepackage{arydshln}
\usepackage{isotope}
\usepackage{lipsum}
\usepackage{rotating}

\usepackage{acronym}

\usepackage{float}

\definecolor{orange}{RGB}{255,165,0}   % Orange color
\definecolor{purple}{RGB}{128,0,128}   % Purple color

\newcommand{\reaction}[6]{\nucc{#1}{#2}(#3,#4)\/\nucc{#5}{#6}}
\newcommand{\nucc}[2]{\ensuremath{^{#1}}{#2}}

\newcommand{\Ercm}[1]{\ensuremath{E_{r}^{\text{c.m.}} = #1}~keV}

\newcommand{\nuc}[2]{$\mathrm{^{#2}#1}$}

\newcommand{\aap}{A\&A}

\newcommand{\apj}{ApJ}
\newcommand{\apjl}{ApJL}

\newcommand{\mnras}{MNRAS}

\newcommand{\nar}{NewAR}
\newcommand{\nat}{Nature}

\newcommand{\prc}{Phys. Rev. C}

\makeatletter
\renewcommand{\@@address}[2][]{%
  \g@addto@macro\auaddress{%
     {\normalsize % <- choose \small, \footnotesize, \scriptsize
     \stepcounter{affn}%
     \xdef\@currentlabel{\theaffn}%
     \jmkLabel{\theaffn}%
     {\textsuperscript{#1}#2.\par}}%
  }%
}

\renewcommand{\@@coraddress}[2][]{%
  \g@addto@macro\auaddress{%
     {\normalsize % same size here
     \stepcounter{affn}%
     \xdef\@currentlabel{\theaffn}%
     \jmkLabel{\theaffn}%
     {\textsuperscript{#1*}#2.\par}}%
  }%
}

\renewcommand{\email}[1]{%
  \global\advance\emailcnt by 1\relax
  \if@corauemail
    \g@addto@macro\corrauthemail{{\small\textcolor{blue}{#1}; }}%
  \else
    \g@addto@macro\authemail{{\small\textcolor{blue}{#1}; }}%
  \fi
}
\makeatother

\makeatother

\begin{document}

\title[NucSyn]{\centering $\gamma$-Ray Lines --\\
Signatures of Nucleosynthesis, Cosmic Rays,\\Positron Annihilation, and Fundamental Physics}

\frenchspacing

\author*[1]{\fnm{Thomas} \sur{Siegert}}\email{thomas.siegert@uni-wuerzburg.de}
\author[2]{\fnm{Francesca} \sur{Calore}}\email{calore@lapth.cnrs.fr}
    \author[3]{\fnm{Pierre} \sur{Jean}}\email{pjean@irap.omp.eu}
\author[4]{\fnm{Mark} \sur{Leising}}\email{lmark@clemson.edu}
\author[5]{\fnm{Nicolas} \sur{de Séréville}}\email{nicolas.de-sereville@ijclab.in2p3.fr}
\author[6,7]{\fnm{Gerald~H.} \sur{Share}}\email{gershare@aol.com}
\author[5]{\fnm{Vincent} \sur{Tatischeff}}\email{vincent.tatischeff@csnsm.in2p3.fr}
\author[8]{\fnm{Wei} \sur{Wang}}\email{wangwei2017@whu.edu.cn}
\author[9,10,11]{\fnm{Meng-Ru} \sur{Wu}}\email{mwu@as.edu.tw}

{
\affil[1]{\orgname{Julius-Maximilians-Universität Würzburg, Institut für Theoretische Physik und Astrophysik, Lehrstuhl für Astronomie}, \orgaddress{\street{Emil-Fischer-Str. 31}, \city{97074 Würzburg}, \country{Germany}}}

\affil[2]{\orgname{LAPTh, CNRS, USMB}, \orgaddress{\city{74940 Annecy}, \country{France}}}

\affil[3]{\orgname{IRAP, Université de Toulouse, CNRS, CNES}, \orgaddress{\street{9 avenue du colonel Roche}, \city{BP 44346, 31028 Toulouse}, \country{France}}} 

\affil[4]{\orgname{Clemson University, Department of Physics and Astronomy}, \orgaddress{\city{Clemson, SC 29634-0978}, \country{USA}}}

\affil[5]{\orgname{Université Paris-Saclay, CNRS/IN2P3, IJCLab}, \orgaddress{\city{91405 Orsay}, \country{France}}}

\affil[6]{\orgname{Astronomy Department, University of Maryland}, \orgaddress{\city{College Park, MD 20740}, \country{USA}}}

\affil[7]{TSC, Resident at the Naval Research Laboratory, \orgaddress{\city{Washington, DC 20375-5352}, \country{USA}}}

\affil[8]{\orgname{School of Physics and Technology, Wuhan University}, \orgaddress{\city{Wuhan 430072}, \country{People's Republic of China}}}

\affil[9]{\orgname{Institute of Physics, Academia Sinica}, \orgaddress{\city{Taipei, 11529}, \country{Taiwan}}}

\affil[10]{Institute of Astronomy and Astrophysics, Academia Sinica, \orgaddress{\city{Taipei, 106319}, \country{Taiwan}}}

\affil[11]{Physics Division, National Center for Theoretical Sciences, \orgaddress{\city{Taipei, 106319}, \country{Taiwan}}}
}

\abstract{
The nuclear \ac{gray} lines in the MeV range of the electromagnetic spectrum hold a vast variety of astrophysical, particle-physical, and fundamental physical information that is otherwise extreme difficult to access.
MeV \ac{gray} line observations provide the most direct evidence for ongoing nucleosynthesis in galaxies by measuring freshly produced radioactive isotopes from massive stars, \acp{SN}, \acp{CN}, or \acp{BNSM}.
Their flux ratios can determine the \ac{LECR} spectrum in different objects and of the Milky Way as a whole.
Different phases of the \ac{ISM} are traced by hot nucleosynthesis ejecta, cooling positrons, or \ac{CR} interactions with molecular clouds.
Positron annihilation itself can be considered as an astrophysical messenger as their production and destruction in typical space environments is inevitable.
Finally, as-of-yet unknown signatures from \ac{BSM} physics might have their elusive imprints in \ac{gray} lines.

\noindent This Chapter gives an overview of historical \ac{gray} line measurements, newest results, and open questions that may only be solved by a new generation of MeV telescopes.
}


\maketitle
%\newpage
\tableofcontents
\newpage

\section*{List of Acronyms}
\addcontentsline{toc}{section}{List of Acronyms}
\setlength{\columnseprule}{0.4pt}
\begin{multicols}{2}
\footnotesize
\raggedright
\begin{acronym}[INTEGRAL]
    \acro{ACS}{anticoincidence system}
    \acro{AGB}{asymptotic giant branch}
    \acro{ALP}{axion-like particle}
    \acroplural{ALP}[ALPs]{axion-like particles}
    \acro{BAT}{Burst Alert Telescope}
    \acro{BATSE}{Burst and Transient Source Experiment}
    \acro{BBN}{Big Bang Nucleosynthesis}
    \acro{BH}{black hole}
%    \acro{BNS}{binary neutron star}
%    \acroplural{BNS}[BNSs]{binary neutron stars}
    \acro{BNSM}{binary neutron star merger}
    \acroplural{BNSM}[BNSMs]{binary neutron star mergers}
    \acro{BSM}{beyond Standard Model}
    \acro{CasA}[Cas\,A]{Cassiopeia A}
    \acro{CGB}{Cosmic Gamma-Ray Background}
    \acro{CMB}{Cosmic Microwave Background}
    \acro{CME}{coronal mass ejection}
    \acroplural{CME}[CMEs]{coronal mass ejections}
    \acro{CXB}{Cosmic X-Ray Background}  
    \acro{CN}{classical nova}
    \acroplural{CN}[CNe]{classical novae}
    \acro{COBE}{Cosmic Background Explorer}
    \acro{CEJSN}{common envelope jet supernova}
    \acroplural{CEJSN}[CEJSNe]{common envelope jet supernovae}
    \acro{COMPTEL}{Imaging Compton Telescope}
    \acro{ccSN}{core-collapse supernova}
    \acroplural{ccSN}[ccSNe]{core-collapse supernova}
    \acro{CGRO}{Compton Gamma Ray Observatory}
    \acro{COSI}{Compton Spectrometer and Imager}
    \acro{CR}{cosmic ray}
    \acroplural{CR}[CRs]{cosmic rays}
    \acro{DM}{dark matter}
    \acro{DMR}{Differential Microwave Radiometer}
    \acro{EC}{electron capture}
%    \acro{EGRET}{Energetic Gamma Ray Experiment Telescope}
    \acro{FIP}{first ionisation potential}
    \acro{FoV}{field of view}
    \acro{FWHM}{full width at half maximum}
%    \acro{GRB}{$\gamma$-ray burst}
    \acroplural{GRB}[GRBs]{$\gamma$-ray bursts}
    \acro{GRS}{Gamma-Ray Spectrometer}
    \acro{GRaND}{Gamma Ray and Neutron Detector}
    \acro{GRIS}{Gamma Ray Imaging Spectrometer}
    \acro{GEANT4}{GEometry ANd Tracking 4}
    \acro{GF}{giant flare}
    \acroplural{GF}[GFs]{giant flares}
    \acro{gray}[$\gamma$-ray]{gamma-ray}
    \acroplural{gray}[$\gamma$-rays]{gamma-rays}
    \acro{HEAO3}[HEAO\,3]{High Energy Astronomy Observatory 3}
    \acro{IBIS}{Imager on Board the INTEGRAL Satellite}
    \acro{IC}{Inverse Compton}
    \acro{IMF}{initial mass function}
    \acro{INTEGRAL}{International Gamma-Ray Astrophysics Laboratory}
    \acro{ISM}{interstellar medium}
    \acro{JT}{Jovian Trojan}
    \acroplural{JT}[JTs]{Jovian Trojans}
    \acro{KBO}{Kuiper Belt Object}
    \acroplural{KBO}[KBOs]{Kuiper Belt Objects}
    \acro{LMC}{Large Magellanic Cloud}
    \acro{LECR}{low-energy cosmic ray}
    \acroplural{LECR}[LECRs]{low-energy cosmic rays}
    \acro{LoS}{line of sight}
    \acro{LP}{Lunar Prospector}
    \acro{MBA}{Main Belt Asteroid}
    \acroplural{MBA}[MBAs]{Main Belt Asteroids}
    \acro{MESA}{Modules for Experiments in Stellar Astrophysics}
    \acro{MRSN}{magneto-rotational supernova}
    \acroplural{MRSN}[MRSNe]{magneto-rotational supernovae}
    \acro{NEAR}{Near Earth Asteroid Rendezvous}
    \acro{NFW}{Navarro-Frenk-White}
    \acro{NS}{neutron star}
    \acro{NT}{Neptunian Trojan}
    \acroplural{NT}[NTs]{Neptunian Trojans}
    \acro{NuSTAR}{Nuclear Spectroscopic Telescope Array}
    \acro{OSSE}{Oriented Scintillation Spectrometer Experiment}
    \acro{PTSN}{phase-transition supernova}
    \acroplural{PTSN}[PTSNe]{phase-transition supernovae}
    \acro{Ps}{Positronium}
    \acro{PSYCO}{population synthesis code}
    \acro{RHESSI}{Reuven Ramaty High Energy Solar Spectroscopic Imager}
    \acro{RMS}{root mean square}
    \acro{SEP}{Solar energetic particle}
    \acroplural{SEP}[SEPs]{Solar energetic particles}
    \acro{SFR}{star formation rate}
    \acro{SMM}{Solar Maximum Mission}
    \acro{SN}{supernova}
    \acroplural{SN}[SNe]{supernovae}
    \acro{SNR}{supernova remnant}
    \acroplural{SNR}[SNRs]{supernova remnants}
    \acro{SNIa}[SN\,Ia]{type Ia supernova}
    \acroplural{SNIa}[SNe\,Ia]{type Ia supernovae}
    \acro{SPI}{Spectrometer onboard INTEGRAL}
    \acro{TGRS}{Transient Gamma-Ray Spectrometer}
    \acro{TS}{test statistic}
    \acro{WD}{white dwarf}
    \acroplural{WD}[WDs]{white dwarfs}
    \acro{WIMP}{weakly interacting massive particle}
    \acroplural{WIMP}[WIMPs]{weakly interacting massive particles}
    \acro{WISP}{weakly interacting slim particle}
    \acroplural{WISP}[WISPs]{weakly interacting slim particles}
    \acro{WR}{Wolf-Rayet}
\end{acronym}
\end{multicols}

% Introduction
\newpage
\section{Introduction}\label{sec:intro_nucsyn}
\vspace{-0.5em}
{\emph{Written by Thomas Siegert}}
\vspace{0.5em}\\
The MeV range is prone to many astrophysical processes that give rise to \ac{gray} lines \citep[e.g.,][]{Ramaty1979_review,Schoenfelder2001_book,Diehl2018_book}.
Here, we define \ac{gray} lines as mono-energetic spectral features, whose detailed analyses inform about the microphysical mechanisms that ultimately lead to observable large-scale phenomena.
In this chapter, we distinguish between decay \acsp{gray} from nucleosynthesis ejecta and the nuclear excitation of circum- and \ac{ISM}, as well as matter of major and minor bodies, by low-energy cosmic rays \acp{LECR} followed by de-excitation, even though the emission mechanism is identical \citep{Ramaty1979_review,Lingenfelter1986_review,Diehl2018_book}.
The possible subsequent annihilation of positrons created in these processes is treated individually as well because several different interaction channels can lead to the production of positrons than just radioactive decay \citep{Lingenfelter1986_review,Prantzos2011}.
Finally, \ac{gray} lines may also be produced through \ac{BSM} processes directly \citep{Lewin1996_review,Bergstroem1998_neutralinos}.

This Chapter will focus on the observational characteristics of \ac{gray} lines.
Details of the individual source types can be found elsewhere in this volume, and we refer the reader to these Chapters.
Typically, Chapters that consider nuclear \acp{gray} start by listing a table of lines with line energies, lifetimes, channels, and if they have been detected.
Here, we are attempting to show a much broader overview of \ac{gray} line observations and their potential for future MeV space missions.
Therefore, we postpone the list of lines to the individual subsections because a single table would be too large and incomprehensive.

In what follows, we outline how \ac{gray} line measurements are to be understood in general.
This means, we introduce the concept of \ac{gray} lines in space and in the measurement, and how they can change due to different (astro-)physical considerations.
The applications to prominent \ac{gray} lines are then found in Sec.\,\ref{sec:observations}.

\subsection{Defining a $\gamma$-Ray Line}\label{sec:gamma-ray-linees}
We define the spectral shape of any \ac{gray} line, $L(E;\vec{\varphi})$, as a function of photon energy, $E$, and a set of parameters, $\vec{\varphi}$, by a convolution of its intrinsic (microphysical) line shape, $I(E;\vec{\vartheta})$, with an astrophysical broadening shape, $A(E;\vec{\varrho})$, as
\begin{equation}
    L(E;\vec{\varphi}) = \int_{-\infty}^{+\infty}\,dE'\,I(E';\vec{\vartheta}) \cdot A(E-E';\vec{\varrho})\mathrm{.}
    \label{eq:definition_line}
\end{equation}
In many cases, $I(E;\vec{\vartheta})$ has the shape of a Lorentz curve with $\vec{\vartheta}$ including the centroid of the line, $E_{\rm lab}$, and its natural broadening \citep[e.g., its lifetime;][]{Krane1987_book}.
Often, the spectral resolution of typical detectors in space applications is much larger than the natural line width, so that in all meaningful studies, $I(E;\vec{\vartheta})$ can be approximated as a $\delta$-function,
\begin{equation}
    I(E;\vec{\vartheta}) \approx I_0 \delta(E-E_{\rm lab})\mathrm{,}
    \label{eq:delta_function}
\end{equation}
with $I_0$ being the amplitude of the line.

\subsubsection{Astrophysical $\gamma$-Ray Lines}\label{sec:astro_lines}
The astrophysical shape, $A(E;\vec{\varrho})$, then describes the entire expectation towards the measurement and can be of any physically allowed form, such as symmetric or asymmetric Gaussians, box-like shapes, multiple components, etc., depending on the actual geometry of the object, its kinematics, and physical effects like absorption.
In most astrophysical studies, $A(E;\vec{\varrho})$ can be approximated by a Gaussian,
\begin{equation}
    A(E;\vec{\varrho}) \approx G(E;\Delta E,\sigma) = \frac{A_0}{\sqrt{2\pi} \cdot \sigma}\exp\left(-\frac{1}{2}\frac{(\Delta E)^2}{\sigma^2}\right)\mathrm{,}
    \label{eq:gaussian}
\end{equation}
where $\Delta E$ describes the line shift with respect to the laboratory energy, $\sigma$ describes the astrophysical line broadening, and $A_0$ the astrophysical line flux.

\paragraph{Doppler Shift}
The line shift can be directly related to the non-relativistic Doppler shift by
\begin{equation}
    \frac{\Delta E}{E_{\rm lab}} = \frac{v}{c}\mathrm{,}
    \label{eq:Doppler-shift}
\end{equation}
where $v$ is the relative velocity of the object to the observer, and $c$ is the speed of light.
A `blue-shift' of the line indicates that $v$ is positive and that object and observer are approaching each other.

\paragraph{Doppler Broadening}\label{sec:Doppler_broadening}
The astrophysical line broadening $\sigma$ can also be linked to a Doppler-broadening velocity of an object, but which depends on the actual situation, and should take into account opacities and density distributions \citep{Ferriere2001_ISMreview}.
On the one hand, in cases in which an observer sees a spatially unresolved expanding sphere, assumed to be homogeneously filled and expanding at the same velocity isotropically with $v_{\rm exp}$, the astrophysical shape is \emph{not} Gaussian, but a boxcar that ranges from $E_{\rm lab}(1-v_{\rm exp}/c)$ to $E_{\rm lab}(1+v_{\rm exp}/c)$ \citep{Chan1987_SNe}.
The same shape applies to a thin shell.
If, nevertheless, this shape is to be approximated as a Gaussian, the \ac{FWHM} of the Gaussian would be
\begin{equation}
    \mathrm{FWHM} = 2 \sqrt{2 \ln2} \cdot \sigma = 2 \frac{v_{\rm exp}}{c} E_{\rm lab} \Leftrightarrow \sigma \approx 0.85 \frac{v_{\rm exp}}{c} E_{\rm lab} \mathrm{.}
    \label{eq:fwhm_shell}
\end{equation}
On the other hand, the $\sigma$ value of the astrophysical Gaussian shape can also be discussed in terms of the turbulent motion or temperature, that is, not bulk motion, or in terms of conditions for annihilation.
We will return to these interpretations in the respective subsections.

\subsubsection{Taking into Account the Instrument}
Beyond the astrophysical line shape, the measured line shape is subject to the spectral response of the detectors used \citep{Knoll2000_book}.
This means the energy resolution of the instrument has to be taken into account to interpret the measurement correctly.
While also here, the line shape can be very asymmetric, for example from incomplete charge collection \citep{Knoll2000_book,Diehl2018}, due to degradation of the detectors due to cosmic-ray bombardment, or simply from the electronics, a symmetric Gaussian is typically sufficient to describe the instrumental resolution.
As a result, Eq.\,(\ref{eq:definition_line}) has to be convolved with the spectral response \citep[e.g.,][]{Pendleton1995_response,Siegert2022_book}.
In a simple measurement of turbulent motion, for example, the measured line width, $\sigma_{\rm meas}$, is then approximately the quadratic sum of the instrumental line width (`resolution'), $\sigma_{\rm inst}$, and the astrophysical line width, $\sigma_{\rm astro}$,
\begin{equation}
    \sigma_{\rm meas}^2(E) = \sigma_{\rm inst}^2(E) + \sigma_{\rm astro}^2(E) 
    %\Leftrightarrow \mathrm{FWHM}_{\rm meas}^2(E) = \mathrm{FWHM}_{\rm inst}^2(E) + \mathrm{FWHM}_{\rm astro}^2(E) \mathrm{.}
    \label{eq:line_widths}
\end{equation}
Taking into account that instrumental resolutions, $\mathrm{FWHM}_{\rm inst}(E)/E$, are typically on the order of 0.1\% (Ge) to 10\% (NaI), and further depend on the photon energy with an approximate square-root dependence, $\mathrm{FWHM}_{\rm inst}(E) \propto \sqrt{E}$, it becomes evident that the spectral resolution in the MeV range is generally improving as a function of energy.
However, due to other constraints, such as detector efficiency and photon count statistics, the `best' measurements are typically achieved between 0.5--2.5\,MeV.
Continuing the example of turbulent motion along the \ac{LoS}, the width of the astronomical component is directly related to the 1D \ac{LoS} \ac{RMS},
\begin{equation}
    \sigma_{\rm astro} = \frac{\left\langle v^2_{\rm los}\right\rangle^{1/2}}{c} E_{\rm lab}\mathrm{.}
\end{equation}
For turbulent motion in the \ac{ISM}, $\sigma_{\rm astro}$ is the about 100\,$\mathrm{km\,s^{-1}}$ \citep{Ferriere2001_ISMreview}, which results in a Doppler broadening of $0.6\,\mathrm{keV}$ at 1809\,keV (radioactive \nuc{Al}{26}, see Sec.\,\ref{sec:Al26}).
Even with Ge detectors and their excellent energy resolution of $\approx 3.1$\,keV (\ac{FWHM} at 1809\,keV), the measured line width would only be 3.4\,keV (\ac{FWHM}), that is, slightly above instrumental resolution.

\subsection{Astrophysical Line Fluxes and Luminosities}\label{sec:line_fluxes}
The astrophysical line flux, $A_0$, together with the amplitude of the intrinsic line, $I_0$, can be combined into a measurable flux, $F$.
Typically, $F$ is in units of $\mathrm{ph\,cm^{-2}\,s^{-1}}$ if integrated over the region of interest, or if the source is point-like, or in units of $\mathrm{ph\,cm^{-2}\,s^{-1}\,sr^{-1}}$ to take into account possible spatial variations.
In general, $F$ depends on the astrophysics studied.
Again, we distinguish between radioactive decay, nuclear excitation, positron annihilation, and \ac{BSM} processes, such as \ac{DM} annihilation and decay.
We can separate the microphysical components from the astrophysical (geometrical) ones by considering
\begin{equation}
    F(\ell,b) = \rho_0 \cdot \frac{1}{4\pi\,\mathrm{sr}}\int_0^\infty\,ds\,\rho\left(\tilde{x}(s,\ell,b),\tilde{y}(s,\ell,b),\tilde{z}(s,\ell,b)\right)\mathrm{,}
    \label{eq:los_integral}
\end{equation}
where $\left(\tilde{x},\tilde{y},\tilde{z}\right)^T$ is the \ac{LoS} vector from the observer ($s=0$) to infinity, $(\ell,b)$ are (Galactic) coordinates, $\rho\left(x,y,z\right)$ is a normalised, unitless emissivity distribution in three dimensions, and $\rho_0$ is the (central) emissivity in units of $\mathrm{ph\,cm^{-3}\,s^{-1}}$.
The microphysical input is completely covered in $\rho_0$.
The factor $(4\pi\,\mathrm{sr})^{-1}$ is explicitly written out for the full differential flux as a function of the chosen coordinates, $F(\ell,b)$.
Integrating over the sphere of the sky results in the total flux,
\begin{equation}
    F = \int_{-\pi}^{+\pi}\,d\ell\,\int_{-\pi/2}^{+\pi/2}\,\sin(b)\,db\,F(\ell,b)\mathrm{,}
    \label{eq:total_flux}
\end{equation}
in units of $\mathrm{ph\,cm^{-2}\,s^{-1}}$.
The volume integral over the emissivity,
\begin{equation}
    L = \rho_0 \cdot \int_{-\pi}^{+\pi}\,d\ell\,\int_{-\pi/2}^{+\pi/2}\,\sin(b)\,db\,\int_0^{\infty}\,ds\,s^2\,\rho\left(\tilde{x}(s,\ell,b),\tilde{y}(s,\ell,b),\tilde{z}(s,\ell,b)\right)\mathrm{,}
    \label{eq:luminosity}
\end{equation}
describes the luminosity, $L$, of the source in units of $\mathrm{ph\,s^{-1}}$.
Since the central emissivity is independent of the coordinates, we can describe the luminosity as $L = \rho_0 \cdot V_{\rm eff}$, with $V_{\rm eff}$ being an `effective volume' of the astrophysical source.
In this regard, the volume integral can also be carried out in more suitable coordinates and does not necessarily need to start from the observer.
This is important to note because it directly relates the microphysics to the shape of the object studied.
Keeping this general description of flux and luminosity, we can also define an `effective distance' to any source type (extended or point-like) by 
\begin{equation}
    d_{\rm eff} := \left[\frac{L}{4\pi F}\right]^{1/2}\mathrm{,}
    %= \left[\frac{\int_{d\Omega}\,d\Omega\,\int_0^{\infty}\,ds\,s^2\,\rho\left(\tilde{x}(s,\ell,b),\tilde{y}(s,\ell,b),\tilde{z}(s,\ell,b)\right)}{\int_{d\Omega}\,d\Omega\,\int_0^\infty\,ds\,\rho\left(\tilde{x}(s,\ell,b),\tilde{y}(s,\ell,b),\tilde{z}(s,\ell,b)\right)}\right]^{1/2}\mathrm{,}
    \label{eq:effective_distance}
\end{equation}
which is simply the weighted emissivity distribution, and does not depend on the microphysics.
For the classical case of point sources (ps) with $\rho(x,y,z) = \delta(x-x_0)\delta(y-y_0)\delta(z-z_0)$, and an absolute distance to the observer of $d = (x_0^2 + y_0^2 + z_0^2)^{1/2}$, the effective distance reduces to the absolute distance, and the general `inverse square law' is recovered,
\begin{equation}
    F_{\rm ps} = \frac{L_{\rm ps}}{4 \pi d_{\rm ps}^2}\mathrm{.}
    \label{eq:spherical_law}
\end{equation}

\subsubsection{Radioactive Decay}
The generic equation of photons from radioactive decays is
\begin{equation}
    X \stackrel{\tau}{\longrightarrow} Y^* + Z;~~~Y^* \stackrel{\mathrm{fs}-\mathrm{ps}}{\longrightarrow} Y + \gamma(E_\gamma)\mathrm{,}
    \label{eq:decay_equation}
\end{equation}
where the parent nucleus, $X$, is decaying into a nuclearly excited state of a daughter nucleus, $Y^*$, within a lifetime $\tau$, and some other particle(s), named $Z$.
This process can happen through all possible decay channels, such as $\alpha$- or $\beta^{\pm}$-decay, \ac{EC}, or similar.
The excited daughter nucleus has a typical lifetime of fs to ps and, on astronomical time scales, immediately de-excites by emitting a photon, $\gamma$, with energy $E_\gamma$.
In an even more general case, there can be cascades or these de-excitations so that one decay can result in several decay \acp{gray} with specific branching ratios.
The fundamental statement for these decay \acp{gray} is that they can only be produced once -- when the parent nucleus decayed, it cannot produce these photons any more.
However, the daughter nucleus might also be radioactive and could itself decay and therefore lead to additional \acp{gray}.
This fact will become important in the case of \acp{SNIa}.

In the case of decaying isotopes \citep[e.g.,][]{Clayton1968_book,Woosley1995}, we consider the total radioactive mass of an object, $M_0$, made of an isotope $i$ with atomic mass $m_i$, that is decaying with a lifetime $\tau_i$.
Then, the activity is 
\begin{equation}
    A_i(t) := \frac{N_i(t)}{\tau_i} = \frac{M(t)}{m_i \tau_i} = \frac{M_0}{m_i \tau_i}\exp\left(-\frac{t-t_0}{\tau_i}\right)\mathrm{,}
    \label{eq:activity}
\end{equation}
with $M(t) = M_0 \exp\left(-\frac{t-t_0}{\tau_i}\right)$ describing the decay of the total mass as a function of time, $t$, so that $M(t_0) = M_0$.
The activity is directly related to the luminosity of this astrophysical source by considering that each decay has a probability of $p_i$ to emit a photon from the excited daughter nucleus.
It follows that
\begin{equation}
    L_i(t) = A_i(t) p_i = \frac{M_0 p_i}{m_i \tau_i}\exp\left(-\frac{t-t_0}{\tau_i}\right)\mathrm{.}
    \label{eq:decay_luminosity}
\end{equation}

\subsubsection{Nuclear Excitation}\label{sec:nuc_ex}
The generic equation of photons from nuclear excitation followed by de-excitation is
\begin{equation}
    X + Y \stackrel{\sigma v}{\longrightarrow} X' + Y^*;~~~Y^* \stackrel{\mathrm{fs}-\mathrm{ps}}{\longrightarrow} Y + \gamma(E_\gamma)\mathrm{,}
    \label{eq:nuc_ex_equation}
\end{equation}
where an energetic particle, $X$, typically a \ac{CR} proton or alpha, interacts with an almost stationary particle, $Y$ \citep{Ramaty1979_review,koz02}.
Then, there is a chance, given by the cross-section, $\sigma$, and the relative velocity, $v$, that the particle $Y$ becomes nuclearly excited to $Y^*$.
Similar to the radioactive decay, the excited state immediately de-excites by emitting a \ac{gray} at a specific energy.
However, opposed to the case of radioactive decay, this typically happens for stable nuclei in the \ac{ISM} so that the excitation and de-excitation process is in balance and reflects the spectrum of \acp{LECR} and the ambient densities.
Also, most of the excitations will lead to the first excited state and only rarely will higher levels be occupied.
This is changed, however, for the case of nuclear excitation of nuclei in planetary atmospheres and solid bodies where the densities are much larger than in the \ac{ISM} and the chances to populate higher levels are sizeable.

In the case of exciting isotopes, we consider the particle number density of a (stable) species $i$ in the \ac{ISM}, $n_i$, in units of $\mathrm{cm^{-3}}$, that interacts with a spectrum of \acp{CR} of the form $dn_{\rm CR}(E)/dE$ in units of $\mathrm{cm^{-3}\,eV^{-1}}$, given the interaction cross-section for nuclear excitation, $\sigma_{\rm ex}(E)$, and the relative velocity, $v_{\rm rel}(E)$.
The emissivity is then calculated as
\begin{equation}
    \rho_0 = \int_{E_{\rm th}}^\infty\,dE\,\frac{dn_{\rm CR}(E)}{dE} n_{\rm ISM} \sigma_{\rm ex}(E) v_{\rm rel}(E)\mathrm{,}
    \label{eq:nuc_ex_emissivity}
\end{equation}
where $E_{\rm th}$ is the threshold energy for the nuclear excitation to a certain level.
Note that in the more general case, the photons created in the process could also be absorbed again, so that additional factors have to be included.
Likewise, there are also competing processes to the excitation, such as spallation, elastic scattering, transmutation, etc., which may also lead to (different) \ac{gray} photons or to none.
An important process in this context is the neutron capture by protons, forming deuterium, and a \ac{gray} line at 2224\,keV \citep[see Secs.\,\ref{sec:SFs} \& \ref{sec:outlook_deuterium}; see also][]{Share1995_flares,hua87}.
In the case of two-plasma interactions, also the shape of the resulting \ac{gray} lines will be influenced by the plasma temperature and asymmetric two-peak structures may emerge due to the relativistic Doppler effect \citep{Aharonian1984_lines,Yoneda2023}.

\subsubsection{Positron Annihilation}
A generic equation of photons from positron annihilation does not exist, because there are several channels that can lead to the \ac{gray} line at 511\,keV photon energies.
We will give three production channels in the following:
\begin{eqnarray}
    &\mathrm{a)}~~~e^+ + e^- &\stackrel{\sigma_{da}}{\longrightarrow} \gamma(511\,\mathrm{keV}) + \gamma(511\,\mathrm{keV})\\
    &\mathrm{b)}~~~e^+ + e^- &\stackrel{\sigma_{rr}}{\longrightarrow}~^1\mathrm{Ps};~~~~^1\mathrm{Ps} \stackrel{\tau_{\rm p-Ps}}{\longrightarrow} \gamma(511\,\mathrm{keV}) + \gamma(511\,\mathrm{keV})\\
    &\mathrm{c)}~~~e^+ + X &\stackrel{\sigma_{cx}}{\longrightarrow}~^1\mathrm{Ps} + X^+;~~~~^1\mathrm{Ps} \stackrel{\tau_{\rm p-Ps}}{\longrightarrow} \gamma(511\,\mathrm{keV}) + \gamma(511\,\mathrm{keV})
    \label{eq:annihilation_equation}
\end{eqnarray}
where the cases a), b), and c), are direct annihilation (da), radiative recombination (rr), and charge exchange (cx), respectively \citep{Guessoum2005_511}.
As will be discussed further in Secs.\,\ref{sec:511} \& \ref{sec:not_a_line}, all these positron annihilation channels do \emph{not} necessarily result in a 511\,keV line \citep{Ore1949,Aharonian1981_positrons,Aharonian2000_aif}.
Case a) only results in a 511\,keV line, if the kinetic energies of the participating particles are small, for example on the order of a few 100\,eV at most.
At higher energies, the relative velocities of the electron and the positron leads to a conversion of also their kinetic energies into photons, resulting in a continuous spectrum from $m_e/2$ to $T+m_e/2$, with $m_e$ being the rest mass energy of the electron and $T$ its kinetic energy \citep{Aharonian2000_aif}.
The direct annihilation spectrum can be single- or double-peaked, and the internal shape depends on the kinetic energies of both particles \citep{Svensson1982_positrons,Svensson1983}.
Case a) competes with case b) in the absence of atoms.
Radiative recombination of free positrons with free electrons is the dominant process below $\sim 30$\,eV \citep{Prantzos2011}.
Then, an intermediate bound state, called \ac{Ps}, is formed \citep{Ore1949,Deutsch1951_Ps}.
The exotic atom of \ac{Ps} behaves similarly to a hydrogen atom, but with a limited lifetime of $\tau_{\rm p-Ps} = 0.124$\,ns in the case of para-\ac{Ps}, i.e. the $^1\mathrm{Ps}$ state.
Para-\ac{Ps} is \emph{decaying} via the electromagnetic force to two photons that carry $511$\,keV each.
Due to its finite lifetime, the para-\ac{Ps} \ac{gray} line has a natural width of $5.3\,\mathrm{\mu eV}$, and is as such always taken as a delta-function.
In three out of four recombinations, however, the $^3\mathrm{Ps}$ state is formed, called ortho-\ac{Ps}, which has a longer lifetime of $\tau_{\rm o-Ps} = 13.9\mathrm{\mu s}$ \citep{Deutsch1951b_Ps}.
In this case, due to parity conservation, ortho-\ac{Ps} is decaying into three photons with a characteristic spectrum that peaks at 511\,keV \citep{Ore1949}, but shows no \ac{gray} line.
In the presence of atoms ($X$), case c) is the dominant channel by far (four to six orders of magnitude in cross-section) above the species-specific threshold \citep{Guessoum1991_511}.
The cross-section of charge exchange typically peaks slightly above threshold, for example around 12\,eV with a threshold of 6.8\,eV in the case of natural hydrogen, and competes with the process of atomic or molecular excitation \citep[see][for a more detailed description]{2005A&A...436..171G,2010MNRAS.402.1171G,Prantzos2011,Siegert2023}.

In any of the three cases, the annihilation rate / \ac{Ps}-formation rate per unit volume can be calculated as
\begin{equation}
    \Gamma_\pm = n_+(E_+) n_x(E_x) \sigma_x(E_+,E_x) v_{\rm rel}(E_+,E_x)\mathrm{,}
    \label{eq:positron_emissivity}
\end{equation}
where $n_+(E_+)$ is the particle number density of positrons in units of $\mathrm{cm^{-3}}$ with kinetic energy $E_+ \lesssim 1$\,keV, $n_x(E_x)$ is the particle number density of species $x$, with either being free electrons or at least one-electron atoms or molecules, having a kinetic energy $E_x \lesssim 1$\,keV, $\sigma_x(E_+,E_x)$ being the corresponding cross-section in units of $\mathrm{cm^{-2}}$, and $v_{\rm rel}(E_+,E_x)$ is the relative velocity \citep{Svensson1982_positrons}.
In this calculation, we only consider non-relativistic particles ($E_+ \lesssim 1$\,keV) to obtain an order of magnitude estimate for the annihilation rate.
\ac{Ps} only forms efficiently for small relative velocities of the participating particles.
In the general case of arbitrary kinetic energies, the distribution functions for the positrons and species $x$ has to be taken into account, in addition to using the differential cross-section $d\sigma_x(E_+,E_x,\mu)/dE$.
In the case of slow participating partners, it is often enough to use non-differential values.
Now, in order to calculate the emissivity of \emph{only} the 511\,keV line, we need to take into account the formation of \ac{Ps} with a fraction $f_{\rm Ps}$, typically called the \ac{Ps}-fraction:
The \ac{Ps}-fraction can be calculated from the statistical weight of \ac{Ps} decaying into two or three photons, and the number of direct annihilations without \ac{Ps} formation.
The multiplicity of a particular spin state is $(2S+1)$, so that para-\ac{Ps} will be formed $1/4$ of the time, and ortho-\ac{Ps} $3/4$ of the time.
While para-\ac{Ps} emits two photons, ortho-\ac{Ps} will emit three photons, so that the emissivity $\rho_{3\gamma}$ is proportional to $3\frac{3}{4}f_{\rm Ps} = \frac{9}{4}f_{\rm Ps}$.
The two-photon emissivity (the 511\,keV line) is then given by the sum over para-\ac{Ps} decays and the direct annihilation emissivity, weighted with the respective \ac{Ps}-fraction, $\rho_{2\gamma} \propto 2\frac{1}{4}f_{\rm Ps} + 2(1-f_{\rm Ps}) = 2 - \frac{3}{2}f_{\rm Ps}$ \citep{Leventhal1978}.
Linking the annihilation rate to the 511\,keV emissivity, we find
\begin{equation}
    \rho_{2\gamma} = \Gamma_{\pm} \left( 2 - \frac{3}{2}f_{\rm Ps} \right)\mathrm{,}
    \label{eq:511_emissivity}
\end{equation}
with $f_{\rm Ps}$ being related to the ratio of the two-photon to three-photon fluxes, $R_{32} = F_{3\gamma}/F_{2\gamma}$, by
\begin{equation}
    f_{\rm Ps} = \frac{8R_{32}}{6R_{32} + 9}\mathrm{.}
    \label{eq:Ps_fraction}
\end{equation}
The \ac{Ps}-fraction is a function of the \ac{ISM} / surrounding conditions, such as ionisation state and temperature \citep{Guessoum1991_511,Guessoum2005_511,Churazov2005,Jean2006}.
Likewise, the line width, and in general the line shape of the 511\,keV from these processes, also depend on the ionisation state and temperature.
This can be used to formulate `annihilation conditions in the \ac{ISM}' if the \ac{Ps}-fraction and the 511\,keV line width can be measured in detail.

As a final remark for positron annihilation physics, there would be also annihilation with bound electrons which is subdominant in the \ac{ISM} case, but can play a major role in the case of annihilation on dust grains and solid bodies \citep[see Secs.\,\ref{sec:sssbs} \& \ref{sec:511}; see also][]{Guessoum2005_511,Siegert2017}.

\subsubsection{Dark Matter Decay and Annihilation}\label{sec:DM_dec_ann}
There are at least two generic equations that may generate \ac{gray} lines from \ac{BSM}-processes, such as \ac{DM} decay or annihilation \citep{Bergstroem1998_neutralinos}:
\begin{eqnarray}
    &\mathrm{a)}~~~\mathrm{DM} &\stackrel{\tau_{x}}{\longrightarrow} \gamma(E_\gamma) + x\\
    &\mathrm{b)}~~~\mathrm{DM} + \mathrm{DM} &\stackrel{\langle \sigma v \rangle}{\longrightarrow} \gamma(E_\gamma) + \gamma(E_\gamma)
\end{eqnarray}
We refer to case a) as \ac{DM} decay, where the second particle, $x$, could either be another photon of same energy $E_\gamma$, or a neutrino, $\nu_a$, which may mix with a fourth generation of (sterile) neutrinos, $\nu_s$ \citep[e.g.,][]{Abazajian2001_sterile}.
In the case of a two-photon decay, the \ac{gray} line would be found at $E_\gamma = m_{\rm DM}/2$ with an amplitude of $2$.
In the mixed final state, the \ac{gray} line would be found at $E_\gamma = m_{\rm \nu_s}/2$ with an amplitude of $1$.
The emissivity of \ac{DM} decay is -- formally -- calculated as 
\begin{equation}
    \rho_{\rm decay} = \frac{\Gamma}{m_{\rm DM}}\rho_s\mathrm{,}
    \label{eq:emissivity_decay}
\end{equation}
where $\Gamma = \tau^{-1}$ is the decay rate, in units of $\mathrm{s^{-1}}$, of the \ac{DM} particle with mass $m_{\rm DM}$ in units of $\mathrm{g}$, and $\rho_s$ is the central mass density of a \ac{DM} density halo profile in units of $\mathrm{g\,cm^{-3}}$.
The decay rate, $\Gamma$, is then model dependent (see Sec.\,\ref{sec:outlook_darkmatter}), and $\rho_s$ has to be inferred from other measurements, such as the rotation curve of a galaxy, to obtain the gravitational potential.
In order to comply with the notation in the literature \citep{Jungman1996_review}, the \ac{DM} flux is typically written in the form
\begin{equation}
    \left(\frac{dN}{dE\,dA\,dt\,d\Omega}\right)_{\rm decay} = \frac{\Gamma}{4\pi\,\mathrm{sr}}\frac{1}{m_{\rm DM}}\left(\frac{dN}{dE}\right)_\gamma \int_0^\infty\,ds\,\rho_s \cdot \rho\left(r(s,\ell,b)\right)\mathrm{,}
    \label{eq:dm_decay_full}
\end{equation}
where the integral term is typically referred to as the `D-factor', and the term $(dN/dE)_\gamma$ is the photon spectrum.
Comparing Eq.\,(\ref{eq:dm_decay_full}) to Eqs.\,(\ref{eq:emissivity_decay}) and (\ref{eq:los_integral}), the individual terms can be identified.

In a similar way, case b), the \ac{DM} annihilation can be described.
Here, the general description of \ac{LoS} integrals and emissivities also works, but is less commonly used,
\begin{equation}
    \rho_{\rm annihilation} = \frac{\langle \sigma v \rangle }{m_{\rm DM}^2}\rho_s^2\mathrm{.}
\end{equation}
Instead, we will use the typical notation in the literature \citep{Jungman1996_review} which calculates the differential flux as
\begin{equation}
\resizebox{0.92\linewidth}{!}{$
    \left(\frac{dN}{dE\,dA\,dt}\right)_{\rm ann} = \frac{\langle \sigma v \rangle}{4\pi\,\mathrm{sr}}\frac{1}{m_{\rm DM}^2}\left(\frac{dN}{dE}\right)_\gamma \times \int_{\Delta\Omega}\,d\Omega\,\int_0^\infty\,ds\,\left[\rho_s \cdot \rho\left(r(s,\ell,b)\right)\right]^2\mathrm{,}
    $}
    \label{eq:dm_annihilation_full}    
\end{equation}
where the double-integral, sometimes normalised by the region of interest, $\Delta\Omega$, is typically referred to as the `J-factor' \citep[e.g.,][]{Evans2012}, often shown in units of $\mathrm{GeV^{2}\,cm^{-5}}$ or $\mathrm{M_\odot^2\,pc^{-5}}$.
The differential flux in units of $\mathrm{ph\,cm^{-2}\,s^{-1}\,sr^{-1}}$ simply uses the \ac{LoS} integral without the solid-angle integration.
Because \ac{DM} annihilation is a two-body process, the density and \ac{DM} mass have to be squared.
Instead of the decay rate, now the velocity-averaged annihilation cross-section, $\langle \sigma v \rangle$ \citep{Steigman2012_sigmav}, in units of $\mathrm{cm^{3}\,s^{-1}}$ dictates the amplitude of the flux.
Similar to case a), $\langle \sigma v \rangle$ is model dependent.
The spectrum -- only considering \ac{gray} lines again, $\left(dN/dE\right)_\gamma$, is now peaking at $E_\gamma = m_{\rm DM}$.

\subsubsection{Opacity to $\gamma$-rays}\label{sec:gamma_opacity}
Gamma–ray photons propagating through matter are generally attenuated by photoelectric absorption, Compton scattering, and pair production, which combine into an energy–dependent opacity, $\kappa(E)$, in units of $\mathrm{cm^{2}\,g^{-1}}$.
The radiative-transfer equation,
\begin{equation}
    \frac{dI\left(E,r(s,\ell,b)\right)}{ds} = -\,\kappa\left(E,r(s,\ell,b)\right)\, I\left(E,r(s,\ell,b)\right)\mathrm{,}
    \label{eq:attenuation}
\end{equation}
connects the specific intensity, $I(E)$, to the local opacity and the current intensity along the \ac{LoS} $s$.
The negative sign indicates that absorption and scattering `remove' photons from the beam as they propagate.
The cumulative effect is expressed through the optical depth, $\tau(E)$, which determines whether photons escape or are absorbed,
\begin{equation}
    \tau(E) = \int \kappa\left(E,r(s,\ell,b)\right)\,\varrho\left(r(s,\ell,b)\right)\, ds\mathrm{.}
\end{equation}
Here, the optical depth is the \ac{LoS} integral of the opacity weighted with the mass density, $\varrho$, in units of $\mathrm{g\,cm^{-3}}$. We note that $\tau$ may either be considered as an intrinsic property of an emission region or a value along a \ac{LoS}, representing the number of mean-free-paths encountered by a photon.
A photon escapes or propagates freely when $\tau\lesssim 1$ and is strongly attenuated when $\tau\gg 1$ \citep{Rybicki1979_book}.
In the diffuse \ac{ISM}, with $n \sim 1\,\mathrm{cm^{-3}}$ and column densities $N \sim 10^{21}\,\mathrm{cm^{-2}}$ \citep{Morrison1983_ISM}, \ac{gray} photons above a few keV experience $\tau \ll 1$ and propagate effectively in an optically thin regime.
In contrast, stellar envelopes, \acp{CN}, and young \acp{SN} with densities $\varrho \sim 10^{-12}$--$10^{-7}\,\mathrm{g\,cm^{-3}}$ and column densities $N \sim 10^{25}$--$10^{29}\,\mathrm{cm^{-2}}$ yield large optical depths at early times \citep{Chan1987_SNe,Pinto1988}.
Typical mass attenuation coefficients for neutral gas are $\kappa = \mu/\varrho \approx \{10^{-1},10^{-2},10^{-3},10^{-4}\}\ \mathrm{cm^{2}\,g^{-1}}$ at photon energies $\{10,100,1000,10000\}\ \mathrm{keV}$, respectively.
These give optical depths in \ac{ISM} of approximately $\tau \sim \{10^{-3},10^{-4},10^{-5},10^{-6}\}$, that is, negligible attenuation.
For an \ac{SNIa} a few days after explosion with $\varrho \sim 10^{-11}\,\mathrm{g\,cm^{-3}}$ and size $R \sim 10^{14}\,\mathrm{cm}$, the corresponding optical depths are $\tau \sim \{10^{2},10^{1},1,10^{-1}\}$ at the same energies \citep{Milne2004}.
This means that only $\gtrsim$MeV photons begin to escape at early times. %
The resulting \ac{gray} line shapes in \acp{SNIa} are therefore altered by Comptonisation and Doppler-broadening within the ejecta (see also Sec.\,\ref{sec:ccSNe}).
After escape, the photons propagate freely through the optically thin Galactic environment.

% Main chapter about observations / predictions
\newpage
\section{Observations and Predictions of $\gamma$-Ray Lines}\label{sec:observations}
\vspace{-0.5em}
{\emph{Written by Thomas Siegert, with contributions from multiple authors (see below)}}
\vspace{0.5em}\\
In this Section, we will summarise different observations and predictions of \ac{gray} lines in the MeV band, ranging from diffuse emission of radioactivities in the Milky Way (e.g., \nuc{Al}{26}, \nuc{Fe}{60}, and \nuc{Na}{22}; Secs.\,\ref{sec:Al26}, \ref{sec:Fe60} \& \ref{sec:CNe}), to point-like sources such as \acp{SNIa} (e.g., \nuc{Ni}{56} decay chain; Sec.\,\ref{sec:SNeIa}), \acp{ccSN} (e.g., \nuc{Ni}{56} and \nuc{Ti}{44} decay chain; Sec.\,\ref{sec:ccSNe}), \acp{CN} (e.g., \nuc{Be}{7}; Sec.\,\ref{sec:CNe}), or \acp{BNSM} ($r$-process elements; Sec.\,\ref{sec:NSM}).
Afterwards, \ac{CR} excitation in the \ac{ISM} and \acp{SNR} (e.g., \nuc{C}{12} and \nuc{O}{16}; Sec.\,\ref{sec:LECRs}) will be outlined, and similar processes are discussed in the context of solar flares (e.g., \nuc{H}{2}; Sec.\,\ref{sec:SFs}) and in planetary atmospheric and solid bodies (e.g., \nuc{Fe}{56}, \nuc{O}{16}; Sec.\,\ref{sec:sssbs}).
The positron annihilation line at 511\,keV, though mentioned indirectly in every subsection as secondary process, will be discussed in more detail in a separate section (Sec.\,\ref{sec:511}).
The importance of understanding lifetimes of radioactive isotopes and cross-sections of nuclear excitation, particle capture, and spallation will be discussed further in the context of an outlook and where \ac{gray} line science should proceed to (Sec.\,\ref{sec:lab}).
Finally, due to the rising interest in \ac{gray} signatures from \ac{DM} decay and annihilation, we devote a small section to \ac{BSM} processes (Sec.\,\ref{sec:outlook_what_now}).

% 26Al
%\newpage
\subsection{Diffuse Emission from Radioactive \nuc{Al}{26}}\label{sec:Al26}
\begin{figure}
    \centering
    \includegraphics[width=0.2318\linewidth]{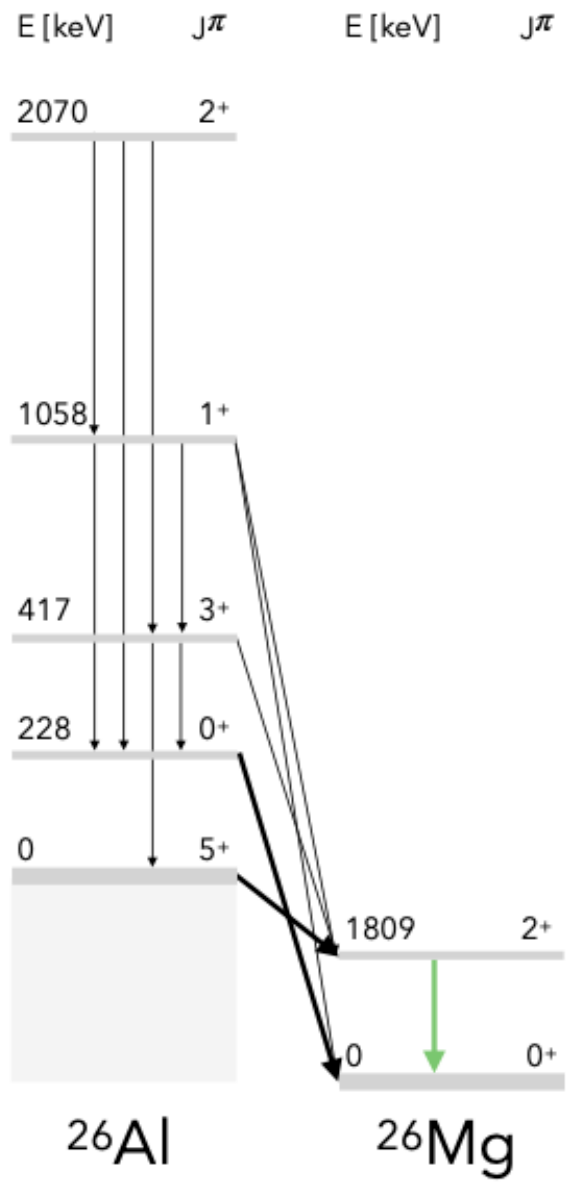}~~~~~~~~~
    \includegraphics[width=0.361\linewidth]{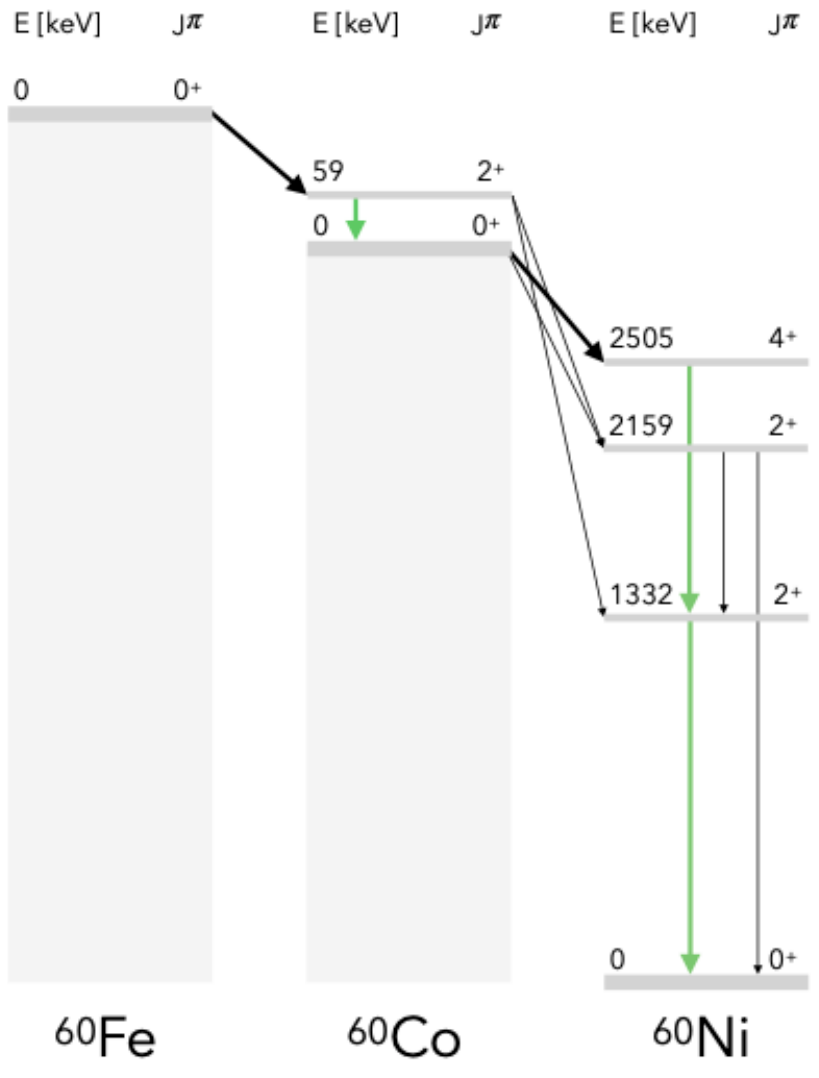}
    \caption{Nuclear energy level schemes. Transitions associated with observationally relevant \ac{gray} emission are marked in green. Left: \nuc{Al}{26} \citep[from][]{Endt1990,Iliadis2011,Pleintinger2020}. Right: \nuc{Fe}{60} \citep[from][]{Rugel2009,Pleintinger2020}.}
    \label{fig:decay_schemes}
\end{figure}

\vspace{-0.5em}
{\emph{Written by Thomas Siegert}}
\vspace{0.5em}\\
Besides the 511\,keV line (Sec.\,\ref{sec:511}), the 1808.65\,keV line from the radioactive decay of \nuc{Al}{26} ($\tau_{26} = 1.04$\,Myr, $p_\gamma = 0.997$, $p_{\beta^+} = 0.82$, see also Fig.\,\ref{fig:decay_schemes}, left) is the best-studied \ac{gray} line in terms of sensitivity (flux), spectral resolution (Doppler shifts and broadening), and angular resolution (imaging).
It was only the second \ac{gray} line detected from outside the Solar System by \ac{HEAO3} observations \citep{Mahoney1984}.
The line was found at $1808.49 \pm 0.41$\,keV, revealing that active nucleosynthesis is ongoing in the Milky Way.
The detection of this line firmly established the presence of freshly synthesised material in the \ac{ISM}.
Subsequent balloon experiments, such as the \ac{GRIS}, reinforced these findings and refined intensity estimates of the line \citep{Teegarden1991}.
Although limited by lower sensitivities and shorter observation times, these early missions paved the way for the more precise studies later conducted by the \ac{COMPTEL} onboard the \ac{CGRO} satellite.

One of the most widely accepted sources of \nuc{Al}{26} are massive stars, in particular during their \ac{WR} phases when intense stellar winds expel freshly synthesised isotopes \citep{Meynet1997}.
Subsequent \ac{ccSN} explosions from these same progenitors also contribute to the overall \nuc{Al}{26} budget through explosive nucleosynthesis \citep{Woosley1995}.
In addition, \acp{CN} on oxygen-neon (ONe) \acp{WD} have long been suggested as potential contributors, although their overall yield remains subject to significant modelling uncertainties \citep[e.g.,][see also Sec.\,\ref{sec:only_massive_stars}]{Jose1998,Vasini2025}.
Another minor channel is found in \ac{AGB} stars, where dredge-up processes can bring \nuc{Al}{26} to the surface and release it via pulsation-driven mass loss \citep{Mowlavi2000}.
\ac{CR} spallation has been examined as a non-negligible mechanism in certain environments, in addition to nuclear excitation of stable \nuc{Mg}{26} (see Sec.\,\ref{sec:LECRs}).
However, its importance relative to stellar production is generally viewed as small \citep{Clayton1984}.
All these processes, to varying degrees, ensure a continuous enrichment of the \ac{ISM}, allowing \nuc{Al}{26} to serve as a unique tracer of recent Galactic nucleosynthesis.

\citet{Knoedlseder1999b} showed that the Galactic distribution of the 1.8\,MeV emission measured with \ac{COMPTEL} strongly correlates with sites of intense star formation, in particular the 53\,GHz free-free emission map obtained by the \ac{DMR} onboard the \ac{COBE}, highlighting massive stars as key producers.
\citet{Diehl2006} then refined these findings using data from the \ac{INTEGRAL}, demonstrating that the large-scale 1.8\,MeV emission pattern traces Galactic spiral arms rich in young, high-mass stellar populations.
These observations underscore the dominance of \ac{WR} phases and subsequent \acp{ccSN} in synthesising  \nuc{Al}{26}.

%In the following, we will outline some prominent observational studies of the 1.8\,MeV \ac{gray} line, give a short overview of the production and destruction channels of \nuc{Al}{26}, and raise some open questions in understanding the measurements.

\subsubsection{Measurements of the 1809\,keV Line}
\begin{figure}[!htp]
    \centering
    \includegraphics[width=0.65\linewidth]{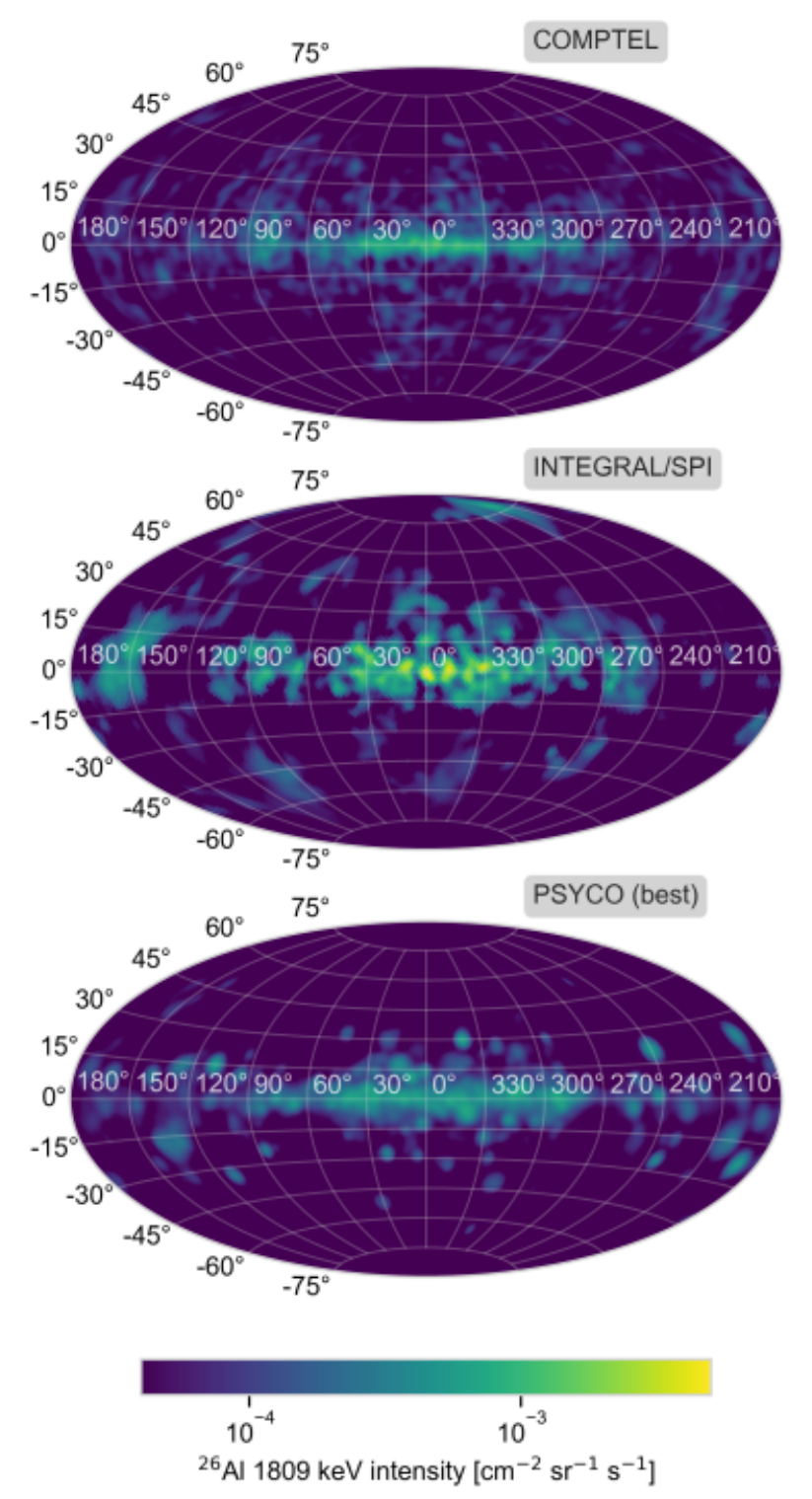}
    \caption{Compilation of observational maps (top: \ac{COMPTEL} \citep{Oberlack1996}; middle: \ac{SPI} \citep{Bouchet2015}, compared to a best-fitting 3D population synthesis model, PSYCO, \citep{Pleintinger2020}), adopted to match the instrument resolution of $3^\circ$. The minimum intensity is set to $5 \times 10^{-5}\,\mathrm{ph\,cm^{-2}\,s^{-1}\,sr^{-1}}$ to mimic potentially observable structures \citep[from][]{Siegert2023b}.}
    \label{fig:26Al_maps}
\end{figure}

\paragraph{Overview of Historic Measurements -- the COMPTEL Era}
With the launch of \ac{CGRO} in 1991, the \ac{COMPTEL} instrument conducted the first detailed all-sky surveys of the \nuc{Al}{26} decay line.
Early analyses \citep[e.g.,][]{Diehl1995} revealed a prominent band of emission along the Galactic plane, confirming massive star regions as principal \nuc{Al}{26} sources.
Quantitatively, \ac{COMPTEL} found a total Inner Galactic flux of $(2.8 \pm 0.6) \times 10^{-4}\,\mathrm{ph\,cm^{-2}\,s^{-1}}$ in the longitude range $-60^\circ \leq \ell \leq +60^\circ$ \citep{Oberlack1996}.
These observations achieved detections at high significance (often $\gtrsim 10\sigma$) and demonstrated that bright emission features correlate with known star-forming complexes, for example in Cygnus \citep{delRio1996}, Carina \citep{Knoedlseder1996}, and Vela \citep{Diehl1995b}.
This reinforced the link to \ac{WR} stars and \acp{ccSN}.
Subsequent work refined these flux estimates while improving imaging techniques, ultimately showing that \nuc{Al}{26} is distributed throughout the Galactic disk in association with regions of recent, high-mass star formation \citep[][see also Fig.\,\ref{fig:26Al_maps}, top]{Oberlack1996,Plueschke2001}.
Overall, \ac{COMPTEL}’s systematic surveys and sensitivity enhancements brought a solid foundation to the 1.8\,MeV \ac{gray} line as a key tracer of ongoing nucleosynthesis in the Milky Way.

\paragraph{Recent Measurements: Imaging Spectroscopy with SPI}
Since the launch of the \ac{INTEGRAL} satellite in 2002, the \ac{SPI} has provided unprecedented sensitivity and spectral resolution ($3.1$\,keV \ac{FWHM} at 1.8\,MeV) for observations of the \nuc{Al}{26} \ac{gray} line.
One of the first major highlights came from \citet{Diehl2006}, who measured a total Inner Galactic flux of $(3.0 \pm 0.3) \times 10^{-4}\,\mathrm{ph\,cm^{-2}\,s^{-1}}$, confirming \ac{COMPTEL}'s result, and found a line centroid of $1808.72 \pm 0.19$\,keV, with an astrophysical line width of $0.4$--$1.9$\,keV (\ac{FWHM}).
These results suggested that \nuc{Al}{26}-enriched material travels at velocities of a few $100\,\mathrm{km\,s^{-1}}$ inside the Milky Way.

Subsequent targeted studies refined these findings in specific star-forming regions:
\citet{Martin2009} focused on the Cygnus OB associations, deriving a local flux of $(3.9 \pm 1.1) \times 10^{-5}\,\mathrm{ph\,cm^{-2}\,s^{-1}}$, underscoring the role of massive stellar clusters in that region.
In \citet{Diehl2010}, the nearby Scorpius-Centaurus association was measured in detail, tying its 1.8\,MeV emission directly to young OB stars at close proximity \citep[see also][for a refined measurement including multiwavelength studies]{Krause2018}.
A breakthrough came in the work by \citet{Kretschmer2013}, who constructed a longitude-velocity diagram by the Doppler shifts in the 1.8\,MeV line.
This showed both, the kinematic imprint of the large-scale Galactic rotation, and a preferential direction of \nuc{Al}{26} ejecta along the spiral arms towards smaller densities \citep{Krause2015}.

The first image reconstruction of the 1.8\,MeV line with \ac{SPI} was performed by \citet{Bouchet2015}.
Their map (Fig.\,\ref{fig:26Al_maps}, middle) showed hotspots in prominent star-forming complexes along the Galactic plane, but could not recover all the suggested sources from \ac{COMPTEL} measurements.
The reason for this is the rather inhomogeneous exposure of \ac{SPI} compared to \ac{COMPTEL}:
\ac{INTEGRAL} performed targeted observations of mainly the Galactic bulge and disk, and other prominent regions, such as the Orion-Eridanus superbubble \citep{Siegert2017b}, have only been visited irregularly during the 22\,yr mission.
More recently, \citet{Siegert2023b} compared \ac{SPI} data with advanced, three dimensional, population synthesis models to link the observed \nuc{Al}{26} distribution to the \ac{SFR}, nucleosynthesis calculations from massive star models, and feedback processes throughout the Galaxy.
Their major finding was an exceptionally high star formation rate of $\gtrsim 5\,\mathrm{M_\odot\,yr^{-1}}$, which will be discussed below in Secs.\,\ref{sec:understanding_26Al} \& \ref{sec:only_massive_stars}.
Finally, \citet{Pleintinger2023} leveraged the capabilities of \ac{SPI} by including ``double event data'', that is, events that trigger two \ac{SPI} detectors simultaneously, to achieve the best constraints yet on the line flux ($(1.84 \pm 0.03) \times 10^{-3}\,\mathrm{ph\,cm^{-2}\,s^{-1}}$), centroid, ($1808.74 \pm 0.04$\,keV), and width ($1.6 \pm 0.3$\,keV \ac{FWHM}).
Over two decades of \ac{SPI} observations have thus firmly established \nuc{Al}{26} as a powerful tracer of massive-star nucleosynthesis and Galactic structure.
However, with more sensitivity comes more questions, which will only be addressed by improved measurements.

\subsubsection{Production and Destruction of \nuc{Al}{26}}
Detailed overviews of the nuclear structure and reaction rates of \nuc{Al}{26} can be found in \citet{Iliadis2011}.
The primary production path for \nuc{Al}{26} is the proton-capture reaction $\,^{25}\mathrm{Mg}(p,\gamma)\,^{26}\mathrm{Al}$ in hydrogen-burning environments (NeNaMgAl-cycle).
Since the seed isotope \nuc{Mg}{25} largely reflects a star’s initial metallicity, the synthesis of \nuc{Al}{26} can be sensitive to the star’s chemical composition.

Despite its robust creation, \nuc{Al}{26} also undergoes destruction, mainly via neutron-capture channels $\,^{26}\mathrm{Al}(n,p)\,^{26}\mathrm{Mg}$ and $\,^{26}\mathrm{Al}(n,\alpha)\,^{23}\mathrm{Na}$.
A lesser contribution to the destruction comes from the proton-capture route $\,^{26}\mathrm{Al}(p,\gamma)\,^{27}\mathrm{Si}$, which is suppressed by the Coulomb barrier \citep{Parikh2014}.
In hot stellar interiors, \nuc{Al}{26} can bypass its otherwise forbidden ground-state decay (from $5^+$ to $0^+$) by populating excited levels such as 228\,keV ($0^+$), 417\,keV ($3^+$), or 1058\,keV ($1^+$).
The equilibrium abundance of the isotope in these environments is strongly temperature-dependent:
For example, the ratio $Y_{26}/Y_{27}$ can vary from $0.03$ to $0.8$ as central stellar temperatures rise from $6 \times 10^7$\,K to $10^8$\,K in intermediate-mass \ac{AGB} stars.
During hot-bottom burning in \ac{AGB} envelopes, \nuc{Al}{26} can be produced in significant quantities, with the final yield depending on thermal pulses and convective mixing processes \citep[see, e.g.,][for \ac{AGB} nucleosynthesis details]{Mowlavi2000}.

\begin{figure}[!ht]
    \centering
    \includegraphics[width=0.459\linewidth]{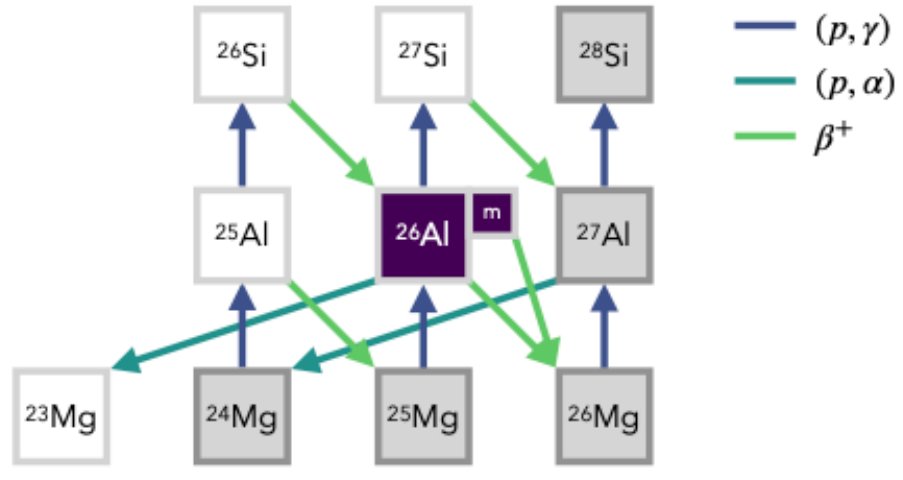}~~~~~~
    \includegraphics[width=0.423\linewidth]{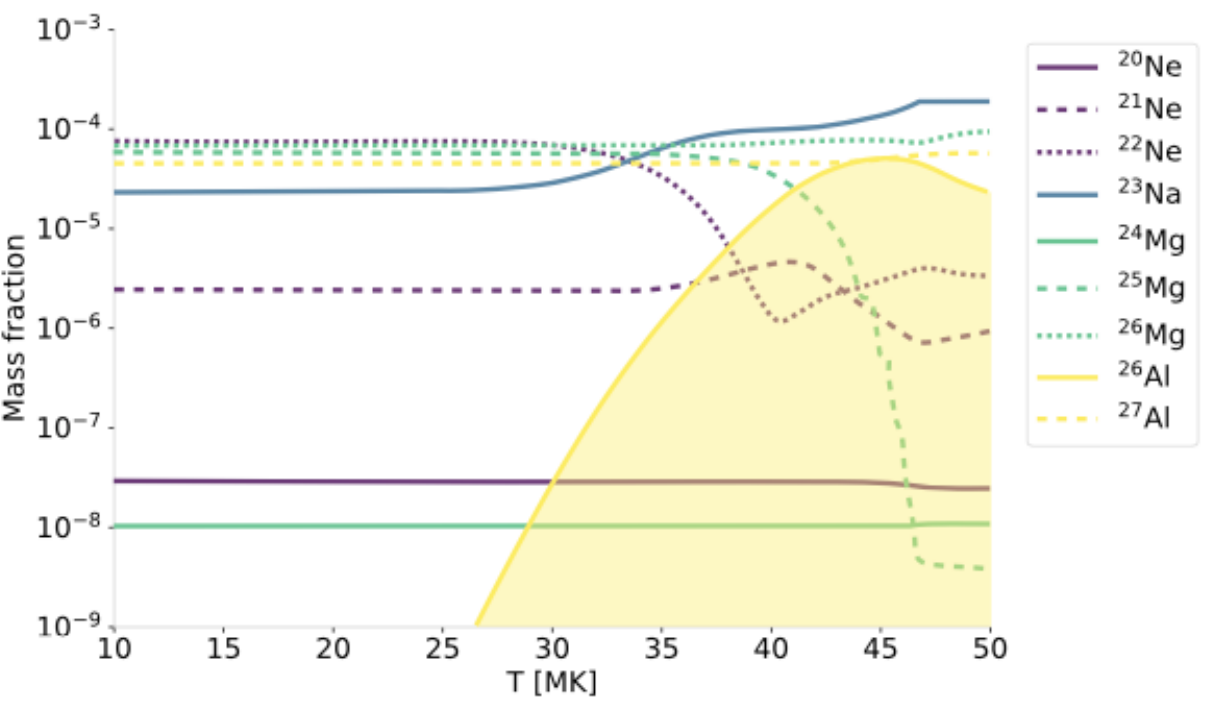}
    \caption{Left: Nuclear production and destruction channels of \nuc{Al}{26}. Stable isotopes are marked in dark grey \citep[from][]{Prantzos1996,Pleintinger2020}. Right: Temperature dependence of element abundances in the NeNaMgAl-cycle in an intermediate-mass \ac{AGB} star. \nuc{Al}{26} is shown as yellow shaded region. The x-axis can also be read as radial information with hotter regions being closer to the stellar centre \citep[from][]{Lugaro2018,Pleintinger2020}.}
    \label{fig:26Al_nuclear}
\end{figure}

Another path arises with the meta-stable isomeric state $\,^{26}\mathrm{Al}^{m}$, which has a half-life of 6.34\,s and becomes significant once temperatures exceed $10^8$\,K.
Above about $4.5 \times 10^8$\,K, ground and isomeric states reach thermal equilibrium, so that rapid convective transport away from high-temperature zones can effectively preserve \nuc{Al}{26}.
In advanced burning stages or \ac{SN} shocks, increased neutron fluxes and high temperatures further influence the \nuc{Al}{26} yields.
Additional explosive contributions arise during \acp{ccSN} and \acp{CN}, where proton captures and neutrino interactions can boost production by 20–-30\% \citep[e.g.,][see also Sec.\,\ref{sec:CNe}]{Sieverding2017}.
Moreover, explosive Ne/C-burning via the chain $\,^{24}\mathrm{Mg}(n,\gamma)\,^{25}\mathrm{Mg}(p,\gamma)\,^{26}\mathrm{Al}$ can produce typical \nuc{Al}{26} mass fractions of about $3 \times 10^{-5}$ in stellar ejecta.

Finally, the typical yield of \nuc{Al}{26} per massive star consequently depends on many factors, of which stellar mass is the dominant one.
The wind yields range from $10^{-10}$--$10^{-7}\,\mathrm{M_\odot}$ for the least-massive stars around 10--15\,$\mathrm{M_\odot}$, up to $10^{-4}$--$10^{-3}\,\mathrm{M_\odot}$ at the high mass end around 60--120\,$\mathrm{M_\odot}$.
The \ac{SN} yields are typically on the order of $10^{-5}$--$10^{-4}\,\mathrm{M_\odot}$.
Weighted with an \ac{IMF}, it becomes evident that most \nuc{Al}{26} should be produced by massive stars in the range 10--30\,$\mathrm{M_\odot}$ \citep{Pleintinger2020}.
The latest model calculations estimate the \nuc{Al}{26} yield in \ac{AGB} stars to be around $10^{-8}$--$10^{-6}\,\mathrm{M_\odot}$ per star \citep{Karakas2014}.
The \ac{CN} yields range from $10^{-9}$--$10^{-7}\,\mathrm{M_\odot}$ per ONe nova event \citep[e.g.,][]{Jose1998,Starrfield2020}.

\subsubsection{Understanding and Interpreting \nuc{Al}{26} Measurements}\label{sec:understanding_26Al}
From the above numbers, it can be easily estimated, how much \nuc{Al}{26} would be present in the Milky Way, given some information on the \ac{SN} rate, for example.
Turning this around, measuring the \nuc{Al}{26} flux and distribution in the Galaxy can provide an estimate of the \ac{SN} rate and the \ac{SFR}.

Assuming one dominant source for the moment (see Sec.\,\ref{sec:only_massive_stars}), massive star and their \acp{SN}, we can convert the measured flux value of $1.8 \times 10^{-3}\,\mathrm{ph\,cm^{-2}\,s^{-1}}$ into a quasi-persistent mass of \nuc{Al}{26} given Eqs.\,(\ref{eq:total_flux},\ref{eq:luminosity},\ref{eq:decay_luminosity}).
For this, one needs to assume a 3D distribution of the emission.
It was found that a doubly-exponential disk with a scale radius of $5.8 \pm 0.6$\,kpc and a scale height of $0.8 \pm 0.2$\,kpc fits the \ac{SPI} \ac{gray} data well \citep{Pleintinger2019}.
From these values, a conversion from flux to luminosity and therefore mass is straightforward as $1\,\mathrm{M_\odot}$ of \nuc{Al}{26} corresponds to a luminosity of $1.4 \times 10^{42}\,\mathrm{s^{-1}}$, and a flux of $3 \times 10^{-4}\,\mathrm{ph\,cm^{-2}\,s^{-1}}$, assuming a distance to the Galactic centre of 8.2\,kpc.
Given the total flux measurement, we would have a total mass of around $6\,\mathrm{M_\odot}$ of \nuc{Al}{26} in the Milky Way.
Now, given the average yield per massive star of $(1$--$2) \times 10^{-4}\,\mathrm{M_\odot}$, we see that on the order of $(3$--$6) \times 10^4$ massive stars contribute to the Galactic 1.8\,MeV line.
The \ac{SN} rate is then the number of contributing stars over the lifetime of \nuc{Al}{26}, that is, 3--6 per century.
Clearly, these values are only rough estimates but give an excellent overview of the concept.
A more complex consideration of the flux values as well as the spatial distribution of the 1.8\,MeV line gives a total \nuc{Al}{26} mass of $1.2$--$2.4\,\mathrm{M_\odot}$ in the Galaxy, which leads to a \ac{SN} rate of $1.8$--$2.8$ per century \citep{Pleintinger2020,Siegert2023}.

The \ac{SN} rate is directly linked to the \ac{SFR} in the Galaxy, roughly by
\begin{equation}
    \mathrm{SFR} = R_{\rm SN} \langle M^* \rangle f_{\rm SN}^{-1}\mathrm{,}
\end{equation}
where $R_{\rm SN}$ is the \ac{SN} rate, $\langle M^* \rangle$ is the average stellar mass in a star formation event, and $f_{\rm SN}$ is the fraction of all stars that actually undergo a \ac{SN} explosion \citep{Diehl2006}.
The average stellar mass is given by the assumed \ac{IMF} and may range around $0.3$--$0.7\,\mathrm{M_\odot}$.
The fraction of stars that undergo a \ac{ccSN} explosion depends on both, a minimum stellar mass, typically chosen as $8\,\mathrm{M_\odot}$, and the assumed ``explodability'', that is, up to which mass stars explode or directly collapse into a \ac{BH}.
The question of the explodability is an important one as it scales the total mass, flux, \ac{SFR} and \ac{SN} rate.
Several different models consider explosions up to $25\,\mathrm{M_\odot}$ \citep[e.g.,][]{Limongi2018}, or up to $100\,\mathrm{M_\odot}$ \citep[e.g.,][]{Janka2012}, or even `islands' of stars that can explode and those that cannot \citep[e.g.,][]{Sukhbold2016}.
The fraction of stars that might explode is then on the order of $(1$--$4) \times 10^{-3}$.
Taking these values into account, a \ac{SFR} between $1$ and $17\,\mathrm{M_\odot\,yr^{-1}}$ is found, with an average around $3.5\,\mathrm{M_\odot\,yr^{-1}}$.
This value, compared to literature values, appears reasonable, although on the high side of estimates \citep[e.g.,][see also Secs.\,\ref{sec:PSYCO} \& \ref{sec:only_massive_stars}]{Licquia2015}.

All these estimates, however, strongly depend on the massive star evolution models, their nucleosynthesis calculations, that is, wind and \ac{SN} yields, the binarity of massive stars, and -- probably most important -- ignore the fact that there are other contributors to \nuc{Al}{26} in the Milky Way.
In addition, an enhanced local production inside the Local Bubble may play a significant role in estimating the actual total flux.
For this reason, these estimates are often based on the central radian of the Milky Way, and not necessarily on the entire sky.
In the following, we will discuss several effects that might impact these measurements, and how they can be addressed in the future.

\begin{figure}
    \centering
    \includegraphics[width=0.45\linewidth]{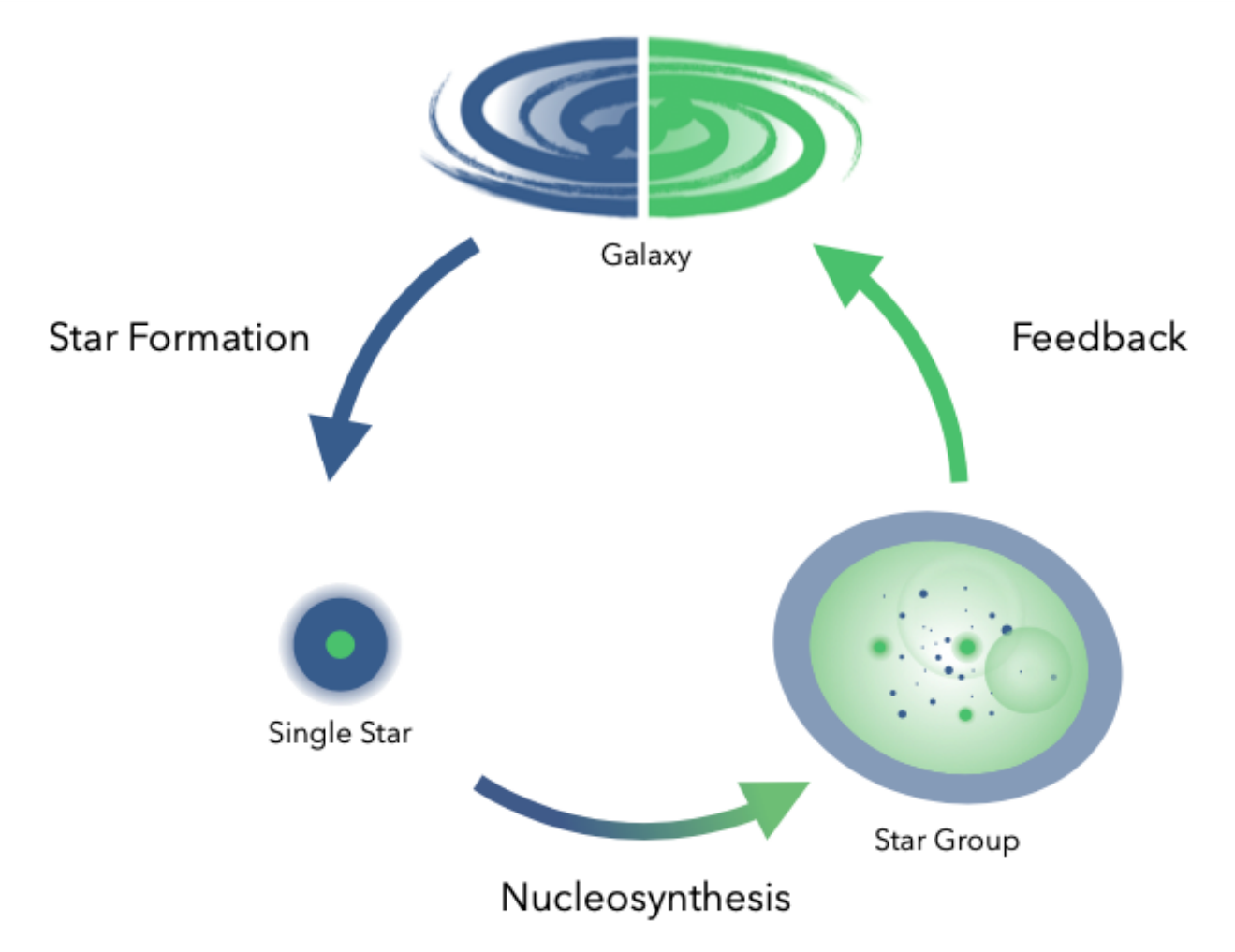}~~~~~~
    \includegraphics[width=0.36\linewidth]{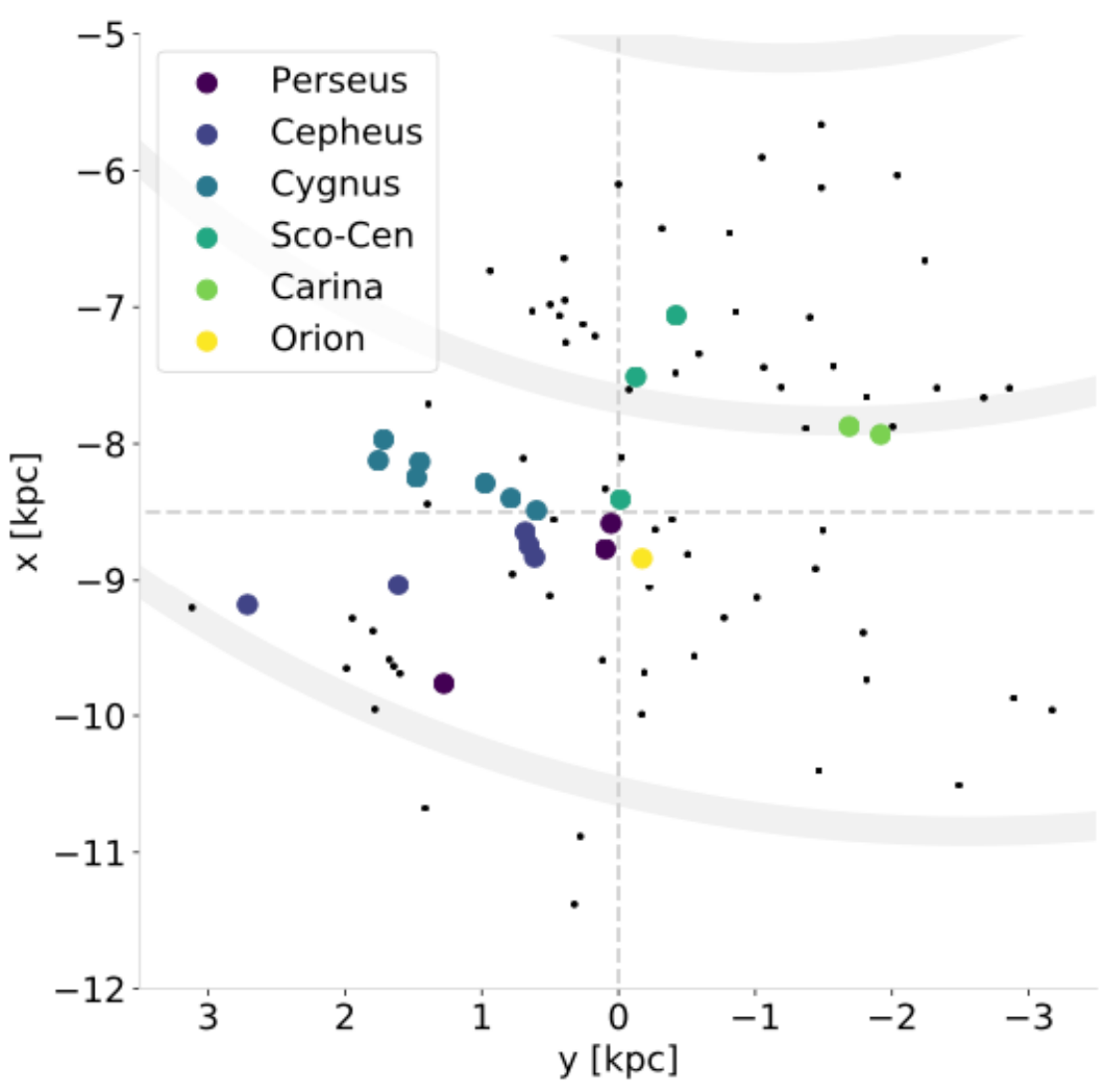}
    \caption{Left: Schematic representation of the cosmic cycle of matter with metallicity symbolically increasing from blue to green. The cycle progresses in three fundamental steps over three major scales \citep[from][]{Pleintinger2020}. Right: Spatial distribution of nearby OB associations within 3.5\,kpc in the Galactic plane. Dots denote the positions of OB associations from the Gaia catalogue \citep{Melnik2017}. Spiral arm tangents are shown in thick grey lines \citep[from][]{Pleintinger2020}.}
    \label{fig:cosmic_matter_cycle}
\end{figure}

\paragraph{Cosmic Cycle of Matter -- Nucleosynthesis Feedback}\label{sec:cycle_of_matter_foreground}
``\emph{[...] feedback is the mechanism by which a process loop alters itself.
While the main system stays the same procedurally with every iteration, feedback describes an inherent sub-loop of self-driven change in process parameters.
In other words, feedback is the mechanism underlying any kind of evolution.
[...]
Such a kind of evolution must have taken place on cosmic scale}'' \citep[][p.\,5]{Pleintinger2020}, as well as happens also locally, and then disturbs the view and interpretation of Galactic measurements.
The cosmic cycle of matter (Fig.\,\ref{fig:cosmic_matter_cycle}, left) happens on all scales so that a local production, for example in one or several recent \ac{ccSN} events inside the Local Bubble \citep[e.g.,][]{Schulreich2023}, would lead to an enhanced flux compared to the Galactic average.
This would explain the large scale height distribution that has been found by \citet{Pleintinger2019}, sometimes reaching 2\,kpc or above, whereas the average was found to be below 300\,pc.

\begin{figure}[!p]
    \centering
    \includegraphics[width=0.75\linewidth]{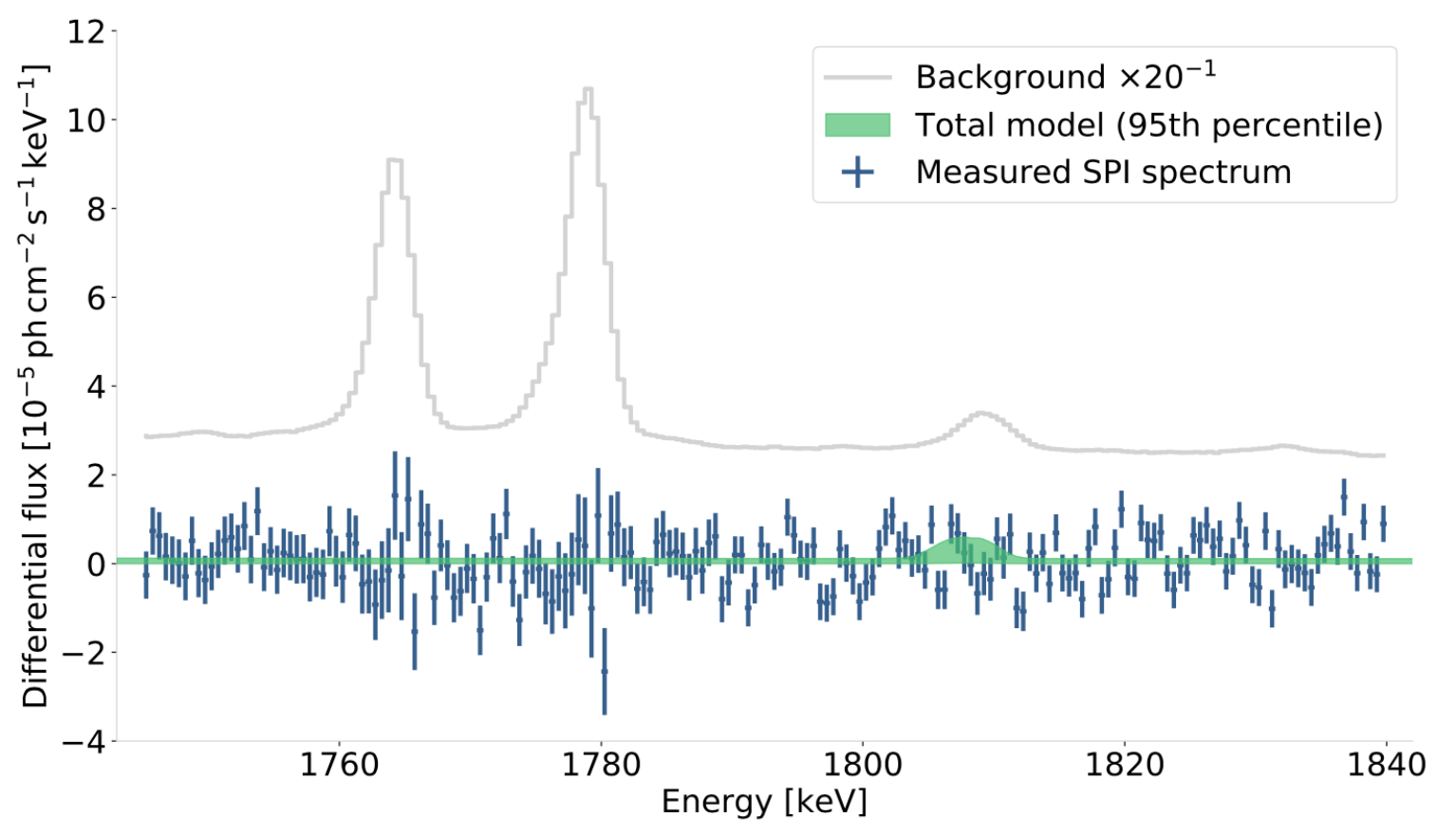}    
    \includegraphics[width=0.75\linewidth]{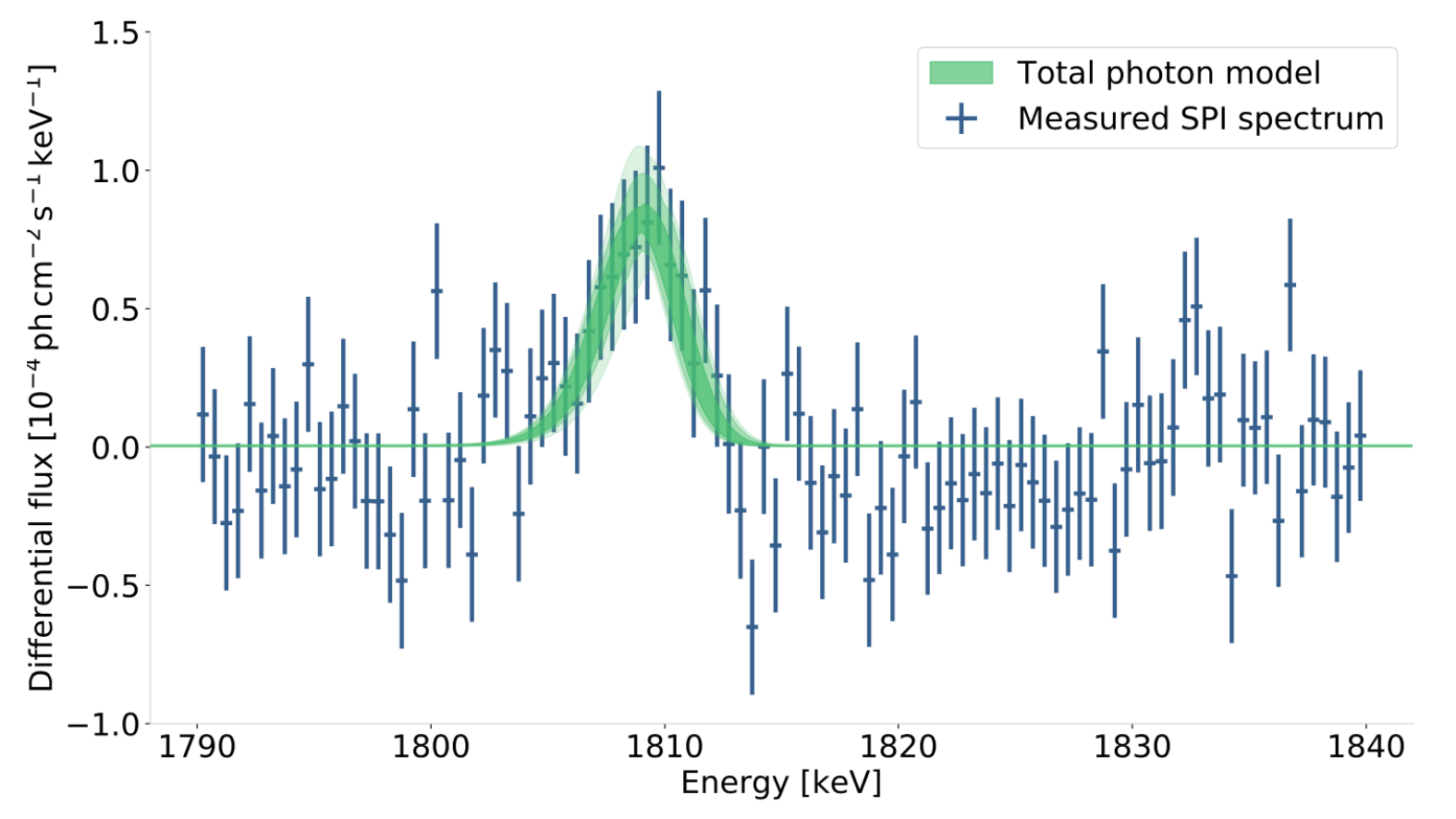}
    \raggedleft
    \includegraphics[width=0.73\linewidth]{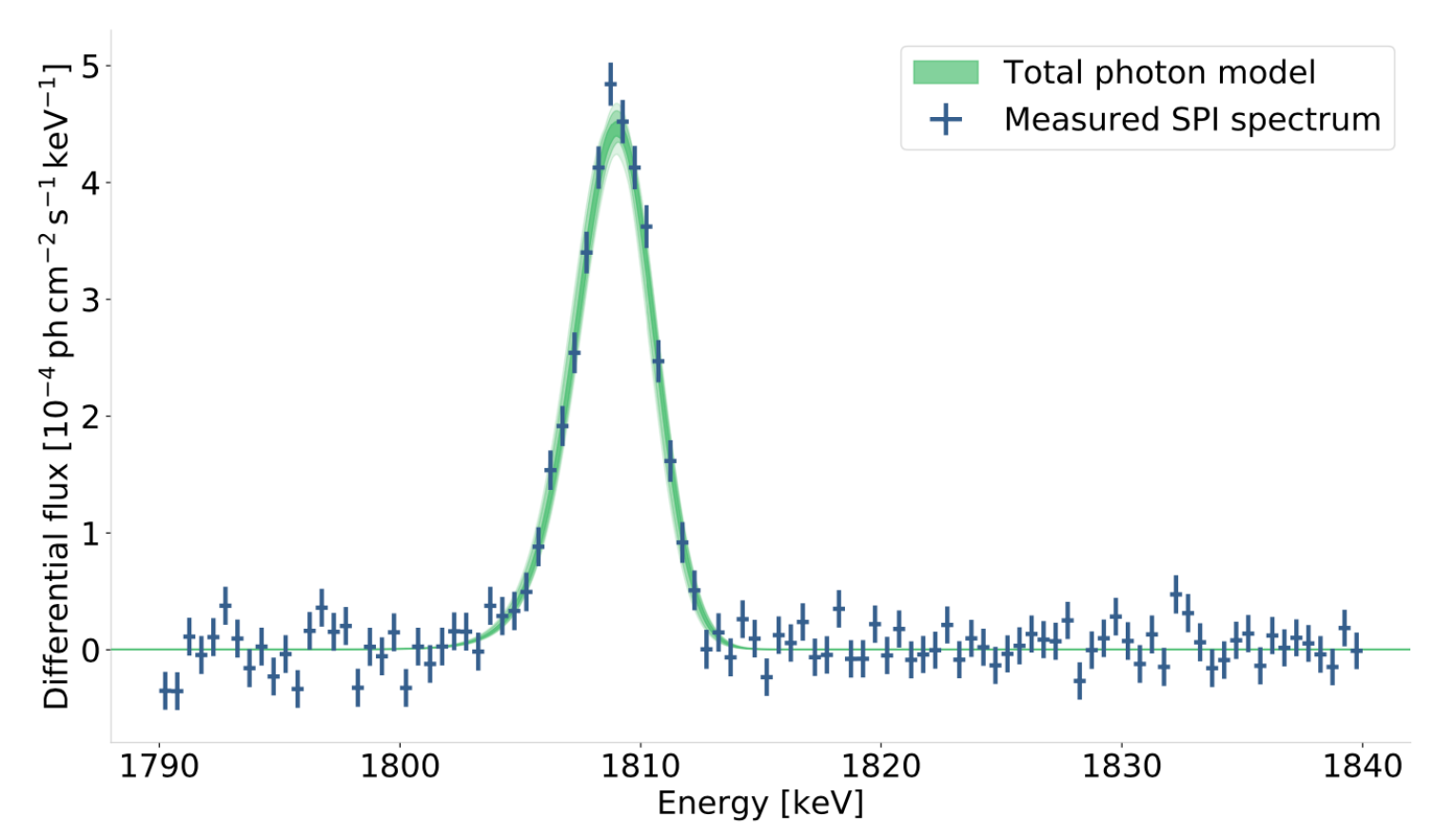}
    \caption{\ac{SPI} spectra (blue data points) around the 1809\,keV line for three different cases. Top: $\gamma^2$ Velorum from 1745--1840\,keV. The line is not detected but a $2\sigma$ upper limit of $1.7 \times 10^{-5}\,\mathrm{ph\,cm^{-2}\,s^{-1}}$ is indicated by the green band. The instrumental background in the same energy range is shown as grey histogram, scaled by a factor of 20 for better comparison. Clearly, the higher the background flux, the larger the error bars per 0.5\,keV bin. Middle: Region around the Perseus OB associations. The line flux is $(3.6 \pm 0.4) \times 10^{-4}\,\mathrm{ph\,cm^{-2}\,s^{-1}}$, broadened by $(1.13 \pm 0.55)$\,keV and shifted by $(0.34 \pm 0.22)$\,keV. Bottom: Galactic-wide spectrum. The line flux is $(1.87 \pm 0.03) \times 10^{-3}\,\mathrm{ph\,cm^{-2}\,s^{-1}}$ with a Doppler shift of $(0.37 \pm 0.04)$\,keV and an astrophysical line width of $(0.62 \pm 0.32)$\,keV which corresponds to a thermal broadening around $80\,\mathrm{km\,s^{-1}}$ \citep[from][]{Pleintinger2020}.}
    \label{fig:26Al_spetra}
\end{figure}

In technical terms, coded aperture mask telescopes, such as \ac{SPI}, have problems in identifying emission regions with small gradients or isotropic emission \citep[e.g.,][]{Siegert2022}.
However, the contrast between different longitudes in the Galaxy, as well as the large scale height from individual doubly-exponential disk fits, could point to a local contribution.
The Local Bubble would therefore lead to an all-sky 1.8\,MeV flux on the order of a few $10^{-6}\,\mathrm{ph\,cm^{-2}\,s^{-1}}$ \citep{Siegert2024} -- too little to explain the large \ac{SFR}, but large enough to increase the scale height to extreme values.
The nearby star-forming regions (Fig.\,\ref{fig:cosmic_matter_cycle}, right) may also show an enhanced nucleosynthesis contribution and lead to a skewed perspective on the entire Galaxy \citep{Pleintinger2020}.
In \citet{Pleintinger2020}, it was found that the nearby ($\lesssim 3.5$\,kpc) groups may reduce the Galactic-wide 1.8\,MeV flux by 20--30\%.
This reduces the \ac{SFR} by the same percentage and alleviates some of the tension.

\begin{figure}[!t]
    \centering
    \includegraphics[width=0.387\linewidth]{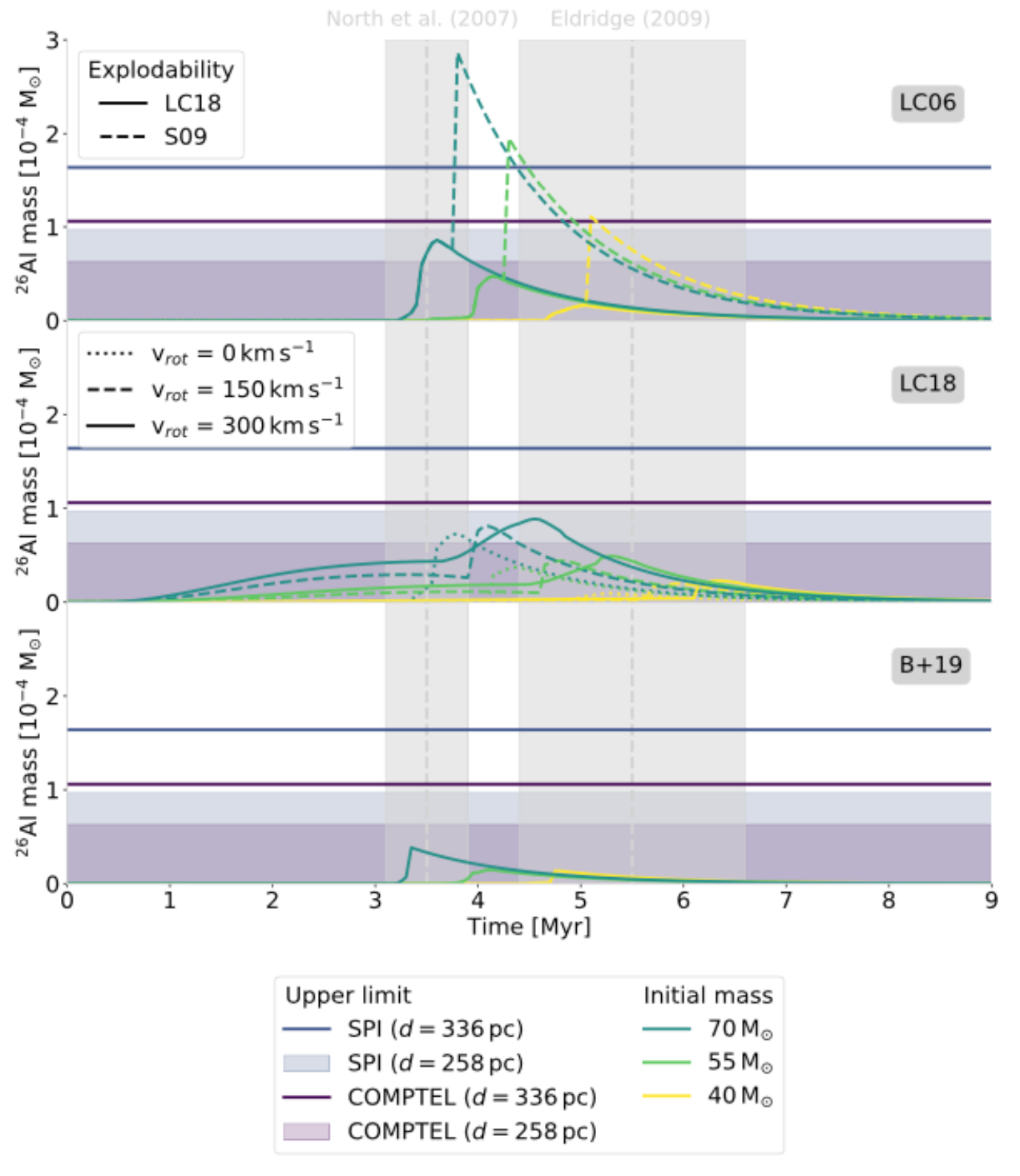}~~~~~~~
    \includegraphics[width=0.486\linewidth]{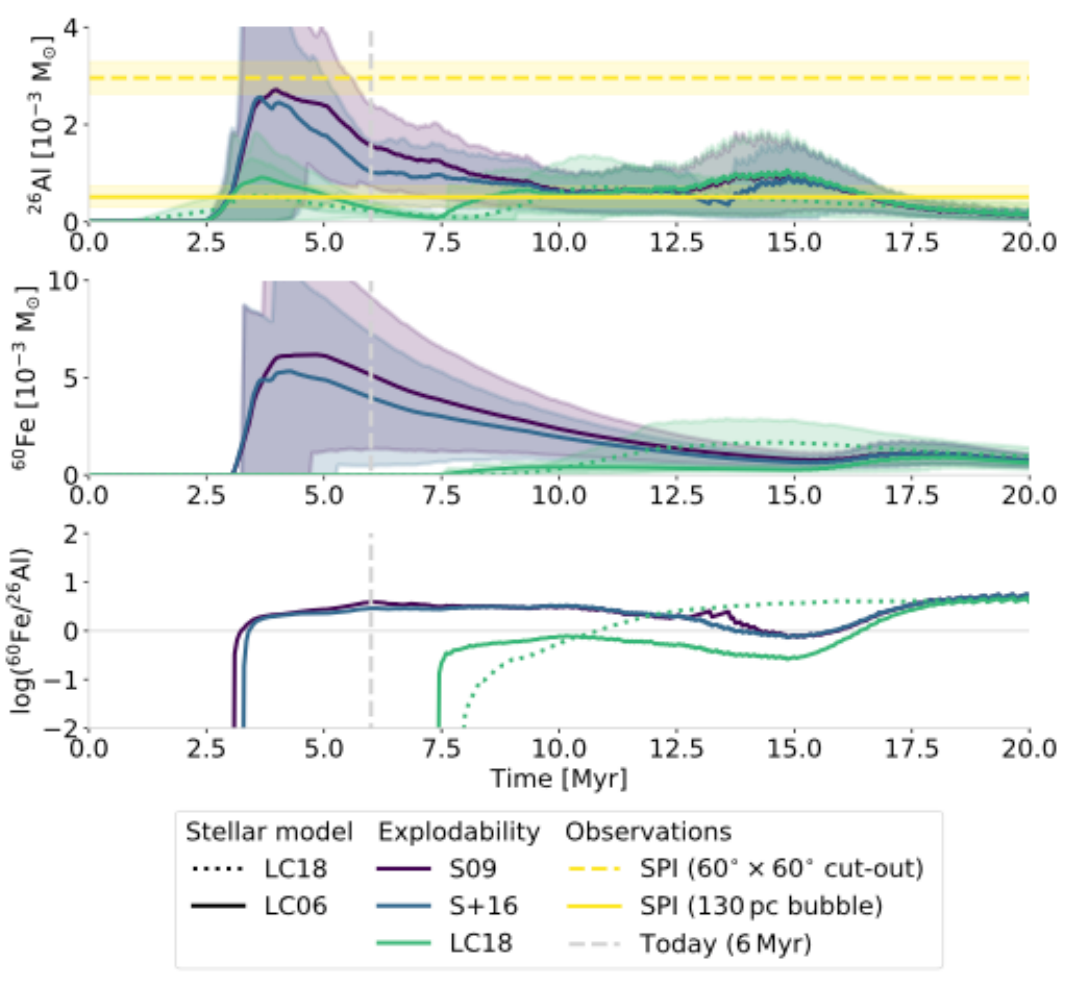}
    \caption{Left: Comparison of single star model predictions with observational limits of $\gamma^2$ Velorum measured with \ac{SPI} and \ac{COMPTEL}. Three different stellar initial masses are show. The panels correspond to single star models \citep[LC06;][]{Limongi2006} (top), \citep[LC18;][]{Limongi2018} LC18 (middle), and binary models \citep[B+19;][]{Brinkman2019} (bottom). The predicted values lie mostly just below the upper limits \citep[from][]{Pleintinger2020}. Right: Population synthesis profiles of \nuc{Al}{26} and \nuc{Fe}{60} from a stellar group representing the Perseus OB2 association. The calculations are based on the evolution stellar models LC06 (solid lines) and LC18 (dotted lines). Different explodability models are applied to the LC06 tracks. Shaded 68\% uncertainty regions are obtained from 1000 Monte Carlo runs each. The two measured values are based on the \ac{SPI} 1.8\,MeV map (dashed line, see also Fig.\,\ref{fig:26Al_spetra}, middle) or a superbubble model with size $R = 130$\,pc at the position of Perseus OB2 (solid line). The group's age is marked at 6 Myr \citep[from][]{Pleintinger2020}.}
    \label{fig:popsyn}
\end{figure}

\paragraph{From Single Stars to the Galaxy}\label{sec:PSYCO}
In order to model the entire Galaxy, \citet{Siegert2023b}, based on the work by \citet{Pleintinger2020}, used a bottom-up, 3D \ac{PSYCO} to place constraints on the \ac{SFR}, the \ac{SN} rate, and the structure of the Galaxy in \nuc{Al}{26} as well as \nuc{Fe}{60}.
The variety of massive star models should be gauged, starting from a single massive star.
However, the 1.8\,MeV \ac{gray} line fluxes from all nearby massive stars are below the sensitivity threshold for \ac{COMPTEL} \citep{Oberlack2000} and \ac{SPI} \citep{Pleintinger2020}.
In the most promising case of the binary star system $\gamma^2$ Velorum, the most recent spectrum (Fig.\,\ref{fig:26Al_spetra}, top) gives an upper limit of $1.7 \times 10^{-5}\,\mathrm{ph\,cm^{-2}\,s^{-1}}$, which is only weakly constraining the initial stellar mass of the star to $\lesssim 50\,\mathrm{M_\odot}$, depending on the nucleosynthesis model, and the distance to the star \citep[see][see also Fig.\,\ref{fig:popsyn}]{Oberlack2000}.
Flux estimates from individual stars would fall into the range of a few $10^{-6}\,\mathrm{ph\,cm^{-2}\,s^{-1}}$ or below, so that with next generation telescopes, it would be possible to calibrate massive star evolution models.
%

%\begin{figure}[!ht]
%    \centering
%    \includegraphics[width=0.65\linewidth]{PSYCO.pdf}
%    \caption{Schematic structure of the \ac{PSYCO} model. See Pleintinger\,\cite[2020;][]{Pleintinger2020} and Siegert et al.\,\cite[2023b;][]{Siegert2023b} for details.}
%    \label{fig:PSYCO}
%\end{figure}

While for now, a single star (binary) is too weak to be detectable, several massive star groups (see above) have been identified as striking \nuc{Al}{26} sources.
In the case of Perseus OB2, for example \citep{Pleintinger2020}, with an age of 6\,Myr and an initial $2 \times 10^4$ stars at a distance of 200--400\,pc, a sizeable amount of \nuc{Al}{26} could be expected.
With more than 15\,Ms of observation time with \ac{SPI}, Perseus OB2 was found at a flux level of $(6.3 \pm 2.8) \times 10^{-5}\,\mathrm{ph\,cm^{-2}\,s^{-1}}$, in agreement with 1.8\,MeV fluxes from the same regions in the \nuc{Al}{26} all-sky maps (Fig.\,\ref{fig:26Al_maps}).
Applying a superbubble expansion model, homogeneously filled with \nuc{Al}{26}, this flux converts to an \nuc{Al}{26} mass of $(5.1 \pm 2.2) \times 10^{-4}\,\mathrm{M_\odot}$ for the entire OB association.
This number can be compared to stellar evolution models in the context of \ac{PSYCO} by calculating the expectation of a population of $2 \times 10^4$ stars as a function of time.
In Fig.\,\ref{fig:popsyn}, right, we show the \nuc{Al}{26} and \nuc{Fe}{60} profiles, as well as their ratio, of a coeval population attempting to mimic Perseus OB2.
It is clear that the stochasticity of the \ac{PSYCO} calculations determine the accuracy of the derived parameters.
In principle it would be possible to determine the age of the population, however only if more details about the remaining stars are known.
If we compare the values at the age of Perseus OB2 at 6\,Myr, we find a generally good agreement with the models from \citet{Limongi2018}, explaining the measurements best.
However, also other models match the measurements, although on the low-side of their expectations, probably because of the enhanced explodability \citep[e.g.,][]{Smartt2009,Sukhbold2016}.

Building the entire Galaxy is also possible with \ac{PSYCO}, based on the embedded cluster mass function and an assumption on the birthplaces of OB associations.
Similar to a Galactic chemical evolution model, an ``empty'' galaxy is filled with supperbubbles, which themselves are filled with \nuc{Al}{26} and \nuc{Fe}{60}, that evolve according to \citet{Weaver1977}.
The initial parameters here are the \ac{SFR}, a specific stellar evolution model that determines the yields per star group, an explodability model, a geometric representation of the Galaxy (spiral arms, scale height), and the metallicity gradient of the Galaxy.
%(see Fig.\,\ref{fig:PSYCO}).
%
At some point during this evolution, the decay and the production of radioactive nuclei are balanced and a snapshot of the simulated galaxy can be taken (see Fig.\,\ref{fig:26Al_maps}, bottom).
Since this is aimed at mimicking the Milky Way, some foreground sources can be introduced.
Repeating this analysis for several thousand Monte Carlo runs leads to estimates of the initial parameters, and parameters linked to them, such as the \ac{SN} rate.
\citet{Siegert2023b} find a \ac{SN} rate of $1.8$--$2.8$ per century, similar to other studies, however only at the expense of a \ac{SFR} of $4$--$8\,\mathrm{M_\odot\,yr^{-1}}$, several factors higher than typically found \citep[e.g.,][with $1$--$2\,\mathrm{M_\odot\,yr^{-1}}$]{Licquia2015}.
The uncertainty range here already includes all possible assumptions as detailed above that actually fit the \ac{SPI} data.
One advantage of this method compared to hydrodymanics simulations \citep[e.g.,][]{Fujimoto2018,RodgersLee2019} is that many realisations of the same assumptions can quickly be calculated, so that a large parameter space can be explored.
Another advantage is that \ac{PSYCO} predictions can be directly fitted to the raw \ac{gray} data and checked for consistency: if one realisation of a parameter set does not fit well, several others might, so that a distribution around the parameters can be found, leading to uncertainty estimates.

%The results, however, some being in agreement with literature values, while others are not, may point to shortcomings of this method.
%
%We will briefly discuss the high \ac{SFR} in the following.

\paragraph{Is it Only Massive Stars?}\label{sec:only_massive_stars}
The high \ac{SFR} measured in the Milky Way from \nuc{Al}{26} \ac{gray} line observations, independent of the method, leads to the question how to significantly scale it down, not by a few per cent, but by several factors.
In most studies, the assumption is that the 1.8\,MeV line is purely from massive star winds and \acp{ccSN}.
With this assumption, it is clear that, depending on the massive star evolution models and nucleosynthesis yields, a high \ac{SFR} is inevitable because these evolution models also need to match many other observations.
Therefore, \nuc{Al}{26} could either be special and all models are wrong, or, more reasonable, the assumptions on the Galactic-wide modelling are lacking input.
Certainly, the \ac{PSYCO} approach handles the unknown unknowns, that is, how many stars and stellar groups are formed where and at what frequency, to the extent that it matches other observations, but always with the free parameter of \ac{SFR}.
The \ac{LoS} integration in \ac{PSYCO} assumes homogeneously filled spheres, which, to a very good approximation, fits several nearby massive star groups.
However, there is the effect of bubble merging \citep[e.g.,][]{Krause2014} which alters the average structure of the bubbles, there is the fact that \nuc{Al}{26} is long-lived enough to also mix with the bubble shells \citep{Schulreich2023}, and there is the density gradient in the Milky Way which can make bubbles appear stretched, such as Orion-Eridanus \citep[e.g.,][]{Burrows1993}.
While on the Galactic scale, these effect may have little to no impact on the total flux because they average out over the contributions of several tens of thousand groups, the nearby associations may lead to an enhanced flux.
While this may make on the order of 20--30\% as described above, one would need to decrease the \ac{SFR} even further by about 50--75\%.

The explodability of massive stars is an open question:
While only $25\,\mathrm{M_\odot}$ stars or below have been observed to actually explode \citep[e.g.,][]{Smartt2015}, a higher contribution of so-called supermassive stars with masses of up to $300\,\mathrm{M_\odot}$ \citep{Martinet2022} might also alleviate some of this tension.
A measurement of this effect might only be possible in extragalactic objects, such as the \ac{LMC}, to prove the enhanced \nuc{Al}{26} production without creating more stars.

The binary evolution of massive stars may also play a bigger role than expected:
About 50--70\% of massive stars are found in binary systems \citep[e.g.,][]{Sana2012}.
\citet{Brinkman2019,Brinkman2021,Brinkman2023} estimated the yield of \nuc{Al}{26} from binary systems using the 1D \ac{MESA} code \citep{Paxton2011}.
The nucleosynthesis is treated `on the fly' so that the stars actually exchange information (e.g., Roche-Lobe overflow, orbital evolution, tidal interactions) but still in a 1D setting.
It has been shown that the yields of binary stars from this approach can be larger, but a full 3D calculation is missing.
Next-generation telescopes could measure the 1.8\,MeV line from individual and binary stars to better gauge the models.

The most interesting factor, however, may be another, long-neglected, source type: \acp{CN}.
Based on a 2D Galactic chemical evolution model, \citet{Vasini2025} found that the fraction of \nuc{Al}{26} from massive stars may be only 25\%, whereas \acp{CN} could make up to 75\%.
Interestingly, these would be the exact numbers required to completely relieve the tension between the 1.8\,MeV \ac{gray} line measurements and other studies.
The \ac{CN} effect will also come back in the measurements of \nuc{Fe}{60}, in that the flux ratio of \nuc{Fe}{60}/\nuc{Al}{26} can hardly be explained by the consensus model of massive star evolution.
Clearly, the \nuc{Al}{26} yields from \acp{CN} can hardly be calibrated because the expected ejecta masses are around $10^{-9}$--$10^{-7}\,\mathrm{M_\odot}$ \citep[e.g.,][]{Jose1998,Starrfield2020}, so that at a distance of 1\,kpc, the 1.8\,MeV flux would be only $10^{-9}\,\mathrm{ph\,cm^{-2}\,s^{-1}}$ at best -- impossible for any \ac{gray} telescope to measure in the near future.
Nevertheless, radioactive \nuc{Al}{26}F molecules have been detected in the old remnant CK Vul (previously believed to be a \ac{CN}, but shown to be probably a stellar merger remnant), proving that, indeed, other objects produce \nuc{Al}{26}, but at a difficult to determine yield \citep{Kaminski2018}.
A reasonable measurement that could be done already today, and which has been attempted with \ac{COMPTEL} \citep{Knoedlseder1999c} as well as \ac{SPI} \citep{Martin2009}, is to disentangle the distribution of \nuc{Al}{26} as a function of Galactocentric radius:
The \ac{CN} distribution would peak around the Galactic centre, whereas there, the massive star contribution is nearly zero, but rises towards a maximum around a few kpc \citep[see also][]{Vasini2025}.
With good angular resolution and deep exposures, a mass profile of \nuc{Al}{26} could be constructed and then again compared to the expected spatial distributions from Galactic chemical evolution models.

% 60Fe
\newpage
%\documentclass{article}
%\usepackage{graphicx} % Required for inserting images

%\title{Chapter $^{60}$Fe}
%\author{Wei Wang }
%\date{February 2025}

\def\ms{$M_\odot$}
\def\Al{$^{26\!}$Al\ }
\def\Fe{$^{60\!}$Fe\ }

%\begin{document}

%\maketitle

\subsection{Diffuse Emission from Radioactive \nuc{Fe}{60}}\label{sec:Fe60}
\vspace{-0.5em}
{\emph{Written by Wei Wang}}
\vspace{0.5em}\\
Similar to the case of \nuc{Al}{26}, \ac{gray} line emission from the radioactive decay of \nuc{Fe}{60} reflects the ongoing nucleosynthesis in our Galaxy.
As opposed to \nuc{Al}{26}, \nuc{Fe}{60} is believed to be only produced in \acp{ccSN} and not in stellar winds.
%is believed to originate from massive stars and their supernovae.

\subsubsection{Production of \nuc{Fe}{60}}
The radioactive isotope \nuc{Fe}{60} can be produced in suitable astrophysical environments through successive neutron captures on pre-existing Fe isotopes such as stable $^{54,56,57,58}$Fe in a neutron-rich environment.
Candidate regions for \nuc{Fe}{60} production are the He and C burning shells inside massive stars, where neutrons are likely to be released from the $^{22}$Ne($\alpha$,n) reaction.
\nuc{Fe}{60} production may occur any time during late evolution of massive stars towards \acp{ccSN} \citep[e.g.,][]{Woosley1995,Limongi2003,Limongi2006,Limongi2018,Pignatari2016,Sukhbold2016,Jones2019b}.
The \ac{EC} variant of such \acp{SN} may be a most-significant producer of \nuc{Fe}{60} in the Galaxy \citep{Wanajo2013,Wanajo2018,Jones2016,Jones2019a}.

There are, however, other possible astrophysical sources of \nuc{Fe}{60}:
From similar considerations, \nuc{Fe}{60} could also be made and released in super-\ac{AGB} stars \citep{Lugaro2012}.
Furthermore, high-density \acp{SNIa} explosions, which would include a deflagration phase \citep{Woosley1997}, could produce even larger amounts per event, and which could form a point-like \ac{gray} emission source.
Much of such newly-produced \nuc{Fe}{60} should be ejected with the \ac{SN} explosion, in particular if produced outside the inner cores of these objects \citep[see][for a discussion of \acp{ccSN} ejecta]{Jones2019b}.

\subsubsection{$\gamma$-Ray Detections and Predictions of \nuc{Fe}{60} in the Galaxy}
\nuc{Fe}{60} has a long lifetime with a radioactive half-life $T_{1/2} \simeq 2.6$\,Myr \citep{Rugel2009,Wallner2015,Ostdiek2016}, it can survive after ejection into the \ac{ISM}.
Several detections of \nuc{Fe}{60}-enriched material in various terrestrial as well as lunar samples \citep{Knie2004,Wallner2016,Wallner2021,Fimiani2016} confirm the evidence for one or more very nearby sources of \nuc{Fe}{60} within several Myr.
The detection of \nuc{Fe}{60} in \acp{CR} also implies that the time required for acceleration and transport to Earth is around several Myr \citep{Binns2016,Schulreich2023}, so that the \nuc{Fe}{60} source distance does not greatly exceed the distance \acp{CR} can diffuse over this time ($\lesssim 1$\,kpc).

\begin{figure}
\centering
\includegraphics[width=0.7\linewidth]{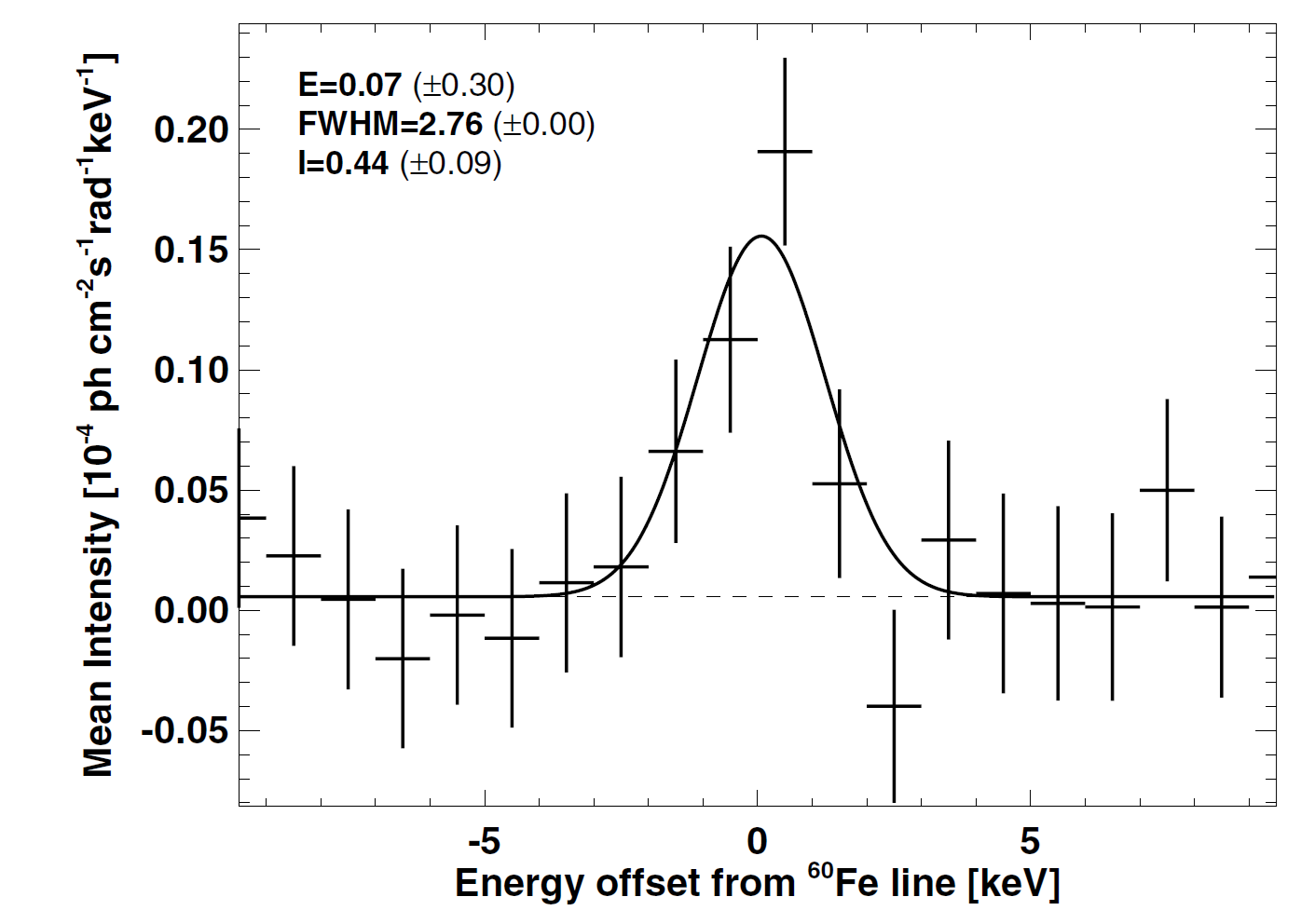}
\vspace*{8pt}
\caption{The combined spectrum of the two \nuc{Fe}{60} \ac{gray} lines in the inner Galaxy observed by \ac{INTEGRAL}/\ac{SPI} \citep[from][]{Wang2007}. In the laboratory, the \nuc{Fe}{60} line energies are 1173.23 and 1332.49\,keV, respectively. Shown here are the two lines superimposed with the centroids at 1173 and 1333\,keV set to a zero shift. The detection significance of the lines combined was $\sim 4.9 \sigma$. The solid line represents a fitted Gaussian profile of fixed instrumental width (2.76\,keV \ac{FWHM}), and a flat continuum. The average line flux is estimated as $(4.4\pm 0.9)\times 10^{-5}\,\mathrm{ph\,cm^{-2}\,s^{-1}\,rad^{-1}}$. \label{f1}}
\end{figure}

\begin{figure}
\centering
\includegraphics[width=0.75\linewidth]{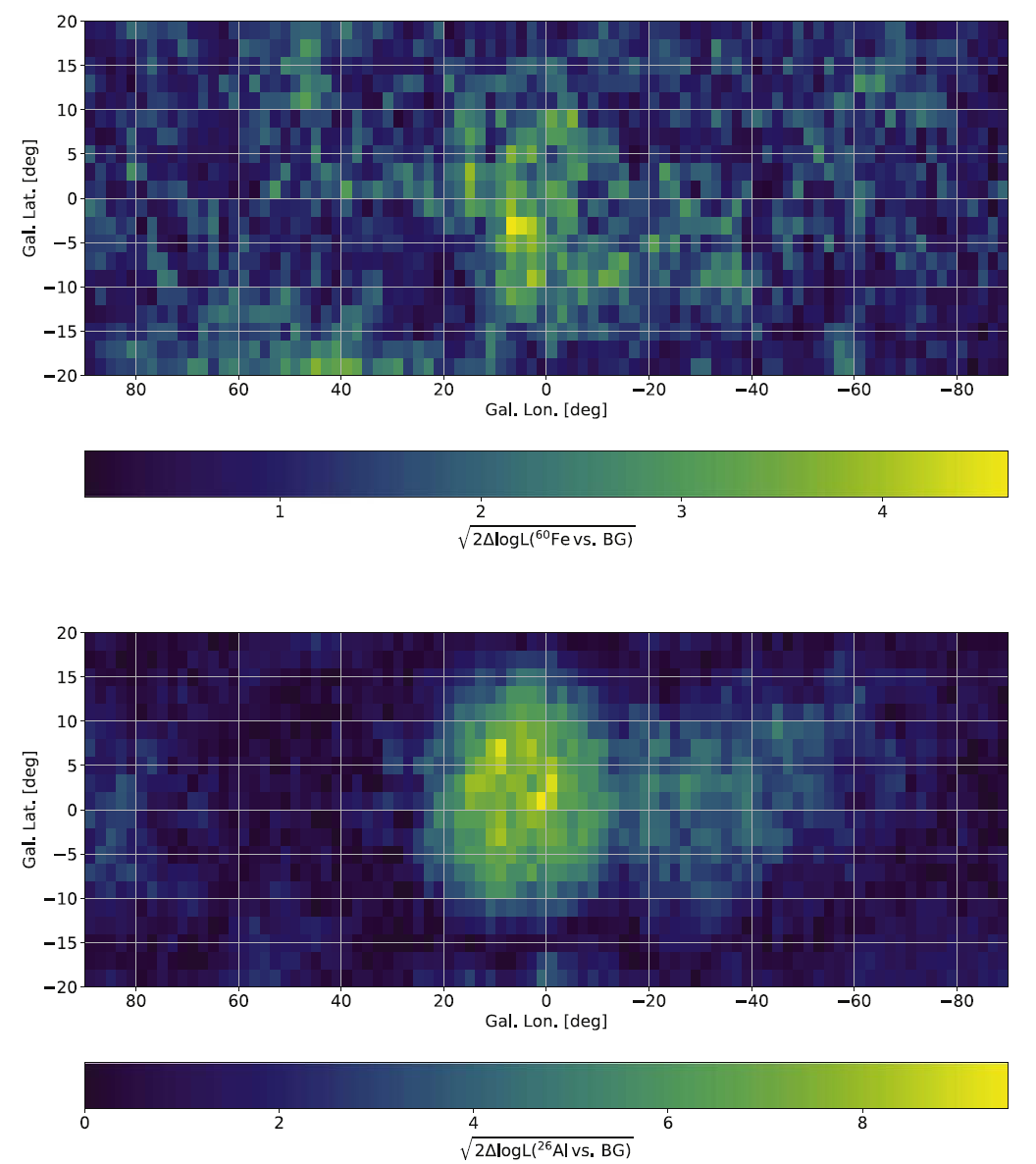}
\vspace*{8pt}
\caption{The distributions of two \ac{gray} line emissions for both \nuc{Fe}{60} (top) and \nuc{Al}{26} (bottom) obtained by \ac{SPI} scanning the inner part of the Galactic plane ($-90^\circ < \ell < 90^\circ; -20^\circ<b<20^\circ$ \citep[from][]{Wang2020}. Each pixel represents an individual point source test to identify excess emission above instrumental background, separated by $2^\circ$ (\ac{TS} map). \nuc{Fe}{60} emission is very weak, and its morphology is still unknown, but it is not a point-like source (such as only the Galactic centre), and probably of diffuse nature. It could be similar to \nuc{Al}{26}, however the uncertainties of also diffuse emission model fits are too large. \label{f2}}
\end{figure}

The \ac{gray} line emission from \nuc{Fe}{60} can directly diagnose the production sites in the Galaxy.
\nuc{Fe}{60} $\beta$-decays to \nuc{Co}{60}, which decays within 5.3\,yr to stable \nuc{Ni}{60} in an excited state that cascades into its ground state by \ac{gray} emission at 1173.23 and 1332.49\,keV, respectively.
The \nuc{Co}{60} ground state transition is accompanied by a \ac{gray} line at 59\,keV, however only in 2\% of all decays.
Thus, observations typically focus on the two high energy \ac{gray} lines.

Before \ac{INTEGRAL}, different \ac{gray} missions have searched for the \ac{gray} lines from the diffuse \nuc{Fe}{60} emission in the Galaxy, however only reported marginal detections or upper limit constraint with a significance level below 3$\sigma$ \citep{Leising1994,Harris1997,Diehl1997,Smith2004}.
This suggested that the \nuc{Fe}{60} emission signal in the Galaxy is very weak.
With nearly 3\,yr of \ac{SPI} observations, both \ac{gray} lines at 1773 and 1332\,keV have been detected for the first time \citep{Wang2007}.
Combining \nuc{Fe}{60} spectra from independent model fits, an \nuc{Fe}{60} signal from the Galaxy with a significance of 4.9$\sigma$ is reported with the average flux of $(4.4\pm 0.9)\times 10^{-5}\,\mathrm{ph\,cm^{-2}\,s^{-1}\,rad^{-1}}$ from the inner Galactic region (see Fig.\,\ref{f1}).
\citet{Wang2020} further determined an \nuc{Fe}{60} line flux of $\sim (3.1\pm 0.5)\times 10^{-4}\,\mathrm{ph\,cm^{-2}\,s^{-1}}$ for the whole sky, and a line flux of $\sim (4.5\pm 0.8)\times 10^{-5}\,\mathrm{ph\,cm^{-2}\,s^{-1}}$ for the inner Galactic region based on 15 \,yr of \ac{SPI} data.
The uncertainties after three and after 15 years of observations with \ac{SPI} are similar due to the radioactive build-up of the instrumental \nuc{Ni}{60} lines at the exact same photon energies \citep{Diehl2018}.
This means that only the first few years of a space mission are potentially useable to firmly detect and study the \nuc{Fe}{60} \ac{gray} lines.
However, other orbit considerations and material choices might alleviate this problem.
Since at least two, but probably more than ten, \acp{ccSN} happened in the vicinity of the moving Solar System, there is also the expectation of \ac{gray} line emission from the Local Bubble in which the Sun is currently located.
\citet{Siegert2024} predict a quasi-isotropic flux of \nuc{Fe}{60} and \nuc{Al}{26} \ac{gray} lines, as well as the associated 511\,keV line, based on a geometric model \citep{Pelgrims2020,Zucker2022} and hydrodynamics simulations \citep{Schulreich2023}.
The total \nuc{Fe}{60} \ac{gray} line flux inside the Local Bubble may range from $(0.5$--$4.2) \times 10^{-5}\,\mathrm{ph\,cm^{-2}\,s^{-1}}$ with an isotropic fraction of $20$--$50\%$.
Likewise, the \nuc{Al}{26} line form the Local Bubble may be $(0.3$--$2.0) \times 10^{-5}\,\mathrm{ph\,cm^{-2}\,s^{-1}}$ across the entire sky, in which the 511\,keV has a typical contribution of 41\%.
Coded mask telescopes, such as \ac{SPI}, cannot measure these isotropic fluxes, but future Compton telescopes could.

The robust detections of \nuc{Fe}{60} by \ac{SPI} can also be used to constrain the morphology of the weak \nuc{Fe}{60} emission in the Galaxy.
By using a so-called \ac{TS} map, testing for excess emission above the instrumental background at different positions in the sky between $-90^\circ<\ell<90^\circ$ and $20^\circ<b<20^\circ$, the sky distribution of \nuc{Fe}{60} can be investigated.
With a $\sim 5\sigma$ detection of \nuc{Fe}{60} in the whole sky, it is possible to test whether \nuc{Fe}{60} is point-like or truly diffuse, and whether \nuc{Fe}{60} and \nuc{Al}{26} share a similar morphology (Fig.\,\ref{f2}).
It was suggested that the \nuc{Fe}{60} and \nuc{Al}{26} emission distributions cannot be attributed to one or several point sources in the Galactic plane region.

%With a catalog of point-source locations with entries between $-90^\circ<l<90^\circ$ and $20^\circ<b<20^\circ$ in 2$^\circ$ pixel resolution, the sky distributions of both two radioactive isotopes \Fe and \Al are obtained (see Figure \ref{f2}). These morphology studies of \Fe and \Al emission distributions suggested that these gamma-ray emissions are not attributed to one
%or several point sources in the Galactic plane region. The \Fe emission morphology would be similar to that of \Al.

\begin{figure}
\centering
\includegraphics[width=0.7\linewidth]{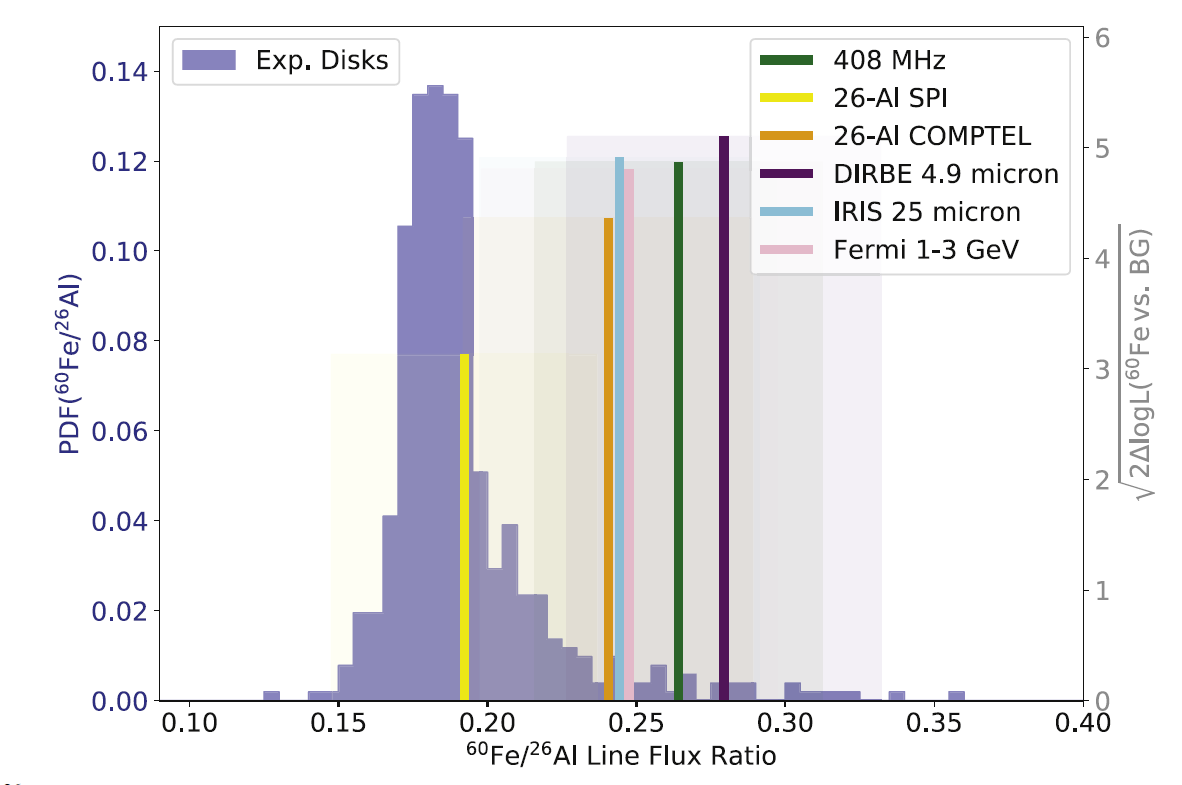}
\vspace*{8pt}
\caption{The flux ratio of \nuc{Fe}{60}/\nuc{Al}{26} determined by 15\,yr of \ac{SPI} observations \citep[from][]{Wang2020} based on model fitting of all-sky distributions, including a series of doubly-exponential disk models (blue) and a set of Galactic emission maps as tracers (vertical lines). The uncertainties are noted as shaded bands. Including the uncertainties of the fluxes from each spectral fit, the estimated \nuc{Fe}{60}/\nuc{Al}{26} flux ratio from the set of exponential disks is $0.18 \pm 0.04$. \label{f3}}
\end{figure}

\subsubsection{Ratio of \nuc{Fe}{60}/\nuc{Al}{26}}\label{sec:60Fe26Alratio}
The long-lived \nuc{Fe}{60} and \nuc{Al}{26} isotopes are expected to accumulate in the \ac{ISM} from many sources if those ejection events occur more often during their radioactive lifetime, leading to diffuse emission of radioactivity \acp{gray}.
Since \nuc{Fe}{60} and \nuc{Al}{26} are expected to have a similar origin, the yield ratio \nuc{Fe}{60}/\nuc{Al}{26} in the Galaxy is expected to notbe dependent on the distance uncertainties.
Therefore, an accurate measurements of the \nuc{Fe}{60}/\nuc{Al}{26} \ac{gray} line flux ratio provides an excellent test of the different models for massive star nucleosynthesis.
The steady-state mass of these radioactive isotopes maintained in the Galaxy through such production counterbalanced by radioactive decay, and thus converts into a ratio for the \ac{gray} line flux in each of the two lines through
\begin{equation}
\frac{ F(\mathrm{^{60}Fe})}{ F(\mathrm{^{26}Al})}  =  0.43 \cdot \frac{\dot{M}(\mathrm{^{60}Fe})}{\dot{M}(\mathrm{^{26}Al})}\mathrm{.}
\label{eq:fluxratio}
\end{equation}

At present, various models have predicted the values of the \nuc{Fe}{60}/\nuc{Al}{26} ratio.
\citet{Timmes1995} carry the massive-star yields into an estimate of chemical evolution for \nuc{Fe}{60} and \nuc{Al}{26} in the Galaxy, predicting a \ac{gray} flux ratio of 0.16.
Further revisions of the model give the different prediction of the \nuc{Fe}{60}/\nuc{Al}{26} ratio from $\sim 0.1$--$1.0$ \citep{Limongi2003,Limongi2006,Limongi2018,Rauscher2002,Prantzos2004,Woosley2007}.
The large uncertainties of the theoretical values come from both astrophysics and nuclear physics.

The yields of these two isotopes depend sensitively on both the stellar evolution details, such as shell burnings and convection, the final stages of the massive star evolution, and also the nuclear reaction rates.
\citet{Tur2010} found that the production of \nuc{Fe}{60} and \nuc{Al}{26} is sensitive to the 3$\alpha$ reaction rates during He burning, that is, the variation of the reaction rate by a factor of two would make a factor of nearly ten change in the \nuc{Fe}{60}/\nuc{Al}{26} ratio.
\nuc{Fe}{60} may be destroyed within its source by further neutron captures: \Fe($n,\gamma$).
Since its closest parent, \nuc{Fe}{59} is unstable, the $^{59}$Fe($n,\gamma$) production process competes with the $^{59}$Fe $\beta$-decay to produce an appreciable amount of \nuc{Fe}{60}:
This reaction pair dominates the nuclear-reaction uncertainties in \nuc{Fe}{60} production, with $^{59}$Fe($n,\gamma$) being difficult to measure in nuclear laboratories due to its long lifetime and multitude of reaction channels \citep{Jones2019b}.
Using an effective He-burning reaction rates can account for correlated behaviour of nuclear reactions and mitigate the overall nuclear uncertainties in these shell burning environments \citep{Austin2017}.

Thus, astronomical observations of the \nuc{Fe}{60}/\nuc{Al}{26} ratio will also help to constrain the nuclear-reaction aspects of massive stars, given the experimental difficulties to measure all reaction channels involved at the astrophysically-relevant energies.
Assuming that the sky distribution of \nuc{Fe}{60} follows the \ac{CGRO}/\ac{COMPTEL} all-sky 1809\,keV emission map of \nuc{Al}{26} \citep[e.g.,][]{Oberlack1996,Plueschke2001}, the early \ac{SPI} observations yield a flux ratio of \nuc{Fe}{60}/\nuc{Al}{26} in the range of $\sim 0.09$--$0.21$ \citep{Wang2007} and $\sim 0.08$--$0.22$ \citep{Bouchet2011,Bouchet2015}.
To cater for the uncertainty of the spatial extent of \nuc{Fe}{60} emission, \citet{Wang2020} performed different tests to avoid possible biases in determining the ratio.
In order to fit the \nuc{Fe}{60} and \nuc{Al}{26} (all-)sky distributions, they first created a grid of of doubly-exponential disk models with different characteristic scale radii from 0.5 to 8.0\,kpc and scale heights from 0.01 to 2.0\,kpc in a grid of 512 possibilities.
The \nuc{Fe}{60}/\nuc{Al}{26} ratio distribution from all 512 exponential-disk configurations is presented in Fig.\,\ref{f3}, including characteristic uncertainties.
The average \nuc{Fe}{60}/\nuc{Al}{26} ratio based on this analysis gives a value of $0.18 \pm 0.04$.
\citet{Wang2020} also fitted the \nuc{Fe}{60} diffuse emissions with a set of maps representing different source tracers (e.g., the 408\,MHz radio map, infrared emission, high-energy \ac{gray} and X-ray sky maps).
These sky maps with significant detections of and adequately describing both \nuc{Al}{26} and \nuc{Fe}{60} emission lines can also constrain the \nuc{Fe}{60}/\nuc{Al}{26} values (Fig.\,\ref{f3}).
The average ratio values distribute around $0.25$, generally with an upper limit range near $0.3$.

The \nuc{Fe}{60}/\nuc{Al}{26} ratio has been promoted as a useful test of stellar evolution and nucleosynthesis models, because the actual source numbers and their distances may cancel out largely in such a ratio.
The present \ac{gray} line studies give an observational ratio of $\sim 0.15$--$0.3$ in the Galaxy, while theoretical predictions seem to produce relatively large values.
Massive star models with a solar composition and a standard stellar mass distribution from 13 to 120\,$\mathrm{M_\odot}$ predict \nuc{Fe}{60} and \nuc{Al}{26} yields with and without rotation effects \citep{Limongi2018}:
With rotation, the flux ratio is $0.8 \pm 0.4$, but the ratio reduces to $\sim 0.2$--$0.6$ in the case of non-rotating models.
In addition, if only the mass range from 13--40$\,\mathrm{M_\odot}$ is considered, the \nuc{Fe}{60}/\nuc{Al}{26} ratio is estimated to be $\sim 0.07$--$0.15$.
The stellar evolution for the stars more massive than $40\,\mathrm{M_\odot}$ has considerable uncertainties, for example from the stellar winds and the effect on the stellar structure and evolution.
Some stars may not explode as \acp{ccSN}, but rather collapse directly to black holes.
Thus, massive stars above $40\,\mathrm{M_\odot}$ may not have the effective contribution to \nuc{Fe}{60} production and could be ignored in a stellar mass-weighted Galactic average.
Compared to the most recent \ac{gray} observations, the present models would either over-estimate \nuc{Fe}{60} yields or under-estimate the \nuc{Al}{26} production \citep{Wang2020}.

This problem, however, may be alleviated if other \nuc{Al}{26} production channels than only massive stars, their winds, and their \acp{ccSN} are taken into account.
\citet{Vasini2025} suggests that up to 75\% of the Galactic \nuc{Al}{26} mass -- and therefore flux if distributed similarly -- may be due to \acp{CN}, based on a Galactic chemical evolution model.
If this scenario is true, as has also been suggested early on in the study of the \nuc{Al}{26} \ac{gray} line \citep[e.g.,][]{Leising1985,Jose1997,Bennett2013}, the \nuc{Fe}{60}/\nuc{Al}{26} flux ratio from only massive stars in the Galaxy may be increased by a factor of four, more in line with massive star model predictions.

% CNe
\newpage
% Introduction
\subsection{$\gamma$-Ray Lines from Classical Novae}\label{sec:CNe}
\vspace{-0.5em}
{\emph{Written by Pierre Jean}}
\vspace{0.5em}\\
%
\begin{comment}
It is commonly accepted that a \ac{CN} originates from the thermonuclear explosion of a layer accumulated by accretion at the surface of a \ac{WD} in a binary system.
%
The thermonuclear runaway converts hydrogen into helium via the CNO process and in the main time synthesises elements with medium atomic numbers.
%
\acp{CN} as sources of nuclear \acp{gray} were first suggested by Clayton \& Hoyle\,\cite[1974;][see also \cite{Clayton1981}]{Clayton1974}.
%
However, no such emission has been observed yet with past and existing \ac{gray} spectrometers.
%
\acp{CN} are good candidates for the observation of their nucleosynthesis activity in \acp{gray} since the freshly produced elements are quickly transported out of their production zone either by convection during the outburst or in the material ejected in the explosion.
%
In these conditions, the unstable nuclei decay in a less dense material, allowing the de-excitation \acp{gray} to escape the system, enabling their observation.
%
The typical ejected masses of $\sim 10^{-6}$--$10^{-4}\,\mathrm{M_{\odot}}$ are low compared to the masses ejected by SNe Ia and ccSNe.
%
These small amounts of ejected unstable nuclei makes the intensity of the expected \ac{gray} lines faint and difficult to detect, taking into account the sensitivity of past and present \ac{gray} spectrometers.
%
On the other hand, the rate of \acp{CN} in our Galaxy is larger than the rate of SNe in the local Universe.
\end{comment}
%
It is commonly accepted that a \ac{CN} arises from the thermonuclear explosion of an accreted layer on a \ac{WD} in a binary system.
The thermonuclear runaway converts hydrogen into helium via the CNO process while synthesizing medium atomic number elements.
\citet{Clayton1974} first proposed \acp{CN} as sources of nuclear \acp{gray}, yet such emission has not been observed with past or current \ac{gray} spectrometers \citep[see also][]{Clayton1981}.
\acp{CN} are promising for observing nucleosynthesis in \acp{gray} because freshly produced elements are rapidly transported from the production zone by convection during the outburst or via ejected material;
under these conditions, unstable nuclei decay in less dense matter, allowing de-excitation \acp{gray} to escape and be detected.
Typical ejected masses of $\sim 10^{-6}$--$10^{-4}\,\mathrm{M_{\odot}}$ are low compared to those from \acp{SNIa} and \acp{ccSN}, and these small amounts of unstable nuclei render the expected \ac{gray} lines faint and difficult to detect without improvements on the sensitivity of \ac{gray} instruments.
%
%On the other hand, the rate of \acp{CN} in our Galaxy is larger than the rate of SNe in the local Universe.

% General properties of novae
\subsubsection{Properties of Nuclear $\gamma$-Rays from Classical Novae}
%
\begin{comment}
The kind of synthesised material during the thermonuclear runaway depends on the composition of the underlying \ac{WD}:
%
carbon-oxygen (CO) or oxygen-neon (ONe).
%
The latter are expected to be more massive than the former according to stellar evolution.
%
The pre-outburst abundance depends on the mixing between the material from the \ac{WD}'s core and the one from the accumulated layer.
%
During their outburst, \acp{CN} from CO-\acp{WD} synthesise light and intermediate-mass elements, mainly of the CNOF group.
%
In addition to elements of the CNOF group, the initial abundance, the higher temperature and densities in ONe novae allow the synthesis of heavier elements, among them nuclei of the NeNa and MgAl groups.
%
The main unstable nuclei that yield to nuclear \ac{gray} emission are listed in Tab.\,\ref{tab:linenova} with their main characteristics.
%
Both kinds of \acp{CN} produce \nuc{Li}{7}, with the mass of \nuc{Li}{7} ejected by CO novae being about ten times larger than by ONe novae \cite{Jose1998}.
%
The isotopes \nuc{Na}{22} and \nuc{Al}{26} belong to the NeNa and MgAl groups, and they are produced in higher abundances in ONe than in CO novae. 
\end{comment}
%
The synthesized material during the thermonuclear runaway depends on the underlying \ac{WD} composition -- carbon-oxygen (CO) or oxygen-neon (ONe), with ONe being more massive.
The pre-outburst abundance is set by mixing core and accreted material.
During the outburst, CO novae mainly synthesise light and intermediate-mass elements of the CNOF group, while ONe novae, thanks to higher temperatures and densities, also produce heavier elements (e.g., from the NeNa and MgAl groups).
Tab.\,\ref{tab:linenova} lists the main unstable nuclei that yield nuclear \ac{gray} emission.
Both \acp{CN} types produce \nuc{Li}{7}, with CO novae ejecting about ten times more than ONe novae \citep{Jose1998}, and the isotopes \nuc{Na}{22} and \nuc{Al}{26} are more abundant in ONe novae.

\begin{table}[!h]
    \centering
    \begin{tabular}{cccc}
        \noalign{\smallskip}
        \hline
        \hline
        \noalign{\smallskip}
        \mbox{Radionuclei(decay mode)}  & \mbox{Main $\gamma$-ray emission} & \mbox{Half-life} & \mbox{Nova type} \\
        \hline
        \noalign{\smallskip}
        \mbox{$^{7}$Be(EC)} & \mbox{478 keV} & \mbox{53 days} & \mbox{CO}  \\
        \noalign{\smallskip}
        \mbox{$^{13}$N($\beta^+$)} & \mbox{$\leq$ 511 keV} & \mbox{10 min} & \mbox{CO \& ONe}  \\
        \noalign{\smallskip}
        \mbox{$^{18}$F($\beta^+$ \& EC)} & \mbox{$\leq$ 511 keV} & \mbox{1.8 h} & \mbox{CO \& ONe}  \\
        \noalign{\smallskip}
        \mbox{$^{22}$Na($\beta^+$ \& EC)} & \mbox{1275 keV} & \mbox{2.6 yr} & \mbox{ONe}  \\
        \noalign{\smallskip}
        \mbox{$^{26}$Al($\beta^+$ \& EC)} & \mbox{1809 keV} & \mbox{0.7 Myr} & \mbox{ONe}  \\
        \noalign{\smallskip}
          \hline
    \end{tabular}
    \caption{List of the main radioactive nuclei with their characteristics, 
     their main \ac{gray} emission (lines) and the type of \acp{CN} in which they are produced \citep[adapted from][]{Hernanz2014}.}
    \label{tab:linenova}
\end{table}

\begin{figure}[!th]
\begin{center}
\includegraphics[width=0.7\linewidth]{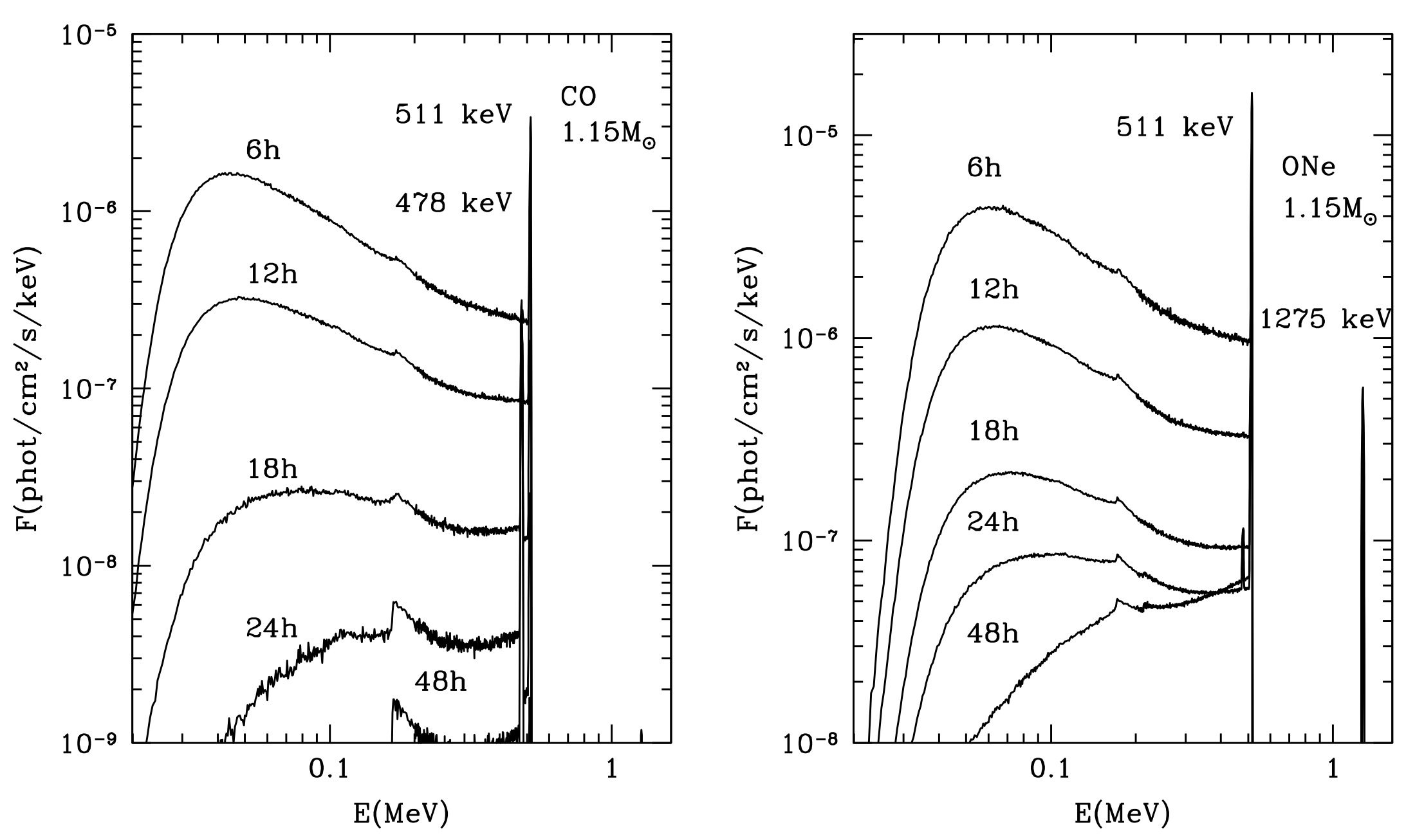}
\caption{Model of early \ac{gray} spectra emitted by a \ac{CN} at a distance of 1\,kpc as a function of time after the peak in temperature of the thermonuclear runaway \citep[from][]{Hernanz2014}.}
\label{fig:specHer14}
\end{center}
\end{figure}

The produced nuclei are proton-rich and unstable, as expected from fast hydrogen burning in the CNO cycle, and decay by $\beta^+$ or \ac{EC}.
Positrons from the short-lived $\beta^+$-decay of \nuc{N}{13} and \nuc{F}{18} rapidly lose kinetic energy and annihilate in the expanding envelope.
The steep temperature gradient in the envelope ensures convection transports unstable nuclei to the surface faster than their lifetime. %
Thus, $\beta^+$-unstable nuclei decay in a less opaque medium, enabling \acp{gray} from positron annihilations to escape.
This yields a 511\,keV line and a continuum $\leq 511$\,keV from ortho-\ac{Ps} annihilation and Compton scattering.
Similarly, positrons from \nuc{Na}{22} decay contribute about three orders of magnitude less at early times than those from \nuc{N}{13} and \nuc{F}{18}.
Furthermore, shortly after the explosion, positrons from \nuc{Na}{22} decay are expected to escape the ejecta (e.g., at $\sim 25$\,d for a \ac{CN} with $1.25\,\mathrm{M_{\odot}}$ \citep{Hernanz2006}).
Due to its lifetime, most positrons from \nuc{Al}{26} decay escape into the \ac{ISM} and likely travel far from the \ac{CN} before annihilation.
\citet{Leising1987} first studied the positron fate in \acp{CN} envelopes and estimated \ac{gray} light curves.
\citet{Gomez-Gomar1998} later analysed the temporal and spectral properties of \ac{gray} line emissions from ONe and CO novae.
They employed a hydrodynamical code with a nuclear reaction network to compute the envelope properties and radioactive synthesis.
Radiative transfer was computed via Monte Carlo simulations.
Fig.\,\ref{fig:specHer14} shows the spectra obtained with this method for \ac{CN} models available in 2014.
The 511\,keV line and continuum are clearly visible.
The low-energy cutoff ($\sim 20$--$30$\,keV) arises from photoelectric absorption, and the 170\,keV feature from 511\,keV photon backscattering in the ejecta.

The light curves of the 478\,keV (\nuc{Be}{7}) and 1275\,keV (\nuc{Na}{22}) lines peak at $5$--$10$\,d post-peak temperature, depending on the \ac{CN} type and ejecta mass.
This maximum occurs when the ejecta become transparent to \acp{gray}, as the lifetimes of \nuc{Be}{7} and \nuc{Na}{22} exceed the ejecta dynamical timescale.
The 511\,keV line and ortho-\ac{Ps} peak is more complex, given the short lifetimes of \nuc{N}{13} and \nuc{F}{18} and the dependence on photon energy and ejecta dynamics.
\citet{Leising1987} modelled 511\,keV emissions peaking at $40$\,min and $6$\,h from \nuc{N}{13} and \nuc{F}{18}, respectively, with fluxes of $5 \times 10^{-3}$ to $10^{-1}\,\mathrm{ph\,cm^{-2}\,s^{-1}}$ for fast \acp{CN} at 1\,kpc.
\citet{Gomez-Gomar1998} obtained similar 511\,keV flux maxima at about $1$\,h and $5$--$8$\,h post-peak temperature \citep[see also][]{Hernanz1999a}.
With revised nuclear rates, \citet{Hernanz2014} reported maximum fluxes of $10^{-2}$ and $8 \times 10^{-4}\,\mathrm{ph\,cm^{-2}\,s^{-1}}$ for CO and ONe novae (both $1.15\,\mathrm{M_{\odot}}$ at 1\,kpc), respectively.
Recently, \citet{Leung2022} performed similar studies;
they found \ac{gray} fluxes over three orders of magnitude lower than those reported by \citet{Hernanz2014} for similar CO and ONe novae.
This is mainly due to their use of lower expansion velocities and peak temperatures than previous studies.
Thus, the ejecta become transparent later ($5$--$10$\,d), by which time most short-lived nuclei have decayed.

\begin{comment}
Since the radioactive nuclei decay in an expanding envelope, the \ac{gray} lines are broadened by Doppler effect.
%
The early emitted 511\,keV line is expected to have a width of $5$--$8$\,keV (\ac{FWHM}), depending on the model, and is slightly blue-shifted, because the only photons that escape toward the observer, due to opacity, are the ones emitted in the external layer of the expanding envelope \cite{Hernanz1999a,Leising1987}.
%
With ejecta velocities ranging between $1300$--$2500\,\mathrm{km\,s^{-1}}$, the width of the 478\,keV line ranges between $3$ and $7$\,keV (\ac{FWHM}) for CO nova models of $0.8\,\mathrm{M_{\odot}}$ and $1.15\,\mathrm{M_{\odot}}$, respectively.
%
For the width of the 1275\,keV line, emitted by ONe novae models (that is, ejecta velocities of $2500\,\mathrm{km\,s^{-1}}$), a broadening of about 20\,keV is expected \cite{Gomez-Gomar1998}.
\end{comment}
%
Since the radioactive nuclei decay in an expanding envelope, the \ac{gray} lines are broadened by the Doppler effect.
The early 511\,keV line is expected to have a width of $5$--$8$\,keV (\ac{FWHM}), model-dependent, and is slightly blue-shifted since only photons from the external envelope escape opacity \citep{Hernanz1999a,Leising1987}.
For CO novae models ($0.8\,\mathrm{M_{\odot}}$ and $1.15\,\mathrm{M_{\odot}}$) with ejecta velocities of $1300$--$2500\,\mathrm{km\,s^{-1}}$, the 478\,keV line has a width of $3$--$7$\,keV (\ac{FWHM}).
For ONe novae with ejecta velocities around $2500\,\mathrm{km\,s^{-1}}$, the 1275\,keV line is broadened by about 20\,keV \citep{Gomez-Gomar1998}.

\begin{comment}
For the diffuse emission of the 511\,keV line in Galaxy, positrons from \acp{CN} will also contribute.
%
The contribution of $\beta^+$-decays of elements synthesised by \acp{CN} alone, however, cannot explain the entire rate of annihilation (see Sec.\,\ref{sec:511}).
%
Indeed, positrons from \nuc{N}{13} and \nuc{F}{18} decays annihilate within the envelope, resulting in a transient emission and the amounts of \nuc{Al}{26} and \nuc{Na}{22} nuclei released by \acp{CN} are probably not enough:
%
If an ONe nova produces up to $10^{-8}\,\mathrm{M_{\odot}}$ of \nuc{Na}{22}, and if the rate of ONe novae in the Galaxy is about $1/3$ of $35\,\mathrm{yr^{-1}}$, Prantzos et al.\,\cite[2011;][]{Prantzos2011} estimated a positron production rate of $\lesssim 1.5 \times 10^{41}\,\mathrm{e^+\,s^{-1}}$ in the Milky Way, which corresponds to less than few percent of the measured annihilation rate.
%
This is also corroborated by the non-detection of diffuse and point-like \nuc{Na}{22} \acp{gray} from the Galaxy (Sec.\,\ref{sec:Na22_obs}).
\end{comment}
%
Positrons from \acp{CN} also contribute to the diffuse 511\,keV emission in the Galaxy.
However, $\beta^+$-decays from \acp{CN} alone cannot account for the full annihilation rate (see Sec.\,\ref{sec:511}).
Positrons from \nuc{N}{13} and \nuc{F}{18} decay within the envelope, causing transient emission, and the \nuc{Al}{26} and \nuc{Na}{22} yields from \acp{CN} are likely insufficient.
Assuming an ONe nova produces up to $10^{-8}\,\mathrm{M_{\odot}}$ of \nuc{Na}{22} and that ONe novae represent about one-third of $35\,\mathrm{yr^{-1}}$, \citet{Prantzos2011} estimated a positron production rate of $\lesssim 1.5 \times 10^{41}\,\mathrm{e^+\,s^{-1}}$, only a few percent of the measured rate.
This is further supported by the non-detection of diffuse and point-like \nuc{Na}{22} \acp{gray} from the Galaxy (Sec.\,\ref{sec:Na22_obs}).

\subsubsection{Observations of $\gamma$-Rays from Classical Novae}
%
% 22Na 
%
\paragraph{$\gamma$-Ray Emission from \nuc{Na}{22} Decays}\label{sec:Na22_obs}
%
% 22Na from individual sources
The 1275\,keV line flux expected from an individual ONe nova is
\begin{equation}
    F_{22} = 3.8 \times 10^{-4} \times \left(\frac{d}{\mathrm{1\,kpc}}\right)^{-2} \times \left(\frac{M_{22}}{10^{-8}\,\mathrm{M_{\odot}}}\right) \times \exp\left(-\frac{\Delta t}{\tau_{22}}\right)\,\mathrm{ph\,cm^{-2}\,s^{-1}}\mathrm{,}
    \label{eq:na22_flux}
\end{equation}
where $M_{22}$ is the ejecta mass of \nuc{Na}{22}, $\Delta t$ is the time since outburst, $\tau_{22}$ the lifetime of \nuc{Na}{22}, and $d$ is the distance to the \ac{CN}.
\citet{Leising1988} undertook the first attempt to search for the 1275\,keV line from individual \acp{CN}.
They analysed data of the \ac{GRS} onboard the \ac{SMM} from 1980 to 1987.
The did not detect the line and thus derived upper limit fluxes of four nearby \acp{CN}, which resulted in a most constraining upper limit of \nuc{Na}{22} ejecta mass of $7 \times 10^{-7}\,\mathrm{M_{\odot}}$ from N Vul 1984b.
Later, \citet{Iyudin1995} obtained an upper limit of $3.7 \times 10^{-8}\,\mathrm{M_{\odot}}$ from the observation of eleven Galactic \acp{CN} exploding between 1991 and 1993, with \ac{COMPTEL}.
The more recent results come from the observations with \ac{SPI}:
\citet{Siegert2021} derive an upper limit of $1.3 \times 10^{-7}\,\mathrm{M_{\odot}}$ from V5115 Sgr, a probable ONe nova that exploded on March 2005, assuming a distance of 3\,kpc.

\begin{figure}[h]
\begin{center}
\includegraphics[scale=0.3,angle=0]{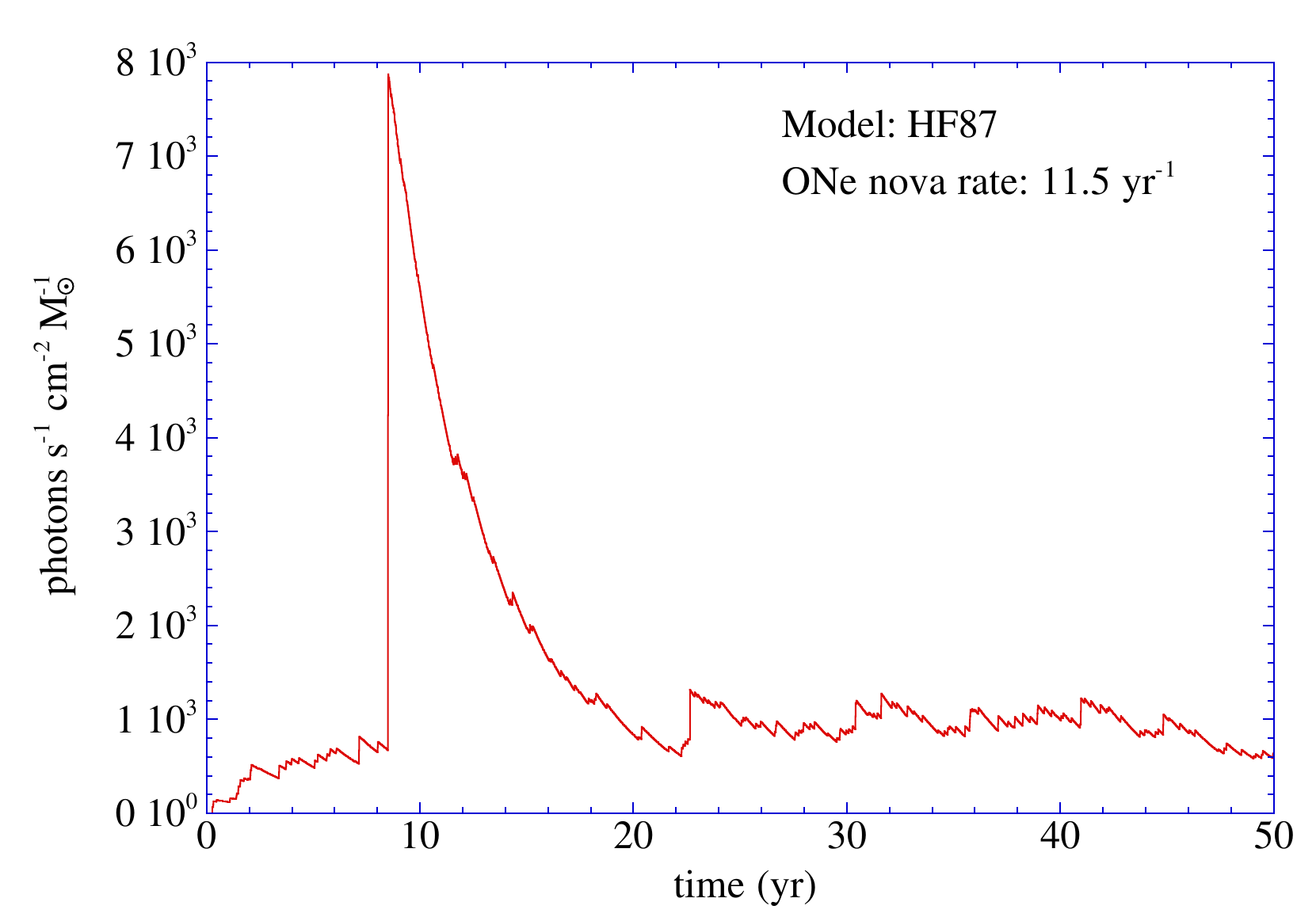}
\caption{Example of the normalised 1275\,keV flux in a $12.5^{\circ}$ region around the Galactic centre as a function of time with a ONe nova rate of $11.5\,\mathrm{yr^{-1}}$. The flux from the Galactic centre direction can be dominated by a single nearby \ac{CN}, such as around year $8.6$.}
\label{fig:timevar}
\end{center}
\end{figure}

% Observation of the cumulative emission of 22Na from ONe novae. 
\begin{comment}
Since the lifetime of \nuc{Na}{22} is larger than the typical period between two ONe novae in the Galaxy ($R_{\rm CNe} \approx 11$--$17\,\mathrm{yr^{-1}} \Leftrightarrow \tau_{\rm ONe} \approx 0.06$--$0.09\,\mathrm{yr}$), the accumulated \nuc{Na}{22} is expected to produce a diffuse 1275\,keV $\gamma$-line in the Milky Way.
%
However, the cumulative flux would be dominated by the contribution of only a few tens of \ac{CN} (e.g., $\tau_{22} \times R_{\rm ONe} \approx 50$ \ac{CN} on average) and the diffuse emission will show fluctuations.
%
Fig.\,\ref{fig:timevar} shows an example of the total 1275\,keV flux variations in a $12.5^{\circ}$ region around the Galactic centre computed with a Monte Carlo simulation as described in Jean et al.\,\cite[2000;][]{Jean2000}.
%
The flux has been normalised to the mass of \nuc{Na}{22} ejected per outburst.
%
In this example, the flux is dominated by a single \ac{CN} at the year $8.6$, which contributes 91\% of the total flux.
%
After year 20, the flux shows fluctuations with a maximum amplitude of $\pm 40\%$ of the mean flux of about $10^{3}\,\mathrm{ph\,cm^{-2}\,s^{-1}M_{\odot}^{-1}}$. 
%
Similar works have also been performed later, taking into account renewed \ac{CN} rates \cite{Siegert2021,Shafter2017}.
\end{comment}
%
Given that the lifetime of \nuc{Na}{22} exceeds the typical interval between ONe novae in the Galaxy ($R_{\rm CNe} \approx 11$--$17\,\mathrm{yr^{-1}} \Leftrightarrow \tau_{\rm ONe} \approx 0.06$--$0.09\,\mathrm{yr}$), the accumulated \nuc{Na}{22} should yield a diffuse 1275\,keV $\gamma$-line in the Milky Way.
However, the cumulative flux is dominated by just a few tens of \acp{CN} (e.g., $\tau_{22} \times R_{\rm ONe} \approx 50$ on average), so the diffuse emission fluctuates.
Fig.\,\ref{fig:timevar} illustrates the total 1275\,keV flux variations over a $12.5^{\circ}$ region around the Galactic centre, computed via Monte Carlo simulation by \citet{Jean2000}.
The flux is normalized to the mass of \nuc{Na}{22} ejected per outburst.
In this case, a single \ac{CN} at year $8.6$ contributes 91\% of the total flux.
After year 20, the flux fluctuates with a maximum amplitude of $\pm 40\%$ around a mean of about $10^{3}\,\mathrm{ph\,cm^{-2}\,s^{-1}M_{\odot}^{-1}}$.
Subsequent studies with updated \ac{CN} rates have also been performed and showed similar features \citep{Siegert2021,Shafter2017}.

\citet{Higdon1987} were the first to search for the Galactic diffuse emission of the \ac{gray} line from the cumulative decay of \nuc{Na}{22} ejected by active ONe novae.
They compared their Monte Carlo emission model with the upper flux limits from \ac{HEAO3} observations by \citet{Mahoney1982}.
Their model, based on M31's stellar distributions scaled to the Milky Way's disk, spheroid, and local \ac{CN} rate, yields an upper mass limit of $5.6 \times 10^{-7}\,\mathrm{M_{\odot}}$ per \ac{CN}.
However, this result is uncertain due to the unknown spatial distribution and rate of \acp{CN}, as interstellar extinction obscures many Galactic \acp{CN} at optical wavelengths.
Analyses of \ac{SMM} data yielded an upper flux limit of $\mathrm{1.2 \times 10^{-4}\,\mathrm{ph\,cm^{-2}\,s^{-1}}}$ from the Galactic centre region; assuming a \ac{CN} rate of $40\,\mathrm{yr^{-1}}$, this implies an upper limit of $2 \times 10^{-8}\,\mathrm{M_{\odot}}$ of \nuc{Na}{22} ejected per \ac{CN} \citep{Leising1988}.
With a more realistic ONe nova rate of $\sim 10\,\mathrm{yr^{-1}}$ \citep{Jean2000}, the upper limit becomes $8 \times 10^{-8}\,\mathrm{M_{\odot}}$.
Using six years of \ac{COMPTEL} data of the Galactic bulge, \citet{Jean2001} derived an upper limit of $6 \times 10^{-7}\,\mathrm{M_{\odot}}$ per \ac{CN}, accounting for uncertainties in the ONe nova rate and spatial distribution.
Fifteen years of \ac{SPI} observations yielded an upper limit of $2.7 \times 10^{-7}\,\mathrm{M_{\odot}}$ from the 1275\,keV diffuse emission \citep{Siegert2021}.

The latest study of the \nuc{Na}{22} amount ejected per ONe nova was conducted by \citet{Fougeres2023} that accounted for updated nuclear rates.
They showed that the \nuc{Na}{22} ejecta depend on the accretion rate and the initial \ac{WD} luminosity.
They calculated \nuc{Na}{22} ejecta masses of $(4$--$8)\times 10^{-9}\,\mathrm{M_{\odot}}$, depending on the \ac{WD} mass for an initial luminosity of $10^{-2}\,\mathrm{L_{\odot}}$ and an accretion rate of $2 \times 10^{-10}\,\mathrm{M_{\odot}\,yr^{-1}}$.
For a $1.2\,\mathrm{M_{\odot}}$ \ac{WD}, they obtained \nuc{Na}{22} ejecta masses of $(0.3$--$8)\times 10^{-9}\,\mathrm{M_{\odot}}$.
This is about one to two orders of magnitude lower than the upper limits obtained with \ac{gray} observations. 

%
% 26Al
%
\paragraph{$\gamma$-Ray Emission from \nuc{Al}{26} Decays}
%
%The diffuse 1809 keV line emission from our Galaxy was observed since the 80's.
The spatial distribution of the 1809\,keV line from \nuc{Al}{26} measured by \ac{COMPTEL} fits with the distribution of the young stellar population rather than with the old star (\ac{CN}) population \citep{Knoedlseder1999}.
Its origin is mainly explained by massive stars that eject freshly produced \nuc{Al}{26} via winds during their Wolf-Rayet phase or when they explode at the end of their life.
Crude estimation of the amount of \nuc{Al}{26} produced by \ac{CN} in the Milky Way was found to be of $\sim 0.4\,\mathrm{M_\odot}$ \citep{Jose1997}.
With updated nuclear reaction rates and a new chemical evolution model, \citet{Canete2023} reached a similar result with an estimation of the maximum contribution of \ac{CN} of $\sim 0.2\,\mathrm{M_\odot}$.
With such a value, the contribution of all \ac{CN} would yield to a total 1809\,keV flux of $\lesssim 4 \times 10^{-5}\,\mathrm{ph\,cm^{-2}\,s^{-1}}$ which is about a factor of nine times less than the flux in the inner Galaxy measured by \ac{SPI} \citep{Diehl2006,Pleintinger2023}.
This is in stark contrast to recent Galactic Chemical Evolution estimates by \citet{Vasini2025} who predict a contribution of up to 75\% of the \nuc{Al}{26} mass in the Galaxy (see also Sec.\,\ref{sec:60Fe26Alratio}).

%
% 7Li
%
\paragraph{$\gamma$-Ray Emission from \nuc{Be}{7} Decays}
The 478\,keV line emission from a \ac{CN} has not been detected yet.
Neglecting absorption at the early stage of the outburst, the instantaneous line flux from the decay of a certain mass of $M_7$ of \nuc{Be}{7} ejected at a distance $d$ by a \ac{CN} \citep{Jose1998} is 
\begin{equation}
    F_{7} = 2.2 \times 10^{-6} \times \left(\frac{d}{\mathrm{1\,kpc}}\right)^{-2} \times \left(\frac{M_{7}}{10^{-10}\,\mathrm{M_{\odot}}}\right) \times \exp\left(-\frac{\Delta t}{\tau_{7}}\right)\,\mathrm{ph\,cm^{-2}\,s^{-1}}\mathrm{.}
    \label{eq:be7_flux}
\end{equation}
Using SMM observations, \citet{Harris1991} reported upper limits on the line flux around $10^{-3}\,\mathrm{ph\,cm^{-2}\,s^{-1}}$ from three Galactic \acp{CN} at distances $\leq 3.5$\,kpc, between 1982 and 1986.
They derived a most constraining upper limit of the \nuc{Be}{7} mass ejected of $5.2 \times 10^{-8}\,\mathrm{M_{\odot}}$ from N Cen 1986, which was assumed to be at a distance of 1.1\,kpc.
More recently, \citet{Siegert2018} obtained an upper limit of $1.2 \times 10^{-8}\,\mathrm{M_{\odot}}$ from the observation with \ac{SPI} of V5668 Sgr in 2015 assuming a distance of 1.6\,kpc.
Combining this result with the \nuc{Be}{7} mass estimated from UV observations, \citet{Siegert2018} concluded that the \ac{CN} should be at a distance larger than 1.2\,kpc.
The observations of the cumulative 478\,keV line emission from the Galactic centre region provide less constraining upper mass limits than the ones obtained from observation of individual \ac{CN} \citep{Harris1991,Siegert2021}.
With a better distance estimation of V5668 Sgr and a revised analysis of \ac{SPI} data, \citet{Siegert2021} obtained an upper limit of $3 \times 10^{-8}\,\mathrm{M_{\odot}}$.
This value is similar to the one obtained by \citet{Izzo2025} with the observation of V1369 Cen in 2013, with \ac{SPI}.
Here again, the upper limit masses derived from \ac{gray} observations are about two orders of magnitude larger than the masses estimated by theoretical studies.
However, some UV observations of \ac{CN} suggest higher \nuc{Be}{7} ejecta masses, such as around $10^{-9}\,\mathrm{M_{\odot}}$ \citep[e.g.,][]{Tajitsu2016,Molaro2020}, based on relative abundance measurements.

%
% Prompt emission
%
\paragraph{Prompt $\gamma$-Ray Emissions}
Several attempts have been made to detect the prompt $\leq 511$\,keV emissions from \ac{CN} using space-borne \ac{gray} detectors with wide fields of view.
The delay between the prompt \ac{gray} emission and the optical maximum being uncertain, the analyses should be performed on long observation periods around the outburst date.
Moreover, the theoretical duration of the brightest $\leq 511$\,keV emission ($\lesssim 1$\,h) is similar to the typical orbital period ($\sim 1.5$\,h) of instrument in low Earth orbit, such as the \ac{BATSE} onboard \ac{CGRO} and the \ac{BAT} onboard Swift.
In these conditions, it is not possible to continuously monitor the sources and, therefore, it is difficult to extract upper limits on fluxes, since the source may have been occulted by Earth during its maximum emission.
Gamma-ray detectors in high excentric orbits, such as the \ac{ACS} of \ac{SPI} \citep{Jean1999} or in interplanetary space such as \ac{TGRS} onboard the Wind probe, do not have such a limitation but they cannot perform images of sources and they experience a higher instrumental background.

\citet{Hernanz2000} did not detect the prompt emission in the $250$--$511$\,keV band of \ac{BATSE} from two nearby \ac{CN} (Nova Cyg 1992 and Nova Vel 1999).
By analysing \ac{TGRS} data within 6\,h intervals, \citet{Harris2000} obtained upper limits on the 511\,keV line of $(2$--$3) \times 10^{-3}\,\mathrm{ph\,cm^{-2}\,s^{-1}}$ from five \ac{CN}, between 1995 and 1997.
\citet{Senziani2008} searched for the emission in the $14$--$200$\,keV data from Swift/\ac{BAT} from 24 \ac{CN} by analysing periods from 20\,d before to 20\,d after their optical discoveries, during the first three years of the Swift mission.
Using models of the prompt emission from \citet{Hernanz1999a} and the Monte Carlo simulation method described in \citet{Jean2000}, these authors estimated a rate of detection of two to five \ac{CN} in ten years of the Swift mission.
The lack of claimed detection up to now with Swift/\ac{BAT} suggests that the model would overestimate the flux in that energy band.
\citet{Siegert2018} did not find significant excess at 511\,keV from V5668 Sgr in 2015 with \ac{SPI} data.
They found six excesses in the \ac{SPI}/\ac{ACS} data during the weeks before the discovery of the \ac{CN} but their temporal signatures do not correspond to the expected prompt emission.
Other retrospective searches with \ac{INTEGRAL} did also not yield any significant detections up to the year 2018 (Urruty, priv. comm.).

%
% Conclusions
%
\subsubsection{Conclusions}
Observations of \ac{gray} lines from \acp{CN} can reveal the underlying \ac{WD} type (CO or ONe) and constrain nucleosynthesis during the thermonuclear runaway, enhancing our understanding of Galactic element origins.
The spectral and temporal features of the early 511\,keV line and \ac{gray} continuum ($\leq 511$\,keV) provide insights into the expanding envelope, its early convection, and opacity through the distribution of $\beta^+$-emitters.
A key expected result is the yet-unknown delay between the thermonuclear runaway's peak temperature and the optical maximum.
No \ac{gray} emission has been detected from \acp{CN}, partly due to the lack of a nearby outburst during the 22-year \ac{INTEGRAL} mission.
Current upper limits remain uncertain because of poorly known distances, rates, and spatial distributions of Galactic \acp{CN}.
Most limits exceed theoretical masses by over an order of magnitude.
A new generation of \ac{gray} instruments with at least an order of magnitude improved sensitivity is needed.
The upcoming \ac{COSI} mission, set to monitor the 0.2--5.0\,MeV band from 2027 \citep{Tomsick2024} with a $\sim 25\%$ wide field of view and imaging capability, will efficiently observe prompt emissions from nearby \acp{CN}.
\ac{COSI} could detect the 1275\,keV line from a $1.25\,\mathrm{M_{\odot}}$ ONe nova within $\lesssim 2.5$\,kpc.

%%%%%%%%%%%%%%%%%%%%%%%%%%%%%%%%%%%%%%%%%%%%%%%%%%%%%%%%%%%%%%%%%%%%%%%%%%
% Bibliography
%%%%%%%%%%%%%%%%%%%%%%%%%%%%%%%%%%%%%%%%%%%%%%%%%%%%%%%%%%%%%%%%%%%%%%%%%%

\begin{comment}

\end{comment}

%\end{document}

% SNe in general
\newpage
\subsection{$\gamma$-Ray Lines from Supernovae}\label{sec:SNe}
\vspace{-0.5em}
{\emph{Written by Mark Leising \& Thomas Siegert}}
\vspace{0.5em}\\
The light of \acp{SN} (both \acp{ccSN} and \acp{SNIa}) is powered by radioactive decays \citep{Burbidge1957,Colgate1966}.
Differences in \ac{gray} line observations arise from the yield per event and the progenitor system's density structure and dynamics.
In general, both \acp{SN} produce large amounts of \nuc{Ni}{56} -- up to $0.1\,\mathrm{M_\odot}$ for \acp{ccSN} \citep[e.g.,][]{Andrews2020} and $0.1$--$1.0\,\mathrm{M_\odot}$ for \acp{SNIa} \citep[e.g.,][]{Stritzinger2006}.
The \nuc{Ni}{56} decay chain produces \ac{gray} lines via the \ac{EC} decay of \nuc{Ni}{56} (6.1\,d half-life) to \nuc{Co}{56} at 158 and 812\,keV, and the \ac{EC}+$\beta^+$ decay of \nuc{Co}{56} (77.2\,d half-life) to \nuc{Fe}{56} at 847 and 1238\,keV.
Other isotopes are produced in yields an order of magnitude lower than \nuc{Ni}{56}, yet they still provide key insights into explosion energy and ejecta distribution.
The long-lived \nuc{Ti}{44} is expected in both \acp{SN} types but has been observed only in \acp{ccSN} \citep[e.g.,][]{Weinberger2020}.
\nuc{Ti}{44} (59\,yr half-life) decays by \ac{EC} to \nuc{Sc}{44}, emitting \ac{gray} lines at 67.9 and 78.3\,keV.
\nuc{Sc}{44} decays in 4\,h to stable \nuc{Ca}{44} (mainly via $\beta^+$-decay), emitting a \ac{gray} photon at 1157\,keV.
Measuring these \ac{gray} line fluxes determines the absolute ejecta masses of these isotopes in \acp{SN}.
The \nuc{Ni}{56} to \nuc{Ti}{44} yield ratio links early nucleosynthesis to late-time observations, constraining progenitor systems even centuries after the event \citep[e.g.][]{Magkotsios2010}.
Another Chapter of this volume details theoretical explosion mechanisms, progenitors, yields, and open questions for \acp{ccSN} and \acp{SNIa}.
This section discusses \ac{gray} line observations from \acp{SN} and their leverage to understanding \ac{SN} explosions, nucleosynthesis, Galactic chemical evolution, massive star evolution, and cosmology.
Sec.\,\ref{sec:SNeIa} focuses on thermonuclear \acp{SN}, while Sec.\,\ref{sec:ccSNe} highlights two \acp{ccSN}: SN\,1987A in the \ac{LMC} and the 350\,yr-old \ac{SNR} \ac{CasA} in the Milky Way.

\begin{table}[!h]
  \centering
  %\scriptsize
  \caption{List of the radioactive nuclei, their half life times, decay modes, prominent \ac{gray} lines and branching ratios in \acp{ccSN} and \acp{SNIa} \citep[adapted from][]{Andrews2020}. Detected \ac{gray} lines in \acp{SN} are shown in bold-face. The \ac{SN} type indicates where the isotopes are \emph{most commonly} found (all isotopes are produced in both types to some extent). The contribution to a possible 511\,keV line and annihilation continuum from $\beta^+$-decays is discussed in Sec.\,\ref{sec:511}.}
  \begin{tabular}{lrc|rr|c}
    \hline
    Isotope & Half-life & Decay Mode(s) & $E_{\gamma}$ [keV] & Percentage & SN type \\
    \hline
    \hline
    \isotope[56]{Ni} $\rightarrow$ \isotope[56]{Co} & $6.075$\,d & EC ($\sim 100\%$) & \textbf{158.4}  & 98.8\% & both \\
    & & & 269.5  & 36.5\% \\
    & & & 480.4  & 36.5\% \\
    & & & 750.0  & 49.5\% \\
    & & & \textbf{811.9}  & 86.0\% \\
    & & & 1561.8  & 14.0\% \\
    \hline
    \isotope[56]{Co} $\rightarrow$ \isotope[56]{Fe} & $77.24$\,d & EC ($81.6\%$), $\beta^+$ ($18.4\%$) & \textbf{846.8}  & 99.9\% & both \\
    & & & 1037.8  & 14.1\% \\
    & & & \textbf{1238.3}  & 66.5\% \\
    & & & 1771.4  & 15.4\% \\
    & & & 2598.5  & 17.0\% \\
    \hline
    \hline
    \isotope[47]{Ca} $\rightarrow$ \isotope[47]{Sc} & $4.536$\,d & $\beta^-$ ($\sim 100\%$) & 1297.1  & 67.0\% & ccSNe \\
    \hline
    \isotope[47]{Sc} $\rightarrow$ \isotope[47]{Ti} & $3.3492$\,d & $\beta^-$ ($\sim 100\%$) & 158.4  & 68.3\% & ccSNe \\
    \hline
    \hline
    \isotope[43]{K} $\rightarrow$ \isotope[43]{Ca} & $22.3$\,h & $\beta^-$ ($\sim 100\%$) & 372.8  & 86.8\% & ccSNe\\
    & & & 396.9  & 11.9\% \\
    & & & 593.4  & 11.3\% \\
    & & & 617.5  & 79.2\% \\
    \hline
    \hline
    \isotope[44]{Ti} $\rightarrow$ \isotope[44]{Sc} & $60.0$\,yr & EC ($\sim 100\%$) & \textbf{67.9}  & 93.0\% & both \\
    & & & \textbf{78.3}  & 96.4\% \\
    \hline
    %\hline
    \isotope[44]{Sc} $\rightarrow$ \isotope[44]{Ca} & $4.0$\,h & EC ($5.7\%$), $\beta^+$ ($94.3\%$) &  \textbf{1157.0}  & 99.9\% & both \\
    \hline
    \hline
    \isotope[48]{Cr} $\rightarrow$ \isotope[48]{V} & $21.56$\,d & EC ($98.5\%$), $\beta^+$ ($1.5\%$) &  112.3  & 96.0\% & both \\
    & & & 308.2  & 100.0\% \\
    \hline
    \isotope[48]{V} $\rightarrow$ \isotope[48]{Ti} & $15.97$\,d & EC ($49.6\%$), $\beta^+$ ($50.4\%$) & 983.5  & 100.0\% & both \\
    & & & 1312.1  & 96.0\% \\
    \hline
    \hline
    %\isotope[51]{Cr} $\rightarrow$ \isotope[51]{V} & 4.952  & 12.9\% \\
    %$T_{1/2}=$27.704 d &  &  \\
    %\hline
    %\hline
    \isotope[52]{Mn} $\rightarrow$ \isotope[52]{Cr} & $21.1$\,min & EC ($70.6\%$), $\beta^+$ ($29.4\%$) & 744.2  & 90.0\% & both \\
    & & & 935.5  & 94.5\% \\
    & & & 1434.1  & 1.0\% \\
    \hline
    \hline
    \isotope[59]{Fe} $\rightarrow$ \isotope[59]{Co} & $44.49$\,d & $\beta^-$ ($\sim 100\%$) & 1099.2  & 56.5\% & both \\
    & & & 1291.6  & 43.2\% \\
    \hline
    \hline
    \isotope[57]{Ni} $\rightarrow$ \isotope[57]{Co} & $35.6$\,h & EC ($64.7\%$), $\beta^+$ ($35.3\%$) & 1377.6  & 81.7\% & both \\
    & & & 1919.5  & 12.3\% \\
    \hline
    \isotope[57]{Co} $\rightarrow$ \isotope[57]{Fe} & $271.74$\,d & EC ($\sim 100\%$) & \textbf{122.1}  & 85.6\% & both \\
    & & & \textbf{136.5}  & 10.7\%  \\
    \hline
    \hline
    \isotope[26]{Al} $\rightarrow$ \isotope[26]{Mg} & $7.17 \times 10^5$\,yr & EC ($18.3\%$), $\beta^+$ ($81.7\%$) & \textbf{1808.7}  & 99.8\% & ccSNe\\
    & & & 1129.7 & 2.5\%\\
    \hline
    \hline
    \isotope[60]{Fe} $\rightarrow$ \isotope[60]{Co} & $2.60 \times 10^6$\,yr & $\beta^-$ ($\sim 100\%$) & 58.6  & 99.8\% & ccSNe \\
    \hline
    \isotope[60]{Co} $\rightarrow$ \isotope[60]{Ni} & $5.27$\,yr & $\beta^-$ ($\sim 100\%$) & \textbf{1332.5} & 99.9\% & ccSNe \\
    & & & \textbf{1173.2} & 99.9\% \\
    \hline
    \hline
  \end{tabular}
  \label{tab:decay}
\end{table}

\begin{comment}
In addition to the isotopes from the \nuc{Ni}{56} and the \nuc{Ti}{44} decay chains, several other radioactive isotopes can lead to \ac{gray} line emission in \acp{ccSN} and \acp{SNIa} .
%
Tab.\,\ref{tab:decay} summarises measured and expected \ac{gray} lines from both \ac{SN} types.
%
While only a few of them have been unambiguously detected in \acp{SN} so far, the discrimination power of measuring these lines is invaluable \cite{Prantzos2011b}.
%
Not only do they contribute to our understanding of single \ac{SN} events, but also the Galactic Chemical Evolution in the entire Milky Way by observing the cumulative effects of isotopes with lifetimes much longer than the waiting times between events, such as \nuc{Al}{26} and \nuc{Fe}{60} \cite[e.g.,][see Secs.\,\ref{sec:Al26} \& \ref{sec:Fe60}]{Timmes1995,Prantzos2008,Kobayashi2020,Vasini2022}.
%
Because \acp{gray} from the \nuc{Ni}{56} and \nuc{Ti}{44} decay chains have been studied in large detail within the last 50 years, we detail out expectations of their fluxes before continuing to the individual \ac{SN} types.
\end{comment}
%
In addition to the \nuc{Ni}{56} and \nuc{Ti}{44} decay chains, other radioactive isotopes can produce \ac{gray} line emission in both \acp{ccSN} and \acp{SNIa}.
Tab.\,\ref{tab:decay} summarises the measured and expected \ac{gray} lines from both \ac{SN} types.
Although only a few \ac{gray} lines have been unambiguously detected, their measurements are invaluable for deciphering \ac{SN} physics \citep{Prantzos2011b}.
They enhance our understanding of individual \ac{SN} events and Galactic chemical evolution by tracking long-lived isotopes such as \nuc{Al}{26} and \nuc{Fe}{60} \citep[e.g.,][see Secs.\,\ref{sec:Al26} \& \ref{sec:Fe60}]{Timmes1995,Prantzos2008,Kobayashi2020,Vasini2022}.
Since \acp{gray} from the \nuc{Ni}{56} and \nuc{Ti}{44} decay chains have been extensively studied over the past 50 years, we first detail out their expected fluxes before discussing individual \ac{SN} types.

\begin{comment}
According to Eq.\,(\ref{eq:decay_luminosity}), the photon flux, $F_{56}$, of a mass, $M_{56}$ of \nuc{Ni}{56} without considering absorption and kinematics in a \ac{SN} envelope, of the major 847\,keV line, at a distance $d$ from the observer is given by
%
\begin{equation}
        F_{56} = 9.3 \times 10^{-3} \times \left(\frac{d}{\mathrm{1\,Mpc}}\right)^{-2} \times \left(\frac{M_{56}}{0.5\,\mathrm{M_{\odot}}}\right) \times \exp\left(-\frac{\Delta t}{\tau_{56}}\right)\,\mathrm{ph\,cm^{-2}\,s^{-1}}\mathrm{.} 
        \label{eq:F_56}
\end{equation}
%
This equation is valid for \acp{SNIa} times typically around $40$--$100$\,d \cite[e.g.,][]{The2014} after the explosion when the envelope is mostly optical thin to these \acp{gray}, and around $1$--$2$\,yr after the explosion for a \ac{ccSN}.
%
This leads to considerably smaller fluxes around the time after the \ac{gray} line peak of $5 \times 10^{-3} \times [d/(1\,\mathrm{Mpc})]^{-2}\,\mathrm{ph\,cm^{-2}\,s^{-1}}$ around day 70 for normal \acp{SNIa} , and $4 \times 10^{-3} \times [d/(50\,\mathrm{kpc})]^{-2}\,\mathrm{ph\,cm^{-2}\,s^{-1}}$ for unobscured \acp{ccSN}.
%
While for \acp{SNIa} , this is a relatively good approximation at later times, the 847\,keV line fluxes for \acp{ccSN} certainly have to be adjusted for the envelopes opacity \cite[e.g.][]{Fu1989,Grebenev1987,Palmer1993}.
\end{comment}
%
According to Eq.\,(\ref{eq:decay_luminosity}), the photon flux, $F_{56}$, of a mass, $M_{56}$, of \nuc{Ni}{56} (for the major 847\,keV line) at a distance, $d$, is given by 
\begin{equation}
        F_{56} = 9.3 \times 10^{-3} \times \left(\frac{d}{\mathrm{1\,Mpc}}\right)^{-2} \times \left(\frac{M_{56}}{0.5\,\mathrm{M_{\odot}}}\right) \times \exp\left(-\frac{\Delta t}{\tau_{56}}\right)\,\mathrm{ph\,cm^{-2}\,s^{-1}}\mathrm{,} 
        \label{eq:F_56}
\end{equation}
which applies around 40--100\,d post-explosion for \acp{SNIa} (when the envelope is mostly optically thin) and around 1--2\,yr for \acp{ccSN}.
This results in considerably lower fluxes after the \ac{gray} line light curve peak, of $5 \times 10^{-3} \times [d/(1\,\mathrm{Mpc})]^{-2}\,\mathrm{ph\,cm^{-2}\,s^{-1}}$ around day 70 for typical \acp{SNIa} and $4 \times 10^{-3} \times [d/(50\,\mathrm{kpc})]^{-2}\,\mathrm{ph\,cm^{-2}\,s^{-1}}$ for unobscured \acp{ccSN}.
While this approximation holds for \acp{SNIa} at later times, the 847\,keV fluxes for \acp{ccSN} must be corrected for the envelope opacity \citep[e.g.][]{Fu1989,Grebenev1987,Palmer1993}.

In both cases, however, the simplest approximation includes a time-, that is, radius-dependent optical depth, $\tau_\gamma(t)$, that takes into account the overlaying total ejecta mass, $M_{\rm ej}$, the expansion velocity of the \ac{SN} envelope, $v_{\rm ej}$, and a \ac{SN}-specific and in general energy-dependent opacity, $\kappa_\gamma$, in units of $\mathrm{cm^2\,g^{-1}}$ \citep{Grebenev1987}.
Then, the optical depth can be approximated as
\begin{equation}
    \tau_\gamma(t) = \frac{\kappa_\gamma M_{\rm ej}}{4\pi v_{\rm ej}^2 t^2}\mathrm{,}
    \label{eq:optical_depth_SN}
\end{equation}
where the general trend, $\tau \propto t^{-2}$, exists in all \ac{SN} models.
By applying the optical depth to the general flux function, Eq.\,(\ref{eq:F_56}), we find
\begin{equation}
\resizebox{0.92\linewidth}{!}{$
    F_{56}(t) = F_{56}^{\rm unabs}(t) \times \exp\left[-636 \left(\frac{\kappa_\gamma}{0.03\,\mathrm{cm^2\,g^{-1}}}\right)\left(\frac{M_{\rm ej}}{1\,\mathrm{M_\odot}}\right) \left(\frac{v_{\rm ej}}{10000\,\mathrm{km\,s^{-1}}}\right)^{-2} \left(\frac{\Delta t}{1\,\mathrm{d}}\right)^{-2} \right]\mathrm{,}
    $}
    \label{eq:opacity_flux_SN}
\end{equation}
where typical values are given in the equation for \acp{SNIa}.
In the case of \acp{ccSN}, the opacity may range within $0.02$--$0.05\,\mathrm{cm^2\,g^{-1}}$, the ejecta mass is much larger with $5$--$20\,\mathrm{M_\odot}$, and the ejecta velocity somewhat smaller $5000$--$10000\,\mathrm{km\,s^{-1}}$.
From these simple considerations, it can be seen that the \acp{ccSN} \ac{gray} line flux can be suppressed by one or two orders of magnitude even within the first year after the explosion, whereas \acp{SNIa} become transparent to \acp{gray} after a few weeks.
The absorbed photons will be re-processed in the envelope and will power the UVOIR lightcurve (see Sec.\,\ref{sec:ccSNe}).
It should be noted that this model can fall short in explaining actual observations of, for example, SN\,1987A \citep[e.g.,][]{Leising1990}.

Considering \nuc{Ti}{44}, the unabsorbed flux is calculated as before,
\begin{equation}
        F_{44} = 7.6 \times 10^{-4} \times \left(\frac{d}{\mathrm{3.3\,kpc}}\right)^{-2} \times \left(\frac{M_{44}}{10^{-4}\,\mathrm{M_{\odot}}}\right) \times \exp\left(-\frac{\Delta t}{\tau_{44}}\right)\,\mathrm{ph\,cm^{-2}\,s^{-1}}\mathrm{,} 
        \label{eq:F_44}
\end{equation}
with typical values given for the \ac{SNR} \ac{CasA}.
For \ac{CasA}'s age of about 350\,yr, the expected flux is on the order of $10^{-5}\,\mathrm{ph\,cm^{-2}\,s^{-1}}$, consistent with previous measurements \citep[e.g.,][see also Sec.\,\ref{sec:ccSNe}]{Iyudin1994,The1996,Vink2001,Renaud2006,Grefenstette2014,Siegert2015,Weinberger2020}.
Within the first two years post-explosion, not all three \ac{gray} lines from the \nuc{Ti}{44} decay chain are visible:
The 1157\,keV line from \nuc{Sc}{44} appears around the same time as the 847 and 1238\,keV \nuc{Co}{56} lines, but the 68 and 78\,keV ``hard X-ray lines'' emerge only about ten years later due to the high opacity ($\kappa_\gamma \approx 1\,\mathrm{cm^{2}\,g^{-1}}$).
However, the expected rise in the low-energy lines has not yet been observed in any \ac{SN} explosion.
Since the late 1960s, the \ac{CasA} \nuc{Ti}{44} \ac{gray} line flux has decreased by a factor of two.
Different instruments, epochs, and sensitivities to various lines (68 \& 78\,keV vs. 1157\,keV) make this decay difficult to probe \citep[see, e.g.,][]{Siegert2015}.
At late times, although line ratios should approach unity, the stable nucleus \nuc{Ca}{44} may be re-excited by \acp{LECR}, enhancing its flux \citep[][see also Sec.\,\ref{sec:LECRs}]{Siegert2015}.

Because radioactive nuclei may decay through several competing channels, the total decay rate is therefore the sum of the individual processes:
\begin{equation}
    \lambda_{\mathrm{tot}} = \lambda_{\mathrm{EC}} + \lambda_{\beta^{+}} + \lambda_{\beta^{-}} + \lambda_{\gamma} + \cdots
    \label{eq:changed_lifetime}
\end{equation}
Thus, the mean lifetime is given by $\tau_{\mathrm{tot}} = 1/\lambda_{\mathrm{tot}}$, so that if one channel is suppressed, the total lifetime increases correspondingly.
In astrophysical plasmas the \ac{EC} component is particularly sensitive to the ionisation state as it requires the presence of bound K–shell\footnote{Also L-shell or free electrons can lead to \ac{EC} on nuclei but at a much smaller probability.} electrons,
\begin{equation}
    \lambda_{\mathrm{EC}} = \lambda_{\mathrm{EC}}^{0}\, f_{e}\mathrm{.}
\end{equation}
Here $\lambda_{\mathrm{EC}}^{0}$ is the laboratory \ac{EC} decay rate and $f_{e}$ is the fraction of occupied inner shells, which can be strongly reduced in highly ionised environments \citep{Motizuki2004_lifetimes}.

For \nuc{Ti}{44} with decays almost entirely through \ac{EC}, even modest stripping of inner electrons can decrease $\lambda_{\mathrm{EC}}$ significantly and thus extend the effective lifetime in young \acp{SNR}.
This effect is relevant for SN\,1987A and \ac{CasA} (see Secs.\,\ref{sec:SN1987A} \& \ref{sec:CasA}), where shock heating and incomplete recombination can maintain ions in highly ionised states, thereby reducing the \ac{gray} and hard X–ray flux expected from the \nuc{Ti}{44} decay chain.
Such environmental dependence introduces uncertainties when comparing theoretical yields to observed line intensities, especially at early times when ionisation is highest.

The same mechanism could influence other isotopes that have an \ac{EC} decay branch, such as \nuc{Al}{26} with its $\sim 18\%$ \ac{EC} component.
In hot regions of the \ac{ISM} or in shocked ejecta, partial ionisation may reduce the \ac{EC} contribution to the total decay rate of \nuc{Al}{26}, increasing its effective lifetime relative to laboratory conditions by at most 22\%.
%

% SNe Ia
%\newpage
\subsubsection{$\gamma$-Ray Lines from the \nuc{Ni}{56} Decay Chain in Type Ia Supernovae}\label{sec:SNeIa}
\paragraph{Measurements of \nuc{Ni}{56} Decay $\gamma$-Rays}
\citet{Clayton1969} recognised decades ago the potential of \ac{gray} line observations for understanding \ac{SN} explosions.
This included especially the lines of the decay of abundant \nuc{Ni}{56} and \nuc{Co}{56}, \nuc{Co}{57}, longer-lived \nuc{Ti}{44} and \nuc{Sc}{44}, as well as \nuc{V}{48} and \nuc{Cr}{48} (see Tab.\,\ref{tab:decay}).
At the time, they were considering ``Type I'' \acp{SN}, before the various cases and progenitors were recognised.
The $A = 56$ isotopes are clearly produced in large amounts in thermonuclear -- or Type Ia -- supernovae (\acp{SNIa}), as they power much of the luminous display, especially after a few weeks of the initial explosion.
Although the \nuc{Co}{56}, \nuc{Sc}{44}, and \nuc{Ti}{44} \ac{gray} line predictions were eventually realised (the latter two only in \acp{ccSN} as of yet; see Sec.\,\ref{sec:ccSNe}), decades of attempts to measure these \ac{gray} lines yielded only upper limits.
One of the more surprising upper limits was obtained for \nuc{Co}{56} emission from SN\,1986G in the nearby galaxy Centaurus\,A \citep{Matz1990}.

\begin{figure}[!h]
	\centering
	\includegraphics[width=1.0\linewidth]{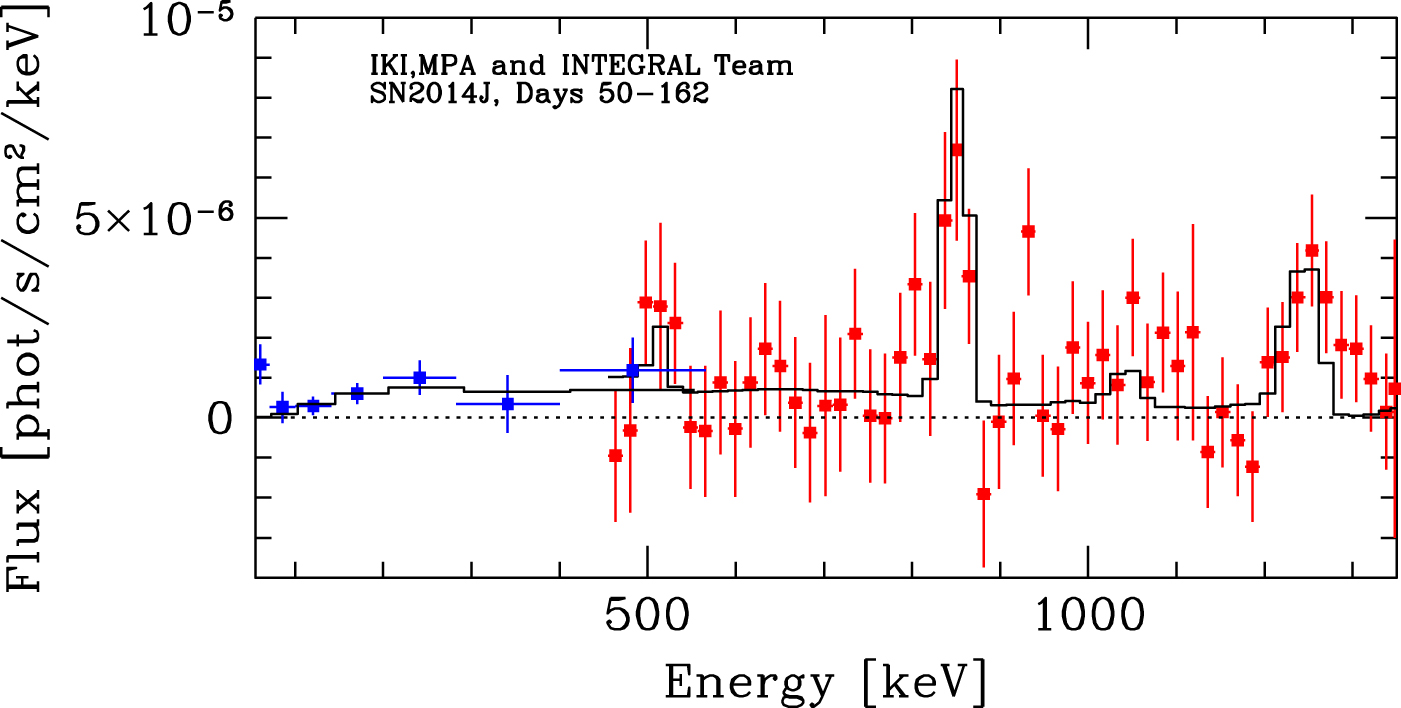}
	\caption{Integrated spectrum of SN\,2014J measured by \ac{INTEGRAL}/\ac{SPI} and \ac{IBIS} \citep[from][]{Churazov2014}. Two \nuc{Co}{56} lines and the scattered continuum are apparent. This is the first \ac{SNIa} firmly detected in \acp{gray}.}
	\label{one}
\end{figure}

\begin{figure}[!t]
    \centering
    \includegraphics[width=0.7\linewidth]{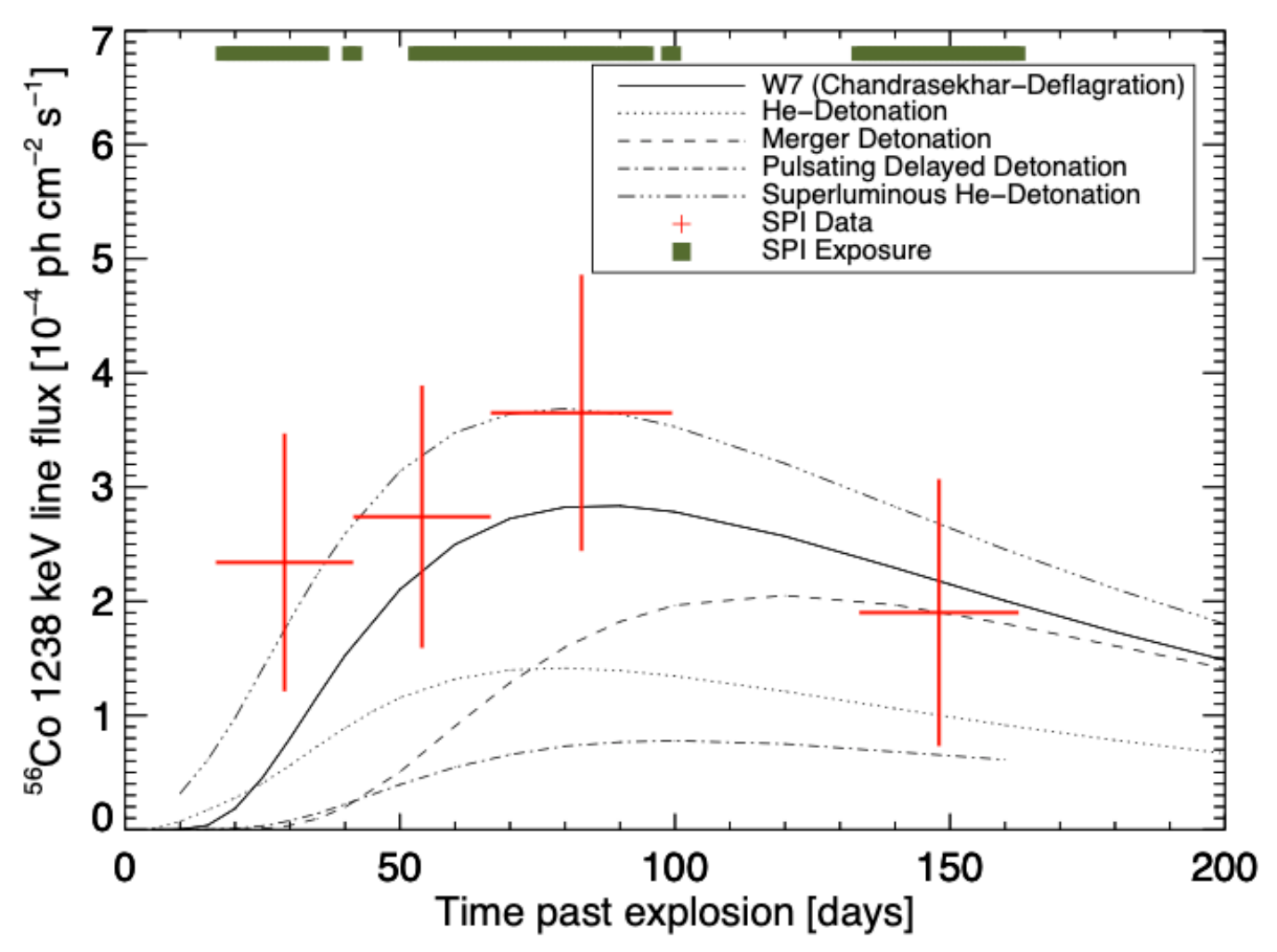}
    \caption{SN\,2014J signal intensity variations for the 847\,keV line (red data points; note that the y-axis label should be 847\,keV, not 1238\,keV) as observed with \ac{SPI} \citep[from][]{Diehl2015}. The fluxes were derived by Gaussians fitted to the spectra of the \nuc{Co}{56} line at four independent epochs. For reference, several candidate model light curves are shown from \citet{The2014}.}
    \label{fig:Diehl2015_lc}
\end{figure}

More distant \acp{SNIa} were observed with instruments with somewhat better sensitivity, but without clear \ac{gray} line detections \citep{Lichti1994,Leising1995,Georgii2002,Isern2013}.
SN\,2014J, at a distance of 3.3\,Mpc in the galaxy M82, was detected in the \nuc{Co}{56} lines at 847 and 1238\,keV at a few $\sigma$ significance in each of four, month-long intervals \citep[][see also Figs.\,\ref{one} \& \ref{fig:Diehl2015_lc}]{Churazov2015,Diehl2015}. 
Weak features corresponding to the \nuc{Ni}{56} line energies at 158 and 812\,keV were also measured from the direction to SN\,2014J, and interpreted as \nuc{Ni}{56} propelled ahead of the main ejecta, perpendicular to the \ac{LoS} \citep{Diehl2014}.
The \nuc{Co}{56} \acp{gray} from that same material are not evident. 
An alternative study by \citet{Isern2016} also found an early \nuc{Ni}{56} signal at a similar flux level, but with lines broadened and redshifted.
It was concluded that the measured signals of the \nuc{Ni}{56} decay lines were probably originating from SN\,2014J because the ejecta masses were estimated to be within each others uncertainties, but the geometrical scenarios have to be left indecisive.

%\paragraph{General Expectations of Line Fluxes}

%flux, masses, etc.

\paragraph{Utilising $\gamma$-Ray Lines to Study Supernova Interiors}
Many calculations of the \nuc{Ni}{56}--\nuc{Co}{56} \ac{gray} line light curves and spectra have been made for many hydrodynamic models of \acp{SNIa} of all types \citep[e.g.,][]{Nomoto1984,Burrows1990,Mueller1991,Hoeflich1998,Gomez-Gomar1998b,Milne2004,Sim2008,Maeda2012,Summa2013,The2014}.
Monte Carlo \ac{gray} transport calculations have been the method of choice and are quite accurate.
In the past, limited Monte Carlo samples yielded limited precision calculations, but that is rarely an issue nowadays.
The resulting light curves for a suite of different models cover the flux-time plane quite thoroughly, making interpretation of observed light curves of limited precision quite challenging, given the great variety of models \citep[see, e.g.,][]{Churazov2015,The2014,Diehl2015}.
Only exquisitely precisely measured line profiles might allow one to favour one model, or one progenitor type or total ejecta mass, based on simple comparisons to spectra or light curves.
However, global comparisons of many observations tied together by the physics of \ac{gray} escape offer better prospects:

The \ac{gray} line transport through the \ac{SN} ejecta is comparably simple, even though that of the scattered continuum and the energy deposition are not.
These photons interact overwhelmingly through Compton scattering on all electrons, bound and free, equally.
Once a photon scatters, it is removed from the line, as in an absorption process.
Most \acp{SNIa} are expected to be quite spherically symmetric, with net polarization of continuum photons usually significantly less than 1\%, indicating geometric asymmetry of the electron scattering photosphere of $\lesssim 10\%$ \citep{Wang2008,Porter2016,Cikota2019,Hoeflich2023}.
Homologous expansion of the ejecta is obtained just seconds after the explosion \citep{Nomoto1984,Roepke2005}, so deeper layers are progressively uncovered over months.
While there might be aspherical effects of the nuclear burning that set in in the first seconds, calculating the mean escape of \ac{gray} lines versus time and energy at later times is rather straightforward.

\begin{figure}[!t]
    \centering
    \includegraphics[width=1.0\linewidth]{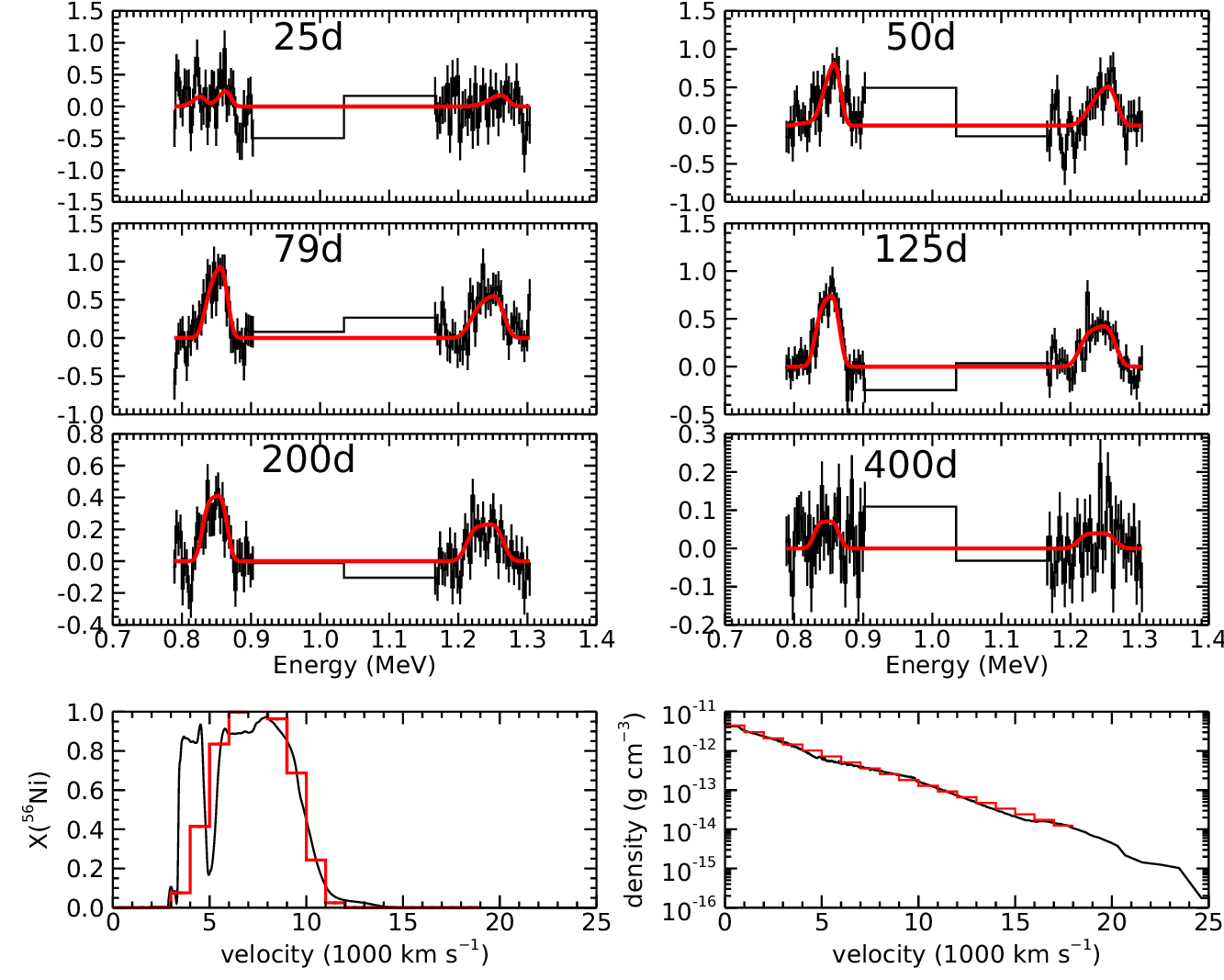}
    \caption{Example fit of the density and \nuc{Ni}{56} mass fraction distributions to line profiles simulated from the model \texttt{7p0z22d20\_27} \citep{Diamond2015} at a distance if 20\,Mpc. This is for an instrument with energy resolution $\Delta E/E = 1\%$ and $3\sigma$ sensitivity to narrow lines of $10^{-6}\,\mathrm{ph\,cm^{-2}\,s^{-1}}$ within 1\,Ms. Twenty observation periods over 400\,d are fitted -- shown are six snapshots. The fitted \acp{SNIa} parameters and resulting line intensity profiles are shown in red. In the bottom panels, the black curves show the model \texttt{7p0z22d20\_27} distributions from which the simulated data were derived \citep[from][]{Leising2022}.}
    \label{two}
\end{figure}

Taking advantage of the assumptions of spherical symmetry and homologous expansion, \citet{Leising2022} showed that the synthesised \nuc{Ni}{56} mass, and the otherwise difficult to obtain total ejecta mass and kinetic energy, could be extracted with good precision from measured \ac{gray} line light curves and/or line profiles.
Further, for sufficiently precise \ac{gray} measurements, the distribution of \nuc{Ni}{56} as a function of the velocity in the ejecta, and density profile parameters can be derived.
Total ejecta mass determinations to $0.1\,\mathrm{M_{\odot}}$ could exclude Chandrasekhar-mass explosions, or merger models, or instead identify them as the progenitor systems.
Total kinetic energy and \nuc{Ni}{56} mass measurements would clarify intermediate mass element nucleosynthesis and gravitational binding energy.
\nuc{Ni}{56} distribution measurements would elucidate nuclear burning sites and flame propagation.

The \ac{gray} line escape calculations are based on the well known fact that a thin expanding spherical shell emits a flat topped line profile (see Sec.\,\ref{sec:Doppler_broadening}).
Then an overlying spherical shell absorbs at all energies according to its optical depth along the observer's line of sight to a source element, and an interior shell absorbs only part of the redshifted side of the line.
For multiple shells, one adds the optical depths at each line of sight velocity.
This is the same for lines of different energies, except for the different Compton optical depths.
As time progresses, the source functions decay -- or, for \nuc{Co}{56}, is fed by the parent decay and decays itself.
The optical depths of all shells fall with time, as $t^{-2}$.
Given the \nuc{Ni}{56} abundance and total electron density versus radial velocity, all \ac{gray} line profiles are easily obtained for all times to first order.
These can be compared directly to a series of measured spectra, or integrated over each line to be compared to line flux data.
While it is not possible to invert the transfer process to get the \nuc{Ni}{56} abundance and density profiles, it is possible to iteratively vary them to fit the observational data quite well.

The precision with which the \ac{SN} properties can be determined is limited entirely by the measurement uncertainties, at least for most events for the foreseeable future.
For these photon-limited measurements, the nearest \acp{SNIa} will be best understood:
For example, for a future wide-field instrument with narrow-line, $3\sigma$ flux sensitivity of $10^{-6}\,\mathrm{ph\,cm^{-2}\,s^{-1}}$ within 1\,Ms of observation time, and energy resolution $\Delta E / E = 1\%$, we can expect to measure the \nuc{Ni}{56} mass to $0.02\,\mathrm{M_{\odot}}$ accuracy, ejecta masses to $0.1\,\mathrm{M_{\odot}}$, and kinetic energies to $0.3 \times 10^{51}\,\mathrm{erg}$, for a normally bright \ac{SNIa} similar to typical delayed detonation models, for a distance of 20\,Mpc \citep{Leising2022}.
The distribution of \nuc{Ni}{56} with radial velocity can also be well recovered in such a measurement.
An example of such a measurement is shown in Fig.\,\ref{two} for a simulated event.

Even for instruments with similar line sensitivity, but several percent energy resolution, such as scintillators, which cannot determine the line profiles to high accuracy, the integrated line flux measurements can still provide nearly the same determination of \acp{SNIa} parameters.
In this case, good coverage of the light curve over at least several months is essential, and suggests that wide-field and/or scanning instruments are necessary.
On the other hand, if \ac{gray} optics can be deployed to provide large collecting area for relatively small detectors, as in other wavelength bands, excellent sensitivity could be achieved.
If combined with good energy resolution, just a few pointings before and after the \ac{gray} peak could reveal the underlying \ac{SNIa} characteristics.

\paragraph{Other $\gamma$-Ray Lines from Type Ia Radioactivity}
Other potentially detectable isotopes from \acp{SNIa} include \nuc{Co}{57} and \nuc{Ti}{44}, which offer key insights into nuclear burning conditions.
\nuc{Co}{57} (half-life 272\,d) is produced as \nuc{Ni}{57} (half-life 36\,h) alongside \nuc{Ni}{56}.
It significantly powers late light curves, and the \nuc{Ni}{57}/\nuc{Ni}{56} ratio is at or above the solar value (0.023) in many models \citep[e.g.,][]{Tiwari2022}.
Measuring \nuc{Ni}{57} from late light curves is challenging and model-dependent.
The \nuc{Co}{57} 122\,keV \ac{gray} line can directly diagnose \ac{SNIa} burning in nearby events.
For example, a \ac{SNIa} at 1\,Mpc ejecting $0.5\,\mathrm{M_{\odot}}$ of \nuc{Ni}{56} with a solar \nuc{Ni}{57}/\nuc{Ni}{56} ratio would yield a flux of $2.4 \times 10^{-5}\,\mathrm{ph\,cm^{-2}\,s^{-1}}$ at one year post-explosion.
Although \nuc{Ti}{44} and \nuc{Sc}{44} \ac{gray} lines are detected in \acp{ccSN}, they remain unconfirmed in thermonuclear remnants.
An excess at 60--85\,keV, possibly from the \nuc{Ti}{44} decay (68 \& 78\,keV) in Tycho's \ac{SNR} \citep{Troja2014}, is unverified by other studies \citep[e.g.,][]{Lopez2015,Weinberger2020}.
The broad range of \nuc{Ti}{44} yields in \ac{SNIa} models makes it a strong diagnostic.
Typically, yields are below $10^{-4}\,\mathrm{M_{\odot}}$ in Chandrasekhar models but can reach up to $10^{-3}\,\mathrm{M_{\odot}}$ in surface helium detonation or \ac{WD} merger models \citep{Woosley1986,Leung2020,Roy2022}.
Besides a few young \ac{SNIa} remnants, other undetected \acp{SNIa} might show measurable \nuc{Ti}{44}.
For instance, $10^{-3}\,\mathrm{M_{\odot}}$ of \nuc{Ti}{44} at 10\,kpc could yield $3 \times 10^{-6}\,\mathrm{ph\,cm^{-2}\,s^{-1}}$ per line even after 500\,yr.
%after the explosion.
%

%
For \nuc{V}{48} (\ac{gray} lines at 983 and 1312\,keV), only a few studies have searched its emission in nearby remnants.
\citet{Panther2021}, following \citet{Sim2012}, set an upper limit on the \nuc{Cr}{48} production in SN\,2014J of $\lesssim 0.1\,\mathrm{M_{\odot}}$, roughly an order of magnitude above expectations.
The \nuc{V}{48} lines might reach $10^{-4}\,\mathrm{ph\,cm^{-2}\,s^{-1}}$ in \acp{SNIa} at approximately 1\,Mpc.
These lines could be detected by next-generation telescopes with proper pointing strategies or a large fields of view (see Sec.\,\ref{sec:future_SNe}).

\paragraph{$\gamma$-Ray Line Contributions to the Cosmic $\gamma$-Ray Background}
We refer to the \ac{CGB} as the diffuse, all-sky, and nearly isotropic emission observed between photon energies of $\sim 100\,\mathrm{keV}$ and several tens of MeV.
Measurements by instruments such as HEAO\,1 \citep{Kinzer1997_CGB,Gruber1999_CGB}, \ac{SMM} \citep{Watanabe2000_CGB}, \ac{COMPTEL} \citep{Weidenspointner2000_CGB}, and \ac{INTEGRAL} \citep{Churazov2007_CGB,Tuerler2010} have shown that the spectrum in this range consists of a smoothly varying continuum that cannot be fully explained by resolved sources alone, implying a substantial contribution from faint or distant populations \citep{1993ApJ...403...32T}.
The origin of the MeV \ac{CGB} remains one of the central questions in high-energy astrophysics because the putative source types trace nucleosynthesis, compact-object formation, cosmic star formation, and the cosmological evolution.
In general, several astrophysical source types may contribute to the \ac{CGB} in this band:
Potential candidates include active galactic nuclei (especially Seyfert galaxies), star-forming galaxies \citep[e.g.,][]{Ueda2003_CGB,Inoue2014_CGB}, \acp{SNIa} \citep[e.g.,][]{1993ApJ...403...32T}, \acp{ccSN} \citep[e.g.,][]{Chugai2023_CGB}, compact object mergers such as \acp{BNSM} \citep[][]{Ruiz-Lapuente2020_CGB}, radioactive isotopes ejected by stellar populations \citep{Lacki2014_CGB}, and more exotic origins such as \ac{DM} annihilation or decay \citep[e.g.,][]{Iguaz2021_CGB}.
Many of these source types produce \ac{gray} lines or Comptonised continuum emission that are redshifted and blended into a diffuse background when integrated over cosmic time.

\acp{SNIa} are of particular interest because they produce large quantities of radioactive \nuc{Ni}{56} and \nuc{Co}{56}, whose \ac{gray} lines dominate the spectrum of young \ac{SNIa}.
Since \acp{SNIa} serve as standardisable candles for cosmology \citep{Arnett1982_SNIa,Phillips1993_SNIa}, their rate evolution is well-studied and their contribution to the \ac{CGB} directly encodes the cosmic history of stellar evolution and chemical enrichment \citep{1993ApJ...403...32T}.
The emitted \acp{gray} undergo Compton scattering and absorption in the ejecta and intergalactic medium, and when redshifted across the Hubble flow they contribute to the MeV-band \ac{CGB}.
In general, the spectrum of the \ac{CGB} from a cosmological source population is
\begin{equation}
    I(E) = \frac{c}{4\pi} \int_0^{z_{\rm max}} \frac{R(z)\,L[E(1+z)]}{H(z)\,(1+z)} \, dz\mathrm{,}
    \label{eq:CGB_calc}
\end{equation}
where $R(z)$ is the comoving volumetric rate density of the source population, $L(E)$ is the rest-frame photon luminosity spectrum per event, and $H(z)$ is the Hubble expansion function \citep{Timmes1993_CGB}.
Eq.\,(\ref{eq:CGB_calc}) applies to any transient population and makes explicit how the redshift distribution and intrinsic spectral models govern the \ac{CGB}.

\begin{figure}[!ht]
    \centering
    \includegraphics[width=0.70\linewidth]{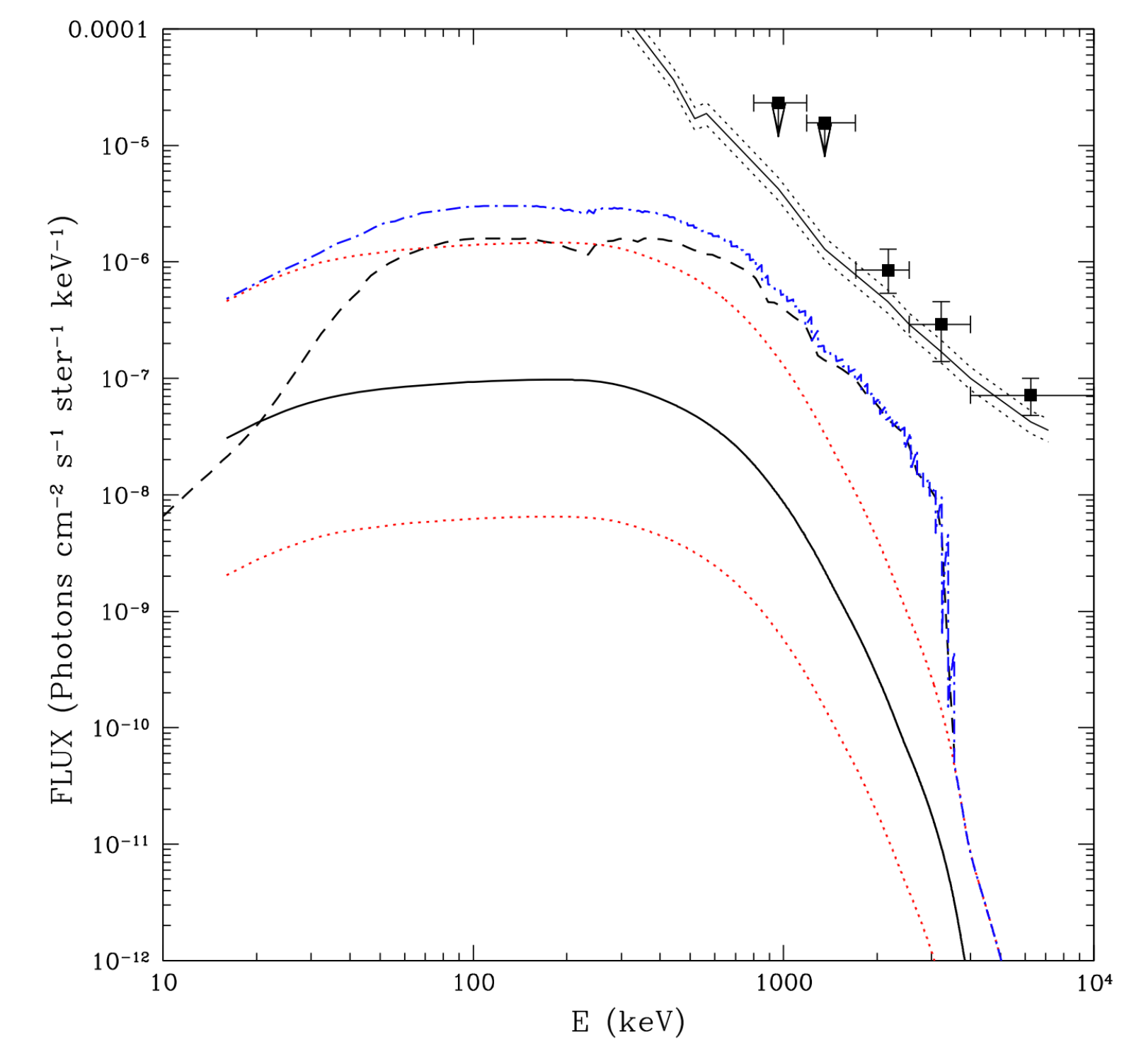}
    \caption{Comparison of the \acp{SNIa} contribution \citep[black dashed lines][]{Ruiz-Lapuente2016_CGB} to the \ac{CGB} (black data points) to the total emission expected from \acp{BNSM} (red dash-dotted), including both the kilonova and late-time remnant phases. The red dotted curves mark the corresponding upper and lower bounds arising from uncertainties in the \ac{BNSM} ejecta masses \citep[from][]{Ruiz-Lapuente2020_CGB}.}
    \label{fig:CGB}
\end{figure}

Historically, the first quantitative estimate of the \ac{SNIa} contribution to the \ac{CGB} was carried out by \citet{1993ApJ...403...32T}, who modelled the \ac{gray} line emission from \nuc{Co}{56} decay and folded these spectra through cosmic \ac{SNIa} rate histories available at the time.
Their result indicated that \acp{SNIa} could provide a non-negligible fraction of the MeV background, although uncertainties in star formation and \ac{SNIa} delay times left a large allowed range.
Subsequent improvements in \ac{SNIa} modelling led to updated \ac{CGB} predictions \citep{Watanabe1999_CGB}.
With improved radioactive-transfer simulations and updated cosmic rate measurements, the authors show that \acp{SNIa} alone likely fall short of explaining the entire MeV background.
Later refinements \citep{Ahn2005_CGB} incorporated better constraints on \ac{SNIa} delay-time distributions and emphasised that contributions peak around a few MeV but still remain below the observed intensity.
Follow-up analyses \citep{Lien2012_CGB} further strengthened this conclusion by combining new \ac{SNIa} rate measurements with modern cosmological parameters.
Recent reviews of \ac{SNIa} progenitors and their environmental signatures \citep{Ruiz-Lapuente2016_CGB} further emphasise that the diversity of explosion channels and circumstellar conditions directly impacts the \ac{gray} signal and thus the integrated \ac{CGB} contribution.
We show the example from \citet{Ruiz-Lapuente2016_CGB,Ruiz-Lapuente2020_CGB} of \ac{SNIa} and other \ac{gray} lines contributions to the \ac{CGB} in Fig.\,\ref{fig:CGB}

In addition to \acp{SNIa}, \acp{ccSN} also produce \acp{gray} through radioactive isotopes such as \nuc{Ni}{56}, \nuc{Co}{56}, \nuc{Ti}{44}, and longer-lived \nuc{Al}{26} and \nuc{Fe}{60}, plus potential $r$-process elements.
However, the \ac{ccSN} contribution to the MeV \ac{CGB} is expected to be smaller than that of \acp{SNIa} due to the heavier obscuration by massive stellar envelopes, which significantly delays and reduces \ac{gray} escape.
Nevertheless, their high rate means \acp{ccSN} still form a baseline contribution to the MeV background across cosmic time \citep{Chugai2023_CGB}.

Compact-object mergers (e.g., \acp{BNSM}) generate radioactive ejecta responsible for kilonova emission, and these ejecta produce \ac{gray} lines from $r$-process nuclei (see Sec.\,\ref{sec:NSM}).
The bulk of this emission peaks in the X-ray and soft \ac{gray} domain and might be highly absorbed.
Recent work by \citet{Ruiz-Lapuente2020_CGB} suggests that \acp{BNSM} may still contribute a small (few percent) amount to the MeV \ac{CGB}.
These contributions remain highly uncertain because $r$-process heating, opacity, and ejecta composition vary significantly between merger models.

Possible contributions from local environments must also be considered in this context.
The Local Bubble \citep{Breitschwerdt1996_LB} contains unresolved radioactive-emission sites such as from \nuc{Al}{26} (1809 \& 511\,keV) or \nuc{Fe}{60} (1173 \& 1332\,keV) whose contributions may be measurable already with next generation telescopes \citep{Schulreich2023,Siegert2024}.
Similarly, a putative \ac{DM} halo could contribute a large amount of positrons from its decay or annihilation, leading to a local 511\,keV line \citep{Iguaz2021_CGB}.
This line would contribute to the isotropic emission and imprint on top of the cosmological \ac{CGB}, similar to the nuclear lines.
Finally, the neutron capture line onto protons at 2224\,keV throughout cosmic time may serve as a `hydrogen calorimeter' if the thermalisation time of spallation neutrons is short enough \citep[][see also Sec.\,\ref{sec:outlook_deuterium}]{Ramaty1980}.
Together, these components highlight that while \acp{SNIa} remain the dominant and best-understood candidate among stellar explosions, a combination of multiple astrophysical and possibly exotic processes is required to explain the full MeV-band \ac{CGB}.
%

%\input{MeV_sniagammas_leising.bbl}

%\bibliography{../main}  
%\bibliographystyle{aasjournal}      



%\end{document}

% ccSNe
%\newpage
% Introduction
\subsubsection{$\gamma$-Ray Lines from Core-Collapse Supernovae}\label{sec:ccSNe}
While there are at least 300 \acp{SNR} inside the Milky Way \citep{Green2019}, only one has been identified as being a \ac{gray} line source so far: the 350\,yr old \ac{SNR} \ac{CasA}.
Several other candidates would match reasonable ages and distances to observe longer-lived isotopes, such as \nuc{Ti}{44}, but which, so far, only resulted in upper limits.
Prompt emission from the \nuc{Ni}{56} decay chain has never been observed in a \ac{SN} inside the Milky Way, but in one of its satellite galaxies: SN\,1987A in the \ac{LMC}.
In SN\,1987A, also the \nuc{Ti}{44} lines have been detected \citep{Boggs2015}, making it a unique object to study over decades.

In this Section, we will discuss the two cases of SN\,1987A (Sec.\,\ref{sec:SN1987A}) and \ac{CasA} (Sec.\,\ref{sec:CasA}), and list the most recent upper limits of other interesting \acp{SNR} according to \citet{Weinberger2020} in Tab.\,\ref{tab:44Ti_limits}.
In the light of these limits, it is clear that more sensitive instruments are required to solve the mystery of why there is only one \nuc{Ti}{44} in the Milky Way so far \citep[e.g.,][see also Sec.\,\ref{sec:future_SNe}]{The2006}.

\begin{figure}[!ht]
    \centering
    \includegraphics[width=0.35\linewidth]{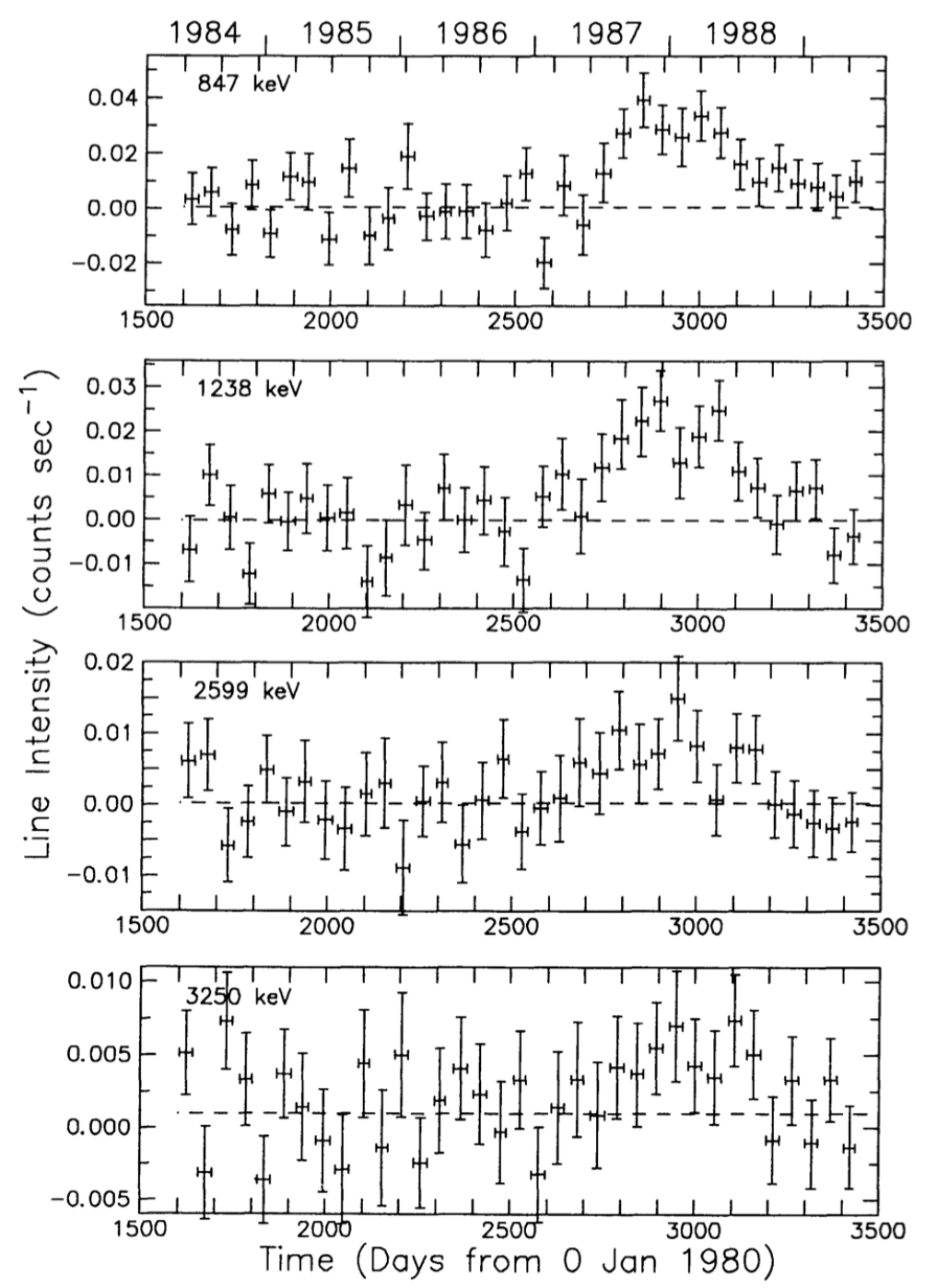}
    \includegraphics[width=0.64\linewidth]{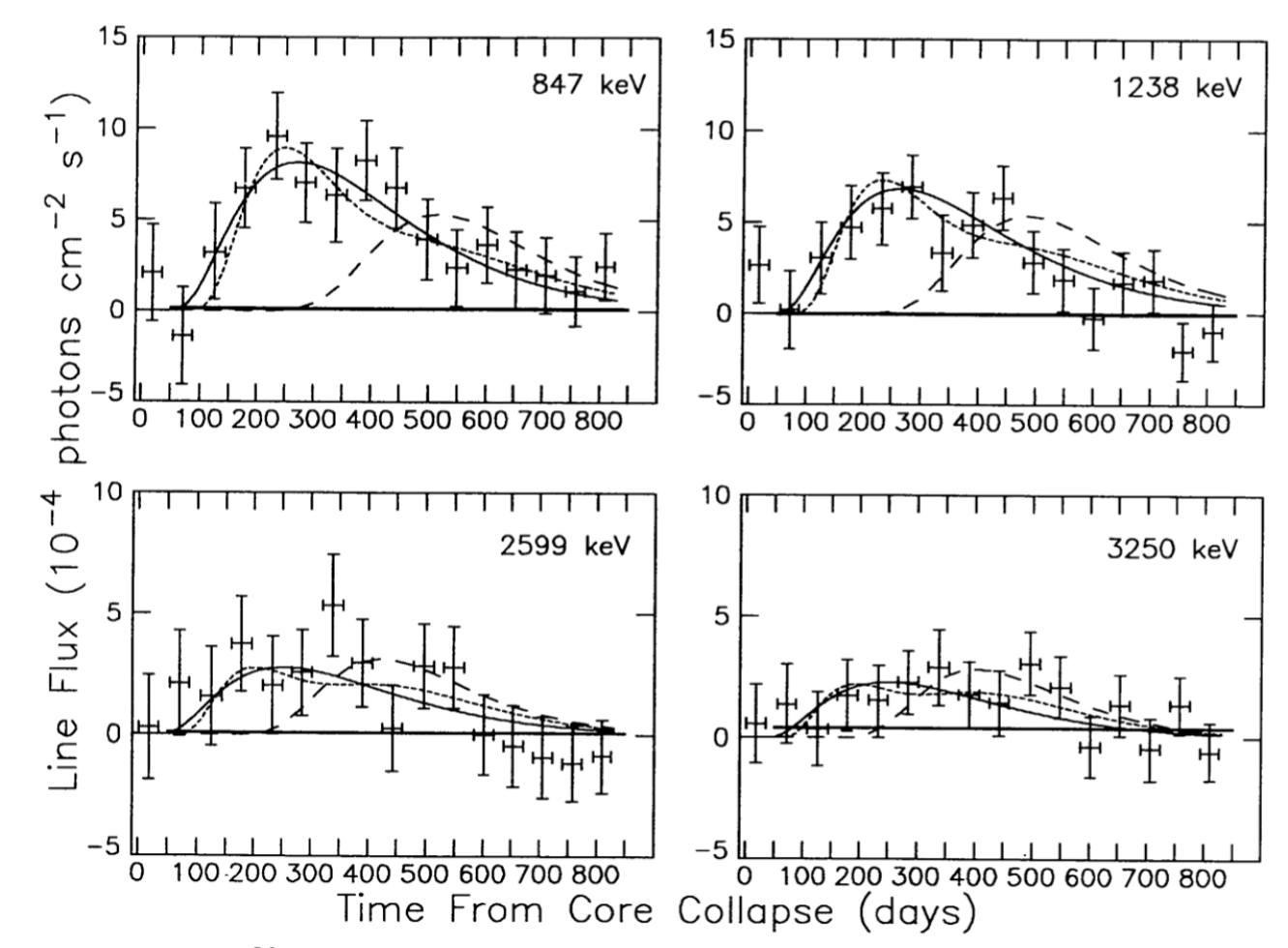}
    \caption{Left: Detection of SN\,1987A in four \ac{gray} lines with \ac{SMM}/\ac{GRS} in counts per second. Right: \ac{gray} line fluxes of the \nuc{Co}{56} decay in SN\,1987A as a function of time, together with three model light curves \citep[from][see also main text]{Leising1990}.}
    \label{fig:87A_lightcurves}
\end{figure}

\paragraph{SN 1987 A in the Large Magellanic Cloud}\label{sec:SN1987A}
There are several studies that tried -- and managed -- to identify \ac{gray} lines (or hard X-ray emission associated with Compton-scattered \ac{gray} lines) from SN\,1987A at early times after the initial explosion \citep[e.g.,][]{Englhauser1989,Sunyayev1990,Matz1988,Sunyayev1987}, together with theoretical expectations for this particular source \citep[e.g.,][including multi-wavelength considerations]{Grebenev1987,Pinto1988,Shigeyama1990,Arnett1989,Bussard1989}.
While it is important to acknowledge the vast ambitions and huge efforts to obtain reasonable measurements, a most comprehensive study of the \nuc{Ni}{56} decay chain, and the associated \ac{gray} line light curves is given in \citet{Leising1990} about three years after the explosion \citep[see also][]{Palmer1993}.

Besides the two prominent lines from the \nuc{Co}{56} decay at 847 (99.9\%) and 1238\,keV (66.5\%), the authors also extracted the light curves of the weaker 2599 (17.0\%) and 3250\,keV (7.9\%) lines\footnote{The 3250\,keV should actually be at a laboratory energy of 3253\,keV but which was not known to this accuracy at the time. Also, this line is not the fourth strongest in the \nuc{Co}{56} decay, see Tab.\,\ref{tab:decay}} using  \ac{SMM}/\ac{GRS}.
The \ac{SMM} observations that include SN\,1987A lasted 1984 until 1989 May, and used a background method based on occultation of the \ac{LMC}, that is, an on-off method \citep{LiMa1983,Vianello2018}.
From August 1987 to May 1988, the \ac{gray} line fluxes of the four lines were $(8.0 \pm 0.9)$, $(5.4 \pm 0.7)$, $(2.6 \pm 0.7)$, and $(2.5 \pm 0.5) \times 10^{-4}\,\mathrm{ph\,cm^{-2}\,s^{-1}}$, respectively.
All lines have been found at the instrumental resolution of \ac{GRS}, and the line centroids are consistent with the expected laboratory energies.
In Fig.\,\ref{fig:87A_lightcurves}, the light curves of the four lines are shown, first as identification in detector count units (left), and then comparing to models (right).

\citet{Leising1990} discuss three theoretical models for the four \ac{gray} line light curves, assuming an initial \nuc{Ni}{56} mass of $0.07\,\mathrm{M_\odot}$ and a distance of 50\,kpc.
\emph{Model 1} follows Eq.\,(\ref{eq:opacity_flux_SN}), where the opacity (or optical depth) decreases as $t^{-2}$.
This model peaks too late and fails to match early measurements.
This discrepancy suggests a range of optical depths rather than a single value.
\emph{Model 2} improves upon Eq.\,(\ref{eq:opacity_flux_SN}) by allowing two distinct \nuc{Co}{56} abundances at different depths.
It fits the light curves well by describing 95\% of \nuc{Co}{56} under an optical depth $\tau_{\rm I} = 200 (100\,\mathrm{d}/t)^2$ and 5\% under $\tau_{\rm II} = 16 (100\,\mathrm{d}/t)^2$, leading to an earlier rise and a longer decay.
The lower $\tau$ corresponds to a hydrogen mass of $1\,\mathrm{M_\odot}$ expanding at $3000\,\mathrm{km\,s^{-1}}$ (cf. Eq.\,(\ref{eq:optical_depth_SN})).
Conceptually, this may represent two distinct \nuc{Ni}{56} clumps displaced by mechanisms such as ``jets'' or ``bubbles'' in the remnant of SN\,1987A.
These 3D structures reappear in the measurements of \ac{CasA} (Sec.\,\ref{sec:CasA}).
\emph{Model 3} assumes homogeneous mixing of \nuc{Co}{56}, meaning outer layers emit earlier than inner ones.
This model also fits the data well, although it may be somewhat unrealistic.
The outer region’s column depth is equivalent to a hydrogen mass of $0.33\,\mathrm{M_\odot}$ expanding at $3000\,\mathrm{km\,s^{-1}}$ (cf. Eq.\,(\ref{eq:optical_depth_SN})).
Based on Fig.\,\ref{fig:87A_lightcurves}, this model requires very low early optical depths for \nuc{Co}{56} and a thick inner region  (massive or slowly moving) that obscures most of it, keeping the \ac{gray} line flux low at later times.

Measurements of the ejected \nuc{Ni}{57} (via \nuc{Co}{57} \ac{gray} line observations) help locate the boundary between ejected and accreted matter -- the ``mass cut'' \citep[e.g.,][]{Thielemann1990,Woosley1991,Wang2024}.
This is because the neutron-to-proton ratio increases toward the star’s centre.
During the first three years, the \nuc{Co}{57} \ac{gray} lines at 122 and 136\,keV from SN\,1987A are undetectable due to high thee opacity, inferred from \nuc{Co}{56} measurements.
Fluxes of a few $10^{-5}\,\mathrm{ph\,cm^{-2}\,s^{-1}}$ are expected around 1000--1200 days post-explosion, corresponding to $1.7 \times 10^{-3}\,\mathrm{M_\odot}$.
\citet{Clayton1992} discuss individual measurements of \nuc{Co}{57} decay \acp{gray} \citep{Kurfess1992}, showing higher fluxes than expected, typically factors of 2--3.
After day 1000, \nuc{Co}{57} becomes the dominant power source for the bolometric light curve until \nuc{Ti}{44} takes over around day 2000.
The \nuc{Ni}{57}/\nuc{Ni}{56} production ratio was found to be about five times the solar value.
This implies roughly $0.009\,\mathrm{M_\odot}$ of \nuc{Ni}{57} in SN\,1987A.
Astrophysical models incorporating opacity, nucleosynthesis, ejecta mass, and velocity yield a \nuc{Ni}{57}/\nuc{Ni}{56} ratio of 1--3.
Although these considerations partly reduce the overproduction, nucleosynthesis and chemical evolution still faced a major problem.

\begin{figure}[!ht]
    \centering
    \includegraphics[width=0.468\linewidth]{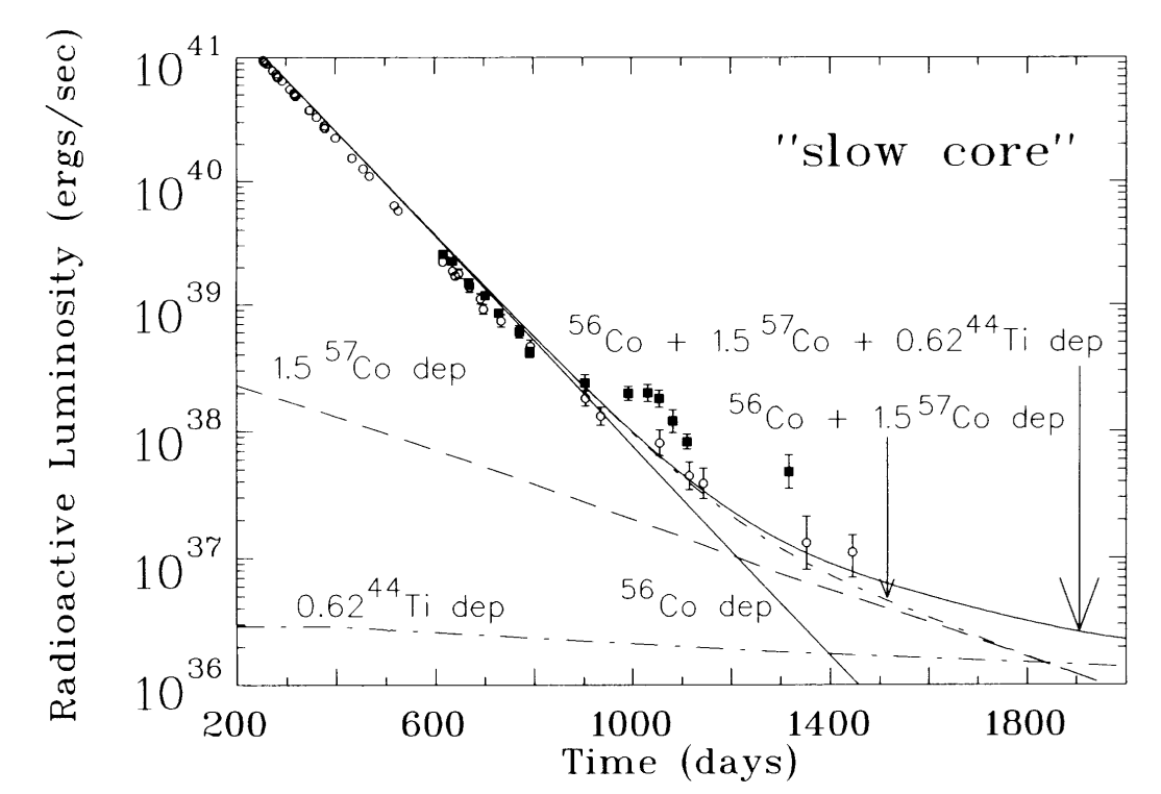}~~~~~~~
    \includegraphics[width=0.423\linewidth]{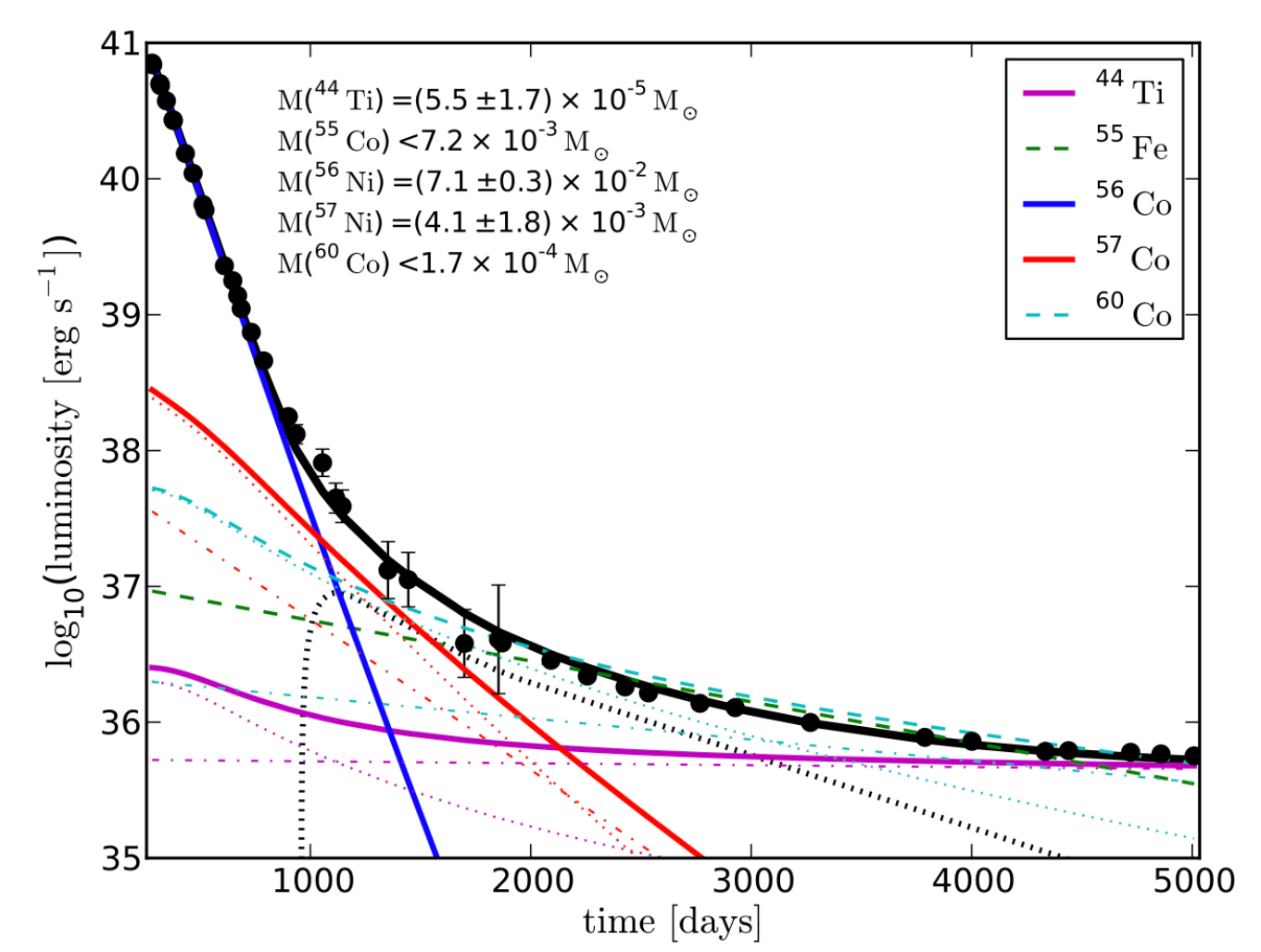}
    \caption{Bolometric light curves of SN\,1987A up to day 1500 \citep[left; from][]{Clayton1992} and up to day 5000 \citep[right; from][]{Seitenzahl2014} after the explosion. Given the updated observations and information, the derived masses of the radioactivities change by factors of 2--3, except for \nuc{Ni}{56}, being consistent with $0.07\,\mathrm{M_\odot}$.} 
    \label{fig:87A_bolo_curves}
\end{figure}

In Fig.\,\ref{fig:87A_bolo_curves}, the bolometric light curves with considerations at early times \citep{Fu1989,Clayton1992} as well as 14 years after the explosion \citep{Seitenzahl2014} are shown\footnote{SN\,1987A shows a re-brightening after that time due to shock heating.}.
It becomes evident that longer observations resulted in updates on the \nuc{Ni}{57} mass, but not on the \nuc{Ni}{56} mass.
\citet{Seitenzahl2014} found a \nuc{Ni}{57} mass of $(4.1 \pm 1.8) \times 10^{-3}\,\mathrm{M_\odot}$, about 2--3 times smaller than the initial estimates.
In addition, they could estimate the mass of \nuc{Ti}{44} to be around $(5.5 \pm 1.7) \times 10^{-5}\,\mathrm{M_\odot}$.
This, however, is about six times smaller than obtained from \ac{INTEGRAL}/\ac{IBIS} \ac{gray} line measurements by \citet{Grebenev2012}, finding $(3.1 \pm 0.8) \times 10^{-4}\,\mathrm{M_\odot}$.
Finally, \citet{Boggs2015}, utilising the \ac{NuSTAR}, found a \nuc{Ti}{44} mass of $(1.5 \pm 0.3) \times 10^{-4}\,\mathrm{M_\odot}$.
All these values do not agree, making it difficult to judge the actual nucleosynthesis happening inside SN\,1987A without improved \ac{gray} line measurements.

The hard X-ray lines from SN\,1987A reported by \citet{Boggs2015} show Doppler-shifts and -broadening corresponding to ejecta velocities of several thousand $\mathrm{km\,s^{-1}}$, revealing high-speed radioactive material.
These features are interpreted as signatures of an asymmetric explosion geometry with clumpy or plume-like structures transporting freshly synthesized \nuc{Ti}{44} outward.
This is similar to the highly solved \ac{NuSTAR} picture from \ac{CasA} (Sec.\,\ref{sec:CasA}), where spatially resolved \nuc{Ti}{44} maps show strong velocity asymmetries that support such explosion dynamics.
Because the 1157\,keV \ac{gray} line of \nuc{Sc}{44} would provide much higher spectral resolving power and therefore more precise constraints on ejecta kinematics, improved measurements of this line remain essential for advancing structural and dynamical studies in combination with nucleosynthesis yields.

\paragraph{Cassiopeia A}\label{sec:CasA}
The \ac{CasA} \ac{SNR} is about 350\,yr old \citep[e.g.,][]{Alarie2014}, so that certainly, no \nuc{Ni}{56}, \nuc{Co}{56}, \nuc{Ni}{57}, or \nuc{Co}{57} can be present in the still expanding remnant.
Given the bolometric light curve, kinematic studies, and light echoes, the \ac{CasA} \nuc{Ni}{56} mass was estimated to be around $0.058$--$0.16\,\mathrm{M_\odot}$ \citep[e.g.,][]{Eriksen2009}.
At this age, however, a sizeable amount of \nuc{Ti}{44} is still left to decay, about 2\% of the initial ejecta mass, so that the flux today should be on the order of $10^{-5}\,\mathrm{ph\,cm^{-2}\,s^{-1}}$ (see Eq.\,(\ref{eq:F_44})).
Indeed, fluxes measured from \ac{CasA} within the last 20 years range from $(1.0$--$12.5) \times 10^{-5}\,\mathrm{ph\,cm^{-2}\,s^{-1}}$ \citep[e.g.,][see also Tab.\,\ref{tab:44Ti_limits}]{Siegert2015,Tsygankov2016,Wang2016,Grefenstette2017,Weinberger2020}.
Within this period, the \nuc{Ti}{44} \ac{gray} line flux should have decreased by 20\%, which is not enough to explain the variations of the line measurements.
It appears that the high-energy \ac{gray} line at 1157\,keV from the \nuc{Sc}{44} decay to stable \nuc{Ca}{44} appears systematically stronger than the \nuc{Ti}{44} decay lines at 68 and 78\,keV to \nuc{Sc}{44} \citep{Siegert2015}.
This can be interpreted in several ways, of which we outline three in the following.

\begin{table}[!ht]
\footnotesize
\caption{Astrophysical parameters for the most promising \acp{SNR} for detections of \nuc{Ti}{44}, together with measured values and upper limits. The age of the remnant is given for 01.01.2011\,AD, which was the average observation date in this study. Age uncertainties of the time bin are included in the mass estimates. Distance to Vela Jr.\ is estimated from \nuc{Ti}{44} yields. Distance to G1.9+0.3 is estimated from absorption towards the Galactic centre. Exposure is dead time corrected, effective \ac{SPI} exposure. Alternative flux measurements with \ac{NuSTAR} and \ac{IBIS} are also shown \citep[adapted from][see also references therein]{Weinberger2020}.}
\centering
\label{tab:44Ti_limits}
\begin{tabular}{c c c c c c c}
\hline\hline
 & Cas A & SN\,1987A & Vela Jr. & Tycho & Kepler & G1.9+0.3 \\
\hline
Distance [kpc]
& $3.3\pm0.1$ & $49.6\pm0.5$ & $0.2$ & $4.1\pm1$ & $5.1^{+0.8}_{-0.7}$ & $8.5$ \\
Explosion Date
& $1681$ & $1987$ & $\sim1320$ & $1572$ & $1604$ & $1890$ \\
Age [yr]
& $330$ & $24$ & $690$ & $438$ & $406$ & $120$ \\
Type
& IIb & II-P & II & Ia & Ia & Ia \\
Exposure [Ms]
& $11.2$ & $7.0$ & $8.3$ & $10.3$ & $29.3$ & $30.6$ \\
\begin{tabular}[c]{@{}c@{}}Gal. Cor.\\$l,b$ [deg]\end{tabular}
& $111.7$,$-2.1$ & $279.7$,$-31.2$ & $266.3$,$-1.21$ & $120.1$,$1.4$ & $4.5$,$6.8$ & $1.9$,$0.3$ \\
\hline
\begin{tabular}[c]{@{}c@{}}Flux $^{44}$Sc\\$[10^{-5}$\,ph\,cm$^{-2}$\,s$^{-1}]$\end{tabular}
& $9.5\pm3.0$ & $\!<\!4.1$ & $\!<\!4.7$ & $\!<\!6.2$ & $\!<\!2.6$ & $\!<\!3.7$ \\
\begin{tabular}[c]{@{}c@{}}Flux $^{44}$Ti\\$[10^{-5}$\,ph\,cm$^{-2}$\,s$^{-1}]$\end{tabular}
& $3.3\pm0.9$ & $\!<\!1.9$ & $\!<\!2.8$ & $\!<\!1.5$ & $\!<\!1.3$ & $\!<\!1.1$ \\
\begin{tabular}[c]{@{}c@{}}Flux Combined\\$[10^{-5}$\,ph\,cm$^{-2}$\,s$^{-1}]$\end{tabular}
& $4.2\pm1.0$ & $\!<\!1.8$ & $\!<\!2.1$ & $\!<\!1.4$ & $\!<\!1.1$ & $\!<\!1.0$ \\
\begin{tabular}[c]{@{}c@{}}Mass Combined\\$[10^{-4}$\,M$_\odot$]\end{tabular}
& $2.6\pm0.6$ & $\!<\!6.9$ & $\!<\!0.3$ & $\!<\!4.8$ & $\!<\!4.0$ & $\!<\!0.3$ \\
\hdashline
\begin{tabular}[c]{@{}c@{}}NuSTAR\\$[10^{-5}$\,ph\,cm$^{-2}$\,s$^{-1}]$\end{tabular}
& $1.8\pm0.3$ & $0.35\pm0.07$ & \textendash & $\!<\!1.0$ & \textendash & $\!<\!1.5$ \\
\begin{tabular}[c]{@{}c@{}}NTEGRAL/IBIS\\$[10^{-5}$\,ph\,cm$^{-2}$\,s$^{-1}]$\end{tabular}
& $1.3\pm0.3$ & $1.7\pm0.4$ & $\!<\!1.8$ & $\!<\!1.5$ & $\!<\!0.63$ & $\!<\!0.9$ \\
\hline\hline
\end{tabular}
\end{table}

\citet{Siegert2015} found an enhanced flux of the 1157\,keV line relative to the 78\,keV line in \ac{SPI} data.
%suggesting reduced systematics.
%
\citet{Weinberger2020} reanalysed over five additional years of data and indicated that the high-energy line was previously misinterpreted, implying an even higher flux.
The 78\,keV line maintained a similar flux, consistent with \ac{NuSTAR} and \ac{IBIS} measurements of $(1.0$--$3.2) \times 10^{-5}\,\mathrm{ph\,cm^{-2}\,s^{-1}}$.
Although \ac{COMPTEL} and \ac{SPI} were the only instruments to measure the high-energy line until 2025, this flux difference hints at possible astrophysical effects:
Stable \nuc{Ca}{44} could be excited by \acp{LECR} and de-excite, emitting only the 1157\,keV line (see also Sec.\,\ref{sec:LECRs}).
Since \nuc{Sc}{44} is very short-lived (4\,h), no de-excitation \acp{gray} lines at 68 or 78\,keV are expected \citep{Siegert2015}.
Although plausible, this scenario would require a \nuc{Ca}{44} density in \ac{CasA} that is orders of magnitude higher than typically observed.
Furthermore, other lines (e.g., from stable \nuc{Fe}{56} in the \nuc{Ni}{56} decay chain) should exhibit a stronger de-excitation flux, but is not observed.
The $3\sigma$ upper limit on the 847\,keV line from \ac{CasA} is approximately $5 \times 10^{-5}\,\mathrm{ph\,cm^{-2}\,s^{-1}}$ for a 4\,keV (\ac{FWHM}) line \citep{Weinberger2021}.
That study also set a flux limit on the \nuc{C}{12} line at 4.4\,MeV (94\,keV \ac{FWHM}) of $5 \times 10^{-5}\,\mathrm{ph\,cm^{-2}\,s^{-1}}$, compared to an expected $\sim 10 \times 10^{-5}\,\mathrm{ph\,cm^{-2}\,s^{-1}}$ \citep[][see, however, \citep{liu2023}]{summa2011}.
This discrepancy has implications for the \ac{LECR} spectrum and the density structure of the \ac{SNR} (see also Sec.\,\ref{sec:LECRs}).
Consequently, the \ac{CR} excitation scenario is ruled out.

Another plausible explanation would be an increased absorption of the lower energy lines.
Two thirds of dust in the Milky Way are probably produced by \acp{ccSN} \citep{McKinnon2016}, so that an enhanced dust production in \ac{CasA} could alter the \ac{LoS} fluxes beyond the expected branching ratios.
The attenuation coefficients for the 1157\,keV line compared to the 68 and 78\,keV lines, considering different dust species, are about a factor of $3$--$12$ smaller \citep{Iyudin2019}. 
This directly impacts the measured flux values, and could lead to a fraction of 16--80\% of the low-energy line fluxes being absorbed by dust \citep{Weinberger2020}.
This scenario appears more realistic, but cannot be proven without further measurements of the high-energy line, for example with \ac{COSI} \citep{Tomsick2024}.

The third, and probably most realistic, scenario is a mixture of systematic uncertainties, wrong assumed line profiles such as the typically used Gaussians vs. the more appropriate tophat function which might overestimate the fluxes \citep{Weinberger2020}, and absorption.
Nuclear excitation of the \nuc{Ca}{44} line does happen, but its \ac{gray} line flux would be orders of magnitude below the measured values.

\begin{figure}[!th]
    \centering
    \includegraphics[width=0.496\linewidth]{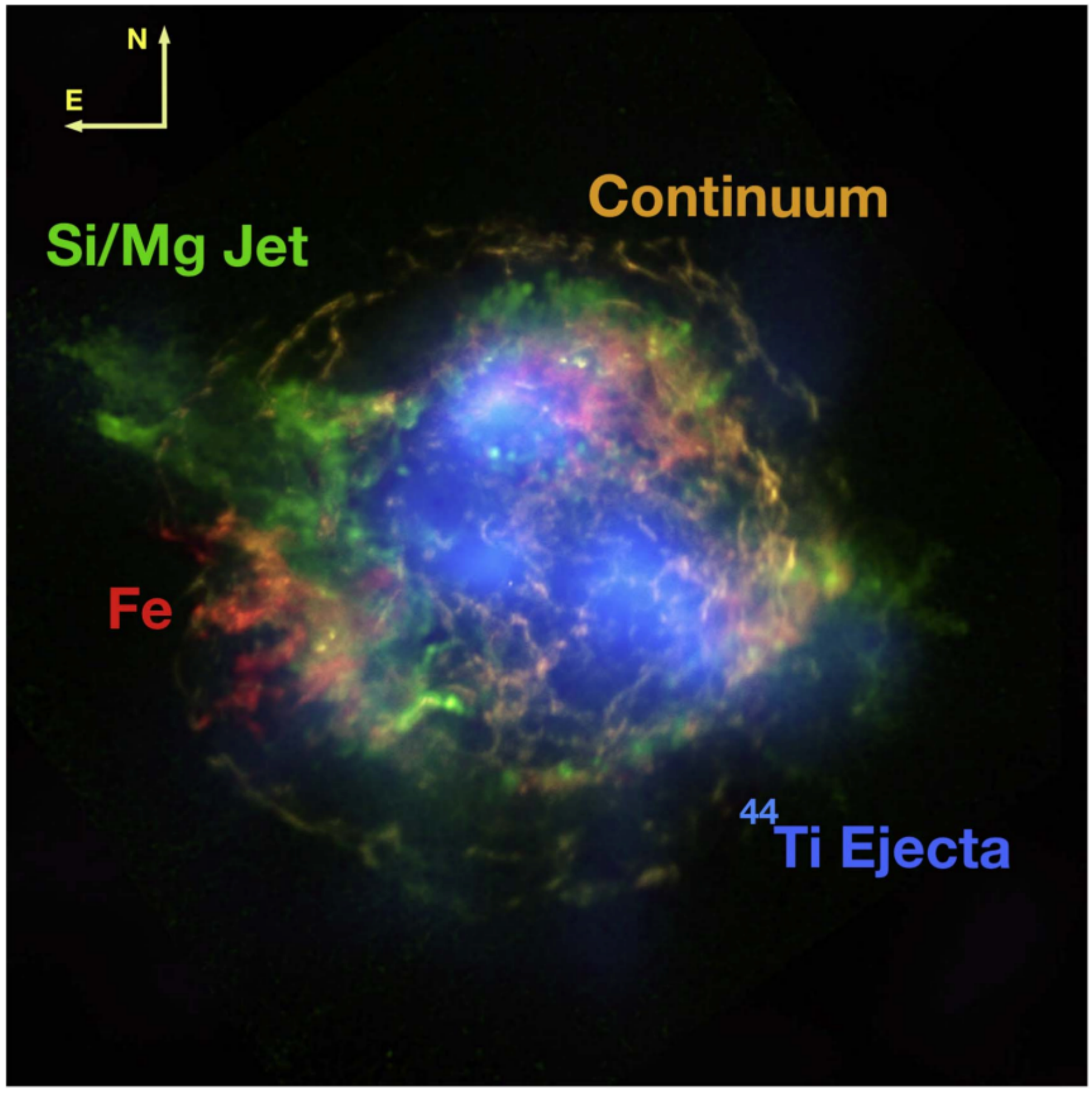}
    \includegraphics[width=0.288\linewidth]{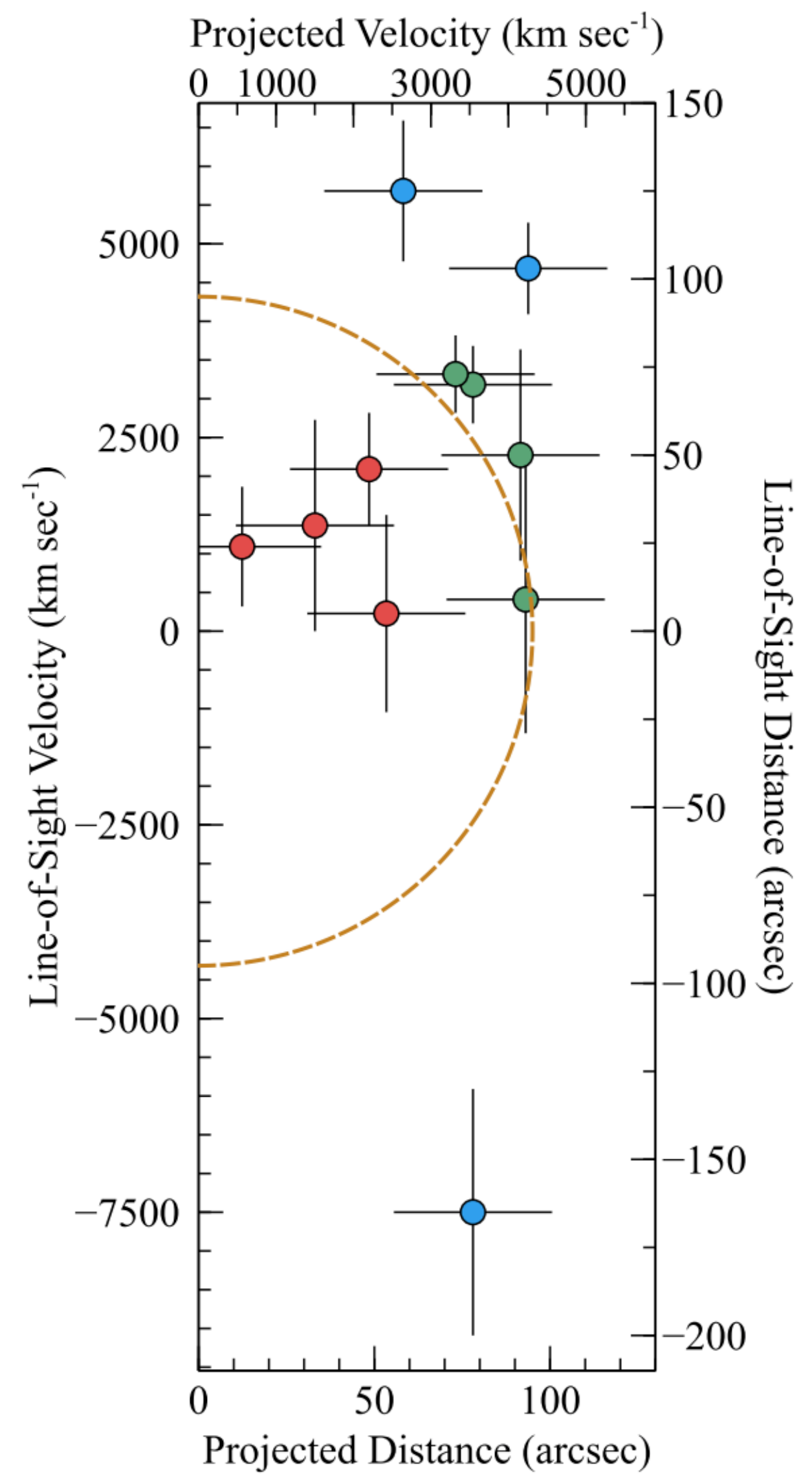}
    \caption{Left: Spatial distribution of \nuc{Ti}{44} (blue) in \ac{CasA} as measured with \ac{NuSTAR}, compared to other X-ray features. The ratio in the Si/Mg band highlights the NW/SW `jet' structure in green. Image credit: Robert Hurt, NASA/JPL-Caltech \citep[from][]{Grefenstette2014,Grefenstette2017}. Right: Doppler velocity measurements of \nuc{Ti}{44} clumps along the \ac{LoS} in \ac{CasA}. The reverse shock radius is marked by the golden dashed curve at 95\,arcsec projected distance \citep[from][]{Grefenstette2017}.}
    \label{fig:CasA_clumps_velocity}
\end{figure}

\begin{comment}
The age and proximity of \ac{CasA} make it possible to not only measure the line fluxes, but also their bulk motion and broadening.
%
In fact, with the focussing optics of \ac{NuSTAR}, it is even possible to map \nuc{Ti}{44} clumps inside the remnant (see Fig.\,\ref{fig:CasA_clumps_velocity}) to find a structured line profile that is not necessarily Gaussian nor tophat-like in total \cite{Grefenstette2017}.
%
The clumpiness in \ac{CasA} is slightly extended along the `jet' axis as seen in the X-ray emission from Si and Mg, as well as the fast moving knots in the optical.
%
The \nuc{Ti}{44} clumps are not uniformly distributed inside the \ac{SNR} and are mostly placed around the centre of the expansion with 80\% being found inside the reverse shock radius \cite{Grefenstette2014}.

Theoretical line profiles of \nuc{Ti}{44} and \nuc{Ni}{56}, for example, can be used to compare to these measurements and find ``jetted'' or ``clumpy'' structures, which may reveal the astrophysics at play that actually made the progenitor star explode \cite[e.g.][]{Vance2020,Orlando2021}.
%
Full 3D modelling of \acp{SN} and in particular \ac{CasA}, such as by Wongwathanarat et al.\,\cite[2017;][]{Wongwathanarat2017}, can be used to identify and falsify the previously conjectured `inverted' nucleosynthesis ejecta \cite[see][]{Hwang2004}.
%
We refer the reader to Chap.\,\ref{chp:XX} for details about \ac{SN} modelling.
\end{comment}
%
\ac{CasA}'s age and proximity enable measurements of line fluxes, bulk motion, and broadening.
With \ac{NuSTAR}'s focussing optics, \nuc{Ti}{44} clumps can be mapped (see Fig.\,\ref{fig:CasA_clumps_velocity}), revealing a structured line profile that is neither purely Gaussian nor tophat-like.
The clumpiness in \ac{CasA} extends along the jet axis, as seen in Si and Mg X-ray emissions and fast optical knots.
Most \nuc{Ti}{44} clumps are concentrated near the expansion center, with 80\% located inside the reverse shock radius \citep{Grefenstette2014}.
Theoretical \nuc{Ti}{44} and \nuc{Ni}{56} line profiles can be compared to observations to identify ``jetted'' or ``clumpy'' structures, revealing the explosion physics \citep[e.g.][]{Vance2020,Orlando2021}.
3D \ac{SN} models, especially for \ac{CasA} \citep[2017;][]{Wongwathanarat2017}, can be used to test the previously conjectured ``inverted'' nucleosynthesis ejecta \citep[see][]{Hwang2004}.
See the Chapter on explosions in this volume for further details on \ac{SN} modelling.

Clearly, \ac{CasA} appears as an exceptional \ac{SNR} because the measured fluxes suggest an otherwise never seen \nuc{Ti}{44} ejecta mass of $(1.0$--$2.0) \times 10^{-4}\,\mathrm{M_\odot}$.
Typical \acp{ccSN} model calculations suggest an order of magnitude less \nuc{Ti}{44} ejecta, and only models with asymmetries show masses of up to $10^{-4}\,\mathrm{M_\odot}$ \citep[e.g.,][also including \acp{SNIa} models]{Wongwathanarat2017,Maeda2003,Seitenzahl2013,Fink2010,Fink2014}.
However, the question ``\emph{Are \nuc{Ti}{44}-producing supernovae exceptional?}'' \citep{The1996} can be answered by Hinchliffe's rule\footnote{According to a large online encyclopedia, the concept known as \emph{Hinchliffe's rule}, after physicist Ian Hinchliffe, states that if a research paper title is in the form of a yes-no question, the answer to that question will probably be ``no''.} because also SN\,1987A appears to have a large \nuc{Ti}{44} mass \citep{Boggs2015}.
On the other hand, the Vela\,Jr. \ac{SNR} appears to have an at least five- to tenfold smaller \nuc{Ti}{44} yield, although the source is much closer than \ac{CasA}, however at an age of almost 700\,yr.
The distance information on Vela\,Jr. is vague and ranges from 0.2\,kpc up to more than 0.5\,kpc, which would change the expected flux from $(1$--$2) \times 10^{-5}\,\mathrm{ph\,cm^{-2}\,s^{-1}}$ to almost $10 \times 10^{-5}\,\mathrm{ph\,cm^{-2}\,s^{-1}}$ (Eq.\,(\ref{eq:F_44})).
The latter is excluded by \ac{INTEGRAL} measurements (Tab.\,\ref{tab:44Ti_limits}), so that either the distance is larger than 0.2\,kpc or the \nuc{Ti}{44} ejecta mass is significantly smaller than for \ac{CasA} and SN\,1987A.
Both scenarios could be investigated by a more sensitive instrument, such as \ac{COSI} \citep{Tomsick2024}.
For more information about \nuc{Ti}{44} nucleosynthesis, including $\alpha$-rich freeze-out, we refer the reader to the Chapter on Explosions.

\begin{figure}
    \centering
    \includegraphics[width=0.46\linewidth]{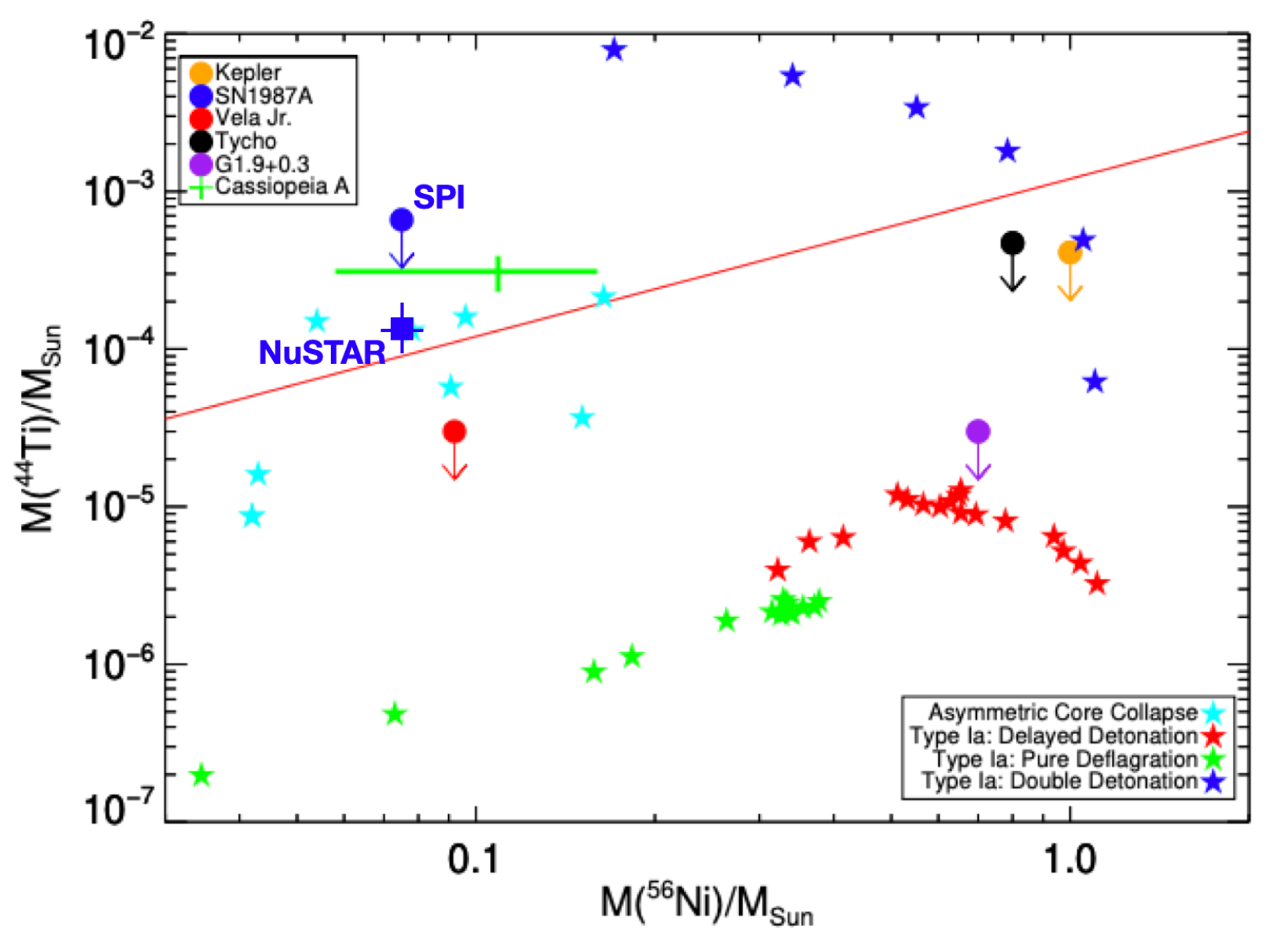}
    \includegraphics[width=0.53\linewidth]{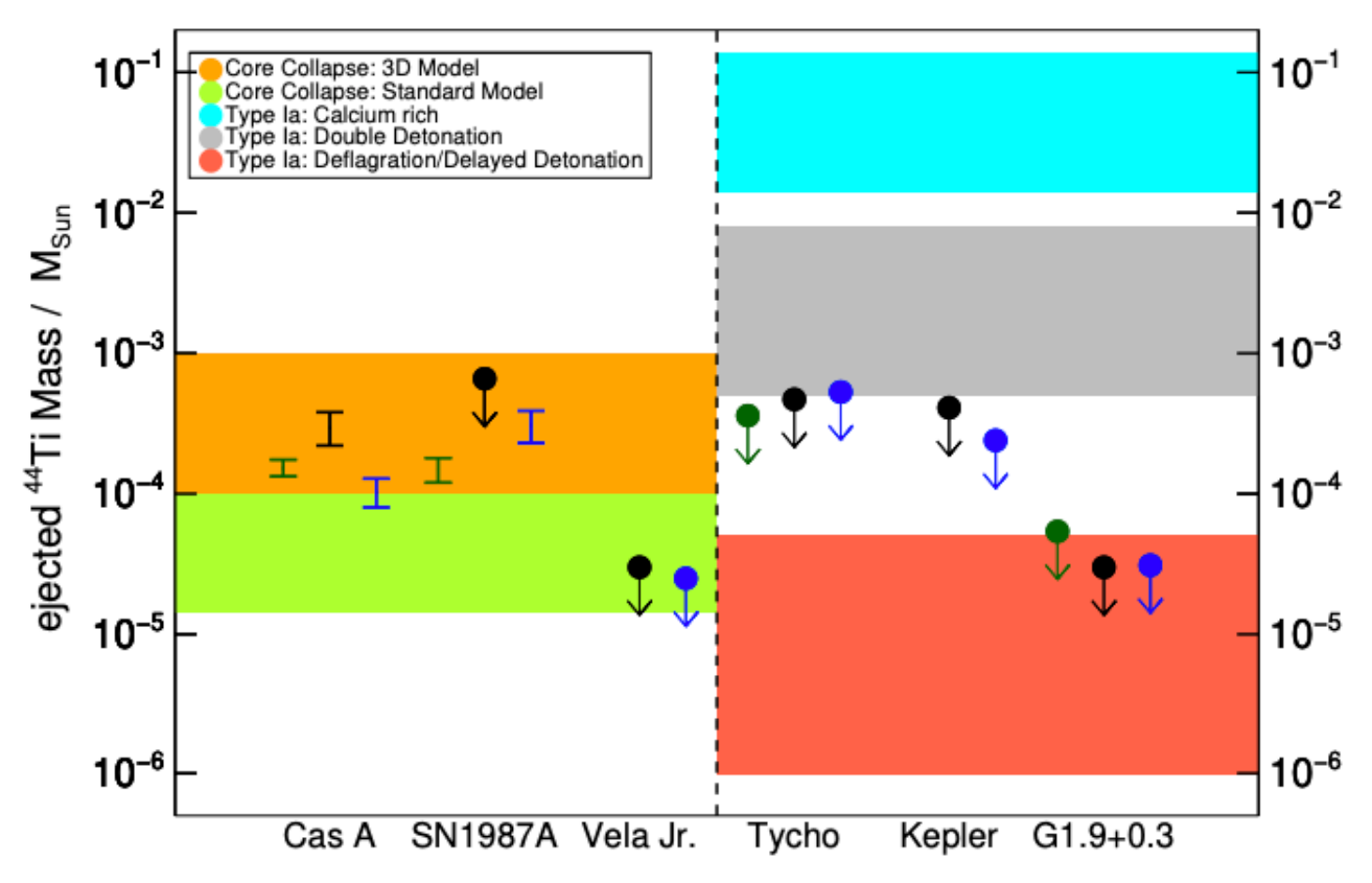}
    \caption{Left: Measured \nuc{Ni}{56} vs. \nuc{Ti}{44} yields for different \acp{SNR} (also \acp{SNIa}), compared to different model calculations. The red line marks the solar [$^{44}\rm Ca /^{56}\rm Fe$] ratio \citep[$1.2\times 10^{-3}$;][]{Anders1989}, which is used as a reference criterion to judge supernova model subtypes. Upper limits from \ac{SPI} are shown as filled circles, measurements (\ac{CasA} only) by the green cross, and \ac{NuSTAR} measurements by a blue square \citep[adapted from][]{Weinberger2020}. Right: Comparison of \nuc{Ti}{44} ejecta masses from different measurements (black: \ac{SPI}; green: \ac{NuSTAR}; blue: \ac{INTEGRAL}/\ac{IBIS}) in \acp{ccSN} (left) and \acp{SNIa} (right) \citep[from][]{Weinberger2020}.}
    \label{fig:enter-label}
\end{figure}

Finally, upper limits obtained for the three youngest \acp{SNIa} exclude the double detonation model and faint SN\,2005E-like scenarios \citep[][see also Sec.\,\ref{sec:SNeIa}]{Weinberger2020}.

\subsubsection{Future Prospects of Supernova Observations}\label{sec:future_SNe}
%
%\paragraph{Future Prospects}\label{sec:future_SNeIa}
%
While numerous \ac{gray} instruments with line sensitivity goals near $10^{-6}\,\mathrm{ph\,cm^{-2}\,s^{-1}}$ within 1\,Ms have been proposed over the past three decades, none have been, or are currently being, realised.
Such an instrument would revolutionise our measurements of the basic quantities defining \acp{SN}.
However, \ac{COSI} \citep{Tomsick2024} is being developed by NASA for launch in 2027.
It is a relatively small instrument whose strength is a wide field and nearly continuous coverage of the entire sky, which are particularly well suited to studying broadly distributed Galactic \ac{gray} lines and continuum, as well as transients.
\ac{COSI}'s sensitivity requirements are $(2$--$3) \times 10^{-5}\,\mathrm{ph\,cm^{-2}\,s^{-1}}$, with energy resolution below 1\% for each of the most prominent \ac{SN} \ac{gray} lines.
With multiple lines and good coverage of the light curve peak, \ac{COSI} should clearly detect \nuc{Co}{56} in approximately one \ac{SNIa} per year.
With some good fortune, \acp{SNIa} at the distances of, e.g., SN\,2011fe and SN\,2014J could be characterised in unprecedented detail.

In the case of \acp{ccSN}, these sensitivities still only reach the Milky Way, plus $100$\,kpc.
For more \nuc{Ti}{44} \ac{gray} line measurements, a sensitivity of $\lesssim 10^{-5}\,\mathrm{ph\,cm^{-2}\,s^{-1}}$ would be required.
Taking into account the line broadening, a `narrow line sensitivity' closer to $10^{-6}\,\mathrm{ph\,cm^{-2}\,s^{-1}}$ is necessary to study more than three \acp{SNR}.
A Galactic \ac{ccSN} would allow us to study not just three isotopes but potentially the whole list in Tab.\,\ref{tab:decay}.
Such observations would be unique and would revolutionise our understanding of \acp{ccSN}.

% NSMs
\newpage
\subsection{Nuclear $\gamma$-Ray Lines from $r$-Process Sources}\label{sec:NSM}
\vspace{-0.5em}
{\emph{Written by Meng-Ru Wu}}
\vspace{0.5em}
\vspace{-12pt}

\subsubsection{$r$-Process Nucleosynthesis} \label{sec:rp_sources}
Nearly half of elements heavier than iron in nature are produced by the so-called rapid neutron capture nucleosynthetic process ($r$ process).
This often occurs in a dynamically evolving environment associated with astrophysical explosions, where matter expands from high density (baryonic mass density $\rho\gtrsim \mathcal{O}(10^{10})\,\mathrm{g\,cm^{-3}}$ and/or high temperature ($k_B T \gtrsim O(1)$\,MeV) \citep{Cowan:2019pkx}.
Due to expansion, the nuclear composition of matter drops out from reaction equilibrium, allowing nuclear transmutation to occur.
At the beginning of the $r$ process, the number density of free neutrons in the environment not only needs to be much higher than that of ``seed nuclei'' (nuclei whose mass number is $A \sim 50$--$100$), but also high enough so that a number of neutron captures by the seed nuclei can take place before the produced unstable nuclei undergo $\beta$-decay.
Through the nuclear transmutation sequence consisting of neutron captures followed by $\beta$-decays, very neutron-rich heavy nuclei up to $A \sim 300$ can be synthesised until nuclear fission prevents further build-up of even heavier nuclei.
The $r$ process ends when the free neutron number density becomes too low, so that neutron captures become inefficient compared to nuclear decays.
From that point onward, the produced $r$-process unstable nuclei gradually decay back toward nuclear stability, forming the heavy isotopes that are eventually identified at Earth, in meteorites, and in stars.

Although the basic theoretical understanding of how an $r$ process occurs outlined above is rather well established, key questions regarding exact astrophysical sites and the associated conditions, as well as the involved nuclear physics properties, remain to be answered \citep{Cowan:2019pkx}.
Among all proposed astrophysical sites (see Tab.\,\ref{tab:source}), including the \acp{BNSM} \citep{Eichler:1989ve}, \ac{NS}--\ac{BH} mergers \citep{Lattimer:1974slx}, neutrino-driven winds from \acp{ccSN} \citep{Woosley:1994ux}, \acp{MRSN} \citep{Winteler:2012hu}, \acp{PTSN} \citep{Fischer:2020xjl}, \acp{CEJSN} \citep{Grichener:2018way}, collapsars \citep{Siegel:2018zxq}, and more recently magnetar \acp{GF} \citep{Cehula:2023hdh}, the \acp{BNSM} are the only unambiguously confirmed sources through the multimessenger detection of gravitational waves, \acp{gray}, and the kilonova emissions associated with the GW170817 event at $\sim 40$\,Mpc away in 2017 \citep{LIGOScientific:2017vwq,LIGOScientific:2017ync}.
This confirmation is primarily based on the large opacity required to account for the infrared kilonova emissions on a timescale of $\sim\mathcal{O}(10)$\,d after the merger.
Advancements in extracting specific emission and absorption features associated with particular $r$ process elements have been made in recent years \citep[e.g.,][]{Watson:2019xjv,Domoto:2022cqp,Sneppen:2023lgo,Gillanders:2023jpd,Hotokezaka:2023aiq}.
However, they are still subject to large theoretical uncertainties originating from the complicated spatial and temporal evolution of the merger ejecta as well as the associated modelling of radiative transfer. 
A potential production of $r$-process elements in the magnetar \ac{GF} SGR\,1806-20 has been recently proposed \citep{Patel:2025frn}.
This claim is based on the observed late-time MeV emission peaked at $\sim 10^3$\,s after the prompt \ac{gray} spike.
The late-time emission may be powered by the \acp{gray} from the radioactive decay of $r$-process nuclei from $\sim 10^{-6}$\,$\mathrm{M_\odot}$ of material with velocities around $\sim 0.15c$ ejected by crustal shocks \citep{Winteler:2012hu}.
If such an association is confirmed in the future, magnetar \acp{GF} may contribute up to $\sim 10\%$ to the Galactic $r$-process inventory.

The \ac{gray} or X-ray lines emitted from decay of $r$ process nuclei in live $r$-process producing events or in their remnants, if detected when the environment become \ac{gray} transparent, will not only be able to offer definitive proof of $r$ process production within these sites, but also provide an independent tool to probe the nucleosynthesis conditions \citep{Meyer:1991xx,Qian:1998cz,Qian:1999vm,Ripley:2013eca,Hotokezaka:2015cma,Li:2018wee,Wu:2019xrq,Korobkin:2019uxw,Wang:2020qkn,Chen:2021tob,Terada:2022hut,Chen:2022nsj,Vassh:2023fnb,Chen:2024nbq,Amend:2024sdm,Patel:2025frn,Liu:2025auu}.
In the following sections, we briefly summarise the most prominent $r$-process \ac{gray} lines from the decay of $r$-process nuclei from different sources, and discuss their implications.

\subsubsection{$\gamma$-Rays from Individual $r$-Process Sources}
For a single $r$-process source located at a distance $d$ from Earth, the \ac{gray} line flux from the decay of a single radioactive nuclear species $i$ with life time $\tau_i$ at a post-production time $t\ll \tau_i$ is given by Eq.\,(\ref{eq:decay_luminosity}), such that
\begin{equation}
%\approx 
%\frac{ I_\gamma N_A M_i}{4\pi d^2 A_i \tau_i} \simeq\\
F^s_\gamma \simeq 10^{-7}\,{\rm ph}\,{\rm cm}^{-2}\,{\rm s}^{-1}\, I_\gamma 
\left( \frac{M_i}{3\times 10^{-4}\,\mathrm{M_\odot}} \right)
%\left( \frac{100}{A_i}\right)
\left( \frac{d}{10\,\rm{kpc}} \right)^{-2}
\left( \frac{\tau_i}{10^5\,\rm{yr}} \right)^{-1}\mathrm{,}
\label{eq:flux_single}
\end{equation}
where $I_\gamma$ is the \ac{gray} emission intensity at a specific energy, $N_A$ is Avogadro's number, and $M_i$ is the mass of the species $i$, assuming $A_i \sim 100$.

\begin{figure}[tbph!]
\centering
\includegraphics[width=0.75\linewidth]{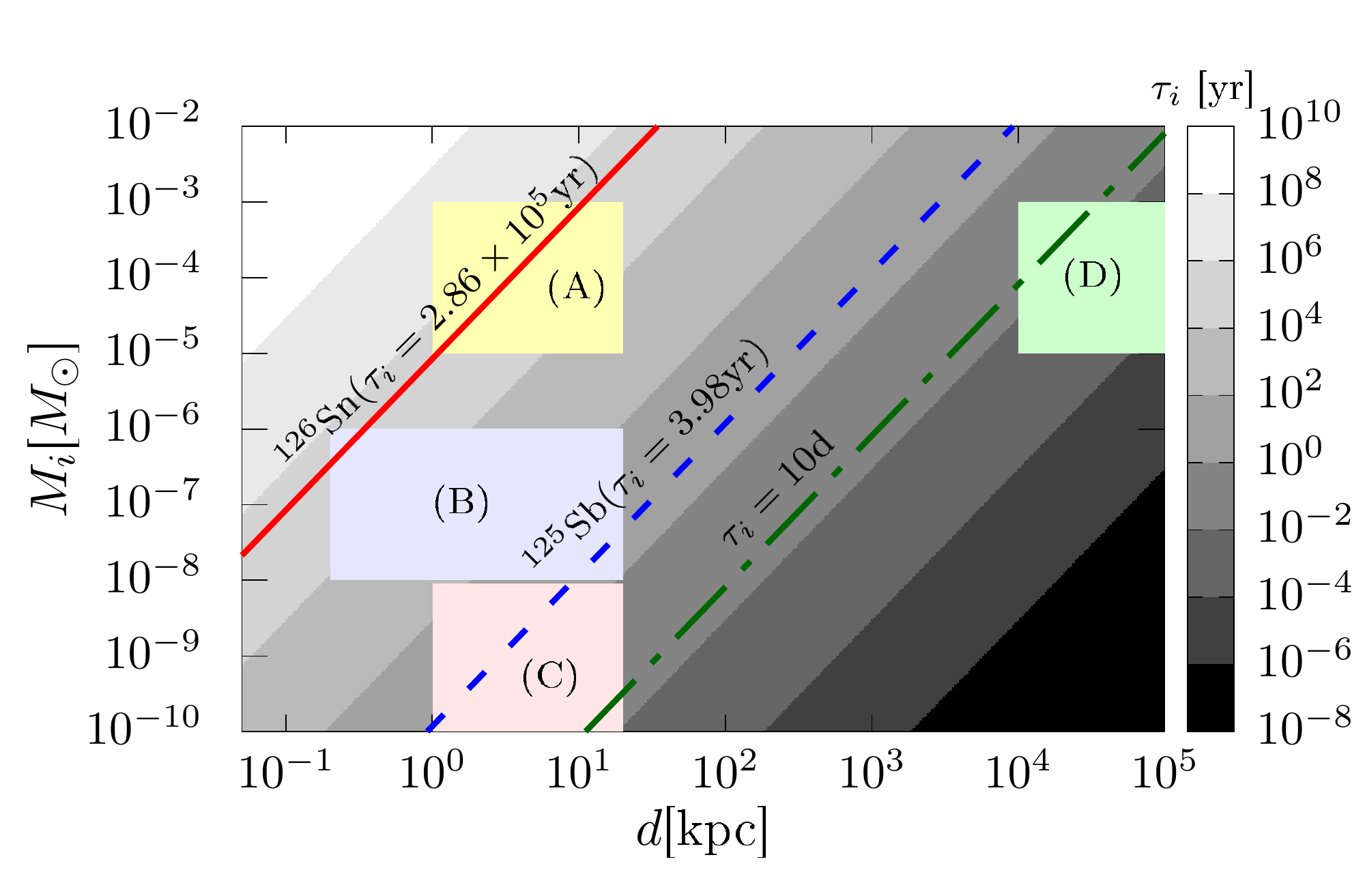}    
\caption{Contour plot of $\tau_i(M_i,d)$ obtained using Eq.\,\eqref{eq:flux_single} for $F_\gamma^s = 10^{-7}\,\mathrm{ph\,cm^{-2}\,s^{-1}}$.
The red solid, blue dashed, and green dash-dotted lines are for the nuclei \nuc{Sn}{126}, \nuc{Sb}{125}, and an assumed one with $\tau_i=10$\,d, respectively. 
The yellow, blue, pink, and green shaded areas indicate the possible range of source distances and the amount of \ac{gray} emitting nuclei produced in (A) Galactic \ac{BNSM} remnants, (B) Galactic \acp{ccSN}, (C) Galactic magnetar \acp{GF}, and (d) extragalactic \acp{BNSM}. 
}
\label{fig:decay_range}
\end{figure}

Fig.\,\ref{fig:decay_range} shows the contour plot of $\tau_i(M_i,d)$ obtained using Eq.\,\eqref{eq:flux_single} for $F_\gamma^s = 10^{-7}\,\mathrm{ph\,cm^{-2}\,s^{-1}}$.
Also plotted are the curves from considering particular \ac{gray} emitting nuclei: \nuc{Sn}{126} ($\tau_i = 2.9\times 10^5$\,yr), \nuc{Sb}{125} ($\tau_i = 4$\,yr), and an assumed nucleus with $\tau_i=10$\,d. 
The shaded rectangle areas indicate the approximate estimated range of source distances and the amount of \ac{gray} emitting nuclei that may be produced in Galactic and extragalactic \acp{BNSM} \citep{Wu:2019xrq}, \acp{ccSN} \citep{Qian:1998cz}, and Galactic magnetar \acp{GF} \citep{Patel:2025tse}.
Clearly, for nuclei with $\tau_i \sim \mathcal{O}(1)$\,yr, it requires a Galactic event that produces $M_i\gtrsim 10^{-8}\,\mathrm{M_\odot}$ to reach a line flux above $10^{-7}\,\mathrm{ph\,cm^{-2}\,s^{-1}}$, overlapping with the potential production sources of \acp{ccSN} and nearby magnetar \acp{GF}.
For nuclei with a longer lifetime of $\tau_i \sim \mathcal{O}(10^{5})$\,yr, the required mass needs to be proportionally higher, reaching $M_i \sim \mathcal{O}(10^{-4})\,\mathrm{M_\odot}$ to satisfy the same criteria. 
Since the maximal value of $M_i$ is typically smaller than $\sim 0.1 M_{\rm ej}^{r}$, only \acp{BNSM}, \ac{NS}-\ac{BH} mergers, \acp{CEJSN}, and collapsars that may produce large enough value of $M_{\rm ej}^{r}$ are good candidates (Tab.\,\ref{tab:source}).

For an extragalactic event with $d \sim \mathcal{O}(10-100)$\,Mpc, \ac{gray} emissions from nuclei with $\tau_i \lesssim \mathcal{O}(10)$\,days in \ac{BNSM} ejecta or \ac{NS}-\ac{BH} mergers may reach the level of $F_\gamma\sim\mathcal{O}(10^{-7})\,\mathrm{ph\,cm^{-2}\,s^{-1}}$.
However, due to the high velocity of merger ejecta reaching $\mathcal{O}(10^{-1})$\,$c$ as well as the many lines from $\mathcal{O}(10$--$30)$ different species of nuclei that can decay to produce \acp{gray} across a wide energy range, the lines are expected to be broadened and mixed into a continuum \citep[e.g.,][]{Hotokezaka:2015cma,Li:2018wee,Korobkin:2019uxw}.
For other sources associated with massive stars that may produce large enough amount of $M_{\rm ej}^{(r)}$, the ejecta environments are not expected to become \ac{gray} transparent yet in days.

After the multimessenger detection of  GW170817 \citep{LIGOScientific:2017vwq,LIGOScientific:2017ync}, confirming that \acp{BNSM} are likely the dominant source for the $r$-process inventory, particular attention has been paid to the potential detectability of \acp{gray} lines from the decay of \nuc{Sn}{126} at $414.7$, $666.3$ and $695.0$\,keV in \ac{BNSM} remnants whose ages are around $\sim\mathcal{O}(10^5)$\,yr in the Milky Way \citep{Wu:2019xrq,Terada:2022hut,Patel:2025frn}.
These studies were done by sampling over assumed occurring frequency of mergers as well as their spatial distributions in the Milky Way (cf. Sec.\,\ref{sec:Al26}).
Despite different assumptions made for the spatial distributions and the nuclear composition of the merger ejecta, they all suggest that it will require a wide field of view \ac{gray} mission with a line sensitivity reaching the level of $10^{-7}\,\mathrm{ph\,cm^{-2}\,s^{-1}}$ in order to have a non-negligible chance to detect the \nuc{Sn}{126} decay \acp{gray} from at least one merger remnant in Milky Way.
However, it should be noted that it may remain possible that one or a few recorded nearby ($\lesssim 3$\,kpc) or young ($\lesssim 10^4$\,yr old) ``\ac{SN} remnants'' are in fact disguised merger remnants. 
In this case, a potential discovery may be possible with a line sensitivity of $\sim 10^{-6}\,\mathrm{ph\,cm^{-2}\,s^{-1}}$ \citep{Wu:2019xrq,Korobkin:2019uxw,Terada:2022hut,Chen:2024nbq}.
The co- or non-detection of the \ac{gray} lines at 351.9 and 609.3\,keV from the decay chain of \nuc{Th}{230} ($\tau_i = 1.1 \times 10^5$\,yr) or the hard X-rays at 74.7\,keV from the decay chain of \nuc{Am}{243} ($\tau_i = 1.1 \times 10^4$\,yr) can possibly offer unique insights on the nucleosynthesis condition associated with the remnant \citep{Wu:2019xrq,Terada:2022hut}.
The amount of the actinide production in any $r$-process sites sensitively depends on the neutron richness of the ejecta.
Moreover, a co- or non-detection of the 1332\,keV line from the decay of \nuc{Co}{60} may further help reveal the nature of the progenitor events \citep{Ripley:2013eca,Liu:2025auu}.

If the line sensitivity of \acp{gray} in future missions can be much improved to reach the level of $10^{-8}\,\mathrm{ph\,cm^{-2}\,s^{-1}}$, then it should be possible to detect the \nuc{Sn}{126} emission from $\gtrsim \mathcal{O}(10)$ merger remnants \citep{Wu:2019xrq}.
Such detections could provide important clues on the property of \ac{NS} binaries that eventually lead to detectable mergers, and probe the variation of the nucleosynthesis conditions in $r$-process sites.
With such a much improved sensitivity, one can expect to answer whether \ac{gray} emitting $r$-process nuclei are produced in the next Galactic \ac{ccSN} explosion \citep{Qian:1998cz} or even in magnetar \acp{GF}.

\begin{figure}[!ht]
  \centering 
  \includegraphics[width=0.49\linewidth]{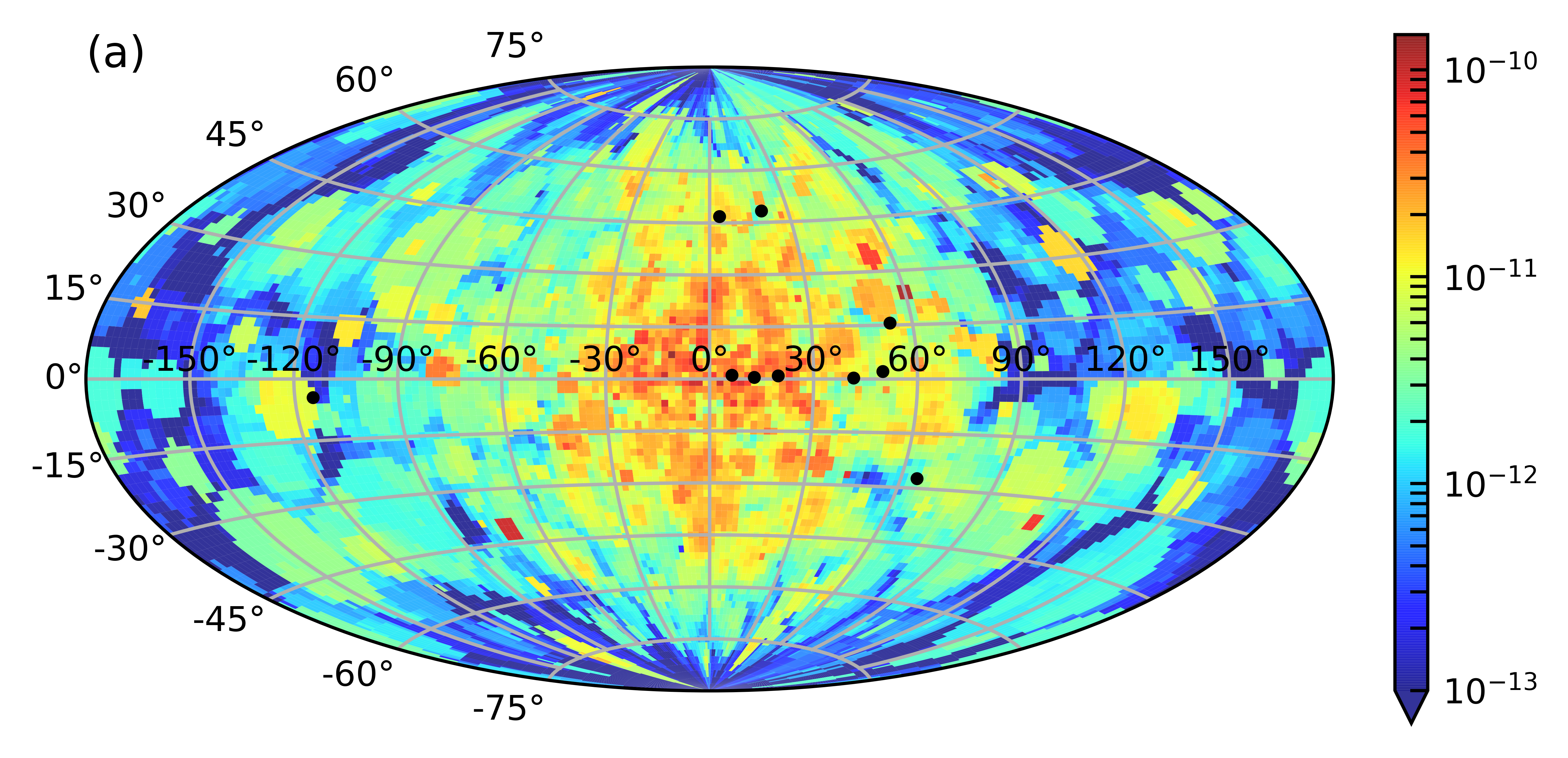}
  \includegraphics[width=0.49\linewidth]{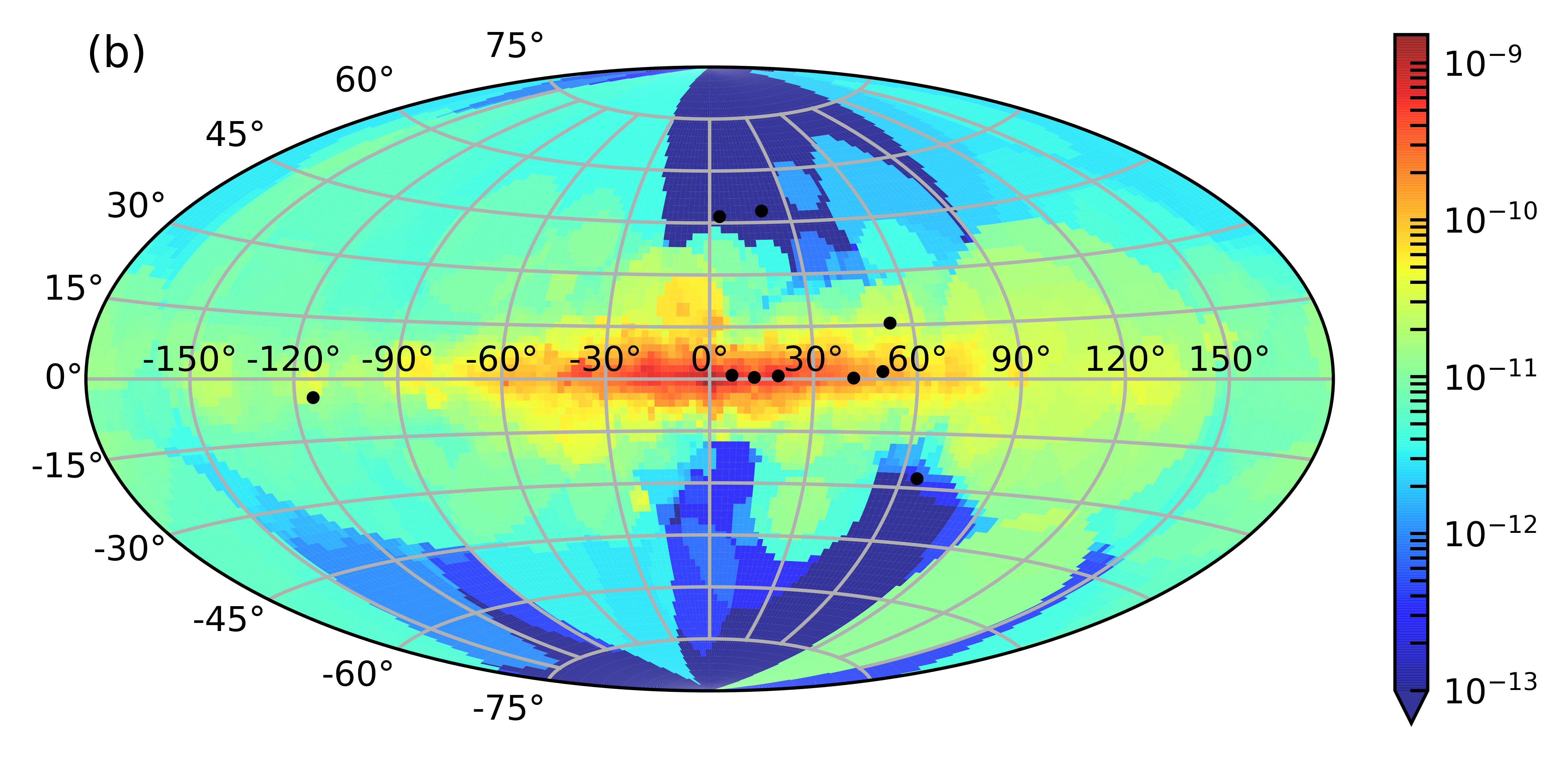}
  \caption{Example flux map from the 1.121\,MeV decay line of \nuc{Hf}{182} (in units of $\mathrm{ph\,cm^{-2}\,s^{-1}\,deg^{-2}}$), shown separately for (a) remnant location sampled by following a drift distance distribution similar to known short \ac{gray} bursts, and (b) following stellar density distributions \citep[from][]{Wu:2019xrq}. 
  }\label{fig:diffuse}
\end{figure}

\subsubsection{Diffuse Extended $r$-Process $\gamma$-Ray Emission}
For the Galactic \ac{gray} sources, if the lifetime of a specific nucleus $\tau_i$ is much longer than the inverse of the occurring frequency of the source in the Milky Way $f_{\rm MW}^{-1}$, emissions from individual sources may no longer be separated, due to the fact that many sources are contributing at the same time, similar to the case of \nuc{Al}{26} in massive stars or \nuc{Na}{22} in \acp{CN}.
In this scenario, the corresponding emission can be considered as ``diffuse'' emission \citep{Qian:1998cz,Wu:2019xrq}.
Since $f_{\rm MW}^{-1}\ll \tau_i$, the total amount of decaying nuclei in the Milky Way at any given moment can be estimated by $N_d \approx {\bar N}_s \tau_i f_{\rm MW}$, where ${\bar N}_s=N_A{\bar M}_i/A_i$ is the average amount produced per source.
Taking $\bar d$ as the average source distance, the diffuse \ac{gray} flux integrated over the full sky can be estimated by 
\begin{equation}
%F^d_\gamma & \approx 
%\frac{I_\gamma N_s}{4\pi {\bar d}^2 \tau_i}
%=\frac{ I_\gamma N_A {\bar M}_i f_{\rm MW}}{4\pi {\bar d}^2 A_i} %\simeq \\
F^d_\gamma \simeq 10^{-7}\,\mathrm{ph\,cm^{-2}\,s^{-1}}\,
I_\gamma 
\left( \frac{{\bar M}_i}{3\times 10^{-4}\,\mathrm{M_\odot}} \right)
\left( \frac{f_{\rm MW}}{10^5\,\rm{yr}}\right)\mathrm{,}
\label{eq:flux_diffuse}
\end{equation}
for $A_i = 100$ and $\bar d=10$\,kpc.

Looking at Eq.\,\eqref{eq:flux_diffuse}, it readily tells us an interesting fact that the diffuse flux does not depend on the lifetime of the nuclei, but is mainly determined by the produced amount per source and the occurring source frequency.
Moreover, it implies an even more interesting point as follows:
Assuming that one unknown single source class dominates the production of the $r$-process nuclei in the Milky Way, but all possible sources' occurring frequency satisfy $\tau_i \gg f_{\rm MW}^{-1}$, then the all-sky diffuse \ac{gray} flux from nucleus $i$ will be nearly independent of the nature of the source.
This is because ${\bar M}_i f_{\rm MW}$ is nearly a constant, which is constrained by the known amount of all $r$-process elements produced over the lifetime of Milky Way.
This is precisely the case for the diffuse \acp{gray} from the decay of \nuc{Hf}{182} with a lifetime of $1.3 \times 10^7$\,yr, much longer than the estimated $f_{\rm MW}^{-1}$ for most of the $r$-process sources listed in Tab.\,\ref{tab:source}.
Thus, the estimated all-sky \nuc{Hf}{182} \ac{gray} line flux from, for example, \acp{ccSN} \citep{Qian:1998cz} and from \acp{BNSM} \citep{Wu:2019xrq} at 0.270 or 1.121\,MeV both range around $10^{-8}\,\mathrm{ph\,cm^{-2}\,s^{-1}}$.

On the other hand, the spatial distribution of the $r$-process sources intuitively affects the angular distribution of the diffuse flux.
For sources that track the stellar density profile of Milky Way, such as \acp{ccSN} and collapsars, the resulting diffuse sky maps should concentrate in the region of the Milky Way's disk, similar to the measured emission from \nuc{Al}{26} and \nuc{Fe}{60}. 
If \acp{BNSM} are the dominant source of the $r$ process and the progenitor \ac{NS} binaries had received substantial kick velocities at birth, then the corresponding diffuse sky map can extend to high latitude (see Fig.\,\ref{fig:diffuse}).

For \nuc{Sn}{126} decay \ac{gray} lines, they are considered the best target to detect individual $r$-process sources if they were produced in \acp{BNSM}.
However, if a much more frequent source such as \acp{ccSN} can produce them with much smaller amount in each, their all-sky \ac{gray} flux may be able to reach the level of $10^{-7}\,\mathrm{ph\,cm^{-2}\,s^{-1}}$ \citep{Qian:1998cz} if \acp{ccSN} account for the majority of $r$-process abundances in the Milky Way.
Although such a scenario is currently disfavoured \citep{Cowan:2019pkx}, this argument indicates that such a consideration may be used to set an independent and meaningful limit on the amount of $r$-process production from \acp{ccSN} by the non-detection of diffuse \ac{gray} line emissions from \nuc{Sn}{126} in the future, before the next Galactic \ac{ccSN} happens.

\begin{table}[!ht]
    \centering
    \caption{Selected list of relevant \ac{gray} or hard X-ray lines from $r$-process nuclei mentioned in this work. For lines produced via shorter-lived nuclei inside the decay sequences from a parent isotope, the name of those nuclei are given inside the parenthesis next to the line energy \citep[see][for complete lists of potentially relevant nuclei]{Qian:1998cz,Wu:2019xrq,Terada:2022hut}.}
    \begin{tabular}{cccrr}
        \hline
        Isotope & Decay channel & $\tau_i$~(yr) & Line energy [keV] & Intensity (\%)\\
          \hline
          \hline
         $^{125}$Sb & $\beta$ to $^{125}$Te & 3.98 & 427.9 & 29.6 
          \\ \hline 
         $^{243}$Am & $\alpha\beta$ to $^{239}$Pu        & $1.06\times 10^4$ & 74.7 & 67.2 \\ \hline
         $^{230}$Th & $\alpha\beta$ to $^{208}$Pb   & $1.09\times 10^5$ & 351.9 ($^{214}$Pb) & 35.6 \\
            & & & 609.3 ($^{214}$Bi)& 45.5 \\ \hline
         $^{126}$Sn & $\beta$ to $^{126}$Te & $2.86\times 10^5$ & 414.7 ($^{126}$Sb) & 98.0 \\ 
         & & & 666.3 ($^{126}$Sb) & 100.0\\ 
         & & & 695.0 ($^{126}$Sb) & 97.0\\ \hline 
         $^{182}$Hf & $\beta$ to $^{182}$W & $1.28\times 10^7$ & 270.4 & 79.0 \\ 
         & & & 1121.3 ($^{182}$Ta)& 35.2 \\ 
         \hline
         \hline
    \end{tabular}

    \label{tab:nuc}
\end{table}

\clearpage
\begin{sidewaystable}[p]
	\centering
	\caption{Possible astrophysical $r$-process sources, the estimated $r$-process masses in the ejecta per event $M_{\rm ej}$, and the estimated occurring rate in the Milky Way (MW) $f_{\rm MW}$. Among these proposed sites, only the \acp{BNSM} are unambiguously confirmed by the GW170817 observation \citep{LIGOScientific:2017vwq,LIGOScientific:2017ync}. \citet{Patel:2025frn} recently proposed the observation of SGR\,1806-20 in 2004 as evidence of $r$ process production in magnetar \acp{GF}. All other sites remain speculative and the quoted amount of $M_{\rm ej}$ should be treated as upper limits.}
	\begin{tabular}{cccc}    
		\hline
		& $M_{\rm ej}$~$[M_\odot]$ & $f_{\rm MW}$~[Myr$^{-1}$] & Confirmation? \\
		\hline
		\hline
		\acp{BNSM} & $\mathcal{O}(10^{-3}-10^{-2})$ \citep{Shibata:2019wef} & $\sim 30$ \citep{Terada:2022hut} & GW170817 \\
		\ac{NS}-\ac{BH} mergers & $\lesssim \mathcal{O}(10^{-1})$ \citep{Shibata:2019wef} & $\lesssim \mathcal{O}(10)$~\citep{Mandel:2021smh} & X \\
		\acp{ccSN} $\nu$ wind & $\lesssim \mathcal{O}(10^{-4})$ \citep{Goriely:2016gfe} & $\sim 10^4$~\citep{Rozwadowska:2020nab} & X \\
		\acp{MRSN} & $\lesssim \mathcal{O}(10^{-3})$ \citep{Winteler:2012hu} & ? & X \\
		\acp{PTSN} & $\lesssim \mathcal{O}(10^{-3})$ \citep{Fischer:2020xjl} & ? & X \\
		\acp{CEJSN} & $\lesssim \mathcal{O}(10^{-2})$ \citep{Grichener:2018way} & $\lesssim \mathcal{O}(10)$ \citep{Grichener:2018way} & X \\
		Collapsars & $\lesssim \mathcal{O}(10^{-1})$ \citep{Siegel:2018zxq} & $\lesssim \mathcal{O}(10)$ \citep{Siegel:2018zxq} & X \\ 
		Magnetar \acp{GF} & $\lesssim \mathcal{O}(10^{-6})$ \citep{Patel:2025tse} & $\sim 10^5$ \citep{Patel:2025tse} & SGR\,1806-20? \\
		\hline
		\hline
	\end{tabular}
	
	\label{tab:source}
\end{sidewaystable}
\clearpage

\begin{comment}
\begin{acknowledgments}
MRW acknowledges support of the National Science and Technology Council, Taiwan under Grant No.~111-2628-M-001-003-MY4, the Academia Sinica under Project No. AS-IV-114-M04, and the Physics Division of the National Center for Theoretical Sciences, Taiwan.
\end{acknowledgments}

\bibliographystyle{apsrev4-1}
% \nocite{*}
\bibliography{ref.bib}

% \onecolumngrid
\appendix

\end{comment}

% LECRs
\newpage
\newpage
\subsection{Nuclear De-Excitation $\gamma$-Ray Lines from Low-Energy Cosmic Rays}\label{sec:LECRs}
\vspace{-0.5em}
{\emph{Written by Vincent Tatischeff}}
\vspace{0.5em}
\vspace{-12pt}

\subsubsection{Utilising Low-Energy Cosmic Rays}
The interaction of accelerated nuclei of kinetic energy $E_{\rm kin} \gtrsim 1$\,MeV/nucleon with ambient matter can produce a wealth of \ac{gray} lines with energies ranging from tens of keV to about 10\,MeV.
This non-thermal \ac{gray} line emission is often observed from the Sun during strong solar flares \citep[e.g.,][see also Sec.\,\ref{sec:SFs}]{smith2003,kiener2006}.
It has furnished valuable information on solar ambient abundances, density and temperature, as well as on accelerated particle composition, spectra and transport in the solar atmosphere \citep[e.g.,][]{turler2021}.
Similarly, interactions of Galactic \ac{CR} ions with interstellar matter should produce a diffuse \ac{gray} line emission that can shed new light on the sources of \acp{LECR} ($E_{\rm kin} \lesssim 1$\,GeV/nucleon) in the Galaxy and the isotopic composition of the \ac{ISM}.

\acp{LECR} are thought to be responsible for the nucleosynthesis of the light elements lithium, beryllium and boron (LiBeB) from spallation nuclear reactions with nuclei of the \ac{ISM}.
This process is important to explain the observed evolution of the abundances of the LiBeB isotopes throughout the lifetime of the Galaxy.
Remarkably, the observed quasi-linear increase of the \nuc{Be}{9} abundances measured in stellar atmospheres with the star metallicity provides evidence for the existence of a significant component of  \ac{LECR} nuclei in the Galaxy.
This has to be in addition to the standard \acp{CR} thought to be produced by diffusive shock acceleration in \ac{SN} remnants \citep{tatischeff2018}.

\acp{LECR} are also thought to play a key role in the chemistry and dynamics of the \ac{ISM}.
They are a primary source of ionisation of heavily shielded, dense molecular clouds and the resulting ionisation fraction conditions both a rich ion-neutral chemistry in these regions and the coupling of the gas with the ambient magnetic field.
\acp{LECR} also represent an important source of heating that contribute to hold molecular cores in equilibrium against gravitational forces.
Hence, \acp{LECR} play a central role in the process of star formation.
Despite \acp{LECR} being a fundamental component of the Galactic ecosystem, their composition and flux are very uncertain Galaxy-wide.
They are better known in the local \ac{ISM} thanks to the valuable \ac{CR} measurements of the Voyager\,1 and Voyager\,2 probes as they crossed the heliopause into interstellar space \citep{cummings2016,stone2019}.
But the total \ac{CR} ionisation rate of atomic hydrogen resulting from the measured \ac{CR} spectra, $\zeta_{\rm H} = (1.51$--$1.64) \times 10^{-17}\,\mathrm{s^{-1}}$, is a factor $>10$ lower than the average \ac{CR} ionisation rate measured in clouds across the Galactic disk using Herschel observations, $\zeta_{\rm H} = 1.78 \times 10^{-16}\,\mathrm{s^{-1}}$ \citep{indriolo2015}.
This suggests that \acp{LECR} are relatively less abundant in the local \ac{ISM} than elsewhere in the Galaxy.
Observations of H$_3^+$ in diffuse clouds show indeed that the density of \acp{LECR} can strongly vary from one region to another in the Galactic disk \citep{indriolo2012}.

\begin{figure}[!ht]
\centering
\includegraphics[width=0.85\textwidth]{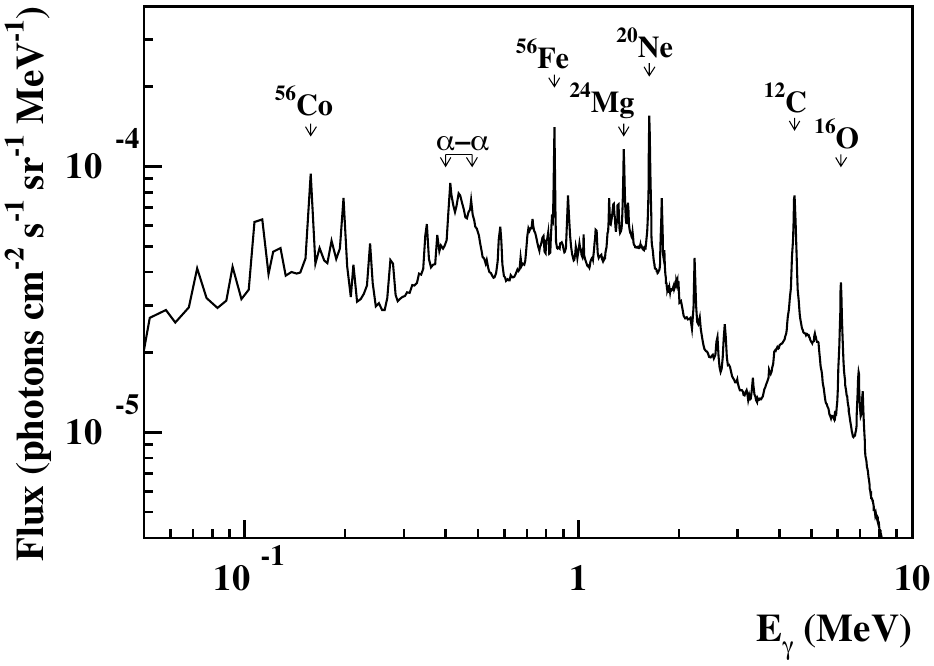}
\caption{Calculated \ac{gray} line emission produced by \acp{LECR} in the inner Galaxy (longitude $300^\circ < \ell < 60^\circ$, latitude $|b| < 10^\circ$). For some of the most significant lines, the emitting nuclei are indicated. The broad feature at $\sim 450$\,keV is called the $\alpha$--$\alpha$ line and is mainly produced by $\alpha$+$^4$He fusion reactions \citep[adapted from][]{tatischeff2011}.}
\label{fig:gray_spectrum}
\end{figure}

\subsubsection{MeV $\gamma$-Ray Line Observations and Predictions}
\paragraph{Diffuse Emission -- The Galactic Plane}
MeV \ac{gray} astronomy can provide the most direct way of studying the density and effects of \acp{LECR} in the \ac{ISM}, through the observation of nuclear de-excitation lines.
Fig.\,\ref{fig:gray_spectrum} shows a calculated \ac{gray} line spectrum from  \ac{LECR} interactions with ambient matter in the inner Galaxy, based on Eq.\,(\ref{eq:nuc_ex_emissivity}).
The density and energy spectrum of \acp{LECR} have been set to produce the mean ionisation rate of diffuse molecular clouds in the inner Galaxy \citep[see][and references therein]{tatischeff2011}.
We can see in Fig.\,\ref{fig:gray_spectrum} a number of relatively narrow lines, which are mainly produced by excitation of abundant heavy nuclei of the \ac{ISM}, such as \nuc{C}{12}, \nuc{O}{16}, \nuc{Ne}{20}, \nuc{Mg}{24}, and \nuc{Fe}{56}, by \ac{CR} protons and alpha particles of kinetic energies between a few MeV/nucleon and a few hundred MeV/nucleon.
The emitting nuclei are mainly excited by inelastic scattering or spallation reactions, but some lines are also produced by charge-exchange reactions, such as the one at 158\,keV produced by the reaction $^{56}$Fe($p$,$n$)$^{56}$Co$^*_{158}$ \citep[see][for a compilation of the most significant \ac{gray} lines produced by nuclear collisions, with the involved nuclear reactions and their cross sections]{ramaty1979,murphy2009}.
Besides strong narrow lines, the total nuclear line emission is also composed of broad lines produced by interaction of \ac{CR} heavy ions with ambient H and He, and of thousands of weaker lines that together form a quasi-continuum in the range $E_\gamma \sim 0.1$--$10$\,MeV \citep{murphy2009,ben13}.
The $\alpha$--$\alpha$ line feature at $\sim 0.45$\,MeV mainly results from the merging of two prompt lines from the reactions $^4$He($\alpha$,$n$$\gamma_{429}$)$^7$Be and $^4$He($\alpha$,$p$$\gamma_{478}$)$^7$Li and the delayed line at 478\,keV from \nuc{Be}{7} decay ($T_{1/2} = 53.3$\,d).

The narrow lines produced in a gaseous ambient medium are generally broadened by the recoil velocity of the excited nucleus and their \ac{FWHM} is about 0.5--5\% of the transition energy.
However, some lines produced in interstellar dust grains can be very narrow, because some of the excited nuclei can stop in solid materials before emitting \acp{gray} \citep{tatischeff2004}.
The most promising of such lines are from the de-excitation of the level of \nuc{Fe}{56} at 847\,keV ($T_{1/2} = 6.1$\,ps), of \nuc{Mg}{24} at 1369\,keV ($T_{1/2} = 1.35$\,ps), of \nuc{Si}{28} at 1.779\,keV ($T_{1/2} = 475$\,fs) and of \nuc{O}{16} at 6.129\,keV ($T_{1/2} = 18.4$\,ps)\footnote{The effect of \ac{ISM} grains is not taken into account in Fig.\,\ref{fig:gray_spectrum}.}.
Most of the interstellar Fe, Mg and Si are contained in dust grains, whereas about 30\% of the interstellar O could be in grains \citep{jones2017}.

\begin{figure}[!ht]
\centering
\includegraphics[width=0.8\textwidth]{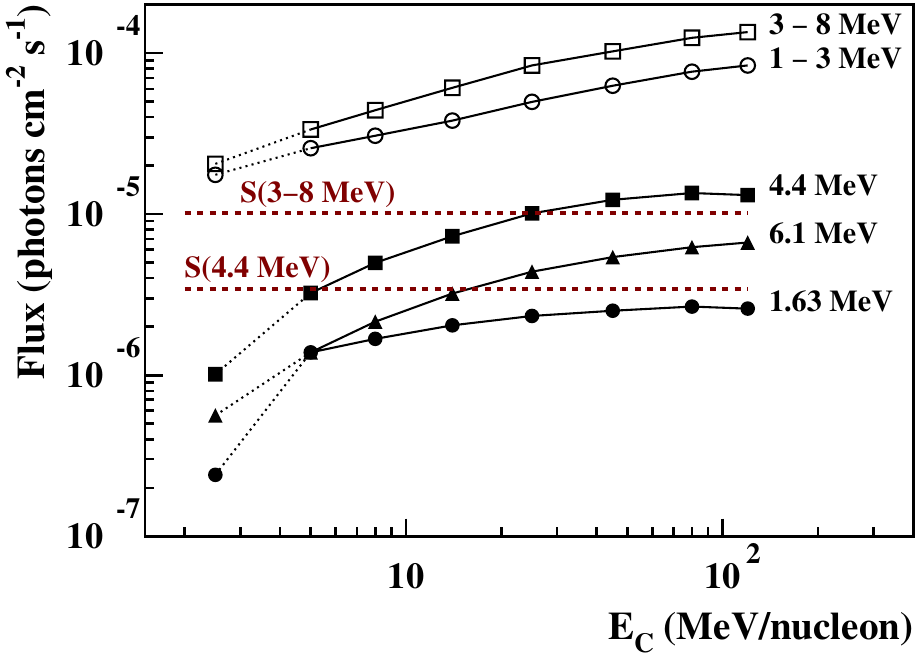}
\caption{Calculated narrow-line fluxes of the 1.63\,MeV (\nuc{Ne}{20}; filled circles), 6.1\,MeV (\nuc{O}{16}; triangles) and 4.4\,MeV lines (\nuc{C}{12}; filled squares), and integrated fluxes in the $1$--$3$\,MeV range (open circles) and $3$--$8$\,MeV range (open squares), from \acp{LECR} in the inner Galaxy (see also Fig.\,\ref{fig:gray_spectrum}), as a function of the cut-off energy of the \ac{CR} source spectrum (Eq.\,(\ref{eq:lecr_spectrum})). The fluxes given at $E_C = 2.5$\,MeV/nucleon are produced by a \ac{CR} population consistent with the measurements of the Voyager probes in the local \ac{ISM}. The other data points assume an additional  \ac{LECR} component accounting for the mean ionisation rate of diffuse clouds in the inner Galaxy deduced from H$_3^+$ observations (see text). The red dashed lines show the e-ASTROGAM sensitivity in two years of survey observations for the diffuse emissions in the 4.4\,MeV line and the $3$--$8$\,MeV range \citep[see][]{eastrogam_science17}.}
\label{fig:line_flux}
\end{figure}

In Fig.\,\ref{fig:gray_spectrum}, the most intense narrow lines, in descending order, are the 4.4\,MeV line mainly from \nuc{C}{12} (with a contribution from \nuc{B}{11} produced by spallation), the 6.1\,MeV line mainly from \nuc{O}{16} (with a contribution from \nuc{N}{15} and \nuc{O}{15}) and the 1.63\,MeV line from \nuc{Ne}{20}.
Calculated fluxes in these three lines are shown in Fig.\,\ref{fig:line_flux}, together with integrated fluxes in the energy ranges 1--3\,MeV and 3--8\,MeV, as a function of the cut-off energy $E_C$ of the  \ac{LECR} source spectrum.
Here, it is assumed to be the same for all species and to result from diffusive shock acceleration \citep[see][]{ben13}:
\begin{equation}
{dN_i \over dt}(E)={C_i R_i^{-s} \over \beta} e^{-E/E_C}\mathrm{,}
\label{eq:lecr_spectrum}
\end{equation}
where $C_i$ is the abundance of nuclei of type $i$ in the  \ac{LECR} composition, $R_i$ is the particle rigidity and $\beta=v/c$ the particle velocity in units of the speed of light.
We see in Fig.\,\ref{fig:line_flux} that a future \ac{gray} observatory such as e-ASTROGAM, which was proposed for ESA's M5 mission \citep{eastrogam_science17}, could detect the characteristic bump between 3 and 8\,MeV (see Fig.\,\ref{fig:gray_spectrum}) independent of the actual density of \acp{LECR}in the inner Galaxy.
The detection of individual \ac{gray} lines, such as the one of \nuc{C}{12} at 4.4\,MeV, also seems feasible if low-energy protons contribute significantly to the ionisation of diffuse molecular clouds.
The diffuse \ac{gray} line emission from \acp{LECR}is expected to be concentrated in the Galactic plane, similar to the high-energy diffuse emission observed by \textit{Fermi}-LAT, of which approximately 60\% of the total intensity is within a band of $\pm 1.5^\circ$ in latitude.
The inverse Compton radiation from accelerated electrons, which is also expected to contribute significantly to the diffuse Galactic emission in the MeV range, is predicted to be much more extended, to latitudes $|b|> 10^\circ$ \citep{ackermann2012}.
Such a difference in spatial distribution should help to identify the nuclear \ac{gray} line component within the total Galactic diffuse emission.
A Compton-imaging instrument such as e-ASTROGAM, having an angular resolution of $0.7^\circ$ \ac{FWHM} at 5\,MeV, seems appropriate for such a detection.

Also here, it is interesting to note that a small fraction, up to several percent, of the 1.8\,MeV \ac{gray} line flux may actually come from \ac{LECR} excitation of stable \nuc{Mg}{26} in the Galaxy \citep{Weinberger2021}.

\paragraph{Point Sources -- Supernova Remnants}
The \ac{CR} ionisation rates in dense molecular clouds close to \ac{SN} remnants can be very high -- more than 100 times the standard value in average molecular clouds.
This has been evidenced by measurements of molecular abundance ratios such as DCO$^+$/HCO$^+$ and HCO$^+$/CO in clouds near the \ac{SN} remnants W51C \citep{ceccarelli2011}, W28 \citep{vaupre2014}, and W49B \citep{zhou2022}.
This suggests that intense fluxes of \acp{LECR} could escape \ac{SN} remnants and interact with neighbouring molecular clouds, thus producing a potentially observable emission of nuclear \ac{gray} lines.
The young \ac{SN} remnant \ac{CasA} also appears to be a promising source, as it is the remnant of a massive star explosion $\sim 350$\,yr ago where accelerated nuclei could still interact with an ambient gas enriched in heavy ions from the progenitor winds and the \ac{SN} ejecta \citep{summa2011}.
Predicted fluxes in the 4.4 and 6.1\,MeV lines strongly depend on model assumptions on the accelerated particle spectrum and ambient medium composition, and they could be within the reach of the next generation of MeV \ac{gray} observatories in certain cases \citep{liu2023}.

% SFs
\newpage
\subsection{Solar Flare $\gamma$-Rays}\label{sec:SFs}
\vspace{-0.5em}
{\emph{Written by Gerald H. Share}}
\vspace{0.5em}
\vspace{-12pt}
\subsubsection{Particle Acceleration in Flares} 
Electron and ion acceleration to MeV energies and above occurs in settings throughout the Universe.
These particles interact with matter, magnetic fields, and electromagnetic radiation to produce \acp{gray} that enable studies of explosive phenomena that occurred up to billions of years ago.
As this volume focuses on the Universe at MeV energies, it is helpful to understand what we have learned when we look at our closest explosive neighbour, the Sun, in this band of the electromagnetic spectrum. 
\begin{figure*}[!ht]
    \centering
    \includegraphics[width=0.42\linewidth,trim=0.005cm 0.3cm 0.0cm 0.2cm,clip]{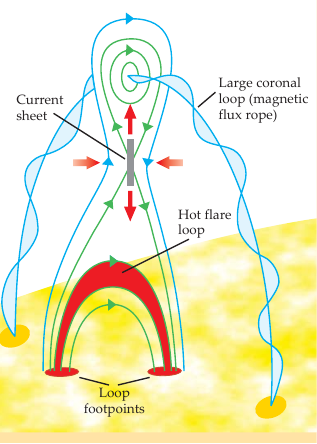} 
    \caption{Cartoon depicting how reconnecting Solar magnetic field lines provide energy to accelerate particles to relativistic energies and release \acp{CME} \citep[from][]{holm12}.  
    \label{holm}}
\end{figure*}
The Sun accelerates particles in eruptions and flares as a result of energy produced by the reconnection of coronal magnetic fields \citep{chen20}.
The highest magnetic field strengths at the Sun are a few thousand Gauss, as compared with the $10^{15}$\,G fields that accelerate particles in pulsars.
Fig.\,\ref{holm} is a simple picture of the Solar phenomenon in which reconnecting magnetic fields energise the coronal plasma to accelerate electrons and ions that become trapped in magnetic loops reaching into the Solar atmosphere where they interact at the flare foot points to produce hard X-rays and \acp{gray} \citep{vilm11}.
Ions and electrons from the same acceleration process can be ejected into interplanetary space as impulsive \acp{SEP} \citep{ream21}. 
The flare energisation process also forms magnetic flux ropes that erupt in mountain-sized \acp{CME}.
Shocks from these $\gtrsim 500\,\mathrm{km\,s^{-1}}$ \acp{CME} can further accelerate supra-thermal ions from the flare and in the Solar wind to GeV energies \citep{ream21}.
These \acp{SEP} also reach interplanetary space and Earth where they can be hazardous for astronauts, spacecraft systems, and passengers on aircraft flying polar routes.

\subsubsection{The Solar $\gamma$-Ray Spectrum} 
\begin{figure*}[ht!]
\centering
\includegraphics[width=0.7\linewidth,trim=0.5cm 0.0cm 0.0cm 0.2cm,clip]{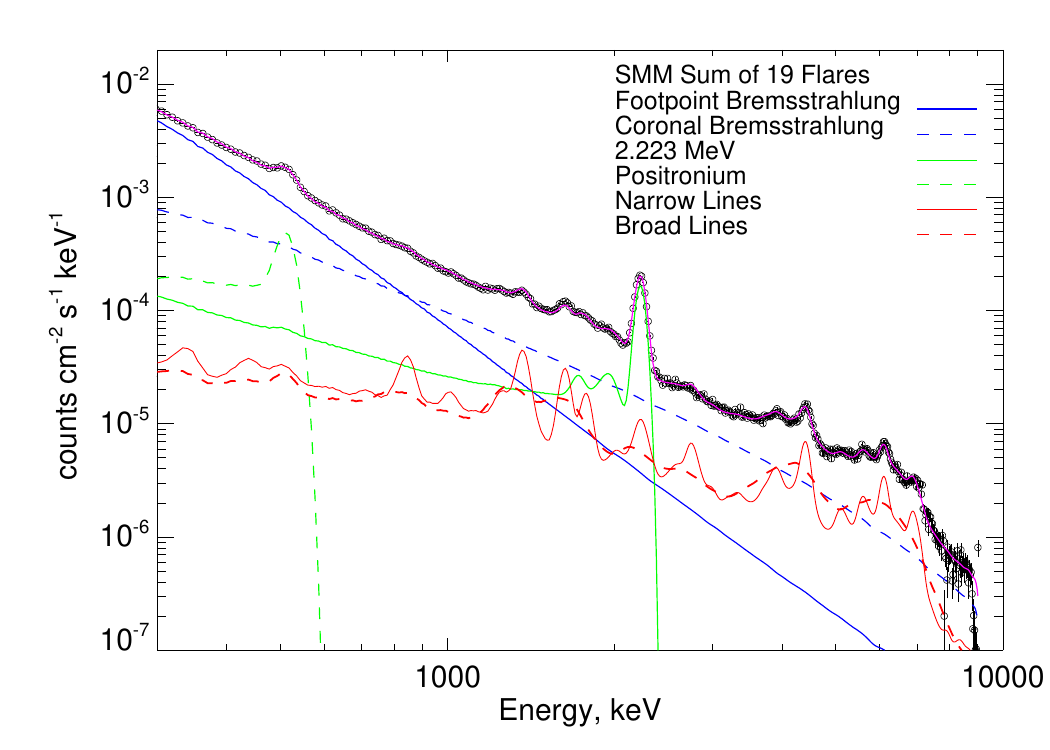} 
\caption{Summed background-subtracted spectrum from 19 Solar flares observed by \ac{SMM}/\ac{GRS}.  
\label{19flr}}
\end{figure*}

To understand the acceleration of electrons and ions with energies higher than a few hundred keV at the Sun, it is necessary to study the \ac{gray} spectra observed during Solar flares.
In addition to the high-resolution \ac{gray} spectra measured by the germanium detectors on the \ac{RHESSI} \citep{smit02}, some of the best Solar flare spectra were made with moderate resolution NaI detectors flown on the \ac{SMM}/\ac{GRS} \citep{forr80}.
%
%This same instrument made pioneering spectral measurements of \acp{SN} \cite{leis90}, high-energy \acp{GRB} \cite{shar86}, the diffuse Galactic \nuc{Al}{26} \cite{shar85}, and the 511\,keV annihilation line \cite{shar88, shar90}.
%
Its excellent spectral stability allowed us to sum up spectra from 19 large Solar flares from 0.3 to 8.5\,MeV that were observed over a nine-year period from 1980 to 1989.
We show this summed spectrum in Fig.\,\ref{19flr}.
It reveals the significant continuum and line features of \acp{gray} emitted in Solar flares.
Note that because this is the raw count spectrum, it contains not only the lines but also the escape peaks and low-energy continua from partial energy losses in the instrument.
Using seven years of data from the same spectrometer, \citet{harr92} showed the Galactic spectrum in the direction of the Galactic Centre, revealing a striking power-law continuum and the annihilation and \nuc{Al}{26} lines.
The spectral index of the continuum plotted in their Fig.\,3 is consistent with more recent measurements \citep[e.g.,][]{sieg22}.

\subsubsection{The $\gamma$-Ray Continuum from Electrons}
The electron-produced Solar \ac{gray} continuum is comprised of at least two components \citep{shar25}:
The solid blue curve represents the bremsstrahlung from the flare foot points produced by a power-law distribution of electrons.
This radiation is anisotropic and can extend up to tens of MeV in some flares.
The dashed blue curve represents the hard emission from a recently discovered population of electrons that peaks at energies between 3--5\,MeV if they radiate by bremsstrahlung.
If the electrons radiate by inverse Compton scattering of soft flare X-ray photons, their spectrum is flat and extends up to 10--20\,MeV.
This hard component of electrons, is observed in flares 10--100 times weaker than the 19 nuclear-line flares whose summed spectrum is plotted in Fig.\,\ref{19flr}.
In contrast to the electrons in the power-law component interacting at the foot points, these electrons appear to be isotropic and radiate from the Solar corona.

\subsubsection{Neutron-Capture and Annihilation Lines} 
The 2.223\,MeV line (green solid curve) is the most striking line feature in the flare spectrum shown in Fig.\,\ref{19flr}.
It is formed when neutrons, produced when flare-accelerated ions interact in the Solar atmosphere, slow down and are captured by hydrogen to produce deuterium with the binding energy appearing as a few eV wide line \citep{hua87}.
The resulting delay in emission of the line is dependent of the abundance of \nuc{He}{3} because it also captures neutrons, but without emitting radiation \citep{wang74}.
Measurement of the \nuc{He}{3}/\nuc{H}{1} ratio in the Sun using the 2.223\,MeV time history is uncertain, ranging from $(0.5$--$10) \times 10^{-5}$ \citep{murp03}, compared with the $1 \times 10^{-5}$ primordial ratio \citep{bani21}.
Because the 2.223\,MeV line is formed near photospheric depths, it is strongly attenuated for flares near the Solar limb.
A Compton-scattered continuum from the \ac{gray} line emission leaving the Solar atmosphere is detectable by instruments with large photopeak efficiencies (Murphy et al., 2025, in prep.).

\ac{RHESSI} imaging of the foot points of the electron bremsstrahlung and the neutron-capture line reveals that flare-accelerated ions and electrons interact at the same location \citep{batt25} in at least one flare.
The close correlation between the $\gtrsim 300$\,keV bremsstrahlung fluence and the 2.223\,MeV line fluence over three orders of magnitude indicates that hundreds of keV electrons and $\gtrsim 5$\,MeV protons have a related origin \citep{shih09}.
\citet{shar25} have used the relatively weak 2.223\,MeV line flux in flares $\sim 100$ times less intense than the nuclear-line flares to conclude that the accelerated ion spectra are significantly softer in these flares.
From upper limits on the quiescent 2.223\,MeV line flux, it was found that the Solar corona is not heated by $\gtrsim 1$\,MeV protons in microflares \citep{harr92}.
The authors set a limit on tritium production at the Sun, and proved that large nuclear line flares are not the result of storage of quiescently accelerated protons in coronal magnetic fields.

The second most striking feature in the flare spectrum is the 511\,keV line and lower energy continuum (green dashed curve) formed when positrons produced in $\beta^+$-decays of excited nuclei and $\pi^+$-decays annihilate with ambient electrons producing two or three \acp{gray} \citep{murp05}.
\ac{RHESSI} high-resolution measurements of the 511\,keV line width and \ac{Ps} fraction provide information on the conditions of the ambient environment where the positrons annihilate \citep{shar04}.
There are times during the flares observed on 2002 July 23, 2003 October 28, 2003 November 2, and 2005 January 20 in which the Solar 511\,keV line is significantly broadened.
These measurements suggest that the ambient chromosphere at densities of $\sim 10^{15}\,\mathrm{cm^{-3}}$ may, at times, be heated to temperatures reaching $\sim 5 \times 10^5$\,K during Solar flares. 
There is also a variable \ac{Ps} continuum observed in flares measured by the \ac{RHESSI} and \ac{SMM}/\ac{GRS} spectrometers.
%
%We note that the latter spectrometer also measured the Galactic \ac{Ps} fraction \cite[see also Sec.\,\ref{sec:511}][]{harr95}. 

\subsubsection{Nuclear De-Excitation Lines} 
A variety of nuclear lines is produced when flare-accelerated protons, $\alpha$-particles, and heavier ions interact with ambient elements in the Sun's chromosphere \citep{rama79}.
The dominant ambient elements responsible for the lines are helium, carbon, nitrogen, oxygen, neon, magnesium, silicon, and iron.
Fusion of $\alpha$-particles and He produce detectable Be and Li lines between 431 and 478\,keV, respectively.
Interactions of protons and $\alpha$-particles with ambient nuclei impart momentum to them producing Doppler-broadened and -shifted narrow lines (solid red curves).
The widths (\ac{FWHM}) of these lines range from $\sim 8$\,keV for the iron line at 847\,keV to $\sim 100$\,keV for the carbon line at 4.43\,MeV.
Interactions of accelerated heavy ions with chromospheric hydrogen and helium produce significantly more shifted and broadened lines (dashed red curves).
The widths of these lines range from $\sim 80$\,keV for iron to $\sim 500$\,keV for carbon.
The most significant narrow lines found in the \ac{SMM} spectrum and shown in Fig.\,\ref{19flr} are listed in Tab.\,\ref{tab:sf_lines} \citep[see also][]{murp09}. 
The interactions producing these lines mostly occur at chromospheric depths.
\begin{table}[!h]
    \centering
    \begin{tabular}{rrr}
        Line Energy [MeV] & Target Nucleus & Excited Nucleus  \\
        \hline
        0.339 & \nuc{Fe}{56} & \nuc{Ni}{59} \\
        0.431 & \nuc{He}{4} & \nuc{Be}{7} \\
        0.478 & \nuc{He}{4} & \nuc{Li}{7} \\
        0.847 & \nuc{Fe}{56} & \nuc{Fe}{56} \\
        1.014 & \nuc{Mg}{24}, \nuc{Si}{28} & \nuc{Al}{27} \\
        1.022 & \nuc{C}{12}, \nuc{O}{16} & \nuc{B}{10} \\
        1.238 & \nuc{Fe}{56} & \nuc{Fe}{56} \\
        1.369 & \nuc{Mg}{24} & \nuc{Mg}{24} \\
        1.634 & \nuc{Ne}{20} & \nuc{Ne}{20} \\
        1.779 & \nuc{Si}{28} & \nuc{Si}{28} \\
        2.313 & \nuc{N}{14} & \nuc{N}{14} \\
        2.754 & \nuc{Mg}{24} & \nuc{Mg}{24} \\
        4.439 & \nuc{C}{12} & \nuc{C}{12} \\
        6.129 & \nuc{O}{16} & \nuc{O}{16} \\
        6.176 & \nuc{O}{16} & \nuc{O}{16} \\
        6.917 & \nuc{O}{16} & \nuc{O}{16} \\
        15.111 & \nuc{C}{12} & \nuc{C}{12} \\
        \hline        
    \end{tabular}
    \caption{Prominant Solar Nuclear De-excitation Lines.}
    \label{tab:sf_lines}
\end{table}
For comparison with celestial line studies, using over nine years of data from \ac{SMM}/\ac{GRS}, \citet{harr95} has set a 3$\sigma$ limit of $8.7 \times 10^{-5}\,\mathrm{ph\,cm^{-2}\,s^{-1}\,rad^{-1}}$ on the summed \nuc{C}{12} plus \nuc{O}{16} narrow-line flux from the Galactic Centre region.
This limit is as much as a factor ten higher than that calculated from \ac{CR} interactions with interstellar carbon \citep{rama79}.

There have been many spectroscopic studies of nuclear de-excitation lines in Solar-flare spectra based on observations by \ac{SMM}/\ac{GRS}, Yohkoh, Konus-Wind, \ac{CGRO}, \ac{RHESSI}, CORONAS-F/SONG, and Fermi/GBM \citep[e.g.,][]{shar95,rama96,murp97,mand99,mand00,smit02,vilm11,acke12,lyse19,yush23}.
These studies addressed ambient abundances where flare-accelerated ions interacted, and the spectra and composition of these ions.
Unfortunately, the results obtained were of limited accuracy because the spectra were typically fit by a single electron-produced continuum and by Gaussian fits to the narrow and broad de-excitation lines.
What is needed to obtain accurate information on the ambient composition where flare-accelerated particles interact and the abundances of these particles are calculated nuclear \ac{gray} line spectra for a variety of accelerated particle spectra and compositions, and ambient abundances.
\citet{murp91} performed a limited analysis of this type for the 1981 April 27 flare observed by \ac{SMM} using the seminal work of \citet{rama79}.

Significant improvement in our knowledge about the Solar nuclear lines has come with detailed studies of particle acceleration onto magnetic loops \citep{hua89,murp07} and from accurate knowledge of the production of broad and narrow nuclear de-excitation lines and unresolved continuum \citep{koz02,murp09,murp16,tuns19}.
The end product of this work is an array of nuclear \ac{gray} spectral templates for each of the processes for different ambient and accelerated particle abundances, flare locations, and ion spectral indices.
Access to these templates is provided by OSPEX\footnote{Object Spectral Executive: \url{https://hesperia.gsfc.nasa.gov/ssw/packages/spex/doc/ospex_explanation.htm}} that is available in the SSW IDL software depository\footnote{\url{http://www.lmsal.com/solarsoft/ssw_whatitis.html}.} 
We have also identified issues with how the original \ac{SMM}/\ac{GRS} and \ac{RHESSI} detector response matrices accounted for \acp{gray} that do not lose all of their energy in the instrument.
The response matrices of both instruments have been corrected and incorporated into OSPEX \citep{shar25}.

There are some studies that were not as dependent on improvements in the analysis techniques discussed above.
\citet{kien06,kien19} analysed \ac{INTEGRAL} high-resolution measurements of the 4.43\,MeV \nuc{C}{12} line in the 2003 October 28 Solar flare and found that relatively narrow-downward directed distributions of accelerated protons and $\alpha$-particles provided the best fits.
Assuming a model where particles are trapped in magnetic loops with mirroring and pitch angle scattering, the best fitting parameter, $\lambda$ (the scattering mean free path divided by the half-length of the coronal segment of the magnetic loop), was consistent with a value of $30$ and not with a value of $300$.
%
%This is a somewhat narrower angular distribution of particles than found in Doppler red-shift measurements of the centroids of the high-\ac{FIP} lines with heliocentric angle in \ac{SMM} flares \cite{shar02} and from \ac{RHESSI} measurements of the $\alpha$--\nuc{He}{4} line in the 2002 July 23 flare \cite{smit03}.
%
This is a somewhat narrower angular distribution of particles than found in Doppler red-shift measurements of the centroids of the high-\ac{FIP} lines with heliocentric angle in \ac{SMM} flares \citep{shar02} and from \ac{RHESSI} measurements of the $\alpha$--\nuc{He}{4} \citep{shar03a} and de-excitation lines \citep{smit03} in the 2002 July 23 flare.
The latter observation and later \ac{INTEGRAL} observations \citep{harr07} suggested that the Solar magnetic loops were tilted by as much as $40^\circ$ from the normal.
\citet{kien19} also derived an accelerated $\alpha$/$p$ ratio of $0.22^{+0.20}_{-0.13}$, consistent with \ac{SMM} measurements \citep{shar97}.
We note that a comparison of the fluxes in the 339 and 847\,keV lines provide a direct measurement of the $\alpha$/$p$ ratio because the 339\,keV line is produced by $\alpha$-particle interactions.

A study of the global energetics of Solar eruptions \citep{emsl12} revealed that  ``\textit{the energy content in flare-accelerated electrons and ions is sufficient to supply the bolometric energy radiated across all wavelengths throughout the event}'' and that ``\textit{the energy contents of flare-accelerated electrons and ions are comparable}''.
This is a significant finding as it shows how important particle acceleration is in the release of energy in flares.
We note that the ion energies were only calculated above $\sim 1$\,MeV, therefore, the estimated energy in ions may be significantly higher if the emission extends to lower energies.
Determining the spectra of interacting flare-accelerated protons below 1\,MeV appears to be a formidable problem \citep{shar01a}.

With our increased understanding of the $\gamma$-radiation produced by ion interactions in the chromosphere, \citet{murp16} has provided an example of what can now be achieved in planned studies.
They used \ac{gray} line flux ratios determined from an archival fit to the summed spectrum plotted in Fig.\,\ref{19flr} to conclude that there is only a $10^{-3}$ probability that the \nuc{He}{3} abundance in flare-accelerated particles could be as small as that found in the Sun's photosphere.
They obtained \nuc{He}{3}/$\alpha$ ratios ranging from 0.05 to 0.3, depending on the assumed accelerated $\alpha$/$p$ ratio.
With the improvements in our analysis of Solar flares discussed above, we have analysed the spectra from the 19 \ac{SMM} flares and the 2002 July 23 flare observed by \ac{RHESSI}.
Planned studies include: (1) elemental composition of high-\ac{FIP} elements, \nuc{C}{12}, \nuc{N}{14}, \nuc{O}{16}, and \nuc{Ne}{20}, (2) composition and variability of the low-\ac{FIP} elements, \nuc{Mg}{24}, \nuc{Si}{28}, and \nuc{Fe}{56}, and (3) spectral variations and abundances of accelerated particles, including \nuc{He}{3}.  

\vspace{1em}
\noindent \textit{My thanks to Brian Dennis and Ronald Murphy for reviewing the manuscript.}

\newpage
\subsection{Solar System Bodies and their $\gamma$-Ray Albedos}\label{sec:sssbs}
\begin{figure}[htbp]
  \centering

  % Left column with two stacked images (image1 and image3)
  \begin{minipage}[t]{0.384\textwidth}
    \vspace{0pt}
    \includegraphics[width=\linewidth]{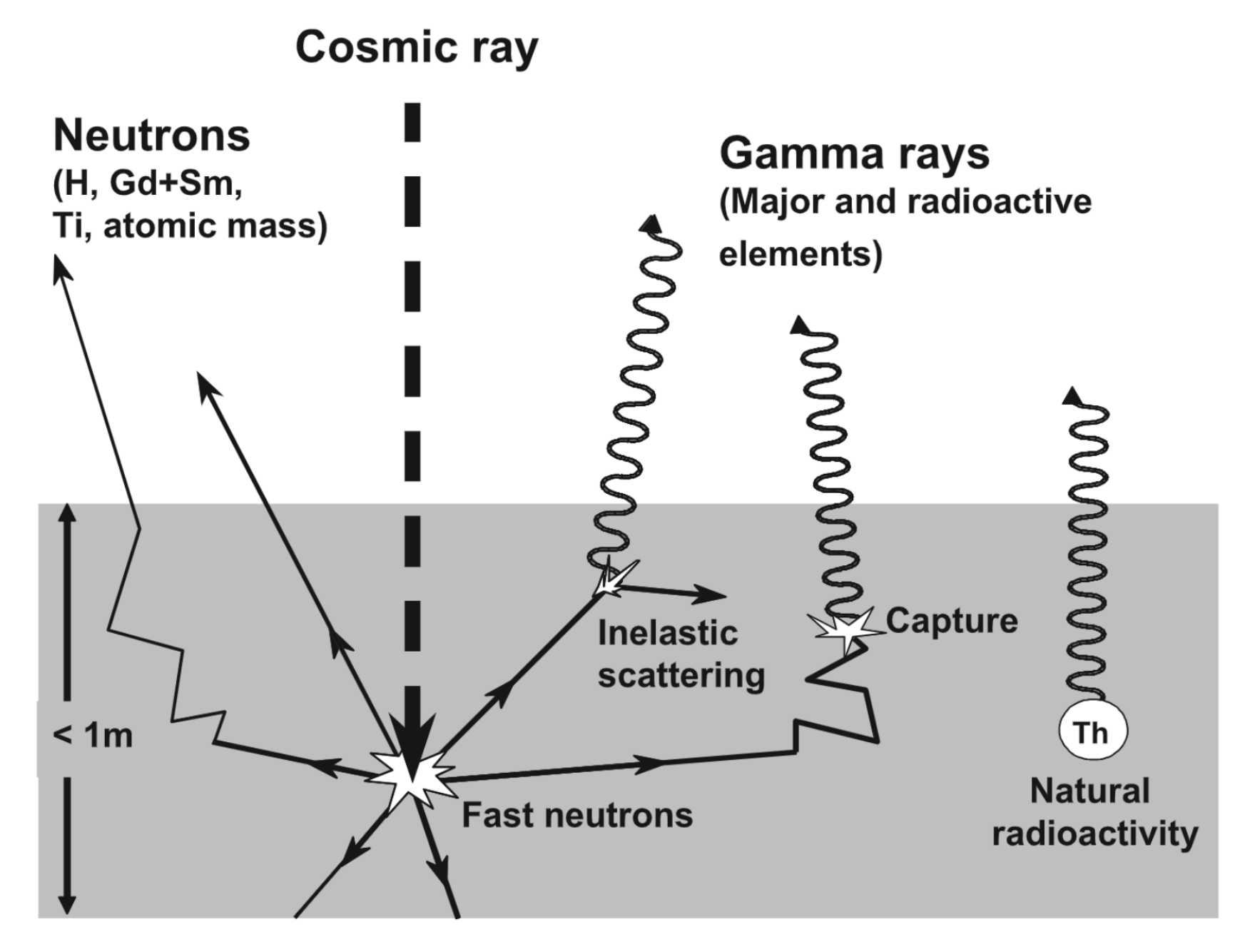} \\[1ex]
    \includegraphics[width=\linewidth]{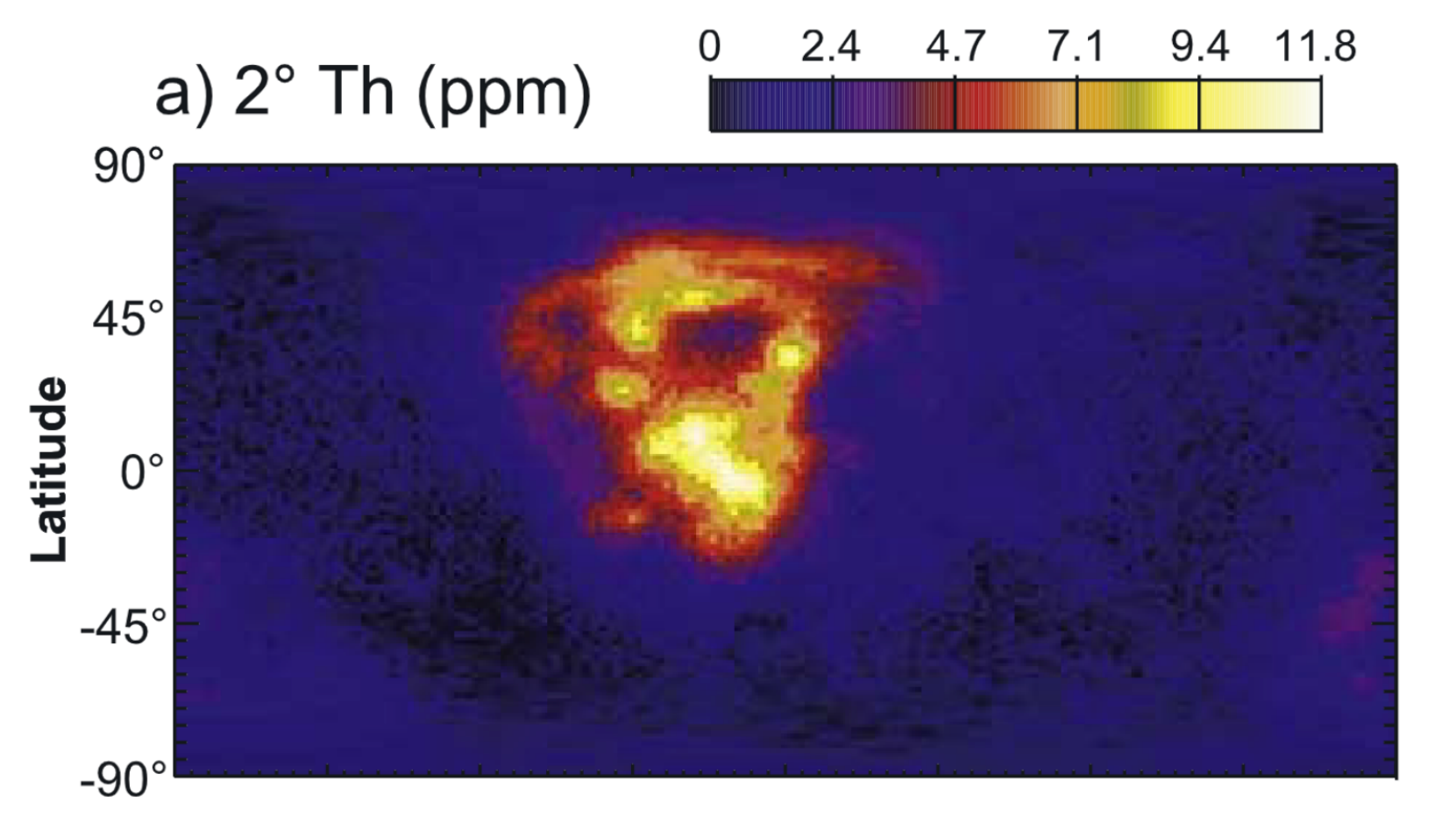}
  \end{minipage}
  %\hfill
  % Right column with one tall image (image2)
  \begin{minipage}[t]{0.384\textwidth}
    \vspace{0pt}
    \includegraphics[width=\linewidth,height=0.98\textheight,keepaspectratio]{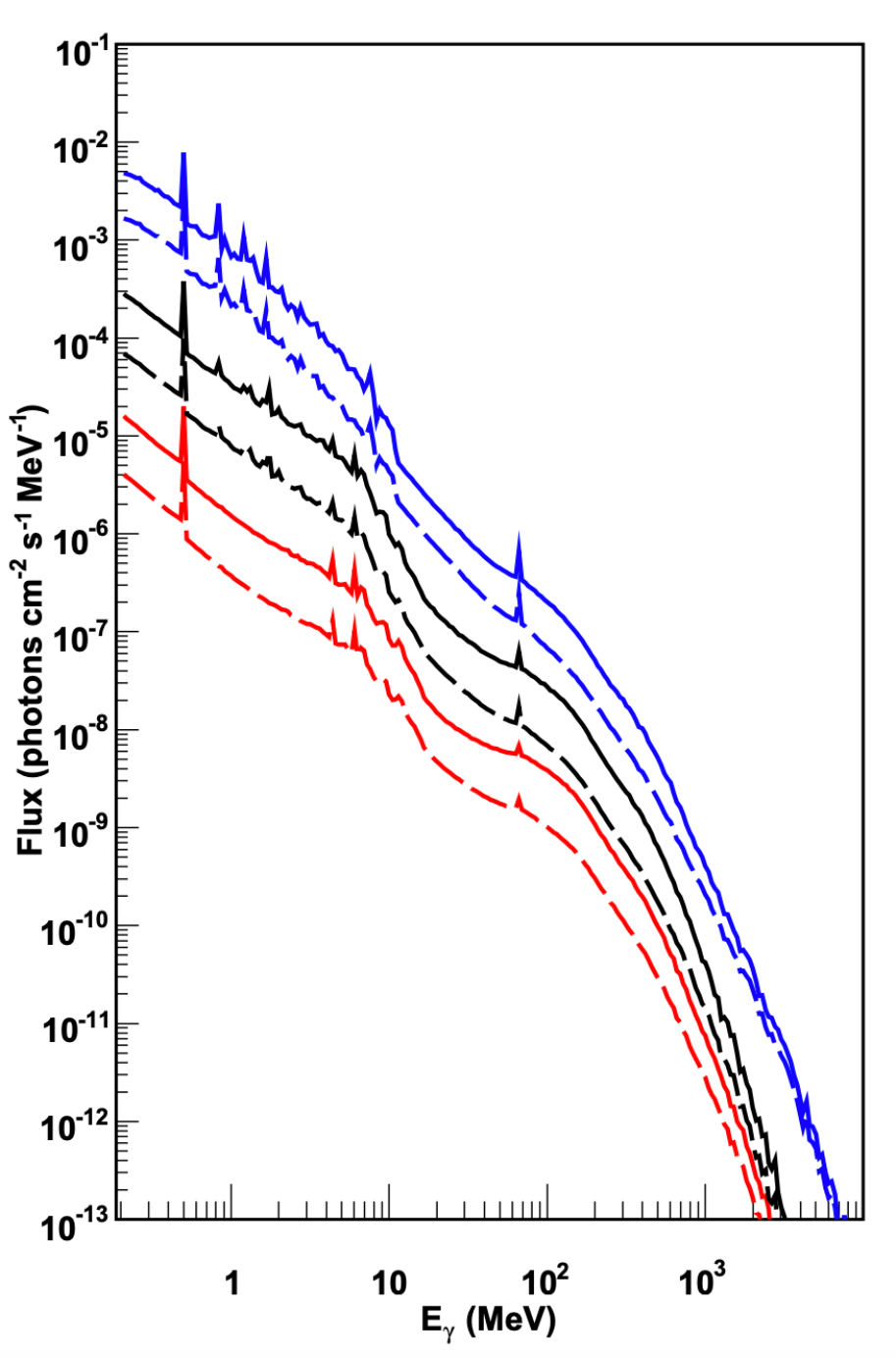}
  \end{minipage}

  \caption{Top left: Expected processes leading to \ac{gray} emission from the Lunar or asteroids surfaces \citep[from][]{Prettyman2006}; in atmospheres, the processes are similar. Bottom left: \emph{Measured} natural radioactivity from the Lunar surface in Th lines from the \ac{LP} mission \citep[from][]{Prettyman2006}. Right: Expected \ac{gray} spectrum from a Moon-sized body at the distance of the Moon with different compositions (black: Moon rock; blue: iron $\times 10$; red: water ice $\times 0.1$) for two different modulation potentials (solid: no modulation; dashed: $1500$\,MV) \citep[from][]{Moskalenko2008}.}
  \label{fig:gamma-ray_albedo}
\end{figure}

\vspace{-0.5em}
{\emph{Written by Thomas Siegert}}
\vspace{0.5em}\\
Solid bodies, with or without atmospheres, acquire a \ac{gray} albedo when bombarded by \acp{CR}.
When \acp{LECR} ($\lesssim 1$\,GeV) interact with nuclei, the nuclei become excited and almost immediately de-excite by emitting characteristic \acp{gray} (see Eq.\,(\ref{eq:nuc_ex_emissivity}) and Sec.\,\ref{sec:nuc_ex}).
In solid bodies, unlike in the \ac{ISM}, the penetration depth of \acp{LECR} and the resulting opacity for the emitted \acp{gray} must be considered.
Thus, not every \ac{LECR} produces a measurable photon, and the overall \ac{gray} spectrum is soft \citep{Moskalenko2007}.
Fig.\,\ref{fig:gamma-ray_albedo} (top left) illustrates the production mechanisms for \ac{gray} photons, particularly \ac{gray} lines.
Natural radioactivity is generally observed from \nuc{K}{40} (1460.8\,keV, $\tau = 1.8$\,Gyr) and, in heavier materials, from actinide alpha decay chains; elements such as U, Th, and Sm and Gd (as fission products) are dominant \ac{gray} line emitters in Moon rock \citep{Prettyman2006}.
Most Th and U lines occur below 3\,MeV.
When a \ac{CR} impacts a solid body, it produces secondary particles such as fast neutrons, pions, and electron-positron pairs.
The pions quickly decay into either \ac{gray} photons or charged muons, which then decay into electrons and positrons, resulting in positron annihilation and bremsstrahlung.
Pair annihilation mainly occurs via \ac{Ps} formation, producing both the 511\,keV line and the ortho-\ac{Ps} continuum.
Fast neutrons may escape the surface, undergo inelastic scattering (nuclear excitation), or be captured by the material, also leading to nuclear excitation.
Thus, measuring the \acp{gray} lines allows us to study the surface composition of solid bodies.
Fig.\,\ref{fig:gamma-ray_albedo} (bottom left) displays the Moon's Th surface concentration as measured by the \ac{LP} mission \citep{Prettyman2006}.
The figures shows the measurement from a \ac{gray} spectrometer orbiting the Moon at 100\,km.
In the following, we briefly describe how these \ac{gray} (line) albedos are calculated, review observed sources, and discuss the cumulative effect of all small bodies (asteroids and trojans) in the Solar System.

\subsubsection{Calculations of $\gamma$-Ray Albedos}
Eq.\,(\ref{eq:nuc_ex_emissivity}) provides the general formula for \ac{gray} production from \acp{CR} interacting with matter.
The challenge lies not in the equation's complexity but in the many particle species, competing processes (nuclear excitation, elastic scattering, spallation, pair production), and secondary interactions involved.
Therefore, Monte Carlo simulations tailored to specific compositions and \ac{CR} spectra are the preferred method.
In detailed 3D models, penetration depth and \ac{gray} escape in solid bodies and atmospheres are automatically accounted for.
Cosmic photon interactions with solids and atmospheres also produce a \ac{gray} albedo, mostly Compton-scattered to lower energies, though if the pair creation threshold is reached, the 511\,keV line reappears in the continuum.
This effect is important for \ac{CGB} measurements since planetary atmospheres and solid bodies both reprocess \ac{gray} emission \citep[e.g.,][]{Siegert2024,Churazov2007,Sazonov2007}.

\citet{Moskalenko2007,Moskalenko2008,Moskalenko2009} employed the \ac{GEANT4} Monte Carlo generator \citep{Agostinelli2003} to calculate the Moon's \ac{gray} albedo, then extrapolated their results to all small Solar System bodies.
%, including the Oort Cloud.
%
Fig.\,\ref{fig:gamma-ray_albedo} (right) shows the simulated \ac{gray} albedo from a Moon-sized body at Lunar distance.
As expected, the spectrum is very soft with an index of $-1.0$ below $\sim 10$\,MeV, and it breaks according to the material's composition, density, and the Solar modulation potential.

\paragraph{Cosmic-Ray Induced $\gamma$-Ray Lines}
On top of the mainly bremsstrahlung continuum, several \acp{gray} lines are visible, depending on the material's composition.
The absolute continuum flux remains similar for comparable modulation potentials, while the \ac{gray} lines appear more sensitive to changes.
By measuring these line fluxes, one can distinguish asteroid compositions: ``dirty snowballs'' \citep[ices from water, carbon dioxide, ammonia;][]{Whipple1950} show strong oxygen lines (red), pure iron bodies (blue) exhibit lines similar to those in \acp{SN} (e.g., 847 and 1238\,keV), and Moon rock–like materials display a blend of lines (Al, O, Fe, Mg, Ti, Si).
The most intriguing aspect is that these \ac{gray} lines and the continuum allow the determination of the \ac{LECR} spectrum in the Solar System as a function of distance, thereby revealing the Solar modulation potential and the Sun's magnetic activity.
Major asteroid populations include the \acp{MBA} between Mars and Jupiter (2.8\,AU), the \acp{JT} at the Jupiter-Sun Lagrange points L4 and L5 (5.2\,AU), the \acp{NT} (28\,AU), and the \acp{KBO} as another belt reaching from 40--100\,AU.
Since the \ac{CR} spectrum and its intensity vary with distance from the Sun, different objects will exhibit distinct \ac{gray} albedos.
De-excitation \ac{gray} line fluxes are expected to be on the order of $\lesssim 10^{-6}\,\mathrm{ph\,cm^{-2}\,s^{-1}}$, as estimated by \citet{Moskalenko2007,Moskalenko2008,Moskalenko2009}.
%
%Next-to-next generation telescopes might be able to resolve these different components.
%
Next generation telescopes should take these lines into account, most certainly the 511\,keV line.

\begin{figure}[!h]
    \centering
    \includegraphics[width=0.65\linewidth]{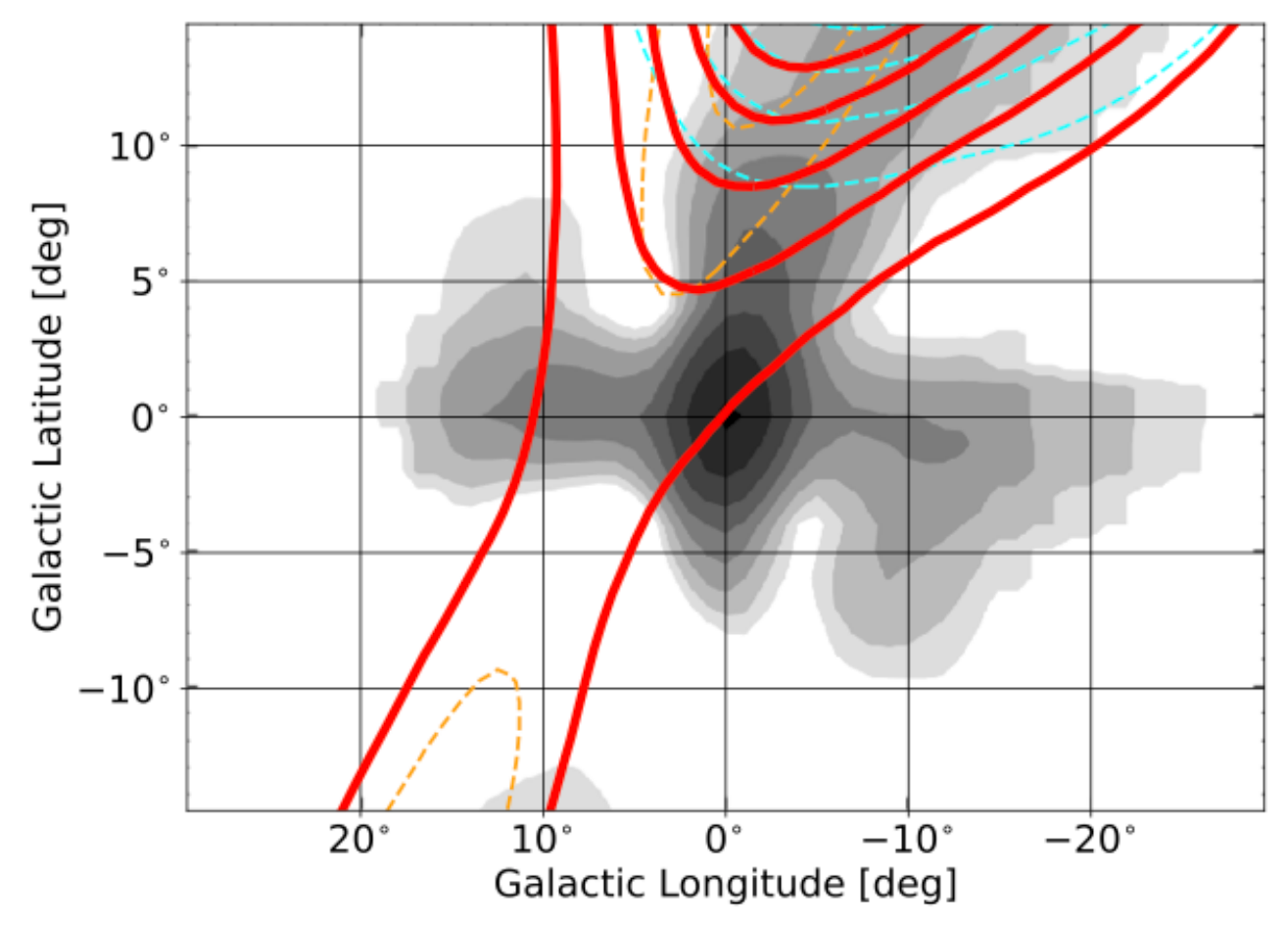}
    \caption{\ac{OSSE} 511\,keV map by \citet{Purcell1997}, together with the cumulative contributions from the \acp{JT} (orange) and \acp{NT} (cyan) from 1990-01-01 to 1997-12-31, and the combined model (red). The ``OSSE fountain'' was a serendipitous and, so far, never observed afterwards, positive latitude enhancement \citep[from][]{Siegert2024}.}
    \label{fig:JT_variation}
\end{figure}

\paragraph{Time-Variable $\gamma$-Ray Foreground}
However, the 511\,keV line from the asteroids is expected to be significantly stronger, with a flux of $(1$--$4) \times 10^{-4}\,\mathrm{ph\,cm^{-2}\,s^{-1}}$ along the Solar System ecliptic.
Since the ecliptic is crossing the Galactic plane around the Galactic bulge region, the contribution of the 511\,keV line flux in the bulge from the asteroids is on the order of $2 \times 10^{-5}\,\mathrm{ph\,cm^{-2}\,s^{-1}}$, that is, about 2\% of the total (see Sec.\,\ref{sec:511}).
The interesting effect with the \acp{JT} and \acp{NT}, and to a lesser extent also with the \acp{MBA}, is that their positions on the sky change with time.
This will lead to an effective time-variable diffuse \ac{gray} foreground to Galactic measurements in the 511\,keV (and other, weaker lines, as well as the continuum).
The ecliptic band in the MeV would then be similar to the zodiac light in the infrared.
However, the \ac{JT} and \ac{NT} regions may be populated by a much larger number of objects than currently known (only km sized objects are known), so that they may stick out of the ecliptic \ac{gray} band.
This might be measurable already now even though the exposure of, for example, \ac{INTEGRAL}/\ac{SPI} is mostly concentrated in the Galactic bulge.
A variation of 2\% of the flux from the spatial variance alone might not be visible, but the effect of the solar cycle in addition might lead to measurable effects on the time scale of several years.
Considering the long exposure times of MeV \ac{gray} missions, the cumulative effect of these trojan asteroid accumulations over time may lead to asymmetric features, such as seen in the ``OSSE 511\,keV Galactic fountain'' \citep{Purcell1997}.
\citet{Siegert2024} attempts to explain the OSSE fountain by the cumulative effect of the \acp{JT} and \acp{NT} and shows that within the observation period of \ac{CGRO}/\ac{OSSE} in this study, between 1991 and 1997, the ``fountain'' matches the cumulative distribution of \acp{JT} and \acp{NT}.
In Fig.\,\ref{fig:JT_variation}, we show the spatial overlap of the ``OSSE 511\,keV fountain'' with the cumulative effect of the \acp{JT}.

\subsubsection{$\gamma$-Ray Measurements of Solar System Objects}
There have been several measurements of \ac{gray} lines from different Solar System objects -- not necessarily with MeV telescopes, but with \ac{gray} spectrometers onboard probes that flew to the objects.
We will give a short overview here and some details for individual objects in the following.

Ceres and Vesta are the largest asteroids in the Solar System with radii of 473 and 263\,km, respectively, and have been
visited by the Dawn spacecraft \citep{Peplowski2013,Lawrence2018}.
Dawn's \ac{GRaND} measured the surface composition of both asteroids, viz. dwarf planets.
Other atmosphere-free objects whose surface composition has been measured by \ac{gray} observations in-situ are Mercury \citep{Evans2012}, the Mars moon Phobos \citep{Lawrence2019}, the asteroid 433 Eros (see Sec.\,\ref{sec:433Eros}), as well as the Moon (see Sec.\,\ref{sec:Moon}, see also Fig.\,\ref{fig:gamma-ray_albedo}, bottom left).
Solar System objects with atmosphere that have \ac{gray} measurements include the Earth (Sec.\,\ref{sec:Earth}) and the Sun.
However, for the quiet Sun, only upper limits in the MeV range exist up to 200\,keV \citep{Hannah2007}.

\paragraph{The Moon}\label{sec:Moon}
Besides the natural radioactivity (Fig.\,\ref{fig:gamma-ray_albedo}, bottom left), \citet{Prettyman2006} disentangled more \ac{gray} lines from the surface survey of the Moon than could be expected from the measurements with BGO crystals and an energy resolution of 12\%.
Besides Th, the \ac{LP} mission created \ac{gray} surface maps of the Moon also in the elements of K, Fe, O, Al, Ti, Mg, and Si (mostly in oxide form) at resolutions of $2$--$20^\circ$. 
In fact, they also measured the 511\,keV line but given the scientific purpose of the \ac{LP} mission, this line was disregarded in their study.
At an orbit of 100\,km above the Lunar surface, their fluxes range from about $10^{-3}\,\mathrm{ph\,cm^{-2}\,s^{-1}}$ at 9\,MeV to more than $10^{-1}\,\mathrm{ph\,cm^{-2}\,s^{-1}}$ around 0.5\,MeV.
Clearly, this is integrating over a large solid angle of $2$--$20^\circ$, and also on top of a 10--90\% background from the instrument and bremsstrahlung component.
\citet{Moskalenko2007} estimated a 511\,keV flux from the Moon of $2.4 \times 10^{-5}\,\mathrm{ph\,cm^{-2}\,s^{-1}}$ at a distance of 384000\,km, which would be equivalent to a differential flux per $(10^\circ)^2$-pixel of $5.2 \times 10^{-2}\,\mathrm{ph\,cm^{-2}\,s^{-1}}$.
This appears reasonable, though is on the high side of the \citet{Prettyman2006} measurements.
Using this value as a reference, we can estimate the fluxes of some other nuclear de-excitation lines from the Moon if it was observed from a satellite around Earth.
For example, the most prominent Fe, Si, and O lines are all be on the order of $(1$--$10) \times 10^{-4}\,\mathrm{ph\,cm^{-2}\,s^{-1}}$ at a distance of \ac{LP} mission, so that the total surface flux at Earth would be at most a few $10^{-7}\,\mathrm{ph\,cm^{-2}\,s^{-1}}$.
While this is not an extremely strong foreground, next-to-next generation telescopes should take also these lines into account -- however certainly the 511\,keV line.

\paragraph{Named Asteroids}\label{sec:433Eros}
Asteroid 433 Eros was measured by the \ac{NEAR} probe which had a closest approach to the asteroid of about 35\,km \citep{Peplowski2016}.
It could measure the nuclear de-excitation of elements like Fe, Mg, K, Si, Al, C, and O, as well as the 511\,keV line.
Extrapolating \ac{NEAR}'s \ac{gray} spectrum from 433 Eros towards the entire population of asteroids in all accumulations in
the Solar System is difficult, if not meaningless, as there are hardly any other measurements in soft \acp{gray} for individual asteroids, and especially not at smaller sizes.
The smaller asteroids would be required because most induced \ac{gray} emission from \ac{LECR} interactions come from grazing angles, that is, the rim of the objects \citep{Moskalenko2008}.
If, however, the objects are too small to develop a full particle cascade when hit, the resulting spectrum might differ from the expectations of \citet[][see also Fig.\,\ref{fig:gamma-ray_albedo}, right]{Moskalenko2008}.
The number of small objects ($\lesssim 1$\,m) in the Solar System is not known but is expected to exceed that of the known ones by several orders of magnitude \citep{Dohnanyi1969}, making them the most luminous component in total, even when considering the incomplete particle cascades.
A measurement of the cumulative effect of asteroids in the Solar System with next-generation telescopes will help to understand the distribution -- and formation -- of small-sized bodies and in how far they contribute to the \ac{gray} foreground albedo.

\paragraph{Earth}\label{sec:Earth}
It appears surprising that \ac{gray} line observations of Earth are rare.
While it is important for, for example, \ac{gray} burst observations to include the reflection from the atmosphere into the satellite instrument \citep[e.g.,][]{Kippen2007}, many observatories strictly avoid observing Earth because it is large (no star tracker information) and bright (detector saturation).
Especially at MeV energies, there are only a few Earth \ac{gray} albedo measurements, of which \citet{Share2001} provide the most useful.
The authors used \ac{SMM}/\ac{GRS} observations of Earth in low-Earth orbit with and without a solar particle event.
Despite the 5\% resolution of the seven NaI detectors, it was possible to identify the major \ac{gray} lines coming from nuclear excitation and spallation of Earth's atmosphere.
Among others, the strongest line blends are around 4.4\,MeV with, for example, \nuc{C}{12} (4439\,keV) and \nuc{B}{11} (4444\,keV), around 7\,MeV with \nuc{O}{16} (6917, 7115\,keV) and \nuc{C}{14} (7010\,keV), and from 3.1--3.7\,MeV with \nuc{N}{14} (3073, 3378, 3384\,keV) and \nuc{C}{13} (3089\,keV) \citep[see][Tab.\,1]{Share2001}.
Most of the lines appear broadened up to several 100\,keV (\ac{FWHM}) and at blend flux levels on the order of $(5$--$10) \times 10^{-3}\,\mathrm{ph\,cm^{-2}\,s^{-1}}$.
The strongest line, however, is at 511\,keV with a flux level of several $10^{-2}\,\mathrm{ph\,cm^{-2}\,s^{-1}}$ and hardly broadened with respect to instrumental resolution.
These measurements are important to gauge simulations of the \ac{gray} albedo of Earth, for example for measurements of the \ac{CGB} that rely on Earth occultation measurements \citep[e.g.,][]{Sazonov2007,Churazov2007,Churazov2008,Tuerler2010}.
Unless these measurements are validated, any measurement of the \ac{CGB} beyond a few 100\,keV from Earth occultation is highly uncertain.
We note that in the case of a Solar particle event, the atmospheric 511\,keV line reaches fluxes levels in low-Earth orbit of $\gtrsim 1\,\mathrm{ph\,cm^{-2}\,s^{-1}}$ \citep{Share2001}.

% 511
\newpage
\subsection{The Positron Annihilation Line at 511\,keV}\label{sec:511}
\vspace{-0.5em}
{\emph{Written by Thomas Siegert}}
\vspace{0.5em}
\vspace{-12pt}

\subsubsection{Overview of 511\,keV Observations}

\paragraph{Historical Observations}
The 511\,keV line from electron positron annihilation has first been detected by \citet{Johnson1973} during balloon flights in 1970--1971.
In fact, the line was already indicated in \citet{Haymes1969} at the $2\sigma$ level during a balloon flight in April 1968.
At first, it was not clear whether the line-like feature was actually due to positron annihilation because the centroid was found to be $476 \pm 24$\,keV, which is closer to the \nuc{Be}{7} decay line at 478\,keV (see Sec.\,\ref{sec:CNe}).
However, with better spectral resolution using Ge detectors on a balloon flight in 1977, \citet{Leventhal1978} unambiguously found that the \ac{gray} line from the direction of the Galactic centre was indeed at 511\,keV and therefore due to the annihilation of positrons.

After this initial detection, several other balloon-borne \ac{gray} detectors as well as satellite-based observations led to a conundrum about the flux of the 511\,keV line:
Even the same instruments \citep[e.g., \ac{HEAO3}, \ac{SMM},][]{Riegler1981,Share1988} found different fluxes, ranging between $(0.4$--$6.0) \times 10^{-3}\,\mathrm{ph\,cm^{-2}\,s^{-1}}$, on timescales of years.
This surprisingly strong variability turned out to be a misconception of what was actually observed -- not a single variable point source, but diffuse extended emission.
After \citet{Lingenfelter1989} found a strong correlation of the observed 511\,keV flux with the \ac{FoV} of the instrument used, \citet{Purcell1997} was finally able to reconstruct the first image of the annihilation line using \ac{CGRO}/\ac{OSSE}.
The \ac{OSSE} image included a strong bulge component with a radial extent of $5^\circ$--$8^\circ$, a weak disk that only extended out to longitudes $|\ell| \lesssim 20^\circ$, and a positive latitude enhancement that was dubbed the ``\ac{OSSE} 511\,keV fountain'' (see Fig.\,\ref{fig:JT_variation}).

With these findings, it was clear that this strongest \ac{gray} line detected was the first one from outside the Solar System.
Furthermore, it was not only the 511\,keV line that was detected, but also the ortho-\ac{Ps} continuum below the line that shows a turn-over of the falling \ac{IC} spectrum around 350\,keV after which the differential spectrum is, in fact, rising as a function of energy \citep[e.g.,][]{Share1988,Purcell1997}.
Beyond the 511\,keV line at which the \ac{Ps}-decay spectrum peaks with the para-\ac{Ps} line, there is a sharp drop and the Galactic ridge continuum spectrum continues.
Newer observations, however, did not find all the structures observed with \ac{OSSE}.

\paragraph{Recent Observations with INTEGRAL/SPI}
With the launch of \ac{INTEGRAL}, the spectrometer \ac{SPI} could do both at unprecedented accuracy -- imaging \emph{and} spectroscopy.
The two seminal papers by \citet{Knoedlseder2005} and \citet{Jean2006} consider the imaging and the spectral analysis of positron annihilation using 1.5\,yr of \ac{SPI} data, respectively.
The central parts of the Galaxy were found to be explained by a 2D Gaussian
%, slightly shifted to negative longitudes
with a spherical shape of radius $8^\circ$ (\ac{FWHM}).
These values are completely consistent with the previous \ac{OSSE} study.
However, neither the disk was seen significantly
%, even though further iterations of the imaging algorithm used made it appear,
nor the positive latitude enhancement.

\begin{figure}[!ht]
    \centering
    \includegraphics[width=1.0\linewidth]{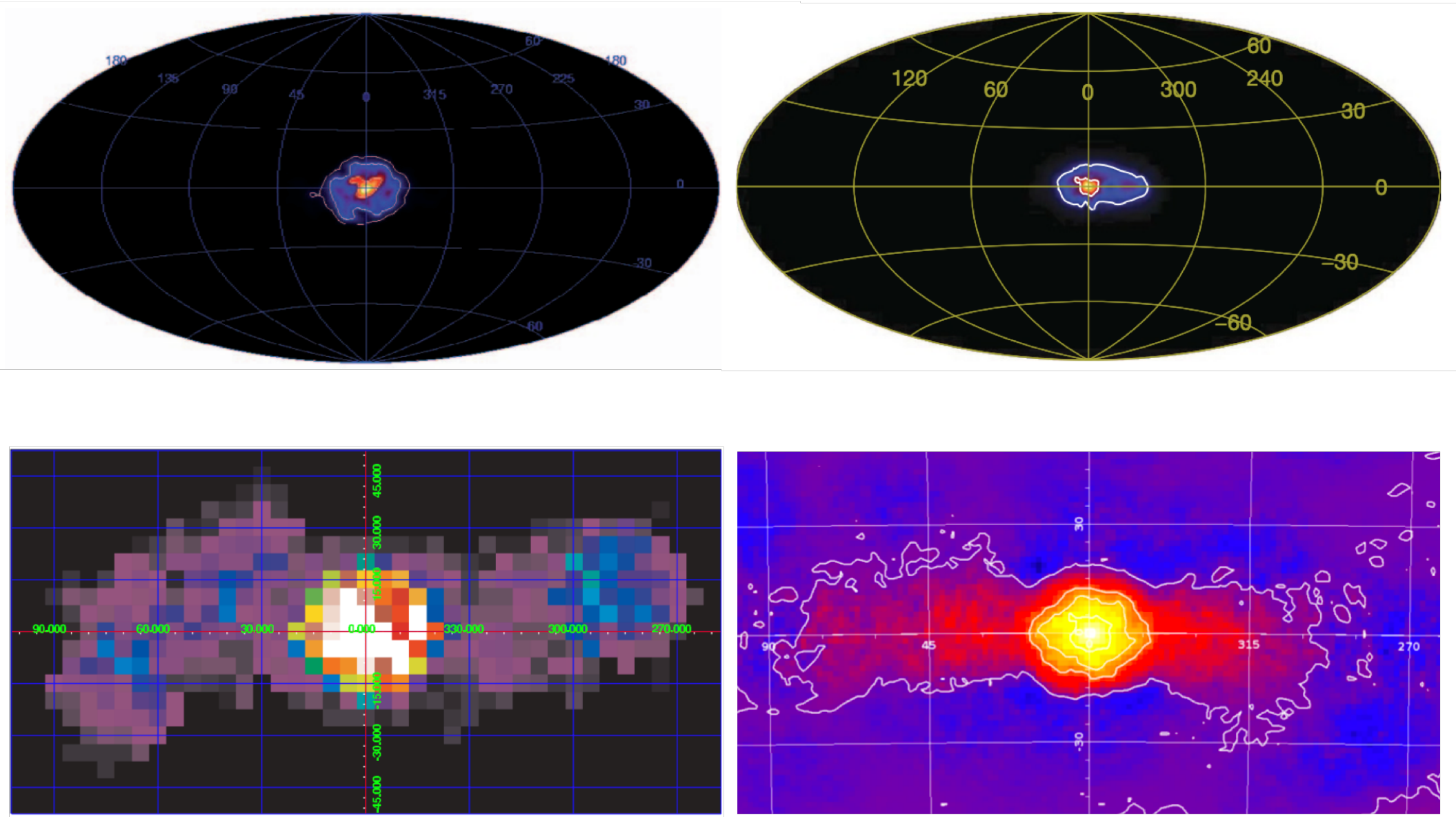}
    \caption{Reconstructed images of the 511\,keV line with \ac{SPI} throughout the \ac{INTEGRAL} mission with different exposure times and algorithms. Top left: 1.5\,yr \citep[from][using Richardson-Lucy]{Knoedlseder2005}. Top right: 3.5\,yr \citep[from][using Richardson-Lucy]{Weidenspointner2008}. Bottom left: 7.5\,yr \citep[from][using Maximum Likelihood]{Bouchet2010}. Bottom right: 10.5\,yr \citep[from][using Maximum Entropy]{Siegert2017}.}
    \label{fig:511keV_maps}
\end{figure}

\begin{figure}
    \centering
    \includegraphics[width=0.75\linewidth]{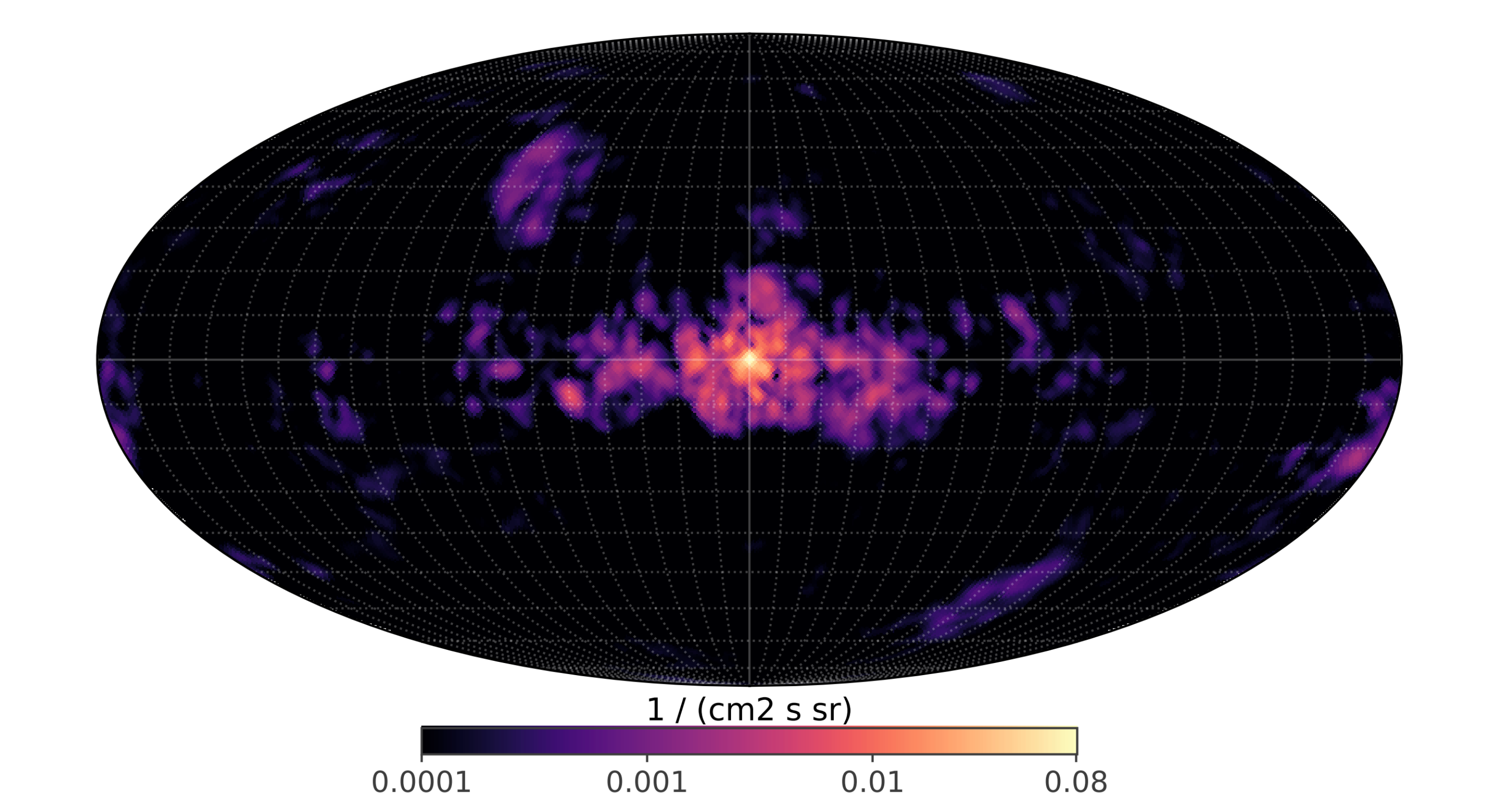}
    \caption{Richardson-Lucy image reconstruction of the 511\,keV line emission from 20.5\,yr of \ac{INTEGRAL}/\ac{SPI} data \citep[from][]{Yoneda2025_511}.}
    \label{fig:yoneda_511}
\end{figure}

Later studies by \citet{Weidenspointner2008}, \citet{Bouchet2010}, \citet{Skinner2014}, and \citet{Siegert2016a,Siegert2019b,Siegert2022a}, all found a disk in 511\,keV with sizes confirming \ac{OSSE}, or even more extended in longitude and latitude.
The initial extremely high bulge-to-disk flux ratios from \ac{SPI} and \ac{OSSE} observations of $0.2$--$3.3$, corresponding to luminosity ratios of up to $9$, were then step-by-step decreased by including more exposure over the timescale of the \ac{INTEGRAL} mission (see Fig.\,\ref{fig:511keV_maps}).
The more or less accepted value for the bulge-to-disk luminosity ratio is now  ``below $1.0$'', which however, depends on the effective distances to bulge and disk, which scale the measured fluxes of $F_{\rm bulge} = (0.9$--$1.2) \times 10^{-3}\,\mathrm{ph\,cm^{-3}\,s^{-1}}$ and $F_{\rm disk} = (1.4$--$2.0) \times 10^{-3}\,\mathrm{ph\,cm^{-3}\,s^{-1}}$.
The most recent image reconstruction from \ac{INTEGRAL}/\ac{SPI} has been performed by \citet{Yoneda2025_511} with more than 20\,yr of data.
This newest image (Fig.\,\ref{fig:yoneda_511}) shows the same regions as before and might finally reveal sub-structures, such as from massive star groups.

\begin{figure}
    \centering
    \includegraphics[width=1.0\linewidth]{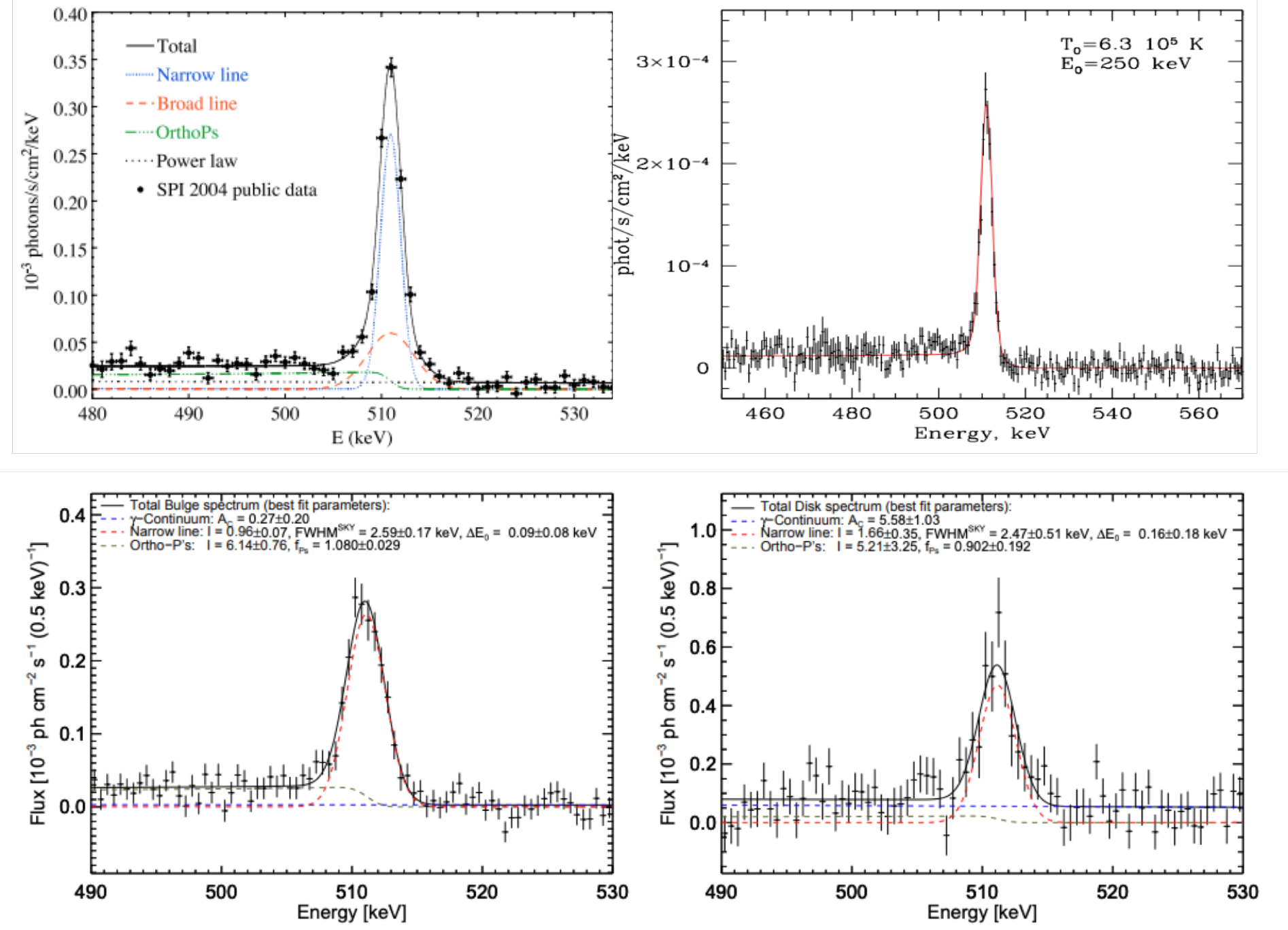}
    \caption{Spectra of the 511\,keV line in the Galaxy with different exposure times and analysis methods (see main text). Top left: Bulge, 1.5\,yr \citep[from][]{Jean2006}. Top right: Bulge, 8.5\,yr \citep[from][]{Churazov2011}. Bottom left: Bulge, 10.5\,yr \citep[from][]{Siegert2016a}. Bottom right: Disk, 10.5\,yr \citep[from][]{Siegert2016a}.}
    \label{fig:511keV_spectra}
\end{figure}

The spectral analysis with \ac{SPI} revealed the positron annihilation \emph{conditions}, that is, not their sources but their sinks.
It was found that positrons mostly annihilate in warm ($7000$--$40000$\,K) and partly ionised ($0.01$--$0.25$\%) phases of the \ac{ISM} \citep{Jean2006,Churazov2005} or in a cooling phase \citep{Churazov2011} (Fig.\,\ref{fig:511keV_spectra}).
Given the spectral resolution of \ac{SPI} of 2.1\,keV at 511\,keV (\ac{FWHM}), it was possible to find both, a narrow line with a width of $0.9$--$1.7$\,keV, and a broad line with $4$--$11$\,keV in the bulge region of the Milky Way \citep{Jean2006,Siegert2019b}.
The annihilation conditions also imprint in the \ac{Ps} fraction, $f_{\rm Ps}$, which, however, is difficult to determine across the Galaxy, or even for different spatial components \citep[e.g.,][]{Siegert2016a}.
In an attempt to construct a 511\,keV longitude-velocity diagram to reveal the internal dynamics of positron annihilation in the Galaxy, similar the \nuc{Al}{26} case \citep[][see also Sec.\,\ref{sec:Al26}]{Kretschmer2013}, \citet{Siegert2019b} found a velocity gradient along the inner $\pm 30^\circ$ of $4 \pm 6\,\mathrm{km\,s^{-1}\,deg^{-1}}$.
This is consistent with zero rotation, consistent with the Galactic rotation speed from CO measurements ($2$--$3\,\mathrm{km\,s^{-1}\,deg^{-1}}$), and also with the 1.8\,MeV \nuc{Al}{26} measurement of $7.5$--$9.5\,\mathrm{km\,s^{-1}\,deg^{-1}}$.
Statistically, it appeared that none of the spectral parameters of the 511\,keV line, nor the \ac{Ps} fraction change as a function of longitude.
However, a slight, insignificant positive latitude enhancement in the line flux was found, reminiscent of the \ac{OSSE} fountain measurement \citep{Purcell1997}.
Systematic, astrophysical variations with longitude are, nevertheless, possible as discussed in much detail in the appendix of \citet{Siegert2019b}.

\begin{figure}
    \centering
    \includegraphics[width=0.49\linewidth]{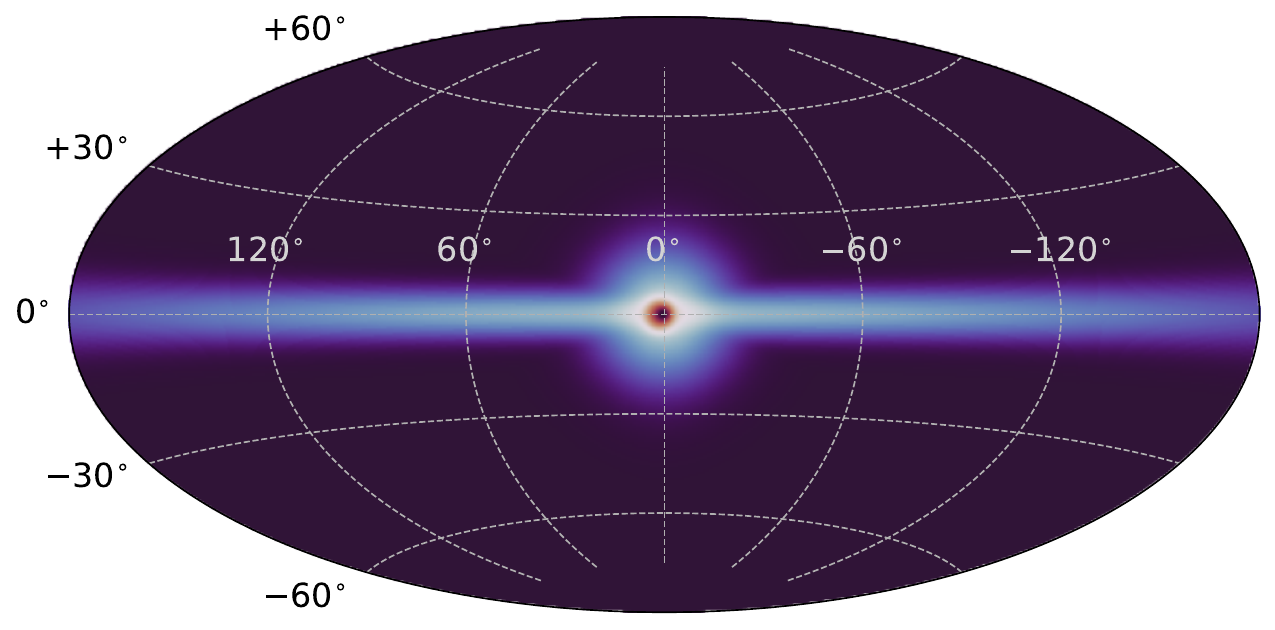}
    \includegraphics[width=0.49\linewidth]{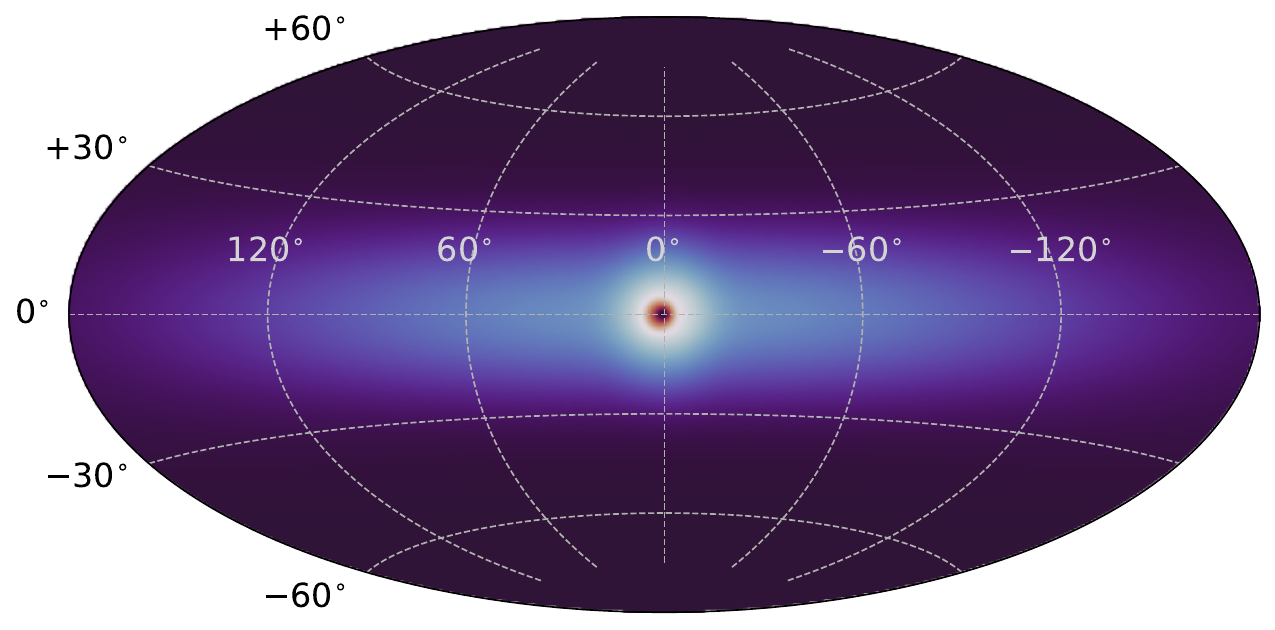}
    \caption{Maximum likelihood solutions of fits to the same 10.5\,yr \ac{INTEGRAL}/\ac{SPI} dataset. Left: Thin disk model with vertical extent of $\sim 3^\circ$ \citep[from][using a single 508--514\,keV bin]{Skinner2014}. Right: Thick disk model with vertical extent of $\sim 10^\circ$ \citep[from][using 0.5\,keV bins from 490--530\,keV]{Siegert2016a}.}
    \label{fig:thin_vs_thick_disk}
\end{figure}

Given 10.5\,yr of observations, \citet{Skinner2014} and \citet{Siegert2016a} performed an analysis to determine the size of the 511\,keV disk (see Fig.\,\ref{fig:thin_vs_thick_disk}).
While \citet{Skinner2014} used a single energy bin from 508--514\,keV, \citet{Siegert2016a} used 0.5\,keV bins from 490--530\,keV to estimate the extent and flux of the disk.
Both works assumed the same empirical three-component model for the bulge.
Here, \citet{Skinner2014} found a point-like source in the 508--514\,keV bin at the position of the Galactic centre, which was then shown to be a 511\,keV line and not just continuum by the spectral analysis in \citet{Siegert2016a}.
For the disk, the two approaches led to similar line fluxes, from $(1.4$--$2.0) \times 10^{-3}\,\mathrm{ph\,cm^{-2}\,s^{-1}}$, but from vastly different scale heights:
The single bin from \citet{Skinner2014} seems to prefer a thin disk ($\sim 3^\circ$), while \citet{Siegert2016a} found a thick disk ($\sim 10^\circ$).
The statistical uncertainties in both cases are rather small, so that the tension between the two studies using the same dataset is about 3--4$\sigma$.
Significant differences in how the instrumental background is treated probably leads to a stark contrast in how the disk is emerging from the likelihood fits, so that most of the difference here is likely of systematic nature.
One method may prefer narrower components, while the other could maybe identify also dimmer, more extended components.
Astrophysically, both scenarios can make sense because a thin disk would represent the dense gas distribution of the Milky Way in which positrons may be expected to annihilate quickly, and a thick disk may show positrons on their way propagating from their sources in a partly ionised \ac{ISM}.
Since both may be true, the 511\,keV disk is likely thin \emph{and} thick.

Finally, from these and similar studies, a Galactic-wide positron annihilation rate can be inferred, given some educated guesses on the effective distances to the observed components.
The observed Milky Way positron annihilation rate is therefore on the order of $5 \times 10^{43}\,\mathrm{e^{+}\,s^{-1}}$, of which 30--40\% can be attributed to the bulge, and the remaining annihilation to the disk.
A fraction of 10\% of the bulge annihilation rate may be assigned to a point-like source in the Galactic centre, consistent with the position of the supermassive \ac{BH} Sagittarius A$^*$, but also consistent with the size of the entire Central Molecular Zone given \ac{SPI}'s angular resolution of $2.7^\circ$.
While the bulge values are less prone to systematics and statistical uncertainties, the disk annihilation rate may be uncertain by a factor of two.
This uncertainty includes the unknown effective distance, the vertical extent, and the possibility that there could be a halo component which is hard to detect for \ac{SPI} \citep[see, e.g.,][]{Purcell1997,Kinzer2001}.
In fact, an (almost) isotropic contribution to the 511\,keV line would be entire invisible for coded mask instruments because it would be disregarded as instrumental background in the data analysis \citep{Siegert2022d}.
Such a component might be very large compared to the Galactic contribution and renewed measurements of the \ac{CGB} are required to fill this information gap.

A focus study of the bulge region to obtain the best-fitting \emph{astrophysical} components rather than empirical ones, found -- and corroborated \citep{Knoedlseder2005} -- that the 511\,keV line and the ortho-\ac{Ps} continuum trace the old stellar population in the Milky Way \citep[][see also Fig.\,\ref{fig:astrophysics_511keV_models}]{Siegert2022a}.
With a 3D smoothing kernel, it was then determined how far from these ``sources'', whatever they are (see Sec.\,\ref{sec:positron_sources}), positrons could actually propagate.
The characteristic length scale is here $150 \pm 50\,\mathrm{pc}$, which would either be close to annihilation \emph{in-situ} at the astrophysical sources which may have been kicked at formation, or the consensus model that requires propagation of positrons away from their sources.
The former would suggest kick velocities on the order of $\gtrsim 50\,\mathrm{km\,s^{-1}}$ within the Galactic nucleus but $\lesssim 15\,\mathrm{km\,s^{-1}}$ in the broad bulge.
Positron propagation could explain the overall picture of the 511\,keV signal, however only if the injection energies are below 1.4\,keV, which may point to a pure nucleosynthesis origin.

Several other authors studied the positron annihilation signal in the Milky Way or tried to set limits on extragalactic objects.
%
%However, this Section intends to give only a brief overview of this strongest extra-solar \ac{gray} line.
%
Non-Galactic observations can be summarised as finding only upper limits, such as for early observations of individual globular clusters and some extragalactic objects \citep{Knoedlseder2005} or dwarf galaxies \citep{Siegert2016c,Siegert2022b}.

\subsubsection{Possible Positron Sources}\label{sec:positron_sources}
\begin{figure}[!ht]
    \centering
    \includegraphics[width=0.85\linewidth]{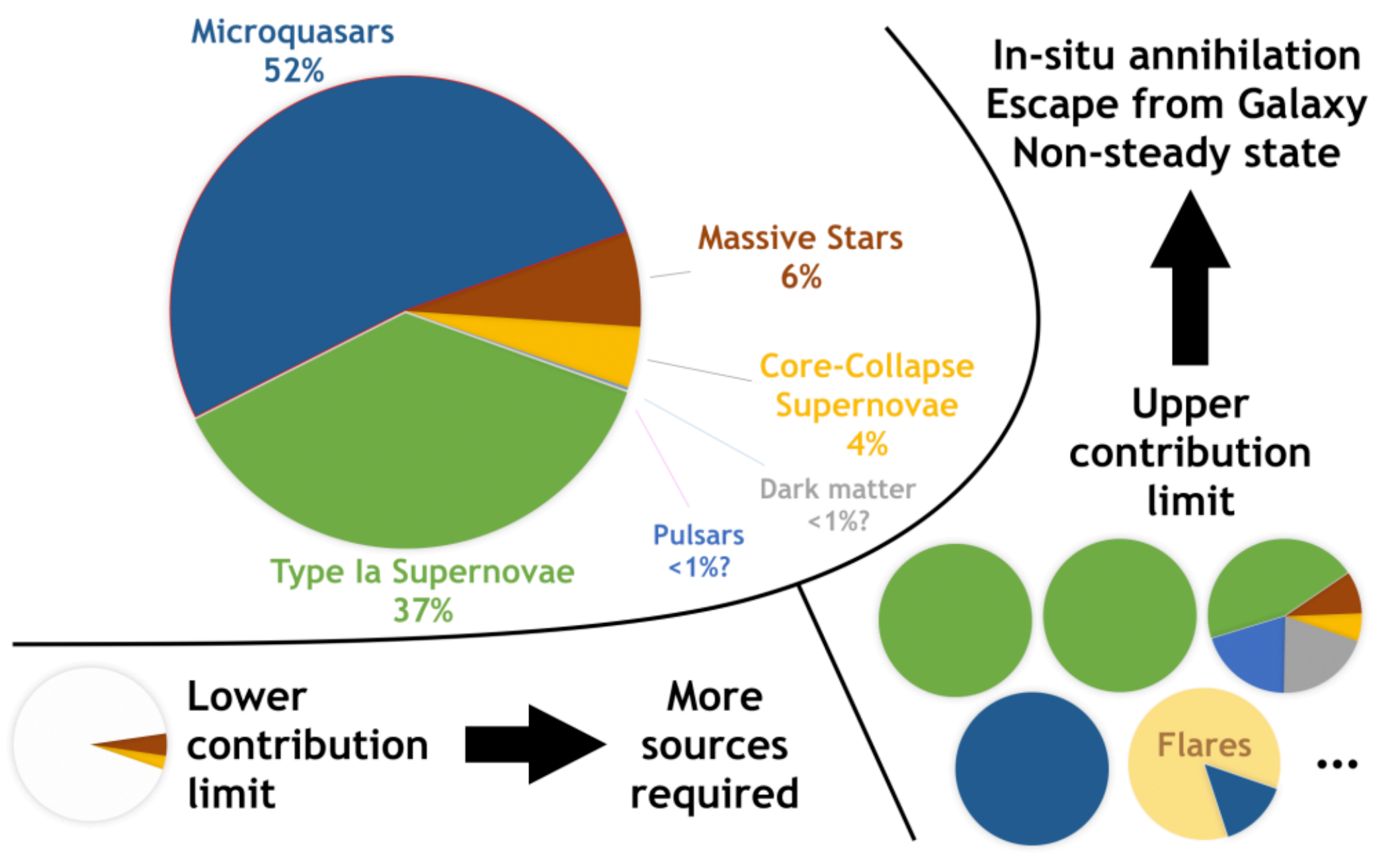}
    \caption{Positron production rate in the Milky Way. Top left: About 105\% of positrons \emph{seen to annihilate} can be explained by nucleosynthesis sources and microquasars. Other sources might not contribute at all to the 511\,keV line (but could contribute to positron annihilation, see Sec.\,\ref{sec:not_a_line}). Bottom left and right: Taking into account measurement uncertainties in production, annihilation, population sizes, and model variants. Both extreme ends would point to a non-steady-state solution of positron annihilation in the Galaxy \citep[from][]{Siegert2023}.}
    \label{fig:sources}
\end{figure}

In the following, we will provide a brief overview of possible and frequently discussed positrons sources.
Individual source types might fill entire chapters of books, so that we will only show the most relevant points in the context of future observations.
A detailed overview can be found in \citet{Prantzos2011} or \citet{Siegert2023}.
A summary of the contribution of \emph{possible} sources, including many uncertainties of both theoretical and observational studies, is shown in Fig.\,\ref{fig:sources}.
It should be noted that this figure is, in some sense, ignorant about circumstantial (and anecdotal) evidence: 
some sources, such as \acp{SNIa}, may be much weaker than shown in Fig.\,\ref{fig:sources} but not necessarily contribute nothing -- the figure is merely a showcase of possibilities.

\paragraph{Massive Stars and Supernovae}
Massive stars and \acp{SN} are the only unambiguously confirmed sources of positrons in the Milky Way.
Given that \acp{gray} of $\beta^+$-unstable radioactive isotopes have been measured (\nuc{Al}{26}, \nuc{Co}{56}, \nuc{Ti}{44}), the expected contribution, that is, the production rate from these isotopes, can be estimated.
However it must be noted that there is so far no one-to-one correspondence of these radioactive isotopes and the 511\,keV line, as could be expected from point-like sources such as \acp{SNR} (\nuc{Co}{56}, \nuc{Ti}{44}) or slightly extended sources such as superbubbles (\nuc{Al}{26}).
These nucleosynthesis positrons are typically ejected with kinetic energies of a few hundred keV up to 1\,MeV, depending on the isotope, at the position of the mother nucleus' decay.
This means that one positron source map would be equivalent to the \nuc{Al}{26} map of the Galaxy, or the \nuc{Ti}{44} hotspots within \acp{SNR}.
But since the ejection energy is mildly relativistic, the sink distribution of positrons, that is, where they annihilate, could be significantly different, given the propagation conditions in the \ac{ISM}.
This applies to all source types described below.

With the measured quasi-persistent \nuc{Al}{26} mass in the Milky Way, the observed \acp{SNR} in \nuc{Ti}{44}, and the \emph{detected} positron annihilation flux, about $11 \pm 3$\% of positrons would originate in these isotopes, taking into account measurement uncertainties in the radioactivity \ac{gray} lines.
In the case of \nuc{Co}{56} as a daughter of \nuc{Ni}{56} in both \acp{ccSN} and \acp{SNIa}, the short lifetime of the isotopes makes it difficult to estimate the escape from these sources.
There are no solid measurements of the 511\,keV line from \acp{ccSN} or \acp{SNIa} directly, but infrared measurements of late lightcurves of \acp{SNIa} suggest an escape fraction of positrons of 1--5\% \citep{Milne2001}.
This would make a contribution of \acp{SNIa} to the positron annihilation budget of the Milky Way of $3$--$17$\% for an \ac{SNIa} rate of 0.25 per century.
Given the measurements of SN2014\,J in M82, \citet{Siegert2017} found an escape fraction of $0.06_{-0.06}^{+0.33}$, which, at face value, might explain even more than the total annihilation budget of the Galaxy.

\paragraph{Classical Novae}
In \acp{CN}, two channels can be expected to produce positrons:
either by the radioactive decays of \nuc{Na}{22} and \nuc{Al}{26} after the material has been ejected, or by the ``annihilation flash'' from \nuc{N}{15} and \nuc{F}{18} at the time of explosive burning on the surface of the \ac{WD} (see Sec.\,\ref{sec:CNe}).
The case of \nuc{Al}{26} is difficult to calibrate because the consensus model of \nuc{Al}{26} only invokes massive stars to explain the 1.8\,MeV signal, so that \ac{CN} contributions are ignored.
They may, however, be on the order of a few per cent, or even dominating the \nuc{Al}{26}.
In case of positron annihilation, this will not change the total budget, because it is the same amount of \nuc{Al}{26} in the Galaxy but only assigned to different sources.
For \nuc{Na}{22}, a measurement of the diffuse 1275\,keV line would directly calibrate the positron production rate from this isotope.
However, the line has never been detected and only upper limits on the line flux, integrated over the whole Galaxy and recent ONe novae, lead to an upper limit on the positron production rate of $\lesssim 10$\% \citep{Siegert2021}.

The case of the ``annihilation flash'' is also difficult to gauge because it has also never been measured.
It is expected that most, if not all, positrons annihilate directly in the expanding \ac{CN} shell, which may result in an approximately one hour long \ac{gray} signal from positron annihilation.
Estimates find a peak 511\,keV flux on the order of $10^{-3}$--$10^{-2}\,\mathrm{ph\,cm^{-2}\,s^{-1}}$ for \acp{CN} at a distance of 1\,kpc.
With an effective distance to \acp{CN} of 5\,kpc, and a \ac{CN} rate of $50\,\mathrm{yr^{-1}}$, one expects to not see a quasi-persistent flux from these sources, but rather individual points from time to time.
If, by some mechanism, the positrons would escape, they would only make a contribution on the order of $10^{-5}$ of the total Galactic annihilation budget.

\paragraph{Cosmic Rays}
The \ac{CR} interactions with molecular clouds lead to the production of positively charged pions and other mesons that, at some point, will decay into positrons, however with larger energies than the nucleosynthesis positrons.
It has been suggested that these positrons do not contribute to the Galactic annihilation signal because their energies are too large to be slowed down efficiently.
Early studies of the in-flight annihilation spectrum in the Milky Way find limits on the injection energy of positrons on the order of 3--7\,MeV \citep[e.g.,][]{Sizun2006,Beacom2006,Churazov2011}.
However, these studies assume knowledge about the underlying Galactic diffuse continuum from \ac{IC} scattering and bremsstrahlung which in themselves are barely understood \citep{Siegert2022c}.
It is clear that if the assumptions on the Galactic diffuse emission are relaxed, much higher injection energies, up to several tens or hundreds of MeV, are possible \citep{Siegert2023}.
By estimates of the average \ac{CR} spectrum throughout the Galaxy by GeV observations, it was found that on the order of 3--10\% of the positrons \emph{seen to annihilate} may originate in \ac{CR} interactions \citep[e.g.,][]{Aharonian2000,Porter2008}.
The GeV map of the Galaxy, showing dense gas regions, would be the starting point of these higher-energy positrons.

\paragraph{Pulsars}
Through interactions of particles and photons with magnetic fields, pulsars, millisecond pulsars, and magnetars, could produce high-energy electron-positron pairs \citep[see][for discussion]{Prantzos2011,Siegert2023}.
Also here, no measurement actually confirms the logical sequence of production in pulsars, followed by slowing down in the \ac{ISM}, and final annihilation at a distance, nor the also plausible path of in-situ annihilation near the pulsar, for example in pulsar wind nebulae.
Given the consensus model of pulsar emission, pair production is inevitable, and the positron production rate of all types of pulsars may be on the order of 10\% for the entire Milky Way.
However, also here, the energy considerations and propagation through the \ac{ISM} with possible in-flight annihilation, must be considered.
Individual pulsars are currently too dim to be observable at 511\,keV.

\paragraph{Black Holes}
Like pulsars, the strong gravity around accreting \acp{BH} may lead to photon-photon pair production, ejection, or in-situ annihilation.
There are three historical measurements that claim to have seen an annihilation feature (\emph{not necessarily} a 511\,keV line) from microquasars \citep{Bouchet1991,Sunyaev1991,Sunyaev1992,Siegert2016b} -- all of which are contested.
Pair production in highly compact objects is inevitable whenever the pair creation threshold is met.
It is then expected that an electron-positron pair-plasma is ejected, which may be related to jets observed at radio frequencies.
The pair-plasma is unstable and would annihilate quickly with itself, resulting in potentially erratic Doppler-blue-shifted and Doppler-broadened features of thermal pair annihilation \citep{Svensson1983,Beloborodov1999}.
However, it may also be the case that the pairs do not annihilate in-situ but rather propagate in the \ac{ISM} and only annihilate later at a distance to their sources, similar to the nucleosynthesis case.
Both scenarios appear plausible and would make microquasars to a strong contributor to the Galactic positron budget, given that each year about one accreting \ac{BH} is in outburst.
The uncertainties in the total population of microquasars make estimates difficult but their total production rate may range around 50\% with large systematics.
Also quiescent or weakly active \acp{BH} may lead to a quasi-steady production of positrons \citep{Bartels2018}, but this population is even more uncertain than the measurable outbursts.

\paragraph{Stellar Flares}
The Sun is the only star for which we have measurements of the 511\,keV line during a flare \citep[e.g.,][see also Sec.\,\ref{sec:SFs}]{Murphy2005}.
The Sun is also expected to have a quasi-steady flux at 511\,keV from \ac{CR} interactions with the Solar atmosphere \citep{Frost1966,Mazziotta2020} and photon-photon pair production followed by \ac{IC} cooling.
With the more than $10^{11}$ stars in the Milky Way, one can therefore expect to see a quasi-persistent flux at 511\,keV that may directly follow the old stellar population.
With rough assumptions, it was estimated that this quasi-persistent flux could make up to 100\% of the bulge luminosity at 511\,keV \cite{Bisnovatyi-Kogan2017}.
However, since the 511\,keV line and the 2.2\,MeV line from neutron capture on protons are almost one-to-one correlated in Solar flares, it would be expected that a diffuse 2.2\,MeV line is visible in the same region with the same flux, but which is not the case \citep[e.g.,][]{Harris1991,McConnell1997}.
Given the upper limits on the 2.2\,MeV line, the quasi-persistent flux of the 511\,keV line from normal flaring stars must be on the order of 10\% or less in the bulge region.

\paragraph{Nearby Objects}
The interactions of \acp{CR} with asteroids in the Solar System also lead to a 511\,keV line \citep[e.g.,][]{Moskalenko2007,Moskalenko2008,Moskalenko2009}.
Individual asteroids have been observed at MeV \ac{gray} energies, but an extrapolation to the entire population of the expectedly more than $10^{20}$ objects inside the Solar System is difficult.
From \ac{GEANT4} simulations, a flux contribution of the asteroids towards the bulge region may be on the order of 2\%, so that the entire ecliptic should show a flux of $4 \times 10^{-4}\,\mathrm{ph\,cm^{-2}\,s^{-1}}$.
This flux is difficult to measure with \ac{INTEGRAL}/\ac{SPI} because it rarely observed the ecliptic.
However, since the signal is expected to be variable in time, because objects and accumulations move with respect to Earth, the bulge signal may be variable in time at this per cent level.
A strong reduction of the total annihilation rate is not plausible because otherwise the asteroids accumulations would have been found already.
In fact, the ``\ac{OSSE} 511\,keV fountain'' may be the cumulative effect of such asteroids integrated over several years \citep{Siegert2024}.
Future observations and analyses are required to gauge the local contribution.

\paragraph{Dark Matter}
The case of \ac{DM} will be discussed in more detail in the outlook Sec.\,\ref{sec:outlook_darkmatter}.
In principle, \ac{DM} can explain the entire 511\,keV signal statistically.
But astrophysically, this would imply that other objects should be much brighter than they are:
the dwarf galaxies should be visible, the halo of M31, and galaxy clusters should all show a sizeable 511\,keV signal, which is not the case, limiting the power of this explanation vastly.

%\subsubsection{Measured Annihilation Rate vs. Production Rates}

\subsubsection{The Positron Puzzle}\label{sec:positron_puzzle}
The description above can be summarised into the ``Positron Puzzle'', which poses one of the biggest unsolved problems in astronomy:
\begin{enumerate}
    \item What do we see?
    \item Why does it look like that?
    \item Where do the positrons come from?
\end{enumerate}
While we believe to understand what we actually observe in terms of positron annihilation, that is, the 511\,keV \ac{gray} line and the ortho-\ac{Ps} continuum, it is, in fact, not clear how the line is formed:
The line shape is the cumulative effect of \ac{LoS} integrals which would weight different phases of the \ac{ISM} with the emissivity from annihilation of different channels, of positrons being slowed down by continuous and catastrophic energy losses, having propagated through the same phases according to a rather unknown diffusion coefficient (tensor) $D_{\mu\mu}$ \citep[e.g.,][]{Spanier2022}.
The lack of angular resolution in combination with MeV instrument sensitivities allows us to only probe an average spectrum across large regions in the sky and of the Milky Way.
In fact, we cannot even say for certain that the observed emission features are Galactic or very nearby -- clearly the spatial coincidence with the Galactic centre and the Galactic plane make this plausible, but no kinematic measurement could prove or provide the assumed distances to the celestial positron annihilation.
It is again the epistemic problem of astronomy that there is no inherent distance information, and detailed modelling always carries an enormous uncertainty.
In what follows, we will discuss different parts of the more extended ``Positron Puzzle'' briefly.

\paragraph{Positron Transport in the Interstellar Medium?}
The distribution of annihilating positrons as a function of time, position and momentum should be described as the steady-state solution of a suitable Fokker-Planck-equation.
Such a diffusion-loss treatment should include drift (re-acceleration, convection, advection, energy losses), diffusion (coefficients due to transport mode and conditions), sources (see above), and sinks (annihilation rates, cross sections, escape, leakage timescales).
The problem in such a treatment is that it requires realistic 3D distributions of both, positron sources and the \ac{ISM}, proper injection spectra and considering turbulent magnetic fields, different transport modes and corresponding diffusion coefficients.
Most of these input parameters and distributions are very uncertain, so that other treatments, such as Monte Carlo simulations of individual positrons experiencing possible interactions, have been used in the past.

\begin{figure}
    \centering
    \includegraphics[width=1.0\linewidth]{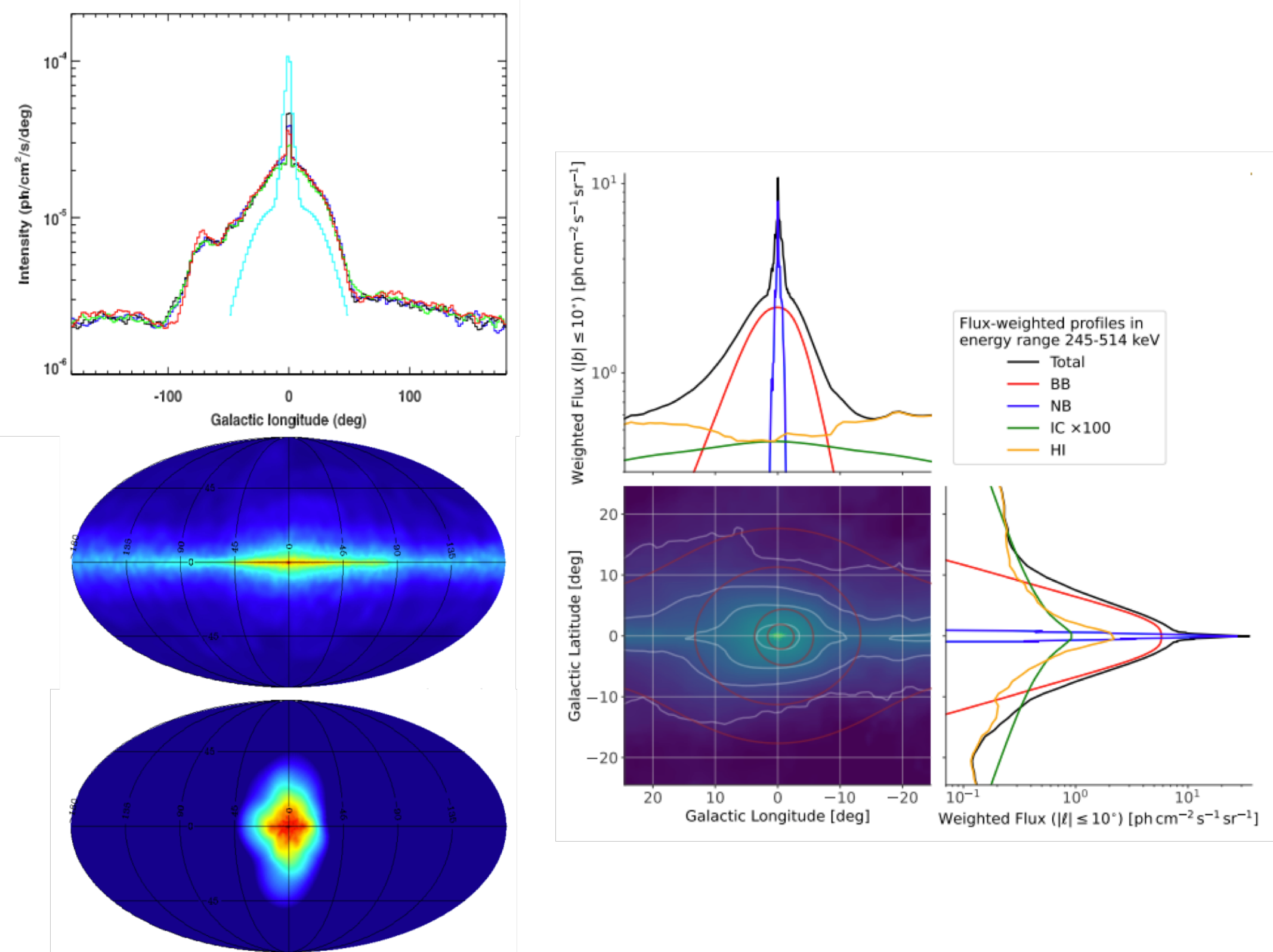}
    \caption{Astrophysical modelling of the 511\,keV line. Left: Monte Carlo simulations of positron transport in the \ac{ISM}. Top: Expected profile from only nucleosynthesis sources (colours) compared to the measured longitude profile (cyan). Middle: Expected sky map from only nucleosynthesis sources. Bottom: Simulated outburst of \nuc{Al}{26} positrons in the Central Molecular Zone after 2\,Myrs \citep[from][]{Alexis2014}. Right: Best-fit model of astrophysical components of positron annihilation (ortho-\ac{Ps} + 511\,keV line) in the bulge region of the Galaxy. Top: Longitudinal profile. Right: Latitudinal profile. Bottom left: Combination of sky maps (coloured) compared to best-fit maximum likelihood fit (red).}
    \label{fig:astrophysics_511keV_models}
\end{figure}

In \citet{Jean2009}, such Monte Carlo simulations have been performed based on a thorough recapitulation of possible transport modes.
In collisional transport, for example, positrons propagate almost ballistically and are guided by the large-scale magnetic field in the Galaxy.
Through pitch-angle scatterings in a turbulent magnetic field, the generic stopping distance for MeV positrons in typical \ac{ISM} conditions is only reduced by 25\%, resulting in an effective distance of $10\, (n/\mathrm{cm^{-3})^{-1}}$\,kpc.
This appears very large compared to the intuitive case of positrons annihilating whenever they enter dense regions.
Such a treatment, however, cannot reproduce the observed features in the Galaxy:
In \citet[][see Fig.\,\ref{fig:astrophysics_511keV_models}, left]{Alexis2014}, first nucleosynthesis sources only have been used as starting distribution to explain emission after propagation, but those fail to explain the large bulge flux, and also predict a spiral arm structure at 511\,keV, but which is not seen.
However, with a single Galactic outburst a few Myr ago from the direction of the Galactic centre, the tension between the production channels, propagation, and annihilation is not as severe.
Certainly, this topics require more attention also for future observations.

\paragraph{Smooth or Structured 511\,keV Image?}
Most analyses of the 511\,keV line as well as the ortho-\ac{Ps} assume spatial models for which the spectra are extracted.
This forward-folding approach has the advantage that the response function is properly included and that the uncertainties of the models used can be easily estimated by likelihood profiles.
As described above, the entire Milky Way in positron annihilation (line + ortho-\ac{Ps}; see also Sec.\,\ref{sec:not_a_line}) can be modelled by 3--4 Gaussian-shaped components -- three for bulge, and one for the disk \citep[e.g.,][]{Skinner2014,Siegert2016a}.
These models can be replaced by three or more astrophysically-motivated components, such as the infrared bulge and the nuclear stellar cluster, the HI or CO disk, or by \ac{DM} annihilation models, on top of some \ac{IC} continuum, which is always present.
Astrophysical models that describe the Galaxy by the old stellar population only require a weak disk component \citep{Siegert2022a}, but avoid the need for a \ac{DM} halo component altogether when both the 511\,keV line and the ortho-\ac{Ps} component is considered (Fig.\,\ref{fig:astrophysics_511keV_models}).
The 511\,keV line alone only makes less than 20\% of positron annihilation in the Galaxy, so that it must not be considered by theoretical models and data analyses separately.
However, due to computational constraints and the fact that the ortho-\ac{Ps} continuum quickly becomes sub-dominant towards lower energies, it is typically treated separately, especially because also the \ac{IC} continuum has its own large uncertainties.

The problem with this approach is that it always includes smooth models without structure or template maps whose meanings can be interpreted in different, non-unique ways.
Therefore, image reconstructions are used to obtain a possibly less-biased view of the MeV sky.
But image reconstruction can come with artefacts due to improper modelling of instrumental background, for example, low-number statistics, small exposure regions, strong gradients in exposure, extremely long datasets, and changes in the imaging and spectral response.
Therefore, one must take a large number of possible reconstruction algorithms and study the emerging features.
The uncertainties per pixel or per region are then not trivially determined -- a possible way is bootstrapping of the dataset so that individual emission features can be validated.
This again is computationally very expensive and only a few tens or hundred realisations of one and the same method can be achieved unless high-performance computers are available.

The image reconstructions of the 511\,keV line \citep[e.g.,][]{Knoedlseder2005,Weidenspointner2008,Bouchet2010} always resulted in the bulge region, plus some more features with increasing exposure, that resulted in the 511\,keV disk.
%
%After 2010, no study tried to reconstruct a 511\,keV image with \ac{INTEGRAL}/\ac{SPI}.
%
An important effect of imaging the sky with an instrument like \ac{SPI} that ``only'' sees about 1\% of the sky at any given time (\ac{FoV}), is that one can only see photons from direction where the instrument was actually pointed to.
This trivial statement, however, has the effect of dark regions in the sky whenever there was only weak exposure.
All-sky maps should therefore always be compared to the exposure maps of the respective instrument, so that also artefacts at exposure edges could easily be identified and not misinterpreted.

A truly smooth image could mean annihilation in gas, nearby objects, or stars.
A structured image could show a link to point sources of 511\,keV emission, similar to the case of the \ac{CXB} that was finally resolved by ROSAT observations to be due to active galactic nuclei.
Only with more than one robustly identified positron annihilation source can this part of the puzzle actually be solved.

\paragraph{More Positrons Injected than Observed?}
Given the large uncertainties in measuring positron sources, and the fact that some sources have not been identified as sources even though it is suggested (\acp{CN}, \acp{SNIa}, etc.), there is a possibiliyy that more positrons are created than observed to annihilate.
This could be due to an observational bias with coded mask telescopes that cannot observe isotropic emission \citep{Siegert2022d}.
As has been suggested by \ac{OSSE} observations, there could be a much stronger halo, which however would be difficult to identify with \ac{SPI}.
In fact, there could be an almost or entirely isotropic component of positron annihilation that would make up some part of the \ac{CGB}.
By previous measurements, such a scenario is not excluded as most \ac{CGB} measurements above 200\,keV and below 30\,MeV are notoriously uncertain and not highly resolved.
With occultation observations or with a dedicated Compton telescope that sees large parts of the sky and with high spectral resolution, such a measurement could be made possible.

The observed \emph{annihilation rate} (of the Milky Way) is on the order of $5 \times 10^{43}\,\mathrm{e^+\,s^{-1}}$ -- the potential positron \emph{production rate}, taking into account uncertainties in models, observations, cross sections, etc., could be a factor of five higher \citep[][see also Fig.\,\ref{fig:sources}]{Siegert2023}.
Observations of another Milky-Way-like galaxy, such as M31, could gauge this number because the entire galaxy would be in the \ac{FoV} and the distance-question would not be much of an issue.
However, if the \emph{production rate} is significantly higher than the \emph{annihilation rate}, where do the remaining positrons go?
A possible solution would be the escape into intergalactic medium in which highly relativistic positrons would potentially take Gyrs to cool to produce a 511\,keV line, so that their contribution to the \ac{CGB} spectrum might be small.
On the other hand, since this effect might have occurred already over the time scale of the Galaxy, the intergalactic positrons may also have relaxed to a sizeable stead-state annihilation rate.

\paragraph{In-Flight Annihilation?}
So far, no firm detection of in-flight annihilation has been detected.
Previous limits for the injection energy are on the order of only 3--7\,MeV \citep[e.g.,][]{Sizun2006,Beacom2006} which is at odds with the high-energy positrons observed at GeV energies.
\citet{Siegert2023} contested these values by re-investigating the data analysis:
Given that the \ac{IC} continuum is uncertain in itself and that the combination of different instruments that measured at two different epochs with different apertures is delicate, and given updated measurements from \ac{SPI} that finally superseded the accuracy of the \ac{COMPTEL} data points \citep{Siegert2022c}, injection energies of up to 100\,MeV suddenly appeared possible.
If this analysis turns out to be accurate, it would open up space for even more sources, which have previously been rejected because of too high injection energies, such as pulsars, \acp{CR}, \acp{WIMP}, or \ac{gray} bursts.
\begin{figure}[!t]
    \centering
    \includegraphics[width=0.60\linewidth]{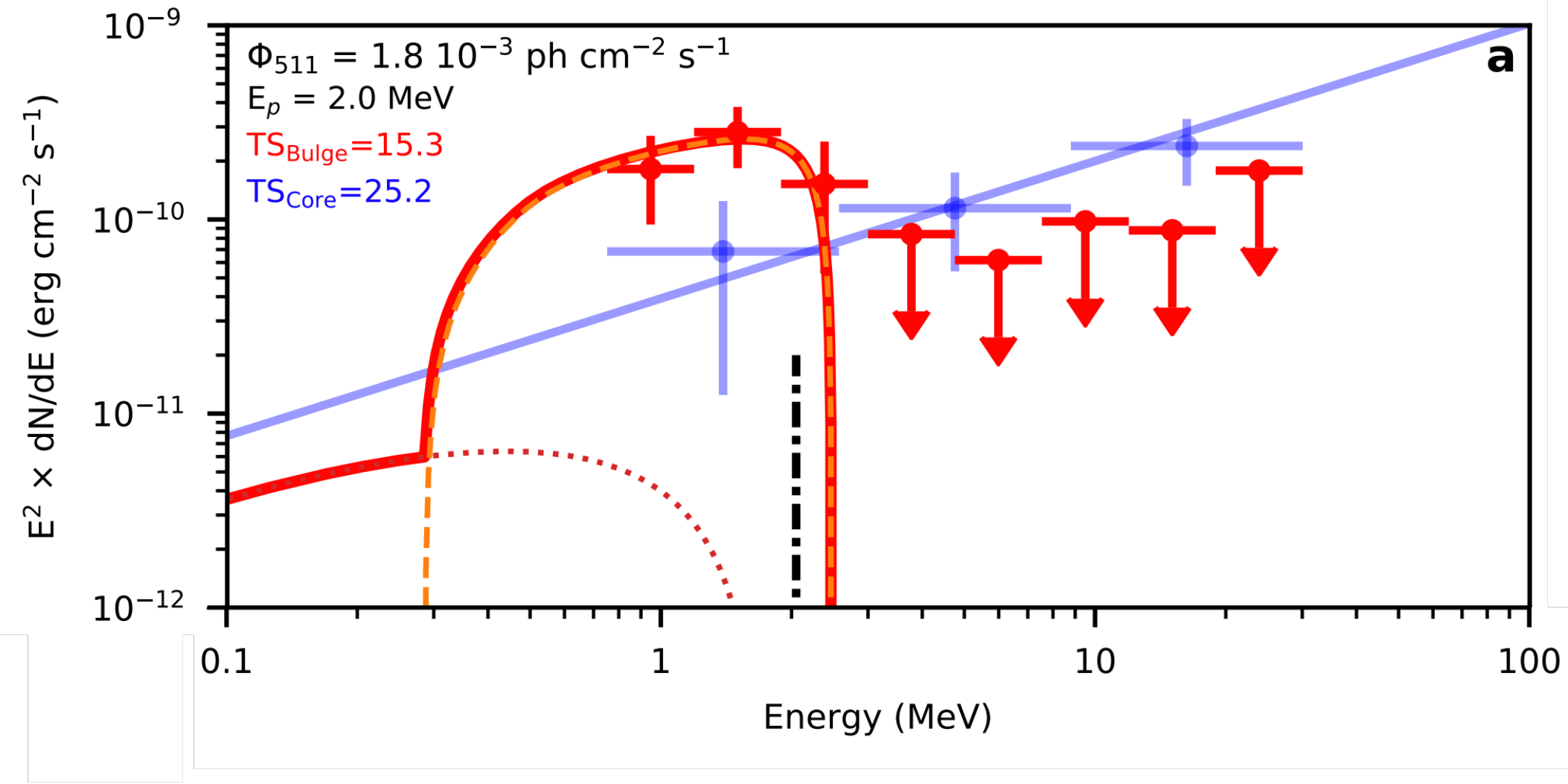}\\
    \includegraphics[width=0.60\linewidth]{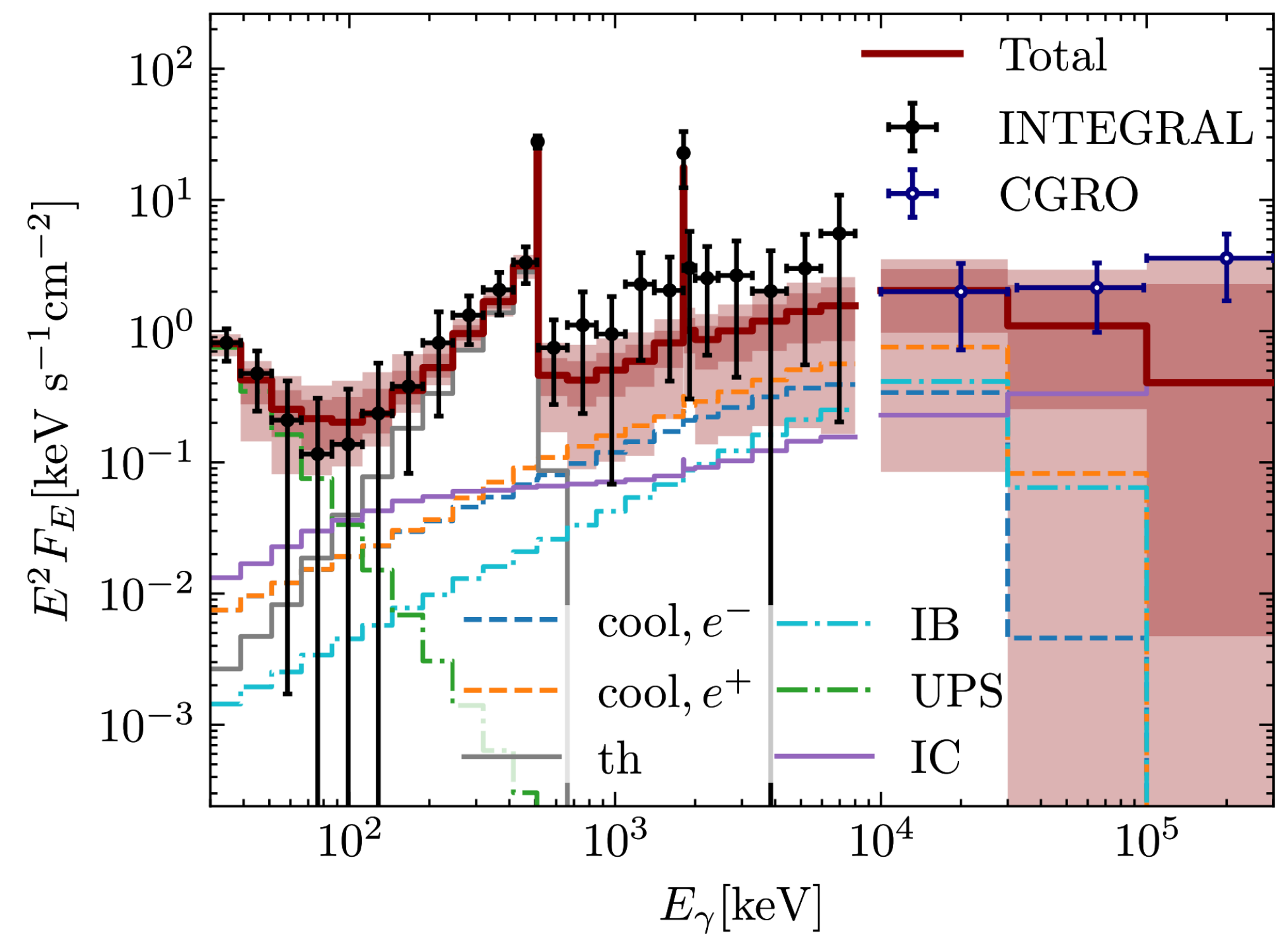}\quad
    \caption{Measured continuum spectrum in the Milky Way bulge region. Top: \ac{CGRO}/\ac{COMPTEL} analysis (red data points: bulge, blue data points: centre) with spectral fit of in-flight annihilation of positrons with an injection energy of 2\,MeV (dashed orange) and bremsstrahlung (dotted) \citep[adapted from][]{Knoedlseder2025_aif}. Bottom: \ac{INTEGRAL} (black data points) and \ac{CGRO} (blue data points) analysis of a circular $9^\circ$-radius region around the Galactic centre, together with a multi-component fit and $1$ and $2\sigma$ (shaded red) bands \citep[from][]{Das2025_aif}.}
    \label{fig:aif}
\end{figure}

Two recent studies investigate the possibility of in-flight annihilation in the Galactic bulge region.
While \citet{Knoedlseder2025_aif} focus on a re-analysis of the energy range of 0.7--30\,MeV with \ac{CGRO}/\ac{COMPTEL}, \citet{Das2025_aif} combined the most recent \ac{INTEGRAL}/\ac{SPI} analysis \citep{Siegert2022c,Berteaud:2022tws}, and data from \ac{CGRO}.
The paper by \citet{Knoedlseder2025_aif} claims to have detected positron annihilation in flight, and sets a strict injection energy of positrons of $\approx 2$\,MeV.
This would rule out positrons from radioactive decays ($\lesssim 1$\,MeV) and from \ac{CR} secondaries ($\gtrsim 100$\,MeV).
%
%This would rule out almost all positrons sources except nucleosynthesis ($\lesssim 1$\,MeV), pair-plasma (jet) ejections from accreting compact-object binaries ($\lesssim$\,few MeV), or \ac{DM} ($\gtrsim$\,MeV).
%
The work by \citet{Das2025_aif} finds less stringent bounds on the injection energy of positrons up to $\sim 100$\,MeV, which would allow more source types to contribute, such as pulsars and \ac{CR} positron production.
\citet{Das2025_aif} make the important remark that the excess found may not necessarily be due to positrons annihilating in flight, but that a population of electrons with a cut-off around an energy of 100\,MeV would show a similar spectrum from bremsstrahlung.
The authors find that the bremsstrahlung component would always be dominant, except for photon energies close to the cut-off.
The spectrum of \citet{Knoedlseder2025_aif} in Fig.\,\ref{fig:aif}, top, shows only the excess component, the authors find above the Galactic ridge, which already includes the bremsstrahlung of \ac{CR} electrons.
The bottom spectrum of Fig.\,\ref{fig:aif} shows the total Galactic spectrum in circular a $9^\circ$-radius region around the Galactic centre from \citet{Das2025_aif}.
While the spectral analyses and models differ in detail, it becomes clear that with improved analysis techniques, a long-sought component could finally be studied in detail.

\paragraph{Similarities to Infrared Emission and Dark Matter}
\subparagraph{Old stars} Fitting full-sky maps of other wavelengths to the raw \ac{gray} data results typically in bad fits.
This is true except for the maps around 1.25--4.9\,micron, i.e. at infrared wavelengths that mostly show starlight.
This points to the or an old population of stars, as opposed to the case of \nuc{Al}{26} from young massive stars, but which has been proven to a positron source.
However, alternative explanations for this match may come from flaring stars.
As described above, \citet{Bisnovatyi-Kogan2017} suggest that all the \emph{annihilating} positrons in the Milky Way come from main sequence and evolved stars.
While this study probably predicts too much 511\,keV flux, it would naturally explain the morphology and the spectrum, and would not require a lot of positron propagation.
Good candidates to further test this scenario are globular clusters to identify point-like enhancements above the diffuse emission.
Alternatively, satellite and other galaxies should then show a 511\,keV flux that only scales with stellar mass.
Only future observations with improved sensitivity or detailed analyses with the 22\,yr of \ac{INTEGRAL}/\ac{SPI} observations could verify such a scenario.

\subparagraph{Dark matter} \ac{NFW} profiles squared can replace the bulge components unless other physically motivated templates are used, again pointing to an old stellar population (see above).
If the 511\,keV emission \emph{and} the ortho-\ac{Ps} continuum \emph{and} the inflight annihilation continuum all point to \ac{DM}, it potentially points to a \ac{DM} particle in the MeV mass range \citep{DelaTorreLuque:2023cef,DelaTorreLuque:2023huu,DelaTorreLuque:2024zsr}.
However then, the annihilation cross section would be extremely large compared to the thermal relic cross section \citep{Slatyer:2016qyl}.
Other sources would then be required to explain the data.
It is not formally ruled out that some positrons are due to \ac{DM}, but there are always other sources to explain the signal.
\ac{DM}, be it in the MeV mass range or GeV--TeV, would always add more difficulties than solutions because suddenly other objects should be visible.
More details about the \ac{DM} case can be found in Sec.\,\ref{sec:outlook_darkmatter}.

\subsubsection{Positron Annihilation is not Always a 511\,keV Line}\label{sec:not_a_line}
As already mentioned several times in the previous sections, \emph{positron annihilation does not always result in a 511\,keV line}.
In fact, the 511\,keV line is the rarer case in the \ac{ISM}, given that most positrons form \ac{Ps} with a fraction of 90--100\%.
This means that at least 3.1, up to maximally 4.5, times the flux of the 511\,keV is found in the ortho-\ac{Ps} continuum.
Solving Eq.\,(\ref{eq:Ps_fraction}) for the ratio $R_{32}$ of ortho-\ac{Ps} continuum vs. 511\,keV line flux sets the \emph{total annihilation flux of thermalised positrons} to
\begin{equation}
    F_{\rm ann}^{\rm th} \approx (1+R_{32}) F_{511}\mathrm{.}
    \label{eq:total_ann_flux_therm}
\end{equation}
Furthermore, the in-flight annihilation spectrum of mildly and strongly relativistic positrons would add a continuum that depends strongly on the injection energy and the exact cooling conditions.
The survival probability,
\begin{equation}
    P \equiv P(E_{\rm kin}, m_e) = \exp\left(-n_{X} \int_{m_e}^{E_{\rm kin}}\,dE \frac{\sigma(E)}{\left|\frac{dE(E,n_X)}{dx}\right|}\right)\,\mathrm{,}
    \label{eq:surv_proba}
\end{equation}
where $\sigma(E)$ is the total annihilation cross section, and $\left|\frac{dE(E,n_X)}{dx}\right|$ is the stopping power (energy loss rate, cooling function) of a positron propagating in a medium with density $n_X$, suggests that the \emph{total annihilation flux} can be expressed as 
\begin{equation}
    F_{\rm ann}^{\rm tot} = (1 + R_{32} + R_{\rm IA}) F_{511}\mathrm{.}
    \label{eq:total_ann_flux}
\end{equation}
Here, the ratio of the in-flight annihilation flux and 511\,keV line flux, $R_{\rm IA}$, is given by
\begin{equation}
    R_{\rm IA} = \frac{1}{1 - \frac{3}{4}f_{\rm Ps}}\frac{1-P}{P} \mathrm{.}
    \label{eq:R_IA}
\end{equation}
With typical values of $P$ in the \ac{ISM} of 80--95\% \citep[e.g.,][]{Beacom2006,Siegert2022b}, the in-flight annihilation continuum is expected to be on the order of 20\%--100\% of the 511\,keV line flux.
However, since the spectrum is spread out to energies equivalent to the injection energies of positrons, the in-flight annihilation spectrum is notoriously difficult to detect.
In fact, so far all studies failed to uniquely identify positron annihilation in flight as an additional component above the Galactic diffuse emission.
Given Eq.\,(\ref{eq:total_ann_flux}), it is clear that the 511\,keV line is just one part of the \emph{positron annihilation spectrum}, and certainly not the strongest one.

It should be noted here that there are also other positron annihilation spectra:
\begin{enumerate}
    \item The ``accelerator case'' in which relativistic positrons interact and potentially annihilate with relativistic electrons can lead to doubly peaked or singly peaked spectra, depending on the kinetic energies.
    \item ``Thermal pair plasma'' annihilation happens when an electron-positron pair-plasma, for example as ejected in jets of microquasars or active galactic nuclei, is relaxing. Here, the relativistic Doppler effects broadens and blue-shifts the annihilation feature \citep{Svensson1983}. \emph{In general, there is no narrow 511\,keV ``line'' in microquasars unless there are very specific circumstances \citep{Mirabel1992}.}
    \item The ``narrow 511\,keV line'' from the Galactic bulge region is actually broadened. The true width depends on the \ac{ISM} conditions such as temperature, composition, and ionisation state. There is also a ``broad 511\,keV line'' from the same region.
    %, but which is often ignored in theoretical papers.
    %
    \item Positron annihilation in dense material, such as atmospheres, rocks, asteroids, dust, etc., will inevitably lead to a ``narrow 511\,keV line''.
\end{enumerate}
%

%BSM
%\newpage
%\input{BSM}

\newpage
\section{Outlook}\label{sec:outlook}
\vspace{-0.5em}
{\emph{Written by Thomas Siegert, with contributions from multiple authors (see below)}}
\vspace{0.5em}
\vspace{-12pt}

% ****** Start of file apssamp.tex ******
%
%   This file is part of the APS files in the REVTeX 4.2 distribution.
%   Version 4.2a of REVTeX, December 2014
%
%   Copyright (c) 2014 The American Physical Society.
%
%   See the REVTeX 4 README file for restrictions and more information.
%
% TeX'ing this file requires that you have AMS-LaTeX 2.0 installed
% as well as the rest of the prerequisites for REVTeX 4.2
%
% See the REVTeX 4 README file
% It also requires running BibTeX. The commands are as follows:
%
%  1)  latex apssamp.tex
%  2)  bibtex apssamp
%  3)  latex apssamp.tex
%  4)  latex apssamp.tex
%
%\documentclass[%
%preprint, linenumbers,
%superscriptaddress,
%groupedaddress,
%unsortedaddress,
%runinaddress,
%frontmatterverbose, 
%preprint,
%preprintnumbers,
%nofootinbib,
%nobibnotes,
%bibnotes,
% amsmath,amssymb,
% aps, physrev,
%]{revtex4-2}

%\usepackage{graphicx}% Include figure files
%\usepackage{dcolumn}% Align table columns on decimal point
%\usepackage{bm}% bold math
%\usepackage{aas_macros}
%\usepackage{hyperref}% add hypertext capabilities
%\usepackage[mathlines]{lineno}% Enable numbering of text and display math
%\linenumbers\relax % Commence numbering lines

%\usepackage[showframe,%Uncomment any one of the following lines to test 
%%scale=0.7, marginratio={1:1, 2:3}, ignoreall,% default settings
%%text={7in,10in},centering,
%%margin=1.5in,
%%total={6.5in,8.75in}, top=1.2in, left=0.9in, includefoot,
%%height=10in,a5paper,hmargin={3cm,0.8in},
%]{geometry}

%\begin{document}

%\preprint{APS/123-QED}

\subsection{Laboratory Measurements for the Production of $\gamma$-Ray Emitters}\label{sec:lab}
\vspace{-0.5em}
{\emph{Written by Nicolas de Séréville}}
\vspace{0.5em}\\
%\vspace{-12pt}
%
%\subsubsection{Nuclear Cross-Sections Relevant For MeV \ac{gray} Astronomy}\label{sec:lab_level1}
%
The MeV band, covering two energy decades, is particularly interesting as it corresponds to the energy range of atomic nuclear transitions, where MeV \acp{gray} are produced either during nuclear processes or interactions.
Observations in this spectral range provide a unique window to directly study nucleosynthesis by measuring the abundance of radionuclides, rather than inferring elemental abundances from observations at other wavelengths.
Comparing observed abundances with those predicted by astrophysical models is essential for understanding astrophysical objects.
These models incorporate multiple ingredients, and abundance predictions are closely tied to thermonuclear reaction rates.

\subsubsection{Reaction Rates, Nuclear Cross-Sections and Their Determination}\label{sec:level2}
In a stellar plasma where the kinetic energy of nuclei is given by the thermal agitation velocity, and for a non-degenerate perfect gas where the velocity follows the Maxwell-Boltzmann distribution, thermonuclear reaction rates are defined per particle pair (in units of $\mathrm{cm^{3}\,s^{-1}}$) as \citep{Iliadis2015}:
\begin{equation}
    \langle\sigma v\rangle = \sqrt{\frac{8}{\pi\mu}} \cdot \frac{1}{(k_{\rm B}T)^{3/2}} \int_{0}^{\infty}\sigma(E) E e^{-E/k_{\rm B}T}dE,
    \label{eq:rate}
\end{equation}
where $\mu$ is the reduced mass of the interacting nuclei, $k_{\rm B}$ is the Boltzmann constant, $T$ is the temperature at which the reaction rate is evaluated, and $\sigma(E)$ is the energy-dependent cross-section of the reaction.
Thermonuclear reaction rates are crucial inputs to stellar models, with cross-sections being the fundamental physical quantities that need to be determined, either through theoretical calculations or experimental measurements.
However, the experimental determination of cross-sections of nuclear reactions relevant to astrophysics presents significant challenges.

\begin{figure}[!ht]
    \centering
    \includegraphics[width=0.65\linewidth]{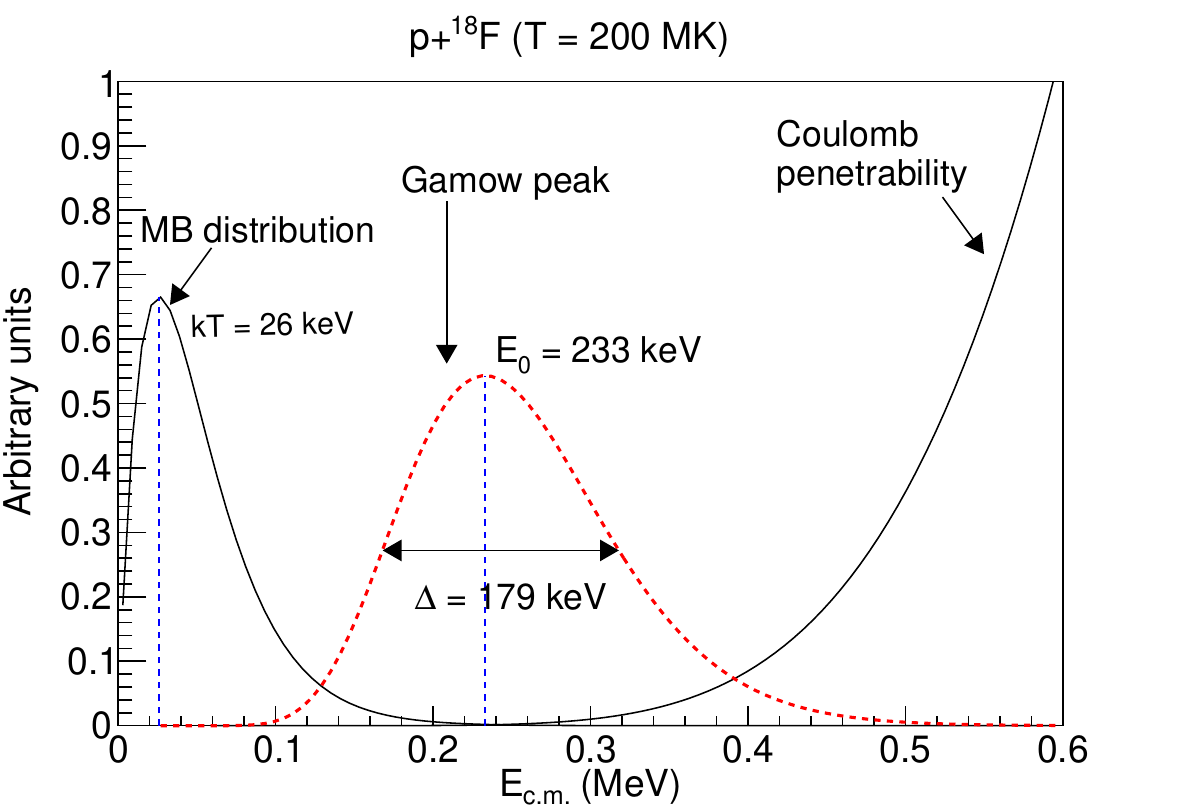}
    \caption{Maxwell-Boltzmann distribution, Coulomb penetrability and Gamow peak for the p + \nucc{18}{F} system at $2 \times 10^8$\,K relevant for the \reaction{18}{F}{p}{$\alpha$}{15}{O} reaction and the prediction of the \ac{gray} emission at \protect{$\leq511$}\,keV in \acp{CN}.}
    \label{fig:gamow}
\end{figure}

In astrophysical environments, and for interacting charged-particles, nuclear reactions occur at low relative energies in the so-called Gamow peak, located around typical energies of a few tens to a few hundreds of keV in quiescent burning environments, and around several hundreds of keV to a few MeV during explosive nucleosynthesis.
The Gamow peak (Fig.\,\ref{fig:gamow}) results from the product of the terms present in the integrand of Eq.\,(\ref{eq:rate}).
Indeed, cross-sections drop exponentially at low energies because of the Coulomb (in case of charged particles) and centrifugal barriers (given by the relative angular momentum of interacting nuclei), while the Maxwell-Boltzmann distribution approaches zero for large energies.

There are two experimental approaches for measuring cross-sections relevant to astrophysics.
In direct measurements, the cross-section of the astrophysical reaction is the one which is measured in the laboratory, at the lowest energy possible.
However, because the cross-sections are extremely small, and well below the Coulomb and centrifugal barriers, dedicated facilities and experimental setups are used to suppress background signals which would otherwise prevent the measurement at the lowest energies.
These measurements include the use of recoil separators (e.g., DRAGON \citep{DRAGON}, St Georges \citep{Couder2008}, SECAR \citep{SECAR}), coincidence detection systems \citep[e.g., STELLA][]{STELLA}, and underground laboratories (e.g., LUNA \citep{LUNA}, JUNA \citep{JUNA}, Felsenkeller \citep{Felsenkeller}), among others.
When direct measurements are not feasible, indirect techniques such as transfer reaction method, Coulomb dissociation method, surrogate reactions and Trojan Horse Method are good alternatives \citep[see][and references therein]{Hammache2021}.
In these approaches the experiments are usually performed at energies above the Coulomb barrier where cross-sections are significantly higher than in direct measurements.
However, since the reactions studied in the laboratory differ from the astrophysically ones, the extraction of the relevant information, and the link with the cross-section of interest, rely on nuclear theory.
Moreover, many astrophysical processes involve unstable nuclei, which presents additional challenges for experimental measurements, further complicating the determination of reliable cross-sections.

\subsubsection{Observed $\gamma$-Ray Emitters}
The \ac{gray} emission of only a few limited number of radionuclides has been observed in the Milky Way by balloon-borne and space-borne instruments so far.
Here, we recapitulate on the detected ones and then give details on their nuclear structure and laboratory measurements.
The isotope \nucc{26}{Al} undergoes $\beta^+$-decay to the first excited state of \nucc{26}{Mg}, followed by de-excitation to the ground state that produces a \ac{gray} line at 1809\,keV (see Sec.\,\ref{sec:Al26}).
This \ac{gray} emission is observed in the Galactic plane and is mainly associated to the population of massive stars during their core-collapse and/or during their Wolf-Rayet phase for the most massive ones.
\ac{AGB} stars and \acp{CN} are also expected to contribute to the \nucc{26}{Al} budget.
In these astrophysical sites, \nucc{26}{Al} is mainly produced by radiative proton capture on \nucc{25}{Mg}.
The most influential resonances above the $p+$\nucc{25}{Mg} threshold are located at $E_r^{c.m.}=59$-, 93-, 191- and 305-keV, the lowest two dominating the \reaction{25}{Mg}{p}{$\gamma$}{26}{Al} reaction rate in the temperature region $(0.02$--$0.15) \times 10^9$\,K, and the other two above.
The strength of the highest three resonances has been measured directly at LUNA \citep{Limata2010,Strieder2012}, and at JUNA for the \Ercm{93} one \citep{Su2022}.
For the lowest resonance at \Ercm{59} its energy is too low for direct measurements and its resonance strength is computed using the proton width deduced from one-proton transfer reactions such as \reaction{}{}{$^3$He}{d}{}{} \citep{Champagne1989,Rollefson1990} and more recently \reaction{}{}{$^7$Li}{$^6$He}{}{} \citep{Li2020}.
The resonance strengths agree within uncertainty when several determinations have been reported, and the thermonuclear \reaction{25}{Mg}{p}{$\gamma$}{26}{Al} reaction rate has in the end an uncertainty smaller than 20\% over the whole temperature range covered by the different astrophysical sites producing \nucc{26}{Al}.

The main destruction mode of \nucc{26}{Al} in stellar environments is either through proton or neutron capture.
During core H-burning in massive stars and in \ac{AGB} stars, \nucc{26}{Al} is destroyed through the \reaction{26}{Al}{p}{$\gamma$}{27}{Si} reaction whose rate is dominated by a single $\ell_p=0$ resonance at \Ercm{128}.
Transfer reactions have been used to determine the single-particle width of its associated state in \nucc{27}{Si} \citep{Vogelaar1996} and of its analogue in \nucc{27}{Al} \citep{Pain2015,Lotay2020}.
These two approaches are in disagreement by a factor of four, which could be solved with a direct measurement of the resonance strength when more intense \nucc{26}{Al} beams become available.
At higher temperatures in \acp{CN}, the destruction of \nucc{26}{Al} is dominated by the \Ercm{188} resonance.
While its spin and parity are now confirmed to be $11/2^-$, its precise energy is still not accurately determined, and differs by 5\,keV between DRAGON measurement \citep{Ruiz2006} and \ac{gray} studies \citep{Lotay2009}.
During hydrostatic C-shell and explosive Ne/C-shell burning phases in massive stars, \nucc{26}{Al} is destroyed by neutron capture.
Spectroscopic information on \nucc{27}{Al} is needed up to 500\,keV above the $n+$\nucc{26}{Al} threshold, but the data remain incomplete.
Direct measurements have been performed using the high neutron flux beam-line EAR-2 at n\_TOF CERN, and at the GELINA facility \citep{Lederer2021a,Lederer2021b} allowing cross-sections determination up to about 150\,keV above the $n+$\nucc{26}{Al} threshold.
Even though indirect measurements identified \nucc{27}{Al} levels up to 500\,keV above the $n+$\nucc{26}{Al} threshold \citep{Benamara2014}, the resonance strength are not yet determined.
The isomeric \nucc{26m}{Al} state has a short lifetime of about 6\,s, allowing rapid thermal equilibration with the ground state under typical stellar temperatures.
Its presence slightly alters the effective destruction and production rates of \nucc{26}{Al} in stellar interiors because thermal coupling can shift population between the two levels.
In laboratory measurements, however, the isomer decays before significant reactions can occur, so experimental cross sections generally probe only the ground-state behaviour.
Further information concerning the nuclear aspects of \nucc{26}{Al} nucleosynthesis including also the isomeric state at 228\,keV can be found in recent reviews \citep[e.g.,][]{Diehl2021, Laird2023}.

The isotope \nucc{60}{Fe} decays via the chain \nucc{60}{Fe}$\rightarrow$\nucc{60}{Co}$\rightarrow$\nucc{60}{Ni}, producing two \ac{gray} lines at 1173 and 1332\,keV that have been observed in the Galactic plane by \ac{INTEGRAL}/\ac{SPI} and \ac{RHESSI} (see Sec.\,\ref{sec:Fe60}).
The \nucc{60}{Fe} radionuclide is produced in massive stars and intermediate-mass thermal-pulsating \ac{AGB} stars during the $s$-process which results from a competition between slow radiative neutron captures and $\beta^-$-decays.
\nucc{60}{Fe} nucleosynthesis occurs for temperatures greater than $5 \times 10^8$\,K and neutron densities on the order of $\sim 10^{10-12}\,\mathrm{cm^{-3}}$ \citep{Limongi2006}.
For temperatures smaller than $2 \times 10^9$\,K, \nucc{60}{Fe} is mainly produced by neutron capture on the unstable nucleus \nucc{59}{Fe} and destroyed by the \reaction{}{}{n}{$\gamma$}{}{} reaction.
Indeed, the destruction of \nucc{60}{Fe} by neutron capture is always faster than the $\beta^-$-decay of \nucc{60}{Fe} for which the terrestrial lilfetime is now well determined as $\tau=(3.78\pm0.06)$\,Myr \citep{Rugel2009}.
The Maxwellian averaged \reaction{60}{Fe}{n}{$\gamma$}{61}{Fe} cross-section at $k_{\rm B}T=25$\,keV has been measured by activation and is $9.9^{+2.8 (\rm syst)}_{-1.4 (\rm stat)}$\,mbarn \citep{Uberseder2009}, confirming that the resonant capture is dominant \citep{Giron2017}.
The main nuclear uncertainty comes from \nucc{60}{Fe} production which requires the nuclear flow to bypass the \nucc{59}{Fe} bottleneck.
The stellar lifetime of \nucc{59}{Fe} is much shorter than its terrestrial lifetime ($\tau=64.2$\,d) because of the thermal excitation of its excited state at 472\,keV which decays more favourably to the ground state of its daughter nucleus.
The effect of this excited state has only been recently studied experimentally at NSCL using the charge exchange \reaction{59}{Co}{t}{$^3$He}{59}{Fe} reaction \citep{Gao2021}.
The obtained stellar lifetime is in agreement with the latest theoretical estimates \citep{Li2016}, but a factor 3.5 larger than the \nucc{59}{Fe} decay rate used in stellar models \citep{Langanke2001}, leading to a decrease by 40\% of the \nucc{60}{Fe} production yield for a 18\,$\mathrm{M}_\odot$ star of Solar metallicity \citep{Gao2021}.
Due to the unstable nature of \nucc{59}{Fe}, the \reaction{59}{Fe}{n}{$\gamma$}{60}{Fe} cross-section has been the focus of several indirect experimental studies.
Earlier studies relying on the Coulomb dissociation method \citep{Uberseder2014} and the surrogate method \citep{Yan2021} agree within uncertainty.
However, a recent work using the $\beta$-Oslo method finds a significant enhancement of the \ac{gray} strength function at low energies, increasing the Maxwellian averaged cross-section by a factor of $1.6$--$2.1$, leading to an increase of almost a factor of two of \nucc{60}{Fe} yields for Solar metallicity stars of $15$--$25\,\mathrm{M}_\odot$ initial mass \citep{Spyrou2024}.
%
%Further details on the different aspects of \nucc{60}{Fe} nucleosynthesis can be in found in~\cite{Diehl2021}.

Other astrophysical \ac{gray} emitters are \nucc{44}{Ti} and \nucc{56}{Co} produced by \acp{SN} (see Sec.\,\ref{sec:SNe}).
\nucc{44}{Ti} has been observed in SN\,1987A \citep{Grebenev2012,Boggs2015} and \ac{CasA} \citep{Grenfenstette2014,Weinberger2020} \acp{ccSN}.
The \ac{gray} lines associated to the \nucc{56}{Ni} decay chain have been observed in the \ac{SNIa} SN\,2014J \citep{Churazov2014,Diehl2015,Diehl2014,Isern2016}.
The nucleosynthesis of \nucc{44}{Ti} and \nucc{56}{Ni} happen during $\alpha$-rich freeze-out in the deepest layers of a massive star during its core-collapse, providing the opportunity to constrain properties of \ac{ccSN} explosions.
Because the temperatures are typically between $4$ and $10 \times10^9$\,K, the nucleosynthesis network is extended and involves many nuclei, charged particle and weak reaction rates.
Several sensitivity studies have been conducted \citep[e.g.,][]{Magkotsios2010, Subedi2020, Hermansen2020} to assess the rates impacting the most \nucc{44}{Ti} and \nucc{56}{Ni} nucleosynthesis, pointing out a dozen of key reactions including ($\alpha$,p) and (p,$\gamma$).

\subsubsection{Yet to be Observed \ac{gray} Emitters}
Several nuclear \ac{gray} lines have yet to be observed.
A few of them originate from \acp{CN} which results from a thermonuclear runaway at the surface of an accreting \ac{WD} powered by explosive hydrogen burning (see Sec.\,\ref{sec:CNe}).
The main \ac{gray} emitters produced are \nucc{7}{Be}, \nucc{13}{N}, \nucc{18}{F}, \nucc{22}{Na} and \nucc{26}{Al}.

Even though the 1.8\,MeV \ac{gray} line is mostly associated with massive stars and \acp{ccSN}, the contribution of \acp{CN} is not well known \citep{Vasini2025} and could be 12\% or higher \citep[e.g.,][]{Laird2023,Vasini2025}.
In this context, the \reaction{25}{Al}{p}{$\gamma$}{26}{Si} reaction producing \nucc{26}{Si}, which decays to \nucc{26}{Al}, is the last reaction for which the uncertainty still needs to be reduced.
The reaction rate uncertainty at $10^8$\,K of a factor of ten is associated with the lack of spectroscopic information for \nucc{26}{Si} states, mainly the \Ercm{163} resonance.
This situation arises from the difficulty to produce intense radioactive \nucc{25}{Al} beams, so that several properties have to be estimated using theoretical shell model calculations or properties of analogous levels in the mirror nucleus \nucc{26}{Mg}.
A recent experiment was performed at FRIB using the GRETINA \ac{gray} array and the S800 magnetic spectrometer with the goal to measure the strength of \nucc{26}{Si} resonant states \citep{Fougeres2023b}.

The 1275\,keV \ac{gray} line associated with the \nucc{22}{Na} decay is particularly interesting since it could be used as a proxy to the \ac{CN} rate in the Milky Way because the lifetime of \nucc{22}{Na} ($\tau = 3.75$\,yr) is much longer than the time interval between two outbursts (\ac{CN} rate $\sim 50^{+31}_{-23}\,\mathrm{yr}^{-1}$ \citep{Shafter2017}).
The main nuclear uncertainty used to come from \reaction{22}{Na}{p}{$\gamma$}{23}{Mg} whose reaction rate is dominated by the \Ercm{204} resonance.
On the one hand, direct determinations of the resonance strength disagree at the $5\sigma$ level by at least a factor of four \citep{Seuthe1990,Stegmuller1996,Sallaska2010}.
On the other hand, indirect measurements give a resonance strength at least five times smaller than the direct ones.
This tension was alleviated by a recent study of the $^{3}$He($^{24}$Mg,$^{4}$He)$^{23}$Mg reaction at GANIL, where the strength of the \Ercm{204} resonance was determined using the AGATA \ac{gray} array together with the VAMOS magnetic spectrometer, and the SPIDER silicon detector \citep{Fougeres2023}.
In this approach, both the lifetime ($\tau = 11^{+7}_{-5}$\,fs) and proton branching ratio ($\mathrm{BR}_p = 0.68 \pm 0.17$) were determined, which, combined with existing measurements, give a consolidated resonance strength $\omega\gamma = 0.24^{+0.11}_{-0.04}$\,meV, consistent with indirect measurements.
The uncertainty on the \reaction{22}{Na}{p}{$\gamma$}{23}{Mg} reaction rate is now reduced to 10--40\% at the \ac{CN} peak temperature, resulting in an uncertainty of 1.4 in the estimated \nucc{22}{Na} yield \citep{Fougeres2023}.

The \ac{gray} emission at and below 511\,keV is produced by positrons, coming mostly from the $\beta^+$-decay of \nucc{18}{F}, annihilating in the material ejected by the outburst.
This emission happens before the visual peak luminosity of the \ac{CN}, which makes it a difficult observational target.
However, its observation would bring very valuable information on the nucleosynthesis and on the dynamic of the ejection since the lifetime of \nucc{18}{F} ($\tau = 159$\,min) is similar to the characteristic time for the envelope to become transparent to \acp{gray}.
The main remaining nuclear uncertainty for \nucc{18}{F} nucleosynthesis comes from its destruction path through the \reaction{18}{F}{p}{$\alpha$}{15}{O} reaction.
This reaction is still the focus of many experimental studies despite efforts over more than two decades \citep{deOliveira2025}.
It represents a complex case where the number and energy of $\ell=0$ \nucc{19}{Ne} states in the vicinity of the $p+$\nucc{18}{F} threshold are not yet firmly established.
Direct measurements focused on the determination of the cross-section at low-energy \citep{deSereville2009,Beer2011}, with the goal to constrain the sign of interferences between $s$-wave states, whose effect is maximum in the Gamow peak.
A wide variety of indirect methods \citep[see, e.g.,][and references therein]{deOliveira2025} have also been used to determine the properties of individual states in the compound nucleus \nucc{19}{Ne} above and below the proton threshold.
However, the current understanding is still incomplete mainly due to missing properties of $\ell=0$ states below the proton threshold, the sign of their interference with other $s$-wave states above the proton threshold, for which their $\alpha$-particle width is unknown.
This situation leads to an uncertainty on the \reaction{18}{F}{p}{$\alpha$}{15}{O} reaction rate about a factor of three, which translates to a \nucc{18}{F} yield uncertain by more than a factor of ten \citep{Kahl2021}.

Massive stars during their core-collapse produce a large number of nuclides, and the characteristic \ac{gray} lines of a few of them have been observed (see previous sections).
With improved line sensitivity in future MeV observatories other \ac{gray} emitters such as \nucc{43}{K}, \nucc{48}{Cr}, \nucc{48}{V}, \nucc{52}{Mn}, \nucc{44,47}{Sc}, \nucc{47}{Ca}, \nucc{51}{Cr} and \nucc{59}{Fe} could potentially become observable \citep{Timmes2019}.
Many of these nuclides have lifetimes from a few hours to a few days, requiring a fast mixing of material to the outer layers where the \acp{gray} can escape.
A recent sensitivity study focusing on the explosive silicon burning phase in a $12\,\mathrm{M}_\odot$ massive star has identified key reaction rates that influence the synthesis of these nuclides \citep{Hermansen2020}.

\subsubsection{Perspectives}
Predictions of elemental and isotopic yields are strongly linked to nuclear cross-sections and thermonuclear reaction rates.
Yet, their determination remains very challenging due to the low, sub-Coulomb reaction energies and the radioactive nature of the species involved.
On the one hand, next generation facilities such as FRIB \citep{FRIB} and RAON \citep{RAON}, equipped with state of the art detection systems, should significantly improve cross-sections determination, leading to more reliable predictions of \ac{gray} emitters production.
On the other hand, future MeV \ac{gray} missions such as \ac{COSI} \citep{Kierans2017}, AMEGO \citep{AMEGO} or newASTROGAM \citep{ASTROGAM} will observe these \ac{gray} lines with unprecedented sensitivity.
Combined with other multi-messenger instruments, they will also be able to detect \acp{gray} from \acp{BNSM} and trace $r$-process nucleosynthesis \citep{Hotokezaka2016} (see Sec.\,\ref{sec:NSM}).
The interplay between laboratory measurements and \ac{gray} observations are here at the heart of nuclear astrophysics and more collaboration should be encouraged on all levels.

%\nocite{*}

%\bibliography{issi_deSereville}% Produces the bibliography via BibTeX.

%\end{document}
%
% ****** End of file apssamp.tex ******

\subsection{Quo Vadis, $\gamma$-Ray Line Science?}\label{sec:outlook_what_now}
%
%\vspace{-0.5em}
%{\emph{Written by Thomas Siegert}}
%\vspace{0.5em}
%\vspace{-12pt}
%\\

\subsubsection{Neutron Capture Line: 2223\,keV}\label{sec:outlook_deuterium}
\vspace{-0.5em}
{\emph{Written by Thomas Siegert}}
\vspace{0.5em}\\
%\vspace{-12pt}
%
The 2223\,keV line from the neutron capture on protons to form deuterium is particularly strong in Solar flares (Sec.\,\ref{sec:SFs}).
But besides the detection from the Sun and a few rare terrestrial cases, the 2.2\,MeV has not been observed from outside the Solar System.
It has, however, the potential to revolutionise MeV \ac{gray} astrophysics because it relates to more than stellar flares.
In the following, we will present three cases for how a detection of the 2.2\,MeV might contribute to our understanding of difference sources.

\begin{figure}[!ht]
    \centering
    \includegraphics[width=0.7\linewidth]{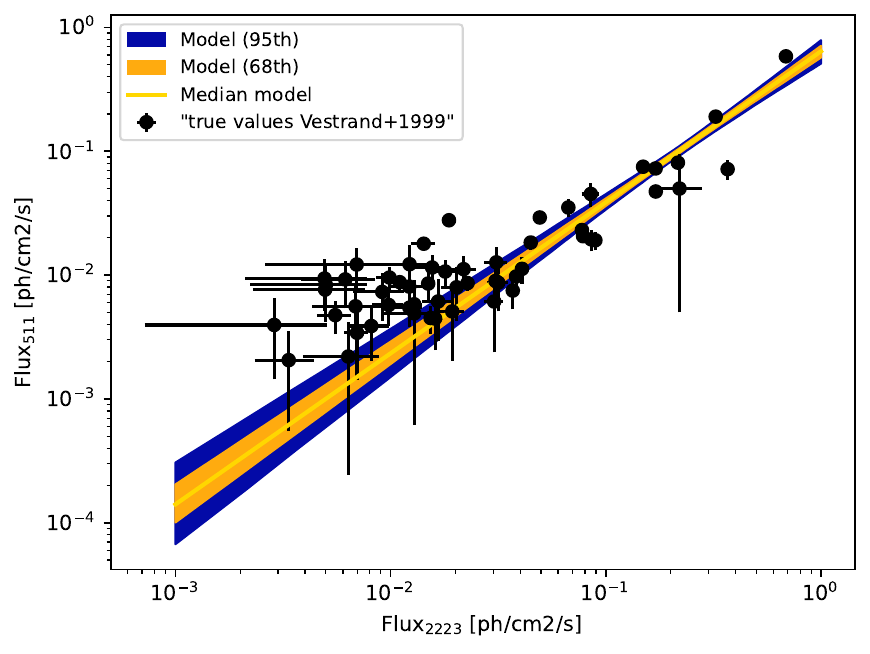}
    \caption{Correlation between the prompt 511\,keV line due to electron-positron annihilation and the 2.2\,MeV line due to neutron capture in Solar flares. Shown are data from \citet{Vestrand1999} (Mittal et al.\,[2025; in prep.]).}
    \label{fig:correlation_511_22}
\end{figure}

\paragraph{Stellar Flares and the 511\,keV Line}
As mentioned before in Secs.\,\ref{sec:SFs} \& \ref{sec:positron_sources}, the Sun is known to flare with a certain flare-energy frequency distribution that often results in two prominent \ac{gray} lines -- the 511\,keV line from positron annihilation and the 2.2\,MeV line from neutron capture.
In fact, there is a strong one-to-one correlation between the fluxes of both lines in the Sun (Fig.\,\ref{fig:correlation_511_22}).
This means that if parts of the Galactic 511\,keV line are due to one hundred billion intermittently flaring stars \citep{Bisnovatyi-Kogan2017}, a similar flux would be expected at 2.2\,MeV.
As it turns out, only up to 10\% of the annihilation line can potentially be explained by flaring stars as otherwise, a diffuse 2.2\,MeV line would have already been detected \citep{McConnell1997}.
From a bottom-up model of flaring stars, extrapolating from the Sun as the best-studied star, a diffuse extended 2.2\,MeV line with a flux level of $10^{-6}$--$10^{-4}\,\mathrm{ph\,cm^{-2}\,s^{-1}}$, distributed identical to the 511\,keV line, would be expected.
This is certainly within the range of future MeV observatories.
A detection of the 2.2\,MeV line co-spatial with the 511\,keV line would be a spectacular result that would solve large parts of the ``Positron Puzzle'', even though only 10\% at most can be explained.
But it would point to a possibility that there is, indeed, in-situ annihilation happening -- and maybe in more than one source type.

\paragraph{A Relic 2.2\,MeV Background? -- The `Hydrogen Calorimeter'}
After early measurement of the \ac{CGB}, such as by the Apollo\,15 mission, a suspicious `bump' around 1--2\,MeV was claimed.
\citet{Leventhal1973} speculated that this emission feature may be due to deuterium formation in a warm Galactic halo.
In fact, \ac{CR} spallation will inevitably lead to free neutrons in the \ac{ISM} and the intergalactic medium.
If the thermalisation time scale of the neutrons is faster, or at least not much longer, than the neutron decay time, they might be captured by protons and emit a 2.2\,MeV photon.
While this process is the general production channel in Solar flares, \citet{Ramaty1980} pointed out that the thermalisation time scale might be on the order of $10^4$\,s for a $1\,\mathrm{cm^{-3}}$ \ac{ISM}, while the neutron decay time is $880$\,s.
This means that only a few per cent of neutrons will be captured.
However, this would mean that the 2.2\,MeV brightness would integrate the \emph{entire} sub-GeV \ac{CR} power and act as a `hydrogen calorimeter' in the Galaxy -- in fact in all galaxies.
A formal estimate of the expected 2.2\,MeV line flux is given nowhere in the literature, but measurements are available:
\ac{COMPTEL} and \ac{SMM} provided narrow line upper limit fluxes of $\lesssim 10^{-4}\,\mathrm{ph\,cm^{-2}\,s^{-1}}$ \citep{Harris1991,McConnell1997}.
It is up to the next generation of soft \ac{gray} telescopes to improve upon these limits and decipher if a diffuse 2.2\,MeV line traces the dense gas in the Milky Way or even its halo.

\paragraph{Accretion Flow Around Compact Objects}
There are several scenarios that could produce a 2.2\,MeV line in accreting \acp{NS} or \acp{BH}.
Neutron capture within the accretion flow is one possibility.
Neutron capture in a \ac{NS}’s atmosphere is another.
Neutrons escaping the accretion flow and being captured in the companion star comprise a third scenario.
Finally, a beam of accelerated particles striking the companion can induce neutron capture, analogous to the 2.2\,MeV line observed in Solar flares.
We will briefly summarise these ideas:

\subparagraph{Neutron capture within the accretion flow}
Accretion onto a compact object heats ions to 100\,MeV, enough to spallate heavy nuclei and release neutrons.
Energetic protons on \nuc{He}{4} dominate neutron production at Solar abundances.
Some neutrons could capture on protons in the flow, yielding a 2.2\,MeV line, but this demands proton densities of $\gtrsim 10^{16}\,\mathrm{cm^{-3}}$.
Even then, neutrons more often escape than capture \citep{Guessoum1988}, and captures in hot plasma produce extremely broad lines \citep{Aharonian1984}.
Therefore, detectable 2.2\,MeV emission from within the accretion flow is improbable.

\subparagraph{Neutron capture in a neutron star atmosphere}
Again, accreting matter onto the \ac{NS} surface can spallate \nuc{He}{4}, releasing neutrons \citep{Shvartsman1972}.
Thermalised neutrons may again recombine radiatively with protons to emit a 2.2\,MeV photon or non-radiatively with \nuc{He}{3}.
Predicted fluxes reach $2 \times 10^{-5}\,\mathrm{cm^{-2}\,s^{-1}}$ for heavy-element-rich accreting sources, such as 4U\,1916-05, 4U\,1626-67 or 4U\,1820-30 \citep{Bildsten1992}.
Since this emission originates near the surface, the 2.2\,MeV line may be gravitationally redshifted to $\gtrsim 1.76$\,MeV.

\subparagraph{Neutron capture in the companion star}
Neutrons produced in the accretion flow can escape the compact object’s gravity and reach the companion’s atmosphere.
Thermalised neutrons captured by protons there emit again a 2.2\,MeV line.
Close binaries are more likely sources, as neutron decay limits travel distance and a small separation increases the companion’s solid angle.
Emission arises from the companion’s irradiated face and is modulated by the orbital period, peaking near the X-ray maximum.
\citet{Guessoum1988} predict fluxes up to $\sim 10^{-5}\,\mathrm{cm^{-2}\,s^{-1}}$ for Cygnus\,X-1.
\citet{2001A&A...378..509J} revised the work of \citet{Guessoum1988} and computed this emission for different accretion disk models and two kinds of companion stars.
Later, \citet{2002A&A...396..157G} calculated the 2.2\,MeV line emission characteristics for several nearby X-ray binary sources, as a function of their accretion rate and accretion disk models.
The authors obtained fluxes of $\lesssim 10^{-5}\,\mathrm{ph\,cm^{-2}\,s^{-1}}$.

\subparagraph{Neutron capture from a beam dump in the companion star}
Very high-energy photons ($E \gtrsim 10^{12}$\,eV) detected from accreting sources like Cygnus\,X-3, Vela\,X-1, and Hercules\,X-1 imply proton beam acceleration.
Beams striking the companion produce neutrons that capture on protons, analogous to Solar flares.
This yields a narrow and unshifted 2.2\,MeV line that varies with orbital phase.
\citet{Vestrand1990} predicts peak fluxes for Cygnus\,X-3 on the order of $\sim 10^{-4}\,\mathrm{cm^{-2}\,s^{-1}}$, but which has already been excluded by \ac{COMPTEL} measurements \citep{McConnell1997}.

\subparagraph{Summary}
In any of these cases, the emission map would probably reveal individual point sources, such as the attempt by \citet{McConnell1997} who found one intriguing hotspot in the 2.2\,MeV map from \ac{COMPTEL}.
Later this point-like emission feature was interpreted as the highly magnetised \ac{WD} RE\,J0317-853 at a distance of only 35\,pc.
However, the gravitational red-shift of a \ac{WD} would result in a line around $\gtrsim 1.9$\,MeV and not necessarily at 2.2\,MeV.
For \acp{NS}, a measurement of a red-shifted neutron capture line would immediately provide an \emph{equation of state} by disentangling the parameter combination $M(R)/R$ as exactly this the gravitational red-shift will be identified.
The need for better measurements is therefore clear: high spectral resolution in combination with at least one order of magnitude better sensitivity.

%\subsubsection{Cyclotron Lines: Magnetars}\label{sec:outlook_magnetars}

%
\subsubsection{Dark Matter and other Beyond-Standard-Model Candidates}\label{sec:outlook_darkmatter}
\vspace{-0.5em}
{\emph{Written by Francesca Calore}}
\vspace{0.5em}\\
%\vspace{-12pt}
%
The MeV domain in \ac{gray} astronomy offers a unique observational window for \ac{BSM} physics, especially in the search for light \ac{DM} candidates and other feebly interacting particles, with masses in the keV to sub-GeV range.
In this energy regime, narrow \ac{gray} lines and distinctive continuum features may act as smoking guns for \ac{DM} decay or annihilation, standing out against smoother astrophysical backgrounds mostly from faint, unresolved point sources and \ac{IC} photons from \ac{CR} electrons interacting with the stellar radiation field of the Galaxy.
Several \ac{BSM} scenarios predict such signals, ranging from thermal relic \ac{DM} below the GeV scale to non-thermal candidates like sterile neutrinos, \acp{ALP}, and other light \acp{WISP}, not necessarily linked to the cosmological \ac{DM}.

Importantly, the cosmological viability of these scenarios -- especially light thermal relics -- is tightly constrained by observations of the \ac{CMB} and by \ac{BBN}.
In general, thermal production through the freeze-out mechanism is possible from eV up to GeV masses.
However, below 1\,MeV, light \ac{DM} modifies predictions for light-element abundances during \ac{BBN} and thermal production is therefore strongly constrained \citep{Sabti:2019mhn}.
Above 1\,MeV, \ac{DM} annihilation during the recombination epoch injects energy that alters the ionisation history, which is stringently constrained by Planck data \citep{Slatyer:2015jla}.
As a result, thermal annihilation cross sections compatible with a relic abundance are often excluded for sub-GeV \ac{DM} with velocity-independent annihilation rate, that is, elastic $s$-wave annihilation into non-neutrino Standard Model final states.
The relevant annihilation cross section is typically the one into electron-positron pairs.
\ac{CMB} and \ac{BBN} constraints leave room for other \ac{DM} types, with late-time velocity suppression or secluded annihilation channels, like the one found in rich dark sectors with vector mediators \citep{Antel:2023hkf}.
Besides, other production mechanisms can be invoked such as freeze-in at low re-heating temperatures, or dark phase transitions at low temperatures \citep{Krnjaic:2019dzc}.

\paragraph{Dark Matter $\gamma$-Ray Lines}
%\subsection{Decaying and Annihilating Dark Matter}
Monochromatic \ac{gray} lines are among the most striking and least ambiguous spectral features that could arise from \ac{DM} decay or annihilation.
In the MeV regime and depending on the centre-of-mass energy\footnote{In the Galactic halo, \ac{DM} is non-relativistic today so that the centre-of-mass energy simply connects to the \ac{DM} mass, $E_{\rm cm} = N m_\chi$, where $N = 1$ for decay and $2$ for annihilation.}, such lines would typically originate from two-body final states like $\chi (\chi) \rightarrow \gamma X$, where  $\chi$ indicates the \ac{DM} candidate, and $X$ is another neutral particle (e.g., another photon, a neutrino or a light boson if kinematically allowed).
%
%General equations for the line signal 
For example, for annihilation into two photons, the \ac{gray} spectrum ${\rm d}N_\gamma/{\rm d}E$ is equal to $2\delta(E - N m_\chi/2)$ (see Sec.\,\ref{sec:DM_dec_ann}).

Sterile neutrinos provide a textbook example \citep{Boyarsky:2018tvu}:
Their radiative decay channel, $\nu_s \rightarrow \nu \gamma$, leads to a sharp line at half the mass of the sterile neutrino.
Though keV-mass sterile neutrinos have received attention as warm \ac{DM} candidates, heavier versions in the MeV regime have been explored as unstable \ac{DM} producing \ac{gray} lines in this range.
However, such candidates are often in tension with cosmological constraints or decay too rapidly to contribute substantially to today’s \ac{DM} density.
\acp{ALP}, on the other hand, may decay to photon pairs ($a \rightarrow \gamma \gamma$), producing a characteristic line at $m_a/2$.
The \acp{ALP} spontaneous decay rate is proportional to $m_a^3 g_{a\gamma}^2$, with $g_{a\gamma}$ the fundamental \ac{ALP}-photon coupling.
This makes MeV-mass \acp{ALP} with weak couplings long-lived on cosmological timescales.

% Expression for the flux from COST school 
The differential \ac{gray} flux from a given direction in the sky, identified by Galactic coordinates $(\ell, b)$, due to \ac{DM} decay or annihilation, has been introduced by Eqs.\,(\ref{eq:emissivity_decay})--(\ref{eq:dm_annihilation_full}), and is repeated here for convenience:
\begin{equation}
    \frac{{\rm d}\Phi_\gamma}{{\rm d}E}(\ell, b)= \mathcal{A}(\theta_{\chi}) \times \frac{{\rm d}N_\gamma}{{\rm d}E} \times \int_{\rm l.o.s.} \rho^N_{\chi}(s, \ell, b) ds\,\mathrm{,}
    \label{eq:calore_dm_equation}
\end{equation}
where $\theta_{\chi} = \{\Gamma_\gamma, m_{\chi}\} $ for decay, and
$\theta_{\chi} = \{\langle \sigma v \rangle, m^2_{\chi}\}$ for annihilation.
The \ac{LoS} integral reflects the \ac{DM} distribution in the target, such as a galactic halo.
In the Milky Way, the \ac{DM} profile is constrained at large radii via the rotation curve, but its inner structure, especially near the Galactic centre, relies on semi-analytic models or cosmological simulations.
Common choices include the \ac{NFW} profile, although theoretical uncertainties near the centre are large.
This significantly impacts flux predictions, particularly for annihilation scenarios, where the flux scales with the square of the \ac{DM} density.
% Discussion of targets 
The \acp{gray} from \ac{DM} can be expected in the continuum, diffuse \ac{gray} emission of the Galaxy, or from specific targets like dwarf galaxies or the Galactic centre.

\subparagraph{Current constraints}
% Comment on sigle target analysis from dwarfs, Reticulum II, not yet competitive
% Diffuse emission from latest Integral analysis (Calore+, Siegert+): report briefly novelty of the analysis, first time \ac{DM} template included in full analysis chain and extraction of spectral points; results superseding CMB bounds for both decay and annihilation demonstrating potential of MeV domain. Comment on reach for $p$-wave: They have the potential for sufficient sensitivity to probe thermal freeze-out even when the dominant annihilation is $p$-wave
%
%
A recent reanalysis of 16\,yr of \ac{INTEGRAL}/\ac{SPI} data has significantly improved our understanding of the Galactic diffuse emission up to 8\,MeV, superseding the two-decade-old \ac{COMPTEL} results \citep{2022A&A...660A.130S}. 
This updated measurement offers critical insights not only into \ac{CR} propagation in the MeV range but also provides sensitivity to exotic emission processes, including those associated with particle and non-particle \ac{DM}.
Currently, these observations set the among some of the highest sensitivities available in this energy band.
For annihilation, the bounds on the two-photon final state are superseding the limits from the \ac{CMB} in the range from 50\,keV up to $\sim$3\,MeV, showing the potential future MeV missions will have in probing light \ac{DM}.
These analyses currently provide also some among the strongest constraints on light decaying \ac{DM} in the 0.1--10\,MeV.

In addition to the diffuse Galactic background, targeted observations of \ac{DM}-dominated systems, such as dwarf spheroidal galaxies, offer complementary constraints.
For instance, \ac{SPI} data from a 1.5\,Ms exposure toward the Reticulum\,II dwarf galaxy has been analysed in \citet{Siegert:2021upf}, setting competitive limits despite the shorter exposure time.

\paragraph{The 511\,keV Line and Dark Matter}
The bright 511\,keV \ac{gray} line from the Galactic bulge (see Sec.\,\ref{sec:511}) has long puzzled astrophysicists.
Its interpretation as arising from positron annihilation implies the existence of a large and concentrated source of low-energy positrons in the inner Galaxy, with little corresponding emission from the disk \citep{Siegert:2015knp,Siegert:2021trw}.
While astrophysical sources remain viable explanations, \ac{DM} has been presented as a compelling alternative, especially given the spatial morphology of the signal \citep{Prantzos:2010wi}.

\subparagraph{A \ac{DM} explanation of the 511\,keV line}
Sub-GeV \ac{DM} annihilating or decaying into $e^+e^-$ pairs has been proposed as a mechanism to generate these positrons.
For example, MeV-scale scalar or vector \ac{DM} particles could annihilate via $\chi \chi \rightarrow e^+ e^-$ or decay through suppressed channels into leptonic final states.
The subsequent positron annihilation with electrons in the \ac{ISM} would produce a 511\,keV line, accompanied by a low-energy continuum from \ac{Ps} formation and in-flight annihilation.
However, such scenarios are constrained by both cosmology and \ac{gray} data.
As seen above, annihilating \ac{DM} in this mass range must have suppressed late-time cross sections (e.g., $p$-wave annihilation) to evade \ac{CMB} constraints on energy injection during recombination.
Similarly, the injection of positrons must be soft and spatially confined to avoid overproducing in-flight \acp{gray} or breaching bounds from diffuse MeV backgrounds \citep[e.g.,][]{DelaTorreLuque:2024zsr}.
Candidates that can potentially match both the morphology and flux of the 511\,keV line include asymmetric \ac{DM} and secluded annihilation models.
In particular, annihilations or decays into intermediate metastable states, which subsequently decay into $e^+ e^-$ with delayed injection, provide a plausible mechanism.
These models can evade stringent cosmological bounds (e.g., allowing for neutrino injection from \ac{DM} annihilation in the early Universe \citep{Sabti:2019mhn}), while naturally explaining the spatial profile of the signal.
A recent model has been proposed by \citet{Aghaie:2025dgl} where MeV-scale Dirac \ac{DM} annihilates into \acp{ALP}:
$p$-wave annihilation into two \acp{ALP} set the relic abundance, whereas $s$-wave annihilation into three \acp{ALP} later decaying into electron-positron pairs is responsible for the line intensity. 

\subparagraph{Current constraints on dark matter}
Sub-GeV \ac{DM} particles producing electron-positron pairs and form \ac{Ps} after thermalisation are strongly constrained by \ac{INTEGRAL}/\ac{SPI} observations.
A first conservative bound can be set by requiring that the \ac{DM} annihilation/decay rate should not overproduce the positrons necessary to sustain the line observation.
This argument gave a bound of $\langle \sigma v \rangle \sim 5 \times 10^{-31}\,\mathrm{cm^{3}\,s^{-1}}$ at 3\,MeV, for an \ac{NFW} profile \ac{DM} density distribution \citep{Vincent:2012an}.
Including the effects of positron propagation in the Galaxy, most notably diffusion, and fully exploiting the spatial distribution 
of the observed signal, \citet{DelaTorreLuque:2023cef} set the currently leading constraints for masses in the range 1--300 (1000)\,MeV for annihilation (decay), touching values of  $\langle \sigma v \rangle \sim 10^{-32}\,\mathrm{cm^{3}\,s^{-1}}$ at 1\,MeV for electron-positron annihilation.
Theoretical uncertainties related to positrons propagation and thermalisation requires further investigation (see Sec.\,\ref{sec:positron_puzzle}).

% $p$-wave
In this context, we stress that for velocity-dependent annihilation scenarios with $p$-wave cross sections (scaling as $v^2$), constraints from the \ac{CMB} are significantly weakened due to the low relative velocities of \ac{DM} at recombination ($v_{\rm CMB} \lesssim 10^{-5}$ c) \citep{Liu:2016cnk}.
In contrast, present-day indirect detection probes remain sensitive, since Galactic \ac{DM} particles today move with $v_0 \lesssim 10^{-3}$\,c.
Consequently, for $p$-wave models, the strongest constraints come from the MeV diffuse \ac{gray} background and 511\,keV line, touching upon the expected $p$-wave vanilla cross section.
The relic abundance condition for $p$-wave annihilating \ac{DM} indeed -- set at freeze-out when $v_{\rm f.o.} \sim 0.15$\,c --requires a cross section of $\sim 10^{-26}\,\mathrm{cm^{3}\,s^{-1}}$ \citep{Diamanti:2013bia}.
Extrapolating to today’s velocities, the expected thermal cross section becomes $\langle \sigma v \rangle_0^{\rm f.o.} \sim 10^{-32}\,\mathrm{cm^{3}\,s^{-1}}$ \citep{Bartels:2017dpb}, potentially in the reach of current indirect detection efforts.

\subparagraph{Current constraints on other \acp{WISP}}
\acp{ALP}, as other \acp{WISP} like dark photons and sterile neutrinos, can also be produced in stars, and especially in \acp{ccSN} \citep{Carenza:2024ehj}.
\acp{ALP} production in \ac{SN} cores is due to the coupling with nucleons, whereas, if coupled also to electrons, for suitable masses and couplings, these particles escape from the \ac{SN} and decay into positrons.
Those then eventually annihilate with electrons in the \ac{ISM}.
The resulting signal is a 511\,keV annihilation line.
However, the spatial distribution of the 511\,keV line signal would be dramatically different than the observed one and correlated with the Galactic disk, allowing to set strong constraints on \ac{ALP}-electron coupling.
In \citet{Calore:2021klc}, using the \ac{INTEGRAL}/\ac{SPI} observation of this \ac{gray} line, a wide range of the axion-electron couplings has been excluded, $10^{-19} \lesssim g_{ae} \lesssim  10^{-11}$, for
$g_{ap} \sim 10^{-9}$.
Also, \acp{ALP} from extragalactic \acp{SN} decaying into electron-positron pairs contribute to the \ac{CXB}, leading to constraints down to $g_{ae} \sim 10^{-20}$.
%
%Beyond ALPs, other WISPs such as dark photons and sterile neutrinos can be produced in a supernova
%core via mixing with photons and ordinary neutrinos,
%respectively, and give rise to 511 keV line which has, likewise, a spatial distribution in strong disagreement with the data~\cite{Calore:2021lih}. 
These argument allow us to set stringent limits on the mixing parameters, in a region not accessible by current and planned laboratory experiments, and competitive with cosmological bounds.
The 511\,keV observations were shown to be the most constraining observable among many others in X- and \acp{gray} \citep[e.g.,][]{DelaTorreLuque:2023huu}.

We stress again that these bounds rely on the modelling of the positrons propagation and thermalisation in the \ac{ISM} where improvement is warranted.
A better treatment of these aspects would strengthen the confidence in the results, and might also allow to tighten the limits via a detailed modelling of the signal morphology.

\paragraph{In-Flight Annihilation of Positrons}
A crucial consequence of positron injection at MeV energies is the generation of ``in-flight annihilation'' \acp{gray}.
These arise when positrons annihilate before thermalising with the ambient \ac{ISM}, producing a broad spectral feature above 511\,keV -- and not a \ac{gray} line (see Sec.\,\ref{sec:not_a_line}).
The resulting emission can conflict with observed diffuse \ac{gray} data if the positron injection energy is too high or the rate too large.
In-flight annihilation constraints, derived from \ac{SPI} and \ac{COMPTEL} data, severely limit models that inject positrons with energies above a few MeV.
For instance, decaying \ac{DM} models with mass $m_\chi \gtrsim 10$\,MeV decaying to $e^+e^-$ pairs are strongly constrained by this effect.
Only scenarios where the positron injection energy is below a few MeV, or where most positrons annihilate after slowing down, remain viable \citep{DelaTorreLuque:2024zsr}.
As a result, viable models must finely balance the need to reproduce the observed 511\,keV signal while suppressing the in-flight annihilation component.
We note that the joint analysis of \ac{gray} instruments decades apart might invoke some bias and that a wrong inter-calibration might lead to stronger effects than what is actually happening \citep{Siegert2023}.

\paragraph{Smoking-Gun Observables and Future Prospects}
Identifying a ``smoking gun'' signature of MeV \ac{DM} requires observations that are both spectrally distinctive and spatially correlated with plausible \ac{DM} distributions.
The most promising signatures include:

\begin{itemize}
    \item Narrow \ac{gray} lines from decaying or annihilating \ac{DM}, e.g., $\chi \rightarrow \gamma \nu$ or $a \rightarrow \gamma \gamma$, especially if observed from regions of high \ac{DM} density like the Galactic centre, dwarf spheroidal galaxies or other nearby galaxies.
    \item The 511\,keV line with a morphology consistent with \ac{DM}-induced positron injection, ideally confirmed with improved angular resolution and spectral modelling.
    \item Possibly suppressed in-flight annihilation emission, confirming that positrons annihilate after thermalisation and thus match the spectral constraints.
\end{itemize}

To detect such signatures, dedicated MeV \ac{gray} telescopes with high spectral and angular resolution are essential.
In the near future, \ac{COSI} is scheduled for launch in 2027 following multiple balloon campaigns.
It will operate in the 0.2--5.0\,MeV range and is expected to significantly enhance current limits on light dark matter decay or annihilation \citep[e.g.,][]{Aramaki:2022zpw, Caputo:2022dkz}.
The proposed GECCO mission (Galactic Explorer with a Coded Aperture Mask Compton Telescope) aims to extend coverage up to 10\,MeV, offering improved angular and energy resolution \citep{Coogan:2021sjs}.
To probe \ac{DM} masses beyond this range, broader-band missions reaching $\sim 100$\,MeV are required, such as GRAMS, APT, AMEGO-X, and all-sky-ASTROGAM.
These instruments would be crucial to close the sensitivity gap at intermediate energies.
For a comprehensive overview of planned MeV \ac{gray} missions, see \citet{Aramaki:2022zpw}. 
Future detectors are projected to tighten existing constraints by at least an order of magnitude.
Combining such observations with \ac{CMB} and \ac{BBN} constraints, and modelling of astrophysical foregrounds, will be key to disentangling \ac{DM} signals from standard astrophysical backgrounds.
Besides the discovery potential, the MeV band holds unique promise to \emph{exclude} some among the most well-motivated extensions of the minimal \ac{DM} model, namely dark sectors with vector- or scalar-type mediators which possess, in most cases, velocity-suppressed cross sections.
Excluding a sizeable part of $p$-wave cross section is in the reach of these future instruments for a large range of \ac{DM} masses.

\subsection{What is Needed? -- A New MeV $\gamma$-Ray Mission}\label{sec:future_mission}
\vspace{-0.5em}
{\emph{Written by Thomas Siegert}}
\vspace{0.5em}
%\vspace{-12pt}
\\
From all the considerations above, the summary of this chapter is the following:
\emph{We need a better soft \ac{gray} telescope.}

But the path is not straight-forward.
Some science cases require accurate spatial resolution but may ignore the spectral requirements of others.
Some need a smooth and all-sky covering exposure with a large \ac{FoV}, others need a focussing aperture to detect extragalactic sources.
For \ac{gray} line science -- as is true for all other observations -- a sensitivity improvement should be the major driver for any new observatory.
Modern data analysis tools can then still be applied to mediocre spatial and spectral resolution, but identifications of more sources and different lines should be in the focus of future developments.
MeV telescopes should be built for dedicated purposes as either of the common apertures, coded masks, Compton telescopes, and Laue lenses (concentrators), have their ups and downs -- \emph{there is no one-fits-it-all solution}.

\newpage
\section*{Declarations}
Conflict of interest statement: Not applicable.

\newpage
\addcontentsline{toc}{section}{References}
\bibliographystyle{sn-mathphys}
\bibliography{bib_60Fe,bib_CNe,bib_SNeIa,bib_SNe,bib_NSM,bib_LECR,bib_intro,bib_ccSNe,bib_SSSBs,bib_26Al,bib_BSM,bib_511,bib_SFs,bib_22,bib_lab}


\begin{thebibliography}{526}
\ifx \bisbn   \undefined \def \bisbn  #1{ISBN #1}\fi
\ifx \binits  \undefined \def \binits#1{#1}\fi
\ifx \bauthor  \undefined \def \bauthor#1{#1}\fi
\ifx \batitle  \undefined \def \batitle#1{#1}\fi
\ifx \bjtitle  \undefined \def \bjtitle#1{#1}\fi
\ifx \bvolume  \undefined \def \bvolume#1{\textbf{#1}}\fi
\ifx \byear  \undefined \def \byear#1{#1}\fi
\ifx \bissue  \undefined \def \bissue#1{#1}\fi
\ifx \bfpage  \undefined \def \bfpage#1{#1}\fi
\ifx \blpage  \undefined \def \blpage #1{#1}\fi
\ifx \burl  \undefined \def \burl#1{\textsf{#1}}\fi
\ifx \doiurl  \undefined \def \doiurl#1{\url{https://doi.org/#1}}\fi
\ifx \betal  \undefined \def \betal{\textit{et al.}}\fi
\ifx \binstitute  \undefined \def \binstitute#1{#1}\fi
\ifx \binstitutionaled  \undefined \def \binstitutionaled#1{#1}\fi
\ifx \bctitle  \undefined \def \bctitle#1{#1}\fi
\ifx \beditor  \undefined \def \beditor#1{#1}\fi
\ifx \bpublisher  \undefined \def \bpublisher#1{#1}\fi
\ifx \bbtitle  \undefined \def \bbtitle#1{#1}\fi
\ifx \bedition  \undefined \def \bedition#1{#1}\fi
\ifx \bseriesno  \undefined \def \bseriesno#1{#1}\fi
\ifx \blocation  \undefined \def \blocation#1{#1}\fi
\ifx \bsertitle  \undefined \def \bsertitle#1{#1}\fi
\ifx \bsnm \undefined \def \bsnm#1{#1}\fi
\ifx \bsuffix \undefined \def \bsuffix#1{#1}\fi
\ifx \bparticle \undefined \def \bparticle#1{#1}\fi
\ifx \barticle \undefined \def \barticle#1{#1}\fi
\bibcommenthead
\ifx \bconfdate \undefined \def \bconfdate #1{#1}\fi
\ifx \botherref \undefined \def \botherref #1{#1}\fi
\ifx \url \undefined \def \url#1{\textsf{#1}}\fi
\ifx \bchapter \undefined \def \bchapter#1{#1}\fi
\ifx \bbook \undefined \def \bbook#1{#1}\fi
\ifx \bcomment \undefined \def \bcomment#1{#1}\fi
\ifx \oauthor \undefined \def \oauthor#1{#1}\fi
\ifx \citeauthoryear \undefined \def \citeauthoryear#1{#1}\fi
\ifx \endbibitem  \undefined \def \endbibitem {}\fi
\ifx \bconflocation  \undefined \def \bconflocation#1{#1}\fi
\ifx \arxivurl  \undefined \def \arxivurl#1{\textsf{#1}}\fi
\csname PreBibitemsHook\endcsname

%%% 1
\bibitem[\protect\citeauthoryear{{Ramaty} and
  {Lingenfelter}}{1979}]{Ramaty1979_review}
\begin{barticle}
\bauthor{\bsnm{{Ramaty}}, \binits{R.}},
\bauthor{\bsnm{{Lingenfelter}}, \binits{R.E.}}:
\batitle{{Gamma-ray line astronomy}}.
\bjtitle{\nat}
\bvolume{278},
\bfpage{127}--\blpage{132}
(\byear{1979})
\doiurl{10.1038/278127a0}
\end{barticle}
\endbibitem

%%% 2
\bibitem[\protect\citeauthoryear{Sch{\"o}nfelder}{2001}]{Schoenfelder2001_book}
\begin{bbook}
\bauthor{\bsnm{Sch{\"o}nfelder}, \binits{V.}}:
\bbtitle{The Universe in Gamma Rays}.
\bsertitle{Astronomy and Astrophysics Library}.
\bpublisher{Springer}, \blocation{???}
(\byear{2001}).
\doiurl{10.1007/978-3-662-04593-0}
\end{bbook}
\endbibitem

%%% 3
\bibitem[\protect\citeauthoryear{Diehl et~al.}{2018}]{Diehl2018_book}
\begin{bbook}
\bauthor{\bsnm{Diehl}, \binits{R.}},
\bauthor{\bsnm{Hartmann}, \binits{D.H.}},
\bauthor{\bsnm{Prantzos}, \binits{N.}}:
\bbtitle{Astronomy with Radioactivities I: Radioactivities and Other Tracers of
  the Stellar Cycle}.
\bsertitle{Space Sciences Series of ISSI}.
\bpublisher{Springer},
\blocation{Cham, Switzerland}
(\byear{2018}).
\doiurl{10.1007/978-94-024-0848-1}
\end{bbook}
\endbibitem

%%% 4
\bibitem[\protect\citeauthoryear{Lingenfelter and
  Ramaty}{1986}]{Lingenfelter1986_review}
\begin{bchapter}
\bauthor{\bsnm{Lingenfelter}, \binits{R.E.}},
\bauthor{\bsnm{Ramaty}, \binits{R.}}:
\bctitle{Gamma‐ray line astrophysics}.
In: \beditor{\bsnm{Shen}, \binits{B.S.P.}} (ed.)
\bbtitle{Theoretical Nuclear Reaction Astrophysics},
p. .
\bpublisher{Benjamin/Cummings}, \blocation{???}
(\byear{1986}).
\bcomment{NASA NTRS Report 19860022013}.
\burl{https://ntrs.nasa.gov/api/citations/19860022013/downloads/19860022013.pdf}
\end{bchapter}
\endbibitem

%%% 5
\bibitem[\protect\citeauthoryear{{Prantzos} et~al.}{2011}]{Prantzos2011}
\begin{barticle}
\bauthor{\bsnm{{Prantzos}}, \binits{N.}},
\bauthor{\bsnm{{Boehm}}, \binits{C.}},
\bauthor{\bsnm{{Bykov}}, \binits{A.M.}},
\bauthor{\bsnm{{Diehl}}, \binits{R.}},
\bauthor{\bsnm{{Ferri{\`e}re}}, \binits{K.}},
\bauthor{\bsnm{{Guessoum}}, \binits{N.}},
\bauthor{\bsnm{{Jean}}, \binits{P.}},
\bauthor{\bsnm{{Kn{\"o}dlseder}}, \binits{J.}},
\bauthor{\bsnm{{Marcowith}}, \binits{A.}},
\bauthor{\bsnm{{Moskalenko}}, \binits{I.V.}},
\bauthor{\bsnm{{Strong}}, \binits{A.}},
\bauthor{\bsnm{{Weidenspointner}}, \binits{G.}}:
\batitle{{The 511 keV emission from positron annihilation in the Galaxy}}.
\bjtitle{Reviews of Modern Physics}
\bvolume{83}(\bissue{3}),
\bfpage{1001}--\blpage{1056}
(\byear{2011})
\doiurl{10.1103/RevModPhys.83.1001}
{\href{https://arxiv.org/abs/1009.4620}{{arXiv:1009.4620}}}
{[astro-ph.HE]}
\end{barticle}
\endbibitem

%%% 6
\bibitem[\protect\citeauthoryear{{Lewin} and {Smith}}{1996}]{Lewin1996_review}
\begin{barticle}
\bauthor{\bsnm{{Lewin}}, \binits{J.D.}},
\bauthor{\bsnm{{Smith}}, \binits{P.F.}}:
\batitle{{Review of mathematics, numerical factors, and corrections for dark
  matter experiments based on elastic nuclear recoil}}.
\bjtitle{Astroparticle Physics}
\bvolume{6}(\bissue{1}),
\bfpage{87}--\blpage{112}
(\byear{1996})
\doiurl{10.1016/S0927-6505(96)00047-3}
\end{barticle}
\endbibitem

%%% 7
\bibitem[\protect\citeauthoryear{{Bergstr{\"o}m}
  et~al.}{1998}]{Bergstroem1998_neutralinos}
\begin{barticle}
\bauthor{\bsnm{{Bergstr{\"o}m}}, \binits{L.}},
\bauthor{\bsnm{{Ullio}}, \binits{P.}},
\bauthor{\bsnm{{Buckley}}, \binits{J.H.}}:
\batitle{{Observability of {\ensuremath{\gamma}} rays from dark matter
  neutralino annihilations in the Milky Way halo}}.
\bjtitle{Astroparticle Physics}
\bvolume{9}(\bissue{2}),
\bfpage{137}--\blpage{162}
(\byear{1998})
\doiurl{10.1016/S0927-6505(98)00015-2}
{\href{https://arxiv.org/abs/astro-ph/9712318}{{arXiv:astro-ph/9712318}}}
{[astro-ph]}
\end{barticle}
\endbibitem

%%% 8
\bibitem[\protect\citeauthoryear{Krane}{1987}]{Krane1987_book}
\begin{bbook}
\bauthor{\bsnm{Krane}, \binits{K.S.}}:
\bbtitle{Introductory Nuclear Physics}.
\bpublisher{John Wiley \& Sons},
\blocation{New York}
(\byear{1987}).
\bcomment{2nd edition}
\end{bbook}
\endbibitem

%%% 9
\bibitem[\protect\citeauthoryear{{Ferri{\`e}re}}{2001}]{Ferriere2001_ISMreview}
\begin{barticle}
\bauthor{\bsnm{{Ferri{\`e}re}}, \binits{K.M.}}:
\batitle{{The interstellar environment of our galaxy}}.
\bjtitle{Reviews of Modern Physics}
\bvolume{73}(\bissue{4}),
\bfpage{1031}--\blpage{1066}
(\byear{2001})
\doiurl{10.1103/RevModPhys.73.1031}
{\href{https://arxiv.org/abs/astro-ph/0106359}{{arXiv:astro-ph/0106359}}}
{[astro-ph]}
\end{barticle}
\endbibitem

%%% 10
\bibitem[\protect\citeauthoryear{{Chan} and
  {Lingenfelter}}{1987}]{Chan1987_SNe}
\begin{barticle}
\bauthor{\bsnm{{Chan}}, \binits{L.W.}},
\bauthor{\bsnm{{Lingenfelter}}, \binits{R.E.}}:
\batitle{{Calculated Gamma-Ray Line Fluxes from the Type II Supernova 1987A}}.
\bjtitle{\apjl}
\bvolume{318},
\bfpage{51}
(\byear{1987})
\doiurl{10.1086/184936}
\end{barticle}
\endbibitem

%%% 11
\bibitem[\protect\citeauthoryear{{Knoll}}{2000}]{Knoll2000_book}
\begin{bbook}
\bauthor{\bsnm{{Knoll}}, \binits{G.F.}}:
\bbtitle{{Radiation Detection and Measurement}},
(\byear{2000})
\end{bbook}
\endbibitem

%%% 12
\bibitem[\protect\citeauthoryear{{Diehl} et~al.}{2018}]{Diehl2018}
\begin{barticle}
\bauthor{\bsnm{{Diehl}}, \binits{R.}},
\bauthor{\bsnm{{Siegert}}, \binits{T.}},
\bauthor{\bsnm{{Greiner}}, \binits{J.}},
\bauthor{\bsnm{{Krause}}, \binits{M.}},
\bauthor{\bsnm{{Kretschmer}}, \binits{K.}},
\bauthor{\bsnm{{Lang}}, \binits{M.}},
\bauthor{\bsnm{{Pleintinger}}, \binits{M.}},
\bauthor{\bsnm{{Strong}}, \binits{A.W.}},
\bauthor{\bsnm{{Weinberger}}, \binits{C.}},
\bauthor{\bsnm{{Zhang}}, \binits{X.}}:
\batitle{{INTEGRAL/SPI {\ensuremath{\gamma}}-ray line spectroscopy. Response
  and background characteristics}}.
\bjtitle{\aap}
\bvolume{611},
\bfpage{12}
(\byear{2018})
\doiurl{10.1051/0004-6361/201731815}
{\href{https://arxiv.org/abs/1710.10139}{{arXiv:1710.10139}}}
{[astro-ph.IM]}
\end{barticle}
\endbibitem

%%% 13
\bibitem[\protect\citeauthoryear{{Pendleton}
  et~al.}{1995}]{Pendleton1995_response}
\begin{barticle}
\bauthor{\bsnm{{Pendleton}}, \binits{G.N.}},
\bauthor{\bsnm{{Paciesas}}, \binits{W.S.}},
\bauthor{\bsnm{{Mallozzi}}, \binits{R.S.}},
\bauthor{\bsnm{{Koshut}}, \binits{T.M.}},
\bauthor{\bsnm{{Fishman}}, \binits{G.J.}},
\bauthor{\bsnm{{Meegan}}, \binits{C.A.}},
\bauthor{\bsnm{{Wilson}}, \binits{R.B.}},
\bauthor{\bsnm{{Horack}}, \binits{J.M.}},
\bauthor{\bsnm{{Lestrade}}, \binits{J.P.}}:
\batitle{{The detector response matrices of the burst and transient source
  experiment (BATSE) on the Compton Gamma Ray Observatory}}.
\bjtitle{Nuclear Instruments and Methods in Physics Research A}
\bvolume{364},
\bfpage{567}--\blpage{577}
(\byear{1995})
\doiurl{10.1016/0168-9002(95)00448-3}
\end{barticle}
\endbibitem

%%% 14
\bibitem[\protect\citeauthoryear{{Siegert} et~al.}{2022}]{Siegert2022_book}
\begin{bchapter}
\bauthor{\bsnm{{Siegert}}, \binits{T.}},
\bauthor{\bsnm{{Horan}}, \binits{D.}},
\bauthor{\bsnm{{Kanbach}}, \binits{G.}}:
\bctitle{{Telescope Concepts in Gamma-Ray Astronomy}}.
In: \beditor{\bsnm{{Bambi}}, \binits{C.}},
\beditor{\bsnm{{Sangangelo}}, \binits{A.}} (eds.)
\bbtitle{Handbook of X-ray and Gamma-ray Astrophysics},
p. \bfpage{80}
(\byear{2022}).
\doiurl{10.1007/978-981-16-4544-0_43-1}
\end{bchapter}
\endbibitem

%%% 15
\bibitem[\protect\citeauthoryear{Clayton}{1968}]{Clayton1968_book}
\begin{bbook}
\bauthor{\bsnm{Clayton}, \binits{D.D.}}:
\bbtitle{Principles of Stellar Evolution and Nucleosynthesis}.
\bpublisher{University of Chicago Press},
\blocation{Chicago}
(\byear{1968}).
\bcomment{Classic textbook on nuclear astrophysics}
\end{bbook}
\endbibitem

%%% 16
\bibitem[\protect\citeauthoryear{{Woosley} and {Weaver}}{1995}]{Woosley1995}
\begin{barticle}
\bauthor{\bsnm{{Woosley}}, \binits{S.E.}},
\bauthor{\bsnm{{Weaver}}, \binits{T.A.}}:
\batitle{{The Evolution and Explosion of Massive Stars. II. Explosive
  Hydrodynamics and Nucleosynthesis}}.
\bjtitle{\apjs}
\bvolume{101},
\bfpage{181}
(\byear{1995})
\doiurl{10.1086/192237}
\end{barticle}
\endbibitem

%%% 17
\bibitem[\protect\citeauthoryear{{Kozlovsky} et~al.}{2002}]{koz02}
\begin{barticle}
\bauthor{\bsnm{{Kozlovsky}}, \binits{B.}},
\bauthor{\bsnm{{Murphy}}, \binits{R.J.}},
\bauthor{\bsnm{{Ramaty}}, \binits{R.}}:
\batitle{{Nuclear Deexcitation Gamma-Ray Lines from Accelerated Particle
  Interactions}}.
\bjtitle{\apjs}
\bvolume{141}(\bissue{2}),
\bfpage{523}--\blpage{541}
(\byear{2002})
\doiurl{10.1086/340545}
\end{barticle}
\endbibitem

%%% 18
\bibitem[\protect\citeauthoryear{{Share} and {Murphy}}{1995}]{Share1995_flares}
\begin{barticle}
\bauthor{\bsnm{{Share}}, \binits{G.H.}},
\bauthor{\bsnm{{Murphy}}, \binits{R.J.}}:
\batitle{{Gamma-Ray Measurements of Flare-to-Flare Variations in Ambient Solar
  Abundances}}.
\bjtitle{\apj}
\bvolume{452},
\bfpage{933}
(\byear{1995})
\doiurl{10.1086/176360}
\end{barticle}
\endbibitem

%%% 19
\bibitem[\protect\citeauthoryear{{Hua} and {Lingenfelter}}{1987}]{hua87}
\begin{barticle}
\bauthor{\bsnm{{Hua}}, \binits{X.-M.}},
\bauthor{\bsnm{{Lingenfelter}}, \binits{R.E.}}:
\batitle{{Solar flare neutron production and the angular dependence of the
  capture gamma-ray emission}}.
\bjtitle{\solphys}
\bvolume{107},
\bfpage{351}--\blpage{383}
(\byear{1987})
\doiurl{10.1007/BF00152031}
\end{barticle}
\endbibitem

%%% 20
\bibitem[\protect\citeauthoryear{{Aharonian} and
  {Sunyaev}}{1984}]{Aharonian1984_lines}
\begin{barticle}
\bauthor{\bsnm{{Aharonian}}, \binits{F.A.}},
\bauthor{\bsnm{{Sunyaev}}, \binits{R.A.}}:
\batitle{{Gamma-ray line emission, nuclear destruction and neutron production
  in hot astrophysical plasmas.The deuterium boiler as a gamma-ray source.}}
\bjtitle{\mnras}
\bvolume{210},
\bfpage{257}--\blpage{277}
(\byear{1984})
\doiurl{10.1093/mnras/210.2.257}
\end{barticle}
\endbibitem

%%% 21
\bibitem[\protect\citeauthoryear{{Yoneda} et~al.}{2023}]{Yoneda2023}
\begin{barticle}
\bauthor{\bsnm{{Yoneda}}, \binits{H.}},
\bauthor{\bsnm{{Aharonian}}, \binits{F.}},
\bauthor{\bsnm{{Coppi}}, \binits{P.}},
\bauthor{\bsnm{{Siegert}}, \binits{T.}},
\bauthor{\bsnm{{Takahashi}}, \binits{T.}}:
\batitle{{Line profile of nuclear de-excitation gamma-ray emission from very
  hot plasma}}.
\bjtitle{\mnras}
\bvolume{526}(\bissue{1}),
\bfpage{1460}--\blpage{1470}
(\byear{2023})
\doiurl{10.1093/mnras/stad2780}
{\href{https://arxiv.org/abs/2309.05426}{{arXiv:2309.05426}}}
{[astro-ph.HE]}
\end{barticle}
\endbibitem

%%% 22
\bibitem[\protect\citeauthoryear{{Guessoum} et~al.}{2005}]{Guessoum2005_511}
\begin{barticle}
\bauthor{\bsnm{{Guessoum}}, \binits{N.}},
\bauthor{\bsnm{{Jean}}, \binits{P.}},
\bauthor{\bsnm{{Gillard}}, \binits{W.}}:
\batitle{{The lives and deaths of positrons in the interstellar medium}}.
\bjtitle{\aap}
\bvolume{436}(\bissue{1}),
\bfpage{171}--\blpage{185}
(\byear{2005})
\doiurl{10.1051/0004-6361:20042454}
{\href{https://arxiv.org/abs/astro-ph/0504186}{{arXiv:astro-ph/0504186}}}
{[astro-ph]}
\end{barticle}
\endbibitem

%%% 23
\bibitem[\protect\citeauthoryear{Ore and Powell}{1949}]{Ore1949}
\begin{barticle}
\bauthor{\bsnm{Ore}, \binits{A.}},
\bauthor{\bsnm{Powell}, \binits{J.L.}}:
\batitle{Three-photon annihilation of an electron-positron pair}.
\bjtitle{Phys. Rev.}
\bvolume{75},
\bfpage{1696}--\blpage{1699}
(\byear{1949})
\doiurl{10.1103/PhysRev.75.1696}
\end{barticle}
\endbibitem

%%% 24
\bibitem[\protect\citeauthoryear{{Aharonian} and
  {Atoyan}}{1981}]{Aharonian1981_positrons}
\begin{barticle}
\bauthor{\bsnm{{Aharonian}}, \binits{F.A.}},
\bauthor{\bsnm{{Atoyan}}, \binits{A.M.}}:
\batitle{{Cosmic gamma-rays associated with annihilation of relativistic e
  $^{+}$ -e $^{-}$ pairs}}.
\bjtitle{Physics Letters B}
\bvolume{99}(\bissue{3}),
\bfpage{301}--\blpage{304}
(\byear{1981})
\doiurl{10.1016/0370-2693(81)91130-8}
\end{barticle}
\endbibitem

%%% 25
\bibitem[\protect\citeauthoryear{{Aharonian} and
  {Atoyan}}{2000}]{Aharonian2000_aif}
\begin{barticle}
\bauthor{\bsnm{{Aharonian}}, \binits{F.A.}},
\bauthor{\bsnm{{Atoyan}}, \binits{A.M.}}:
\batitle{{Broad-band diffuse gamma ray emission of the galactic disk}}.
\bjtitle{\aap}
\bvolume{362},
\bfpage{937}--\blpage{952}
(\byear{2000})
\doiurl{10.48550/arXiv.astro-ph/0009009}
{\href{https://arxiv.org/abs/astro-ph/0009009}{{arXiv:astro-ph/0009009}}}
{[astro-ph]}
\end{barticle}
\endbibitem

%%% 26
\bibitem[\protect\citeauthoryear{{Svensson}}{1982}]{Svensson1982_positrons}
\begin{barticle}
\bauthor{\bsnm{{Svensson}}, \binits{R.}}:
\batitle{{Electron-Positron Pair Equilibria in Relativistic Plasmas}}.
\bjtitle{\apj}
\bvolume{258},
\bfpage{335}
(\byear{1982})
\doiurl{10.1086/160082}
\end{barticle}
\endbibitem

%%% 27
\bibitem[\protect\citeauthoryear{{Svensson}}{1983}]{Svensson1983}
\begin{barticle}
\bauthor{\bsnm{{Svensson}}, \binits{R.}}:
\batitle{{The thermal pair annihilation spectrum - A detailed balance
  approach}}.
\bjtitle{\apj}
\bvolume{270},
\bfpage{300}--\blpage{304}
(\byear{1983})
\doiurl{10.1086/161122}
\end{barticle}
\endbibitem

%%% 28
\bibitem[\protect\citeauthoryear{Deutsch}{1951a}]{Deutsch1951_Ps}
\begin{barticle}
\bauthor{\bsnm{Deutsch}, \binits{M.}}:
\batitle{{Evidence for the Formation of Positronium in Gases}}.
\bjtitle{Physical Review}
\bvolume{82},
\bfpage{455}--\blpage{456}
(\byear{1951})
\doiurl{10.1103/PhysRev.82.455}
\end{barticle}
\endbibitem

%%% 29
\bibitem[\protect\citeauthoryear{Deutsch}{1951b}]{Deutsch1951b_Ps}
\begin{barticle}
\bauthor{\bsnm{Deutsch}, \binits{M.}}:
\batitle{{Observation of the Long‐Lived Positronium State}}.
\bjtitle{Physical Review}
\bvolume{83},
\bfpage{866}--\blpage{868}
(\byear{1951})
\doiurl{10.1103/PhysRev.83.866}
\end{barticle}
\endbibitem

%%% 30
\bibitem[\protect\citeauthoryear{{Guessoum} et~al.}{1991}]{Guessoum1991_511}
\begin{barticle}
\bauthor{\bsnm{{Guessoum}}, \binits{N.}},
\bauthor{\bsnm{{Ramaty}}, \binits{R.}},
\bauthor{\bsnm{{Lingenfelter}}, \binits{R.E.}}:
\batitle{{Positron Annihilation in the Interstellar Medium}}.
\bjtitle{\apj}
\bvolume{378},
\bfpage{170}
(\byear{1991})
\doiurl{10.1086/170417}
\end{barticle}
\endbibitem

%%% 31
\bibitem[\protect\citeauthoryear{{Guessoum} et~al.}{2005}]{2005A&A...436..171G}
\begin{barticle}
\bauthor{\bsnm{{Guessoum}}, \binits{N.}},
\bauthor{\bsnm{{Jean}}, \binits{P.}},
\bauthor{\bsnm{{Gillard}}, \binits{W.}}:
\batitle{{The lives and deaths of positrons in the interstellar medium}}.
\bjtitle{\aap}
\bvolume{436}(\bissue{1}),
\bfpage{171}--\blpage{185}
(\byear{2005})
\doiurl{10.1051/0004-6361:20042454}
{\href{https://arxiv.org/abs/astro-ph/0504186}{{arXiv:astro-ph/0504186}}}
{[astro-ph]}
\end{barticle}
\endbibitem

%%% 32
\bibitem[\protect\citeauthoryear{{Guessoum} et~al.}{2010}]{2010MNRAS.402.1171G}
\begin{barticle}
\bauthor{\bsnm{{Guessoum}}, \binits{N.}},
\bauthor{\bsnm{{Jean}}, \binits{P.}},
\bauthor{\bsnm{{Gillard}}, \binits{W.}}:
\batitle{{Positron annihilation on polycyclic aromatic hydrocarbon molecules in
  the interstellar medium}}.
\bjtitle{\mnras}
\bvolume{402}(\bissue{2}),
\bfpage{1171}--\blpage{1178}
(\byear{2010})
\doiurl{10.1111/j.1365-2966.2009.15954.x}
{\href{https://arxiv.org/abs/0911.0582}{{arXiv:0911.0582}}}
{[astro-ph.HE]}
\end{barticle}
\endbibitem

%%% 33
\bibitem[\protect\citeauthoryear{{Siegert}}{2023}]{Siegert2023}
\begin{barticle}
\bauthor{\bsnm{{Siegert}}, \binits{T.}}:
\batitle{{The Positron Puzzle}}.
\bjtitle{\apss}
\bvolume{368}(\bissue{4}),
\bfpage{27}
(\byear{2023})
\doiurl{10.1007/s10509-023-04184-4}
{\href{https://arxiv.org/abs/2303.15582}{{arXiv:2303.15582}}}
{[astro-ph.HE]}
\end{barticle}
\endbibitem

%%% 34
\bibitem[\protect\citeauthoryear{{Leventhal} et~al.}{1978}]{Leventhal1978}
\begin{barticle}
\bauthor{\bsnm{{Leventhal}}, \binits{M.}},
\bauthor{\bsnm{{MacCallum}}, \binits{C.J.}},
\bauthor{\bsnm{{Stang}}, \binits{P.D.}}:
\batitle{{Detection of 511 keV positron annihilation radiation from the
  galactic center direction.}}
\bjtitle{\apjl}
\bvolume{225},
\bfpage{11}--\blpage{14}
(\byear{1978})
\doiurl{10.1086/182782}
\end{barticle}
\endbibitem

%%% 35
\bibitem[\protect\citeauthoryear{{Churazov} et~al.}{2005}]{Churazov2005}
\begin{barticle}
\bauthor{\bsnm{{Churazov}}, \binits{E.}},
\bauthor{\bsnm{{Sunyaev}}, \binits{R.}},
\bauthor{\bsnm{{Sazonov}}, \binits{S.}},
\bauthor{\bsnm{{Revnivtsev}}, \binits{M.}},
\bauthor{\bsnm{{Varshalovich}}, \binits{D.}}:
\batitle{{Positron annihilation spectrum from the Galactic Centre region
  observed by SPI/INTEGRAL}}.
\bjtitle{\mnras}
\bvolume{357}(\bissue{4}),
\bfpage{1377}--\blpage{1386}
(\byear{2005})
\doiurl{10.1111/j.1365-2966.2005.08757.x}
{\href{https://arxiv.org/abs/astro-ph/0411351}{{arXiv:astro-ph/0411351}}}
{[astro-ph]}
\end{barticle}
\endbibitem

%%% 36
\bibitem[\protect\citeauthoryear{{Jean} et~al.}{2006}]{Jean2006}
\begin{barticle}
\bauthor{\bsnm{{Jean}}, \binits{P.}},
\bauthor{\bsnm{{Kn{\"o}dlseder}}, \binits{J.}},
\bauthor{\bsnm{{Gillard}}, \binits{W.}},
\bauthor{\bsnm{{Guessoum}}, \binits{N.}},
\bauthor{\bsnm{{Ferri{\`e}re}}, \binits{K.}},
\bauthor{\bsnm{{Marcowith}}, \binits{A.}},
\bauthor{\bsnm{{Lonjou}}, \binits{V.}},
\bauthor{\bsnm{{Roques}}, \binits{J.P.}}:
\batitle{{Spectral analysis of the Galactic e$^{+}$e$^{-}$ annihilation
  emission}}.
\bjtitle{\aap}
\bvolume{445}(\bissue{2}),
\bfpage{579}--\blpage{589}
(\byear{2006})
\doiurl{10.1051/0004-6361:20053765}
{\href{https://arxiv.org/abs/astro-ph/0509298}{{arXiv:astro-ph/0509298}}}
{[astro-ph]}
\end{barticle}
\endbibitem

%%% 37
\bibitem[\protect\citeauthoryear{{Siegert}}{2017}]{Siegert2017}
\begin{botherref}
\oauthor{\bsnm{{Siegert}}, \binits{T.}}:
{Positron-Annihilation Spectroscopy throughout the Milky Way}.
PhD thesis,
Max-Planck-Institute for Extraterrestrial Physics, Garching
(February 2017)
\end{botherref}
\endbibitem

%%% 38
\bibitem[\protect\citeauthoryear{{Abazajian}
  et~al.}{2001}]{Abazajian2001_sterile}
\begin{barticle}
\bauthor{\bsnm{{Abazajian}}, \binits{K.}},
\bauthor{\bsnm{{Fuller}}, \binits{G.M.}},
\bauthor{\bsnm{{Patel}}, \binits{M.}}:
\batitle{{Sterile neutrino hot, warm, and cold dark matter}}.
\bjtitle{\prd}
\bvolume{64}(\bissue{2}),
\bfpage{023501}
(\byear{2001})
\doiurl{10.1103/PhysRevD.64.023501}
{\href{https://arxiv.org/abs/astro-ph/0101524}{{arXiv:astro-ph/0101524}}}
{[astro-ph]}
\end{barticle}
\endbibitem

%%% 39
\bibitem[\protect\citeauthoryear{{Jungman} et~al.}{1996}]{Jungman1996_review}
\begin{barticle}
\bauthor{\bsnm{{Jungman}}, \binits{G.}},
\bauthor{\bsnm{{Kamionkowski}}, \binits{M.}},
\bauthor{\bsnm{{Griest}}, \binits{K.}}:
\batitle{{Supersymmetric dark matter}}.
\bjtitle{\physrep}
\bvolume{267},
\bfpage{195}--\blpage{373}
(\byear{1996})
\doiurl{10.1016/0370-1573(95)00058-5}
{\href{https://arxiv.org/abs/hep-ph/9506380}{{arXiv:hep-ph/9506380}}}
{[hep-ph]}
\end{barticle}
\endbibitem

%%% 40
\bibitem[\protect\citeauthoryear{{Evans} et~al.}{2012}]{Evans2012}
\begin{barticle}
\bauthor{\bsnm{{Evans}}, \binits{L.G.}},
\bauthor{\bsnm{{Peplowski}}, \binits{P.N.}},
\bauthor{\bsnm{{Rhodes}}, \binits{E.A.}},
\bauthor{\bsnm{{Lawrence}}, \binits{D.J.}},
\bauthor{\bsnm{{McCoy}}, \binits{T.J.}},
\bauthor{\bsnm{{Nittler}}, \binits{L.R.}},
\bauthor{\bsnm{{Solomon}}, \binits{S.C.}},
\bauthor{\bsnm{{Sprague}}, \binits{A.L.}},
\bauthor{\bsnm{{Stockstill-Cahill}}, \binits{K.R.}},
\bauthor{\bsnm{{Starr}}, \binits{R.D.}},
\bauthor{\bsnm{{Weider}}, \binits{S.Z.}},
\bauthor{\bsnm{{Boynton}}, \binits{W.V.}},
\bauthor{\bsnm{{Hamara}}, \binits{D.K.}},
\bauthor{\bsnm{{Goldsten}}, \binits{J.O.}}:
\batitle{{Major-element abundances on the surface of Mercury: Results from the
  MESSENGER Gamma-Ray Spectrometer}}.
\bjtitle{Journal of Geophysical Research (Planets)}
\bvolume{117},
\bfpage{00}--\blpage{07}
(\byear{2012})
\doiurl{10.1029/2012JE004178}
\end{barticle}
\endbibitem

%%% 41
\bibitem[\protect\citeauthoryear{{Steigman} et~al.}{2012}]{Steigman2012_sigmav}
\begin{barticle}
\bauthor{\bsnm{{Steigman}}, \binits{G.}},
\bauthor{\bsnm{{Dasgupta}}, \binits{B.}},
\bauthor{\bsnm{{Beacom}}, \binits{J.F.}}:
\batitle{{Precise relic WIMP abundance and its impact on searches for dark
  matter annihilation}}.
\bjtitle{\prd}
\bvolume{86}(\bissue{2}),
\bfpage{023506}
(\byear{2012})
\doiurl{10.1103/PhysRevD.86.023506}
{\href{https://arxiv.org/abs/1204.3622}{{arXiv:1204.3622}}}
{[hep-ph]}
\end{barticle}
\endbibitem

%%% 42
\bibitem[\protect\citeauthoryear{{Rybicki} and
  {Lightman}}{1979}]{Rybicki1979_book}
\begin{bbook}
\bauthor{\bsnm{{Rybicki}}, \binits{G.B.}},
\bauthor{\bsnm{{Lightman}}, \binits{A.P.}}:
\bbtitle{{Radiative Processes in Astrophysics}},
(\byear{1979})
\end{bbook}
\endbibitem

%%% 43
\bibitem[\protect\citeauthoryear{{Morrison} and
  {McCammon}}{1983}]{Morrison1983_ISM}
\begin{barticle}
\bauthor{\bsnm{{Morrison}}, \binits{R.}},
\bauthor{\bsnm{{McCammon}}, \binits{D.}}:
\batitle{{Interstellar photoelectric absorption cross sections, 0.03-10 keV.}}
\bjtitle{\apj}
\bvolume{270},
\bfpage{119}--\blpage{122}
(\byear{1983})
\doiurl{10.1086/161102}
\end{barticle}
\endbibitem

%%% 44
\bibitem[\protect\citeauthoryear{{Pinto} and {Woosley}}{1988}]{Pinto1988}
\begin{barticle}
\bauthor{\bsnm{{Pinto}}, \binits{P.A.}},
\bauthor{\bsnm{{Woosley}}, \binits{S.E.}}:
\batitle{{The theory of gamma-ray emergence in supernova 1987A}}.
\bjtitle{\nat}
\bvolume{333}(\bissue{6173}),
\bfpage{534}--\blpage{537}
(\byear{1988})
\doiurl{10.1038/333534a0}
\end{barticle}
\endbibitem

%%% 45
\bibitem[\protect\citeauthoryear{{Milne} et~al.}{2004}]{Milne2004}
\begin{barticle}
\bauthor{\bsnm{{Milne}}, \binits{P.A.}},
\bauthor{\bsnm{{Hungerford}}, \binits{A.L.}},
\bauthor{\bsnm{{Fryer}}, \binits{C.L.}},
\bauthor{\bsnm{{Evans}}, \binits{T.M.}},
\bauthor{\bsnm{{Urbatsch}}, \binits{T.J.}},
\bauthor{\bsnm{{Boggs}}, \binits{S.E.}},
\bauthor{\bsnm{{Isern}}, \binits{J.}},
\bauthor{\bsnm{{Bravo}}, \binits{E.}},
\bauthor{\bsnm{{Hirschmann}}, \binits{A.}},
\bauthor{\bsnm{{Kumagai}}, \binits{S.}},
\bauthor{\bsnm{{Pinto}}, \binits{P.A.}},
\bauthor{\bsnm{{The}}, \binits{L.-S.}}:
\batitle{{Unified One-Dimensional Simulations of Gamma-Ray Line Emission from
  Type Ia Supernovae}}.
\bjtitle{\apj}
\bvolume{613}(\bissue{2}),
\bfpage{1101}--\blpage{1119}
(\byear{2004})
\doiurl{10.1086/423235}
{\href{https://arxiv.org/abs/astro-ph/0406173}{{arXiv:astro-ph/0406173}}}
{[astro-ph]}
\end{barticle}
\endbibitem

%%% 46
\bibitem[\protect\citeauthoryear{{Mahoney} et~al.}{1984}]{Mahoney1984}
\begin{barticle}
\bauthor{\bsnm{{Mahoney}}, \binits{W.A.}},
\bauthor{\bsnm{{Ling}}, \binits{J.C.}},
\bauthor{\bsnm{{Wheaton}}, \binits{W.A.}},
\bauthor{\bsnm{{Jacobson}}, \binits{A.S.}}:
\batitle{{HEAO 3 discovery of Al-26 in the interstellar medium}}.
\bjtitle{\apj}
\bvolume{286},
\bfpage{578}--\blpage{585}
(\byear{1984})
\doiurl{10.1086/162632}
\end{barticle}
\endbibitem

%%% 47
\bibitem[\protect\citeauthoryear{{Teegarden} et~al.}{1991}]{Teegarden1991}
\begin{barticle}
\bauthor{\bsnm{{Teegarden}}, \binits{B.J.}},
\bauthor{\bsnm{{Barthelmy}}, \binits{S.D.}},
\bauthor{\bsnm{{Gehrels}}, \binits{N.}},
\bauthor{\bsnm{{Tueller}}, \binits{J.}},
\bauthor{\bsnm{{Leventhal}}, \binits{M.}},
\bauthor{\bsnm{{MacCallum}}, \binits{C.J.}}:
\batitle{{GRIS Observations of 26Al Gamma-Ray Line Emission from Two Points in
  the Galactic Plane}}.
\bjtitle{\apjl}
\bvolume{375},
\bfpage{9}
(\byear{1991})
\doiurl{10.1086/186076}
\end{barticle}
\endbibitem

%%% 48
\bibitem[\protect\citeauthoryear{{Endt}}{1990}]{Endt1990}
\begin{barticle}
\bauthor{\bsnm{{Endt}}, \binits{P.M.}}:
\batitle{{Energy levels of A = 21-44 nuclei (VII)}}.
\bjtitle{\nphysa}
\bvolume{521},
\bfpage{1}--\blpage{400}
(\byear{1990})
\doiurl{10.1016/0375-9474(90)90598-G}
\end{barticle}
\endbibitem

%%% 49
\bibitem[\protect\citeauthoryear{{Iliadis} et~al.}{2011}]{Iliadis2011}
\begin{barticle}
\bauthor{\bsnm{{Iliadis}}, \binits{C.}},
\bauthor{\bsnm{{Champagne}}, \binits{A.}},
\bauthor{\bsnm{{Chieffi}}, \binits{A.}},
\bauthor{\bsnm{{Limongi}}, \binits{M.}}:
\batitle{{The Effects of Thermonuclear Reaction Rate Variations on $^{26}$Al
  Production in Massive Stars: A Sensitivity Study}}.
\bjtitle{\apjs}
\bvolume{193}(\bissue{1}),
\bfpage{16}
(\byear{2011})
\doiurl{10.1088/0067-0049/193/1/16}
{\href{https://arxiv.org/abs/1101.5553}{{arXiv:1101.5553}}}
{[astro-ph.SR]}
\end{barticle}
\endbibitem

%%% 50
\bibitem[\protect\citeauthoryear{{Pleintinger}}{2020}]{Pleintinger2020}
\begin{botherref}
\oauthor{\bsnm{{Pleintinger}}, \binits{M.M.M.}}:
{Star Groups and their Nucleosynthesis}.
PhD thesis,
Max-Planck-Institute for Extraterrestrial Physics, Garching
(November 2020)
\end{botherref}
\endbibitem

%%% 51
\bibitem[\protect\citeauthoryear{{Rugel} et~al.}{2009}]{Rugel2009}
\begin{barticle}
\bauthor{\bsnm{{Rugel}}, \binits{G.}},
\bauthor{\bsnm{{F{\"a}stermann}}, \binits{T.}},
\bauthor{\bsnm{{Knie}}, \binits{K.}},
\bauthor{\bsnm{{Korschinek}}, \binits{G.}},
\bauthor{\bsnm{{Poutivtsev}}, \binits{M.}},
\bauthor{\bsnm{{Schumann}}, \binits{D.}},
\bauthor{\bsnm{{Kivel}}, \binits{N.}},
\bauthor{\bsnm{{G{\"u}nther-Leopold}}, \binits{I.}},
\bauthor{\bsnm{{Weinreich}}, \binits{R.}},
\bauthor{\bsnm{{Wohlmuther}}, \binits{M.}}:
\batitle{{New Measurement of the Fe60 Half-Life}}.
\bjtitle{\prl}
\bvolume{103}(\bissue{7}),
\bfpage{072502}
(\byear{2009})
\doiurl{10.1103/PhysRevLett.103.072502}
\end{barticle}
\endbibitem

%%% 52
\bibitem[\protect\citeauthoryear{{Meynet} et~al.}{1997}]{Meynet1997}
\begin{barticle}
\bauthor{\bsnm{{Meynet}}, \binits{G.}},
\bauthor{\bsnm{{Arnould}}, \binits{M.}},
\bauthor{\bsnm{{Prantzos}}, \binits{N.}},
\bauthor{\bsnm{{Paulus}}, \binits{G.}}:
\batitle{{Contribution of Wolf-Rayet stars to the synthesis of \^26\^Al. I. The
  {\ensuremath{\gamma}}-ray connection.}}
\bjtitle{\aap}
\bvolume{320},
\bfpage{460}--\blpage{468}
(\byear{1997})
\end{barticle}
\endbibitem

%%% 53
\bibitem[\protect\citeauthoryear{{Jos{\'e}} and {Hernanz}}{1998}]{Jose1998}
\begin{barticle}
\bauthor{\bsnm{{Jos{\'e}}}, \binits{J.}},
\bauthor{\bsnm{{Hernanz}}, \binits{M.}}:
\batitle{{Nucleosynthesis in Classical Novae: CO versus ONe White Dwarfs}}.
\bjtitle{\apj}
\bvolume{494}(\bissue{2}),
\bfpage{680}--\blpage{690}
(\byear{1998})
\doiurl{10.1086/305244}
{\href{https://arxiv.org/abs/astro-ph/9709153}{{arXiv:astro-ph/9709153}}}
{[astro-ph]}
\end{barticle}
\endbibitem

%%% 54
\bibitem[\protect\citeauthoryear{{Vasini} et~al.}{2025}]{Vasini2025}
\begin{barticle}
\bauthor{\bsnm{{Vasini}}, \binits{A.}},
\bauthor{\bsnm{{Spitoni}}, \binits{E.}},
\bauthor{\bsnm{{Matteucci}}, \binits{F.}},
\bauthor{\bsnm{{Cescutti}}, \binits{G.}},
\bauthor{\bsnm{{Della Valle}}, \binits{M.}}:
\batitle{{Tracing the Milky Way spiral arms with $^{26}$Al: The role of nova
  systems in the 2D distribution of $^{26}$Al}}.
\bjtitle{\aap}
\bvolume{693},
\bfpage{37}
(\byear{2025})
\doiurl{10.1051/0004-6361/202451630}
{\href{https://arxiv.org/abs/2407.16765}{{arXiv:2407.16765}}}
{[astro-ph.GA]}
\end{barticle}
\endbibitem

%%% 55
\bibitem[\protect\citeauthoryear{{Mowlavi} and {Meynet}}{2000}]{Mowlavi2000}
\begin{barticle}
\bauthor{\bsnm{{Mowlavi}}, \binits{N.}},
\bauthor{\bsnm{{Meynet}}, \binits{G.}}:
\batitle{{Aluminum 26 production in asymptotic giant branch stars}}.
\bjtitle{\aap}
\bvolume{361},
\bfpage{959}--\blpage{976}
(\byear{2000})
\end{barticle}
\endbibitem

%%% 56
\bibitem[\protect\citeauthoryear{{Clayton}}{1984}]{Clayton1984}
\begin{barticle}
\bauthor{\bsnm{{Clayton}}, \binits{D.D.}}:
\batitle{{Al-26 in the interstellar medium}}.
\bjtitle{\apj}
\bvolume{280},
\bfpage{144}--\blpage{149}
(\byear{1984})
\doiurl{10.1086/161978}
\end{barticle}
\endbibitem

%%% 57
\bibitem[\protect\citeauthoryear{{Kn{\"o}dlseder}
  et~al.}{1999}]{Knoedlseder1999b}
\begin{barticle}
\bauthor{\bsnm{{Kn{\"o}dlseder}}, \binits{J.}},
\bauthor{\bsnm{{Bennett}}, \binits{K.}},
\bauthor{\bsnm{{Bloemen}}, \binits{H.}},
\bauthor{\bsnm{{Diehl}}, \binits{R.}},
\bauthor{\bsnm{{Hermsen}}, \binits{W.}},
\bauthor{\bsnm{{Oberlack}}, \binits{U.}},
\bauthor{\bsnm{{Ryan}}, \binits{J.}},
\bauthor{\bsnm{{Sch{\"o}nfelder}}, \binits{V.}},
\bauthor{\bsnm{{von Ballmoos}}, \binits{P.}}:
\batitle{{A multiwavelength comparison of COMPTEL 1.8 MeV $^{26}$Al line
  data}}.
\bjtitle{\aap}
\bvolume{344},
\bfpage{68}--\blpage{82}
(\byear{1999})
\end{barticle}
\endbibitem

%%% 58
\bibitem[\protect\citeauthoryear{{Diehl} et~al.}{2006}]{Diehl2006}
\begin{barticle}
\bauthor{\bsnm{{Diehl}}, \binits{R.}},
\bauthor{\bsnm{{Halloin}}, \binits{H.}},
\bauthor{\bsnm{{Kretschmer}}, \binits{K.}},
\bauthor{\bsnm{{Lichti}}, \binits{G.G.}},
\bauthor{\bsnm{{Sch{\"o}nfelder}}, \binits{V.}},
\bauthor{\bsnm{{Strong}}, \binits{A.W.}},
\bauthor{\bsnm{{von Kienlin}}, \binits{A.}},
\bauthor{\bsnm{{Wang}}, \binits{W.}},
\bauthor{\bsnm{{Jean}}, \binits{P.}},
\bauthor{\bsnm{{Kn{\"o}dlseder}}, \binits{J.}},
\bauthor{\bsnm{{Roques}}, \binits{J.-P.}},
\bauthor{\bsnm{{Weidenspointner}}, \binits{G.}},
\bauthor{\bsnm{{Schanne}}, \binits{S.}},
\bauthor{\bsnm{{Hartmann}}, \binits{D.H.}},
\bauthor{\bsnm{{Winkler}}, \binits{C.}},
\bauthor{\bsnm{{Wunderer}}, \binits{C.}}:
\batitle{{Radioactive $^{26}$Al from massive stars in the Galaxy}}.
\bjtitle{\nat}
\bvolume{439}(\bissue{7072}),
\bfpage{45}--\blpage{47}
(\byear{2006})
\doiurl{10.1038/nature04364}
{\href{https://arxiv.org/abs/astro-ph/0601015}{{arXiv:astro-ph/0601015}}}
{[astro-ph]}
\end{barticle}
\endbibitem

%%% 59
\bibitem[\protect\citeauthoryear{{Diehl} et~al.}{1995}]{Diehl1995}
\begin{barticle}
\bauthor{\bsnm{{Diehl}}, \binits{R.}},
\bauthor{\bsnm{{Dupraz}}, \binits{C.}},
\bauthor{\bsnm{{Bennett}}, \binits{K.}},
\bauthor{\bsnm{{Bloemen}}, \binits{H.}},
\bauthor{\bsnm{{Hermsen}}, \binits{W.}},
\bauthor{\bsnm{{Kn{\"o}dlseder}}, \binits{J.}},
\bauthor{\bsnm{{Lichti}}, \binits{G.}},
\bauthor{\bsnm{{Morris}}, \binits{D.}},
\bauthor{\bsnm{{Ryan}}, \binits{J.}},
\bauthor{\bsnm{{Sch{\"o}nfelder}}, \binits{V.}},
\bauthor{\bsnm{{Steinle}}, \binits{H.}},
\bauthor{\bsnm{{Strong}}, \binits{A.}},
\bauthor{\bsnm{{Swanenburg}}, \binits{B.}},
\bauthor{\bsnm{{Varendorff}}, \binits{M.}},
\bauthor{\bsnm{{Winkler}}, \binits{C.}}:
\batitle{{COMPTEL observations of Galactic \^26\^Al emission.}}
\bjtitle{\aap}
\bvolume{298},
\bfpage{445}
(\byear{1995})
\end{barticle}
\endbibitem

%%% 60
\bibitem[\protect\citeauthoryear{{Oberlack} et~al.}{1996}]{Oberlack1996}
\begin{barticle}
\bauthor{\bsnm{{Oberlack}}, \binits{U.}},
\bauthor{\bsnm{{Bennett}}, \binits{K.}},
\bauthor{\bsnm{{Bloemen}}, \binits{H.}},
\bauthor{\bsnm{{Diehl}}, \binits{R.}},
\bauthor{\bsnm{{Dupraz}}, \binits{C.}},
\bauthor{\bsnm{{Hermsen}}, \binits{W.}},
\bauthor{\bsnm{{Kn{\"o}dlseder}}, \binits{J.}},
\bauthor{\bsnm{{Morris}}, \binits{D.}},
\bauthor{\bsnm{{Sch{\"o}nfelder}}, \binits{V.}},
\bauthor{\bsnm{{Strong}}, \binits{A.}},
\bauthor{\bsnm{{Winkler}}, \binits{C.}}:
\batitle{{The COMPTEL 1.809MeV all-sky image.}}
\bjtitle{\aaps}
\bvolume{120},
\bfpage{311}--\blpage{314}
(\byear{1996})
\end{barticle}
\endbibitem

%%% 61
\bibitem[\protect\citeauthoryear{{del Rio} et~al.}{1996}]{delRio1996}
\begin{barticle}
\bauthor{\bsnm{{del Rio}}, \binits{E.}},
\bauthor{\bsnm{{von Ballmoos}}, \binits{P.}},
\bauthor{\bsnm{{Bennett}}, \binits{K.}},
\bauthor{\bsnm{{Bloemen}}, \binits{H.}},
\bauthor{\bsnm{{Diehl}}, \binits{R.}},
\bauthor{\bsnm{{Hermsen}}, \binits{W.}},
\bauthor{\bsnm{{Kn{\"o}dlseder}}, \binits{J.}},
\bauthor{\bsnm{{Oberlack}}, \binits{U.}},
\bauthor{\bsnm{{Ryan}}, \binits{J.}},
\bauthor{\bsnm{{Sch{\"o}nfelder}}, \binits{V.}},
\bauthor{\bsnm{{Winkler}}, \binits{C.}}:
\batitle{{1.8 MeV line emission from the Cygnus region.}}
\bjtitle{\aap}
\bvolume{315},
\bfpage{237}--\blpage{242}
(\byear{1996})
\end{barticle}
\endbibitem

%%% 62
\bibitem[\protect\citeauthoryear{{Kn{\"o}dlseder}
  et~al.}{1996}]{Knoedlseder1996}
\begin{barticle}
\bauthor{\bsnm{{Kn{\"o}dlseder}}, \binits{J.}},
\bauthor{\bsnm{{Bennett}}, \binits{K.}},
\bauthor{\bsnm{{Bloemen}}, \binits{H.}},
\bauthor{\bsnm{{Diehl}}, \binits{R.}},
\bauthor{\bsnm{{Hermsen}}, \binits{W.}},
\bauthor{\bsnm{{Oberlack}}, \binits{U.}},
\bauthor{\bsnm{{Ryan}}, \binits{J.}},
\bauthor{\bsnm{{Sch{\"o}nfelder}}, \binits{V.}}:
\batitle{{1.8 MeV emission from the Carina region.}}
\bjtitle{\aaps}
\bvolume{120},
\bfpage{327}--\blpage{330}
(\byear{1996})
\doiurl{10.48550/arXiv.astro-ph/9604054}
{\href{https://arxiv.org/abs/astro-ph/9604054}{{arXiv:astro-ph/9604054}}}
{[astro-ph]}
\end{barticle}
\endbibitem

%%% 63
\bibitem[\protect\citeauthoryear{{Diehl} et~al.}{1995}]{Diehl1995b}
\begin{barticle}
\bauthor{\bsnm{{Diehl}}, \binits{R.}},
\bauthor{\bsnm{{Bennett}}, \binits{K.}},
\bauthor{\bsnm{{Bloemen}}, \binits{H.}},
\bauthor{\bsnm{{Dupraz}}, \binits{C.}},
\bauthor{\bsnm{{Hermsen}}, \binits{W.}},
\bauthor{\bsnm{{Kn{\"o}dlseder}}, \binits{J.}},
\bauthor{\bsnm{{Lichti}}, \binits{G.}},
\bauthor{\bsnm{{Morris}}, \binits{D.}},
\bauthor{\bsnm{{Oberlack}}, \binits{U.}},
\bauthor{\bsnm{{Ryan}}, \binits{J.}},
\bauthor{\bsnm{{Sch{\"o}nfelder}}, \binits{V.}},
\bauthor{\bsnm{{Steinle}}, \binits{H.}},
\bauthor{\bsnm{{Varendorff}}, \binits{M.}},
\bauthor{\bsnm{{Winkler}}, \binits{C.}}:
\batitle{{1.809 MeV gamma-rays from the VELA region.}}
\bjtitle{\aap}
\bvolume{298},
\bfpage{25}
(\byear{1995})
\end{barticle}
\endbibitem

%%% 64
\bibitem[\protect\citeauthoryear{{Pl{\"u}schke} et~al.}{2001}]{Plueschke2001}
\begin{bchapter}
\bauthor{\bsnm{{Pl{\"u}schke}}, \binits{S.}},
\bauthor{\bsnm{{Diehl}}, \binits{R.}},
\bauthor{\bsnm{{Sch{\"o}nfelder}}, \binits{V.}},
\bauthor{\bsnm{{Bloemen}}, \binits{H.}},
\bauthor{\bsnm{{Hermsen}}, \binits{W.}},
\bauthor{\bsnm{{Bennett}}, \binits{K.}},
\bauthor{\bsnm{{Winkler}}, \binits{C.}},
\bauthor{\bsnm{{McConnell}}, \binits{M.}},
\bauthor{\bsnm{{Ryan}}, \binits{J.}},
\bauthor{\bsnm{{Oberlack}}, \binits{U.}},
\bauthor{\bsnm{{Kn{\"o}dlseder}}, \binits{J.}}:
\bctitle{{The COMPTEL 1.809 MeV survey}}.
In: \beditor{\bsnm{{Gimenez}}, \binits{A.}},
\beditor{\bsnm{{Reglero}}, \binits{V.}},
\beditor{\bsnm{{Winkler}}, \binits{C.}} (eds.)
\bbtitle{Exploring the Gamma-Ray Universe}.
\bsertitle{ESA Special Publication},
vol. \bseriesno{459},
pp. \bfpage{55}--\blpage{58}
(\byear{2001}).
\doiurl{10.48550/arXiv.astro-ph/0104047}
\end{bchapter}
\endbibitem

%%% 65
\bibitem[\protect\citeauthoryear{{Bouchet} et~al.}{2015}]{Bouchet2015}
\begin{barticle}
\bauthor{\bsnm{{Bouchet}}, \binits{L.}},
\bauthor{\bsnm{{Jourdain}}, \binits{E.}},
\bauthor{\bsnm{{Roques}}, \binits{J.-P.}}:
\batitle{{The Galactic $^{26}$Al Emission Map as Revealed by INTEGRAL SPI}}.
\bjtitle{\apj}
\bvolume{801}(\bissue{2}),
\bfpage{142}
(\byear{2015})
\doiurl{10.1088/0004-637X/801/2/142}
{\href{https://arxiv.org/abs/1501.05247}{{arXiv:1501.05247}}}
{[astro-ph.HE]}
\end{barticle}
\endbibitem

%%% 66
\bibitem[\protect\citeauthoryear{{Siegert} et~al.}{2023}]{Siegert2023b}
\begin{barticle}
\bauthor{\bsnm{{Siegert}}, \binits{T.}},
\bauthor{\bsnm{{Pleintinger}}, \binits{M.M.M.}},
\bauthor{\bsnm{{Diehl}}, \binits{R.}},
\bauthor{\bsnm{{Krause}}, \binits{M.G.H.}},
\bauthor{\bsnm{{Greiner}}, \binits{J.}},
\bauthor{\bsnm{{Weinberger}}, \binits{C.}}:
\batitle{{Galactic population synthesis of radioactive nucleosynthesis
  ejecta}}.
\bjtitle{\aap}
\bvolume{672},
\bfpage{54}
(\byear{2023})
\doiurl{10.1051/0004-6361/202244457}
{\href{https://arxiv.org/abs/2301.10192}{{arXiv:2301.10192}}}
{[astro-ph.GA]}
\end{barticle}
\endbibitem

%%% 67
\bibitem[\protect\citeauthoryear{{Martin} et~al.}{2009}]{Martin2009}
\begin{barticle}
\bauthor{\bsnm{{Martin}}, \binits{P.}},
\bauthor{\bsnm{{Kn{\"o}dlseder}}, \binits{J.}},
\bauthor{\bsnm{{Diehl}}, \binits{R.}},
\bauthor{\bsnm{{Meynet}}, \binits{G.}}:
\batitle{{New estimates of the gamma-ray line emission of the Cygnus region
  from INTEGRAL/SPI observations}}.
\bjtitle{\aap}
\bvolume{506}(\bissue{2}),
\bfpage{703}--\blpage{710}
(\byear{2009})
\doiurl{10.1051/0004-6361/200912178}
{\href{https://arxiv.org/abs/1001.1521}{{arXiv:1001.1521}}}
{[astro-ph.HE]}
\end{barticle}
\endbibitem

%%% 68
\bibitem[\protect\citeauthoryear{{Diehl} et~al.}{2010}]{Diehl2010}
\begin{barticle}
\bauthor{\bsnm{{Diehl}}, \binits{R.}},
\bauthor{\bsnm{{Lang}}, \binits{M.G.}},
\bauthor{\bsnm{{Martin}}, \binits{P.}},
\bauthor{\bsnm{{Ohlendorf}}, \binits{H.}},
\bauthor{\bsnm{{Preibisch}}, \binits{T.}},
\bauthor{\bsnm{{Voss}}, \binits{R.}},
\bauthor{\bsnm{{Jean}}, \binits{P.}},
\bauthor{\bsnm{{Roques}}, \binits{J.-P.}},
\bauthor{\bsnm{{von Ballmoos}}, \binits{P.}},
\bauthor{\bsnm{{Wang}}, \binits{W.}}:
\batitle{{Radioactive $^{26}$Al from the Scorpius-Centaurus association}}.
\bjtitle{\aap}
\bvolume{522},
\bfpage{51}
(\byear{2010})
\doiurl{10.1051/0004-6361/201014302}
{\href{https://arxiv.org/abs/1007.4462}{{arXiv:1007.4462}}}
{[astro-ph.HE]}
\end{barticle}
\endbibitem

%%% 69
\bibitem[\protect\citeauthoryear{{Krause} et~al.}{2018}]{Krause2018}
\begin{barticle}
\bauthor{\bsnm{{Krause}}, \binits{M.G.H.}},
\bauthor{\bsnm{{Burkert}}, \binits{A.}},
\bauthor{\bsnm{{Diehl}}, \binits{R.}},
\bauthor{\bsnm{{Fierlinger}}, \binits{K.}},
\bauthor{\bsnm{{Gaczkowski}}, \binits{B.}},
\bauthor{\bsnm{{Kr{\"o}ll}}, \binits{D.}},
\bauthor{\bsnm{{Ngoumou}}, \binits{J.}},
\bauthor{\bsnm{{Roccatagliata}}, \binits{V.}},
\bauthor{\bsnm{{Siegert}}, \binits{T.}},
\bauthor{\bsnm{{Preibisch}}, \binits{T.}}:
\batitle{{Surround and Squash: the impact of superbubbles on the interstellar
  medium in Scorpius-Centaurus OB2}}.
\bjtitle{\aap}
\bvolume{619},
\bfpage{120}
(\byear{2018})
\doiurl{10.1051/0004-6361/201732416}
{\href{https://arxiv.org/abs/1808.04788}{{arXiv:1808.04788}}}
{[astro-ph.GA]}
\end{barticle}
\endbibitem

%%% 70
\bibitem[\protect\citeauthoryear{{Kretschmer} et~al.}{2013}]{Kretschmer2013}
\begin{barticle}
\bauthor{\bsnm{{Kretschmer}}, \binits{K.}},
\bauthor{\bsnm{{Diehl}}, \binits{R.}},
\bauthor{\bsnm{{Krause}}, \binits{M.}},
\bauthor{\bsnm{{Burkert}}, \binits{A.}},
\bauthor{\bsnm{{Fierlinger}}, \binits{K.}},
\bauthor{\bsnm{{Gerhard}}, \binits{O.}},
\bauthor{\bsnm{{Greiner}}, \binits{J.}},
\bauthor{\bsnm{{Wang}}, \binits{W.}}:
\batitle{{Kinematics of massive star ejecta in the Milky Way as traced by
  $^{26}$Al}}.
\bjtitle{\aap}
\bvolume{559},
\bfpage{99}
(\byear{2013})
\doiurl{10.1051/0004-6361/201322563}
{\href{https://arxiv.org/abs/1309.4980}{{arXiv:1309.4980}}}
{[astro-ph.HE]}
\end{barticle}
\endbibitem

%%% 71
\bibitem[\protect\citeauthoryear{{Krause} et~al.}{2015}]{Krause2015}
\begin{barticle}
\bauthor{\bsnm{{Krause}}, \binits{M.G.H.}},
\bauthor{\bsnm{{Diehl}}, \binits{R.}},
\bauthor{\bsnm{{Bagetakos}}, \binits{Y.}},
\bauthor{\bsnm{{Brinks}}, \binits{E.}},
\bauthor{\bsnm{{Burkert}}, \binits{A.}},
\bauthor{\bsnm{{Gerhard}}, \binits{O.}},
\bauthor{\bsnm{{Greiner}}, \binits{J.}},
\bauthor{\bsnm{{Kretschmer}}, \binits{K.}},
\bauthor{\bsnm{{Siegert}}, \binits{T.}}:
\batitle{{$^{26}$Al kinematics: superbubbles following the spiral arms?.
  Constraints from the statistics of star clusters and HI supershells}}.
\bjtitle{\aap}
\bvolume{578},
\bfpage{113}
(\byear{2015})
\doiurl{10.1051/0004-6361/201525847}
{\href{https://arxiv.org/abs/1504.03120}{{arXiv:1504.03120}}}
{[astro-ph.GA]}
\end{barticle}
\endbibitem

%%% 72
\bibitem[\protect\citeauthoryear{{Siegert} and {Diehl}}{2017}]{Siegert2017b}
\begin{bchapter}
\bauthor{\bsnm{{Siegert}}, \binits{T.}},
\bauthor{\bsnm{{Diehl}}, \binits{R.}}:
\bctitle{{The $^{26}$Al Gamma-ray Line from Massive-Star Regions}}.
In: \beditor{\bsnm{{Kubono}}, \binits{S.}},
\beditor{\bsnm{{Kajino}}, \binits{T.}},
\beditor{\bsnm{{Nishimura}}, \binits{S.}},
\beditor{\bsnm{{Isobe}}, \binits{T.}},
\beditor{\bsnm{{Nagataki}}, \binits{S.}},
\beditor{\bsnm{{Shima}}, \binits{T.}},
\beditor{\bsnm{{Takeda}}, \binits{Y.}} (eds.)
\bbtitle{14th International Symposium on Nuclei in the Cosmos (NIC2016)},
p. \bfpage{020305}
(\byear{2017}).
\doiurl{10.7566/JPSCP.14.020305}
\end{bchapter}
\endbibitem

%%% 73
\bibitem[\protect\citeauthoryear{{Pleintinger} et~al.}{2023}]{Pleintinger2023}
\begin{barticle}
\bauthor{\bsnm{{Pleintinger}}, \binits{M.M.M.}},
\bauthor{\bsnm{{Diehl}}, \binits{R.}},
\bauthor{\bsnm{{Siegert}}, \binits{T.}},
\bauthor{\bsnm{{Greiner}}, \binits{J.}},
\bauthor{\bsnm{{Krause}}, \binits{M.G.H.}}:
\batitle{{$^{26}$Al gamma rays from the Galaxy with INTEGRAL/SPI}}.
\bjtitle{\aap}
\bvolume{672},
\bfpage{53}
(\byear{2023})
\doiurl{10.1051/0004-6361/202245069}
{\href{https://arxiv.org/abs/2212.11228}{{arXiv:2212.11228}}}
{[astro-ph.HE]}
\end{barticle}
\endbibitem

%%% 74
\bibitem[\protect\citeauthoryear{{Parikh} et~al.}{2014}]{Parikh2014}
\begin{barticle}
\bauthor{\bsnm{{Parikh}}, \binits{A.}},
\bauthor{\bsnm{{Jos{\'e}}}, \binits{J.}},
\bauthor{\bsnm{{Karakas}}, \binits{A.}},
\bauthor{\bsnm{{Ruiz}}, \binits{C.}},
\bauthor{\bsnm{{Wimmer}}, \binits{K.}}:
\batitle{{Strength of the E$_{R}$=127 keV, Al26(p,{\ensuremath{\gamma}})Si27
  resonance}}.
\bjtitle{\prc}
\bvolume{90}(\bissue{3}),
\bfpage{038801}
(\byear{2014})
\doiurl{10.1103/PhysRevC.90.038801}
{\href{https://arxiv.org/abs/1408.5227}{{arXiv:1408.5227}}}
{[astro-ph.SR]}
\end{barticle}
\endbibitem

%%% 75
\bibitem[\protect\citeauthoryear{{Prantzos} and {Diehl}}{1996}]{Prantzos1996}
\begin{barticle}
\bauthor{\bsnm{{Prantzos}}, \binits{N.}},
\bauthor{\bsnm{{Diehl}}, \binits{R.}}:
\batitle{{Radioactive 26Al in the galaxy: observations versus theory}}.
\bjtitle{\physrep}
\bvolume{267},
\bfpage{1}--\blpage{69}
(\byear{1996})
\doiurl{10.1016/0370-1573(95)00055-0}
\end{barticle}
\endbibitem

%%% 76
\bibitem[\protect\citeauthoryear{{Lugaro} et~al.}{2018}]{Lugaro2018}
\begin{barticle}
\bauthor{\bsnm{{Lugaro}}, \binits{M.}},
\bauthor{\bsnm{{Ott}}, \binits{U.}},
\bauthor{\bsnm{{Kereszturi}}, \binits{{\'A}.}}:
\batitle{{Radioactive nuclei from cosmochronology to habitability}}.
\bjtitle{Progress in Particle and Nuclear Physics}
\bvolume{102},
\bfpage{1}--\blpage{47}
(\byear{2018})
\doiurl{10.1016/j.ppnp.2018.05.002}
{\href{https://arxiv.org/abs/1808.00233}{{arXiv:1808.00233}}}
{[astro-ph.SR]}
\end{barticle}
\endbibitem

%%% 77
\bibitem[\protect\citeauthoryear{{Sieverding} et~al.}{2017}]{Sieverding2017}
\begin{bchapter}
\bauthor{\bsnm{{Sieverding}}, \binits{A.}},
\bauthor{\bsnm{{Mart{\'\i}nez-Pinedo}}, \binits{G.}},
\bauthor{\bsnm{{Langanke}}, \binits{K.}},
\bauthor{\bsnm{{Heger}}, \binits{A.}}:
\bctitle{{Neutrino Induced Nucleosynthesis of Radioactive Nuclei in
  Core-Collapse Supernovae}}.
In: \beditor{\bsnm{{Kubono}}, \binits{S.}},
\beditor{\bsnm{{Kajino}}, \binits{T.}},
\beditor{\bsnm{{Nishimura}}, \binits{S.}},
\beditor{\bsnm{{Isobe}}, \binits{T.}},
\beditor{\bsnm{{Nagataki}}, \binits{S.}},
\beditor{\bsnm{{Shima}}, \binits{T.}},
\beditor{\bsnm{{Takeda}}, \binits{Y.}} (eds.)
\bbtitle{14th International Symposium on Nuclei in the Cosmos (NIC2016)},
p. \bfpage{020701}
(\byear{2017}).
\doiurl{10.7566/JPSCP.14.020701}
\end{bchapter}
\endbibitem

%%% 78
\bibitem[\protect\citeauthoryear{{Karakas} and {Lattanzio}}{2014}]{Karakas2014}
\begin{barticle}
\bauthor{\bsnm{{Karakas}}, \binits{A.I.}},
\bauthor{\bsnm{{Lattanzio}}, \binits{J.C.}}:
\batitle{{The Dawes Review 2: Nucleosynthesis and Stellar Yields of Low- and
  Intermediate-Mass Single Stars}}.
\bjtitle{\pasa}
\bvolume{31},
\bfpage{030}
(\byear{2014})
\doiurl{10.1017/pasa.2014.21}
{\href{https://arxiv.org/abs/1405.0062}{{arXiv:1405.0062}}}
{[astro-ph.SR]}
\end{barticle}
\endbibitem

%%% 79
\bibitem[\protect\citeauthoryear{{Starrfield} et~al.}{2020}]{Starrfield2020}
\begin{barticle}
\bauthor{\bsnm{{Starrfield}}, \binits{S.}},
\bauthor{\bsnm{{Bose}}, \binits{M.}},
\bauthor{\bsnm{{Iliadis}}, \binits{C.}},
\bauthor{\bsnm{{Hix}}, \binits{W.R.}},
\bauthor{\bsnm{{Woodward}}, \binits{C.E.}},
\bauthor{\bsnm{{Wagner}}, \binits{R.M.}}:
\batitle{{Carbon-Oxygen Classical Novae Are Galactic $^{7}$Li Producers as well
  as Potential Supernova Ia Progenitors}}.
\bjtitle{\apj}
\bvolume{895}(\bissue{1}),
\bfpage{70}
(\byear{2020})
\doiurl{10.3847/1538-4357/ab8d23}
{\href{https://arxiv.org/abs/1910.00575}{{arXiv:1910.00575}}}
{[astro-ph.SR]}
\end{barticle}
\endbibitem

%%% 80
\bibitem[\protect\citeauthoryear{{Pleintinger} et~al.}{2019}]{Pleintinger2019}
\begin{barticle}
\bauthor{\bsnm{{Pleintinger}}, \binits{M.M.M.}},
\bauthor{\bsnm{{Siegert}}, \binits{T.}},
\bauthor{\bsnm{{Diehl}}, \binits{R.}},
\bauthor{\bsnm{{Fujimoto}}, \binits{Y.}},
\bauthor{\bsnm{{Greiner}}, \binits{J.}},
\bauthor{\bsnm{{Krause}}, \binits{M.G.H.}},
\bauthor{\bsnm{{Krumholz}}, \binits{M.R.}}:
\batitle{{Comparing simulated $^{26}$Al maps to gamma-ray measurements}}.
\bjtitle{\aap}
\bvolume{632},
\bfpage{73}
(\byear{2019})
\doiurl{10.1051/0004-6361/201935911}
{\href{https://arxiv.org/abs/1910.06112}{{arXiv:1910.06112}}}
{[astro-ph.HE]}
\end{barticle}
\endbibitem

%%% 81
\bibitem[\protect\citeauthoryear{{Limongi} and {Chieffi}}{2018}]{Limongi2018}
\begin{barticle}
\bauthor{\bsnm{{Limongi}}, \binits{M.}},
\bauthor{\bsnm{{Chieffi}}, \binits{A.}}:
\batitle{{Presupernova Evolution and Explosive Nucleosynthesis of Rotating
  Massive Stars in the Metallicity Range -3 {\ensuremath{\leq}} [Fe/H]
  {\ensuremath{\leq}} 0}}.
\bjtitle{\apjs}
\bvolume{237}(\bissue{1}),
\bfpage{13}
(\byear{2018})
\doiurl{10.3847/1538-4365/aacb24}
{\href{https://arxiv.org/abs/1805.09640}{{arXiv:1805.09640}}}
{[astro-ph.SR]}
\end{barticle}
\endbibitem

%%% 82
\bibitem[\protect\citeauthoryear{{Janka}}{2012}]{Janka2012}
\begin{barticle}
\bauthor{\bsnm{{Janka}}, \binits{H.-T.}}:
\batitle{{Explosion Mechanisms of Core-Collapse Supernovae}}.
\bjtitle{Annual Review of Nuclear and Particle Science}
\bvolume{62}(\bissue{1}),
\bfpage{407}--\blpage{451}
(\byear{2012})
\doiurl{10.1146/annurev-nucl-102711-094901}
{\href{https://arxiv.org/abs/1206.2503}{{arXiv:1206.2503}}}
{[astro-ph.SR]}
\end{barticle}
\endbibitem

%%% 83
\bibitem[\protect\citeauthoryear{{Sukhbold} et~al.}{2016}]{Sukhbold2016}
\begin{barticle}
\bauthor{\bsnm{{Sukhbold}}, \binits{T.}},
\bauthor{\bsnm{{Ertl}}, \binits{T.}},
\bauthor{\bsnm{{Woosley}}, \binits{S.E.}},
\bauthor{\bsnm{{Brown}}, \binits{J.M.}},
\bauthor{\bsnm{{Janka}}, \binits{H.-T.}}:
\batitle{{Core-collapse Supernovae from 9 to 120 Solar Masses Based on
  Neutrino-powered Explosions}}.
\bjtitle{\apj}
\bvolume{821}(\bissue{1}),
\bfpage{38}
(\byear{2016})
\doiurl{10.3847/0004-637X/821/1/38}
{\href{https://arxiv.org/abs/1510.04643}{{arXiv:1510.04643}}}
{[astro-ph.HE]}
\end{barticle}
\endbibitem

%%% 84
\bibitem[\protect\citeauthoryear{{Licquia} and {Newman}}{2015}]{Licquia2015}
\begin{barticle}
\bauthor{\bsnm{{Licquia}}, \binits{T.C.}},
\bauthor{\bsnm{{Newman}}, \binits{J.A.}}:
\batitle{{Improved Estimates of the Milky Way's Stellar Mass and Star Formation
  Rate from Hierarchical Bayesian Meta-Analysis}}.
\bjtitle{\apj}
\bvolume{806}(\bissue{1}),
\bfpage{96}
(\byear{2015})
\doiurl{10.1088/0004-637X/806/1/96}
{\href{https://arxiv.org/abs/1407.1078}{{arXiv:1407.1078}}}
{[astro-ph.GA]}
\end{barticle}
\endbibitem

%%% 85
\bibitem[\protect\citeauthoryear{{Mel'nik} and {Dambis}}{2017}]{Melnik2017}
\begin{barticle}
\bauthor{\bsnm{{Mel'nik}}, \binits{A.M.}},
\bauthor{\bsnm{{Dambis}}, \binits{A.K.}}:
\batitle{{Kinematics of OB-associations in Gaia epoch}}.
\bjtitle{\mnras}
\bvolume{472}(\bissue{4}),
\bfpage{3887}--\blpage{3904}
(\byear{2017})
\doiurl{10.1093/mnras/stx2225}
{\href{https://arxiv.org/abs/1708.08337}{{arXiv:1708.08337}}}
{[astro-ph.GA]}
\end{barticle}
\endbibitem

%%% 86
\bibitem[\protect\citeauthoryear{{Schulreich} et~al.}{2023}]{Schulreich2023}
\begin{barticle}
\bauthor{\bsnm{{Schulreich}}, \binits{M.M.}},
\bauthor{\bsnm{{Feige}}, \binits{J.}},
\bauthor{\bsnm{{Breitschwerdt}}, \binits{D.}}:
\batitle{{Numerical studies on the link between radioisotopic signatures on
  Earth and the formation of the Local Bubble. II. Advanced modelling of
  interstellar $^{26}$Al, $^{53}$Mn, $^{60}$Fe, and $^{244}$Pu influxes as
  traces of past supernova activity in the solar neighbourhood}}.
\bjtitle{\aap}
\bvolume{680},
\bfpage{39}
(\byear{2023})
\doiurl{10.1051/0004-6361/202347532}
{\href{https://arxiv.org/abs/2309.13983}{{arXiv:2309.13983}}}
{[astro-ph.SR]}
\end{barticle}
\endbibitem

%%% 87
\bibitem[\protect\citeauthoryear{{Siegert} et~al.}{2022}]{Siegert2022}
\begin{bchapter}
\bauthor{\bsnm{{Siegert}}, \binits{T.}},
\bauthor{\bsnm{{Horan}}, \binits{D.}},
\bauthor{\bsnm{{Kanbach}}, \binits{G.}}:
\bctitle{{Telescope Concepts in Gamma-Ray Astronomy}}.
In: \beditor{\bsnm{{Bambi}}, \binits{C.}},
\beditor{\bsnm{{Sangangelo}}, \binits{A.}} (eds.)
\bbtitle{Handbook of X-ray and Gamma-ray Astrophysics},
p. \bfpage{80}
(\byear{2022}).
\doiurl{10.1007/978-981-16-4544-0_43-1}
\end{bchapter}
\endbibitem

%%% 88
\bibitem[\protect\citeauthoryear{{Siegert} et~al.}{2024}]{Siegert2024}
\begin{barticle}
\bauthor{\bsnm{{Siegert}}, \binits{T.}},
\bauthor{\bsnm{{Schulreich}}, \binits{M.M.}},
\bauthor{\bsnm{{Bauer}}, \binits{N.}},
\bauthor{\bsnm{{Reinhardt}}, \binits{R.}},
\bauthor{\bsnm{{Mittal}}, \binits{S.}},
\bauthor{\bsnm{{Yoneda}}, \binits{H.}}:
\batitle{{Gamma-ray line emission from the Local Bubble}}.
\bjtitle{\aap}
\bvolume{689},
\bfpage{2}
(\byear{2024})
\doiurl{10.1051/0004-6361/202450310}
{\href{https://arxiv.org/abs/2405.15262}{{arXiv:2405.15262}}}
{[astro-ph.HE]}
\end{barticle}
\endbibitem

%%% 89
\bibitem[\protect\citeauthoryear{{Limongi} and {Chieffi}}{2006}]{Limongi2006}
\begin{barticle}
\bauthor{\bsnm{{Limongi}}, \binits{M.}},
\bauthor{\bsnm{{Chieffi}}, \binits{A.}}:
\batitle{{The Nucleosynthesis of $^{26}$Al and $^{60}$Fe in Solar Metallicity
  Stars Extending in Mass from 11 to 120 M$_{solar}$: The Hydrostatic and
  Explosive Contributions}}.
\bjtitle{\apj}
\bvolume{647}(\bissue{1}),
\bfpage{483}--\blpage{500}
(\byear{2006})
\doiurl{10.1086/505164}
{\href{https://arxiv.org/abs/astro-ph/0604297}{{arXiv:astro-ph/0604297}}}
{[astro-ph]}
\end{barticle}
\endbibitem

%%% 90
\bibitem[\protect\citeauthoryear{{Brinkman} et~al.}{2019}]{Brinkman2019}
\begin{barticle}
\bauthor{\bsnm{{Brinkman}}, \binits{H.E.}},
\bauthor{\bsnm{{Doherty}}, \binits{C.L.}},
\bauthor{\bsnm{{Pols}}, \binits{O.R.}},
\bauthor{\bsnm{{Li}}, \binits{E.T.}},
\bauthor{\bsnm{{C{\^o}t{\'e}}}, \binits{B.}},
\bauthor{\bsnm{{Lugaro}}, \binits{M.}}:
\batitle{{Aluminium-26 from Massive Binary Stars. I. Nonrotating Models}}.
\bjtitle{\apj}
\bvolume{884}(\bissue{1}),
\bfpage{38}
(\byear{2019})
\doiurl{10.3847/1538-4357/ab40ae}
{\href{https://arxiv.org/abs/1909.04433}{{arXiv:1909.04433}}}
{[astro-ph.SR]}
\end{barticle}
\endbibitem

%%% 91
\bibitem[\protect\citeauthoryear{{Oberlack} et~al.}{2000}]{Oberlack2000}
\begin{barticle}
\bauthor{\bsnm{{Oberlack}}, \binits{U.}},
\bauthor{\bsnm{{Wessolowski}}, \binits{U.}},
\bauthor{\bsnm{{Diehl}}, \binits{R.}},
\bauthor{\bsnm{{Bennett}}, \binits{K.}},
\bauthor{\bsnm{{Bloemen}}, \binits{H.}},
\bauthor{\bsnm{{Hermsen}}, \binits{W.}},
\bauthor{\bsnm{{Kn{\"o}dlseder}}, \binits{J.}},
\bauthor{\bsnm{{Morris}}, \binits{D.}},
\bauthor{\bsnm{{Sch{\"o}nfelder}}, \binits{V.}},
\bauthor{\bsnm{{von Ballmoos}}, \binits{P.}}:
\batitle{{COMPTEL limits on $^{26}$Al 1.809 MeV line emission from gamma$^{2}$
  Velorum}}.
\bjtitle{\aap}
\bvolume{353},
\bfpage{715}--\blpage{721}
(\byear{2000})
\doiurl{10.48550/arXiv.astro-ph/9910555}
{\href{https://arxiv.org/abs/astro-ph/9910555}{{arXiv:astro-ph/9910555}}}
{[astro-ph]}
\end{barticle}
\endbibitem

%%% 92
\bibitem[\protect\citeauthoryear{{Smartt}}{2009}]{Smartt2009}
\begin{barticle}
\bauthor{\bsnm{{Smartt}}, \binits{S.J.}}:
\batitle{{Progenitors of Core-Collapse Supernovae}}.
\bjtitle{\araa}
\bvolume{47}(\bissue{1}),
\bfpage{63}--\blpage{106}
(\byear{2009})
\doiurl{10.1146/annurev-astro-082708-101737}
{\href{https://arxiv.org/abs/0908.0700}{{arXiv:0908.0700}}}
{[astro-ph.SR]}
\end{barticle}
\endbibitem

%%% 93
\bibitem[\protect\citeauthoryear{{Weaver} et~al.}{1977}]{Weaver1977}
\begin{barticle}
\bauthor{\bsnm{{Weaver}}, \binits{R.}},
\bauthor{\bsnm{{McCray}}, \binits{R.}},
\bauthor{\bsnm{{Castor}}, \binits{J.}},
\bauthor{\bsnm{{Shapiro}}, \binits{P.}},
\bauthor{\bsnm{{Moore}}, \binits{R.}}:
\batitle{{Interstellar bubbles. II. Structure and evolution.}}
\bjtitle{\apj}
\bvolume{218},
\bfpage{377}--\blpage{395}
(\byear{1977})
\doiurl{10.1086/155692}
\end{barticle}
\endbibitem

%%% 94
\bibitem[\protect\citeauthoryear{{Fujimoto} et~al.}{2018}]{Fujimoto2018}
\begin{barticle}
\bauthor{\bsnm{{Fujimoto}}, \binits{Y.}},
\bauthor{\bsnm{{Krumholz}}, \binits{M.R.}},
\bauthor{\bsnm{{Tachibana}}, \binits{S.}}:
\batitle{{Short-lived radioisotopes in meteorites from Galactic-scale
  correlated star formation}}.
\bjtitle{\mnras}
\bvolume{480}(\bissue{3}),
\bfpage{4025}--\blpage{4039}
(\byear{2018})
\doiurl{10.1093/mnras/sty2132}
{\href{https://arxiv.org/abs/1802.08695}{{arXiv:1802.08695}}}
{[astro-ph.GA]}
\end{barticle}
\endbibitem

%%% 95
\bibitem[\protect\citeauthoryear{{Rodgers-Lee} et~al.}{2019}]{RodgersLee2019}
\begin{barticle}
\bauthor{\bsnm{{Rodgers-Lee}}, \binits{D.}},
\bauthor{\bsnm{{Krause}}, \binits{M.G.H.}},
\bauthor{\bsnm{{Dale}}, \binits{J.}},
\bauthor{\bsnm{{Diehl}}, \binits{R.}}:
\batitle{{Synthetic $^{26}$Al emission from galactic-scale superbubble
  simulations}}.
\bjtitle{\mnras}
\bvolume{490}(\bissue{2}),
\bfpage{1894}--\blpage{1912}
(\byear{2019})
\doiurl{10.1093/mnras/stz2708}
{\href{https://arxiv.org/abs/1909.10978}{{arXiv:1909.10978}}}
{[astro-ph.GA]}
\end{barticle}
\endbibitem

%%% 96
\bibitem[\protect\citeauthoryear{{Krause} et~al.}{2014}]{Krause2014}
\begin{barticle}
\bauthor{\bsnm{{Krause}}, \binits{M.}},
\bauthor{\bsnm{{Diehl}}, \binits{R.}},
\bauthor{\bsnm{{B{\"o}hringer}}, \binits{H.}},
\bauthor{\bsnm{{Freyberg}}, \binits{M.}},
\bauthor{\bsnm{{Lubos}}, \binits{D.}}:
\batitle{{Feedback by massive stars and the emergence of superbubbles. II.
  X-ray properties}}.
\bjtitle{\aap}
\bvolume{566},
\bfpage{94}
(\byear{2014})
\doiurl{10.1051/0004-6361/201423871}
{\href{https://arxiv.org/abs/1405.0037}{{arXiv:1405.0037}}}
{[astro-ph.GA]}
\end{barticle}
\endbibitem

%%% 97
\bibitem[\protect\citeauthoryear{{Burrows} et~al.}{1993}]{Burrows1993}
\begin{barticle}
\bauthor{\bsnm{{Burrows}}, \binits{D.N.}},
\bauthor{\bsnm{{Singh}}, \binits{K.P.}},
\bauthor{\bsnm{{Nousek}}, \binits{J.A.}},
\bauthor{\bsnm{{Garmire}}, \binits{G.P.}},
\bauthor{\bsnm{{Good}}, \binits{J.}}:
\batitle{{A Multiwavelength Study of the Eridanus Soft X-Ray Enhancement}}.
\bjtitle{\apj}
\bvolume{406},
\bfpage{97}
(\byear{1993})
\doiurl{10.1086/172423}
\end{barticle}
\endbibitem

%%% 98
\bibitem[\protect\citeauthoryear{{Smartt}}{2015}]{Smartt2015}
\begin{barticle}
\bauthor{\bsnm{{Smartt}}, \binits{S.J.}}:
\batitle{{Observational Constraints on the Progenitors of Core-Collapse
  Supernovae: The Case for Missing High-Mass Stars}}.
\bjtitle{\pasa}
\bvolume{32},
\bfpage{016}
(\byear{2015})
\doiurl{10.1017/pasa.2015.17}
{\href{https://arxiv.org/abs/1504.02635}{{arXiv:1504.02635}}}
{[astro-ph.SR]}
\end{barticle}
\endbibitem

%%% 99
\bibitem[\protect\citeauthoryear{{Martinet} et~al.}{2022}]{Martinet2022}
\begin{barticle}
\bauthor{\bsnm{{Martinet}}, \binits{S.}},
\bauthor{\bsnm{{Meynet}}, \binits{G.}},
\bauthor{\bsnm{{Nandal}}, \binits{D.}},
\bauthor{\bsnm{{Ekstr{\"o}m}}, \binits{S.}},
\bauthor{\bsnm{{Georgy}}, \binits{C.}},
\bauthor{\bsnm{{Haemmerl{\'e}}}, \binits{L.}},
\bauthor{\bsnm{{Hirschi}}, \binits{R.}},
\bauthor{\bsnm{{Yusof}}, \binits{N.}},
\bauthor{\bsnm{{Gounelle}}, \binits{M.}},
\bauthor{\bsnm{{Dwarkadas}}, \binits{V.}}:
\batitle{{Very massive star winds as sources of the short-lived radioactive
  isotope $^{26}$Al}}.
\bjtitle{\aap}
\bvolume{664},
\bfpage{181}
(\byear{2022})
\doiurl{10.1051/0004-6361/202243474}
{\href{https://arxiv.org/abs/2205.15184}{{arXiv:2205.15184}}}
{[astro-ph.SR]}
\end{barticle}
\endbibitem

%%% 100
\bibitem[\protect\citeauthoryear{{Sana} et~al.}{2012}]{Sana2012}
\begin{barticle}
\bauthor{\bsnm{{Sana}}, \binits{H.}},
\bauthor{\bsnm{{de Mink}}, \binits{S.E.}},
\bauthor{\bsnm{{de Koter}}, \binits{A.}},
\bauthor{\bsnm{{Langer}}, \binits{N.}},
\bauthor{\bsnm{{Evans}}, \binits{C.J.}},
\bauthor{\bsnm{{Gieles}}, \binits{M.}},
\bauthor{\bsnm{{Gosset}}, \binits{E.}},
\bauthor{\bsnm{{Izzard}}, \binits{R.G.}},
\bauthor{\bsnm{{Le Bouquin}}, \binits{J.-B.}},
\bauthor{\bsnm{{Schneider}}, \binits{F.R.N.}}:
\batitle{{Binary Interaction Dominates the Evolution of Massive Stars}}.
\bjtitle{Science}
\bvolume{337}(\bissue{6093}),
\bfpage{444}
(\byear{2012})
\doiurl{10.1126/science.1223344}
{\href{https://arxiv.org/abs/1207.6397}{{arXiv:1207.6397}}}
{[astro-ph.SR]}
\end{barticle}
\endbibitem

%%% 101
\bibitem[\protect\citeauthoryear{{Brinkman} et~al.}{2021}]{Brinkman2021}
\begin{barticle}
\bauthor{\bsnm{{Brinkman}}, \binits{H.E.}},
\bauthor{\bsnm{{den Hartogh}}, \binits{J.W.}},
\bauthor{\bsnm{{Doherty}}, \binits{C.L.}},
\bauthor{\bsnm{{Pignatari}}, \binits{M.}},
\bauthor{\bsnm{{Lugaro}}, \binits{M.}}:
\batitle{{$^{26}$Aluminum from Massive Binary Stars. II. Rotating Single Stars
  Up to Core Collapse and Their Impact on the Early Solar System}}.
\bjtitle{\apj}
\bvolume{923}(\bissue{1}),
\bfpage{47}
(\byear{2021})
\doiurl{10.3847/1538-4357/ac25ea}
{\href{https://arxiv.org/abs/2109.05842}{{arXiv:2109.05842}}}
{[astro-ph.SR]}
\end{barticle}
\endbibitem

%%% 102
\bibitem[\protect\citeauthoryear{{Brinkman} et~al.}{2023}]{Brinkman2023}
\begin{barticle}
\bauthor{\bsnm{{Brinkman}}, \binits{H.E.}},
\bauthor{\bsnm{{Doherty}}, \binits{C.}},
\bauthor{\bsnm{{Pignatari}}, \binits{M.}},
\bauthor{\bsnm{{Pols}}, \binits{O.}},
\bauthor{\bsnm{{Lugaro}}, \binits{M.}}:
\batitle{{Aluminium-26 from Massive Binary Stars. III. Binary Stars up to Core
  Collapse and Their Impact on the Early Solar System}}.
\bjtitle{\apj}
\bvolume{951}(\bissue{2}),
\bfpage{110}
(\byear{2023})
\doiurl{10.3847/1538-4357/acd7ea}
{\href{https://arxiv.org/abs/2305.16787}{{arXiv:2305.16787}}}
{[astro-ph.SR]}
\end{barticle}
\endbibitem

%%% 103
\bibitem[\protect\citeauthoryear{{Paxton} et~al.}{2011}]{Paxton2011}
\begin{barticle}
\bauthor{\bsnm{{Paxton}}, \binits{B.}},
\bauthor{\bsnm{{Bildsten}}, \binits{L.}},
\bauthor{\bsnm{{Dotter}}, \binits{A.}},
\bauthor{\bsnm{{Herwig}}, \binits{F.}},
\bauthor{\bsnm{{Lesaffre}}, \binits{P.}},
\bauthor{\bsnm{{Timmes}}, \binits{F.}}:
\batitle{{Modules for Experiments in Stellar Astrophysics (MESA)}}.
\bjtitle{\apjs}
\bvolume{192}(\bissue{1}),
\bfpage{3}
(\byear{2011})
\doiurl{10.1088/0067-0049/192/1/3}
{\href{https://arxiv.org/abs/1009.1622}{{arXiv:1009.1622}}}
{[astro-ph.SR]}
\end{barticle}
\endbibitem

%%% 104
\bibitem[\protect\citeauthoryear{{Kami{\'n}ski} et~al.}{2018}]{Kaminski2018}
\begin{barticle}
\bauthor{\bsnm{{Kami{\'n}ski}}, \binits{T.}},
\bauthor{\bsnm{{Tylenda}}, \binits{R.}},
\bauthor{\bsnm{{Menten}}, \binits{K.M.}},
\bauthor{\bsnm{{Karakas}}, \binits{A.}},
\bauthor{\bsnm{{Winters}}, \binits{J.M.}},
\bauthor{\bsnm{{Breier}}, \binits{A.A.}},
\bauthor{\bsnm{{Wong}}, \binits{K.T.}},
\bauthor{\bsnm{{Giesen}}, \binits{T.F.}},
\bauthor{\bsnm{{Patel}}, \binits{N.A.}}:
\batitle{{Astronomical detection of radioactive molecule $^{26}$AlF in the
  remnant of an ancient explosion}}.
\bjtitle{Nature Astronomy}
\bvolume{2},
\bfpage{778}--\blpage{783}
(\byear{2018})
\doiurl{10.1038/s41550-018-0541-x}
{\href{https://arxiv.org/abs/1807.10647}{{arXiv:1807.10647}}}
{[astro-ph.SR]}
\end{barticle}
\endbibitem

%%% 105
\bibitem[\protect\citeauthoryear{{Kn{\"o}dlseder}}{1999}]{Knoedlseder1999c}
\begin{bchapter}
\bauthor{\bsnm{{Kn{\"o}dlseder}}, \binits{J.}}:
\bctitle{{$^{26}$Al Sources in the Galaxy as Seen in the 1. 809 MeV Gamma-Ray
  Line}}.
In: \beditor{\bsnm{{Diehl}}, \binits{R.}},
\beditor{\bsnm{{Hartmann}}, \binits{D.}} (eds.)
\bbtitle{Astronomy with Radioactivities},
vol. \bseriesno{274},
p. \bfpage{43}
(\byear{1999}).
\doiurl{10.48550/arXiv.astro-ph/9912132}
\end{bchapter}
\endbibitem

%%% 106
\bibitem[\protect\citeauthoryear{{Limongi} and {Chieffi}}{2003}]{Limongi2003}
\begin{barticle}
\bauthor{\bsnm{{Limongi}}, \binits{M.}},
\bauthor{\bsnm{{Chieffi}}, \binits{A.}}:
\batitle{{Evolution, Explosion, and Nucleosynthesis of Core-Collapse
  Supernovae}}.
\bjtitle{\apj}
\bvolume{592}(\bissue{1}),
\bfpage{404}--\blpage{433}
(\byear{2003})
\doiurl{10.1086/375703}
{\href{https://arxiv.org/abs/astro-ph/0304185}{{arXiv:astro-ph/0304185}}}
{[astro-ph]}
\end{barticle}
\endbibitem

%%% 107
\bibitem[\protect\citeauthoryear{{Pignatari} et~al.}{2016}]{Pignatari2016}
\begin{barticle}
\bauthor{\bsnm{{Pignatari}}, \binits{M.}},
\bauthor{\bsnm{{Herwig}}, \binits{F.}},
\bauthor{\bsnm{{Hirschi}}, \binits{R.}},
\bauthor{\bsnm{{Bennett}}, \binits{M.}},
\bauthor{\bsnm{{Rockefeller}}, \binits{G.}},
\bauthor{\bsnm{{Fryer}}, \binits{C.}},
\bauthor{\bsnm{{Timmes}}, \binits{F.X.}},
\bauthor{\bsnm{{Ritter}}, \binits{C.}},
\bauthor{\bsnm{{Heger}}, \binits{A.}},
\bauthor{\bsnm{{Jones}}, \binits{S.}},
\bauthor{\bsnm{{Battino}}, \binits{U.}},
\bauthor{\bsnm{{Dotter}}, \binits{A.}},
\bauthor{\bsnm{{Trappitsch}}, \binits{R.}},
\bauthor{\bsnm{{Diehl}}, \binits{S.}},
\bauthor{\bsnm{{Frischknecht}}, \binits{U.}},
\bauthor{\bsnm{{Hungerford}}, \binits{A.}},
\bauthor{\bsnm{{Magkotsios}}, \binits{G.}},
\bauthor{\bsnm{{Travaglio}}, \binits{C.}},
\bauthor{\bsnm{{Young}}, \binits{P.}}:
\batitle{{NuGrid Stellar Data Set. I.Stellar Yields from H to Bi for Stars with
  Metallicities Z = 0.02 and Z = 0.01}}.
\bjtitle{\apjs}
\bvolume{225}(\bissue{2}),
\bfpage{24}
(\byear{2016})
\doiurl{10.3847/0067-0049/225/2/24}
{\href{https://arxiv.org/abs/1307.6961}{{arXiv:1307.6961}}}
{[astro-ph.SR]}
\end{barticle}
\endbibitem

%%% 108
\bibitem[\protect\citeauthoryear{{Jones} et~al.}{2019}]{Jones2019b}
\begin{barticle}
\bauthor{\bsnm{{Jones}}, \binits{S.W.}},
\bauthor{\bsnm{{M{\"o}ller}}, \binits{H.}},
\bauthor{\bsnm{{Fryer}}, \binits{C.L.}},
\bauthor{\bsnm{{Fontes}}, \binits{C.J.}},
\bauthor{\bsnm{{Trappitsch}}, \binits{R.}},
\bauthor{\bsnm{{Even}}, \binits{W.P.}},
\bauthor{\bsnm{{Couture}}, \binits{A.}},
\bauthor{\bsnm{{Mumpower}}, \binits{M.R.}},
\bauthor{\bsnm{{Safi-Harb}}, \binits{S.}}:
\batitle{{$^{60}$Fe in core-collapse supernovae and prospects for X-ray and
  gamma-ray detection in supernova remnants}}.
\bjtitle{\mnras}
\bvolume{485}(\bissue{3}),
\bfpage{4287}--\blpage{4310}
(\byear{2019})
\doiurl{10.1093/mnras/stz536}
{\href{https://arxiv.org/abs/1902.05980}{{arXiv:1902.05980}}}
{[astro-ph.SR]}
\end{barticle}
\endbibitem

%%% 109
\bibitem[\protect\citeauthoryear{{Wanajo} et~al.}{2013}]{Wanajo2013}
\begin{barticle}
\bauthor{\bsnm{{Wanajo}}, \binits{S.}},
\bauthor{\bsnm{{Janka}}, \binits{H.-T.}},
\bauthor{\bsnm{{M{\"u}ller}}, \binits{B.}}:
\batitle{{Electron-capture Supernovae as Sources of $^{60}$Fe}}.
\bjtitle{\apjl}
\bvolume{774}(\bissue{1}),
\bfpage{6}
(\byear{2013})
\doiurl{10.1088/2041-8205/774/1/L6}
{\href{https://arxiv.org/abs/1307.3319}{{arXiv:1307.3319}}}
{[astro-ph.SR]}
\end{barticle}
\endbibitem

%%% 110
\bibitem[\protect\citeauthoryear{{Wanajo} et~al.}{2018}]{Wanajo2018}
\begin{barticle}
\bauthor{\bsnm{{Wanajo}}, \binits{S.}},
\bauthor{\bsnm{{M{\"u}ller}}, \binits{B.}},
\bauthor{\bsnm{{Janka}}, \binits{H.-T.}},
\bauthor{\bsnm{{Heger}}, \binits{A.}}:
\batitle{{Nucleosynthesis in the Innermost Ejecta of Neutrino-driven Supernova
  Explosions in Two Dimensions}}.
\bjtitle{\apj}
\bvolume{852}(\bissue{1}),
\bfpage{40}
(\byear{2018})
\doiurl{10.3847/1538-4357/aa9d97}
{\href{https://arxiv.org/abs/1701.06786}{{arXiv:1701.06786}}}
{[astro-ph.SR]}
\end{barticle}
\endbibitem

%%% 111
\bibitem[\protect\citeauthoryear{{Jones} et~al.}{2016}]{Jones2016}
\begin{barticle}
\bauthor{\bsnm{{Jones}}, \binits{S.}},
\bauthor{\bsnm{{R{\"o}pke}}, \binits{F.K.}},
\bauthor{\bsnm{{Pakmor}}, \binits{R.}},
\bauthor{\bsnm{{Seitenzahl}}, \binits{I.R.}},
\bauthor{\bsnm{{Ohlmann}}, \binits{S.T.}},
\bauthor{\bsnm{{Edelmann}}, \binits{P.V.F.}}:
\batitle{{Do electron-capture supernovae make neutron stars?. First
  multidimensional hydrodynamic simulations of the oxygen deflagration}}.
\bjtitle{\aap}
\bvolume{593},
\bfpage{72}
(\byear{2016})
\doiurl{10.1051/0004-6361/201628321}
{\href{https://arxiv.org/abs/1602.05771}{{arXiv:1602.05771}}}
{[astro-ph.SR]}
\end{barticle}
\endbibitem

%%% 112
\bibitem[\protect\citeauthoryear{{Jones} et~al.}{2019}]{Jones2019a}
\begin{barticle}
\bauthor{\bsnm{{Jones}}, \binits{S.}},
\bauthor{\bsnm{{R{\"o}pke}}, \binits{F.K.}},
\bauthor{\bsnm{{Fryer}}, \binits{C.}},
\bauthor{\bsnm{{Ruiter}}, \binits{A.J.}},
\bauthor{\bsnm{{Seitenzahl}}, \binits{I.R.}},
\bauthor{\bsnm{{Nittler}}, \binits{L.R.}},
\bauthor{\bsnm{{Ohlmann}}, \binits{S.T.}},
\bauthor{\bsnm{{Reifarth}}, \binits{R.}},
\bauthor{\bsnm{{Pignatari}}, \binits{M.}},
\bauthor{\bsnm{{Belczynski}}, \binits{K.}}:
\batitle{{Remnants and ejecta of thermonuclear electron-capture supernovae.
  Constraining oxygen-neon deflagrations in high-density white dwarfs}}.
\bjtitle{\aap}
\bvolume{622},
\bfpage{74}
(\byear{2019})
\doiurl{10.1051/0004-6361/201834381}
{\href{https://arxiv.org/abs/1812.08230}{{arXiv:1812.08230}}}
{[astro-ph.SR]}
\end{barticle}
\endbibitem

%%% 113
\bibitem[\protect\citeauthoryear{{Lugaro} et~al.}{2012}]{Lugaro2012}
\begin{barticle}
\bauthor{\bsnm{{Lugaro}}, \binits{M.}},
\bauthor{\bsnm{{Doherty}}, \binits{C.L.}},
\bauthor{\bsnm{{Karakas}}, \binits{A.I.}},
\bauthor{\bsnm{{Maddison}}, \binits{S.T.}},
\bauthor{\bsnm{{Liffman}}, \binits{K.}},
\bauthor{\bsnm{{Garc{\'\i}a-Hern{\'a}ndez}}, \binits{D.A.}},
\bauthor{\bsnm{{Siess}}, \binits{L.}},
\bauthor{\bsnm{{Lattanzio}}, \binits{J.C.}}:
\batitle{{Short-lived radioactivity in the early solar system: The Super-AGB
  star hypothesis}}.
\bjtitle{\maps}
\bvolume{47}(\bissue{12}),
\bfpage{1998}--\blpage{2012}
(\byear{2012})
\doiurl{10.1111/j.1945-5100.2012.01411.x}
{\href{https://arxiv.org/abs/1208.5816}{{arXiv:1208.5816}}}
{[astro-ph.SR]}
\end{barticle}
\endbibitem

%%% 114
\bibitem[\protect\citeauthoryear{{Woosley}}{1997}]{Woosley1997}
\begin{barticle}
\bauthor{\bsnm{{Woosley}}, \binits{S.E.}}:
\batitle{{Neutron-rich Nucleosynthesis in Carbon Deflagration Supernovae}}.
\bjtitle{\apj}
\bvolume{476}(\bissue{2}),
\bfpage{801}--\blpage{810}
(\byear{1997})
\doiurl{10.1086/303650}
\end{barticle}
\endbibitem

%%% 115
\bibitem[\protect\citeauthoryear{{Wallner} et~al.}{2015}]{Wallner2015}
\begin{barticle}
\bauthor{\bsnm{{Wallner}}, \binits{A.}},
\bauthor{\bsnm{{Bichler}}, \binits{M.}},
\bauthor{\bsnm{{Buczak}}, \binits{K.}},
\bauthor{\bsnm{{Dressler}}, \binits{R.}},
\bauthor{\bsnm{{Fifield}}, \binits{L.K.}},
\bauthor{\bsnm{{Schumann}}, \binits{D.}},
\bauthor{\bsnm{{Sterba}}, \binits{J.H.}},
\bauthor{\bsnm{{Tims}}, \binits{S.G.}},
\bauthor{\bsnm{{Wallner}}, \binits{G.}},
\bauthor{\bsnm{{Kutschera}}, \binits{W.}}:
\batitle{{Settling the Half-Life of $^{60}$Fe: Fundamental for a Versatile
  Astrophysical Chronometer}}.
\bjtitle{\prl}
\bvolume{114}(\bissue{4}),
\bfpage{041101}
(\byear{2015})
\doiurl{10.1103/PhysRevLett.114.041101}
\end{barticle}
\endbibitem

%%% 116
\bibitem[\protect\citeauthoryear{{Ostdiek}}{2016}]{Ostdiek2016}
\begin{botherref}
\oauthor{\bsnm{{Ostdiek}}, \binits{K.M.C.}}:
{Measurement of the half-life of $^{60}$Fe for stellar and early solar system
  models using the direct decay of $^{60m}$Co and accelerator mass
  spectrometry}.
PhD thesis,
University of Notre Dame, Indiana
(January 2016)
\end{botherref}
\endbibitem

%%% 117
\bibitem[\protect\citeauthoryear{{Knie} et~al.}{2004}]{Knie2004}
\begin{barticle}
\bauthor{\bsnm{{Knie}}, \binits{K.}},
\bauthor{\bsnm{{Korschinek}}, \binits{G.}},
\bauthor{\bsnm{{F{\"a}stermann}}, \binits{T.}},
\bauthor{\bsnm{{Dorfi}}, \binits{E.A.}},
\bauthor{\bsnm{{Rugel}}, \binits{G.}},
\bauthor{\bsnm{{Wallner}}, \binits{A.}}:
\batitle{{$^{60}$Fe Anomaly in a Deep-Sea Manganese Crust and Implications for
  a Nearby Supernova Source}}.
\bjtitle{\prl}
\bvolume{93}(\bissue{17}),
\bfpage{171103}
(\byear{2004})
\doiurl{10.1103/PhysRevLett.93.171103}
\end{barticle}
\endbibitem

%%% 118
\bibitem[\protect\citeauthoryear{{Wallner} et~al.}{2016}]{Wallner2016}
\begin{barticle}
\bauthor{\bsnm{{Wallner}}, \binits{A.}},
\bauthor{\bsnm{{Feige}}, \binits{J.}},
\bauthor{\bsnm{{Kinoshita}}, \binits{N.}},
\bauthor{\bsnm{{Paul}}, \binits{M.}},
\bauthor{\bsnm{{Fifield}}, \binits{L.K.}},
\bauthor{\bsnm{{Golser}}, \binits{R.}},
\bauthor{\bsnm{{Honda}}, \binits{M.}},
\bauthor{\bsnm{{Linnemann}}, \binits{U.}},
\bauthor{\bsnm{{Matsuzaki}}, \binits{H.}},
\bauthor{\bsnm{{Merchel}}, \binits{S.}},
\bauthor{\bsnm{{Rugel}}, \binits{G.}},
\bauthor{\bsnm{{Tims}}, \binits{S.G.}},
\bauthor{\bsnm{{Steier}}, \binits{P.}},
\bauthor{\bsnm{{Yamagata}}, \binits{T.}},
\bauthor{\bsnm{{Winkler}}, \binits{S.R.}}:
\batitle{{Recent near-Earth supernovae probed by global deposition of
  interstellar radioactive $^{60}$Fe}}.
\bjtitle{\nat}
\bvolume{532}(\bissue{7597}),
\bfpage{69}--\blpage{72}
(\byear{2016})
\doiurl{10.1038/nature17196}
\end{barticle}
\endbibitem

%%% 119
\bibitem[\protect\citeauthoryear{{Wallner} et~al.}{2021}]{Wallner2021}
\begin{barticle}
\bauthor{\bsnm{{Wallner}}, \binits{A.}},
\bauthor{\bsnm{{Fr{\"o}hlich}}, \binits{M.B.}},
\bauthor{\bsnm{{Hotchkis}}, \binits{M.A.C.}},
\bauthor{\bsnm{{Kinoshita}}, \binits{N.}},
\bauthor{\bsnm{{Paul}}, \binits{M.}},
\bauthor{\bsnm{{Martschini}}, \binits{M.}},
\bauthor{\bsnm{{Pavetich}}, \binits{S.}},
\bauthor{\bsnm{{Tims}}, \binits{S.G.}},
\bauthor{\bsnm{{Kivel}}, \binits{N.}},
\bauthor{\bsnm{{Schumann}}, \binits{D.}},
\bauthor{\bsnm{{Honda}}, \binits{M.}},
\bauthor{\bsnm{{Matsuzaki}}, \binits{H.}},
\bauthor{\bsnm{{Yamagata}}, \binits{T.}}:
\batitle{{$^{60}$Fe and $^{244}$Pu deposited on Earth constrain the r-process
  yields of recent nearby supernovae}}.
\bjtitle{Science}
\bvolume{372}(\bissue{6543}),
\bfpage{742}--\blpage{745}
(\byear{2021})
\doiurl{10.1126/science.aax3972}
\end{barticle}
\endbibitem

%%% 120
\bibitem[\protect\citeauthoryear{{Fimiani} et~al.}{2016}]{Fimiani2016}
\begin{barticle}
\bauthor{\bsnm{{Fimiani}}, \binits{L.}},
\bauthor{\bsnm{{Cook}}, \binits{D.L.}},
\bauthor{\bsnm{{F{\"a}stermann}}, \binits{T.}},
\bauthor{\bsnm{{G{\'o}mez-Guzm{\'a}n}}, \binits{J.M.}},
\bauthor{\bsnm{{Hain}}, \binits{K.}},
\bauthor{\bsnm{{Herzog}}, \binits{G.}},
\bauthor{\bsnm{{Knie}}, \binits{K.}},
\bauthor{\bsnm{{Korschinek}}, \binits{G.}},
\bauthor{\bsnm{{Ludwig}}, \binits{P.}},
\bauthor{\bsnm{{Park}}, \binits{J.}},
\bauthor{\bsnm{{Reedy}}, \binits{R.C.}},
\bauthor{\bsnm{{Rugel}}, \binits{G.}}:
\batitle{{Interstellar <mml:mmultiscripts>Fe 60 </mml:mmultiscripts> on the
  Surface of the Moon}}.
\bjtitle{\prl}
\bvolume{116}(\bissue{15}),
\bfpage{151104}
(\byear{2016})
\doiurl{10.1103/PhysRevLett.116.151104}
\end{barticle}
\endbibitem

%%% 121
\bibitem[\protect\citeauthoryear{{Binns} et~al.}{2016}]{Binns2016}
\begin{barticle}
\bauthor{\bsnm{{Binns}}, \binits{W.R.}},
\bauthor{\bsnm{{Israel}}, \binits{M.H.}},
\bauthor{\bsnm{{Christian}}, \binits{E.R.}},
\bauthor{\bsnm{{Cummings}}, \binits{A.C.}},
\bauthor{\bsnm{{de Nolfo}}, \binits{G.A.}},
\bauthor{\bsnm{{Lave}}, \binits{K.A.}},
\bauthor{\bsnm{{Leske}}, \binits{R.A.}},
\bauthor{\bsnm{{Mewaldt}}, \binits{R.A.}},
\bauthor{\bsnm{{Stone}}, \binits{E.C.}},
\bauthor{\bsnm{{von Rosenvinge}}, \binits{T.T.}},
\bauthor{\bsnm{{Wiedenbeck}}, \binits{M.E.}}:
\batitle{{Observation of the $^{60}$Fe nucleosynthesis-clock isotope in
  galactic cosmic rays}}.
\bjtitle{Science}
\bvolume{352}(\bissue{6286}),
\bfpage{677}--\blpage{680}
(\byear{2016})
\doiurl{10.1126/science.aad6004}
\end{barticle}
\endbibitem

%%% 122
\bibitem[\protect\citeauthoryear{{Wang} et~al.}{2007}]{Wang2007}
\begin{barticle}
\bauthor{\bsnm{{Wang}}, \binits{W.}},
\bauthor{\bsnm{{Harris}}, \binits{M.J.}},
\bauthor{\bsnm{{Diehl}}, \binits{R.}},
\bauthor{\bsnm{{Halloin}}, \binits{H.}},
\bauthor{\bsnm{{Cordier}}, \binits{B.}},
\bauthor{\bsnm{{Strong}}, \binits{A.W.}},
\bauthor{\bsnm{{Kretschmer}}, \binits{K.}},
\bauthor{\bsnm{{Kn{\"o}dlseder}}, \binits{J.}},
\bauthor{\bsnm{{Jean}}, \binits{P.}},
\bauthor{\bsnm{{Lichti}}, \binits{G.G.}},
\bauthor{\bsnm{{Roques}}, \binits{J.P.}},
\bauthor{\bsnm{{Schanne}}, \binits{S.}},
\bauthor{\bsnm{{von Kienlin}}, \binits{A.}},
\bauthor{\bsnm{{Weidenspointner}}, \binits{G.}},
\bauthor{\bsnm{{Wunderer}}, \binits{C.}}:
\batitle{{SPI observations of the diffuse $^{60}$Fe emission in the Galaxy}}.
\bjtitle{\aap}
\bvolume{469}(\bissue{3}),
\bfpage{1005}--\blpage{1012}
(\byear{2007})
\doiurl{10.1051/0004-6361:20066982}
{\href{https://arxiv.org/abs/0704.3895}{{arXiv:0704.3895}}}
{[astro-ph]}
\end{barticle}
\endbibitem

%%% 123
\bibitem[\protect\citeauthoryear{{Wang} et~al.}{2020}]{Wang2020}
\begin{barticle}
\bauthor{\bsnm{{Wang}}, \binits{W.}},
\bauthor{\bsnm{{Siegert}}, \binits{T.}},
\bauthor{\bsnm{{Dai}}, \binits{Z.G.}},
\bauthor{\bsnm{{Diehl}}, \binits{R.}},
\bauthor{\bsnm{{Greiner}}, \binits{J.}},
\bauthor{\bsnm{{Heger}}, \binits{A.}},
\bauthor{\bsnm{{Krause}}, \binits{M.}},
\bauthor{\bsnm{{Lang}}, \binits{M.}},
\bauthor{\bsnm{{Pleintinger}}, \binits{M.M.M.}},
\bauthor{\bsnm{{Zhang}}, \binits{X.L.}}:
\batitle{{Gamma-Ray Emission of $^{60}$Fe and $^{26}$Al Radioactivity in Our
  Galaxy}}.
\bjtitle{\apj}
\bvolume{889}(\bissue{2}),
\bfpage{169}
(\byear{2020})
\doiurl{10.3847/1538-4357/ab6336}
{\href{https://arxiv.org/abs/1912.07874}{{arXiv:1912.07874}}}
{[astro-ph.HE]}
\end{barticle}
\endbibitem

%%% 124
\bibitem[\protect\citeauthoryear{{Leising} and {Share}}{1994}]{Leising1994}
\begin{barticle}
\bauthor{\bsnm{{Leising}}, \binits{M.D.}},
\bauthor{\bsnm{{Share}}, \binits{G.H.}}:
\batitle{{Gamma-Ray Limits on Galactic 60Fe and 44Ti Nucleosynthesis}}.
\bjtitle{\apj}
\bvolume{424},
\bfpage{200}
(\byear{1994})
\doiurl{10.1086/173883}
\end{barticle}
\endbibitem

%%% 125
\bibitem[\protect\citeauthoryear{{Harris} et~al.}{1997}]{Harris1997}
\begin{bchapter}
\bauthor{\bsnm{{Harris}}, \binits{M.J.}},
\bauthor{\bsnm{{Purcell}}, \binits{W.R.}},
\bauthor{\bsnm{{McNaron-Brown}}, \binits{K.}},
\bauthor{\bsnm{{Murphy}}, \binits{R.J.}},
\bauthor{\bsnm{{Grove}}, \binits{J.E.}},
\bauthor{\bsnm{{Johnson}}, \binits{W.N.}},
\bauthor{\bsnm{{Kinzer}}, \binits{R.L.}},
\bauthor{\bsnm{{Kurfess}}, \binits{J.D.}},
\bauthor{\bsnm{{Share}}, \binits{G.H.}},
\bauthor{\bsnm{{Jung}}, \binits{G.V.}}:
\bctitle{{OSSE results on Galactic {\ensuremath{\gamma}}-ray line emission}}.
In: \beditor{\bsnm{{Dermer}}, \binits{C.D.}},
\beditor{\bsnm{{Strickman}}, \binits{M.S.}},
\beditor{\bsnm{{Kurfess}}, \binits{J.D.}} (eds.)
\bbtitle{Proceedings of the Fourth Compton Symposium}.
\bsertitle{American Institute of Physics Conference Series},
vol. \bseriesno{410},
pp. \bfpage{1079}--\blpage{1083}.
\bpublisher{AIP}, \blocation{???}
(\byear{1997}).
\doiurl{10.1063/1.54086}
\end{bchapter}
\endbibitem

%%% 126
\bibitem[\protect\citeauthoryear{{Diehl} et~al.}{1997}]{Diehl1997}
\begin{bchapter}
\bauthor{\bsnm{{Diehl}}, \binits{R.}},
\bauthor{\bsnm{{Wessolowski}}, \binits{U.}},
\bauthor{\bsnm{{Oberlack}}, \binits{U.}},
\bauthor{\bsnm{{Bloemen}}, \binits{H.}},
\bauthor{\bsnm{{Georgii}}, \binits{R.}},
\bauthor{\bsnm{{Iyudin}}, \binits{A.}},
\bauthor{\bsnm{{Kn{\"o}dlseder}}, \binits{J.}},
\bauthor{\bsnm{{Lichti}}, \binits{G.}},
\bauthor{\bsnm{{Hermsen}}, \binits{W.}},
\bauthor{\bsnm{{Morris}}, \binits{D.}},
\bauthor{\bsnm{{Ryan}}, \binits{J.}},
\bauthor{\bsnm{{Sch{\"o}nfelder}}, \binits{V.}},
\bauthor{\bsnm{{Strong}}, \binits{A.}},
\bauthor{\bsnm{{von Ballmoos}}, \binits{P.}},
\bauthor{\bsnm{{Winkler}}, \binits{C.}}:
\bctitle{{$^{26}$Al and the COMPTEL $^{60}$Fe data}}.
In: \beditor{\bsnm{{Dermer}}, \binits{C.D.}},
\beditor{\bsnm{{Strickman}}, \binits{M.S.}},
\beditor{\bsnm{{Kurfess}}, \binits{J.D.}} (eds.)
\bbtitle{Proceedings of the Fourth Compton Symposium}.
\bsertitle{American Institute of Physics Conference Series},
vol. \bseriesno{410},
pp. \bfpage{1109}--\blpage{1113}.
\bpublisher{AIP}, \blocation{???}
(\byear{1997}).
\doiurl{10.1063/1.54176}
\end{bchapter}
\endbibitem

%%% 127
\bibitem[\protect\citeauthoryear{{Smith}}{2004}]{Smith2004}
\begin{bchapter}
\bauthor{\bsnm{{Smith}}, \binits{D.M.}}:
\bctitle{{Gamma-Ray Line Observations with RHESSI}}.
In: \beditor{\bsnm{{Sch{\"o}nfelder}}, \binits{V.}},
\beditor{\bsnm{{Lichti}}, \binits{G.}},
\beditor{\bsnm{{Winkler}}, \binits{C.}} (eds.)
\bbtitle{5th INTEGRAL Workshop on the INTEGRAL Universe}.
\bsertitle{ESA Special Publication},
vol. \bseriesno{552},
p. \bfpage{45}
(\byear{2004}).
\doiurl{10.48550/arXiv.astro-ph/0404594}
\end{bchapter}
\endbibitem

%%% 128
\bibitem[\protect\citeauthoryear{{Pelgrims} et~al.}{2020}]{Pelgrims2020}
\begin{barticle}
\bauthor{\bsnm{{Pelgrims}}, \binits{V.}},
\bauthor{\bsnm{{Ferri{\`e}re}}, \binits{K.}},
\bauthor{\bsnm{{Boulanger}}, \binits{F.}},
\bauthor{\bsnm{{Lallement}}, \binits{R.}},
\bauthor{\bsnm{{Montier}}, \binits{L.}}:
\batitle{{Modeling the magnetized Local Bubble from dust data}}.
\bjtitle{\aap}
\bvolume{636},
\bfpage{17}
(\byear{2020})
\doiurl{10.1051/0004-6361/201937157}
{\href{https://arxiv.org/abs/1911.09691}{{arXiv:1911.09691}}}
{[astro-ph.GA]}
\end{barticle}
\endbibitem

%%% 129
\bibitem[\protect\citeauthoryear{{Zucker} et~al.}{2022}]{Zucker2022}
\begin{barticle}
\bauthor{\bsnm{{Zucker}}, \binits{C.}},
\bauthor{\bsnm{{Goodman}}, \binits{A.A.}},
\bauthor{\bsnm{{Alves}}, \binits{J.}},
\bauthor{\bsnm{{Bialy}}, \binits{S.}},
\bauthor{\bsnm{{Foley}}, \binits{M.}},
\bauthor{\bsnm{{Speagle}}, \binits{J.S.}},
\bauthor{\bsnm{{Gro{\^I}{\texttwosuperior}schedl}}, \binits{J.}},
\bauthor{\bsnm{{Finkbeiner}}, \binits{D.P.}},
\bauthor{\bsnm{{Burkert}}, \binits{A.}},
\bauthor{\bsnm{{Khimey}}, \binits{D.}},
\bauthor{\bsnm{{Swiggum}}, \binits{C.}}:
\batitle{{Star formation near the Sun is driven by expansion of the Local
  Bubble}}.
\bjtitle{\nat}
\bvolume{601}(\bissue{7893}),
\bfpage{334}--\blpage{337}
(\byear{2022})
\doiurl{10.1038/s41586-021-04286-5}
{\href{https://arxiv.org/abs/2201.05124}{{arXiv:2201.05124}}}
{[astro-ph.GA]}
\end{barticle}
\endbibitem

%%% 130
\bibitem[\protect\citeauthoryear{{Timmes} et~al.}{1995}]{Timmes1995}
\begin{barticle}
\bauthor{\bsnm{{Timmes}}, \binits{F.X.}},
\bauthor{\bsnm{{Woosley}}, \binits{S.E.}},
\bauthor{\bsnm{{Hartmann}}, \binits{D.H.}},
\bauthor{\bsnm{{Hoffman}}, \binits{R.D.}},
\bauthor{\bsnm{{Weaver}}, \binits{T.A.}},
\bauthor{\bsnm{{Matteucci}}, \binits{F.}}:
\batitle{{26Al and 60Fe from Supernova Explosions}}.
\bjtitle{\apj}
\bvolume{449},
\bfpage{204}
(\byear{1995})
\doiurl{10.1086/176046}
{\href{https://arxiv.org/abs/astro-ph/9503120}{{arXiv:astro-ph/9503120}}}
{[astro-ph]}
\end{barticle}
\endbibitem

%%% 131
\bibitem[\protect\citeauthoryear{{Rauscher} et~al.}{2002}]{Rauscher2002}
\begin{barticle}
\bauthor{\bsnm{{Rauscher}}, \binits{T.}},
\bauthor{\bsnm{{Heger}}, \binits{A.}},
\bauthor{\bsnm{{Hoffman}}, \binits{R.D.}},
\bauthor{\bsnm{{Woosley}}, \binits{S.E.}}:
\batitle{{Nucleosynthesis in Massive Stars with Improved Nuclear and Stellar
  Physics}}.
\bjtitle{\apj}
\bvolume{576}(\bissue{1}),
\bfpage{323}--\blpage{348}
(\byear{2002})
\doiurl{10.1086/341728}
{\href{https://arxiv.org/abs/astro-ph/0112478}{{arXiv:astro-ph/0112478}}}
{[astro-ph]}
\end{barticle}
\endbibitem

%%% 132
\bibitem[\protect\citeauthoryear{{Prantzos}}{2004}]{Prantzos2004}
\begin{barticle}
\bauthor{\bsnm{{Prantzos}}, \binits{N.}}:
\batitle{{Radioactive $^{26}$Al and $^{60}$Fe in the Milky Way: Implications of
  the RHESSI detection of $^{60}$Fe}}.
\bjtitle{\aap}
\bvolume{420},
\bfpage{1033}--\blpage{1037}
(\byear{2004})
\doiurl{10.1051/0004-6361:20035766}
{\href{https://arxiv.org/abs/astro-ph/0402198}{{arXiv:astro-ph/0402198}}}
{[astro-ph]}
\end{barticle}
\endbibitem

%%% 133
\bibitem[\protect\citeauthoryear{{Woosley} and {Heger}}{2007}]{Woosley2007}
\begin{barticle}
\bauthor{\bsnm{{Woosley}}, \binits{S.E.}},
\bauthor{\bsnm{{Heger}}, \binits{A.}}:
\batitle{{Nucleosynthesis and remnants in massive stars of solar metallicity}}.
\bjtitle{\physrep}
\bvolume{442}(\bissue{1-6}),
\bfpage{269}--\blpage{283}
(\byear{2007})
\doiurl{10.1016/j.physrep.2007.02.009}
{\href{https://arxiv.org/abs/astro-ph/0702176}{{arXiv:astro-ph/0702176}}}
{[astro-ph]}
\end{barticle}
\endbibitem

%%% 134
\bibitem[\protect\citeauthoryear{{Tur} et~al.}{2010}]{Tur2010}
\begin{barticle}
\bauthor{\bsnm{{Tur}}, \binits{C.}},
\bauthor{\bsnm{{Heger}}, \binits{A.}},
\bauthor{\bsnm{{Austin}}, \binits{S.M.}}:
\batitle{{Production of $^{26}$Al, $^{44}$Ti, and $^{60}$Fe in Core-collapse
  Supernovae: Sensitivity to the Rates of the Triple Alpha and
  $^{12}$C({\ensuremath{\alpha}}, {\ensuremath{\gamma}})$^{16}$O Reactions}}.
\bjtitle{\apj}
\bvolume{718}(\bissue{1}),
\bfpage{357}--\blpage{367}
(\byear{2010})
\doiurl{10.1088/0004-637X/718/1/357}
{\href{https://arxiv.org/abs/0908.4283}{{arXiv:0908.4283}}}
{[astro-ph.SR]}
\end{barticle}
\endbibitem

%%% 135
\bibitem[\protect\citeauthoryear{{Austin} et~al.}{2017}]{Austin2017}
\begin{barticle}
\bauthor{\bsnm{{Austin}}, \binits{S.M.}},
\bauthor{\bsnm{{West}}, \binits{C.}},
\bauthor{\bsnm{{Heger}}, \binits{A.}}:
\batitle{{Reducing Uncertainties in the Production of the Gamma-emitting Nuclei
  $^{26}$Al, $^{44}$Ti, and $^{60}$Fe in Core-collapse Supernovae by Using
  Effective Helium Burning Rates}}.
\bjtitle{\apjl}
\bvolume{839}(\bissue{1}),
\bfpage{9}
(\byear{2017})
\doiurl{10.3847/2041-8213/aa68e7}
{\href{https://arxiv.org/abs/1704.01240}{{arXiv:1704.01240}}}
{[astro-ph.SR]}
\end{barticle}
\endbibitem

%%% 136
\bibitem[\protect\citeauthoryear{{Bouchet} et~al.}{2011}]{Bouchet2011}
\begin{barticle}
\bauthor{\bsnm{{Bouchet}}, \binits{L.}},
\bauthor{\bsnm{{Strong}}, \binits{A.W.}},
\bauthor{\bsnm{{Porter}}, \binits{T.A.}},
\bauthor{\bsnm{{Moskalenko}}, \binits{I.V.}},
\bauthor{\bsnm{{Jourdain}}, \binits{E.}},
\bauthor{\bsnm{{Roques}}, \binits{J.-P.}}:
\batitle{{Diffuse Emission Measurement with the SPectrometer on INTEGRAL as an
  Indirect Probe of Cosmic-Ray Electrons and Positrons}}.
\bjtitle{\apj}
\bvolume{739}(\bissue{1}),
\bfpage{29}
(\byear{2011})
\doiurl{10.1088/0004-637X/739/1/29}
{\href{https://arxiv.org/abs/1107.0200}{{arXiv:1107.0200}}}
{[astro-ph.HE]}
\end{barticle}
\endbibitem

%%% 137
\bibitem[\protect\citeauthoryear{{Leising} and {Clayton}}{1985}]{Leising1985}
\begin{barticle}
\bauthor{\bsnm{{Leising}}, \binits{M.D.}},
\bauthor{\bsnm{{Clayton}}, \binits{D.D.}}:
\batitle{{Angular distribution of interstellar Al-26}}.
\bjtitle{\apj}
\bvolume{294},
\bfpage{591}--\blpage{598}
(\byear{1985})
\doiurl{10.1086/163326}
\end{barticle}
\endbibitem

%%% 138
\bibitem[\protect\citeauthoryear{{Jos{\'e}} et~al.}{1997}]{Jose1997}
\begin{barticle}
\bauthor{\bsnm{{Jos{\'e}}}, \binits{J.}},
\bauthor{\bsnm{{Hernanz}}, \binits{M.}},
\bauthor{\bsnm{{Coc}}, \binits{A.}}:
\batitle{{New Results on $^{26}$Al Production in Classical Novae}}.
\bjtitle{\apjl}
\bvolume{479}(\bissue{1}),
\bfpage{55}--\blpage{58}
(\byear{1997})
\doiurl{10.1086/310575}
{\href{https://arxiv.org/abs/astro-ph/9701181}{{arXiv:astro-ph/9701181}}}
{[astro-ph]}
\end{barticle}
\endbibitem

%%% 139
\bibitem[\protect\citeauthoryear{{Bennett} et~al.}{2013}]{Bennett2013}
\begin{barticle}
\bauthor{\bsnm{{Bennett}}, \binits{M.B.}},
\bauthor{\bsnm{{Wrede}}, \binits{C.}},
\bauthor{\bsnm{{Chipps}}, \binits{K.A.}},
\bauthor{\bsnm{{Jos{\'e}}}, \binits{J.}},
\bauthor{\bsnm{{Liddick}}, \binits{S.N.}},
\bauthor{\bsnm{{Santia}}, \binits{M.}},
\bauthor{\bsnm{{Bowe}}, \binits{A.}},
\bauthor{\bsnm{{Chen}}, \binits{A.A.}},
\bauthor{\bsnm{{Cooper}}, \binits{N.}},
\bauthor{\bsnm{{Irvine}}, \binits{D.}},
\bauthor{\bsnm{{McNeice}}, \binits{E.}},
\bauthor{\bsnm{{Montes}}, \binits{F.}},
\bauthor{\bsnm{{Naqvi}}, \binits{F.}},
\bauthor{\bsnm{{Ortez}}, \binits{R.}},
\bauthor{\bsnm{{Pain}}, \binits{S.D.}},
\bauthor{\bsnm{{Pereira}}, \binits{J.}},
\bauthor{\bsnm{{Prokop}}, \binits{C.}},
\bauthor{\bsnm{{Quaglia}}, \binits{J.}},
\bauthor{\bsnm{{Quinn}}, \binits{S.J.}},
\bauthor{\bsnm{{Schwartz}}, \binits{S.B.}},
\bauthor{\bsnm{{Shanab}}, \binits{S.}},
\bauthor{\bsnm{{Simon}}, \binits{A.}},
\bauthor{\bsnm{{Spyrou}}, \binits{A.}},
\bauthor{\bsnm{{Thiagalingam}}, \binits{E.}}:
\batitle{{Classical-Nova Contribution to the Milky Way's Al26 Abundance: Exit
  Channel of the Key Al25(p,{\ensuremath{\gamma}})Si26 Resonance}}.
\bjtitle{\prl}
\bvolume{111}(\bissue{23}),
\bfpage{232503}
(\byear{2013})
\doiurl{10.1103/PhysRevLett.111.232503}
{\href{https://arxiv.org/abs/1312.3668}{{arXiv:1312.3668}}}
{[nucl-ex]}
\end{barticle}
\endbibitem

%%% 140
\bibitem[\protect\citeauthoryear{{Clayton} and {Hoyle}}{1974}]{Clayton1974}
\begin{barticle}
\bauthor{\bsnm{{Clayton}}, \binits{D.D.}},
\bauthor{\bsnm{{Hoyle}}, \binits{F.}}:
\batitle{{Gamma-Ray Lines from Novae}}.
\bjtitle{\apjl}
\bvolume{187},
\bfpage{101}
(\byear{1974})
\doiurl{10.1086/181406}
\end{barticle}
\endbibitem

%%% 141
\bibitem[\protect\citeauthoryear{{Clayton}}{1981}]{Clayton1981}
\begin{barticle}
\bauthor{\bsnm{{Clayton}}, \binits{D.D.}}:
\batitle{{Li-7 gamma-ray lines from novae}}.
\bjtitle{\apjl}
\bvolume{244},
\bfpage{97}
(\byear{1981})
\doiurl{10.1086/183488}
\end{barticle}
\endbibitem

%%% 142
\bibitem[\protect\citeauthoryear{{Hernanz}}{2014}]{Hernanz2014}
\begin{bchapter}
\bauthor{\bsnm{{Hernanz}}, \binits{M.}}:
\bctitle{{Gamma-ray Emission from Nova Outbursts}}.
In: \beditor{\bsnm{{Woudt}}, \binits{P.A.}},
\beditor{\bsnm{{Ribeiro}}, \binits{V.A.R.M.}} (eds.)
\bbtitle{Stellar Novae: Past and Future Decades}.
\bsertitle{Astronomical Society of the Pacific Conference Series},
vol. \bseriesno{490},
p. \bfpage{319}
(\byear{2014}).
\doiurl{10.48550/arXiv.1305.0769}
\end{bchapter}
\endbibitem

%%% 143
\bibitem[\protect\citeauthoryear{{Hernanz} and {Jos{\'e}}}{2006}]{Hernanz2006}
\begin{barticle}
\bauthor{\bsnm{{Hernanz}}, \binits{M.}},
\bauthor{\bsnm{{Jos{\'e}}}, \binits{J.}}:
\batitle{{Radioactivities from novae}}.
\bjtitle{\nar}
\bvolume{50}(\bissue{7-8}),
\bfpage{504}--\blpage{508}
(\byear{2006})
\doiurl{10.1016/j.newar.2006.06.012}
\end{barticle}
\endbibitem

%%% 144
\bibitem[\protect\citeauthoryear{{Leising} and {Clayton}}{1987}]{Leising1987}
\begin{barticle}
\bauthor{\bsnm{{Leising}}, \binits{M.D.}},
\bauthor{\bsnm{{Clayton}}, \binits{D.D.}}:
\batitle{{Positron Annihilation Gamma Rays from Novae}}.
\bjtitle{\apj}
\bvolume{323},
\bfpage{159}
(\byear{1987})
\doiurl{10.1086/165816}
\end{barticle}
\endbibitem

%%% 145
\bibitem[\protect\citeauthoryear{{Gomez-Gomar} et~al.}{1998}]{Gomez-Gomar1998}
\begin{barticle}
\bauthor{\bsnm{{Gomez-Gomar}}, \binits{J.}},
\bauthor{\bsnm{{Hernanz}}, \binits{M.}},
\bauthor{\bsnm{{Jose}}, \binits{J.}},
\bauthor{\bsnm{{Isern}}, \binits{J.}}:
\batitle{{Gamma-ray emission from individual classical novae}}.
\bjtitle{\mnras}
\bvolume{296}(\bissue{4}),
\bfpage{913}--\blpage{920}
(\byear{1998})
\doiurl{10.1046/j.1365-8711.1998.01421.x}
{\href{https://arxiv.org/abs/astro-ph/9711322}{{arXiv:astro-ph/9711322}}}
{[astro-ph]}
\end{barticle}
\endbibitem

%%% 146
\bibitem[\protect\citeauthoryear{{Hernanz} et~al.}{1999}]{Hernanz1999a}
\begin{barticle}
\bauthor{\bsnm{{Hernanz}}, \binits{M.}},
\bauthor{\bsnm{{Jos{\'e}}}, \binits{J.}},
\bauthor{\bsnm{{Coc}}, \binits{A.}},
\bauthor{\bsnm{{G{\'o}mez-Gomar}}, \binits{J.}},
\bauthor{\bsnm{{Isern}}, \binits{J.}}:
\batitle{{Gamma-Ray Emission from Novae Related to Positron Annihilation:
  Constraints on its Observability Posed by New Experimental Nuclear Data}}.
\bjtitle{\apjl}
\bvolume{526}(\bissue{2}),
\bfpage{97}--\blpage{100}
(\byear{1999})
\doiurl{10.1086/312372}
{\href{https://arxiv.org/abs/astro-ph/9910111}{{arXiv:astro-ph/9910111}}}
{[astro-ph]}
\end{barticle}
\endbibitem

%%% 147
\bibitem[\protect\citeauthoryear{{Leung} and {Siegert}}{2022}]{Leung2022}
\begin{barticle}
\bauthor{\bsnm{{Leung}}, \binits{S.-C.}},
\bauthor{\bsnm{{Siegert}}, \binits{T.}}:
\batitle{{Gamma-ray light curves and spectra of classical novae}}.
\bjtitle{\mnras}
\bvolume{516}(\bissue{1}),
\bfpage{1008}--\blpage{1021}
(\byear{2022})
\doiurl{10.1093/mnras/stac1672}
{\href{https://arxiv.org/abs/2112.06893}{{arXiv:2112.06893}}}
{[astro-ph.HE]}
\end{barticle}
\endbibitem

%%% 148
\bibitem[\protect\citeauthoryear{{Leising} et~al.}{1988}]{Leising1988}
\begin{barticle}
\bauthor{\bsnm{{Leising}}, \binits{M.D.}},
\bauthor{\bsnm{{Share}}, \binits{G.H.}},
\bauthor{\bsnm{{Chupp}}, \binits{E.L.}},
\bauthor{\bsnm{{Kanbach}}, \binits{G.}}:
\batitle{{Gamma-Ray Limits on 22Na Production in Novae}}.
\bjtitle{\apj}
\bvolume{328},
\bfpage{755}
(\byear{1988})
\doiurl{10.1086/166334}
\end{barticle}
\endbibitem

%%% 149
\bibitem[\protect\citeauthoryear{{Iyudin} et~al.}{1995}]{Iyudin1995}
\begin{barticle}
\bauthor{\bsnm{{Iyudin}}, \binits{A.F.}},
\bauthor{\bsnm{{Bennett}}, \binits{K.}},
\bauthor{\bsnm{{Bloemen}}, \binits{H.}},
\bauthor{\bsnm{{Diehl}}, \binits{R.}},
\bauthor{\bsnm{{Hermsen}}, \binits{W.}},
\bauthor{\bsnm{{Lichti}}, \binits{G.G.}},
\bauthor{\bsnm{{Morris}}, \binits{D.}},
\bauthor{\bsnm{{Ryan}}, \binits{J.}},
\bauthor{\bsnm{{Sch{\"o}nfelder}}, \binits{V.}},
\bauthor{\bsnm{{Steinle}}, \binits{H.}},
\bauthor{\bsnm{{Strong}}, \binits{A.}},
\bauthor{\bsnm{{Varendorff}}, \binits{M.}},
\bauthor{\bsnm{{Winkler}}, \binits{C.}}:
\batitle{{COMPTEL search for \^22\^Na line emission from recent novae.}}
\bjtitle{\aap}
\bvolume{300},
\bfpage{422}
(\byear{1995})
\end{barticle}
\endbibitem

%%% 150
\bibitem[\protect\citeauthoryear{{Siegert} et~al.}{2021}]{Siegert2021}
\begin{barticle}
\bauthor{\bsnm{{Siegert}}, \binits{T.}},
\bauthor{\bsnm{{Ghosh}}, \binits{S.}},
\bauthor{\bsnm{{Mathur}}, \binits{K.}},
\bauthor{\bsnm{{Spraggon}}, \binits{E.}},
\bauthor{\bsnm{{Yeddanapudi}}, \binits{A.}}:
\batitle{{Nucleosynthesis constraints through {\ensuremath{\gamma}}-ray line
  measurements from classical novae. Hierarchical model for the ejecta of
  $^{22}$Na and $^{7}$Be}}.
\bjtitle{\aap}
\bvolume{650},
\bfpage{187}
(\byear{2021})
\doiurl{10.1051/0004-6361/202140300}
\end{barticle}
\endbibitem

%%% 151
\bibitem[\protect\citeauthoryear{{Jean} et~al.}{2000}]{Jean2000}
\begin{barticle}
\bauthor{\bsnm{{Jean}}, \binits{P.}},
\bauthor{\bsnm{{Hernanz}}, \binits{M.}},
\bauthor{\bsnm{{G{\'o}mez-Gomar}}, \binits{J.}},
\bauthor{\bsnm{{Jos{\'e}}}, \binits{J.}}:
\batitle{{Galactic 1.275-MeV emission from ONe novae and its detectability by
  INTEGRAL/SPI}}.
\bjtitle{\mnras}
\bvolume{319}(\bissue{2}),
\bfpage{350}--\blpage{364}
(\byear{2000})
\doiurl{10.1046/j.1365-8711.2000.03587.x}
{\href{https://arxiv.org/abs/astro-ph/0004126}{{arXiv:astro-ph/0004126}}}
{[astro-ph]}
\end{barticle}
\endbibitem

%%% 152
\bibitem[\protect\citeauthoryear{{Shafter}}{2017}]{Shafter2017}
\begin{barticle}
\bauthor{\bsnm{{Shafter}}, \binits{A.W.}}:
\batitle{{The Galactic Nova Rate Revisited}}.
\bjtitle{\apj}
\bvolume{834}(\bissue{2}),
\bfpage{196}
(\byear{2017})
\doiurl{10.3847/1538-4357/834/2/196}
{\href{https://arxiv.org/abs/1606.02358}{{arXiv:1606.02358}}}
{[astro-ph.SR]}
\end{barticle}
\endbibitem

%%% 153
\bibitem[\protect\citeauthoryear{{Higdon} and {Fowler}}{1987}]{Higdon1987}
\begin{barticle}
\bauthor{\bsnm{{Higdon}}, \binits{J.C.}},
\bauthor{\bsnm{{Fowler}}, \binits{W.A.}}:
\batitle{{Gamma-Ray Constraints on 22Na Yields in Nova Explosions}}.
\bjtitle{\apj}
\bvolume{317},
\bfpage{710}
(\byear{1987})
\doiurl{10.1086/165317}
\end{barticle}
\endbibitem

%%% 154
\bibitem[\protect\citeauthoryear{{Mahoney} et~al.}{1982}]{Mahoney1982}
\begin{barticle}
\bauthor{\bsnm{{Mahoney}}, \binits{W.A.}},
\bauthor{\bsnm{{Ling}}, \binits{J.C.}},
\bauthor{\bsnm{{Jacobson}}, \binits{A.S.}},
\bauthor{\bsnm{{Lingenfelter}}, \binits{R.E.}}:
\batitle{{Diffuse galactic gamma-ray line emission from nucleosynthetic Fe-60,
  Al-26, and Na-22 - Preliminary limits from HEAO 3.}}
\bjtitle{\apj}
\bvolume{262},
\bfpage{742}--\blpage{748}
(\byear{1982})
\doiurl{10.1086/160469}
\end{barticle}
\endbibitem

%%% 155
\bibitem[\protect\citeauthoryear{{Jean} et~al.}{2001}]{Jean2001}
\begin{bchapter}
\bauthor{\bsnm{{Jean}}, \binits{P.}},
\bauthor{\bsnm{{Kn{\"o}dlseder}}, \binits{J.}},
\bauthor{\bsnm{{von Ballmoos}}, \binits{P.}},
\bauthor{\bsnm{{G{\'o}mez-Gomar}}, \binits{J.}},
\bauthor{\bsnm{{Hernanz}}, \binits{M.}},
\bauthor{\bsnm{{Jos{\'e}}}, \binits{J.}}:
\bctitle{{Upper limits of the $^{22}$Na yield from O-Ne nova}}.
In: \beditor{\bsnm{{Gimenez}}, \binits{A.}},
\beditor{\bsnm{{Reglero}}, \binits{V.}},
\beditor{\bsnm{{Winkler}}, \binits{C.}} (eds.)
\bbtitle{Exploring the Gamma-Ray Universe}.
\bsertitle{ESA Special Publication},
vol. \bseriesno{459},
pp. \bfpage{73}--\blpage{77}
(\byear{2001}).
\doiurl{10.48550/arXiv.astro-ph/0106340}
\end{bchapter}
\endbibitem

%%% 156
\bibitem[\protect\citeauthoryear{{Foug{\`e}res} et~al.}{2023}]{Fougeres2023}
\begin{barticle}
\bauthor{\bsnm{{Foug{\`e}res}}, \binits{C.}},
\bauthor{\bsnm{{de Oliveira Santos}}, \binits{F.}},
\bauthor{\bsnm{{Jos{\'e}}}, \binits{J.}},
\bauthor{\bsnm{{Michelagnoli}}, \binits{C.}},
\bauthor{\bsnm{{Cl{\'e}ment}}, \binits{E.}},
\bauthor{\bsnm{{Kim}}, \binits{Y.H.}},
\bauthor{\bsnm{{Lemasson}}, \binits{A.}},
\bauthor{\bsnm{{Guimar{\~a}es}}, \binits{V.}},
\bauthor{\bsnm{{Barrientos}}, \binits{D.}},
\bauthor{\bsnm{{Bemmerer}}, \binits{D.}},
\bauthor{\bsnm{{Benzoni}}, \binits{G.}},
\bauthor{\bsnm{{Boston}}, \binits{A.J.}},
\bauthor{\bsnm{{B{\"o}ttger}}, \binits{R.}},
\bauthor{\bsnm{{Boulay}}, \binits{F.}},
\bauthor{\bsnm{{Bracco}}, \binits{A.}},
\bauthor{\bsnm{{{\v{C}}elikovi{\'c}}}, \binits{I.}},
\bauthor{\bsnm{{Cederwall}}, \binits{B.}},
\bauthor{\bsnm{{Ciemala}}, \binits{M.}},
\bauthor{\bsnm{{Delafosse}}, \binits{C.}},
\bauthor{\bsnm{{Domingo-Pardo}}, \binits{C.}},
\bauthor{\bsnm{{Dudouet}}, \binits{J.}},
\bauthor{\bsnm{{Eberth}}, \binits{J.}},
\bauthor{\bsnm{{F{\"u}l{\"o}p}}, \binits{Z.}},
\bauthor{\bsnm{{Gonz{\'a}lez}}, \binits{V.}},
\bauthor{\bsnm{{Gottardo}}, \binits{A.}},
\bauthor{\bsnm{{Goupil}}, \binits{J.}},
\bauthor{\bsnm{{Hess}}, \binits{H.}},
\bauthor{\bsnm{{Jungclaus}}, \binits{A.}},
\bauthor{\bsnm{{Ka{\c{s}}ka{\c{s}}}}, \binits{A.}},
\bauthor{\bsnm{{Korichi}}, \binits{A.}},
\bauthor{\bsnm{{Lenzi}}, \binits{S.M.}},
\bauthor{\bsnm{{Leoni}}, \binits{S.}},
\bauthor{\bsnm{{Li}}, \binits{H.}},
\bauthor{\bsnm{{Ljungvall}}, \binits{J.}},
\bauthor{\bsnm{{Lopez-Martens}}, \binits{A.}},
\bauthor{\bsnm{{Menegazzo}}, \binits{R.}},
\bauthor{\bsnm{{Mengoni}}, \binits{D.}},
\bauthor{\bsnm{{Million}}, \binits{B.}},
\bauthor{\bsnm{{Mr{\'a}zek}}, \binits{J.}},
\bauthor{\bsnm{{Napoli}}, \binits{D.R.}},
\bauthor{\bsnm{{Navin}}, \binits{A.}},
\bauthor{\bsnm{{Nyberg}}, \binits{J.}},
\bauthor{\bsnm{{Podoly{\'a}k}}, \binits{Z.}},
\bauthor{\bsnm{{Pullia}}, \binits{A.}},
\bauthor{\bsnm{{Quintana}}, \binits{B.}},
\bauthor{\bsnm{{Ralet}}, \binits{D.}},
\bauthor{\bsnm{{Redon}}, \binits{N.}},
\bauthor{\bsnm{{Reiter}}, \binits{P.}},
\bauthor{\bsnm{{Rezynkina}}, \binits{K.}},
\bauthor{\bsnm{{Saillant}}, \binits{F.}},
\bauthor{\bsnm{{Salsac}}, \binits{M.-D.}},
\bauthor{\bsnm{{S{\'a}nchez-Ben{\'\i}tez}}, \binits{A.M.}},
\bauthor{\bsnm{{Sanchis}}, \binits{E.}},
\bauthor{\bsnm{{{\c{S}}enyi{\v{g}}it}}, \binits{M.}},
\bauthor{\bsnm{{Siciliano}}, \binits{M.}},
\bauthor{\bsnm{{Smirnova}}, \binits{N.A.}},
\bauthor{\bsnm{{Sohler}}, \binits{D.}},
\bauthor{\bsnm{{Stanoiu}}, \binits{M.}},
\bauthor{\bsnm{{Theisen}}, \binits{C.}},
\bauthor{\bsnm{{Valiente-Dob{\'o}n}}, \binits{J.J.}},
\bauthor{\bsnm{{Uji{\'c}}}, \binits{P.}},
\bauthor{\bsnm{{Zieli{\'n}ska}}, \binits{M.}}:
\batitle{{Search for $^{22}$Na in novae supported by a novel method for
  measuring femtosecond nuclear lifetimes}}.
\bjtitle{Nature Communications}
\bvolume{14},
\bfpage{4536}
(\byear{2023})
\doiurl{10.1038/s41467-023-40121-3}
{\href{https://arxiv.org/abs/2212.06302}{{arXiv:2212.06302}}}
{[nucl-ex]}
\end{barticle}
\endbibitem

%%% 157
\bibitem[\protect\citeauthoryear{{Kn{\"o}dlseder}
  et~al.}{1999}]{Knoedlseder1999}
\begin{barticle}
\bauthor{\bsnm{{Kn{\"o}dlseder}}, \binits{J.}},
\bauthor{\bsnm{{Bennett}}, \binits{K.}},
\bauthor{\bsnm{{Bloemen}}, \binits{H.}},
\bauthor{\bsnm{{Diehl}}, \binits{R.}},
\bauthor{\bsnm{{Hermsen}}, \binits{W.}},
\bauthor{\bsnm{{Oberlack}}, \binits{U.}},
\bauthor{\bsnm{{Ryan}}, \binits{J.}},
\bauthor{\bsnm{{Sch{\"o}nfelder}}, \binits{V.}},
\bauthor{\bsnm{{von Ballmoos}}, \binits{P.}}:
\batitle{{A multiwavelength comparison of COMPTEL 1.8 MeV \{(26) \} line
  data}}.
\bjtitle{\aap}
\bvolume{344},
\bfpage{68}--\blpage{82}
(\byear{1999})
\end{barticle}
\endbibitem

%%% 158
\bibitem[\protect\citeauthoryear{{Canete} et~al.}{2023}]{Canete2023}
\begin{barticle}
\bauthor{\bsnm{{Canete}}, \binits{L.}},
\bauthor{\bsnm{{Doherty}}, \binits{D.T.}},
\bauthor{\bsnm{{Lotay}}, \binits{G.}},
\bauthor{\bsnm{{Seweryniak}}, \binits{D.}},
\bauthor{\bsnm{{Campbell}}, \binits{C.M.}},
\bauthor{\bsnm{{Carpenter}}, \binits{M.P.}},
\bauthor{\bsnm{{Catford}}, \binits{W.N.}},
\bauthor{\bsnm{{Chipps}}, \binits{K.A.}},
\bauthor{\bsnm{{Henderson}}, \binits{J.}},
\bauthor{\bsnm{{Izzard}}, \binits{R.G.}},
\bauthor{\bsnm{{Janssens}}, \binits{R.V.F.}},
\bauthor{\bsnm{{Jayatissa}}, \binits{H.}},
\bauthor{\bsnm{{Jos{\'e}}}, \binits{J.}},
\bauthor{\bsnm{{Kennington}}, \binits{A.R.L.}},
\bauthor{\bsnm{{Kondev}}, \binits{F.G.}},
\bauthor{\bsnm{{Korichi}}, \binits{A.}},
\bauthor{\bsnm{{Lauritsen}}, \binits{T.}},
\bauthor{\bsnm{{M{\"u}ller-Gatermann}}, \binits{C.}},
\bauthor{\bsnm{{Paxman}}, \binits{C.}},
\bauthor{\bsnm{{Podoly{\'a}k}}, \binits{Z.}},
\bauthor{\bsnm{{Reed}}, \binits{B.J.}},
\bauthor{\bsnm{{Regan}}, \binits{P.H.}},
\bauthor{\bsnm{{Reviol}}, \binits{W.}},
\bauthor{\bsnm{{Siciliano}}, \binits{M.}},
\bauthor{\bsnm{{Wilson}}, \binits{G.L.}},
\bauthor{\bsnm{{Yates}}, \binits{R.}},
\bauthor{\bsnm{{Zhu}}, \binits{S.}}:
\batitle{{Confirmation of a new resonance in $^{26}$Si and contribution of
  classical novae to the galactic abundance of $^{26}$Al}}.
\bjtitle{\prc}
\bvolume{108}(\bissue{3}),
\bfpage{035807}
(\byear{2023})
\doiurl{10.1103/PhysRevC.108.035807}
\end{barticle}
\endbibitem

%%% 159
\bibitem[\protect\citeauthoryear{{Harris} et~al.}{1991}]{Harris1991}
\begin{barticle}
\bauthor{\bsnm{{Harris}}, \binits{M.J.}},
\bauthor{\bsnm{{Leising}}, \binits{M.D.}},
\bauthor{\bsnm{{Share}}, \binits{G.H.}}:
\batitle{{A Search for the 478 keV Line from the Decay of Nucleosynthetic
  7Be}}.
\bjtitle{\apj}
\bvolume{375},
\bfpage{216}
(\byear{1991})
\doiurl{10.1086/170183}
\end{barticle}
\endbibitem

%%% 160
\bibitem[\protect\citeauthoryear{{Siegert} et~al.}{2018}]{Siegert2018}
\begin{barticle}
\bauthor{\bsnm{{Siegert}}, \binits{T.}},
\bauthor{\bsnm{{Coc}}, \binits{A.}},
\bauthor{\bsnm{{Delgado}}, \binits{L.}},
\bauthor{\bsnm{{Diehl}}, \binits{R.}},
\bauthor{\bsnm{{Greiner}}, \binits{J.}},
\bauthor{\bsnm{{Hernanz}}, \binits{M.}},
\bauthor{\bsnm{{Jean}}, \binits{P.}},
\bauthor{\bsnm{{Jos{\'e}}}, \binits{J.}},
\bauthor{\bsnm{{Molaro}}, \binits{P.}},
\bauthor{\bsnm{{Pleintinger}}, \binits{M.M.M.}},
\bauthor{\bsnm{{Savchenko}}, \binits{V.}},
\bauthor{\bsnm{{Starrfield}}, \binits{S.}},
\bauthor{\bsnm{{Tatischeff}}, \binits{V.}},
\bauthor{\bsnm{{Weinberger}}, \binits{C.}}:
\batitle{{Gamma-ray observations of Nova Sgr 2015 No. 2 with INTEGRAL}}.
\bjtitle{\aap}
\bvolume{615},
\bfpage{107}
(\byear{2018})
\doiurl{10.1051/0004-6361/201732514}
{\href{https://arxiv.org/abs/1803.06888}{{arXiv:1803.06888}}}
{[astro-ph.HE]}
\end{barticle}
\endbibitem

%%% 161
\bibitem[\protect\citeauthoryear{{Izzo} et~al.}{2025}]{Izzo2025}
\begin{botherref}
\oauthor{\bsnm{{Izzo}}, \binits{L.}},
\oauthor{\bsnm{{Siegert}}, \binits{T.}},
\oauthor{\bsnm{{Jean}}, \binits{P.}},
\oauthor{\bsnm{{Molaro}}, \binits{P.}},
\oauthor{\bsnm{{Bonifacio}}, \binits{P.}},
\oauthor{\bsnm{{Della Valle}}, \binits{M.}},
\oauthor{\bsnm{{Parsotan}}, \binits{T.}}:
{Possible evidence for the 478 keV emission line from $^7$Be decay during the
  outburst phases of V1369 Cen}.
arXiv e-prints,
2504--20866
(2025)
\doiurl{10.48550/arXiv.2504.20866}
{\href{https://arxiv.org/abs/2504.20866}{{arXiv:2504.20866}}}
{[astro-ph.HE]}
\end{botherref}
\endbibitem

%%% 162
\bibitem[\protect\citeauthoryear{{Tajitsu} et~al.}{2016}]{Tajitsu2016}
\begin{barticle}
\bauthor{\bsnm{{Tajitsu}}, \binits{A.}},
\bauthor{\bsnm{{Sadakane}}, \binits{K.}},
\bauthor{\bsnm{{Naito}}, \binits{H.}},
\bauthor{\bsnm{{Arai}}, \binits{A.}},
\bauthor{\bsnm{{Kawakita}}, \binits{H.}},
\bauthor{\bsnm{{Aoki}}, \binits{W.}}:
\batitle{{The $^{7}$Be II Resonance Lines in Two Classical Novae V5668 Sgr and
  V2944 Oph}}.
\bjtitle{\apj}
\bvolume{818}(\bissue{2}),
\bfpage{191}
(\byear{2016})
\doiurl{10.3847/0004-637X/818/2/191}
{\href{https://arxiv.org/abs/1601.05168}{{arXiv:1601.05168}}}
{[astro-ph.SR]}
\end{barticle}
\endbibitem

%%% 163
\bibitem[\protect\citeauthoryear{{Molaro} et~al.}{2020}]{Molaro2020}
\begin{barticle}
\bauthor{\bsnm{{Molaro}}, \binits{P.}},
\bauthor{\bsnm{{Izzo}}, \binits{L.}},
\bauthor{\bsnm{{Bonifacio}}, \binits{P.}},
\bauthor{\bsnm{{Hernanz}}, \binits{M.}},
\bauthor{\bsnm{{Selvelli}}, \binits{P.}},
\bauthor{\bsnm{{della Valle}}, \binits{M.}}:
\batitle{{Search for $^{7}$Be in the outbursts of four recent novae}}.
\bjtitle{\mnras}
\bvolume{492}(\bissue{4}),
\bfpage{4975}--\blpage{4985}
(\byear{2020})
\doiurl{10.1093/mnras/stz3587}
{\href{https://arxiv.org/abs/1912.13281}{{arXiv:1912.13281}}}
{[astro-ph.GA]}
\end{barticle}
\endbibitem

%%% 164
\bibitem[\protect\citeauthoryear{{Jean} et~al.}{1999}]{Jean1999}
\begin{barticle}
\bauthor{\bsnm{{Jean}}, \binits{P.}},
\bauthor{\bsnm{{G{\'o}mez-Gomar}}, \binits{J.}},
\bauthor{\bsnm{{Hernanz}}, \binits{M.}},
\bauthor{\bsnm{{Jos{\'e}}}, \binits{J.}},
\bauthor{\bsnm{{Isern}}, \binits{J.}},
\bauthor{\bsnm{{Vedrenne}}, \binits{G.}},
\bauthor{\bsnm{{Mandrou}}, \binits{P.}},
\bauthor{\bsnm{{Sch{\"o}nfelder}}, \binits{V.}},
\bauthor{\bsnm{{Lichti}}, \binits{G.G.}},
\bauthor{\bsnm{{Georgii}}, \binits{R.}}:
\batitle{{Possibility of the Detection of Classical Novae with the Shield of
  the Integral Spectrometer SPI}}.
\bjtitle{Astrophysical Letters and Communications}
\bvolume{38},
\bfpage{421}
(\byear{1999})
\doiurl{10.48550/arXiv.astro-ph/9903015}
{\href{https://arxiv.org/abs/astro-ph/9903015}{{arXiv:astro-ph/9903015}}}
{[astro-ph]}
\end{barticle}
\endbibitem

%%% 165
\bibitem[\protect\citeauthoryear{{Hernanz} et~al.}{2000}]{Hernanz2000}
\begin{bchapter}
\bauthor{\bsnm{{Hernanz}}, \binits{M.}},
\bauthor{\bsnm{{Smith}}, \binits{D.M.}},
\bauthor{\bsnm{{Fishman}}, \binits{J.}},
\bauthor{\bsnm{{Harmon}}, \binits{A.}},
\bauthor{\bsnm{{G{\'o}mez-Gomar}}, \binits{J.}},
\bauthor{\bsnm{{Jos{\'e}}}, \binits{J.}},
\bauthor{\bsnm{{Isern}}, \binits{J.}},
\bauthor{\bsnm{{Jean}}, \binits{P.}}:
\bctitle{{BATSE observations of classical novae}}.
In: \beditor{\bsnm{{McConnell}}, \binits{M.L.}},
\beditor{\bsnm{{Ryan}}, \binits{J.M.}} (eds.)
\bbtitle{The Fifth Compton Symposium}.
\bsertitle{American Institute of Physics Conference Series},
vol. \bseriesno{510},
pp. \bfpage{82}--\blpage{86}.
\bpublisher{AIP}, \blocation{???}
(\byear{2000}).
\doiurl{10.1063/1.1303179}
\end{bchapter}
\endbibitem

%%% 166
\bibitem[\protect\citeauthoryear{{Harris} et~al.}{2000}]{Harris2000}
\begin{barticle}
\bauthor{\bsnm{{Harris}}, \binits{M.J.}},
\bauthor{\bsnm{{Teegarden}}, \binits{B.J.}},
\bauthor{\bsnm{{Cline}}, \binits{T.L.}},
\bauthor{\bsnm{{Gehrels}}, \binits{N.}},
\bauthor{\bsnm{{Palmer}}, \binits{D.M.}},
\bauthor{\bsnm{{Ramaty}}, \binits{R.}},
\bauthor{\bsnm{{Seifert}}, \binits{H.}}:
\batitle{{Transient Gamma-Ray Spectrometer Observations of Gamma-Ray Lines from
  Novae. II. Constraining the Galactic Nova Rate from a Survey of the Southern
  Sky during 1995-1997}}.
\bjtitle{\apj}
\bvolume{542}(\bissue{2}),
\bfpage{1057}--\blpage{1063}
(\byear{2000})
\doiurl{10.1086/317022}
{\href{https://arxiv.org/abs/astro-ph/0004167}{{arXiv:astro-ph/0004167}}}
{[astro-ph]}
\end{barticle}
\endbibitem

%%% 167
\bibitem[\protect\citeauthoryear{{Senziani} et~al.}{2008}]{Senziani2008}
\begin{barticle}
\bauthor{\bsnm{{Senziani}}, \binits{F.}},
\bauthor{\bsnm{{Skinner}}, \binits{G.K.}},
\bauthor{\bsnm{{Jean}}, \binits{P.}},
\bauthor{\bsnm{{Hernanz}}, \binits{M.}}:
\batitle{{Detectability of gamma-ray emission from classical novae with
  Swift/BAT}}.
\bjtitle{\aap}
\bvolume{485}(\bissue{1}),
\bfpage{223}--\blpage{231}
(\byear{2008})
\doiurl{10.1051/0004-6361:200809863}
{\href{https://arxiv.org/abs/0804.4791}{{arXiv:0804.4791}}}
{[astro-ph]}
\end{barticle}
\endbibitem

%%% 168
\bibitem[\protect\citeauthoryear{{Tomsick} et~al.}{2024}]{Tomsick2024}
\begin{bchapter}
\bauthor{\bsnm{{Tomsick}}, \binits{J.}},
\bauthor{\bsnm{{Boggs}}, \binits{S.}},
\bauthor{\bsnm{{Zoglauer}}, \binits{A.}},
\bauthor{\bsnm{{Hartmann}}, \binits{D.H.}},
\bauthor{\bsnm{{Ajello}}, \binits{M.}},
\bauthor{\bsnm{{Burns}}, \binits{E.}},
\bauthor{\bsnm{{Fryer}}, \binits{C.}},
\bauthor{\bsnm{{Karwin}}, \binits{C.}},
\bauthor{\bsnm{{Kierans}}, \binits{C.}},
\bauthor{\bsnm{{Lowell}}, \binits{A.}},
\bauthor{\bsnm{{Malzac}}, \binits{J.}},
\bauthor{\bsnm{{Roberts}}, \binits{J.}},
\bauthor{\bsnm{{Saint-Hilaire}}, \binits{P.}},
\bauthor{\bsnm{{Shih}}, \binits{A.}},
\bauthor{\bsnm{{Siegert}}, \binits{T.}},
\bauthor{\bsnm{{Sleator}}, \binits{C.}},
\bauthor{\bsnm{{Takahashi}}, \binits{T.}},
\bauthor{\bsnm{{Tavecchio}}, \binits{F.}},
\bauthor{\bsnm{{Wulf}}, \binits{E.}},
\bauthor{\bsnm{{Beechert}}, \binits{J.}},
\bauthor{\bsnm{{Gulick}}, \binits{H.}},
\bauthor{\bsnm{{Joens}}, \binits{A.}},
\bauthor{\bsnm{{Lazar}}, \binits{H.}},
\bauthor{\bsnm{{Neights}}, \binits{E.}},
\bauthor{\bsnm{{Martinez Oliveros}}, \binits{J.C.}},
\bauthor{\bsnm{{Matsumoto}}, \binits{S.}},
\bauthor{\bsnm{{Melia}}, \binits{T.}},
\bauthor{\bsnm{{Yoneda}}, \binits{H.}},
\bauthor{\bsnm{{Amman}}, \binits{M.}},
\bauthor{\bsnm{{Bal}}, \binits{D.}},
\bauthor{\bsnm{{von Ballmoos}}, \binits{P.}},
\bauthor{\bsnm{{Bates}}, \binits{H.}},
\bauthor{\bsnm{{B{\"o}ttcher}}, \binits{M.}},
\bauthor{\bsnm{{Bulgarelli}}, \binits{A.}},
\bauthor{\bsnm{{Cavazzuti}}, \binits{E.}},
\bauthor{\bsnm{{Chang}}, \binits{H.K.}},
\bauthor{\bsnm{{Chen}}, \binits{C.}},
\bauthor{\bsnm{{Chu}}, \binits{C.Y.}},
\bauthor{\bsnm{{Ciabattoni}}, \binits{A.}},
\bauthor{\bsnm{{Costamante}}, \binits{L.}},
\bauthor{\bsnm{{Dreyer}}, \binits{L.}},
\bauthor{\bsnm{{Fioretti}}, \binits{V.}},
\bauthor{\bsnm{{Fenu}}, \binits{F.}},
\bauthor{\bsnm{{Gallego}}, \binits{S.}},
\bauthor{\bsnm{{Ghirlanda}}, \binits{G.}},
\bauthor{\bsnm{{Grove}}, \binits{E.}},
\bauthor{\bsnm{{Huang}}, \binits{C.Y.}},
\bauthor{\bsnm{{Jean}}, \binits{P.}},
\bauthor{\bsnm{{Khatiya}}, \binits{N.}},
\bauthor{\bsnm{{Kn{\"o}dlseder}}, \binits{J.}},
\bauthor{\bsnm{{Kraus}}, \binits{M.}},
\bauthor{\bsnm{{Leising}}, \binits{M.}},
\bauthor{\bsnm{{Lewis}}, \binits{T.}},
\bauthor{\bsnm{{Lommler}}, \binits{J.}},
\bauthor{\bsnm{{Marcotulli}}, \binits{L.}},
\bauthor{\bsnm{{Martinez Castellanos}}, \binits{I.}},
\bauthor{\bsnm{{Mittal}}, \binits{S.}},
\bauthor{\bsnm{{Negro}}, \binits{M.}},
\bauthor{\bsnm{{Al Nussirat}}, \binits{S.}},
\bauthor{\bsnm{{Nakazawa}}, \binits{K.}},
\bauthor{\bsnm{{Oberlack}}, \binits{U.}},
\bauthor{\bsnm{{Palmore}}, \binits{D.}},
\bauthor{\bsnm{{Panebianco}}, \binits{G.}},
\bauthor{\bsnm{{Parmiggiani}}, \binits{N.}},
\bauthor{\bsnm{{Pike}}, \binits{S.}},
\bauthor{\bsnm{{Rogers}}, \binits{F.}},
\bauthor{\bsnm{{Schutte}}, \binits{H.}},
\bauthor{\bsnm{{Sheng}}, \binits{Y.}},
\bauthor{\bsnm{{Smale}}, \binits{A.}},
\bauthor{\bsnm{{Smith}}, \binits{J.R.}},
\bauthor{\bsnm{{Trigg}}, \binits{A.}},
\bauthor{\bsnm{{Venters}}, \binits{T.}},
\bauthor{\bsnm{{Watanabe}}, \binits{Y.}},
\bauthor{\bsnm{{Zhang}}, \binits{H.}}:
\bctitle{{The Compton Spectrometer and Imager}}.
In: \bbtitle{38th International Cosmic Ray Conference},
p. \bfpage{745}
(\byear{2024}).
\doiurl{10.48550/arXiv.2308.12362}
\end{bchapter}
\endbibitem

%%% 169
\bibitem[\protect\citeauthoryear{{Burbidge} et~al.}{1957}]{Burbidge1957}
\begin{barticle}
\bauthor{\bsnm{{Burbidge}}, \binits{E.M.}},
\bauthor{\bsnm{{Burbidge}}, \binits{G.R.}},
\bauthor{\bsnm{{Fowler}}, \binits{W.A.}},
\bauthor{\bsnm{{Hoyle}}, \binits{F.}}:
\batitle{{Synthesis of the Elements in Stars}}.
\bjtitle{Reviews of Modern Physics}
\bvolume{29}(\bissue{4}),
\bfpage{547}--\blpage{650}
(\byear{1957})
\doiurl{10.1103/RevModPhys.29.547}
\end{barticle}
\endbibitem

%%% 170
\bibitem[\protect\citeauthoryear{{Colgate} and {White}}{1966}]{Colgate1966}
\begin{barticle}
\bauthor{\bsnm{{Colgate}}, \binits{S.A.}},
\bauthor{\bsnm{{White}}, \binits{R.H.}}:
\batitle{{The Hydrodynamic Behavior of Supernovae Explosions}}.
\bjtitle{\apj}
\bvolume{143},
\bfpage{626}
(\byear{1966})
\doiurl{10.1086/148549}
\end{barticle}
\endbibitem

%%% 171
\bibitem[\protect\citeauthoryear{{Andrews} et~al.}{2020}]{Andrews2020}
\begin{barticle}
\bauthor{\bsnm{{Andrews}}, \binits{S.}},
\bauthor{\bsnm{{Fryer}}, \binits{C.}},
\bauthor{\bsnm{{Even}}, \binits{W.}},
\bauthor{\bsnm{{Jones}}, \binits{S.}},
\bauthor{\bsnm{{Pignatari}}, \binits{M.}}:
\batitle{{The Nucleosynthetic Yields of Core-collapse Supernovae: Prospects for
  the Next Generation of Gamma-Ray Astronomy}}.
\bjtitle{\apj}
\bvolume{890}(\bissue{1}),
\bfpage{35}
(\byear{2020})
\doiurl{10.3847/1538-4357/ab64f8}
{\href{https://arxiv.org/abs/1912.10542}{{arXiv:1912.10542}}}
{[astro-ph.HE]}
\end{barticle}
\endbibitem

%%% 172
\bibitem[\protect\citeauthoryear{{Stritzinger} et~al.}{2006}]{Stritzinger2006}
\begin{barticle}
\bauthor{\bsnm{{Stritzinger}}, \binits{M.}},
\bauthor{\bsnm{{Mazzali}}, \binits{P.A.}},
\bauthor{\bsnm{{Sollerman}}, \binits{J.}},
\bauthor{\bsnm{{Benetti}}, \binits{S.}}:
\batitle{{Consistent estimates of $^{56}$Ni yields for type Ia supernovae}}.
\bjtitle{\aap}
\bvolume{460}(\bissue{3}),
\bfpage{793}--\blpage{798}
(\byear{2006})
\doiurl{10.1051/0004-6361:20065514}
{\href{https://arxiv.org/abs/astro-ph/0609232}{{arXiv:astro-ph/0609232}}}
{[astro-ph]}
\end{barticle}
\endbibitem

%%% 173
\bibitem[\protect\citeauthoryear{{Weinberger} et~al.}{2020}]{Weinberger2020}
\begin{barticle}
\bauthor{\bsnm{{Weinberger}}, \binits{C.}},
\bauthor{\bsnm{{Diehl}}, \binits{R.}},
\bauthor{\bsnm{{Pleintinger}}, \binits{M.M.M.}},
\bauthor{\bsnm{{Siegert}}, \binits{T.}},
\bauthor{\bsnm{{Greiner}}, \binits{J.}}:
\batitle{{$^{44}$Ti ejecta in young supernova remnants}}.
\bjtitle{\aap}
\bvolume{638},
\bfpage{83}
(\byear{2020})
\doiurl{10.1051/0004-6361/202037536}
{\href{https://arxiv.org/abs/2004.12688}{{arXiv:2004.12688}}}
{[astro-ph.HE]}
\end{barticle}
\endbibitem

%%% 174
\bibitem[\protect\citeauthoryear{{Magkotsios} et~al.}{2010}]{Magkotsios2010}
\begin{barticle}
\bauthor{\bsnm{{Magkotsios}}, \binits{G.}},
\bauthor{\bsnm{{Timmes}}, \binits{F.X.}},
\bauthor{\bsnm{{Hungerford}}, \binits{A.L.}},
\bauthor{\bsnm{{Fryer}}, \binits{C.L.}},
\bauthor{\bsnm{{Young}}, \binits{P.A.}},
\bauthor{\bsnm{{Wiescher}}, \binits{M.}}:
\batitle{{Trends in $^{44}$Ti and $^{56}$Ni from Core-collapse Supernovae}}.
\bjtitle{\apjs}
\bvolume{191}(\bissue{1}),
\bfpage{66}--\blpage{95}
(\byear{2010})
\doiurl{10.1088/0067-0049/191/1/66}
{\href{https://arxiv.org/abs/1009.3175}{{arXiv:1009.3175}}}
{[astro-ph.SR]}
\end{barticle}
\endbibitem

%%% 175
\bibitem[\protect\citeauthoryear{{Prantzos}}{2011}]{Prantzos2011b}
\begin{botherref}
\oauthor{\bsnm{{Prantzos}}, \binits{N.}}:
{Nucleosynthesis and gamma-ray lines}.
arXiv e-prints,
1101--2112
(2011)
\doiurl{10.48550/arXiv.1101.2112}
{\href{https://arxiv.org/abs/1101.2112}{{arXiv:1101.2112}}}
{[astro-ph.HE]}
\end{botherref}
\endbibitem

%%% 176
\bibitem[\protect\citeauthoryear{{Prantzos}}{2008}]{Prantzos2008}
\begin{bchapter}
\bauthor{\bsnm{{Prantzos}}, \binits{N.}}:
\bctitle{{An Introduction to Galactic Chemical Evolution}}.
In: \beditor{\bsnm{{Charbonnel}}, \binits{C.}},
\beditor{\bsnm{{Zahn}}, \binits{J.-P.}} (eds.)
\bbtitle{EAS Publications Series}.
\bsertitle{EAS Publications Series},
vol. \bseriesno{32},
pp. \bfpage{311}--\blpage{356}
(\byear{2008}).
\doiurl{10.1051/eas:0832009}
\end{bchapter}
\endbibitem

%%% 177
\bibitem[\protect\citeauthoryear{{Kobayashi} et~al.}{2020}]{Kobayashi2020}
\begin{barticle}
\bauthor{\bsnm{{Kobayashi}}, \binits{C.}},
\bauthor{\bsnm{{Karakas}}, \binits{A.I.}},
\bauthor{\bsnm{{Lugaro}}, \binits{M.}}:
\batitle{{The Origin of Elements from Carbon to Uranium}}.
\bjtitle{\apj}
\bvolume{900}(\bissue{2}),
\bfpage{179}
(\byear{2020})
\doiurl{10.3847/1538-4357/abae65}
{\href{https://arxiv.org/abs/2008.04660}{{arXiv:2008.04660}}}
{[astro-ph.GA]}
\end{barticle}
\endbibitem

%%% 178
\bibitem[\protect\citeauthoryear{{Vasini} et~al.}{2022}]{Vasini2022}
\begin{barticle}
\bauthor{\bsnm{{Vasini}}, \binits{A.}},
\bauthor{\bsnm{{Matteucci}}, \binits{F.}},
\bauthor{\bsnm{{Spitoni}}, \binits{E.}}:
\batitle{{Chemical evolution of $^{26}$Al and $^{60}$Fe in the Milky Way}}.
\bjtitle{\mnras}
\bvolume{517}(\bissue{3}),
\bfpage{4256}--\blpage{4264}
(\byear{2022})
\doiurl{10.1093/mnras/stac2981}
{\href{https://arxiv.org/abs/2204.00510}{{arXiv:2204.00510}}}
{[astro-ph.GA]}
\end{barticle}
\endbibitem

%%% 179
\bibitem[\protect\citeauthoryear{{Fu} and {Arnett}}{1989}]{Fu1989}
\begin{barticle}
\bauthor{\bsnm{{Fu}}, \binits{A.}},
\bauthor{\bsnm{{Arnett}}, \binits{W.D.}}:
\batitle{{The Late Behavior of Supernova 1987A. II. Gamma-Ray Transparency of
  the Ejecta}}.
\bjtitle{\apj}
\bvolume{340},
\bfpage{414}
(\byear{1989})
\doiurl{10.1086/167403}
\end{barticle}
\endbibitem

%%% 180
\bibitem[\protect\citeauthoryear{{Grebenev} and
  {Syunyaev}}{1987}]{Grebenev1987}
\begin{barticle}
\bauthor{\bsnm{{Grebenev}}, \binits{S.A.}},
\bauthor{\bsnm{{Syunyaev}}, \binits{R.A.}}:
\batitle{{The Expected X-Ray Emission from Supernova 1987A - Monte-Carlo
  Calculations}}.
\bjtitle{Soviet Astronomy Letters}
\bvolume{13},
\bfpage{397}
(\byear{1987})
\end{barticle}
\endbibitem

%%% 181
\bibitem[\protect\citeauthoryear{{Palmer} et~al.}{1993}]{Palmer1993}
\begin{barticle}
\bauthor{\bsnm{{Palmer}}, \binits{D.M.}},
\bauthor{\bsnm{{Schindler}}, \binits{S.M.}},
\bauthor{\bsnm{{Cook}}, \binits{W.R.}},
\bauthor{\bsnm{{Grunsfeld}}, \binits{J.M.}},
\bauthor{\bsnm{{Heindl}}, \binits{W.A.}},
\bauthor{\bsnm{{Prince}}, \binits{T.A.}},
\bauthor{\bsnm{{Stone}}, \binits{E.C.}}:
\batitle{{Gamma-Ray Continuum and Line Observations of SN 1987A}}.
\bjtitle{\apj}
\bvolume{412},
\bfpage{203}
(\byear{1993})
\doiurl{10.1086/172912}
\end{barticle}
\endbibitem

%%% 182
\bibitem[\protect\citeauthoryear{{Leising} and {Share}}{1990}]{Leising1990}
\begin{barticle}
\bauthor{\bsnm{{Leising}}, \binits{M.D.}},
\bauthor{\bsnm{{Share}}, \binits{G.H.}}:
\batitle{{The Gamma-Ray Light Curves of SN 1987A}}.
\bjtitle{\apj}
\bvolume{357},
\bfpage{638}
(\byear{1990})
\doiurl{10.1086/168952}
\end{barticle}
\endbibitem

%%% 183
\bibitem[\protect\citeauthoryear{{Iyudin} et~al.}{1994}]{Iyudin1994}
\begin{barticle}
\bauthor{\bsnm{{Iyudin}}, \binits{A.F.}},
\bauthor{\bsnm{{Diehl}}, \binits{R.}},
\bauthor{\bsnm{{Bloemen}}, \binits{H.}},
\bauthor{\bsnm{{Hermsen}}, \binits{W.}},
\bauthor{\bsnm{{Lichti}}, \binits{G.G.}},
\bauthor{\bsnm{{Morris}}, \binits{D.}},
\bauthor{\bsnm{{Ryan}}, \binits{J.}},
\bauthor{\bsnm{{Sch{\"o}nfelder}}, \binits{V.}},
\bauthor{\bsnm{{Steinle}}, \binits{H.}},
\bauthor{\bsnm{{Varendorff}}, \binits{M.}},
\bauthor{\bsnm{{de Vries}}, \binits{C.}},
\bauthor{\bsnm{{Winkler}}, \binits{C.}}:
\batitle{{COMPTEL observations of 44Ti gamma-ray line emission form CAS A.}}
\bjtitle{\aap}
\bvolume{284},
\bfpage{1}--\blpage{4}
(\byear{1994})
\end{barticle}
\endbibitem

%%% 184
\bibitem[\protect\citeauthoryear{{The} et~al.}{1996}]{The1996}
\begin{barticle}
\bauthor{\bsnm{{The}}, \binits{L.-S.}},
\bauthor{\bsnm{{Leising}}, \binits{M.D.}},
\bauthor{\bsnm{{Kurfess}}, \binits{J.D.}},
\bauthor{\bsnm{{Johnson}}, \binits{W.N.}},
\bauthor{\bsnm{{Hartmann}}, \binits{D.H.}},
\bauthor{\bsnm{{Gehrels}}, \binits{N.}},
\bauthor{\bsnm{{Grove}}, \binits{J.E.}},
\bauthor{\bsnm{{Purcell}}, \binits{W.R.}}:
\batitle{{CGRO/OSSE observations of the Cassiopeia A SNR.}}
\bjtitle{\aaps}
\bvolume{120},
\bfpage{357}--\blpage{360}
(\byear{1996})
\end{barticle}
\endbibitem

%%% 185
\bibitem[\protect\citeauthoryear{{Vink} et~al.}{2001}]{Vink2001}
\begin{barticle}
\bauthor{\bsnm{{Vink}}, \binits{J.}},
\bauthor{\bsnm{{Laming}}, \binits{J.M.}},
\bauthor{\bsnm{{Kaastra}}, \binits{J.S.}},
\bauthor{\bsnm{{Bleeker}}, \binits{J.A.M.}},
\bauthor{\bsnm{{Bloemen}}, \binits{H.}},
\bauthor{\bsnm{{Oberlack}}, \binits{U.}}:
\batitle{{Detection of the 67.9 and 78.4 keV Lines Associated with the
  Radioactive Decay of $^{44}$Ti in Cassiopeia A}}.
\bjtitle{\apjl}
\bvolume{560}(\bissue{1}),
\bfpage{79}--\blpage{82}
(\byear{2001})
\doiurl{10.1086/324172}
{\href{https://arxiv.org/abs/astro-ph/0107468}{{arXiv:astro-ph/0107468}}}
{[astro-ph]}
\end{barticle}
\endbibitem

%%% 186
\bibitem[\protect\citeauthoryear{{Renaud} et~al.}{2006}]{Renaud2006}
\begin{barticle}
\bauthor{\bsnm{{Renaud}}, \binits{M.}},
\bauthor{\bsnm{{Vink}}, \binits{J.}},
\bauthor{\bsnm{{Decourchelle}}, \binits{A.}},
\bauthor{\bsnm{{Lebrun}}, \binits{F.}},
\bauthor{\bsnm{{den Hartog}}, \binits{P.R.}},
\bauthor{\bsnm{{Terrier}}, \binits{R.}},
\bauthor{\bsnm{{Couvreur}}, \binits{C.}},
\bauthor{\bsnm{{Kn{\"o}dlseder}}, \binits{J.}},
\bauthor{\bsnm{{Martin}}, \binits{P.}},
\bauthor{\bsnm{{Prantzos}}, \binits{N.}},
\bauthor{\bsnm{{Bykov}}, \binits{A.M.}},
\bauthor{\bsnm{{Bloemen}}, \binits{H.}}:
\batitle{{The Signature of $^{44}$Ti in Cassiopeia A Revealed by IBIS/ISGRI on
  INTEGRAL}}.
\bjtitle{\apjl}
\bvolume{647}(\bissue{1}),
\bfpage{41}--\blpage{44}
(\byear{2006})
\doiurl{10.1086/507300}
{\href{https://arxiv.org/abs/astro-ph/0606736}{{arXiv:astro-ph/0606736}}}
{[astro-ph]}
\end{barticle}
\endbibitem

%%% 187
\bibitem[\protect\citeauthoryear{{Grefenstette}
  et~al.}{2014}]{Grefenstette2014}
\begin{barticle}
\bauthor{\bsnm{{Grefenstette}}, \binits{B.W.}},
\bauthor{\bsnm{{Harrison}}, \binits{F.A.}},
\bauthor{\bsnm{{Boggs}}, \binits{S.E.}},
\bauthor{\bsnm{{Reynolds}}, \binits{S.P.}},
\bauthor{\bsnm{{Fryer}}, \binits{C.L.}},
\bauthor{\bsnm{{Madsen}}, \binits{K.K.}},
\bauthor{\bsnm{{Wik}}, \binits{D.R.}},
\bauthor{\bsnm{{Zoglauer}}, \binits{A.}},
\bauthor{\bsnm{{Ellinger}}, \binits{C.I.}},
\bauthor{\bsnm{{Alexander}}, \binits{D.M.}},
\bauthor{\bsnm{{An}}, \binits{H.}},
\bauthor{\bsnm{{Barret}}, \binits{D.}},
\bauthor{\bsnm{{Christensen}}, \binits{F.E.}},
\bauthor{\bsnm{{Craig}}, \binits{W.W.}},
\bauthor{\bsnm{{Forster}}, \binits{K.}},
\bauthor{\bsnm{{Giommi}}, \binits{P.}},
\bauthor{\bsnm{{Hailey}}, \binits{C.J.}},
\bauthor{\bsnm{{Hornstrup}}, \binits{A.}},
\bauthor{\bsnm{{Kaspi}}, \binits{V.M.}},
\bauthor{\bsnm{{Kitaguchi}}, \binits{T.}},
\bauthor{\bsnm{{Koglin}}, \binits{J.E.}},
\bauthor{\bsnm{{Mao}}, \binits{P.H.}},
\bauthor{\bsnm{{Miyasaka}}, \binits{H.}},
\bauthor{\bsnm{{Mori}}, \binits{K.}},
\bauthor{\bsnm{{Perri}}, \binits{M.}},
\bauthor{\bsnm{{Pivovaroff}}, \binits{M.J.}},
\bauthor{\bsnm{{Puccetti}}, \binits{S.}},
\bauthor{\bsnm{{Rana}}, \binits{V.}},
\bauthor{\bsnm{{Stern}}, \binits{D.}},
\bauthor{\bsnm{{Westergaard}}, \binits{N.J.}},
\bauthor{\bsnm{{Zhang}}, \binits{W.W.}}:
\batitle{{Asymmetries in core-collapse supernovae from maps of radioactive
  $^{44}$Ti in CassiopeiaA}}.
\bjtitle{\nat}
\bvolume{506}(\bissue{7488}),
\bfpage{339}--\blpage{342}
(\byear{2014})
\doiurl{10.1038/nature12997}
{\href{https://arxiv.org/abs/1403.4978}{{arXiv:1403.4978}}}
{[astro-ph.HE]}
\end{barticle}
\endbibitem

%%% 188
\bibitem[\protect\citeauthoryear{{Siegert} et~al.}{2015}]{Siegert2015}
\begin{barticle}
\bauthor{\bsnm{{Siegert}}, \binits{T.}},
\bauthor{\bsnm{{Diehl}}, \binits{R.}},
\bauthor{\bsnm{{Krause}}, \binits{M.G.H.}},
\bauthor{\bsnm{{Greiner}}, \binits{J.}}:
\batitle{{Revisiting INTEGRAL/SPI observations of $^{44}$Ti from Cassiopeia
  A}}.
\bjtitle{\aap}
\bvolume{579},
\bfpage{124}
(\byear{2015})
\doiurl{10.1051/0004-6361/201525877}
{\href{https://arxiv.org/abs/1505.05999}{{arXiv:1505.05999}}}
{[astro-ph.HE]}
\end{barticle}
\endbibitem

%%% 189
\bibitem[\protect\citeauthoryear{{Motizuki} and
  {Kumagai}}{2004}]{Motizuki2004_lifetimes}
\begin{barticle}
\bauthor{\bsnm{{Motizuki}}, \binits{Y.}},
\bauthor{\bsnm{{Kumagai}}, \binits{S.}}:
\batitle{{$^{44}$Ti radioactivity in young supernova remnants: Cas A and SN
  1987A}}.
\bjtitle{\nar}
\bvolume{48}(\bissue{1-4}),
\bfpage{69}--\blpage{73}
(\byear{2004})
\doiurl{10.1016/j.newar.2003.11.009}
{\href{https://arxiv.org/abs/astro-ph/0311080}{{arXiv:astro-ph/0311080}}}
{[astro-ph]}
\end{barticle}
\endbibitem

%%% 190
\bibitem[\protect\citeauthoryear{{Clayton} et~al.}{1969}]{Clayton1969}
\begin{barticle}
\bauthor{\bsnm{{Clayton}}, \binits{D.D.}},
\bauthor{\bsnm{{Colgate}}, \binits{S.A.}},
\bauthor{\bsnm{{Fishman}}, \binits{G.J.}}:
\batitle{{Gamma-Ray Lines from Young Supernova Remnants}}.
\bjtitle{\apj}
\bvolume{155},
\bfpage{75}
(\byear{1969})
\doiurl{10.1086/149849}
\end{barticle}
\endbibitem

%%% 191
\bibitem[\protect\citeauthoryear{{Matz} and {Share}}{1990}]{Matz1990}
\begin{barticle}
\bauthor{\bsnm{{Matz}}, \binits{S.M.}},
\bauthor{\bsnm{{Share}}, \binits{G.H.}}:
\batitle{{A Limit on the Production of 56Ni in a Type I Supernova}}.
\bjtitle{\apj}
\bvolume{362},
\bfpage{235}
(\byear{1990})
\doiurl{10.1086/169259}
\end{barticle}
\endbibitem

%%% 192
\bibitem[\protect\citeauthoryear{{Churazov} et~al.}{2014}]{Churazov2014}
\begin{barticle}
\bauthor{\bsnm{{Churazov}}, \binits{E.}},
\bauthor{\bsnm{{Sunyaev}}, \binits{R.}},
\bauthor{\bsnm{{Isern}}, \binits{J.}},
\bauthor{\bsnm{{Kn{\"o}dlseder}}, \binits{J.}},
\bauthor{\bsnm{{Jean}}, \binits{P.}},
\bauthor{\bsnm{{Lebrun}}, \binits{F.}},
\bauthor{\bsnm{{Chugai}}, \binits{N.}},
\bauthor{\bsnm{{Grebenev}}, \binits{S.}},
\bauthor{\bsnm{{Bravo}}, \binits{E.}},
\bauthor{\bsnm{{Sazonov}}, \binits{S.}},
\bauthor{\bsnm{{Renaud}}, \binits{M.}}:
\batitle{{Cobalt-56 {\ensuremath{\gamma}}-ray emission lines from the type Ia
  supernova 2014J}}.
\bjtitle{\nat}
\bvolume{512}(\bissue{7515}),
\bfpage{406}--\blpage{408}
(\byear{2014})
\doiurl{10.1038/nature13672}
{\href{https://arxiv.org/abs/1405.3332}{{arXiv:1405.3332}}}
{[astro-ph.HE]}
\end{barticle}
\endbibitem

%%% 193
\bibitem[\protect\citeauthoryear{{Lichti} et~al.}{1994}]{Lichti1994}
\begin{barticle}
\bauthor{\bsnm{{Lichti}}, \binits{G.G.}},
\bauthor{\bsnm{{Bennett}}, \binits{K.}},
\bauthor{\bsnm{{den Herder}}, \binits{J.W.}},
\bauthor{\bsnm{{Diehl}}, \binits{R.}},
\bauthor{\bsnm{{Morris}}, \binits{D.}},
\bauthor{\bsnm{{Ryan}}, \binits{J.}},
\bauthor{\bsnm{{Sch{\"o}nfelder}}, \binits{V.}},
\bauthor{\bsnm{{Steinle}}, \binits{H.}},
\bauthor{\bsnm{{Strong}}, \binits{A.W.}},
\bauthor{\bsnm{{Winkler}}, \binits{C.}}:
\batitle{{COMPTEL upper limits on gamma-ray line emission from Supernova
  1991T.}}
\bjtitle{\aap}
\bvolume{292},
\bfpage{569}
(\byear{1994})
\end{barticle}
\endbibitem

%%% 194
\bibitem[\protect\citeauthoryear{{Leising} et~al.}{1995}]{Leising1995}
\begin{barticle}
\bauthor{\bsnm{{Leising}}, \binits{M.D.}},
\bauthor{\bsnm{{Johnson}}, \binits{W.N.}},
\bauthor{\bsnm{{Kurfess}}, \binits{J.D.}},
\bauthor{\bsnm{{Clayton}}, \binits{D.D.}},
\bauthor{\bsnm{{Grabelsky}}, \binits{D.A.}},
\bauthor{\bsnm{{Jung}}, \binits{G.V.}},
\bauthor{\bsnm{{Kinzer}}, \binits{R.L.}},
\bauthor{\bsnm{{Purcell}}, \binits{W.R.}},
\bauthor{\bsnm{{Strickman}}, \binits{M.S.}},
\bauthor{\bsnm{{The}}, \binits{L.-S.}},
\bauthor{\bsnm{{Ulmer}}, \binits{M.P.}}:
\batitle{{Compton Gamma Ray Observatory OSSE Observations of SN 1991T}}.
\bjtitle{\apj}
\bvolume{450},
\bfpage{805}
(\byear{1995})
\doiurl{10.1086/176185}
\end{barticle}
\endbibitem

%%% 195
\bibitem[\protect\citeauthoryear{{Georgii} et~al.}{2002}]{Georgii2002}
\begin{barticle}
\bauthor{\bsnm{{Georgii}}, \binits{R.}},
\bauthor{\bsnm{{Pl{\"u}schke}}, \binits{S.}},
\bauthor{\bsnm{{Diehl}}, \binits{R.}},
\bauthor{\bsnm{{Lichti}}, \binits{G.G.}},
\bauthor{\bsnm{{Sch{\"o}nfelder}}, \binits{V.}},
\bauthor{\bsnm{{Bloemen}}, \binits{H.}},
\bauthor{\bsnm{{Hermsen}}, \binits{W.}},
\bauthor{\bsnm{{Ryan}}, \binits{J.}},
\bauthor{\bsnm{{Bennett}}, \binits{K.}}:
\batitle{{COMPTEL upper limits for the $^{56}$Co gamma -ray emission from
  SN1998bu}}.
\bjtitle{\aap}
\bvolume{394},
\bfpage{517}--\blpage{523}
(\byear{2002})
\doiurl{10.1051/0004-6361:20021133}
{\href{https://arxiv.org/abs/astro-ph/0208152}{{arXiv:astro-ph/0208152}}}
{[astro-ph]}
\end{barticle}
\endbibitem

%%% 196
\bibitem[\protect\citeauthoryear{{Isern} et~al.}{2013}]{Isern2013}
\begin{barticle}
\bauthor{\bsnm{{Isern}}, \binits{J.}},
\bauthor{\bsnm{{Jean}}, \binits{P.}},
\bauthor{\bsnm{{Bravo}}, \binits{E.}},
\bauthor{\bsnm{{Diehl}}, \binits{R.}},
\bauthor{\bsnm{{Kn{\"o}dlseder}}, \binits{J.}},
\bauthor{\bsnm{{Domingo}}, \binits{A.}},
\bauthor{\bsnm{{Hirschmann}}, \binits{A.}},
\bauthor{\bsnm{{H{\"o}flich}}, \binits{P.}},
\bauthor{\bsnm{{Lebrun}}, \binits{F.}},
\bauthor{\bsnm{{Renaud}}, \binits{M.}},
\bauthor{\bsnm{{Soldi}}, \binits{S.}},
\bauthor{\bsnm{{Elias-Rosa}}, \binits{N.}},
\bauthor{\bsnm{{Hernanz}}, \binits{M.}},
\bauthor{\bsnm{{Kulebi}}, \binits{B.}},
\bauthor{\bsnm{{Zhang}}, \binits{X.}},
\bauthor{\bsnm{{Badenes}}, \binits{C.}},
\bauthor{\bsnm{{Dom{\'\i}nguez}}, \binits{I.}},
\bauthor{\bsnm{{Garcia-Senz}}, \binits{D.}},
\bauthor{\bsnm{{Jordi}}, \binits{C.}},
\bauthor{\bsnm{{Lichti}}, \binits{G.}},
\bauthor{\bsnm{{Vedrenne}}, \binits{G.}},
\bauthor{\bsnm{{Von Ballmoos}}, \binits{P.}}:
\batitle{{Observation of SN2011fe with INTEGRAL. I. Pre-maximum phase}}.
\bjtitle{\aap}
\bvolume{552},
\bfpage{97}
(\byear{2013})
\doiurl{10.1051/0004-6361/201220303}
{\href{https://arxiv.org/abs/1302.3381}{{arXiv:1302.3381}}}
{[astro-ph.HE]}
\end{barticle}
\endbibitem

%%% 197
\bibitem[\protect\citeauthoryear{{Churazov} et~al.}{2015}]{Churazov2015}
\begin{barticle}
\bauthor{\bsnm{{Churazov}}, \binits{E.}},
\bauthor{\bsnm{{Sunyaev}}, \binits{R.}},
\bauthor{\bsnm{{Isern}}, \binits{J.}},
\bauthor{\bsnm{{Bikmaev}}, \binits{I.}},
\bauthor{\bsnm{{Bravo}}, \binits{E.}},
\bauthor{\bsnm{{Chugai}}, \binits{N.}},
\bauthor{\bsnm{{Grebenev}}, \binits{S.}},
\bauthor{\bsnm{{Jean}}, \binits{P.}},
\bauthor{\bsnm{{Kn{\"o}dlseder}}, \binits{J.}},
\bauthor{\bsnm{{Lebrun}}, \binits{F.}},
\bauthor{\bsnm{{Kuulkers}}, \binits{E.}}:
\batitle{{Gamma-rays from Type Ia Supernova SN2014J}}.
\bjtitle{\apj}
\bvolume{812}(\bissue{1}),
\bfpage{62}
(\byear{2015})
\doiurl{10.1088/0004-637X/812/1/62}
{\href{https://arxiv.org/abs/1502.00255}{{arXiv:1502.00255}}}
{[astro-ph.HE]}
\end{barticle}
\endbibitem

%%% 198
\bibitem[\protect\citeauthoryear{{Diehl} et~al.}{2015}]{Diehl2015}
\begin{barticle}
\bauthor{\bsnm{{Diehl}}, \binits{R.}},
\bauthor{\bsnm{{Siegert}}, \binits{T.}},
\bauthor{\bsnm{{Hillebrandt}}, \binits{W.}},
\bauthor{\bsnm{{Krause}}, \binits{M.}},
\bauthor{\bsnm{{Greiner}}, \binits{J.}},
\bauthor{\bsnm{{Maeda}}, \binits{K.}},
\bauthor{\bsnm{{R{\"o}pke}}, \binits{F.K.}},
\bauthor{\bsnm{{Sim}}, \binits{S.A.}},
\bauthor{\bsnm{{Wang}}, \binits{W.}},
\bauthor{\bsnm{{Zhang}}, \binits{X.}}:
\batitle{{SN2014J gamma rays from the $^{56}$Ni decay chain}}.
\bjtitle{\aap}
\bvolume{574},
\bfpage{72}
(\byear{2015})
\doiurl{10.1051/0004-6361/201424991}
{\href{https://arxiv.org/abs/1409.5477}{{arXiv:1409.5477}}}
{[astro-ph.HE]}
\end{barticle}
\endbibitem

%%% 199
\bibitem[\protect\citeauthoryear{{Diehl} et~al.}{2014}]{Diehl2014}
\begin{barticle}
\bauthor{\bsnm{{Diehl}}, \binits{R.}},
\bauthor{\bsnm{{Siegert}}, \binits{T.}},
\bauthor{\bsnm{{Hillebrandt}}, \binits{W.}},
\bauthor{\bsnm{{Grebenev}}, \binits{S.A.}},
\bauthor{\bsnm{{Greiner}}, \binits{J.}},
\bauthor{\bsnm{{Krause}}, \binits{M.}},
\bauthor{\bsnm{{Kromer}}, \binits{M.}},
\bauthor{\bsnm{{Maeda}}, \binits{K.}},
\bauthor{\bsnm{{R{\"o}pke}}, \binits{F.}},
\bauthor{\bsnm{{Taubenberger}}, \binits{S.}}:
\batitle{{Early $^{56}$Ni decay gamma rays from SN2014J suggest an unusual
  explosion}}.
\bjtitle{Science}
\bvolume{345}(\bissue{6201}),
\bfpage{1162}--\blpage{1165}
(\byear{2014})
\doiurl{10.1126/science.1254738}
{\href{https://arxiv.org/abs/1407.3061}{{arXiv:1407.3061}}}
{[astro-ph.HE]}
\end{barticle}
\endbibitem

%%% 200
\bibitem[\protect\citeauthoryear{{Isern} et~al.}{2016}]{Isern2016}
\begin{barticle}
\bauthor{\bsnm{{Isern}}, \binits{J.}},
\bauthor{\bsnm{{Jean}}, \binits{P.}},
\bauthor{\bsnm{{Bravo}}, \binits{E.}},
\bauthor{\bsnm{{Kn{\"o}dlseder}}, \binits{J.}},
\bauthor{\bsnm{{Lebrun}}, \binits{F.}},
\bauthor{\bsnm{{Churazov}}, \binits{E.}},
\bauthor{\bsnm{{Sunyaev}}, \binits{R.}},
\bauthor{\bsnm{{Domingo}}, \binits{A.}},
\bauthor{\bsnm{{Badenes}}, \binits{C.}},
\bauthor{\bsnm{{Hartmann}}, \binits{D.H.}},
\bauthor{\bsnm{{H{\"o}flich}}, \binits{P.}},
\bauthor{\bsnm{{Renaud}}, \binits{M.}},
\bauthor{\bsnm{{Soldi}}, \binits{S.}},
\bauthor{\bsnm{{Elias-Rosa}}, \binits{N.}},
\bauthor{\bsnm{{Hernanz}}, \binits{M.}},
\bauthor{\bsnm{{Dom{\'\i}nguez}}, \binits{I.}},
\bauthor{\bsnm{{Garc{\'\i}a-Senz}}, \binits{D.}},
\bauthor{\bsnm{{Lichti}}, \binits{G.G.}},
\bauthor{\bsnm{{Vedrenne}}, \binits{G.}},
\bauthor{\bsnm{{Von Ballmoos}}, \binits{P.}}:
\batitle{{Gamma-ray emission from SN2014J near maximum optical light}}.
\bjtitle{\aap}
\bvolume{588},
\bfpage{67}
(\byear{2016})
\doiurl{10.1051/0004-6361/201526941}
{\href{https://arxiv.org/abs/1602.02918}{{arXiv:1602.02918}}}
{[astro-ph.HE]}
\end{barticle}
\endbibitem

%%% 201
\bibitem[\protect\citeauthoryear{{The} and {Burrows}}{2014}]{The2014}
\begin{barticle}
\bauthor{\bsnm{{The}}, \binits{L.-S.}},
\bauthor{\bsnm{{Burrows}}, \binits{A.}}:
\batitle{{Expectations for the Hard X-Ray Continuum and Gamma-Ray Line Fluxes
  from the Type Ia Supernova SN 2014J in M82}}.
\bjtitle{\apj}
\bvolume{786}(\bissue{2}),
\bfpage{141}
(\byear{2014})
\doiurl{10.1088/0004-637X/786/2/141}
{\href{https://arxiv.org/abs/1402.4806}{{arXiv:1402.4806}}}
{[astro-ph.HE]}
\end{barticle}
\endbibitem

%%% 202
\bibitem[\protect\citeauthoryear{{Nomoto} et~al.}{1984}]{Nomoto1984}
\begin{barticle}
\bauthor{\bsnm{{Nomoto}}, \binits{K.}},
\bauthor{\bsnm{{Thielemann}}, \binits{F.-K.}},
\bauthor{\bsnm{{Yokoi}}, \binits{K.}}:
\batitle{{Accreting white dwarf models for type I supernovae. III. Carbon
  deflagration supernovae.}}
\bjtitle{\apj}
\bvolume{286},
\bfpage{644}--\blpage{658}
(\byear{1984})
\doiurl{10.1086/162639}
\end{barticle}
\endbibitem

%%% 203
\bibitem[\protect\citeauthoryear{{Burrows} and {The}}{1990}]{Burrows1990}
\begin{barticle}
\bauthor{\bsnm{{Burrows}}, \binits{A.}},
\bauthor{\bsnm{{The}}, \binits{L.-S.}}:
\batitle{{X- and Gamma-Ray Signatures of Type IA Supernovae}}.
\bjtitle{\apj}
\bvolume{360},
\bfpage{626}
(\byear{1990})
\doiurl{10.1086/169150}
\end{barticle}
\endbibitem

%%% 204
\bibitem[\protect\citeauthoryear{{M{\"u}ller} et~al.}{1991}]{Mueller1991}
\begin{barticle}
\bauthor{\bsnm{{M{\"u}ller}}, \binits{E.}},
\bauthor{\bsnm{{H{\"o}flich}}, \binits{P.}},
\bauthor{\bsnm{{Khokhlov}}, \binits{A.}}:
\batitle{{Type IA supernovae : gamma-rays as predicted by delayed detonation
  modelsand SN 1991T.}}
\bjtitle{\aap}
\bvolume{249},
\bfpage{1}
(\byear{1991})
\end{barticle}
\endbibitem

%%% 205
\bibitem[\protect\citeauthoryear{{H{\"o}flich} et~al.}{1998}]{Hoeflich1998}
\begin{barticle}
\bauthor{\bsnm{{H{\"o}flich}}, \binits{P.}},
\bauthor{\bsnm{{Wheeler}}, \binits{J.C.}},
\bauthor{\bsnm{{Khokhlov}}, \binits{A.}}:
\batitle{{Hard X-Rays and Gamma Rays from Type IA Supernovae}}.
\bjtitle{\apj}
\bvolume{492}(\bissue{1}),
\bfpage{228}--\blpage{245}
(\byear{1998})
\doiurl{10.1086/305018}
{\href{https://arxiv.org/abs/astro-ph/9709033}{{arXiv:astro-ph/9709033}}}
{[astro-ph]}
\end{barticle}
\endbibitem

%%% 206
\bibitem[\protect\citeauthoryear{{G{\'o}mez-Gomar}
  et~al.}{1998}]{Gomez-Gomar1998b}
\begin{barticle}
\bauthor{\bsnm{{G{\'o}mez-Gomar}}, \binits{J.}},
\bauthor{\bsnm{{Isern}}, \binits{J.}},
\bauthor{\bsnm{{Jean}}, \binits{P.}}:
\batitle{{Prospects for Type IA supernova explosion mechanism identification
  with gama-rays.}}
\bjtitle{\mnras}
\bvolume{295},
\bfpage{1}--\blpage{9}
(\byear{1998})
\doiurl{10.1046/j.1365-8711.1998.29511115.x}
{\href{https://arxiv.org/abs/astro-ph/9709048}{{arXiv:astro-ph/9709048}}}
{[astro-ph]}
\end{barticle}
\endbibitem

%%% 207
\bibitem[\protect\citeauthoryear{{Sim} and {Mazzali}}{2008}]{Sim2008}
\begin{barticle}
\bauthor{\bsnm{{Sim}}, \binits{S.A.}},
\bauthor{\bsnm{{Mazzali}}, \binits{P.A.}}:
\batitle{{On the {\ensuremath{\gamma}}-ray emission of Type Ia supernovae}}.
\bjtitle{\mnras}
\bvolume{385}(\bissue{4}),
\bfpage{1681}--\blpage{1690}
(\byear{2008})
\doiurl{10.1111/j.1365-2966.2008.12600.x}
{\href{https://arxiv.org/abs/0710.3313}{{arXiv:0710.3313}}}
{[astro-ph]}
\end{barticle}
\endbibitem

%%% 208
\bibitem[\protect\citeauthoryear{{Maeda} et~al.}{2012}]{Maeda2012}
\begin{barticle}
\bauthor{\bsnm{{Maeda}}, \binits{K.}},
\bauthor{\bsnm{{Terada}}, \binits{Y.}},
\bauthor{\bsnm{{Kasen}}, \binits{D.}},
\bauthor{\bsnm{{R{\"o}pke}}, \binits{F.K.}},
\bauthor{\bsnm{{Bamba}}, \binits{A.}},
\bauthor{\bsnm{{Diehl}}, \binits{R.}},
\bauthor{\bsnm{{Nomoto}}, \binits{K.}},
\bauthor{\bsnm{{Kromer}}, \binits{M.}},
\bauthor{\bsnm{{Seitenzahl}}, \binits{I.R.}},
\bauthor{\bsnm{{Yamaguchi}}, \binits{H.}},
\bauthor{\bsnm{{Tamagawa}}, \binits{T.}},
\bauthor{\bsnm{{Hillebrandt}}, \binits{W.}}:
\batitle{{Prospect of Studying Hard X- and Gamma-Rays from Type Ia
  Supernovae}}.
\bjtitle{\apj}
\bvolume{760}(\bissue{1}),
\bfpage{54}
(\byear{2012})
\doiurl{10.1088/0004-637X/760/1/54}
{\href{https://arxiv.org/abs/1208.2094}{{arXiv:1208.2094}}}
{[astro-ph.HE]}
\end{barticle}
\endbibitem

%%% 209
\bibitem[\protect\citeauthoryear{{Summa} et~al.}{2013}]{Summa2013}
\begin{barticle}
\bauthor{\bsnm{{Summa}}, \binits{A.}},
\bauthor{\bsnm{{Ulyanov}}, \binits{A.}},
\bauthor{\bsnm{{Kromer}}, \binits{M.}},
\bauthor{\bsnm{{Boyer}}, \binits{S.}},
\bauthor{\bsnm{{R{\"o}pke}}, \binits{F.K.}},
\bauthor{\bsnm{{Sim}}, \binits{S.A.}},
\bauthor{\bsnm{{Seitenzahl}}, \binits{I.R.}},
\bauthor{\bsnm{{Fink}}, \binits{M.}},
\bauthor{\bsnm{{Mannheim}}, \binits{K.}},
\bauthor{\bsnm{{Pakmor}}, \binits{R.}},
\bauthor{\bsnm{{Ciaraldi-Schoolmann}}, \binits{F.}},
\bauthor{\bsnm{{Diehl}}, \binits{R.}},
\bauthor{\bsnm{{Maeda}}, \binits{K.}},
\bauthor{\bsnm{{Hillebrandt}}, \binits{W.}}:
\batitle{{Gamma-ray diagnostics of Type Ia supernovae. Predictions of
  observables from three-dimensional modeling}}.
\bjtitle{\aap}
\bvolume{554},
\bfpage{67}
(\byear{2013})
\doiurl{10.1051/0004-6361/201220972}
{\href{https://arxiv.org/abs/1304.2777}{{arXiv:1304.2777}}}
{[astro-ph.SR]}
\end{barticle}
\endbibitem

%%% 210
\bibitem[\protect\citeauthoryear{{Wang} and {Wheeler}}{2008}]{Wang2008}
\begin{barticle}
\bauthor{\bsnm{{Wang}}, \binits{L.}},
\bauthor{\bsnm{{Wheeler}}, \binits{J.C.}}:
\batitle{{Spectropolarimetry of supernovae.}}
\bjtitle{\araa}
\bvolume{46},
\bfpage{433}--\blpage{474}
(\byear{2008})
\doiurl{10.1146/annurev.astro.46.060407.145139}
{\href{https://arxiv.org/abs/0811.1054}{{arXiv:0811.1054}}}
{[astro-ph]}
\end{barticle}
\endbibitem

%%% 211
\bibitem[\protect\citeauthoryear{{Porter} et~al.}{2016}]{Porter2016}
\begin{barticle}
\bauthor{\bsnm{{Porter}}, \binits{A.L.}},
\bauthor{\bsnm{{Leising}}, \binits{M.D.}},
\bauthor{\bsnm{{Williams}}, \binits{G.G.}},
\bauthor{\bsnm{{Milne}}, \binits{P.}},
\bauthor{\bsnm{{Smith}}, \binits{P.}},
\bauthor{\bsnm{{Smith}}, \binits{N.}},
\bauthor{\bsnm{{Bilinski}}, \binits{C.}},
\bauthor{\bsnm{{Hoffman}}, \binits{J.L.}},
\bauthor{\bsnm{{Huk}}, \binits{L.}},
\bauthor{\bsnm{{Leonard}}, \binits{D.C.}}:
\batitle{{Asymmetries in SN 2014J near Maximum Light Revealed through
  Spectropolarimetry}}.
\bjtitle{\apj}
\bvolume{828}(\bissue{1}),
\bfpage{24}
(\byear{2016})
\doiurl{10.3847/0004-637X/828/1/24}
{\href{https://arxiv.org/abs/1605.03994}{{arXiv:1605.03994}}}
{[astro-ph.HE]}
\end{barticle}
\endbibitem

%%% 212
\bibitem[\protect\citeauthoryear{{Cikota} et~al.}{2019}]{Cikota2019}
\begin{barticle}
\bauthor{\bsnm{{Cikota}}, \binits{A.}},
\bauthor{\bsnm{{Patat}}, \binits{F.}},
\bauthor{\bsnm{{Wang}}, \binits{L.}},
\bauthor{\bsnm{{Wheeler}}, \binits{J.C.}},
\bauthor{\bsnm{{Bulla}}, \binits{M.}},
\bauthor{\bsnm{{Baade}}, \binits{D.}},
\bauthor{\bsnm{{H{\"o}flich}}, \binits{P.}},
\bauthor{\bsnm{{Cikota}}, \binits{S.}},
\bauthor{\bsnm{{Clocchiatti}}, \binits{A.}},
\bauthor{\bsnm{{Maund}}, \binits{J.R.}},
\bauthor{\bsnm{{Stevance}}, \binits{H.F.}},
\bauthor{\bsnm{{Yang}}, \binits{Y.}}:
\batitle{{Linear spectropolarimetry of 35 Type Ia supernovae with VLT/FORS: an
  analysis of the Si II line polarization}}.
\bjtitle{\mnras}
\bvolume{490}(\bissue{1}),
\bfpage{578}--\blpage{599}
(\byear{2019})
\doiurl{10.1093/mnras/stz2322}
{\href{https://arxiv.org/abs/1908.07526}{{arXiv:1908.07526}}}
{[astro-ph.HE]}
\end{barticle}
\endbibitem

%%% 213
\bibitem[\protect\citeauthoryear{{H{\"o}flich} et~al.}{2023}]{Hoeflich2023}
\begin{barticle}
\bauthor{\bsnm{{H{\"o}flich}}, \binits{P.}},
\bauthor{\bsnm{{Yang}}, \binits{Y.}},
\bauthor{\bsnm{{Baade}}, \binits{D.}},
\bauthor{\bsnm{{Cikota}}, \binits{A.}},
\bauthor{\bsnm{{Maund}}, \binits{J.R.}},
\bauthor{\bsnm{{Mishra}}, \binits{D.}},
\bauthor{\bsnm{{Patat}}, \binits{F.}},
\bauthor{\bsnm{{Patra}}, \binits{K.C.}},
\bauthor{\bsnm{{Wang}}, \binits{L.}},
\bauthor{\bsnm{{Wheeler}}, \binits{J.C.}},
\bauthor{\bsnm{{Filippenko}}, \binits{A.V.}},
\bauthor{\bsnm{{Gal-Yam}}, \binits{A.}},
\bauthor{\bsnm{{Schulze}}, \binits{S.}}:
\batitle{{The core normal Type Ia supernova 2019np - an overall spherical
  explosion with an aspherical surface layer and an aspherical $^{56}$Ni
  core}}.
\bjtitle{\mnras}
\bvolume{520}(\bissue{1}),
\bfpage{560}--\blpage{582}
(\byear{2023})
\doiurl{10.1093/mnras/stad172}
{\href{https://arxiv.org/abs/2301.04721}{{arXiv:2301.04721}}}
{[astro-ph.SR]}
\end{barticle}
\endbibitem

%%% 214
\bibitem[\protect\citeauthoryear{{R{\"o}pke}}{2005}]{Roepke2005}
\begin{barticle}
\bauthor{\bsnm{{R{\"o}pke}}, \binits{F.K.}}:
\batitle{{Following multi-dimensional type Ia supernova explosion models to
  homologous expansion}}.
\bjtitle{\aap}
\bvolume{432}(\bissue{3}),
\bfpage{969}--\blpage{983}
(\byear{2005})
\doiurl{10.1051/0004-6361:20041700}
{\href{https://arxiv.org/abs/astro-ph/0408296}{{arXiv:astro-ph/0408296}}}
{[astro-ph]}
\end{barticle}
\endbibitem

%%% 215
\bibitem[\protect\citeauthoryear{{Diamond} et~al.}{2015}]{Diamond2015}
\begin{barticle}
\bauthor{\bsnm{{Diamond}}, \binits{T.R.}},
\bauthor{\bsnm{{H{\"o}flich}}, \binits{P.}},
\bauthor{\bsnm{{Gerardy}}, \binits{C.L.}}:
\batitle{{Late-time Near-infrared Observations of SN 2005df}}.
\bjtitle{\apj}
\bvolume{806}(\bissue{1}),
\bfpage{107}
(\byear{2015})
\doiurl{10.1088/0004-637X/806/1/107}
{\href{https://arxiv.org/abs/1410.6759}{{arXiv:1410.6759}}}
{[astro-ph.SR]}
\end{barticle}
\endbibitem

%%% 216
\bibitem[\protect\citeauthoryear{{Leising}}{2022}]{Leising2022}
\begin{barticle}
\bauthor{\bsnm{{Leising}}, \binits{M.D.}}:
\batitle{{Deriving Thermonuclear Supernova Properties from Gamma-Ray Line
  Measurements}}.
\bjtitle{\apj}
\bvolume{932}(\bissue{1}),
\bfpage{63}
(\byear{2022})
\doiurl{10.3847/1538-4357/ac6efa}
{\href{https://arxiv.org/abs/2205.06348}{{arXiv:2205.06348}}}
{[astro-ph.HE]}
\end{barticle}
\endbibitem

%%% 217
\bibitem[\protect\citeauthoryear{{Tiwari} et~al.}{2022}]{Tiwari2022}
\begin{barticle}
\bauthor{\bsnm{{Tiwari}}, \binits{V.}},
\bauthor{\bsnm{{Graur}}, \binits{O.}},
\bauthor{\bsnm{{Fisher}}, \binits{R.}},
\bauthor{\bsnm{{Seitenzahl}}, \binits{I.}},
\bauthor{\bsnm{{Leung}}, \binits{S.-C.}},
\bauthor{\bsnm{{Nomoto}}, \binits{K.}},
\bauthor{\bsnm{{Perets}}, \binits{H.B.}},
\bauthor{\bsnm{{Shen}}, \binits{K.}}:
\batitle{{The late-time light curves of Type Ia supernovae: confronting models
  with observations}}.
\bjtitle{\mnras}
\bvolume{515}(\bissue{3}),
\bfpage{3703}--\blpage{3715}
(\byear{2022})
\doiurl{10.1093/mnras/stac1618}
{\href{https://arxiv.org/abs/2206.02812}{{arXiv:2206.02812}}}
{[astro-ph.HE]}
\end{barticle}
\endbibitem

%%% 218
\bibitem[\protect\citeauthoryear{{Troja} et~al.}{2014}]{Troja2014}
\begin{barticle}
\bauthor{\bsnm{{Troja}}, \binits{E.}},
\bauthor{\bsnm{{Segreto}}, \binits{A.}},
\bauthor{\bsnm{{La Parola}}, \binits{V.}},
\bauthor{\bsnm{{Hartmann}}, \binits{D.}},
\bauthor{\bsnm{{Baumgartner}}, \binits{W.}},
\bauthor{\bsnm{{Markwardt}}, \binits{C.}},
\bauthor{\bsnm{{Barthelmy}}, \binits{S.}},
\bauthor{\bsnm{{Cusumano}}, \binits{G.}},
\bauthor{\bsnm{{Gehrels}}, \binits{N.}}:
\batitle{{Swift/BAT Detection of Hard X-Rays from Tycho's Supernova Remnant:
  Evidence for Titanium-44}}.
\bjtitle{\apjl}
\bvolume{797}(\bissue{1}),
\bfpage{6}
(\byear{2014})
\doiurl{10.1088/2041-8205/797/1/L6}
{\href{https://arxiv.org/abs/1411.0991}{{arXiv:1411.0991}}}
{[astro-ph.HE]}
\end{barticle}
\endbibitem

%%% 219
\bibitem[\protect\citeauthoryear{{Lopez} et~al.}{2015}]{Lopez2015}
\begin{barticle}
\bauthor{\bsnm{{Lopez}}, \binits{L.A.}},
\bauthor{\bsnm{{Grefenstette}}, \binits{B.W.}},
\bauthor{\bsnm{{Reynolds}}, \binits{S.P.}},
\bauthor{\bsnm{{An}}, \binits{H.}},
\bauthor{\bsnm{{Boggs}}, \binits{S.E.}},
\bauthor{\bsnm{{Christensen}}, \binits{F.E.}},
\bauthor{\bsnm{{Craig}}, \binits{W.W.}},
\bauthor{\bsnm{{Eriksen}}, \binits{K.A.}},
\bauthor{\bsnm{{Fryer}}, \binits{C.L.}},
\bauthor{\bsnm{{Hailey}}, \binits{C.J.}},
\bauthor{\bsnm{{Harrison}}, \binits{F.A.}},
\bauthor{\bsnm{{Madsen}}, \binits{K.K.}},
\bauthor{\bsnm{{Stern}}, \binits{D.K.}},
\bauthor{\bsnm{{Zhang}}, \binits{W.W.}},
\bauthor{\bsnm{{Zoglauer}}, \binits{A.}}:
\batitle{{A Spatially Resolved Study of the Synchrotron Emission and Titanium
  in Tycho{\textquoteright}s Supernova Remnant Using NuSTAR}}.
\bjtitle{\apj}
\bvolume{814}(\bissue{2}),
\bfpage{132}
(\byear{2015})
\doiurl{10.1088/0004-637X/814/2/132}
{\href{https://arxiv.org/abs/1504.07238}{{arXiv:1504.07238}}}
{[astro-ph.HE]}
\end{barticle}
\endbibitem

%%% 220
\bibitem[\protect\citeauthoryear{{Woosley} et~al.}{1986}]{Woosley1986}
\begin{barticle}
\bauthor{\bsnm{{Woosley}}, \binits{S.E.}},
\bauthor{\bsnm{{Taam}}, \binits{R.E.}},
\bauthor{\bsnm{{Weaver}}, \binits{T.A.}}:
\batitle{{Models for Type I Supernova. I. Detonations in White Dwarfs}}.
\bjtitle{\apj}
\bvolume{301},
\bfpage{601}
(\byear{1986})
\doiurl{10.1086/163926}
\end{barticle}
\endbibitem

%%% 221
\bibitem[\protect\citeauthoryear{{Leung} and {Nomoto}}{2020}]{Leung2020}
\begin{barticle}
\bauthor{\bsnm{{Leung}}, \binits{S.-C.}},
\bauthor{\bsnm{{Nomoto}}, \binits{K.}}:
\batitle{{Explosive Nucleosynthesis in Sub-Chandrasekhar-mass White Dwarf
  Models for Type Ia Supernovae: Dependence on Model Parameters}}.
\bjtitle{\apj}
\bvolume{888}(\bissue{2}),
\bfpage{80}
(\byear{2020})
\doiurl{10.3847/1538-4357/ab5c1f}
{\href{https://arxiv.org/abs/1901.10007}{{arXiv:1901.10007}}}
{[astro-ph.HE]}
\end{barticle}
\endbibitem

%%% 222
\bibitem[\protect\citeauthoryear{{Roy} et~al.}{2022}]{Roy2022}
\begin{barticle}
\bauthor{\bsnm{{Roy}}, \binits{N.C.}},
\bauthor{\bsnm{{Tiwari}}, \binits{V.}},
\bauthor{\bsnm{{Bobrick}}, \binits{A.}},
\bauthor{\bsnm{{Kosakowski}}, \binits{D.}},
\bauthor{\bsnm{{Fisher}}, \binits{R.}},
\bauthor{\bsnm{{Perets}}, \binits{H.B.}},
\bauthor{\bsnm{{Kashyap}}, \binits{R.}},
\bauthor{\bsnm{{Lor{\'e}n-Aguilar}}, \binits{P.}},
\bauthor{\bsnm{{Garc{\'\i}a-Berro}}, \binits{E.}}:
\batitle{{3D Hydrodynamical Simulations of Helium-ignited Double-degenerate
  White Dwarf Mergers}}.
\bjtitle{\apjl}
\bvolume{932}(\bissue{2}),
\bfpage{24}
(\byear{2022})
\doiurl{10.3847/2041-8213/ac75e7}
{\href{https://arxiv.org/abs/2204.09683}{{arXiv:2204.09683}}}
{[astro-ph.SR]}
\end{barticle}
\endbibitem

%%% 223
\bibitem[\protect\citeauthoryear{{Panther} et~al.}{2021}]{Panther2021}
\begin{barticle}
\bauthor{\bsnm{{Panther}}, \binits{F.H.}},
\bauthor{\bsnm{{Seitenzahl}}, \binits{I.R.}},
\bauthor{\bsnm{{Ruiter}}, \binits{A.J.}},
\bauthor{\bsnm{{Siegert}}, \binits{T.}},
\bauthor{\bsnm{{Sim}}, \binits{S.}},
\bauthor{\bsnm{{Crocker}}, \binits{R.M.}}:
\batitle{{Prospects of direct detection of $^{48}$V gamma-rays from
  thermonuclear supernovae}}.
\bjtitle{\mnras}
\bvolume{508}(\bissue{2}),
\bfpage{1590}--\blpage{1598}
(\byear{2021})
\doiurl{10.1093/mnras/stab2701}
{\href{https://arxiv.org/abs/2103.16840}{{arXiv:2103.16840}}}
{[astro-ph.HE]}
\end{barticle}
\endbibitem

%%% 224
\bibitem[\protect\citeauthoryear{{Sim} et~al.}{2012}]{Sim2012}
\begin{barticle}
\bauthor{\bsnm{{Sim}}, \binits{S.A.}},
\bauthor{\bsnm{{Fink}}, \binits{M.}},
\bauthor{\bsnm{{Kromer}}, \binits{M.}},
\bauthor{\bsnm{{R{\"o}pke}}, \binits{F.K.}},
\bauthor{\bsnm{{Ruiter}}, \binits{A.J.}},
\bauthor{\bsnm{{Hillebrandt}}, \binits{W.}}:
\batitle{{2D simulations of the double-detonation model for thermonuclear
  transients from low-mass carbon-oxygen white dwarfs}}.
\bjtitle{\mnras}
\bvolume{420}(\bissue{4}),
\bfpage{3003}--\blpage{3016}
(\byear{2012})
\doiurl{10.1111/j.1365-2966.2011.20162.x}
{\href{https://arxiv.org/abs/1111.2117}{{arXiv:1111.2117}}}
{[astro-ph.HE]}
\end{barticle}
\endbibitem

%%% 225
\bibitem[\protect\citeauthoryear{{Kinzer} et~al.}{1997}]{Kinzer1997_CGB}
\begin{barticle}
\bauthor{\bsnm{{Kinzer}}, \binits{R.L.}},
\bauthor{\bsnm{{Jung}}, \binits{G.V.}},
\bauthor{\bsnm{{Gruber}}, \binits{D.E.}},
\bauthor{\bsnm{{Matteson}}, \binits{J.L.}},
\bauthor{\bsnm{{Peterson}}},
\bauthor{\bsnm{{L.~E.}}}:
\batitle{{Diffuse Cosmic Gamma Radiation Measured by HEAO 1}}.
\bjtitle{\apj}
\bvolume{475}(\bissue{1}),
\bfpage{361}--\blpage{372}
(\byear{1997})
\doiurl{10.1086/303507}
\end{barticle}
\endbibitem

%%% 226
\bibitem[\protect\citeauthoryear{{Gruber} et~al.}{1999}]{Gruber1999_CGB}
\begin{barticle}
\bauthor{\bsnm{{Gruber}}, \binits{D.E.}},
\bauthor{\bsnm{{Matteson}}, \binits{J.L.}},
\bauthor{\bsnm{{Peterson}}, \binits{L.E.}},
\bauthor{\bsnm{{Jung}}, \binits{G.V.}}:
\batitle{{The Spectrum of Diffuse Cosmic Hard X-Rays Measured with HEAO-1}}.
\bjtitle{\apj}
\bvolume{520},
\bfpage{124}--\blpage{129}
(\byear{1999})
\doiurl{10.1086/307450}
\end{barticle}
\endbibitem

%%% 227
\bibitem[\protect\citeauthoryear{{Watanabe} et~al.}{2000}]{Watanabe2000_CGB}
\begin{barticle}
\bauthor{\bsnm{{Watanabe}}, \binits{K.}},
\bauthor{\bsnm{{Leising}}, \binits{M.D.}},
\bauthor{\bsnm{{Share}}, \binits{G.H.}},
\bauthor{\bsnm{{Kinzer}}, \binits{R.L.}}:
\batitle{{The MeV cosmic gamma‐ray background measured with SMM}}.
\bjtitle{AIP Conference Proceedings}
\bvolume{510},
\bfpage{471}--\blpage{475}
(\byear{2000})
\doiurl{10.1063/1.1290203}
\end{barticle}
\endbibitem

%%% 228
\bibitem[\protect\citeauthoryear{{Weidenspointner}
  et~al.}{2000}]{Weidenspointner2000_CGB}
\begin{barticle}
\bauthor{\bsnm{{Weidenspointner}}, \binits{G.}},
\bauthor{\bsnm{{Varendorff}}, \binits{M.}},
\bauthor{\bsnm{{Oberlack}}, \binits{U.}},
\bauthor{\bsnm{{Morris}}, \binits{D.}},
\bauthor{\bsnm{{Pl{\"u}schke}}, \binits{S.}},
\bauthor{\bsnm{{Diehl}}, \binits{R.}},
\bauthor{\bsnm{{Kappadath}}, \binits{S.C.}},
\bauthor{\bsnm{{McConnell}}, \binits{M.L.}},
\bauthor{\bsnm{{Ryan}}, \binits{J.}},
\bauthor{\bsnm{{Sch{\"o}nfelder}}, \binits{V.}},
\bauthor{\bsnm{{Steinle}}, \binits{H.}}:
\batitle{{The cosmic diffuse gamma‐ray background spectrum between 0.8 and 30
  MeV measured with COMPTEL}}.
\bjtitle{AIP Conference Proceedings}
\bvolume{510},
\bfpage{467}--\blpage{471}
(\byear{2000})
\doiurl{10.1063/1.1290201}
\end{barticle}
\endbibitem

%%% 229
\bibitem[\protect\citeauthoryear{{Churazov} et~al.}{2007}]{Churazov2007_CGB}
\begin{barticle}
\bauthor{\bsnm{{Churazov}}, \binits{E.}},
\bauthor{\bsnm{{Sunyaev}}, \binits{R.}},
\bauthor{\bsnm{{Revnivtsev}}, \binits{M.}},
\bauthor{\bsnm{{Sazonov}}, \binits{S.}},
\bauthor{\bsnm{{Yao}}, \binits{Y.}},
\bauthor{\bsnm{{Ubertini}}, \binits{P.}},
\bauthor{\bsnm{{Lebrun}}, \binits{F.}},
\bauthor{\bsnm{{Terrier}}, \binits{R.}},
\bauthor{\bsnm{{Broderick}}, \binits{J.W.}},
\bauthor{\bsnm{{Shtykovskiy}}, \binits{P.}},
\bauthor{\bsnm{{Jourdain}}, \binits{E.}},
\bauthor{\bsnm{{Natalucci}}, \binits{L.}}:
\batitle{{INTEGRAL observations of the cosmic X-ray background in the 5--100
  keV range via occultation by the Earth}}.
\bjtitle{A\&A}
\bvolume{467},
\bfpage{529}--\blpage{540}
(\byear{2007})
\doiurl{10.1051/0004-6361:20066230}
\end{barticle}
\endbibitem

%%% 230
\bibitem[\protect\citeauthoryear{{T{\"u}rler} et~al.}{2010}]{Tuerler2010}
\begin{barticle}
\bauthor{\bsnm{{T{\"u}rler}}, \binits{M.}},
\bauthor{\bsnm{{Chernyakova}}, \binits{M.}},
\bauthor{\bsnm{{Courvoisier}}, \binits{T.J.-L.}},
\bauthor{\bsnm{{Lubi{\'n}ski}}, \binits{P.}},
\bauthor{\bsnm{{Neronov}}, \binits{A.}},
\bauthor{\bsnm{{Produit}}, \binits{N.}},
\bauthor{\bsnm{{Walter}}, \binits{R.}}:
\batitle{{INTEGRAL hard X-ray spectra of the cosmic X-ray background and
  Galactic ridge emission}}.
\bjtitle{\aap}
\bvolume{512},
\bfpage{49}
(\byear{2010})
\doiurl{10.1051/0004-6361/200913072}
{\href{https://arxiv.org/abs/1001.2110}{{arXiv:1001.2110}}}
{[astro-ph.CO]}
\end{barticle}
\endbibitem

%%% 231
\bibitem[\protect\citeauthoryear{{The} et~al.}{1993}]{1993ApJ...403...32T}
\begin{barticle}
\bauthor{\bsnm{{The}}, \binits{L.-S.}},
\bauthor{\bsnm{{Leising}}, \binits{M.D.}},
\bauthor{\bsnm{{Clayton}}, \binits{D.D.}}:
\batitle{{The Cosmic Gamma-Ray Background from Type IA Supernovae}}.
\bjtitle{\apj}
\bvolume{403},
\bfpage{32}
(\byear{1993})
\doiurl{10.1086/172180}
\end{barticle}
\endbibitem

%%% 232
\bibitem[\protect\citeauthoryear{{Ueda} et~al.}{2003}]{Ueda2003_CGB}
\begin{barticle}
\bauthor{\bsnm{{Ueda}}, \binits{Y.}},
\bauthor{\bsnm{{Akiyama}}, \binits{M.}},
\bauthor{\bsnm{{Ohta}}, \binits{K.}},
\bauthor{\bsnm{{Miyaji}}, \binits{T.}}:
\batitle{{Cosmological Evolution of the Hard X-Ray Active Galactic Nucleus
  Luminosity Function and the Origin of the Hard X-Ray Background}}.
\bjtitle{\apj}
\bvolume{598}(\bissue{2}),
\bfpage{886}--\blpage{908}
(\byear{2003})
\doiurl{10.1086/378940}
{\href{https://arxiv.org/abs/astro-ph/0308140}{{arXiv:astro-ph/0308140}}}
{[astro-ph]}
\end{barticle}
\endbibitem

%%% 233
\bibitem[\protect\citeauthoryear{{Inoue}}{2014}]{Inoue2014_CGB}
\begin{botherref}
\oauthor{\bsnm{{Inoue}}, \binits{Y.}}:
{Cosmic Gamma-ray Background Radiation}.
arXiv e-prints,
1412--3886
(2014)
\doiurl{10.48550/arXiv.1412.3886}
{\href{https://arxiv.org/abs/1412.3886}{{arXiv:1412.3886}}}
{[astro-ph.HE]}
\end{botherref}
\endbibitem

%%% 234
\bibitem[\protect\citeauthoryear{{Chugai}}{2023}]{Chugai2023_CGB}
\begin{barticle}
\bauthor{\bsnm{{Chugai}}, \binits{N.N.}}:
\batitle{{Cosmic Abundance of Iron}}.
\bjtitle{Astronomy Letters}
\bvolume{49}(\bissue{5}),
\bfpage{209}--\blpage{215}
(\byear{2023})
\doiurl{10.1134/S1063773723050018}
{\href{https://arxiv.org/abs/2307.00944}{{arXiv:2307.00944}}}
{[astro-ph.HE]}
\end{barticle}
\endbibitem

%%% 235
\bibitem[\protect\citeauthoryear{{Ruiz-Lapuente} and
  {Korobkin}}{2020}]{Ruiz-Lapuente2020_CGB}
\begin{barticle}
\bauthor{\bsnm{{Ruiz-Lapuente}}, \binits{P.}},
\bauthor{\bsnm{{Korobkin}}, \binits{O.}}:
\batitle{{Gamma-Rays from Kilonovae and the Cosmic Gamma-Ray Background}}.
\bjtitle{\apj}
\bvolume{892}(\bissue{1}),
\bfpage{45}
(\byear{2020})
\doiurl{10.3847/1538-4357/ab744e}
{\href{https://arxiv.org/abs/1912.11974}{{arXiv:1912.11974}}}
{[astro-ph.HE]}
\end{barticle}
\endbibitem

%%% 236
\bibitem[\protect\citeauthoryear{{Lacki} et~al.}{2014}]{Lacki2014_CGB}
\begin{barticle}
\bauthor{\bsnm{{Lacki}}, \binits{B.C.}},
\bauthor{\bsnm{{Horiuchi}}, \binits{S.}},
\bauthor{\bsnm{{Beacom}}, \binits{J.F.}}:
\batitle{{The Star-forming Galaxy Contribution to the Cosmic MeV and GeV
  Gamma-Ray Background}}.
\bjtitle{\apj}
\bvolume{786}(\bissue{1}),
\bfpage{40}
(\byear{2014})
\doiurl{10.1088/0004-637X/786/1/40}
{\href{https://arxiv.org/abs/1206.0772}{{arXiv:1206.0772}}}
{[astro-ph.HE]}
\end{barticle}
\endbibitem

%%% 237
\bibitem[\protect\citeauthoryear{{Iguaz} et~al.}{2021}]{Iguaz2021_CGB}
\begin{barticle}
\bauthor{\bsnm{{Iguaz}}, \binits{J.}},
\bauthor{\bsnm{{Serpico}}, \binits{P.D.}},
\bauthor{\bsnm{{Siegert}}, \binits{T.}}:
\batitle{{Isotropic x-ray bound on primordial black hole dark matter}}.
\bjtitle{\prd}
\bvolume{103}(\bissue{10}),
\bfpage{103025}
(\byear{2021})
\doiurl{10.1103/PhysRevD.103.103025}
{\href{https://arxiv.org/abs/2104.03145}{{arXiv:2104.03145}}}
{[astro-ph.CO]}
\end{barticle}
\endbibitem

%%% 238
\bibitem[\protect\citeauthoryear{Arnett}{1982}]{Arnett1982_SNIa}
\begin{barticle}
\bauthor{\bsnm{Arnett}, \binits{W.D.}}:
\batitle{{Analytic solutions for the early part of the light curve of Type I
  supernovae}}.
\bjtitle{Astrophysical Journal}
\bvolume{253},
\bfpage{785}--\blpage{797}
(\byear{1982})
\doiurl{10.1086/159681}
\end{barticle}
\endbibitem

%%% 239
\bibitem[\protect\citeauthoryear{Phillips}{1993}]{Phillips1993_SNIa}
\begin{barticle}
\bauthor{\bsnm{Phillips}, \binits{M.M.}}:
\batitle{{The Absolute Magnitudes of Type Ia Supernovae}}.
\bjtitle{Astrophysical Journal Letters}
\bvolume{413},
\bfpage{105}--\blpage{108}
(\byear{1993})
\doiurl{10.1086/186970}
\end{barticle}
\endbibitem

%%% 240
\bibitem[\protect\citeauthoryear{{Timmes} and {Woosley}}{1993}]{Timmes1993_CGB}
\begin{barticle}
\bauthor{\bsnm{{Timmes}}, \binits{F.X.}},
\bauthor{\bsnm{{Woosley}}, \binits{S.E.}}:
\batitle{{The metagalactic gamma-ray background from Type Ia supernovae}}.
\bjtitle{\apj}
\bvolume{403},
\bfpage{32}--\blpage{46}
(\byear{1993})
\doiurl{10.1086/172185}
\end{barticle}
\endbibitem

%%% 241
\bibitem[\protect\citeauthoryear{{Watanabe} et~al.}{1999}]{Watanabe1999_CGB}
\begin{barticle}
\bauthor{\bsnm{{Watanabe}}, \binits{K.}},
\bauthor{\bsnm{{Hartmann}}, \binits{D.H.}},
\bauthor{\bsnm{{Leising}}, \binits{M.D.}},
\bauthor{\bsnm{{The}}, \binits{L.-S.}}:
\batitle{{The Diffuse Gamma-Ray Background from Supernovae}}.
\bjtitle{\apj}
\bvolume{516}(\bissue{1}),
\bfpage{285}--\blpage{296}
(\byear{1999})
\doiurl{10.1086/307110}
{\href{https://arxiv.org/abs/astro-ph/9809197}{{arXiv:astro-ph/9809197}}}
{[astro-ph]}
\end{barticle}
\endbibitem

%%% 242
\bibitem[\protect\citeauthoryear{{Ahn} et~al.}{2005}]{Ahn2005_CGB}
\begin{barticle}
\bauthor{\bsnm{{Ahn}}, \binits{K.}},
\bauthor{\bsnm{{Komatsu}}, \binits{E.}},
\bauthor{\bsnm{{H{\"o}flich}}, \binits{P.}}:
\batitle{{Cosmic gamma-ray background from type Ia supernovae reexamined:
  Evidence for missing gamma rays at MeV energy}}.
\bjtitle{\prd}
\bvolume{71}(\bissue{12}),
\bfpage{121301}
(\byear{2005})
\doiurl{10.1103/PhysRevD.71.121301}
{\href{https://arxiv.org/abs/astro-ph/0506126}{{arXiv:astro-ph/0506126}}}
{[astro-ph]}
\end{barticle}
\endbibitem

%%% 243
\bibitem[\protect\citeauthoryear{{Lien} and {Fields}}{2012}]{Lien2012_CGB}
\begin{barticle}
\bauthor{\bsnm{{Lien}}, \binits{A.}},
\bauthor{\bsnm{{Fields}}, \binits{B.D.}}:
\batitle{{The Diffuse Gamma-Ray Background from Type Ia Supernovae}}.
\bjtitle{\apj}
\bvolume{747}(\bissue{2}),
\bfpage{120}
(\byear{2012})
\doiurl{10.1088/0004-637X/747/2/120}
{\href{https://arxiv.org/abs/1201.3447}{{arXiv:1201.3447}}}
{[astro-ph.CO]}
\end{barticle}
\endbibitem

%%% 244
\bibitem[\protect\citeauthoryear{{Ruiz-Lapuente}
  et~al.}{2016}]{Ruiz-Lapuente2016_CGB}
\begin{barticle}
\bauthor{\bsnm{{Ruiz-Lapuente}}, \binits{P.}},
\bauthor{\bsnm{{The}}, \binits{L.-S.}},
\bauthor{\bsnm{{Hartmann}}, \binits{D.H.}},
\bauthor{\bsnm{{Ajello}}, \binits{M.}},
\bauthor{\bsnm{{Canal}}, \binits{R.}},
\bauthor{\bsnm{{R{\"o}pke}}, \binits{F.K.}},
\bauthor{\bsnm{{Ohlmann}}, \binits{S.T.}},
\bauthor{\bsnm{{Hillebrandt}}, \binits{W.}}:
\batitle{{The Origin of the Cosmic Gamma-ray Background in the MeV Range}}.
\bjtitle{\apj}
\bvolume{820}(\bissue{2}),
\bfpage{142}
(\byear{2016})
\doiurl{10.3847/0004-637X/820/2/142}
{\href{https://arxiv.org/abs/1502.06116}{{arXiv:1502.06116}}}
{[astro-ph.HE]}
\end{barticle}
\endbibitem

%%% 245
\bibitem[\protect\citeauthoryear{{Breitschwerdt}
  et~al.}{1996}]{Breitschwerdt1996_LB}
\begin{barticle}
\bauthor{\bsnm{{Breitschwerdt}}, \binits{D.}},
\bauthor{\bsnm{{Egger}}, \binits{R.}},
\bauthor{\bsnm{{Freyberg}}, \binits{M.J.}},
\bauthor{\bsnm{{Frisch}}, \binits{P.C.}},
\bauthor{\bsnm{{Vallerga}}, \binits{J.V.}}:
\batitle{{The Local Bubble Origin and Evolution}}.
\bjtitle{\ssr}
\bvolume{78}(\bissue{1-2}),
\bfpage{183}--\blpage{198}
(\byear{1996})
\doiurl{10.1007/BF00170805}
\end{barticle}
\endbibitem

%%% 246
\bibitem[\protect\citeauthoryear{{Ramaty} and
  {Lingenfelter}}{1980}]{Ramaty1980}
\begin{bbook}
\bauthor{\bsnm{{Ramaty}}, \binits{R.}},
\bauthor{\bsnm{{Lingenfelter}}, \binits{R.E.}}:
\bbtitle{{Interpretations and Implications of Gamma Ray Lines from Solar
  Flares, the Galactic Center and Gamma Ray Transients.}},
(\byear{1980})
\end{bbook}
\endbibitem

%%% 247
\bibitem[\protect\citeauthoryear{{Green}}{2019}]{Green2019}
\begin{barticle}
\bauthor{\bsnm{{Green}}, \binits{D.A.}}:
\batitle{{A revised catalogue of 294 Galactic supernova remnants}}.
\bjtitle{Journal of Astrophysics and Astronomy}
\bvolume{40}(\bissue{4}),
\bfpage{36}
(\byear{2019})
\doiurl{10.1007/s12036-019-9601-6}
{\href{https://arxiv.org/abs/1907.02638}{{arXiv:1907.02638}}}
{[astro-ph.GA]}
\end{barticle}
\endbibitem

%%% 248
\bibitem[\protect\citeauthoryear{{Boggs} et~al.}{2015}]{Boggs2015}
\begin{barticle}
\bauthor{\bsnm{{Boggs}}, \binits{S.E.}},
\bauthor{\bsnm{{Harrison}}, \binits{F.A.}},
\bauthor{\bsnm{{Miyasaka}}, \binits{H.}},
\bauthor{\bsnm{{Grefenstette}}, \binits{B.W.}},
\bauthor{\bsnm{{Zoglauer}}, \binits{A.}},
\bauthor{\bsnm{{Fryer}}, \binits{C.L.}},
\bauthor{\bsnm{{Reynolds}}, \binits{S.P.}},
\bauthor{\bsnm{{Alexander}}, \binits{D.M.}},
\bauthor{\bsnm{{An}}, \binits{H.}},
\bauthor{\bsnm{{Barret}}, \binits{D.}},
\bauthor{\bsnm{{Christensen}}, \binits{F.E.}},
\bauthor{\bsnm{{Craig}}, \binits{W.W.}},
\bauthor{\bsnm{{Forster}}, \binits{K.}},
\bauthor{\bsnm{{Giommi}}, \binits{P.}},
\bauthor{\bsnm{{Hailey}}, \binits{C.J.}},
\bauthor{\bsnm{{Hornstrup}}, \binits{A.}},
\bauthor{\bsnm{{Kitaguchi}}, \binits{T.}},
\bauthor{\bsnm{{Koglin}}, \binits{J.E.}},
\bauthor{\bsnm{{Madsen}}, \binits{K.K.}},
\bauthor{\bsnm{{Mao}}, \binits{P.H.}},
\bauthor{\bsnm{{Mori}}, \binits{K.}},
\bauthor{\bsnm{{Perri}}, \binits{M.}},
\bauthor{\bsnm{{Pivovaroff}}, \binits{M.J.}},
\bauthor{\bsnm{{Puccetti}}, \binits{S.}},
\bauthor{\bsnm{{Rana}}, \binits{V.}},
\bauthor{\bsnm{{Stern}}, \binits{D.}},
\bauthor{\bsnm{{Westergaard}}, \binits{N.J.}},
\bauthor{\bsnm{{Zhang}}, \binits{W.W.}}:
\batitle{{$^{44}$Ti gamma-ray emission lines from SN1987A reveal an asymmetric
  explosion}}.
\bjtitle{Science}
\bvolume{348}(\bissue{6235}),
\bfpage{670}--\blpage{671}
(\byear{2015})
\doiurl{10.1126/science.aaa2259}
\end{barticle}
\endbibitem

%%% 249
\bibitem[\protect\citeauthoryear{{The} et~al.}{2006}]{The2006}
\begin{barticle}
\bauthor{\bsnm{{The}}, \binits{L.-S.}},
\bauthor{\bsnm{{Clayton}}, \binits{D.D.}},
\bauthor{\bsnm{{Diehl}}, \binits{R.}},
\bauthor{\bsnm{{Hartmann}}, \binits{D.H.}},
\bauthor{\bsnm{{Iyudin}}, \binits{A.F.}},
\bauthor{\bsnm{{Leising}}, \binits{M.D.}},
\bauthor{\bsnm{{Meyer}}, \binits{B.S.}},
\bauthor{\bsnm{{Motizuki}}, \binits{Y.}},
\bauthor{\bsnm{{Sch{\"o}nfelder}}, \binits{V.}}:
\batitle{{Are $^{44}$Ti-producing supernovae exceptional?}}
\bjtitle{\aap}
\bvolume{450}(\bissue{3}),
\bfpage{1037}--\blpage{1050}
(\byear{2006})
\doiurl{10.1051/0004-6361:20054626}
{\href{https://arxiv.org/abs/astro-ph/0601039}{{arXiv:astro-ph/0601039}}}
{[astro-ph]}
\end{barticle}
\endbibitem

%%% 250
\bibitem[\protect\citeauthoryear{{Englhauser} et~al.}{1989}]{Englhauser1989}
\begin{bchapter}
\bauthor{\bsnm{{Englhauser}}, \binits{J.}},
\bauthor{\bsnm{{Dobereiner}}, \binits{S.}},
\bauthor{\bsnm{{Pietsch}}, \binits{W.}},
\bauthor{\bsnm{{Reppin}}, \binits{C.}},
\bauthor{\bsnm{{Trumper}}, \binits{J.}},
\bauthor{\bsnm{{Kendziorra}}, \binits{E.}},
\bauthor{\bsnm{{Maisack}}, \binits{M.}},
\bauthor{\bsnm{{Mony}}, \binits{B.}},
\bauthor{\bsnm{{Staubert}}, \binits{R.}},
\bauthor{\bsnm{{Efremov}}, \binits{V.}},
\bauthor{\bsnm{{Kaniovsky}}, \binits{A.}},
\bauthor{\bsnm{{Kuznetsov}}, \binits{A.}},
\bauthor{\bsnm{{Sunyaev}}, \binits{R.}}:
\bctitle{{Mir-Kvant HEXE Hard X-Ray Lightcurve of Supernova 1987A}}.
In: \beditor{\bsnm{{Hunt}}, \binits{J.}},
\beditor{\bsnm{{Battrick}}, \binits{B.}} (eds.)
\bbtitle{Two Topics in X-Ray Astronomy, Volume 1: X Ray Binaries. Volume 2: AGN
  and the X Ray Background}.
\bsertitle{ESA Special Publication},
vol. \bseriesno{1},
p. \bfpage{397}
(\byear{1989})
\end{bchapter}
\endbibitem

%%% 251
\bibitem[\protect\citeauthoryear{Sunyaev et~al.}{1990}]{Sunyayev1990}
\begin{barticle}
\bauthor{\bsnm{Sunyaev}, \binits{R.A.}},
\bauthor{\bsnm{Efremov}, \binits{V.}},
\bauthor{\bsnm{Kaniovsky}, \binits{A.}},
\bauthor{\bsnm{Stepanov}, \binits{D.}},
\bauthor{\bsnm{Unin}, \binits{S.}},
\bauthor{\bsnm{Kuznetsov}, \binits{A.}},
\bauthor{\bsnm{Loznikov}, \binits{V.}},
\bauthor{\bsnm{Mclioransky}, \binits{A.}},
\bauthor{\bsnm{Prudkoglyad}, \binits{A.}},
\bauthor{\bsnm{Grebenev}, \binits{S.}},
\bauthor{\bsnm{Reppin}, \binits{C.}},
\bauthor{\bsnm{Pietsch}, \binits{W.}},
\bauthor{\bsnm{Englhauser}, \binits{J.}},
\bauthor{\bsnm{Trümper}, \binits{J.}},
\bauthor{\bsnm{Voges}, \binits{W.}},
\bauthor{\bsnm{Kendziorra}, \binits{E.}},
\bauthor{\bsnm{Bezler}, \binits{M.}},
\bauthor{\bsnm{Staubert}, \binits{R.}}:
\batitle{Hard x-rays from supernova 1987a}.
\bjtitle{Advances in Space Research}
\bvolume{10}(\bissue{2}),
\bfpage{47}--\blpage{53}
(\byear{1990})
\doiurl{10.1016/0273-1177(90)90117-I}
\end{barticle}
\endbibitem

%%% 252
\bibitem[\protect\citeauthoryear{{Matz} et~al.}{1988}]{Matz1988}
\begin{barticle}
\bauthor{\bsnm{{Matz}}, \binits{S.M.}},
\bauthor{\bsnm{{Share}}, \binits{G.H.}},
\bauthor{\bsnm{{Leising}}, \binits{M.D.}},
\bauthor{\bsnm{{Chupp}}, \binits{E.L.}},
\bauthor{\bsnm{{Vestrand}}, \binits{W.T.}},
\bauthor{\bsnm{{Purcell}}, \binits{W.R.}},
\bauthor{\bsnm{{Strickman}}, \binits{M.S.}},
\bauthor{\bsnm{{Reppin}}, \binits{C.}}:
\batitle{{Gamma-ray line emission from SN1987A}}.
\bjtitle{\nat}
\bvolume{331}(\bissue{6155}),
\bfpage{416}--\blpage{418}
(\byear{1988})
\doiurl{10.1038/331416a0}
\end{barticle}
\endbibitem

%%% 253
\bibitem[\protect\citeauthoryear{{Sunyaev} et~al.}{1987}]{Sunyayev1987}
\begin{barticle}
\bauthor{\bsnm{{Sunyaev}}, \binits{R.}},
\bauthor{\bsnm{{Kaniovsky}}, \binits{A.}},
\bauthor{\bsnm{{Efremov}}, \binits{V.}},
\bauthor{\bsnm{{Gilfanov}}, \binits{M.}},
\bauthor{\bsnm{{Churazov}}, \binits{E.}},
\bauthor{\bsnm{{Grebenev}}, \binits{S.}},
\bauthor{\bsnm{{Kuznetsov}}, \binits{A.}},
\bauthor{\bsnm{{Melioranskiy}}, \binits{A.}},
\bauthor{\bsnm{{Yamburenko}}, \binits{N.}},
\bauthor{\bsnm{{Yunin}}, \binits{S.}},
\bauthor{\bsnm{{Stepanov}}, \binits{D.}},
\bauthor{\bsnm{{Chulkov}}, \binits{I.}},
\bauthor{\bsnm{{Pappe}}, \binits{N.}},
\bauthor{\bsnm{{Boyarskiy}}, \binits{M.}},
\bauthor{\bsnm{{Gavrilova}}, \binits{E.}},
\bauthor{\bsnm{{Loznikov}}, \binits{V.}},
\bauthor{\bsnm{{Prudkoglyad}}, \binits{A.}},
\bauthor{\bsnm{{Rodin}}, \binits{V.}},
\bauthor{\bsnm{{Reppin}}, \binits{C.}},
\bauthor{\bsnm{{Pietsch}}, \binits{W.}},
\bauthor{\bsnm{{Engelhauser}}, \binits{J.}},
\bauthor{\bsnm{{Tr{\"u}mper}}, \binits{J.}},
\bauthor{\bsnm{{Voges}}, \binits{W.}},
\bauthor{\bsnm{{Kendziorra}}, \binits{E.}},
\bauthor{\bsnm{{Bezler}}, \binits{M.}},
\bauthor{\bsnm{{Staubert}}, \binits{R.}},
\bauthor{\bsnm{{Brinkman}}, \binits{A.C.}},
\bauthor{\bsnm{{Heise}}, \binits{J.}},
\bauthor{\bsnm{{Mels}}, \binits{W.A.}},
\bauthor{\bsnm{{Jager}}, \binits{R.}},
\bauthor{\bsnm{{Skinner}}, \binits{G.K.}},
\bauthor{\bsnm{{Al-Emam}}, \binits{O.}},
\bauthor{\bsnm{{Patterson}}, \binits{T.G.}},
\bauthor{\bsnm{{Willmore}}, \binits{A.P.}},
\bauthor{\bsnm{{Gilfanov}}, \binits{M.}},
\bauthor{\bsnm{{Churazov}}, \binits{E.}}:
\batitle{{Discovery of hard X-ray emission from supernova 1987A.}}
\bjtitle{\nat}
\bvolume{330},
\bfpage{227}--\blpage{229}
(\byear{1987})
\doiurl{10.1038/330227a0}
\end{barticle}
\endbibitem

%%% 254
\bibitem[\protect\citeauthoryear{{Shigeyama} and
  {Nomoto}}{1990}]{Shigeyama1990}
\begin{barticle}
\bauthor{\bsnm{{Shigeyama}}, \binits{T.}},
\bauthor{\bsnm{{Nomoto}}, \binits{K.}}:
\batitle{{Theoretical Light Curve of SN 1987A and Mixing of Hydrogen and Nickel
  in the Ejecta}}.
\bjtitle{\apj}
\bvolume{360},
\bfpage{242}
(\byear{1990})
\doiurl{10.1086/169114}
\end{barticle}
\endbibitem

%%% 255
\bibitem[\protect\citeauthoryear{{Arnett} et~al.}{1989}]{Arnett1989}
\begin{barticle}
\bauthor{\bsnm{{Arnett}}, \binits{W.D.}},
\bauthor{\bsnm{{Bahcall}}, \binits{J.N.}},
\bauthor{\bsnm{{Kirshner}}, \binits{R.P.}},
\bauthor{\bsnm{{Woosley}}, \binits{S.E.}}:
\batitle{{Supernova 1987A.}}
\bjtitle{\araa}
\bvolume{27},
\bfpage{629}--\blpage{700}
(\byear{1989})
\doiurl{10.1146/annurev.aa.27.090189.003213}
\end{barticle}
\endbibitem

%%% 256
\bibitem[\protect\citeauthoryear{{Bussard} et~al.}{1989}]{Bussard1989}
\begin{barticle}
\bauthor{\bsnm{{Bussard}}, \binits{R.W.}},
\bauthor{\bsnm{{Burrows}}, \binits{A.}},
\bauthor{\bsnm{{The}}, \binits{L.S.}}:
\batitle{{SN 1987A Gamma-Ray Line Profiles and Fluxes}}.
\bjtitle{\apj}
\bvolume{341},
\bfpage{401}
(\byear{1989})
\doiurl{10.1086/167503}
\end{barticle}
\endbibitem

%%% 257
\bibitem[\protect\citeauthoryear{{Li} and {Ma}}{1983}]{LiMa1983}
\begin{barticle}
\bauthor{\bsnm{{Li}}, \binits{T.-P.}},
\bauthor{\bsnm{{Ma}}, \binits{Y.-Q.}}:
\batitle{{Analysis methods for results in gamma-ray astronomy.}}
\bjtitle{\apj}
\bvolume{272},
\bfpage{317}--\blpage{324}
(\byear{1983})
\doiurl{10.1086/161295}
\end{barticle}
\endbibitem

%%% 258
\bibitem[\protect\citeauthoryear{{Vianello}}{2018}]{Vianello2018}
\begin{barticle}
\bauthor{\bsnm{{Vianello}}, \binits{G.}}:
\batitle{{The Significance of an Excess in a Counting Experiment: Assessing the
  Impact of Systematic Uncertainties and the Case with a Gaussian Background}}.
\bjtitle{\apjs}
\bvolume{236}(\bissue{1}),
\bfpage{17}
(\byear{2018})
\doiurl{10.3847/1538-4365/aab780}
{\href{https://arxiv.org/abs/1712.00118}{{arXiv:1712.00118}}}
{[physics.data-an]}
\end{barticle}
\endbibitem

%%% 259
\bibitem[\protect\citeauthoryear{{Thielemann} et~al.}{1990}]{Thielemann1990}
\begin{barticle}
\bauthor{\bsnm{{Thielemann}}, \binits{F.-K.}},
\bauthor{\bsnm{{Hashimoto}}, \binits{M.-A.}},
\bauthor{\bsnm{{Nomoto}}, \binits{K.}}:
\batitle{{Explosive Nucleosynthesis in SN 1987A. II. Composition,
  Radioactivities, and the Neutron Star Mass}}.
\bjtitle{\apj}
\bvolume{349},
\bfpage{222}
(\byear{1990})
\doiurl{10.1086/168308}
\end{barticle}
\endbibitem

%%% 260
\bibitem[\protect\citeauthoryear{{Woosley} and {Hoffman}}{1991}]{Woosley1991}
\begin{barticle}
\bauthor{\bsnm{{Woosley}}, \binits{S.E.}},
\bauthor{\bsnm{{Hoffman}}, \binits{R.D.}}:
\batitle{{57Co and 44Ti Production in SN 1987A}}.
\bjtitle{\apjl}
\bvolume{368},
\bfpage{31}
(\byear{1991})
\doiurl{10.1086/185941}
\end{barticle}
\endbibitem

%%% 261
\bibitem[\protect\citeauthoryear{{Wang} and {Burrows}}{2024}]{Wang2024}
\begin{barticle}
\bauthor{\bsnm{{Wang}}, \binits{T.}},
\bauthor{\bsnm{{Burrows}}, \binits{A.}}:
\batitle{{Insights into the Production of $^{44}$Ti and Nickel Isotopes in
  Core-collapse Supernovae}}.
\bjtitle{\apj}
\bvolume{974}(\bissue{1}),
\bfpage{39}
(\byear{2024})
\doiurl{10.3847/1538-4357/ad6983}
{\href{https://arxiv.org/abs/2406.13746}{{arXiv:2406.13746}}}
{[astro-ph.HE]}
\end{barticle}
\endbibitem

%%% 262
\bibitem[\protect\citeauthoryear{{Clayton} et~al.}{1992}]{Clayton1992}
\begin{barticle}
\bauthor{\bsnm{{Clayton}}, \binits{D.D.}},
\bauthor{\bsnm{{Leising}}, \binits{M.D.}},
\bauthor{\bsnm{{The}}, \binits{L.-S.}},
\bauthor{\bsnm{{Johnson}}, \binits{W.N.}},
\bauthor{\bsnm{{Kurfess}}, \binits{J.D.}}:
\batitle{{The 57 CO Abundance in SN 1987A}}.
\bjtitle{\apjl}
\bvolume{399},
\bfpage{141}
(\byear{1992})
\doiurl{10.1086/186627}
\end{barticle}
\endbibitem

%%% 263
\bibitem[\protect\citeauthoryear{{Kurfess} et~al.}{1992}]{Kurfess1992}
\begin{barticle}
\bauthor{\bsnm{{Kurfess}}, \binits{J.D.}},
\bauthor{\bsnm{{Johnson}}, \binits{W.N.}},
\bauthor{\bsnm{{Kinzer}}, \binits{R.L.}},
\bauthor{\bsnm{{Kroeger}}, \binits{R.A.}},
\bauthor{\bsnm{{Strickman}}, \binits{M.S.}},
\bauthor{\bsnm{{Grove}}, \binits{J.E.}},
\bauthor{\bsnm{{Leising}}, \binits{M.D.}},
\bauthor{\bsnm{{Clayton}}, \binits{D.D.}},
\bauthor{\bsnm{{Grabelsky}}, \binits{D.A.}},
\bauthor{\bsnm{{Purcell}}, \binits{W.R.}},
\bauthor{\bsnm{{Ulmer}}, \binits{M.P.}},
\bauthor{\bsnm{{Cameron}}, \binits{R.A.}},
\bauthor{\bsnm{{Jung}}, \binits{G.V.}}:
\batitle{{Oriented Scintillation Spectrometer Experiment Observations of 57Co
  in SN 1987A}}.
\bjtitle{\apjl}
\bvolume{399},
\bfpage{137}
(\byear{1992})
\doiurl{10.1086/186626}
\end{barticle}
\endbibitem

%%% 264
\bibitem[\protect\citeauthoryear{{Seitenzahl} et~al.}{2014}]{Seitenzahl2014}
\begin{barticle}
\bauthor{\bsnm{{Seitenzahl}}, \binits{I.R.}},
\bauthor{\bsnm{{Timmes}}, \binits{F.X.}},
\bauthor{\bsnm{{Magkotsios}}, \binits{G.}}:
\batitle{{The Light Curve of SN 1987A Revisited: Constraining Production Masses
  of Radioactive Nuclides}}.
\bjtitle{\apj}
\bvolume{792}(\bissue{1}),
\bfpage{10}
(\byear{2014})
\doiurl{10.1088/0004-637X/792/1/10}
{\href{https://arxiv.org/abs/1408.5986}{{arXiv:1408.5986}}}
{[astro-ph.SR]}
\end{barticle}
\endbibitem

%%% 265
\bibitem[\protect\citeauthoryear{{Grebenev} et~al.}{2012}]{Grebenev2012}
\begin{barticle}
\bauthor{\bsnm{{Grebenev}}, \binits{S.A.}},
\bauthor{\bsnm{{Lutovinov}}, \binits{A.A.}},
\bauthor{\bsnm{{Tsygankov}}, \binits{S.S.}},
\bauthor{\bsnm{{Winkler}}, \binits{C.}}:
\batitle{{Hard-X-ray emission lines from the decay of $^{44}$Ti in the remnant
  of supernova 1987A}}.
\bjtitle{\nat}
\bvolume{490}(\bissue{7420}),
\bfpage{373}--\blpage{375}
(\byear{2012})
\doiurl{10.1038/nature11473}
{\href{https://arxiv.org/abs/1211.2656}{{arXiv:1211.2656}}}
{[astro-ph.HE]}
\end{barticle}
\endbibitem

%%% 266
\bibitem[\protect\citeauthoryear{{Alarie} et~al.}{2014}]{Alarie2014}
\begin{barticle}
\bauthor{\bsnm{{Alarie}}, \binits{A.}},
\bauthor{\bsnm{{Bilodeau}}, \binits{A.}},
\bauthor{\bsnm{{Drissen}}, \binits{L.}}:
\batitle{{A hyperspectral view of Cassiopeia A}}.
\bjtitle{\mnras}
\bvolume{441}(\bissue{4}),
\bfpage{2996}--\blpage{3008}
(\byear{2014})
\doiurl{10.1093/mnras/stu774}
\end{barticle}
\endbibitem

%%% 267
\bibitem[\protect\citeauthoryear{{Eriksen} et~al.}{2009}]{Eriksen2009}
\begin{barticle}
\bauthor{\bsnm{{Eriksen}}, \binits{K.A.}},
\bauthor{\bsnm{{Arnett}}, \binits{D.}},
\bauthor{\bsnm{{McCarthy}}, \binits{D.W.}},
\bauthor{\bsnm{{Young}}, \binits{P.}}:
\batitle{{The Reddening Toward Cassiopeia A's Supernova: Constraining the
  $^{56}$Ni Yield}}.
\bjtitle{\apj}
\bvolume{697}(\bissue{1}),
\bfpage{29}--\blpage{36}
(\byear{2009})
\doiurl{10.1088/0004-637X/697/1/29}
{\href{https://arxiv.org/abs/0902.4029}{{arXiv:0902.4029}}}
{[astro-ph.SR]}
\end{barticle}
\endbibitem

%%% 268
\bibitem[\protect\citeauthoryear{{Tsygankov} et~al.}{2016}]{Tsygankov2016}
\begin{barticle}
\bauthor{\bsnm{{Tsygankov}}, \binits{S.S.}},
\bauthor{\bsnm{{Krivonos}}, \binits{R.A.}},
\bauthor{\bsnm{{Lutovinov}}, \binits{A.A.}},
\bauthor{\bsnm{{Revnivtsev}}, \binits{M.G.}},
\bauthor{\bsnm{{Churazov}}, \binits{E.M.}},
\bauthor{\bsnm{{Sunyaev}}, \binits{R.A.}},
\bauthor{\bsnm{{Grebenev}}, \binits{S.A.}}:
\batitle{{Galactic survey of $^{44}$Ti sources with the IBIS telescope onboard
  INTEGRAL}}.
\bjtitle{\mnras}
\bvolume{458}(\bissue{4}),
\bfpage{3411}--\blpage{3419}
(\byear{2016})
\doiurl{10.1093/mnras/stw549}
{\href{https://arxiv.org/abs/1603.01264}{{arXiv:1603.01264}}}
{[astro-ph.HE]}
\end{barticle}
\endbibitem

%%% 269
\bibitem[\protect\citeauthoryear{{Wang} and {Li}}{2016}]{Wang2016}
\begin{barticle}
\bauthor{\bsnm{{Wang}}, \binits{W.}},
\bauthor{\bsnm{{Li}}, \binits{Z.}}:
\batitle{{Hard X-Ray Emissions from Cassiopeia A Observed by INTEGRAL}}.
\bjtitle{\apj}
\bvolume{825}(\bissue{2}),
\bfpage{102}
(\byear{2016})
\doiurl{10.3847/0004-637X/825/2/102}
{\href{https://arxiv.org/abs/1605.00360}{{arXiv:1605.00360}}}
{[astro-ph.HE]}
\end{barticle}
\endbibitem

%%% 270
\bibitem[\protect\citeauthoryear{{Grefenstette}
  et~al.}{2017}]{Grefenstette2017}
\begin{barticle}
\bauthor{\bsnm{{Grefenstette}}, \binits{B.W.}},
\bauthor{\bsnm{{Fryer}}, \binits{C.L.}},
\bauthor{\bsnm{{Harrison}}, \binits{F.A.}},
\bauthor{\bsnm{{Boggs}}, \binits{S.E.}},
\bauthor{\bsnm{{DeLaney}}, \binits{T.}},
\bauthor{\bsnm{{Laming}}, \binits{J.M.}},
\bauthor{\bsnm{{Reynolds}}, \binits{S.P.}},
\bauthor{\bsnm{{Alexander}}, \binits{D.M.}},
\bauthor{\bsnm{{Barret}}, \binits{D.}},
\bauthor{\bsnm{{Christensen}}, \binits{F.E.}},
\bauthor{\bsnm{{Craig}}, \binits{W.W.}},
\bauthor{\bsnm{{Forster}}, \binits{K.}},
\bauthor{\bsnm{{Giommi}}, \binits{P.}},
\bauthor{\bsnm{{Hailey}}, \binits{C.J.}},
\bauthor{\bsnm{{Hornstrup}}, \binits{A.}},
\bauthor{\bsnm{{Kitaguchi}}, \binits{T.}},
\bauthor{\bsnm{{Koglin}}, \binits{J.E.}},
\bauthor{\bsnm{{Lopez}}, \binits{L.}},
\bauthor{\bsnm{{Mao}}, \binits{P.H.}},
\bauthor{\bsnm{{Madsen}}, \binits{K.K.}},
\bauthor{\bsnm{{Miyasaka}}, \binits{H.}},
\bauthor{\bsnm{{Mori}}, \binits{K.}},
\bauthor{\bsnm{{Perri}}, \binits{M.}},
\bauthor{\bsnm{{Pivovaroff}}, \binits{M.J.}},
\bauthor{\bsnm{{Puccetti}}, \binits{S.}},
\bauthor{\bsnm{{Rana}}, \binits{V.}},
\bauthor{\bsnm{{Stern}}, \binits{D.}},
\bauthor{\bsnm{{Westergaard}}, \binits{N.J.}},
\bauthor{\bsnm{{Wik}}, \binits{D.R.}},
\bauthor{\bsnm{{Zhang}}, \binits{W.W.}},
\bauthor{\bsnm{{Zoglauer}}, \binits{A.}}:
\batitle{{The Distribution of Radioactive $^{44}$Ti in Cassiopeia A}}.
\bjtitle{\apj}
\bvolume{834}(\bissue{1}),
\bfpage{19}
(\byear{2017})
\doiurl{10.3847/1538-4357/834/1/19}
{\href{https://arxiv.org/abs/1612.02774}{{arXiv:1612.02774}}}
{[astro-ph.HE]}
\end{barticle}
\endbibitem

%%% 271
\bibitem[\protect\citeauthoryear{{Weinberger}}{2021}]{Weinberger2021}
\begin{botherref}
\oauthor{\bsnm{{Weinberger}}, \binits{C.}}:
{Supernova Diagnostics from Gamma-Ray Lines in the Young Remnant Phase}.
PhD thesis,
Max-Planck-Institute for Extraterrestrial Physics, Garching
(February 2021).
\url{https://mediatum.ub.tum.de/1609856}
\end{botherref}
\endbibitem

%%% 272
\bibitem[\protect\citeauthoryear{{Summa} et~al.}{2011}]{summa2011}
\begin{barticle}
\bauthor{\bsnm{{Summa}}, \binits{A.}},
\bauthor{\bsnm{{Els{\"a}sser}}, \binits{D.}},
\bauthor{\bsnm{{Mannheim}}, \binits{K.}}:
\batitle{{Nuclear de-excitation line spectrum of Cassiopeia A}}.
\bjtitle{\aap}
\bvolume{533},
\bfpage{13}
(\byear{2011})
\doiurl{10.1051/0004-6361/201117267}
{\href{https://arxiv.org/abs/1107.4331}{{arXiv:1107.4331}}}
{[astro-ph.HE]}
\end{barticle}
\endbibitem

%%% 273
\bibitem[\protect\citeauthoryear{{Liu} et~al.}{2023}]{liu2023}
\begin{barticle}
\bauthor{\bsnm{{Liu}}, \binits{B.}},
\bauthor{\bsnm{{Yang}}, \binits{R.-z.}},
\bauthor{\bsnm{{He}}, \binits{X.-y.}},
\bauthor{\bsnm{{Aharonian}}, \binits{F.}}:
\batitle{{New estimation of the nuclear de-excitation line emission from the
  supernova remnant Cassiopeia A}}.
\bjtitle{\mnras}
\bvolume{524}(\bissue{4}),
\bfpage{5248}--\blpage{5253}
(\byear{2023})
\doiurl{10.1093/mnras/stad2165}
{\href{https://arxiv.org/abs/2307.08967}{{arXiv:2307.08967}}}
{[astro-ph.HE]}
\end{barticle}
\endbibitem

%%% 274
\bibitem[\protect\citeauthoryear{{McKinnon} et~al.}{2016}]{McKinnon2016}
\begin{barticle}
\bauthor{\bsnm{{McKinnon}}, \binits{R.}},
\bauthor{\bsnm{{Torrey}}, \binits{P.}},
\bauthor{\bsnm{{Vogelsberger}}, \binits{M.}}:
\batitle{{Dust formation in Milky Way-like galaxies}}.
\bjtitle{\mnras}
\bvolume{457}(\bissue{4}),
\bfpage{3775}--\blpage{3800}
(\byear{2016})
\doiurl{10.1093/mnras/stw253}
{\href{https://arxiv.org/abs/1505.04792}{{arXiv:1505.04792}}}
{[astro-ph.GA]}
\end{barticle}
\endbibitem

%%% 275
\bibitem[\protect\citeauthoryear{{Iyudin} et~al.}{2019}]{Iyudin2019}
\begin{barticle}
\bauthor{\bsnm{{Iyudin}}, \binits{A.F.}},
\bauthor{\bsnm{{M{\"u}ller}}, \binits{E.}},
\bauthor{\bsnm{{Obergaulinger}}, \binits{M.}}:
\batitle{{Titanium hidden in dust}}.
\bjtitle{\mnras}
\bvolume{485}(\bissue{3}),
\bfpage{3288}--\blpage{3295}
(\byear{2019})
\doiurl{10.1093/mnras/stz419}
{\href{https://arxiv.org/abs/1902.02249}{{arXiv:1902.02249}}}
{[astro-ph.HE]}
\end{barticle}
\endbibitem

%%% 276
\bibitem[\protect\citeauthoryear{{Vance} et~al.}{2020}]{Vance2020}
\begin{barticle}
\bauthor{\bsnm{{Vance}}, \binits{G.S.}},
\bauthor{\bsnm{{Young}}, \binits{P.A.}},
\bauthor{\bsnm{{Fryer}}, \binits{C.L.}},
\bauthor{\bsnm{{Ellinger}}, \binits{C.I.}}:
\batitle{{Titanium and Iron in the Cassiopeia A Supernova Remnant}}.
\bjtitle{\apj}
\bvolume{895}(\bissue{2}),
\bfpage{82}
(\byear{2020})
\doiurl{10.3847/1538-4357/ab8ade}
{\href{https://arxiv.org/abs/2005.03777}{{arXiv:2005.03777}}}
{[astro-ph.HE]}
\end{barticle}
\endbibitem

%%% 277
\bibitem[\protect\citeauthoryear{{Orlando} et~al.}{2021}]{Orlando2021}
\begin{barticle}
\bauthor{\bsnm{{Orlando}}, \binits{S.}},
\bauthor{\bsnm{{Wongwathanarat}}, \binits{A.}},
\bauthor{\bsnm{{Janka}}, \binits{H.-T.}},
\bauthor{\bsnm{{Miceli}}, \binits{M.}},
\bauthor{\bsnm{{Ono}}, \binits{M.}},
\bauthor{\bsnm{{Nagataki}}, \binits{S.}},
\bauthor{\bsnm{{Bocchino}}, \binits{F.}},
\bauthor{\bsnm{{Peres}}, \binits{G.}}:
\batitle{{The fully developed remnant of a neutrino-driven supernova. Evolution
  of ejecta structure and asymmetries in SNR Cassiopeia A}}.
\bjtitle{\aap}
\bvolume{645},
\bfpage{66}
(\byear{2021})
\doiurl{10.1051/0004-6361/202039335}
{\href{https://arxiv.org/abs/2009.01789}{{arXiv:2009.01789}}}
{[astro-ph.HE]}
\end{barticle}
\endbibitem

%%% 278
\bibitem[\protect\citeauthoryear{{Wongwathanarat}
  et~al.}{2017}]{Wongwathanarat2017}
\begin{barticle}
\bauthor{\bsnm{{Wongwathanarat}}, \binits{A.}},
\bauthor{\bsnm{{Janka}}, \binits{H.-T.}},
\bauthor{\bsnm{{M{\"u}ller}}, \binits{E.}},
\bauthor{\bsnm{{Pllumbi}}, \binits{E.}},
\bauthor{\bsnm{{Wanajo}}, \binits{S.}}:
\batitle{{Production and Distribution of $^{44}$Ti and $^{56}$Ni in a
  Three-dimensional Supernova Model Resembling Cassiopeia A}}.
\bjtitle{\apj}
\bvolume{842}(\bissue{1}),
\bfpage{13}
(\byear{2017})
\doiurl{10.3847/1538-4357/aa72de}
{\href{https://arxiv.org/abs/1610.05643}{{arXiv:1610.05643}}}
{[astro-ph.HE]}
\end{barticle}
\endbibitem

%%% 279
\bibitem[\protect\citeauthoryear{{Hwang} et~al.}{2004}]{Hwang2004}
\begin{barticle}
\bauthor{\bsnm{{Hwang}}, \binits{U.}},
\bauthor{\bsnm{{Laming}}, \binits{J.M.}},
\bauthor{\bsnm{{Badenes}}, \binits{C.}},
\bauthor{\bsnm{{Berendse}}, \binits{F.}},
\bauthor{\bsnm{{Blondin}}, \binits{J.}},
\bauthor{\bsnm{{Cioffi}}, \binits{D.}},
\bauthor{\bsnm{{DeLaney}}, \binits{T.}},
\bauthor{\bsnm{{Dewey}}, \binits{D.}},
\bauthor{\bsnm{{Fesen}}, \binits{R.}},
\bauthor{\bsnm{{Flanagan}}, \binits{K.A.}},
\bauthor{\bsnm{{Fryer}}, \binits{C.L.}},
\bauthor{\bsnm{{Ghavamian}}, \binits{P.}},
\bauthor{\bsnm{{Hughes}}, \binits{J.P.}},
\bauthor{\bsnm{{Morse}}, \binits{J.A.}},
\bauthor{\bsnm{{Plucinsky}}, \binits{P.P.}},
\bauthor{\bsnm{{Petre}}, \binits{R.}},
\bauthor{\bsnm{{Pohl}}, \binits{M.}},
\bauthor{\bsnm{{Rudnick}}, \binits{L.}},
\bauthor{\bsnm{{Sankrit}}, \binits{R.}},
\bauthor{\bsnm{{Slane}}, \binits{P.O.}},
\bauthor{\bsnm{{Smith}}, \binits{R.K.}},
\bauthor{\bsnm{{Vink}}, \binits{J.}},
\bauthor{\bsnm{{Warren}}, \binits{J.S.}}:
\batitle{{A Million Second Chandra View of Cassiopeia A}}.
\bjtitle{\apjl}
\bvolume{615}(\bissue{2}),
\bfpage{117}--\blpage{120}
(\byear{2004})
\doiurl{10.1086/426186}
{\href{https://arxiv.org/abs/astro-ph/0409760}{{arXiv:astro-ph/0409760}}}
{[astro-ph]}
\end{barticle}
\endbibitem

%%% 280
\bibitem[\protect\citeauthoryear{{Maeda} and {Nomoto}}{2003}]{Maeda2003}
\begin{barticle}
\bauthor{\bsnm{{Maeda}}, \binits{K.}},
\bauthor{\bsnm{{Nomoto}}, \binits{K.}}:
\batitle{{Bipolar Supernova Explosions: Nucleosynthesis and Implications for
  Abundances in Extremely Metal-Poor Stars}}.
\bjtitle{\apj}
\bvolume{598}(\bissue{2}),
\bfpage{1163}--\blpage{1200}
(\byear{2003})
\doiurl{10.1086/378948}
{\href{https://arxiv.org/abs/astro-ph/0304172}{{arXiv:astro-ph/0304172}}}
{[astro-ph]}
\end{barticle}
\endbibitem

%%% 281
\bibitem[\protect\citeauthoryear{{Seitenzahl} et~al.}{2013}]{Seitenzahl2013}
\begin{barticle}
\bauthor{\bsnm{{Seitenzahl}}, \binits{I.R.}},
\bauthor{\bsnm{{Ciaraldi-Schoolmann}}, \binits{F.}},
\bauthor{\bsnm{{R{\"o}pke}}, \binits{F.K.}},
\bauthor{\bsnm{{Fink}}, \binits{M.}},
\bauthor{\bsnm{{Hillebrandt}}, \binits{W.}},
\bauthor{\bsnm{{Kromer}}, \binits{M.}},
\bauthor{\bsnm{{Pakmor}}, \binits{R.}},
\bauthor{\bsnm{{Ruiter}}, \binits{A.J.}},
\bauthor{\bsnm{{Sim}}, \binits{S.A.}},
\bauthor{\bsnm{{Taubenberger}}, \binits{S.}}:
\batitle{{Three-dimensional delayed-detonation models with nucleosynthesis for
  Type Ia supernovae}}.
\bjtitle{\mnras}
\bvolume{429}(\bissue{2}),
\bfpage{1156}--\blpage{1172}
(\byear{2013})
\doiurl{10.1093/mnras/sts402}
{\href{https://arxiv.org/abs/1211.3015}{{arXiv:1211.3015}}}
{[astro-ph.SR]}
\end{barticle}
\endbibitem

%%% 282
\bibitem[\protect\citeauthoryear{{Fink} et~al.}{2010}]{Fink2010}
\begin{barticle}
\bauthor{\bsnm{{Fink}}, \binits{M.}},
\bauthor{\bsnm{{R{\"o}pke}}, \binits{F.K.}},
\bauthor{\bsnm{{Hillebrandt}}, \binits{W.}},
\bauthor{\bsnm{{Seitenzahl}}, \binits{I.R.}},
\bauthor{\bsnm{{Sim}}, \binits{S.A.}},
\bauthor{\bsnm{{Kromer}}, \binits{M.}}:
\batitle{{Double-detonation sub-Chandrasekhar supernovae: can minimum helium
  shell masses detonate the core?}}
\bjtitle{\aap}
\bvolume{514},
\bfpage{53}
(\byear{2010})
\doiurl{10.1051/0004-6361/200913892}
{\href{https://arxiv.org/abs/1002.2173}{{arXiv:1002.2173}}}
{[astro-ph.SR]}
\end{barticle}
\endbibitem

%%% 283
\bibitem[\protect\citeauthoryear{{Fink} et~al.}{2014}]{Fink2014}
\begin{barticle}
\bauthor{\bsnm{{Fink}}, \binits{M.}},
\bauthor{\bsnm{{Kromer}}, \binits{M.}},
\bauthor{\bsnm{{Seitenzahl}}, \binits{I.R.}},
\bauthor{\bsnm{{Ciaraldi-Schoolmann}}, \binits{F.}},
\bauthor{\bsnm{{R{\"o}pke}}, \binits{F.K.}},
\bauthor{\bsnm{{Sim}}, \binits{S.A.}},
\bauthor{\bsnm{{Pakmor}}, \binits{R.}},
\bauthor{\bsnm{{Ruiter}}, \binits{A.J.}},
\bauthor{\bsnm{{Hillebrandt}}, \binits{W.}}:
\batitle{{Three-dimensional pure deflagration models with nucleosynthesis and
  synthetic observables for Type Ia supernovae}}.
\bjtitle{\mnras}
\bvolume{438}(\bissue{2}),
\bfpage{1762}--\blpage{1783}
(\byear{2014})
\doiurl{10.1093/mnras/stt2315}
{\href{https://arxiv.org/abs/1308.3257}{{arXiv:1308.3257}}}
{[astro-ph.SR]}
\end{barticle}
\endbibitem

%%% 284
\bibitem[\protect\citeauthoryear{{Anders} and {Grevesse}}{1989}]{Anders1989}
\begin{barticle}
\bauthor{\bsnm{{Anders}}, \binits{E.}},
\bauthor{\bsnm{{Grevesse}}, \binits{N.}}:
\batitle{{Abundances of the elements: Meteoritic and solar}}.
\bjtitle{\gca}
\bvolume{53}(\bissue{1}),
\bfpage{197}--\blpage{214}
(\byear{1989})
\doiurl{10.1016/0016-7037(89)90286-X}
\end{barticle}
\endbibitem

%%% 285
\bibitem[\protect\citeauthoryear{Cowan et~al.}{2021}]{Cowan:2019pkx}
\begin{barticle}
\bauthor{\bsnm{Cowan}, \binits{J.J.}},
\bauthor{\bsnm{Sneden}, \binits{C.}},
\bauthor{\bsnm{Lawler}, \binits{J.E.}},
\bauthor{\bsnm{Aprahamian}, \binits{A.}},
\bauthor{\bsnm{Wiescher}, \binits{M.}},
\bauthor{\bsnm{Langanke}, \binits{K.}},
\bauthor{\bsnm{Mart\'\i{}nez-Pinedo}, \binits{G.}},
\bauthor{\bsnm{Thielemann}, \binits{F.-K.}}:
\batitle{{Origin of the heaviest elements: The rapid neutron-capture process}}.
\bjtitle{Rev. Mod. Phys.}
\bvolume{93}(\bissue{1}),
\bfpage{15002}
(\byear{2021})
\doiurl{10.1103/RevModPhys.93.015002}
{\href{https://arxiv.org/abs/1901.01410}{{arXiv:1901.01410}}}
{[astro-ph.HE]}
\end{barticle}
\endbibitem

%%% 286
\bibitem[\protect\citeauthoryear{Eichler et~al.}{1989}]{Eichler:1989ve}
\begin{barticle}
\bauthor{\bsnm{Eichler}, \binits{D.}},
\bauthor{\bsnm{Livio}, \binits{M.}},
\bauthor{\bsnm{Piran}, \binits{T.}},
\bauthor{\bsnm{Schramm}, \binits{D.N.}}:
\batitle{{Nucleosynthesis, Neutrino Bursts and Gamma-Rays from Coalescing
  Neutron Stars}}.
\bjtitle{Nature}
\bvolume{340},
\bfpage{126}--\blpage{128}
(\byear{1989})
\doiurl{10.1038/340126a0}
\end{barticle}
\endbibitem

%%% 287
\bibitem[\protect\citeauthoryear{Lattimer and Schramm}{1974}]{Lattimer:1974slx}
\begin{barticle}
\bauthor{\bsnm{Lattimer}, \binits{J.M.}},
\bauthor{\bsnm{Schramm}, \binits{D.N.}}:
\batitle{{Black-hole-neutron-star collisions}}.
\bjtitle{Astrophys. J. Lett.}
\bvolume{192},
\bfpage{145}
(\byear{1974})
\doiurl{10.1086/181612}
\end{barticle}
\endbibitem

%%% 288
\bibitem[\protect\citeauthoryear{Woosley et~al.}{1994}]{Woosley:1994ux}
\begin{barticle}
\bauthor{\bsnm{Woosley}, \binits{S.E.}},
\bauthor{\bsnm{Wilson}, \binits{J.R.}},
\bauthor{\bsnm{Mathews}, \binits{G.J.}},
\bauthor{\bsnm{Hoffman}, \binits{R.D.}},
\bauthor{\bsnm{Meyer}, \binits{B.S.}}:
\batitle{{The r process and neutrino heated supernova ejecta}}.
\bjtitle{Astrophys. J.}
\bvolume{433},
\bfpage{229}--\blpage{246}
(\byear{1994})
\doiurl{10.1086/174638}
\end{barticle}
\endbibitem

%%% 289
\bibitem[\protect\citeauthoryear{Winteler et~al.}{2012}]{Winteler:2012hu}
\begin{barticle}
\bauthor{\bsnm{Winteler}, \binits{C.}},
\bauthor{\bsnm{Kaeppeli}, \binits{R.}},
\bauthor{\bsnm{Perego}, \binits{A.}},
\bauthor{\bsnm{Arcones}, \binits{A.}},
\bauthor{\bsnm{Vasset}, \binits{N.}},
\bauthor{\bsnm{Nishimura}, \binits{N.}},
\bauthor{\bsnm{Liebend{\"o}rfer}, \binits{M.}},
\bauthor{\bsnm{Thielemann}, \binits{F.-K.}}:
\batitle{{Magneto-rotationally driven Supernovae as the origin of early galaxy
  r-process elements?}}
\bjtitle{Astrophys. J. Lett.}
\bvolume{750},
\bfpage{22}
(\byear{2012})
\doiurl{10.1088/2041-8205/750/1/L22}
{\href{https://arxiv.org/abs/1203.0616}{{arXiv:1203.0616}}}
{[astro-ph.SR]}
\end{barticle}
\endbibitem

%%% 290
\bibitem[\protect\citeauthoryear{Fischer et~al.}{2020}]{Fischer:2020xjl}
\begin{barticle}
\bauthor{\bsnm{Fischer}, \binits{T.}},
\bauthor{\bsnm{Wu}, \binits{M.-R.}},
\bauthor{\bsnm{Wehmeyer}, \binits{B.}},
\bauthor{\bsnm{Bastian}, \binits{N.-U.F.}},
\bauthor{\bsnm{Mart\'\i{}nez-Pinedo}, \binits{G.}},
\bauthor{\bsnm{Thielemann}, \binits{F.-K.}}:
\batitle{{Core-collapse Supernova Explosions Driven by the Hadron-quark Phase
  Transition as a Rare $r$-process Site}}.
\bjtitle{Astrophys. J.}
\bvolume{894}(\bissue{1}),
\bfpage{9}
(\byear{2020})
\doiurl{10.3847/1538-4357/ab86b0}
{\href{https://arxiv.org/abs/2003.00972}{{arXiv:2003.00972}}}
{[astro-ph.HE]}
\end{barticle}
\endbibitem

%%% 291
\bibitem[\protect\citeauthoryear{Grichener and Soker}{2019}]{Grichener:2018way}
\begin{barticle}
\bauthor{\bsnm{Grichener}, \binits{A.}},
\bauthor{\bsnm{Soker}, \binits{N.}}:
\batitle{{The Common Envelope Jet Supernova (CEJSN) r-process Scenario}}.
\bjtitle{Astrophys. J.}
\bvolume{878}(\bissue{1}),
\bfpage{24}
(\byear{2019})
\doiurl{10.3847/1538-4357/ab1d5d}
{\href{https://arxiv.org/abs/1810.03889}{{arXiv:1810.03889}}}
{[astro-ph.SR]}
\end{barticle}
\endbibitem

%%% 292
\bibitem[\protect\citeauthoryear{Siegel et~al.}{2019}]{Siegel:2018zxq}
\begin{barticle}
\bauthor{\bsnm{Siegel}, \binits{D.M.}},
\bauthor{\bsnm{Barnes}, \binits{J.}},
\bauthor{\bsnm{Metzger}, \binits{B.D.}}:
\batitle{{Collapsars as a major source of r-process elements}}.
\bjtitle{Nature}
\bvolume{569},
\bfpage{241}
(\byear{2019})
\doiurl{10.1038/s41586-019-1136-0}
{\href{https://arxiv.org/abs/1810.00098}{{arXiv:1810.00098}}}
{[astro-ph.HE]}
\end{barticle}
\endbibitem

%%% 293
\bibitem[\protect\citeauthoryear{Cehula et~al.}{2024}]{Cehula:2023hdh}
\begin{barticle}
\bauthor{\bsnm{Cehula}, \binits{J.}},
\bauthor{\bsnm{Thompson}, \binits{T.A.}},
\bauthor{\bsnm{Metzger}, \binits{B.D.}}:
\batitle{{Dynamics of baryon ejection in magnetar giant flares: implications
  for radio afterglows, r-process nucleosynthesis, and fast radio bursts}}.
\bjtitle{Mon. Not. Roy. Astron. Soc.}
\bvolume{528}(\bissue{3}),
\bfpage{5323}--\blpage{5345}
(\byear{2024})
\doiurl{10.1093/mnras/stae358}
{\href{https://arxiv.org/abs/2311.05681}{{arXiv:2311.05681}}}
{[astro-ph.HE]}
\end{barticle}
\endbibitem

%%% 294
\bibitem[\protect\citeauthoryear{Abbott et~al.}{2017a}]{LIGOScientific:2017vwq}
\begin{barticle}
\bauthor{\bsnm{Abbott}, \binits{B.P.}}, \betal:
\batitle{{GW170817: Observation of Gravitational Waves from a Binary Neutron
  Star Inspiral}}.
\bjtitle{Phys. Rev. Lett.}
\bvolume{119}(\bissue{16}),
\bfpage{161101}
(\byear{2017})
\doiurl{10.1103/PhysRevLett.119.161101}
{\href{https://arxiv.org/abs/1710.05832}{{arXiv:1710.05832}}}
{[gr-qc]}
\end{barticle}
\endbibitem

%%% 295
\bibitem[\protect\citeauthoryear{Abbott et~al.}{2017b}]{LIGOScientific:2017ync}
\begin{barticle}
\bauthor{\bsnm{Abbott}, \binits{B.P.}}, \betal:
\batitle{{Multi-messenger Observations of a Binary Neutron Star Merger}}.
\bjtitle{Astrophys. J. Lett.}
\bvolume{848}(\bissue{2}),
\bfpage{12}
(\byear{2017})
\doiurl{10.3847/2041-8213/aa91c9}
{\href{https://arxiv.org/abs/1710.05833}{{arXiv:1710.05833}}}
{[astro-ph.HE]}
\end{barticle}
\endbibitem

%%% 296
\bibitem[\protect\citeauthoryear{Watson et~al.}{2019}]{Watson:2019xjv}
\begin{barticle}
\bauthor{\bsnm{Watson}, \binits{D.}}, \betal:
\batitle{{Identification of strontium in the merger of two neutron stars}}.
\bjtitle{Nature}
\bvolume{574}(\bissue{7779}),
\bfpage{497}--\blpage{500}
(\byear{2019})
\doiurl{10.1038/s41586-019-1676-3}
{\href{https://arxiv.org/abs/1910.10510}{{arXiv:1910.10510}}}
{[astro-ph.HE]}
\end{barticle}
\endbibitem

%%% 297
\bibitem[\protect\citeauthoryear{Domoto et~al.}{2022}]{Domoto:2022cqp}
\begin{barticle}
\bauthor{\bsnm{Domoto}, \binits{N.}},
\bauthor{\bsnm{Tanaka}, \binits{M.}},
\bauthor{\bsnm{Kato}, \binits{D.}},
\bauthor{\bsnm{Kawaguchi}, \binits{K.}},
\bauthor{\bsnm{Hotokezaka}, \binits{K.}},
\bauthor{\bsnm{Wanajo}, \binits{S.}}:
\batitle{{Lanthanide Features in Near-infrared Spectra of Kilonovae}}.
\bjtitle{Astrophys. J.}
\bvolume{939}(\bissue{1}),
\bfpage{8}
(\byear{2022})
\doiurl{10.3847/1538-4357/ac8c36}
{\href{https://arxiv.org/abs/2206.04232}{{arXiv:2206.04232}}}
{[astro-ph.HE]}
\end{barticle}
\endbibitem

%%% 298
\bibitem[\protect\citeauthoryear{Sneppen and Watson}{2023}]{Sneppen:2023lgo}
\begin{barticle}
\bauthor{\bsnm{Sneppen}, \binits{A.}},
\bauthor{\bsnm{Watson}, \binits{D.}}:
\batitle{{Discovery of a 760 nm P Cygni line in AT2017gfo: Identification of
  yttrium in the kilonova photosphere}}.
\bjtitle{Astron. Astrophys.}
\bvolume{675},
\bfpage{194}
(\byear{2023})
\doiurl{10.1051/0004-6361/202346421}
{\href{https://arxiv.org/abs/2306.14942}{{arXiv:2306.14942}}}
{[astro-ph.HE]}
\end{barticle}
\endbibitem

%%% 299
\bibitem[\protect\citeauthoryear{Gillanders et~al.}{2024}]{Gillanders:2023jpd}
\begin{barticle}
\bauthor{\bsnm{Gillanders}, \binits{J.H.}},
\bauthor{\bsnm{Sim}, \binits{S.A.}},
\bauthor{\bsnm{Smartt}, \binits{S.J.}},
\bauthor{\bsnm{Goriely}, \binits{S.}},
\bauthor{\bsnm{Bauswein}, \binits{A.}}:
\batitle{{Modelling the spectra of the kilonova AT2017gfo \textendash{} II.
  Beyond the photospheric epochs}}.
\bjtitle{Mon. Not. Roy. Astron. Soc.}
\bvolume{529}(\bissue{3}),
\bfpage{2918}--\blpage{2945}
(\byear{2024})
\doiurl{10.1093/mnras/stad3688}
{\href{https://arxiv.org/abs/2306.15055}{{arXiv:2306.15055}}}
{[astro-ph.HE]}
\end{barticle}
\endbibitem

%%% 300
\bibitem[\protect\citeauthoryear{Hotokezaka et~al.}{2023}]{Hotokezaka:2023aiq}
\begin{barticle}
\bauthor{\bsnm{Hotokezaka}, \binits{K.}},
\bauthor{\bsnm{Tanaka}, \binits{M.}},
\bauthor{\bsnm{Kato}, \binits{D.}},
\bauthor{\bsnm{Gaigalas}, \binits{G.}}:
\batitle{{Tellurium emission line in kilonova AT 2017gfo}}.
\bjtitle{Mon. Not. Roy. Astron. Soc.}
\bvolume{526}(\bissue{1}),
\bfpage{155}--\blpage{159}
(\byear{2023})
\doiurl{10.1093/mnrasl/slad128}
{\href{https://arxiv.org/abs/2307.00988}{{arXiv:2307.00988}}}
{[astro-ph.HE]}
\end{barticle}
\endbibitem

%%% 301
\bibitem[\protect\citeauthoryear{Patel et~al.}{2025}]{Patel:2025frn}
\begin{botherref}
\oauthor{\bsnm{Patel}, \binits{A.}},
\oauthor{\bsnm{Metzger}, \binits{B.D.}},
\oauthor{\bsnm{Cehula}, \binits{J.}},
\oauthor{\bsnm{Burns}, \binits{E.}},
\oauthor{\bsnm{Goldberg}, \binits{J.A.}},
\oauthor{\bsnm{Thompson}, \binits{T.A.}}:
{Direct evidence for r-process nucleosynthesis in delayed MeV emission from the
  SGR 1806-20 magnetar giant flare}
(2025)
{\href{https://arxiv.org/abs/2501.09181}{{arXiv:2501.09181}}}
{[astro-ph.HE]}
\end{botherref}
\endbibitem

%%% 302
\bibitem[\protect\citeauthoryear{Meyer and Howard}{1991}]{Meyer:1991xx}
\begin{bchapter}
\bauthor{\bsnm{Meyer}, \binits{B.S.}},
\bauthor{\bsnm{Howard}, \binits{W.M.}}:
\bctitle{Possible gamma-ray signatures of an r-process event}.
In: \beditor{\bsnm{Woosley}, \binits{S.E.}} (ed.)
\bbtitle{Supernovae},
pp. \bfpage{630}--\blpage{632}.
\bpublisher{Springer},
\blocation{New York, NY}
(\byear{1991})
\end{bchapter}
\endbibitem

%%% 303
\bibitem[\protect\citeauthoryear{Qian et~al.}{1998}]{Qian:1998cz}
\begin{barticle}
\bauthor{\bsnm{Qian}, \binits{Y.Z.}},
\bauthor{\bsnm{Vogel}, \binits{P.}},
\bauthor{\bsnm{Wasserburg}, \binits{G.J.}}:
\batitle{{Supernovae as the site of the r process: Implications for gamma-ray
  astronomy}}.
\bjtitle{Astrophys. J.}
\bvolume{506},
\bfpage{868}--\blpage{873}
(\byear{1998})
\doiurl{10.1086/306285}
{\href{https://arxiv.org/abs/astro-ph/9803300}{{arXiv:astro-ph/9803300}}}
\end{barticle}
\endbibitem

%%% 304
\bibitem[\protect\citeauthoryear{Qian et~al.}{1999}]{Qian:1999vm}
\begin{barticle}
\bauthor{\bsnm{Qian}, \binits{Y.Z.}},
\bauthor{\bsnm{Vogel}, \binits{P.}},
\bauthor{\bsnm{Wasserburg}, \binits{G.J.}}:
\batitle{{Probing r-process production of nuclei beyond bi209 with
  gamma-rays}}.
\bjtitle{Astrophys. J.}
\bvolume{524},
\bfpage{213}--\blpage{219}
(\byear{1999})
\doiurl{10.1086/307805}
{\href{https://arxiv.org/abs/astro-ph/9905387}{{arXiv:astro-ph/9905387}}}
\end{barticle}
\endbibitem

%%% 305
\bibitem[\protect\citeauthoryear{Ripley et~al.}{2014}]{Ripley:2013eca}
\begin{barticle}
\bauthor{\bsnm{Ripley}, \binits{J.L.}},
\bauthor{\bsnm{Metzger}, \binits{B.D.}},
\bauthor{\bsnm{Arcones}, \binits{A.}},
\bauthor{\bsnm{Martinez-Pinedo}, \binits{G.}}:
\batitle{{X-ray Decay Lines from Heavy Nuclei in Supernova Remnants as a Probe
  of the r-Process Origin and the Birth Periods of Magnetars}}.
\bjtitle{Mon. Not. Roy. Astron. Soc.}
\bvolume{438}(\bissue{4}),
\bfpage{3243}--\blpage{3254}
(\byear{2014})
\doiurl{10.1093/mnras/stt2434}
{\href{https://arxiv.org/abs/1310.2950}{{arXiv:1310.2950}}}
{[astro-ph.HE]}
\end{barticle}
\endbibitem

%%% 306
\bibitem[\protect\citeauthoryear{Hotokezaka et~al.}{2016}]{Hotokezaka:2015cma}
\begin{barticle}
\bauthor{\bsnm{Hotokezaka}, \binits{K.}},
\bauthor{\bsnm{Wanajo}, \binits{S.}},
\bauthor{\bsnm{Tanaka}, \binits{M.}},
\bauthor{\bsnm{Bamba}, \binits{A.}},
\bauthor{\bsnm{Terada}, \binits{Y.}},
\bauthor{\bsnm{Piran}, \binits{T.}}:
\batitle{{Radioactive decay products in neutron star merger ejecta: heating
  efficiency and \ensuremath{\gamma}-ray emission}}.
\bjtitle{Mon. Not. Roy. Astron. Soc.}
\bvolume{459}(\bissue{1}),
\bfpage{35}--\blpage{43}
(\byear{2016})
\doiurl{10.1093/mnras/stw404}
{\href{https://arxiv.org/abs/1511.05580}{{arXiv:1511.05580}}}
{[astro-ph.HE]}
\end{barticle}
\endbibitem

%%% 307
\bibitem[\protect\citeauthoryear{Li}{2019}]{Li:2018wee}
\begin{barticle}
\bauthor{\bsnm{Li}, \binits{L.-X.}}:
\batitle{{Radioactive Gamma-Ray Emissions from Neutron Star Mergers}}.
\bjtitle{Astrophys. J.}
\bvolume{872}(\bissue{1}),
\bfpage{19}
(\byear{2019})
\doiurl{10.3847/1538-4357/aaf961}
{\href{https://arxiv.org/abs/1808.09833}{{arXiv:1808.09833}}}
{[astro-ph.HE]}
\end{barticle}
\endbibitem

%%% 308
\bibitem[\protect\citeauthoryear{Wu et~al.}{2019}]{Wu:2019xrq}
\begin{barticle}
\bauthor{\bsnm{Wu}, \binits{M.-R.}},
\bauthor{\bsnm{Banerjee}, \binits{P.}},
\bauthor{\bsnm{Metzger}, \binits{B.D.}},
\bauthor{\bsnm{Mart\'\i{}nez-Pinedo}, \binits{G.}},
\bauthor{\bsnm{Aramaki}, \binits{T.}},
\bauthor{\bsnm{Burns}, \binits{E.}},
\bauthor{\bsnm{Hailey}, \binits{C.J.}},
\bauthor{\bsnm{Barnes}, \binits{J.}},
\bauthor{\bsnm{Karagiorgi}, \binits{G.}}:
\batitle{{Finding the remnants of the Milky Way's last neutron star mergers}}.
\bjtitle{Astrophys. J.}
\bvolume{880}(\bissue{1}),
\bfpage{23}
(\byear{2019})
\doiurl{10.3847/1538-4357/ab2593}
{\href{https://arxiv.org/abs/1905.03793}{{arXiv:1905.03793}}}
{[astro-ph.HE]}
\end{barticle}
\endbibitem

%%% 309
\bibitem[\protect\citeauthoryear{Korobkin et~al.}{2019}]{Korobkin:2019uxw}
\begin{botherref}
\oauthor{\bsnm{Korobkin}, \binits{O.}}, et al.:
{Gamma-rays from kilonova: a potential probe of r-process nucleosynthesis}
(2019)
\doiurl{10.3847/1538-4357/ab64d8}
{\href{https://arxiv.org/abs/1905.05089}{{arXiv:1905.05089}}}
{[astro-ph.HE]}
\end{botherref}
\endbibitem

%%% 310
\bibitem[\protect\citeauthoryear{Wang et~al.}{2020}]{Wang:2020qkn}
\begin{barticle}
\bauthor{\bsnm{Wang}, \binits{X.}},
\bauthor{\bsnm{Vassh}, \binits{N.}},
\bauthor{\bsnm{Sprouse}, \binits{T.}},
\bauthor{\bsnm{Mumpower}, \binits{M.}},
\bauthor{\bsnm{Vogt}, \binits{R.}},
\bauthor{\bsnm{Randrup}, \binits{J.}},
\bauthor{\bsnm{Surman}, \binits{R.}}:
\batitle{{MeV Gamma Rays from Fission: A Distinct Signature of Actinide
  Production in Neutron Star Mergers}}.
\bjtitle{Astrophys. J. Lett.}
\bvolume{903}(\bissue{1}),
\bfpage{3}
(\byear{2020})
\doiurl{10.3847/2041-8213/abbe18}
{\href{https://arxiv.org/abs/2008.03335}{{arXiv:2008.03335}}}
{[astro-ph.HE]}
\end{barticle}
\endbibitem

%%% 311
\bibitem[\protect\citeauthoryear{Chen et~al.}{2021}]{Chen:2021tob}
\begin{barticle}
\bauthor{\bsnm{Chen}, \binits{M.-H.}},
\bauthor{\bsnm{Li}, \binits{L.-X.}},
\bauthor{\bsnm{Lin}, \binits{D.-B.}},
\bauthor{\bsnm{Liang}, \binits{E.-W.}}:
\batitle{{Gamma-Ray Emission Produced by r-process Elements from Neutron Star
  Mergers}}.
\bjtitle{Astrophys. J.}
\bvolume{919}(\bissue{1}),
\bfpage{59}
(\byear{2021})
\doiurl{10.3847/1538-4357/ac1267}
{\href{https://arxiv.org/abs/2107.02982}{{arXiv:2107.02982}}}
{[astro-ph.HE]}
\end{barticle}
\endbibitem

%%% 312
\bibitem[\protect\citeauthoryear{Terada et~al.}{2022}]{Terada:2022hut}
\begin{barticle}
\bauthor{\bsnm{Terada}, \binits{Y.}},
\bauthor{\bsnm{Miwa}, \binits{Y.}},
\bauthor{\bsnm{Ohsumi}, \binits{H.}},
\bauthor{\bsnm{Fujimoto}, \binits{S.-i.}},
\bauthor{\bsnm{Katsuda}, \binits{S.}},
\bauthor{\bsnm{Bamba}, \binits{A.}},
\bauthor{\bsnm{Yamazaki}, \binits{R.}}:
\batitle{{Gamma-ray Diagnostics of r-process Nucleosynthesis in the Remnants of
  Galactic Binary Neutron-Star Mergers}}.
\bjtitle{Astrophys. J.}
\bvolume{933}(\bissue{1}),
\bfpage{111}
(\byear{2022})
\doiurl{10.3847/1538-4357/ac721f}
{\href{https://arxiv.org/abs/2205.05407}{{arXiv:2205.05407}}}
{[astro-ph.HE]}
\end{barticle}
\endbibitem

%%% 313
\bibitem[\protect\citeauthoryear{Chen et~al.}{2022}]{Chen:2022nsj}
\begin{barticle}
\bauthor{\bsnm{Chen}, \binits{M.-H.}},
\bauthor{\bsnm{Hu}, \binits{R.-C.}},
\bauthor{\bsnm{Liang}, \binits{E.-W.}}:
\batitle{{Radioactively Powered Gamma-Ray Transient Associated with a Kilonova
  from Neutron Star Merger}}.
\bjtitle{Astrophys. J. Lett.}
\bvolume{932}(\bissue{1}),
\bfpage{7}
(\byear{2022})
\doiurl{10.3847/2041-8213/ac7470}
{\href{https://arxiv.org/abs/2204.13269}{{arXiv:2204.13269}}}
{[astro-ph.HE]}
\end{barticle}
\endbibitem

%%% 314
\bibitem[\protect\citeauthoryear{Vassh et~al.}{2024}]{Vassh:2023fnb}
\begin{barticle}
\bauthor{\bsnm{Vassh}, \binits{N.}},
\bauthor{\bsnm{Wang}, \binits{X.}},
\bauthor{\bsnm{Lariviere}, \binits{M.}},
\bauthor{\bsnm{Sprouse}, \binits{T.}},
\bauthor{\bsnm{Mumpower}, \binits{M.R.}},
\bauthor{\bsnm{Surman}, \binits{R.}},
\bauthor{\bsnm{Liu}, \binits{Z.}},
\bauthor{\bsnm{McLaughlin}, \binits{G.C.}},
\bauthor{\bsnm{Denissenkov}, \binits{P.}},
\bauthor{\bsnm{Herwig}, \binits{F.}}:
\batitle{{Thallium-208: A Beacon of In~Situ Neutron Capture Nucleosynthesis}}.
\bjtitle{Phys. Rev. Lett.}
\bvolume{132}(\bissue{5}),
\bfpage{052701}
(\byear{2024})
\doiurl{10.1103/PhysRevLett.132.052701}
{\href{https://arxiv.org/abs/2311.10895}{{arXiv:2311.10895}}}
{[nucl-th]}
\end{barticle}
\endbibitem

%%% 315
\bibitem[\protect\citeauthoryear{Chen et~al.}{2024}]{Chen:2024nbq}
\begin{barticle}
\bauthor{\bsnm{Chen}, \binits{M.-H.}},
\bauthor{\bsnm{Li}, \binits{L.-X.}},
\bauthor{\bsnm{Liang}, \binits{E.-W.}}:
\batitle{{Radioactive Gamma-Ray Lines from Long-lived Neutron Star Merger
  Remnants}}.
\bjtitle{Astrophys. J.}
\bvolume{971}(\bissue{2}),
\bfpage{143}
(\byear{2024})
\doiurl{10.3847/1538-4357/ad65ec}
{\href{https://arxiv.org/abs/2407.14762}{{arXiv:2407.14762}}}
{[astro-ph.HE]}
\end{barticle}
\endbibitem

%%% 316
\bibitem[\protect\citeauthoryear{{Amend} et~al.}{2025}]{Amend:2024sdm}
\begin{barticle}
\bauthor{\bsnm{{Amend}}, \binits{B.}},
\bauthor{\bsnm{{Fryer}}, \binits{C.L.}},
\bauthor{\bsnm{{Mumpower}}, \binits{M.R.}},
\bauthor{\bsnm{{Korobkin}}, \binits{O.}}:
\batitle{{Spatial Models of R-process Remnants and their
  {\ensuremath{\gamma}}-Ray Detectability}}.
\bjtitle{\apj}
\bvolume{991}(\bissue{2}),
\bfpage{216}
(\byear{2025})
\doiurl{10.3847/1538-4357/adfdde}
{\href{https://arxiv.org/abs/2412.05424}{{arXiv:2412.05424}}}
{[astro-ph.HE]}
\end{barticle}
\endbibitem

%%% 317
\bibitem[\protect\citeauthoryear{Liu et~al.}{2025}]{Liu:2025auu}
\begin{botherref}
\oauthor{\bsnm{Liu}, \binits{Z.}},
\oauthor{\bsnm{Grohs}, \binits{E.}},
\oauthor{\bsnm{Lund}, \binits{K.A.}},
\oauthor{\bsnm{McLaughlin}, \binits{G.C.}},
\oauthor{\bsnm{Reichert}, \binits{M.}},
\oauthor{\bsnm{R{\"o}derer}, \binits{I.U.}},
\oauthor{\bsnm{Surman}, \binits{R.}},
\oauthor{\bsnm{Wang}, \binits{X.}}:
{Gamma rays as a signature of r-process producing supernovae: remnants and
  future Galactic explosions}
(2025)
{\href{https://arxiv.org/abs/2506.14991}{{arXiv:2506.14991}}}
{[astro-ph.HE]}
\end{botherref}
\endbibitem

%%% 318
\bibitem[\protect\citeauthoryear{Patel et~al.}{2025}]{Patel:2025tse}
\begin{botherref}
\oauthor{\bsnm{Patel}, \binits{A.}},
\oauthor{\bsnm{Metzger}, \binits{B.D.}},
\oauthor{\bsnm{Goldberg}, \binits{J.A.}},
\oauthor{\bsnm{Cehula}, \binits{J.}},
\oauthor{\bsnm{Thompson}, \binits{T.A.}},
\oauthor{\bsnm{Renzo}, \binits{M.}}:
{r-Process Nucleosynthesis and Radioactively Powered Transients from Magnetar
  Giant Flares}
(2025)
{\href{https://arxiv.org/abs/2501.17253}{{arXiv:2501.17253}}}
{[astro-ph.HE]}
\end{botherref}
\endbibitem

%%% 319
\bibitem[\protect\citeauthoryear{Shibata and
  Hotokezaka}{2019}]{Shibata:2019wef}
\begin{barticle}
\bauthor{\bsnm{Shibata}, \binits{M.}},
\bauthor{\bsnm{Hotokezaka}, \binits{K.}}:
\batitle{{Merger and Mass Ejection of Neutron-Star Binaries}}.
\bjtitle{Ann. Rev. Nucl. Part. Sci.}
\bvolume{69},
\bfpage{41}--\blpage{64}
(\byear{2019})
\doiurl{10.1146/annurev-nucl-101918-023625}
{\href{https://arxiv.org/abs/1908.02350}{{arXiv:1908.02350}}}
{[astro-ph.HE]}
\end{barticle}
\endbibitem

%%% 320
\bibitem[\protect\citeauthoryear{Mandel and
  Broekgaarden}{2022}]{Mandel:2021smh}
\begin{barticle}
\bauthor{\bsnm{Mandel}, \binits{I.}},
\bauthor{\bsnm{Broekgaarden}, \binits{F.S.}}:
\batitle{{Rates of compact object coalescences}}.
\bjtitle{Living Rev. Rel.}
\bvolume{25}(\bissue{1}),
\bfpage{1}
(\byear{2022})
\doiurl{10.1007/s41114-021-00034-3}
{\href{https://arxiv.org/abs/2107.14239}{{arXiv:2107.14239}}}
{[astro-ph.HE]}
\end{barticle}
\endbibitem

%%% 321
\bibitem[\protect\citeauthoryear{Goriely and Janka}{2016}]{Goriely:2016gfe}
\begin{barticle}
\bauthor{\bsnm{Goriely}, \binits{S.}},
\bauthor{\bsnm{Janka}, \binits{H.-T.}}:
\batitle{{Solar r-process-constrained actinide production in neutrino-driven
  winds of supernovae}}.
\bjtitle{Mon. Not. Roy. Astron. Soc.}
\bvolume{459}(\bissue{4}),
\bfpage{4174}--\blpage{4182}
(\byear{2016})
\doiurl{10.1093/mnras/stw946}
{\href{https://arxiv.org/abs/1603.04282}{{arXiv:1603.04282}}}
{[astro-ph.SR]}
\end{barticle}
\endbibitem

%%% 322
\bibitem[\protect\citeauthoryear{Rozwadowska
  et~al.}{2021}]{Rozwadowska:2020nab}
\begin{barticle}
\bauthor{\bsnm{Rozwadowska}, \binits{K.}},
\bauthor{\bsnm{Vissani}, \binits{F.}},
\bauthor{\bsnm{Cappellaro}, \binits{E.}}:
\batitle{{On the rate of core collapse supernovae in the milky way}}.
\bjtitle{New Astron.}
\bvolume{83},
\bfpage{101498}
(\byear{2021})
\doiurl{10.1016/j.newast.2020.101498}
{\href{https://arxiv.org/abs/2009.03438}{{arXiv:2009.03438}}}
{[astro-ph.HE]}
\end{barticle}
\endbibitem

%%% 323
\bibitem[\protect\citeauthoryear{{Smith} et~al.}{2003}]{smith2003}
\begin{barticle}
\bauthor{\bsnm{{Smith}}, \binits{D.M.}},
\bauthor{\bsnm{{Share}}, \binits{G.H.}},
\bauthor{\bsnm{{Murphy}}, \binits{R.J.}},
\bauthor{\bsnm{{Schwartz}}, \binits{R.A.}},
\bauthor{\bsnm{{Shih}}, \binits{A.Y.}},
\bauthor{\bsnm{{Lin}}, \binits{R.P.}}:
\batitle{{High-Resolution Spectroscopy of Gamma-Ray Lines from the X-Class
  Solar Flare of 2002 July 23}}.
\bjtitle{\apjl}
\bvolume{595}(\bissue{2}),
\bfpage{81}--\blpage{84}
(\byear{2003})
\doiurl{10.1086/378173}
{\href{https://arxiv.org/abs/astro-ph/0306292}{{arXiv:astro-ph/0306292}}}
{[astro-ph]}
\end{barticle}
\endbibitem

%%% 324
\bibitem[\protect\citeauthoryear{{Kiener} et~al.}{2006}]{kiener2006}
\begin{barticle}
\bauthor{\bsnm{{Kiener}}, \binits{J.}},
\bauthor{\bsnm{{Gros}}, \binits{M.}},
\bauthor{\bsnm{{Tatischeff}}, \binits{V.}},
\bauthor{\bsnm{{Weidenspointner}}, \binits{G.}}:
\batitle{{Properties of the energetic particle distributions during the October
  28, 2003 solar flare from INTEGRAL/SPI observations}}.
\bjtitle{\aap}
\bvolume{445}(\bissue{2}),
\bfpage{725}--\blpage{733}
(\byear{2006})
\doiurl{10.1051/0004-6361:20053665}
{\href{https://arxiv.org/abs/astro-ph/0511091}{{arXiv:astro-ph/0511091}}}
{[astro-ph]}
\end{barticle}
\endbibitem

%%% 325
\bibitem[\protect\citeauthoryear{{T{\"u}rler} et~al.}{2021}]{turler2021}
\begin{barticle}
\bauthor{\bsnm{{T{\"u}rler}}, \binits{M.}},
\bauthor{\bsnm{{Tatischeff}}, \binits{V.}},
\bauthor{\bsnm{{Beckmann}}, \binits{V.}},
\bauthor{\bsnm{{Churazov}}, \binits{E.}}:
\batitle{{INTEGRAL serendipitous observations of solar and terrestrial X-rays
  and gamma rays}}.
\bjtitle{\nar}
\bvolume{93},
\bfpage{101616}
(\byear{2021})
\doiurl{10.1016/j.newar.2021.101616}
{\href{https://arxiv.org/abs/2104.06073}{{arXiv:2104.06073}}}
{[astro-ph.HE]}
\end{barticle}
\endbibitem

%%% 326
\bibitem[\protect\citeauthoryear{{Tatischeff} and
  {Gabici}}{2018}]{tatischeff2018}
\begin{barticle}
\bauthor{\bsnm{{Tatischeff}}, \binits{V.}},
\bauthor{\bsnm{{Gabici}}, \binits{S.}}:
\batitle{{Particle Acceleration by Supernova Shocks and Spallogenic
  Nucleosynthesis of Light Elements}}.
\bjtitle{Annual Review of Nuclear and Particle Science}
\bvolume{68}(\bissue{1}),
\bfpage{377}--\blpage{404}
(\byear{2018})
\doiurl{10.1146/annurev-nucl-101917-021151}
{\href{https://arxiv.org/abs/1803.01794}{{arXiv:1803.01794}}}
{[astro-ph.HE]}
\end{barticle}
\endbibitem

%%% 327
\bibitem[\protect\citeauthoryear{{Cummings} et~al.}{2016}]{cummings2016}
\begin{barticle}
\bauthor{\bsnm{{Cummings}}, \binits{A.C.}},
\bauthor{\bsnm{{Stone}}, \binits{E.C.}},
\bauthor{\bsnm{{Heikkila}}, \binits{B.C.}},
\bauthor{\bsnm{{Lal}}, \binits{N.}},
\bauthor{\bsnm{{Webber}}, \binits{W.R.}},
\bauthor{\bsnm{{J{\'o}hannesson}}, \binits{G.}},
\bauthor{\bsnm{{Moskalenko}}, \binits{I.V.}},
\bauthor{\bsnm{{Orlando}}, \binits{E.}},
\bauthor{\bsnm{{Porter}}, \binits{T.A.}}:
\batitle{{Galactic Cosmic Rays in the Local Interstellar Medium: Voyager 1
  Observations and Model Results}}.
\bjtitle{\apj}
\bvolume{831}(\bissue{1}),
\bfpage{18}
(\byear{2016})
\doiurl{10.3847/0004-637X/831/1/18}
\end{barticle}
\endbibitem

%%% 328
\bibitem[\protect\citeauthoryear{{Stone} et~al.}{2019}]{stone2019}
\begin{barticle}
\bauthor{\bsnm{{Stone}}, \binits{E.C.}},
\bauthor{\bsnm{{Cummings}}, \binits{A.C.}},
\bauthor{\bsnm{{Heikkila}}, \binits{B.C.}},
\bauthor{\bsnm{{Lal}}, \binits{N.}}:
\batitle{{Cosmic ray measurements from Voyager 2 as it crossed into
  interstellar space}}.
\bjtitle{Nature Astronomy}
\bvolume{3},
\bfpage{1013}--\blpage{1018}
(\byear{2019})
\doiurl{10.1038/s41550-019-0928-3}
\end{barticle}
\endbibitem

%%% 329
\bibitem[\protect\citeauthoryear{{Indriolo} et~al.}{2015}]{indriolo2015}
\begin{barticle}
\bauthor{\bsnm{{Indriolo}}, \binits{N.}},
\bauthor{\bsnm{{Neufeld}}, \binits{D.A.}},
\bauthor{\bsnm{{Gerin}}, \binits{M.}},
\bauthor{\bsnm{{Schilke}}, \binits{P.}},
\bauthor{\bsnm{{Benz}}, \binits{A.O.}},
\bauthor{\bsnm{{Winkel}}, \binits{B.}},
\bauthor{\bsnm{{Menten}}, \binits{K.M.}},
\bauthor{\bsnm{{Chambers}}, \binits{E.T.}},
\bauthor{\bsnm{{Black}}, \binits{J.H.}},
\bauthor{\bsnm{{Bruderer}}, \binits{S.}},
\bauthor{\bsnm{{Falgarone}}, \binits{E.}},
\bauthor{\bsnm{{Godard}}, \binits{B.}},
\bauthor{\bsnm{{Goicoechea}}, \binits{J.R.}},
\bauthor{\bsnm{{Gupta}}, \binits{H.}},
\bauthor{\bsnm{{Lis}}, \binits{D.C.}},
\bauthor{\bsnm{{Ossenkopf}}, \binits{V.}},
\bauthor{\bsnm{{Persson}}, \binits{C.M.}},
\bauthor{\bsnm{{Sonnentrucker}}, \binits{P.}},
\bauthor{\bsnm{{van der Tak}}, \binits{F.F.S.}},
\bauthor{\bsnm{{van Dishoeck}}, \binits{E.F.}},
\bauthor{\bsnm{{Wolfire}}, \binits{M.G.}},
\bauthor{\bsnm{{Wyrowski}}, \binits{F.}}:
\batitle{{Herschel Survey of Galactic OH$^{+}$, H$_{2}$O$^{+}$, and
  H$_{3}$O$^{+}$: Probing the Molecular Hydrogen Fraction and Cosmic-Ray
  Ionization Rate}}.
\bjtitle{\apj}
\bvolume{800}(\bissue{1}),
\bfpage{40}
(\byear{2015})
\doiurl{10.1088/0004-637X/800/1/40}
{\href{https://arxiv.org/abs/1412.1106}{{arXiv:1412.1106}}}
{[astro-ph.GA]}
\end{barticle}
\endbibitem

%%% 330
\bibitem[\protect\citeauthoryear{{Indriolo} and {McCall}}{2012}]{indriolo2012}
\begin{barticle}
\bauthor{\bsnm{{Indriolo}}, \binits{N.}},
\bauthor{\bsnm{{McCall}}, \binits{B.J.}}:
\batitle{{Investigating the Cosmic-Ray Ionization Rate in the Galactic Diffuse
  Interstellar Medium through Observations of H$^{+}$ $_{3}$}}.
\bjtitle{\apj}
\bvolume{745}(\bissue{1}),
\bfpage{91}
(\byear{2012})
\doiurl{10.1088/0004-637X/745/1/91}
{\href{https://arxiv.org/abs/1111.6936}{{arXiv:1111.6936}}}
{[astro-ph.GA]}
\end{barticle}
\endbibitem

%%% 331
\bibitem[\protect\citeauthoryear{{Tatischeff} and
  {Kiener}}{2011}]{tatischeff2011}
\begin{barticle}
\bauthor{\bsnm{{Tatischeff}}, \binits{V.}},
\bauthor{\bsnm{{Kiener}}, \binits{J.}}:
\batitle{{Nuclear interactions of low-energy cosmic rays with the interstellar
  medium .}}
\bjtitle{\memsai}
\bvolume{82},
\bfpage{903}
(\byear{2011})
\doiurl{10.48550/arXiv.1109.1672}
{\href{https://arxiv.org/abs/1109.1672}{{arXiv:1109.1672}}}
{[astro-ph.HE]}
\end{barticle}
\endbibitem

%%% 332
\bibitem[\protect\citeauthoryear{{Ramaty} et~al.}{1979}]{ramaty1979}
\begin{barticle}
\bauthor{\bsnm{{Ramaty}}, \binits{R.}},
\bauthor{\bsnm{{Kozlovsky}}, \binits{B.}},
\bauthor{\bsnm{{Lingenfelter}}, \binits{R.E.}}:
\batitle{{Nuclear gamma-rays from energetic particle interactions.}}
\bjtitle{\apjs}
\bvolume{40},
\bfpage{487}--\blpage{526}
(\byear{1979})
\doiurl{10.1086/190596}
\end{barticle}
\endbibitem

%%% 333
\bibitem[\protect\citeauthoryear{{Murphy} et~al.}{2009}]{murphy2009}
\begin{barticle}
\bauthor{\bsnm{{Murphy}}, \binits{R.J.}},
\bauthor{\bsnm{{Kozlovsky}}, \binits{B.}},
\bauthor{\bsnm{{Kiener}}, \binits{J.}},
\bauthor{\bsnm{{Share}}, \binits{G.H.}}:
\batitle{{Nuclear Gamma-Ray De-Excitation Lines and Continuum from
  Accelerated-Particle Interactions in Solar Flares}}.
\bjtitle{\apjs}
\bvolume{183}(\bissue{1}),
\bfpage{142}--\blpage{155}
(\byear{2009})
\doiurl{10.1088/0067-0049/183/1/142}
\end{barticle}
\endbibitem

%%% 334
\bibitem[\protect\citeauthoryear{{Benhabiles-Mezhoud} et~al.}{2013}]{ben13}
\begin{barticle}
\bauthor{\bsnm{{Benhabiles-Mezhoud}}, \binits{H.}},
\bauthor{\bsnm{{Kiener}}, \binits{J.}},
\bauthor{\bsnm{{Tatischeff}}, \binits{V.}},
\bauthor{\bsnm{{Strong}}, \binits{A.W.}}:
\batitle{{De-excitation Nuclear Gamma-Ray Line Emission from Low-energy Cosmic
  Rays in the Inner Galaxy}}.
\bjtitle{\apj}
\bvolume{763}(\bissue{2}),
\bfpage{98}
(\byear{2013})
\doiurl{10.1088/0004-637X/763/2/98}
{\href{https://arxiv.org/abs/1212.1622}{{arXiv:1212.1622}}}
{[astro-ph.HE]}
\end{barticle}
\endbibitem

%%% 335
\bibitem[\protect\citeauthoryear{{Tatischeff} and
  {Kiener}}{2004}]{tatischeff2004}
\begin{barticle}
\bauthor{\bsnm{{Tatischeff}}, \binits{V.}},
\bauthor{\bsnm{{Kiener}}, \binits{J.}}:
\batitle{{{\ensuremath{\gamma}}-ray lines from cosmic-ray interactions with
  interstellar dust grains}}.
\bjtitle{\nar}
\bvolume{48}(\bissue{1-4}),
\bfpage{99}--\blpage{103}
(\byear{2004})
\doiurl{10.1016/j.newar.2003.11.013}
{\href{https://arxiv.org/abs/astro-ph/0310827}{{arXiv:astro-ph/0310827}}}
{[astro-ph]}
\end{barticle}
\endbibitem

%%% 336
\bibitem[\protect\citeauthoryear{{Jones} et~al.}{2017}]{jones2017}
\begin{barticle}
\bauthor{\bsnm{{Jones}}, \binits{A.P.}},
\bauthor{\bsnm{{K{\"o}hler}}, \binits{M.}},
\bauthor{\bsnm{{Ysard}}, \binits{N.}},
\bauthor{\bsnm{{Bocchio}}, \binits{M.}},
\bauthor{\bsnm{{Verstraete}}, \binits{L.}}:
\batitle{{The global dust modelling framework THEMIS}}.
\bjtitle{\aap}
\bvolume{602},
\bfpage{46}
(\byear{2017})
\doiurl{10.1051/0004-6361/201630225}
{\href{https://arxiv.org/abs/1703.00775}{{arXiv:1703.00775}}}
{[astro-ph.GA]}
\end{barticle}
\endbibitem

%%% 337
\bibitem[\protect\citeauthoryear{{de Angelis}
  et~al.}{2018}]{eastrogam_science17}
\begin{barticle}
\bauthor{\bsnm{{de Angelis}}, \binits{A.}},
\bauthor{\bsnm{{Tatischeff}}, \binits{V.}},
\bauthor{\bsnm{{Grenier}}, \binits{I.A.}},
\bauthor{\bsnm{{McEnery}}, \binits{J.}},
\bauthor{\bsnm{{Mallamaci}}, \binits{M.}},
\bauthor{\bsnm{{Tavani}}, \binits{M.}},
\bauthor{\bsnm{{Oberlack}}, \binits{U.}},
\bauthor{\bsnm{{Hanlon}}, \binits{L.}},
\bauthor{\bsnm{{Walter}}, \binits{R.}},
\bauthor{\bsnm{{Argan}}, \binits{A.}},
\bauthor{\bsnm{{von Ballmoos}}, \binits{P.}},
\bauthor{\bsnm{{Bulgarelli}}, \binits{A.}},
\bauthor{\bsnm{{Bykov}}, \binits{A.}},
\bauthor{\bsnm{{Hernanz}}, \binits{M.}},
\bauthor{\bsnm{{Kanbach}}, \binits{G.}},
\bauthor{\bsnm{{Kuvvetli}}, \binits{I.}},
\bauthor{\bsnm{{Pearce}}, \binits{M.}},
\bauthor{\bsnm{{Zdziarski}}, \binits{A.}},
\bauthor{\bsnm{{Conrad}}, \binits{J.}},
\bauthor{\bsnm{{Ghisellini}}, \binits{G.}},
\bauthor{\bsnm{{Harding}}, \binits{A.}},
\bauthor{\bsnm{{Isern}}, \binits{J.}},
\bauthor{\bsnm{{Leising}}, \binits{M.}},
\bauthor{\bsnm{{Longo}}, \binits{F.}},
\bauthor{\bsnm{{Madejski}}, \binits{G.}},
\bauthor{\bsnm{{Martinez}}, \binits{M.}},
\bauthor{\bsnm{{Mazziotta}}, \binits{M.N.}},
\bauthor{\bsnm{{Paredes}}, \binits{J.M.}},
\bauthor{\bsnm{{Pohl}}, \binits{M.}},
\bauthor{\bsnm{{Rando}}, \binits{R.}},
\bauthor{\bsnm{{Razzano}}, \binits{M.}},
\bauthor{\bsnm{{Aboudan}}, \binits{A.}},
\bauthor{\bsnm{{Ackermann}}, \binits{M.}},
\bauthor{\bsnm{{Addazi}}, \binits{A.}},
\bauthor{\bsnm{{Ajello}}, \binits{M.}},
\bauthor{\bsnm{{Albertus}}, \binits{C.}},
\bauthor{\bsnm{{{\'A}lvarez}}, \binits{J.M.}},
\bauthor{\bsnm{{Ambrosi}}, \binits{G.}},
\bauthor{\bsnm{{Ant{\'o}n}}, \binits{S.}},
\bauthor{\bsnm{{Antonelli}}, \binits{L.A.}},
\bauthor{\bsnm{{Babic}}, \binits{A.}},
\bauthor{\bsnm{{Baibussinov}}, \binits{B.}},
\bauthor{\bsnm{{Balbo}}, \binits{M.}},
\bauthor{\bsnm{{Baldini}}, \binits{L.}},
\bauthor{\bsnm{{Balman}}, \binits{S.}},
\bauthor{\bsnm{{Bambi}}, \binits{C.}},
\bauthor{\bsnm{{Barres de Almeida}}, \binits{U.}},
\bauthor{\bsnm{{Barrio}}, \binits{J.A.}},
\bauthor{\bsnm{{Bartels}}, \binits{R.}},
\bauthor{\bsnm{{Bastieri}}, \binits{D.}},
\bauthor{\bsnm{{Bednarek}}, \binits{W.}},
\bauthor{\bsnm{{Bernard}}, \binits{D.}},
\bauthor{\bsnm{{Bernardini}}, \binits{E.}},
\bauthor{\bsnm{{Bernasconi}}, \binits{T.}},
\bauthor{\bsnm{{Bertucci}}, \binits{B.}},
\bauthor{\bsnm{{Biland}}, \binits{A.}},
\bauthor{\bsnm{{Bissaldi}}, \binits{E.}},
\bauthor{\bsnm{{B{\"o}ttcher}}, \binits{M.}},
\bauthor{\bsnm{{Bonvicini}}, \binits{V.}},
\bauthor{\bsnm{{Bosch-Ramon}}, \binits{V.}},
\bauthor{\bsnm{{Bottacini}}, \binits{E.}},
\bauthor{\bsnm{{Bozhilov}}, \binits{V.}},
\bauthor{\bsnm{{Bretz}}, \binits{T.}},
\bauthor{\bsnm{{Branchesi}}, \binits{M.}},
\bauthor{\bsnm{{Brdar}}, \binits{V.}},
\bauthor{\bsnm{{Bringmann}}, \binits{T.}},
\bauthor{\bsnm{{Brogna}}, \binits{A.}},
\bauthor{\bsnm{{Budtz J{\o}rgensen}}, \binits{C.}},
\bauthor{\bsnm{{Busetto}}, \binits{G.}},
\bauthor{\bsnm{{Buson}}, \binits{S.}},
\bauthor{\bsnm{{Busso}}, \binits{M.}},
\bauthor{\bsnm{{Caccianiga}}, \binits{A.}},
\bauthor{\bsnm{{Camera}}, \binits{S.}},
\bauthor{\bsnm{{Campana}}, \binits{R.}},
\bauthor{\bsnm{{Caraveo}}, \binits{P.}},
\bauthor{\bsnm{{Cardillo}}, \binits{M.}},
\bauthor{\bsnm{{Carlson}}, \binits{P.}},
\bauthor{\bsnm{{Celestin}}, \binits{S.}},
\bauthor{\bsnm{{Cerme{\~n}o}}, \binits{M.}},
\bauthor{\bsnm{{Chen}}, \binits{A.}},
\bauthor{\bsnm{{Cheung}}, \binits{C.C.}},
\bauthor{\bsnm{{Churazov}}, \binits{E.}},
\bauthor{\bsnm{{Ciprini}}, \binits{S.}},
\bauthor{\bsnm{{Coc}}, \binits{A.}},
\bauthor{\bsnm{{Colafrancesco}}, \binits{S.}},
\bauthor{\bsnm{{Coleiro}}, \binits{A.}},
\bauthor{\bsnm{{Collmar}}, \binits{W.}},
\bauthor{\bsnm{{Coppi}}, \binits{P.}},
\bauthor{\bsnm{{Curado da Silva}}, \binits{R.}},
\bauthor{\bsnm{{Cutini}}, \binits{S.}},
\bauthor{\bsnm{{D'Ammando}}, \binits{F.}},
\bauthor{\bsnm{{de Lotto}}, \binits{B.}},
\bauthor{\bsnm{{de Martino}}, \binits{D.}},
\bauthor{\bsnm{{De Rosa}}, \binits{A.}},
\bauthor{\bsnm{{Del Santo}}, \binits{M.}},
\bauthor{\bsnm{{Delgado}}, \binits{L.}},
\bauthor{\bsnm{{Diehl}}, \binits{R.}},
\bauthor{\bsnm{{Dietrich}}, \binits{S.}},
\bauthor{\bsnm{{Dolgov}}, \binits{A.D.}},
\bauthor{\bsnm{{Dom{\'\i}nguez}}, \binits{A.}},
\bauthor{\bsnm{{Dominis Prester}}, \binits{D.}},
\bauthor{\bsnm{{Donnarumma}}, \binits{I.}},
\bauthor{\bsnm{{Dorner}}, \binits{D.}},
\bauthor{\bsnm{{Doro}}, \binits{M.}},
\bauthor{\bsnm{{Dutra}}, \binits{M.}},
\bauthor{\bsnm{{Els{\"a}sser}}, \binits{D.}},
\bauthor{\bsnm{{Fabrizio}}, \binits{M.}},
\bauthor{\bsnm{{Fern{\'a}ndez-Barral}}, \binits{A.}},
\bauthor{\bsnm{{Fioretti}}, \binits{V.}},
\bauthor{\bsnm{{Foffano}}, \binits{L.}},
\bauthor{\bsnm{{Formato}}, \binits{V.}},
\bauthor{\bsnm{{Fornengo}}, \binits{N.}},
\bauthor{\bsnm{{Foschini}}, \binits{L.}},
\bauthor{\bsnm{{Franceschini}}, \binits{A.}},
\bauthor{\bsnm{{Franckowiak}}, \binits{A.}},
\bauthor{\bsnm{{Funk}}, \binits{S.}},
\bauthor{\bsnm{{Fuschino}}, \binits{F.}},
\bauthor{\bsnm{{Gaggero}}, \binits{D.}},
\bauthor{\bsnm{{Galanti}}, \binits{G.}},
\bauthor{\bsnm{{Gargano}}, \binits{F.}},
\bauthor{\bsnm{{Gasparrini}}, \binits{D.}},
\bauthor{\bsnm{{Gehrz}}, \binits{R.}},
\bauthor{\bsnm{{Giammaria}}, \binits{P.}},
\bauthor{\bsnm{{Giglietto}}, \binits{N.}},
\bauthor{\bsnm{{Giommi}}, \binits{P.}},
\bauthor{\bsnm{{Giordano}}, \binits{F.}},
\bauthor{\bsnm{{Giroletti}}, \binits{M.}},
\bauthor{\bsnm{{Ghirlanda}}, \binits{G.}},
\bauthor{\bsnm{{Godinovic}}, \binits{N.}},
\bauthor{\bsnm{{Gouiff{\'e}s}}, \binits{C.}},
\bauthor{\bsnm{{Grove}}, \binits{J.E.}},
\bauthor{\bsnm{{Hamadache}}, \binits{C.}},
\bauthor{\bsnm{{Hartmann}}, \binits{D.H.}},
\bauthor{\bsnm{{Hayashida}}, \binits{M.}},
\bauthor{\bsnm{{Hryczuk}}, \binits{A.}},
\bauthor{\bsnm{{Jean}}, \binits{P.}},
\bauthor{\bsnm{{Johnson}}, \binits{T.}},
\bauthor{\bsnm{{Jos{\'e}}}, \binits{J.}},
\bauthor{\bsnm{{Kaufmann}}, \binits{S.}},
\bauthor{\bsnm{{Khelifi}}, \binits{B.}},
\bauthor{\bsnm{{Kiener}}, \binits{J.}},
\bauthor{\bsnm{{Kn{\"o}dlseder}}, \binits{J.}},
\bauthor{\bsnm{{Kole}}, \binits{M.}},
\bauthor{\bsnm{{Kopp}}, \binits{J.}},
\bauthor{\bsnm{{Kozhuharov}}, \binits{V.}},
\bauthor{\bsnm{{Labanti}}, \binits{C.}},
\bauthor{\bsnm{{Lalkovski}}, \binits{S.}},
\bauthor{\bsnm{{Laurent}}, \binits{P.}},
\bauthor{\bsnm{{Limousin}}, \binits{O.}},
\bauthor{\bsnm{{Linares}}, \binits{M.}},
\bauthor{\bsnm{{Lindfors}}, \binits{E.}},
\bauthor{\bsnm{{Lindner}}, \binits{M.}},
\bauthor{\bsnm{{Liu}}, \binits{J.}},
\bauthor{\bsnm{{Lombardi}}, \binits{S.}},
\bauthor{\bsnm{{Loparco}}, \binits{F.}},
\bauthor{\bsnm{{L{\'o}pez-Coto}}, \binits{R.}},
\bauthor{\bsnm{{L{\'o}pez Moya}}, \binits{M.}},
\bauthor{\bsnm{{Lott}}, \binits{B.}},
\bauthor{\bsnm{{Lubrano}}, \binits{P.}},
\bauthor{\bsnm{{Malyshev}}, \binits{D.}},
\bauthor{\bsnm{{Mankuzhiyil}}, \binits{N.}},
\bauthor{\bsnm{{Mannheim}}, \binits{K.}},
\bauthor{\bsnm{{March{\~a}}}, \binits{M.J.}},
\bauthor{\bsnm{{Marcian{\`o}}}, \binits{A.}},
\bauthor{\bsnm{{Marcote}}, \binits{B.}},
\bauthor{\bsnm{{Mariotti}}, \binits{M.}},
\bauthor{\bsnm{{Marisaldi}}, \binits{M.}},
\bauthor{\bsnm{{McBreen}}, \binits{S.}},
\bauthor{\bsnm{{Mereghetti}}, \binits{S.}},
\bauthor{\bsnm{{Merle}}, \binits{A.}},
\bauthor{\bsnm{{Mignani}}, \binits{R.}},
\bauthor{\bsnm{{Minervini}}, \binits{G.}},
\bauthor{\bsnm{{Moiseev}}, \binits{A.}},
\bauthor{\bsnm{{Morselli}}, \binits{A.}},
\bauthor{\bsnm{{Moura}}, \binits{F.}},
\bauthor{\bsnm{{Nakazawa}}, \binits{K.}},
\bauthor{\bsnm{{Nava}}, \binits{L.}},
\bauthor{\bsnm{{Nieto}}, \binits{D.}},
\bauthor{\bsnm{{Orienti}}, \binits{M.}},
\bauthor{\bsnm{{Orio}}, \binits{M.}},
\bauthor{\bsnm{{Orlando}}, \binits{E.}},
\bauthor{\bsnm{{Orleanski}}, \binits{P.}},
\bauthor{\bsnm{{Paiano}}, \binits{S.}},
\bauthor{\bsnm{{Paoletti}}, \binits{R.}},
\bauthor{\bsnm{{Papitto}}, \binits{A.}},
\bauthor{\bsnm{{Pasquato}}, \binits{M.}},
\bauthor{\bsnm{{Patricelli}}, \binits{B.}},
\bauthor{\bsnm{{P{\'e}rez-Garc{\'\i}a}}, \binits{M.{\'A}.}},
\bauthor{\bsnm{{Persic}}, \binits{M.}},
\bauthor{\bsnm{{Piano}}, \binits{G.}},
\bauthor{\bsnm{{Pichel}}, \binits{A.}},
\bauthor{\bsnm{{Pimenta}}, \binits{M.}},
\bauthor{\bsnm{{Pittori}}, \binits{C.}},
\bauthor{\bsnm{{Porter}}, \binits{T.}},
\bauthor{\bsnm{{Poutanen}}, \binits{J.}},
\bauthor{\bsnm{{Prandini}}, \binits{E.}},
\bauthor{\bsnm{{Prantzos}}, \binits{N.}},
\bauthor{\bsnm{{Produit}}, \binits{N.}},
\bauthor{\bsnm{{Profumo}}, \binits{S.}},
\bauthor{\bsnm{{Queiroz}}, \binits{F.S.}}:
\batitle{{Science with e-ASTROGAM. A space mission for MeV-GeV gamma-ray
  astrophysics}}.
\bjtitle{Journal of High Energy Astrophysics}
\bvolume{19},
\bfpage{1}--\blpage{106}
(\byear{2018})
\doiurl{10.1016/j.jheap.2018.07.001}
{\href{https://arxiv.org/abs/1711.01265}{{arXiv:1711.01265}}}
{[astro-ph.HE]}
\end{barticle}
\endbibitem

%%% 338
\bibitem[\protect\citeauthoryear{{Ackermann} et~al.}{2012}]{ackermann2012}
\begin{barticle}
\bauthor{\bsnm{{Ackermann}}, \binits{M.}},
\bauthor{\bsnm{{Ajello}}, \binits{M.}},
\bauthor{\bsnm{{Atwood}}, \binits{W.B.}},
\bauthor{\bsnm{{Baldini}}, \binits{L.}},
\bauthor{\bsnm{{Ballet}}, \binits{J.}},
\bauthor{\bsnm{{Barbiellini}}, \binits{G.}},
\bauthor{\bsnm{{Bastieri}}, \binits{D.}},
\bauthor{\bsnm{{Bechtol}}, \binits{K.}},
\bauthor{\bsnm{{Bellazzini}}, \binits{R.}},
\bauthor{\bsnm{{Berenji}}, \binits{B.}},
\bauthor{\bsnm{{Blandford}}, \binits{R.D.}},
\bauthor{\bsnm{{Bloom}}, \binits{E.D.}},
\bauthor{\bsnm{{Bonamente}}, \binits{E.}},
\bauthor{\bsnm{{Borgland}}, \binits{A.W.}},
\bauthor{\bsnm{{Brandt}}, \binits{T.J.}},
\bauthor{\bsnm{{Bregeon}}, \binits{J.}},
\bauthor{\bsnm{{Brigida}}, \binits{M.}},
\bauthor{\bsnm{{Bruel}}, \binits{P.}},
\bauthor{\bsnm{{Buehler}}, \binits{R.}},
\bauthor{\bsnm{{Buson}}, \binits{S.}},
\bauthor{\bsnm{{Caliandro}}, \binits{G.A.}},
\bauthor{\bsnm{{Cameron}}, \binits{R.A.}},
\bauthor{\bsnm{{Caraveo}}, \binits{P.A.}},
\bauthor{\bsnm{{Cavazzuti}}, \binits{E.}},
\bauthor{\bsnm{{Cecchi}}, \binits{C.}},
\bauthor{\bsnm{{Charles}}, \binits{E.}},
\bauthor{\bsnm{{Chekhtman}}, \binits{A.}},
\bauthor{\bsnm{{Chiang}}, \binits{J.}},
\bauthor{\bsnm{{Ciprini}}, \binits{S.}},
\bauthor{\bsnm{{Claus}}, \binits{R.}},
\bauthor{\bsnm{{Cohen-Tanugi}}, \binits{J.}},
\bauthor{\bsnm{{Conrad}}, \binits{J.}},
\bauthor{\bsnm{{Cutini}}, \binits{S.}},
\bauthor{\bsnm{{de Angelis}}, \binits{A.}},
\bauthor{\bsnm{{de Palma}}, \binits{F.}},
\bauthor{\bsnm{{Dermer}}, \binits{C.D.}},
\bauthor{\bsnm{{Digel}}, \binits{S.W.}},
\bauthor{\bsnm{{Silva}}, \binits{E.d.C.e.}},
\bauthor{\bsnm{{Drell}}, \binits{P.S.}},
\bauthor{\bsnm{{Drlica-Wagner}}, \binits{A.}},
\bauthor{\bsnm{{Falletti}}, \binits{L.}},
\bauthor{\bsnm{{Favuzzi}}, \binits{C.}},
\bauthor{\bsnm{{Fegan}}, \binits{S.J.}},
\bauthor{\bsnm{{Ferrara}}, \binits{E.C.}},
\bauthor{\bsnm{{Focke}}, \binits{W.B.}},
\bauthor{\bsnm{{Fortin}}, \binits{P.}},
\bauthor{\bsnm{{Fukazawa}}, \binits{Y.}},
\bauthor{\bsnm{{Funk}}, \binits{S.}},
\bauthor{\bsnm{{Fusco}}, \binits{P.}},
\bauthor{\bsnm{{Gaggero}}, \binits{D.}},
\bauthor{\bsnm{{Gargano}}, \binits{F.}},
\bauthor{\bsnm{{Germani}}, \binits{S.}},
\bauthor{\bsnm{{Giglietto}}, \binits{N.}},
\bauthor{\bsnm{{Giordano}}, \binits{F.}},
\bauthor{\bsnm{{Giroletti}}, \binits{M.}},
\bauthor{\bsnm{{Glanzman}}, \binits{T.}},
\bauthor{\bsnm{{Godfrey}}, \binits{G.}},
\bauthor{\bsnm{{Grove}}, \binits{J.E.}},
\bauthor{\bsnm{{Guiriec}}, \binits{S.}},
\bauthor{\bsnm{{Gustafsson}}, \binits{M.}},
\bauthor{\bsnm{{Hadasch}}, \binits{D.}},
\bauthor{\bsnm{{Hanabata}}, \binits{Y.}},
\bauthor{\bsnm{{Harding}}, \binits{A.K.}},
\bauthor{\bsnm{{Hayashida}}, \binits{M.}},
\bauthor{\bsnm{{Hays}}, \binits{E.}},
\bauthor{\bsnm{{Horan}}, \binits{D.}},
\bauthor{\bsnm{{Hou}}, \binits{X.}},
\bauthor{\bsnm{{Hughes}}, \binits{R.E.}},
\bauthor{\bsnm{{J{\'o}hannesson}}, \binits{G.}},
\bauthor{\bsnm{{Johnson}}, \binits{A.S.}},
\bauthor{\bsnm{{Johnson}}, \binits{R.P.}},
\bauthor{\bsnm{{Kamae}}, \binits{T.}},
\bauthor{\bsnm{{Katagiri}}, \binits{H.}},
\bauthor{\bsnm{{Kataoka}}, \binits{J.}},
\bauthor{\bsnm{{Kn{\"o}dlseder}}, \binits{J.}},
\bauthor{\bsnm{{Kuss}}, \binits{M.}},
\bauthor{\bsnm{{Lande}}, \binits{J.}},
\bauthor{\bsnm{{Latronico}}, \binits{L.}},
\bauthor{\bsnm{{Lee}}, \binits{S.-H.}},
\bauthor{\bsnm{{Lemoine-Goumard}}, \binits{M.}},
\bauthor{\bsnm{{Longo}}, \binits{F.}},
\bauthor{\bsnm{{Loparco}}, \binits{F.}},
\bauthor{\bsnm{{Lott}}, \binits{B.}},
\bauthor{\bsnm{{Lovellette}}, \binits{M.N.}},
\bauthor{\bsnm{{Lubrano}}, \binits{P.}},
\bauthor{\bsnm{{Mazziotta}}, \binits{M.N.}},
\bauthor{\bsnm{{McEnery}}, \binits{J.E.}},
\bauthor{\bsnm{{Michelson}}, \binits{P.F.}},
\bauthor{\bsnm{{Mitthumsiri}}, \binits{W.}},
\bauthor{\bsnm{{Mizuno}}, \binits{T.}},
\bauthor{\bsnm{{Monte}}, \binits{C.}},
\bauthor{\bsnm{{Monzani}}, \binits{M.E.}},
\bauthor{\bsnm{{Morselli}}, \binits{A.}},
\bauthor{\bsnm{{Moskalenko}}, \binits{I.V.}},
\bauthor{\bsnm{{Murgia}}, \binits{S.}},
\bauthor{\bsnm{{Naumann-Godo}}, \binits{M.}},
\bauthor{\bsnm{{Norris}}, \binits{J.P.}},
\bauthor{\bsnm{{Nuss}}, \binits{E.}},
\bauthor{\bsnm{{Ohsugi}}, \binits{T.}},
\bauthor{\bsnm{{Okumura}}, \binits{A.}},
\bauthor{\bsnm{{Omodei}}, \binits{N.}},
\bauthor{\bsnm{{Orlando}}, \binits{E.}},
\bauthor{\bsnm{{Ormes}}, \binits{J.F.}},
\bauthor{\bsnm{{Paneque}}, \binits{D.}},
\bauthor{\bsnm{{Panetta}}, \binits{J.H.}},
\bauthor{\bsnm{{Parent}}, \binits{D.}},
\bauthor{\bsnm{{Pesce-Rollins}}, \binits{M.}},
\bauthor{\bsnm{{Pierbattista}}, \binits{M.}},
\bauthor{\bsnm{{Piron}}, \binits{F.}},
\bauthor{\bsnm{{Pivato}}, \binits{G.}},
\bauthor{\bsnm{{Porter}}, \binits{T.A.}},
\bauthor{\bsnm{{Rain{\`o}}}, \binits{S.}},
\bauthor{\bsnm{{Rando}}, \binits{R.}},
\bauthor{\bsnm{{Razzano}}, \binits{M.}},
\bauthor{\bsnm{{Razzaque}}, \binits{S.}},
\bauthor{\bsnm{{Reimer}}, \binits{A.}},
\bauthor{\bsnm{{Reimer}}, \binits{O.}},
\bauthor{\bsnm{{Sadrozinski}}, \binits{H.F.-W.}},
\bauthor{\bsnm{{Sgr{\`o}}}, \binits{C.}},
\bauthor{\bsnm{{Siskind}}, \binits{E.J.}},
\bauthor{\bsnm{{Spandre}}, \binits{G.}},
\bauthor{\bsnm{{Spinelli}}, \binits{P.}},
\bauthor{\bsnm{{Strong}}, \binits{A.W.}},
\bauthor{\bsnm{{Suson}}, \binits{D.J.}},
\bauthor{\bsnm{{Takahashi}}, \binits{H.}},
\bauthor{\bsnm{{Tanaka}}, \binits{T.}},
\bauthor{\bsnm{{Thayer}}, \binits{J.G.}},
\bauthor{\bsnm{{Thayer}}, \binits{J.B.}},
\bauthor{\bsnm{{Thompson}}, \binits{D.J.}},
\bauthor{\bsnm{{Tibaldo}}, \binits{L.}},
\bauthor{\bsnm{{Tinivella}}, \binits{M.}},
\bauthor{\bsnm{{Torres}}, \binits{D.F.}},
\bauthor{\bsnm{{Tosti}}, \binits{G.}},
\bauthor{\bsnm{{Troja}}, \binits{E.}},
\bauthor{\bsnm{{Usher}}, \binits{T.L.}},
\bauthor{\bsnm{{Vandenbroucke}}, \binits{J.}},
\bauthor{\bsnm{{Vasileiou}}, \binits{V.}},
\bauthor{\bsnm{{Vianello}}, \binits{G.}},
\bauthor{\bsnm{{Vitale}}, \binits{V.}},
\bauthor{\bsnm{{Waite}}, \binits{A.P.}},
\bauthor{\bsnm{{Wang}}, \binits{P.}},
\bauthor{\bsnm{{Winer}}, \binits{B.L.}},
\bauthor{\bsnm{{Wood}}, \binits{K.S.}},
\bauthor{\bsnm{{Wood}}, \binits{M.}},
\bauthor{\bsnm{{Yang}}, \binits{Z.}},
\bauthor{\bsnm{{Ziegler}}, \binits{M.}},
\bauthor{\bsnm{{Zimmer}}, \binits{S.}}:
\batitle{{Fermi-LAT Observations of the Diffuse {\ensuremath{\gamma}}-Ray
  Emission: Implications for Cosmic Rays and the Interstellar Medium}}.
\bjtitle{\apj}
\bvolume{750}(\bissue{1}),
\bfpage{3}
(\byear{2012})
\doiurl{10.1088/0004-637X/750/1/3}
{\href{https://arxiv.org/abs/1202.4039}{{arXiv:1202.4039}}}
{[astro-ph.HE]}
\end{barticle}
\endbibitem

%%% 339
\bibitem[\protect\citeauthoryear{{Ceccarelli} et~al.}{2011}]{ceccarelli2011}
\begin{barticle}
\bauthor{\bsnm{{Ceccarelli}}, \binits{C.}},
\bauthor{\bsnm{{Hily-Blant}}, \binits{P.}},
\bauthor{\bsnm{{Montmerle}}, \binits{T.}},
\bauthor{\bsnm{{Dubus}}, \binits{G.}},
\bauthor{\bsnm{{Gallant}}, \binits{Y.}},
\bauthor{\bsnm{{Fiasson}}, \binits{A.}}:
\batitle{{Supernova-enhanced Cosmic-Ray Ionization and Induced Chemistry in a
  Molecular Cloud of W51C}}.
\bjtitle{\apjl}
\bvolume{740}(\bissue{1}),
\bfpage{4}
(\byear{2011})
\doiurl{10.1088/2041-8205/740/1/L4}
{\href{https://arxiv.org/abs/1108.3600}{{arXiv:1108.3600}}}
{[astro-ph.GA]}
\end{barticle}
\endbibitem

%%% 340
\bibitem[\protect\citeauthoryear{{Vaupr{\'e}} et~al.}{2014}]{vaupre2014}
\begin{barticle}
\bauthor{\bsnm{{Vaupr{\'e}}}, \binits{S.}},
\bauthor{\bsnm{{Hily-Blant}}, \binits{P.}},
\bauthor{\bsnm{{Ceccarelli}}, \binits{C.}},
\bauthor{\bsnm{{Dubus}}, \binits{G.}},
\bauthor{\bsnm{{Gabici}}, \binits{S.}},
\bauthor{\bsnm{{Montmerle}}, \binits{T.}}:
\batitle{{Cosmic ray induced ionisation of a molecular cloud shocked by the W28
  supernova remnant}}.
\bjtitle{\aap}
\bvolume{568},
\bfpage{50}
(\byear{2014})
\doiurl{10.1051/0004-6361/201424036}
{\href{https://arxiv.org/abs/1407.0205}{{arXiv:1407.0205}}}
{[astro-ph.GA]}
\end{barticle}
\endbibitem

%%% 341
\bibitem[\protect\citeauthoryear{{Zhou} et~al.}{2022}]{zhou2022}
\begin{barticle}
\bauthor{\bsnm{{Zhou}}, \binits{P.}},
\bauthor{\bsnm{{Zhang}}, \binits{G.-Y.}},
\bauthor{\bsnm{{Zhou}}, \binits{X.}},
\bauthor{\bsnm{{Arias}}, \binits{M.}},
\bauthor{\bsnm{{Koo}}, \binits{B.-C.}},
\bauthor{\bsnm{{Vink}}, \binits{J.}},
\bauthor{\bsnm{{Zhang}}, \binits{Z.-Y.}},
\bauthor{\bsnm{{Sun}}, \binits{L.}},
\bauthor{\bsnm{{Du}}, \binits{F.-J.}},
\bauthor{\bsnm{{Zhu}}, \binits{H.}},
\bauthor{\bsnm{{Chen}}, \binits{Y.}},
\bauthor{\bsnm{{Bovino}}, \binits{S.}},
\bauthor{\bsnm{{Lee}}, \binits{Y.-H.}}:
\batitle{{Unusually High HCO$^{+}$/CO Ratios in and outside Supernova Remnant
  W49B}}.
\bjtitle{\apj}
\bvolume{931}(\bissue{2}),
\bfpage{144}
(\byear{2022})
\doiurl{10.3847/1538-4357/ac63b5}
{\href{https://arxiv.org/abs/2203.13111}{{arXiv:2203.13111}}}
{[astro-ph.GA]}
\end{barticle}
\endbibitem

%%% 342
\bibitem[\protect\citeauthoryear{{Holman}}{2012}]{holm12}
\begin{barticle}
\bauthor{\bsnm{{Holman}}, \binits{G.D.}}:
\batitle{{Solar eruptive events}}.
\bjtitle{Physics Today}
\bvolume{65}(\bissue{4}),
\bfpage{56}
(\byear{2012})
\doiurl{10.1063/PT.3.1520}
\end{barticle}
\endbibitem

%%% 343
\bibitem[\protect\citeauthoryear{{Chen} et~al.}{2020}]{chen20}
\begin{barticle}
\bauthor{\bsnm{{Chen}}, \binits{B.}},
\bauthor{\bsnm{{Shen}}, \binits{C.}},
\bauthor{\bsnm{{Gary}}, \binits{D.E.}},
\bauthor{\bsnm{{Reeves}}, \binits{K.K.}},
\bauthor{\bsnm{{Fleishman}}, \binits{G.D.}},
\bauthor{\bsnm{{Yu}}, \binits{S.}},
\bauthor{\bsnm{{Guo}}, \binits{F.}},
\bauthor{\bsnm{{Krucker}}, \binits{S.}},
\bauthor{\bsnm{{Lin}}, \binits{J.}},
\bauthor{\bsnm{{Nita}}, \binits{G.M.}},
\bauthor{\bsnm{{Kong}}, \binits{X.}}:
\batitle{{Measurement of magnetic field and relativistic electrons along a
  solar flare current sheet}}.
\bjtitle{Nature Astronomy}
\bvolume{4},
\bfpage{1140}--\blpage{1147}
(\byear{2020})
\doiurl{10.1038/s41550-020-1147-7}
{\href{https://arxiv.org/abs/2005.12757}{{arXiv:2005.12757}}}
{[astro-ph.SR]}
\end{barticle}
\endbibitem

%%% 344
\bibitem[\protect\citeauthoryear{{Vilmer} et~al.}{2011}]{vilm11}
\begin{barticle}
\bauthor{\bsnm{{Vilmer}}, \binits{N.}},
\bauthor{\bsnm{{MacKinnon}}, \binits{A.L.}},
\bauthor{\bsnm{{Hurford}}, \binits{G.J.}}:
\batitle{{Properties of Energetic Ions in the Solar Atmosphere from
  {$\gamma$}-Ray and Neutron Observations}}.
\bjtitle{\ssr}
\bvolume{159},
\bfpage{167}--\blpage{224}
(\byear{2011})
\doiurl{10.1007/s11214-010-9728-x}
{\href{https://arxiv.org/abs/1110.2432}{{arXiv:1110.2432}}}
{[astro-ph.SR]}
\end{barticle}
\endbibitem

%%% 345
\bibitem[\protect\citeauthoryear{{Reames}}{2021}]{ream21}
\begin{bbook}
\bauthor{\bsnm{{Reames}}, \binits{D.V.}}:
\bbtitle{{Solar Energetic Particles. A Modern Primer on Understanding Sources,
  Acceleration and Propagation}}
vol. \bseriesno{978},
(\byear{2021}).
\doiurl{10.1007/978-3-030-66402-2}
\end{bbook}
\endbibitem

%%% 346
\bibitem[\protect\citeauthoryear{{Smith} et~al.}{2002}]{smit02}
\begin{barticle}
\bauthor{\bsnm{{Smith}}, \binits{D.M.}},
\bauthor{\bsnm{{Lin}}, \binits{R.P.}},
\bauthor{\bsnm{{Turin}}, \binits{P.}},
\bauthor{\bsnm{{Curtis}}, \binits{D.W.}},
\bauthor{\bsnm{{Primbsch}}, \binits{J.H.}},
\bauthor{\bsnm{{Campbell}}, \binits{R.D.}},
\bauthor{\bsnm{{Abiad}}, \binits{R.}},
\bauthor{\bsnm{{Schroeder}}, \binits{P.}},
\bauthor{\bsnm{{Cork}}, \binits{C.P.}},
\bauthor{\bsnm{{Hull}}, \binits{E.L.}},
\bauthor{\bsnm{{Landis}}, \binits{D.A.}},
\bauthor{\bsnm{{Madden}}, \binits{N.W.}},
\bauthor{\bsnm{{Malone}}, \binits{D.}},
\bauthor{\bsnm{{Pehl}}, \binits{R.H.}},
\bauthor{\bsnm{{Raudorf}}, \binits{T.}},
\bauthor{\bsnm{{Sangsingkeow}}, \binits{P.}},
\bauthor{\bsnm{{Boyle}}, \binits{R.}},
\bauthor{\bsnm{{Banks}}, \binits{I.S.}},
\bauthor{\bsnm{{Shirey}}, \binits{K.}},
\bauthor{\bsnm{{Schwartz}}, \binits{R.}}:
\batitle{{The RHESSI Spectrometer}}.
\bjtitle{\solphys}
\bvolume{210}(\bissue{1}),
\bfpage{33}--\blpage{60}
(\byear{2002})
\doiurl{10.1023/A:1022400716414}
\end{barticle}
\endbibitem

%%% 347
\bibitem[\protect\citeauthoryear{{Forrest} et~al.}{1980}]{forr80}
\begin{barticle}
\bauthor{\bsnm{{Forrest}}, \binits{D.J.}},
\bauthor{\bsnm{{Chupp}}, \binits{E.L.}},
\bauthor{\bsnm{{Ryan}}, \binits{J.M.}},
\bauthor{\bsnm{{Cherry}}, \binits{M.L.}},
\bauthor{\bsnm{{Gleske}}, \binits{I.U.}},
\bauthor{\bsnm{{Reppin}}, \binits{C.}},
\bauthor{\bsnm{{Pinkau}}, \binits{K.}},
\bauthor{\bsnm{{Rieger}}, \binits{E.}},
\bauthor{\bsnm{{Kanbach}}, \binits{G.}},
\bauthor{\bsnm{{Kinzer}}, \binits{R.L.}},
\bauthor{\bsnm{{Share}}, \binits{G.}},
\bauthor{\bsnm{{Johnson}}, \binits{W.N.}},
\bauthor{\bsnm{{Kurfess}}, \binits{J.D.}}:
\batitle{{The gamma ray spectrometer for the Solar Maximum Mission.}}
\bjtitle{\solphys}
\bvolume{65}(\bissue{1}),
\bfpage{15}--\blpage{23}
(\byear{1980})
\doiurl{10.1007/BF00151381}
\end{barticle}
\endbibitem

%%% 348
\bibitem[\protect\citeauthoryear{{Harris} et~al.}{1992}]{harr92}
\begin{barticle}
\bauthor{\bsnm{{Harris}}, \binits{M.J.}},
\bauthor{\bsnm{{Share}}, \binits{G.H.}},
\bauthor{\bsnm{{Beall}}, \binits{J.H.}},
\bauthor{\bsnm{{Murphy}}, \binits{R.J.}}:
\batitle{{Upper limit on the steady emission of the 2.223 MeV neutron capture
  gamma-ray line from the sun}}.
\bjtitle{Solar Physics}
\bvolume{142},
\bfpage{171}--\blpage{185}
(\byear{1992})
\doiurl{10.1007/BF00156640}
\end{barticle}
\endbibitem

%%% 349
\bibitem[\protect\citeauthoryear{{Siegert} et~al.}{2022}]{sieg22}
\begin{barticle}
\bauthor{\bsnm{{Siegert}}, \binits{T.}},
\bauthor{\bsnm{{Berteaud}}, \binits{J.}},
\bauthor{\bsnm{{Calore}}, \binits{F.}},
\bauthor{\bsnm{{Serpico}}, \binits{P.D.}},
\bauthor{\bsnm{{Weinberger}}, \binits{C.}}:
\batitle{{Diffuse Galactic emission spectrum between 0.5 and 8.0 MeV}}.
\bjtitle{\aap}
\bvolume{660},
\bfpage{130}
(\byear{2022})
\doiurl{10.1051/0004-6361/202142639}
{\href{https://arxiv.org/abs/2202.04574}{{arXiv:2202.04574}}}
{[astro-ph.HE]}
\end{barticle}
\endbibitem

%%% 350
\bibitem[\protect\citeauthoryear{{Share} et~al.}{2025}]{shar25}
\begin{barticle}
\bauthor{\bsnm{{Share}}, \binits{G.H.}},
\bauthor{\bsnm{{Murphy}}, \binits{R.J.}},
\bauthor{\bsnm{{Dennis}}, \binits{B.R.}},
\bauthor{\bsnm{{Finke}}, \binits{J.D.}}:
\batitle{{Solar Gamma-Ray Evidence for a Distinct Population of $>1$ MeV
  Flare-accelerated Electrons}}.
\bjtitle{\apj}
\bvolume{981}(\bissue{1}),
\bfpage{11}
(\byear{2025})
\doiurl{10.3847/1538-4357/adac60}
{\href{https://arxiv.org/abs/2412.19586}{{arXiv:2412.19586}}}
{[astro-ph.SR]}
\end{barticle}
\endbibitem

%%% 351
\bibitem[\protect\citeauthoryear{{Wang} and {Ramaty}}{1974}]{wang74}
\begin{barticle}
\bauthor{\bsnm{{Wang}}, \binits{H.T.}},
\bauthor{\bsnm{{Ramaty}}, \binits{R.}}:
\batitle{{Neutron Propagation and 2.2 MeV Gamma-Ray Line Production in the
  Solar Atmosphere}}.
\bjtitle{\solphys}
\bvolume{36}(\bissue{1}),
\bfpage{129}--\blpage{137}
(\byear{1974})
\doiurl{10.1007/BF00151553}
\end{barticle}
\endbibitem

%%% 352
\bibitem[\protect\citeauthoryear{{Murphy} et~al.}{2003}]{murp03}
\begin{barticle}
\bauthor{\bsnm{{Murphy}}, \binits{R.J.}},
\bauthor{\bsnm{{Share}}, \binits{G.H.}},
\bauthor{\bsnm{{Hua}}, \binits{X.-M.}},
\bauthor{\bsnm{{Lin}}, \binits{R.P.}},
\bauthor{\bsnm{{Smith}}, \binits{D.M.}},
\bauthor{\bsnm{{Schwartz}}, \binits{R.A.}}:
\batitle{{Physical Implications of RHESSI Neutron-Capture Line Measurements}}.
\bjtitle{\apjl}
\bvolume{595}(\bissue{2}),
\bfpage{93}--\blpage{96}
(\byear{2003})
\doiurl{10.1086/378175}
\end{barticle}
\endbibitem

%%% 353
\bibitem[\protect\citeauthoryear{{Bania} and {Balser}}{2021}]{bani21}
\begin{barticle}
\bauthor{\bsnm{{Bania}}, \binits{T.M.}},
\bauthor{\bsnm{{Balser}}, \binits{D.S.}}:
\batitle{{Green Bank Telescope Observations of $^{3}$He$^{+}$: Planetary
  Nebulae}}.
\bjtitle{\apj}
\bvolume{910}(\bissue{1}),
\bfpage{73}
(\byear{2021})
\doiurl{10.3847/1538-4357/abd543}
{\href{https://arxiv.org/abs/2012.11707}{{arXiv:2012.11707}}}
{[astro-ph.GA]}
\end{barticle}
\endbibitem

%%% 354
\bibitem[\protect\citeauthoryear{{Battaglia} and {Krucker}}{2025}]{batt25}
\begin{barticle}
\bauthor{\bsnm{{Battaglia}}, \binits{A.F.}},
\bauthor{\bsnm{{Krucker}}, \binits{S.}}:
\batitle{{New insights into the proton precipitation sites in solar flares}}.
\bjtitle{\aap}
\bvolume{694},
\bfpage{58}
(\byear{2025})
\doiurl{10.1051/0004-6361/202453144}
{\href{https://arxiv.org/abs/2412.11490}{{arXiv:2412.11490}}}
{[astro-ph.SR]}
\end{barticle}
\endbibitem

%%% 355
\bibitem[\protect\citeauthoryear{{Shih} et~al.}{2009}]{shih09}
\begin{barticle}
\bauthor{\bsnm{{Shih}}, \binits{A.Y.}},
\bauthor{\bsnm{{Lin}}, \binits{R.P.}},
\bauthor{\bsnm{{Smith}}, \binits{D.M.}}:
\batitle{{RHESSI Observations of the Proportional Acceleration of Relativistic
  $>$0.3 MeV Electrons and $>$30 MeV Protons in Solar Flares}}.
\bjtitle{\apjl}
\bvolume{698}(\bissue{2}),
\bfpage{152}--\blpage{157}
(\byear{2009})
\doiurl{10.1088/0004-637X/698/2/L152}
\end{barticle}
\endbibitem

%%% 356
\bibitem[\protect\citeauthoryear{{Murphy} et~al.}{2005}]{murp05}
\begin{barticle}
\bauthor{\bsnm{{Murphy}}, \binits{R.J.}},
\bauthor{\bsnm{{Share}}, \binits{G.H.}},
\bauthor{\bsnm{{Skibo}}, \binits{J.G.}},
\bauthor{\bsnm{{Kozlovsky}}, \binits{B.}}:
\batitle{{The Physics of Positron Annihilation in the Solar Atmosphere}}.
\bjtitle{\apjs}
\bvolume{161},
\bfpage{495}--\blpage{519}
(\byear{2005})
\doiurl{10.1086/452634}
\end{barticle}
\endbibitem

%%% 357
\bibitem[\protect\citeauthoryear{{Share} et~al.}{2004}]{shar04}
\begin{barticle}
\bauthor{\bsnm{{Share}}, \binits{G.H.}},
\bauthor{\bsnm{{Murphy}}, \binits{R.J.}},
\bauthor{\bsnm{{Smith}}, \binits{D.M.}},
\bauthor{\bsnm{{Schwartz}}, \binits{R.A.}},
\bauthor{\bsnm{{Lin}}, \binits{R.P.}}:
\batitle{{RHESSI e$^{+}$-e$^{-}$ Annihilation Radiation Observations:
  Implications for Conditions in the Flaring Solar Chromosphere}}.
\bjtitle{\apjl}
\bvolume{615}(\bissue{2}),
\bfpage{169}--\blpage{172}
(\byear{2004})
\doiurl{10.1086/426478}
\end{barticle}
\endbibitem

%%% 358
\bibitem[\protect\citeauthoryear{{Ramaty} et~al.}{1979}]{rama79}
\begin{barticle}
\bauthor{\bsnm{{Ramaty}}, \binits{R.}},
\bauthor{\bsnm{{Kozlovsky}}, \binits{B.}},
\bauthor{\bsnm{{Lingenfelter}}, \binits{R.E.}}:
\batitle{{Nuclear gamma-rays from energetic particle interactions}}.
\bjtitle{\apjs}
\bvolume{40},
\bfpage{487}--\blpage{526}
(\byear{1979})
\doiurl{10.1086/190596}
\end{barticle}
\endbibitem

%%% 359
\bibitem[\protect\citeauthoryear{{Murphy} et~al.}{2009}]{murp09}
\begin{barticle}
\bauthor{\bsnm{{Murphy}}, \binits{R.J.}},
\bauthor{\bsnm{{Kozlovsky}}, \binits{B.}},
\bauthor{\bsnm{{Kiener}}, \binits{J.}},
\bauthor{\bsnm{{Share}}, \binits{G.H.}}:
\batitle{{Nuclear Gamma-Ray De-Excitation Lines and Continuum from
  Accelerated-Particle Interactions in Solar Flares}}.
\bjtitle{\apjs}
\bvolume{183},
\bfpage{142}--\blpage{155}
(\byear{2009})
\doiurl{10.1088/0067-0049/183/1/142}
\end{barticle}
\endbibitem

%%% 360
\bibitem[\protect\citeauthoryear{{Harris} et~al.}{1995}]{harr95}
\begin{barticle}
\bauthor{\bsnm{{Harris}}, \binits{M.J.}},
\bauthor{\bsnm{{Share}}, \binits{G.H.}},
\bauthor{\bsnm{{Messina}}, \binits{D.C.}}:
\batitle{{Limits on Galactic Gamma-Ray Lines at 4.44 MeV and 6.13 MeV from
  Nuclear De-Excitation}}.
\bjtitle{\apj}
\bvolume{448},
\bfpage{157}
(\byear{1995})
\doiurl{10.1086/175948}
\end{barticle}
\endbibitem

%%% 361
\bibitem[\protect\citeauthoryear{{Share} and {Murphy}}{1995}]{shar95}
\begin{barticle}
\bauthor{\bsnm{{Share}}, \binits{G.H.}},
\bauthor{\bsnm{{Murphy}}, \binits{R.J.}}:
\batitle{{Gamma-Ray Measurements of Flare-to-Flare Variations in Ambient Solar
  Abundances}}.
\bjtitle{\apj}
\bvolume{452},
\bfpage{933}
(\byear{1995})
\doiurl{10.1086/176360}
\end{barticle}
\endbibitem

%%% 362
\bibitem[\protect\citeauthoryear{{Ramaty} et~al.}{1996}]{rama96}
\begin{bchapter}
\bauthor{\bsnm{{Ramaty}}, \binits{R.}},
\bauthor{\bsnm{{Mandzhavidze}}, \binits{N.}},
\bauthor{\bsnm{{Kozlovsky}}, \binits{B.}}:
\bctitle{{Solar atmospheric abundances from gamma ray spectroscopy}}.
In: \beditor{\bsnm{{Ramaty}}, \binits{R.}},
\beditor{\bsnm{{Mandzhavidze}}, \binits{N.}},
\beditor{\bsnm{{Hua}}, \binits{X.-M.}} (eds.)
\bbtitle{American Institute of Physics Conference Series}.
\bsertitle{American Institute of Physics Conference Series},
vol. \bseriesno{374},
pp. \bfpage{172}--\blpage{183}
(\byear{1996}).
\doiurl{10.1063/1.50953}
\end{bchapter}
\endbibitem

%%% 363
\bibitem[\protect\citeauthoryear{{Murphy} et~al.}{1997}]{murp97}
\begin{barticle}
\bauthor{\bsnm{{Murphy}}, \binits{R.J.}},
\bauthor{\bsnm{{Share}}, \binits{G.H.}},
\bauthor{\bsnm{{Grove}}, \binits{J.E.}},
\bauthor{\bsnm{{Johnson}}, \binits{W.N.}},
\bauthor{\bsnm{{Kinzer}}, \binits{R.L.}},
\bauthor{\bsnm{{Kurfess}}, \binits{J.D.}},
\bauthor{\bsnm{{Strickman}}, \binits{M.S.}},
\bauthor{\bsnm{{Jung}}, \binits{G.V.}}:
\batitle{{Accelerated Particle Composition and Energetics and Ambient
  Abundances from Gamma-Ray Spectroscopy of the 1991 June 4 Solar Flare}}.
\bjtitle{\apj}
\bvolume{490},
\bfpage{883}
(\byear{1997})
\doiurl{10.1086/304902}
\end{barticle}
\endbibitem

%%% 364
\bibitem[\protect\citeauthoryear{{Mandzhavidze} et~al.}{1999}]{mand99}
\begin{barticle}
\bauthor{\bsnm{{Mandzhavidze}}, \binits{N.}},
\bauthor{\bsnm{{Ramaty}}, \binits{R.}},
\bauthor{\bsnm{{Kozlovsky}}, \binits{B.}}:
\batitle{{Determination of the Abundances of Subcoronal $^{4}$He and of Solar
  Flare-accelerated $^{3}$He and $^{4}$He from Gamma-Ray Spectroscopy}}.
\bjtitle{\apj}
\bvolume{518}(\bissue{2}),
\bfpage{918}--\blpage{925}
(\byear{1999})
\doiurl{10.1086/307321}
\end{barticle}
\endbibitem

%%% 365
\bibitem[\protect\citeauthoryear{{Mandzhavidze} and {Ramaty}}{2000}]{mand00}
\begin{bchapter}
\bauthor{\bsnm{{Mandzhavidze}}, \binits{N.}},
\bauthor{\bsnm{{Ramaty}}, \binits{R.}}:
\bctitle{{Particle Acceleration and Abundances from Gamma-Ray Line
  Spectroscopy}}.
In: \beditor{\bsnm{{Ramaty}}, \binits{R.}},
\beditor{\bsnm{{Mandzhavidze}}, \binits{N.}} (eds.)
\bbtitle{High Energy Solar Physics Workshop - Anticipating Hess!}
\bsertitle{Astronomical Society of the Pacific Conference Series},
vol. \bseriesno{206},
p. \bfpage{64}
(\byear{2000})
\end{bchapter}
\endbibitem

%%% 366
\bibitem[\protect\citeauthoryear{{Ackermann} et~al.}{2012}]{acke12}
\begin{barticle}
\bauthor{\bsnm{{Ackermann}}, \binits{M.}},
\bauthor{\bsnm{{Ajello}}, \binits{M.}},
\bauthor{\bsnm{{Allafort}}, \binits{A.}},
\bauthor{\bsnm{{Atwood}}, \binits{W.B.}},
\bauthor{\bsnm{{Baldini}}, \binits{L.}},
\bauthor{\bsnm{{Barbiellini}}, \binits{G.}},
\bauthor{\bsnm{{Bastieri}}, \binits{D.}},
\bauthor{\bsnm{al.}}:
\batitle{{Fermi Detection of {$\gamma$}-Ray Emission from the M2 Soft X-Ray
  Flare on 2010 June 12}}.
\bjtitle{\apj}
\bvolume{745},
\bfpage{144}
(\byear{2012})
\doiurl{10.1088/0004-637X/745/2/144}
\end{barticle}
\endbibitem

%%% 367
\bibitem[\protect\citeauthoryear{{Lysenko} et~al.}{2019}]{lyse19}
\begin{barticle}
\bauthor{\bsnm{{Lysenko}}, \binits{A.L.}},
\bauthor{\bsnm{{Anfinogentov}}, \binits{S.A.}},
\bauthor{\bsnm{{Svinkin}}, \binits{D.S.}},
\bauthor{\bsnm{{Frederiks}}, \binits{D.D.}},
\bauthor{\bsnm{{Fleishman}}, \binits{G.D.}}:
\batitle{{Gamma-Ray Emission from the Impulsive Phase of the 2017 September 6
  X9.3 Flare}}.
\bjtitle{\apj}
\bvolume{877}(\bissue{2}),
\bfpage{145}
(\byear{2019})
\doiurl{10.3847/1538-4357/ab1be0}
{\href{https://arxiv.org/abs/1904.10017}{{arXiv:1904.10017}}}
{[astro-ph.HE]}
\end{barticle}
\endbibitem

%%% 368
\bibitem[\protect\citeauthoryear{{Yushkov} et~al.}{2023}]{yush23}
\begin{barticle}
\bauthor{\bsnm{{Yushkov}}, \binits{B.Y.}},
\bauthor{\bsnm{{Kurt}}, \binits{V.G.}},
\bauthor{\bsnm{{Galkin}}, \binits{V.I.}}:
\batitle{{High-Energy Emissions Observed in the Impulsive Phase of the 2001
  August 25 Eruptive Flare}}.
\bjtitle{\solphys}
\bvolume{298}(\bissue{2}),
\bfpage{31}
(\byear{2023})
\doiurl{10.1007/s11207-023-02123-8}
\end{barticle}
\endbibitem

%%% 369
\bibitem[\protect\citeauthoryear{{Murphy} et~al.}{1991}]{murp91}
\begin{barticle}
\bauthor{\bsnm{{Murphy}}, \binits{R.J.}},
\bauthor{\bsnm{{Ramaty}}, \binits{R.}},
\bauthor{\bsnm{{Reames}}, \binits{D.V.}},
\bauthor{\bsnm{{Kozlovsky}}, \binits{B.}}:
\batitle{{Solar abundances from gamma-ray spectroscopy - Comparisons with
  energetic particle, photospheric, and coronal abundances}}.
\bjtitle{\apj}
\bvolume{371},
\bfpage{793}--\blpage{803}
(\byear{1991})
\doiurl{10.1086/169944}
\end{barticle}
\endbibitem

%%% 370
\bibitem[\protect\citeauthoryear{{Hua} et~al.}{1989}]{hua89}
\begin{barticle}
\bauthor{\bsnm{{Hua}}, \binits{X.-M.}},
\bauthor{\bsnm{{Ramaty}}, \binits{R.}},
\bauthor{\bsnm{{Lingenfelter}}, \binits{R.E.}}:
\batitle{{Deexcitation gamma-ray line emission from solar flare magnetic
  loops}}.
\bjtitle{\apj}
\bvolume{341},
\bfpage{516}--\blpage{532}
(\byear{1989})
\doiurl{10.1086/167513}
\end{barticle}
\endbibitem

%%% 371
\bibitem[\protect\citeauthoryear{{Murphy} et~al.}{2007}]{murp07}
\begin{barticle}
\bauthor{\bsnm{{Murphy}}, \binits{R.J.}},
\bauthor{\bsnm{{Kozlovsky}}, \binits{B.}},
\bauthor{\bsnm{{Share}}, \binits{G.H.}},
\bauthor{\bsnm{{Hua}}, \binits{X.}},
\bauthor{\bsnm{{Lingenfelter}}, \binits{R.E.}}:
\batitle{{Using Gamma-Ray and Neutron Emission to Determine Solar Flare
  Accelerated Particle Spectra and Composition and the Conditions within the
  Flare Magnetic Loop}}.
\bjtitle{\apjs}
\bvolume{168},
\bfpage{167}--\blpage{194}
(\byear{2007})
\doiurl{10.1086/509637}
\end{barticle}
\endbibitem

%%% 372
\bibitem[\protect\citeauthoryear{{Murphy} et~al.}{2016}]{murp16}
\begin{barticle}
\bauthor{\bsnm{{Murphy}}, \binits{R.J.}},
\bauthor{\bsnm{{Kozlovsky}}, \binits{B.}},
\bauthor{\bsnm{{Share}}, \binits{G.H.}}:
\batitle{{Evidence for Enhanced $^{3}$He in Flare-accelerated Particles Based
  on New Calculations of the Gamma-Ray Line Spectrum}}.
\bjtitle{\apj}
\bvolume{833},
\bfpage{196}
(\byear{2016})
\doiurl{10.3847/1538-4357/833/2/196}
\end{barticle}
\endbibitem

%%% 373
\bibitem[\protect\citeauthoryear{{Tusnski} et~al.}{2019}]{tuns19}
\begin{barticle}
\bauthor{\bsnm{{Tusnski}}, \binits{D.S.}},
\bauthor{\bsnm{{Szpigel}}, \binits{S.}},
\bauthor{\bsnm{{Gim{\'e}nez de Castro}}, \binits{C.G.}},
\bauthor{\bsnm{{MacKinnon}}, \binits{A.L.}},
\bauthor{\bsnm{{Sim{\~o}es}}, \binits{P.J.A.}}:
\batitle{{Self-consistent Modeling of Gamma-ray Spectra from Solar Flares with
  the Monte Carlo Simulation Package FLUKA}}.
\bjtitle{\solphys}
\bvolume{294}(\bissue{8}),
\bfpage{103}
(\byear{2019})
\doiurl{10.1007/s11207-019-1499-2}
{\href{https://arxiv.org/abs/1907.11575}{{arXiv:1907.11575}}}
{[astro-ph.HE]}
\end{barticle}
\endbibitem

%%% 374
\bibitem[\protect\citeauthoryear{{Kiener} et~al.}{2006}]{kien06}
\begin{barticle}
\bauthor{\bsnm{{Kiener}}, \binits{J.}},
\bauthor{\bsnm{{Gros}}, \binits{M.}},
\bauthor{\bsnm{{Tatischeff}}, \binits{V.}},
\bauthor{\bsnm{{Weidenspointner}}, \binits{G.}}:
\batitle{{Properties of the energetic particle distributions during the October
  28, 2003 solar flare from INTEGRAL/SPI observations}}.
\bjtitle{\aap}
\bvolume{445}(\bissue{2}),
\bfpage{725}--\blpage{733}
(\byear{2006})
\doiurl{10.1051/0004-6361:20053665}
{\href{https://arxiv.org/abs/astro-ph/0511091}{{arXiv:astro-ph/0511091}}}
{[astro-ph]}
\end{barticle}
\endbibitem

%%% 375
\bibitem[\protect\citeauthoryear{{Kiener}}{2019}]{kien19}
\begin{barticle}
\bauthor{\bsnm{{Kiener}}, \binits{J.}}:
\batitle{{Shape and angular distribution of the 4.439-MeV {\ensuremath{\gamma}}
  -ray line from proton inelastic scattering off $^{12}$C}}.
\bjtitle{\prc}
\bvolume{99}(\bissue{1}),
\bfpage{014605}
(\byear{2019})
\doiurl{10.1103/PhysRevC.99.014605}
{\href{https://arxiv.org/abs/1802.00658}{{arXiv:1802.00658}}}
{[nucl-ex]}
\end{barticle}
\endbibitem

%%% 376
\bibitem[\protect\citeauthoryear{{Share} et~al.}{2002}]{shar02}
\begin{barticle}
\bauthor{\bsnm{{Share}}, \binits{G.H.}},
\bauthor{\bsnm{{Murphy}}, \binits{R.J.}},
\bauthor{\bsnm{{Kiener}}, \binits{J.}},
\bauthor{\bsnm{{de S{\'e}r{\'e}ville}}, \binits{N.}}:
\batitle{{Directionality of Solar Flare-accelerated Protons and
  {$\alpha$}-Particles from {$\gamma$}-Ray Line Measurements}}.
\bjtitle{\apj}
\bvolume{573},
\bfpage{464}--\blpage{470}
(\byear{2002})
\doiurl{10.1086/340595}
{\href{https://arxiv.org/abs/astro-ph/0203215}{{astro-ph/0203215}}}
\end{barticle}
\endbibitem

%%% 377
\bibitem[\protect\citeauthoryear{{Share} et~al.}{2003}]{shar03a}
\begin{barticle}
\bauthor{\bsnm{{Share}}, \binits{G.H.}},
\bauthor{\bsnm{{Murphy}}, \binits{R.J.}},
\bauthor{\bsnm{{Smith}}, \binits{D.M.}},
\bauthor{\bsnm{{Lin}}, \binits{R.P.}},
\bauthor{\bsnm{{Dennis}}, \binits{B.R.}},
\bauthor{\bsnm{{Schwartz}}, \binits{R.A.}}:
\batitle{{Directionality of Flare-accelerated {\ensuremath{\alpha}}-Particles
  at the Sun}}.
\bjtitle{\apjl}
\bvolume{595}(\bissue{2}),
\bfpage{89}--\blpage{92}
(\byear{2003})
\doiurl{10.1086/378176}
\end{barticle}
\endbibitem

%%% 378
\bibitem[\protect\citeauthoryear{{Smith} et~al.}{2003}]{smit03}
\begin{barticle}
\bauthor{\bsnm{{Smith}}, \binits{D.M.}},
\bauthor{\bsnm{{Share}}, \binits{G.H.}},
\bauthor{\bsnm{{Murphy}}, \binits{R.J.}},
\bauthor{\bsnm{{Schwartz}}, \binits{R.A.}},
\bauthor{\bsnm{{Shih}}, \binits{A.Y.}},
\bauthor{\bsnm{{Lin}}, \binits{R.P.}}:
\batitle{{High-Resolution Spectroscopy of Gamma-Ray Lines from the X-Class
  Solar Flare of 2002 July 23}}.
\bjtitle{\apjl}
\bvolume{595}(\bissue{2}),
\bfpage{81}--\blpage{84}
(\byear{2003})
\doiurl{10.1086/378173}
{\href{https://arxiv.org/abs/astro-ph/0306292}{{arXiv:astro-ph/0306292}}}
{[astro-ph]}
\end{barticle}
\endbibitem

%%% 379
\bibitem[\protect\citeauthoryear{{Harris} et~al.}{2007}]{harr07}
\begin{barticle}
\bauthor{\bsnm{{Harris}}, \binits{M.J.}},
\bauthor{\bsnm{{Tatischeff}}, \binits{V.}},
\bauthor{\bsnm{{Kiener}}, \binits{J.}},
\bauthor{\bsnm{{Gros}}, \binits{M.}},
\bauthor{\bsnm{{Weidenspointner}}, \binits{G.}}:
\batitle{{High resolution {\ensuremath{\gamma}}-ray spectroscopy of flares on
  the east and west limbs of the Sun}}.
\bjtitle{\aap}
\bvolume{461}(\bissue{2}),
\bfpage{723}--\blpage{729}
(\byear{2007})
\doiurl{10.1051/0004-6361:20066084}
{\href{https://arxiv.org/abs/astro-ph/0610859}{{arXiv:astro-ph/0610859}}}
{[astro-ph]}
\end{barticle}
\endbibitem

%%% 380
\bibitem[\protect\citeauthoryear{{Share} and {Murphy}}{1997}]{shar97}
\begin{barticle}
\bauthor{\bsnm{{Share}}, \binits{G.H.}},
\bauthor{\bsnm{{Murphy}}, \binits{R.J.}}:
\batitle{{Intensity and Directionality of Flare-accelerated
  {\ensuremath{\alpha}}-Particles at the Sun}}.
\bjtitle{\apj}
\bvolume{485}(\bissue{1}),
\bfpage{409}--\blpage{418}
(\byear{1997})
\doiurl{10.1086/304407}
\end{barticle}
\endbibitem

%%% 381
\bibitem[\protect\citeauthoryear{{Emslie} et~al.}{2012}]{emsl12}
\begin{barticle}
\bauthor{\bsnm{{Emslie}}, \binits{A.G.}},
\bauthor{\bsnm{{Dennis}}, \binits{B.R.}},
\bauthor{\bsnm{{Shih}}, \binits{A.Y.}},
\bauthor{\bsnm{{Chamberlin}}, \binits{P.C.}},
\bauthor{\bsnm{{Mewaldt}}, \binits{R.A.}},
\bauthor{\bsnm{{Moore}}, \binits{C.S.}},
\bauthor{\bsnm{{Share}}, \binits{G.H.}},
\bauthor{\bsnm{{Vourlidas}}, \binits{A.}},
\bauthor{\bsnm{{Welsch}}, \binits{B.T.}}:
\batitle{{Global Energetics of Thirty-eight Large Solar Eruptive Events}}.
\bjtitle{\apj}
\bvolume{759}(\bissue{1}),
\bfpage{71}
(\byear{2012})
\doiurl{10.1088/0004-637X/759/1/71}
{\href{https://arxiv.org/abs/1209.2654}{{arXiv:1209.2654}}}
{[astro-ph.SR]}
\end{barticle}
\endbibitem

%%% 382
\bibitem[\protect\citeauthoryear{{Share} et~al.}{2001}]{shar01a}
\begin{barticle}
\bauthor{\bsnm{{Share}}, \binits{G.H.}},
\bauthor{\bsnm{{Murphy}}, \binits{R.J.}},
\bauthor{\bsnm{{Newton}}, \binits{E.K.}}:
\batitle{{Limits on Radiative Capture {\ensuremath{\gamma}}-Ray Lines and
  Implications for Energy Content in Flare-Accelerated Protons}}.
\bjtitle{\solphys}
\bvolume{201}(\bissue{1}),
\bfpage{191}--\blpage{200}
(\byear{2001})
\doiurl{10.1023/A:1010333807867}
\end{barticle}
\endbibitem

%%% 383
\bibitem[\protect\citeauthoryear{{Moskalenko} and
  {Porter}}{2007}]{Moskalenko2007}
\begin{barticle}
\bauthor{\bsnm{{Moskalenko}}, \binits{I.V.}},
\bauthor{\bsnm{{Porter}}, \binits{T.A.}}:
\batitle{{The Gamma-Ray Albedo of the Moon}}.
\bjtitle{\apj}
\bvolume{670}(\bissue{2}),
\bfpage{1467}--\blpage{1472}
(\byear{2007})
\doiurl{10.1086/522828}
{\href{https://arxiv.org/abs/0708.2742}{{arXiv:0708.2742}}}
{[astro-ph]}
\end{barticle}
\endbibitem

%%% 384
\bibitem[\protect\citeauthoryear{{Prettyman} et~al.}{2006}]{Prettyman2006}
\begin{barticle}
\bauthor{\bsnm{{Prettyman}}, \binits{T.H.}},
\bauthor{\bsnm{{Hagerty}}, \binits{J.J.}},
\bauthor{\bsnm{{Elphic}}, \binits{R.C.}},
\bauthor{\bsnm{{Feldman}}, \binits{W.C.}},
\bauthor{\bsnm{{Lawrence}}, \binits{D.J.}},
\bauthor{\bsnm{{McKinney}}, \binits{G.W.}},
\bauthor{\bsnm{{Vaniman}}, \binits{D.T.}}:
\batitle{{Elemental composition of the lunar surface: Analysis of gamma ray
  spectroscopy data from Lunar Prospector}}.
\bjtitle{Journal of Geophysical Research (Planets)}
\bvolume{111}(\bissue{E12}),
\bfpage{12007}
(\byear{2006})
\doiurl{10.1029/2005JE002656}
\end{barticle}
\endbibitem

%%% 385
\bibitem[\protect\citeauthoryear{{Moskalenko} et~al.}{2008}]{Moskalenko2008}
\begin{barticle}
\bauthor{\bsnm{{Moskalenko}}, \binits{I.V.}},
\bauthor{\bsnm{{Porter}}, \binits{T.A.}},
\bauthor{\bsnm{{Digel}}, \binits{S.W.}},
\bauthor{\bsnm{{Michelson}}, \binits{P.F.}},
\bauthor{\bsnm{{Ormes}}, \binits{J.F.}}:
\batitle{{A Celestial Gamma-Ray Foreground Due to the Albedo of Small Solar
  System Bodies and a Remote Probe of the Interstellar Cosmic-Ray Spectrum}}.
\bjtitle{\apj}
\bvolume{681}(\bissue{2}),
\bfpage{1708}--\blpage{1716}
(\byear{2008})
\doiurl{10.1086/588425}
{\href{https://arxiv.org/abs/0712.2015}{{arXiv:0712.2015}}}
{[astro-ph]}
\end{barticle}
\endbibitem

%%% 386
\bibitem[\protect\citeauthoryear{{Churazov} et~al.}{2007}]{Churazov2007}
\begin{barticle}
\bauthor{\bsnm{{Churazov}}, \binits{E.}},
\bauthor{\bsnm{{Sunyaev}}, \binits{R.}},
\bauthor{\bsnm{{Revnivtsev}}, \binits{M.}},
\bauthor{\bsnm{{Sazonov}}, \binits{S.}},
\bauthor{\bsnm{{Molkov}}, \binits{S.}},
\bauthor{\bsnm{{Grebenev}}, \binits{S.}},
\bauthor{\bsnm{{Winkler}}, \binits{C.}},
\bauthor{\bsnm{{Parmar}}, \binits{A.}},
\bauthor{\bsnm{{Bazzano}}, \binits{A.}},
\bauthor{\bsnm{{Falanga}}, \binits{M.}},
\bauthor{\bsnm{{Gros}}, \binits{A.}},
\bauthor{\bsnm{{Lebrun}}, \binits{F.}},
\bauthor{\bsnm{{Natalucci}}, \binits{L.}},
\bauthor{\bsnm{{Ubertini}}, \binits{P.}},
\bauthor{\bsnm{{Roques}}, \binits{J.-P.}},
\bauthor{\bsnm{{Bouchet}}, \binits{L.}},
\bauthor{\bsnm{{Jourdain}}, \binits{E.}},
\bauthor{\bsnm{{Kn{\"o}dlseder}}, \binits{J.}},
\bauthor{\bsnm{{Diehl}}, \binits{R.}},
\bauthor{\bsnm{{Budtz-Jorgensen}}, \binits{C.}},
\bauthor{\bsnm{{Brandt}}, \binits{S.}},
\bauthor{\bsnm{{Lund}}, \binits{N.}},
\bauthor{\bsnm{{Westergaard}}, \binits{N.J.}},
\bauthor{\bsnm{{Neronov}}, \binits{A.}},
\bauthor{\bsnm{{T{\"u}rler}}, \binits{M.}},
\bauthor{\bsnm{{Chernyakova}}, \binits{M.}},
\bauthor{\bsnm{{Walter}}, \binits{R.}},
\bauthor{\bsnm{{Produit}}, \binits{N.}},
\bauthor{\bsnm{{Mowlavi}}, \binits{N.}},
\bauthor{\bsnm{{Mas-Hesse}}, \binits{J.M.}},
\bauthor{\bsnm{{Domingo}}, \binits{A.}},
\bauthor{\bsnm{{Gehrels}}, \binits{N.}},
\bauthor{\bsnm{{Kuulkers}}, \binits{E.}},
\bauthor{\bsnm{{Kretschmar}}, \binits{P.}},
\bauthor{\bsnm{{Schmidt}}, \binits{M.}}:
\batitle{{INTEGRAL observations of the cosmic X-ray background in the 5-100 keV
  range via occultation by the Earth}}.
\bjtitle{\aap}
\bvolume{467}(\bissue{2}),
\bfpage{529}--\blpage{540}
(\byear{2007})
\doiurl{10.1051/0004-6361:20066230}
{\href{https://arxiv.org/abs/astro-ph/0608250}{{arXiv:astro-ph/0608250}}}
{[astro-ph]}
\end{barticle}
\endbibitem

%%% 387
\bibitem[\protect\citeauthoryear{{Sazonov} et~al.}{2007}]{Sazonov2007}
\begin{barticle}
\bauthor{\bsnm{{Sazonov}}, \binits{S.}},
\bauthor{\bsnm{{Churazov}}, \binits{E.}},
\bauthor{\bsnm{{Sunyaev}}, \binits{R.}},
\bauthor{\bsnm{{Revnivtsev}}, \binits{M.}}:
\batitle{{Hard X-ray emission of the Earth's atmosphere: Monte Carlo
  simulations}}.
\bjtitle{\mnras}
\bvolume{377}(\bissue{4}),
\bfpage{1726}--\blpage{1736}
(\byear{2007})
\doiurl{10.1111/j.1365-2966.2007.11746.x}
{\href{https://arxiv.org/abs/astro-ph/0608253}{{arXiv:astro-ph/0608253}}}
{[astro-ph]}
\end{barticle}
\endbibitem

%%% 388
\bibitem[\protect\citeauthoryear{{Moskalenko} and
  {Porter}}{2009}]{Moskalenko2009}
\begin{barticle}
\bauthor{\bsnm{{Moskalenko}}, \binits{I.V.}},
\bauthor{\bsnm{{Porter}}, \binits{T.A.}}:
\batitle{{Isotropic Gamma-Ray Background: Cosmic-Ray-Induced Albedo from Debris
  in the Solar System?}}
\bjtitle{\apjl}
\bvolume{692}(\bissue{1}),
\bfpage{54}--\blpage{57}
(\byear{2009})
\doiurl{10.1088/0004-637X/692/1/L54}
{\href{https://arxiv.org/abs/0901.0304}{{arXiv:0901.0304}}}
{[astro-ph.HE]}
\end{barticle}
\endbibitem

%%% 389
\bibitem[\protect\citeauthoryear{{Agostinelli} et~al.}{2003}]{Agostinelli2003}
\begin{barticle}
\bauthor{\bsnm{{Agostinelli}}, \binits{S.}},
\bauthor{\bsnm{{Allison}}, \binits{J.}},
\bauthor{\bsnm{{Amako}}, \binits{K.}},
\bauthor{\bsnm{{Apostolakis}}, \binits{J.}},
\bauthor{\bsnm{{Araujo}}, \binits{H.}},
\bauthor{\bsnm{{Arce}}, \binits{P.}},
\bauthor{\bsnm{{Asai}}, \binits{M.}},
\bauthor{\bsnm{{Axen}}, \binits{D.}},
\bauthor{\bsnm{{Banerjee}}, \binits{S.}},
\bauthor{\bsnm{{Barrand}}, \binits{G.}},
\bauthor{\bsnm{{Behner}}, \binits{F.}},
\bauthor{\bsnm{{Bellagamba}}, \binits{L.}},
\bauthor{\bsnm{{Boudreau}}, \binits{J.}},
\bauthor{\bsnm{{Broglia}}, \binits{L.}},
\bauthor{\bsnm{{Brunengo}}, \binits{A.}},
\bauthor{\bsnm{{Burkhardt}}, \binits{H.}},
\bauthor{\bsnm{{Chauvie}}, \binits{S.}},
\bauthor{\bsnm{{Chuma}}, \binits{J.}},
\bauthor{\bsnm{{Chytracek}}, \binits{R.}},
\bauthor{\bsnm{{Cooperman}}, \binits{G.}},
\bauthor{\bsnm{{Cosmo}}, \binits{G.}},
\bauthor{\bsnm{{Degtyarenko}}, \binits{P.}},
\bauthor{\bsnm{{Dell'Acqua}}, \binits{A.}},
\bauthor{\bsnm{{Depaola}}, \binits{G.}},
\bauthor{\bsnm{{Dietrich}}, \binits{D.}},
\bauthor{\bsnm{{Enami}}, \binits{R.}},
\bauthor{\bsnm{{Feliciello}}, \binits{A.}},
\bauthor{\bsnm{{Ferguson}}, \binits{C.}},
\bauthor{\bsnm{{Fesefeldt}}, \binits{H.}},
\bauthor{\bsnm{{Folger}}, \binits{G.}},
\bauthor{\bsnm{{Foppiano}}, \binits{F.}},
\bauthor{\bsnm{{Forti}}, \binits{A.}},
\bauthor{\bsnm{{Garelli}}, \binits{S.}},
\bauthor{\bsnm{{Giani}}, \binits{S.}},
\bauthor{\bsnm{{Giannitrapani}}, \binits{R.}},
\bauthor{\bsnm{{Gibin}}, \binits{D.}},
\bauthor{\bsnm{{G{\'o}mez Cadenas}}, \binits{J.J.}},
\bauthor{\bsnm{{Gonz{\'a}lez}}, \binits{I.}},
\bauthor{\bsnm{{Gracia Abril}}, \binits{G.}},
\bauthor{\bsnm{{Greeniaus}}, \binits{G.}},
\bauthor{\bsnm{{Greiner}}, \binits{W.}},
\bauthor{\bsnm{{Grichine}}, \binits{V.}},
\bauthor{\bsnm{{Grossheim}}, \binits{A.}},
\bauthor{\bsnm{{Guatelli}}, \binits{S.}},
\bauthor{\bsnm{{Gumplinger}}, \binits{P.}},
\bauthor{\bsnm{{Hamatsu}}, \binits{R.}},
\bauthor{\bsnm{{Hashimoto}}, \binits{K.}},
\bauthor{\bsnm{{Hasui}}, \binits{H.}},
\bauthor{\bsnm{{Heikkinen}}, \binits{A.}},
\bauthor{\bsnm{{Howard}}, \binits{A.}},
\bauthor{\bsnm{{Ivanchenko}}, \binits{V.}},
\bauthor{\bsnm{{Johnson}}, \binits{A.}},
\bauthor{\bsnm{{Jones}}, \binits{F.W.}},
\bauthor{\bsnm{{Kallenbach}}, \binits{J.}},
\bauthor{\bsnm{{Kanaya}}, \binits{N.}},
\bauthor{\bsnm{{Kawabata}}, \binits{M.}},
\bauthor{\bsnm{{Kawabata}}, \binits{Y.}},
\bauthor{\bsnm{{Kawaguti}}, \binits{M.}},
\bauthor{\bsnm{{Kelner}}, \binits{S.}},
\bauthor{\bsnm{{Kent}}, \binits{P.}},
\bauthor{\bsnm{{Kimura}}, \binits{A.}},
\bauthor{\bsnm{{Kodama}}, \binits{T.}},
\bauthor{\bsnm{{Kokoulin}}, \binits{R.}},
\bauthor{\bsnm{{Kossov}}, \binits{M.}},
\bauthor{\bsnm{{Kurashige}}, \binits{H.}},
\bauthor{\bsnm{{Lamanna}}, \binits{E.}},
\bauthor{\bsnm{{Lamp{\'e}n}}, \binits{T.}},
\bauthor{\bsnm{{Lara}}, \binits{V.}},
\bauthor{\bsnm{{Lefebure}}, \binits{V.}},
\bauthor{\bsnm{{Lei}}, \binits{F.}},
\bauthor{\bsnm{{Liendl}}, \binits{M.}},
\bauthor{\bsnm{{Lockman}}, \binits{W.}},
\bauthor{\bsnm{{Longo}}, \binits{F.}},
\bauthor{\bsnm{{Magni}}, \binits{S.}},
\bauthor{\bsnm{{Maire}}, \binits{M.}},
\bauthor{\bsnm{{Medernach}}, \binits{E.}},
\bauthor{\bsnm{{Minamimoto}}, \binits{K.}},
\bauthor{\bsnm{{Mora de Freitas}}, \binits{P.}},
\bauthor{\bsnm{{Morita}}, \binits{Y.}},
\bauthor{\bsnm{{Murakami}}, \binits{K.}},
\bauthor{\bsnm{{Nagamatu}}, \binits{M.}},
\bauthor{\bsnm{{Nartallo}}, \binits{R.}},
\bauthor{\bsnm{{Nieminen}}, \binits{P.}},
\bauthor{\bsnm{{Nishimura}}, \binits{T.}},
\bauthor{\bsnm{{Ohtsubo}}, \binits{K.}},
\bauthor{\bsnm{{Okamura}}, \binits{M.}},
\bauthor{\bsnm{{O'Neale}}, \binits{S.}},
\bauthor{\bsnm{{Oohata}}, \binits{Y.}},
\bauthor{\bsnm{{Paech}}, \binits{K.}},
\bauthor{\bsnm{{Perl}}, \binits{J.}},
\bauthor{\bsnm{{Pfeiffer}}, \binits{A.}},
\bauthor{\bsnm{{Pia}}, \binits{M.G.}},
\bauthor{\bsnm{{Ranjard}}, \binits{F.}},
\bauthor{\bsnm{{Rybin}}, \binits{A.}},
\bauthor{\bsnm{{Sadilov}}, \binits{S.}},
\bauthor{\bsnm{{Di Salvo}}, \binits{E.}},
\bauthor{\bsnm{{Santin}}, \binits{G.}},
\bauthor{\bsnm{{Sasaki}}, \binits{T.}},
\bauthor{\bsnm{{Savvas}}, \binits{N.}},
\bauthor{\bsnm{{Sawada}}, \binits{Y.}},
\bauthor{\bsnm{{Scherer}}, \binits{S.}},
\bauthor{\bsnm{{Sei}}, \binits{S.}},
\bauthor{\bsnm{{Sirotenko}}, \binits{V.}},
\bauthor{\bsnm{{Smith}}, \binits{D.}},
\bauthor{\bsnm{{Starkov}}, \binits{N.}},
\bauthor{\bsnm{{Stoecker}}, \binits{H.}},
\bauthor{\bsnm{{Sulkimo}}, \binits{J.}},
\bauthor{\bsnm{{Takahata}}, \binits{M.}},
\bauthor{\bsnm{{Tanaka}}, \binits{S.}},
\bauthor{\bsnm{{Tcherniaev}}, \binits{E.}},
\bauthor{\bsnm{{Safai Tehrani}}, \binits{E.}},
\bauthor{\bsnm{{Tropeano}}, \binits{M.}},
\bauthor{\bsnm{{Truscott}}, \binits{P.}},
\bauthor{\bsnm{{Uno}}, \binits{H.}},
\bauthor{\bsnm{{Urban}}, \binits{L.}},
\bauthor{\bsnm{{Urban}}, \binits{P.}},
\bauthor{\bsnm{{Verderi}}, \binits{M.}},
\bauthor{\bsnm{{Walkden}}, \binits{A.}},
\bauthor{\bsnm{{Wander}}, \binits{W.}},
\bauthor{\bsnm{{Weber}}, \binits{H.}},
\bauthor{\bsnm{{Wellisch}}, \binits{J.P.}},
\bauthor{\bsnm{{Wenaus}}, \binits{T.}},
\bauthor{\bsnm{{Williams}}, \binits{D.C.}},
\bauthor{\bsnm{{Wright}}, \binits{D.}},
\bauthor{\bsnm{{Yamada}}, \binits{T.}},
\bauthor{\bsnm{{Yoshida}}, \binits{H.}},
\bauthor{\bsnm{{Zschiesche}}, \binits{D.}},
\bauthor{\bsnm{{G EANT4 Collaboration}}}:
\batitle{{G EANT4{\textemdash}a simulation toolkit}}.
\bjtitle{Nuclear Instruments and Methods in Physics Research A}
\bvolume{506}(\bissue{3}),
\bfpage{250}--\blpage{303}
(\byear{2003})
\doiurl{10.1016/S0168-9002(03)01368-8}
\end{barticle}
\endbibitem

%%% 390
\bibitem[\protect\citeauthoryear{{Whipple}}{1950}]{Whipple1950}
\begin{barticle}
\bauthor{\bsnm{{Whipple}}, \binits{F.L.}}:
\batitle{{A comet model. I. The acceleration of Comet Encke}}.
\bjtitle{\apj}
\bvolume{111},
\bfpage{375}--\blpage{394}
(\byear{1950})
\doiurl{10.1086/145272}
\end{barticle}
\endbibitem

%%% 391
\bibitem[\protect\citeauthoryear{{Purcell} et~al.}{1997}]{Purcell1997}
\begin{barticle}
\bauthor{\bsnm{{Purcell}}, \binits{W.R.}},
\bauthor{\bsnm{{Cheng}}, \binits{L.-X.}},
\bauthor{\bsnm{{Dixon}}, \binits{D.D.}},
\bauthor{\bsnm{{Kinzer}}, \binits{R.L.}},
\bauthor{\bsnm{{Kurfess}}, \binits{J.D.}},
\bauthor{\bsnm{{Leventhal}}, \binits{M.}},
\bauthor{\bsnm{{Saunders}}, \binits{M.A.}},
\bauthor{\bsnm{{Skibo}}, \binits{J.G.}},
\bauthor{\bsnm{{Smith}}, \binits{D.M.}},
\bauthor{\bsnm{{Tueller}}, \binits{J.}}:
\batitle{{OSSE Mapping of Galactic 511 keV Positron Annihilation Line
  Emission}}.
\bjtitle{\apj}
\bvolume{491}(\bissue{2}),
\bfpage{725}--\blpage{748}
(\byear{1997})
\doiurl{10.1086/304994}
\end{barticle}
\endbibitem

%%% 392
\bibitem[\protect\citeauthoryear{{Peplowski} et~al.}{2013}]{Peplowski2013}
\begin{barticle}
\bauthor{\bsnm{{Peplowski}}, \binits{P.N.}},
\bauthor{\bsnm{{Lawrence}}, \binits{D.J.}},
\bauthor{\bsnm{{Prettyman}}, \binits{T.H.}},
\bauthor{\bsnm{{Yamashita}}, \binits{N.}},
\bauthor{\bsnm{{Bazell}}, \binits{D.}},
\bauthor{\bsnm{{Feldman}}, \binits{W.C.}},
\bauthor{\bsnm{{Le Corre}}, \binits{L.}},
\bauthor{\bsnm{{McCoy}}, \binits{T.J.}},
\bauthor{\bsnm{{Reddy}}, \binits{V.}},
\bauthor{\bsnm{{Reedy}}, \binits{R.C.}},
\bauthor{\bsnm{{Russell}}, \binits{C.T.}},
\bauthor{\bsnm{{Toplis}}, \binits{M.J.}}:
\batitle{{Compositional variability on the surface of 4 Vesta revealed through
  GRaND measurements of high-energy gamma rays}}.
\bjtitle{\maps}
\bvolume{48}(\bissue{11}),
\bfpage{2252}--\blpage{2270}
(\byear{2013})
\doiurl{10.1111/maps.12176}
\end{barticle}
\endbibitem

%%% 393
\bibitem[\protect\citeauthoryear{{Lawrence} et~al.}{2018}]{Lawrence2018}
\begin{barticle}
\bauthor{\bsnm{{Lawrence}}, \binits{D.J.}},
\bauthor{\bsnm{{Peplowski}}, \binits{P.N.}},
\bauthor{\bsnm{{Beck}}, \binits{A.W.}},
\bauthor{\bsnm{{Feldman}}, \binits{W.C.}},
\bauthor{\bsnm{{Prettyman}}, \binits{T.H.}},
\bauthor{\bsnm{{Russell}}, \binits{C.T.}},
\bauthor{\bsnm{{Toplis}}, \binits{M.J.}},
\bauthor{\bsnm{{Wilson}}, \binits{J.T.}},
\bauthor{\bsnm{{Ammannito}}, \binits{E.}},
\bauthor{\bsnm{{Castillo-Rogez}}, \binits{J.C.}},
\bauthor{\bsnm{{Desanctis}}, \binits{M.C.}},
\bauthor{\bsnm{{Mest}}, \binits{S.C.}},
\bauthor{\bsnm{{Neesemann}}, \binits{A.}}:
\batitle{{Compositional variability on the surface of 1 Ceres revealed through
  GRaND measurements of high-energy gamma rays}}.
\bjtitle{\maps}
\bvolume{53}(\bissue{9}),
\bfpage{1805}--\blpage{1819}
(\byear{2018})
\doiurl{10.1111/maps.13124}
\end{barticle}
\endbibitem

%%% 394
\bibitem[\protect\citeauthoryear{{Lawrence} et~al.}{2019}]{Lawrence2019}
\begin{barticle}
\bauthor{\bsnm{{Lawrence}}, \binits{D.J.}},
\bauthor{\bsnm{{Peplowski}}, \binits{P.N.}},
\bauthor{\bsnm{{Beck}}, \binits{A.W.}},
\bauthor{\bsnm{{Burks}}, \binits{M.T.}},
\bauthor{\bsnm{{Chabot}}, \binits{N.L.}},
\bauthor{\bsnm{{Cully}}, \binits{M.J.}},
\bauthor{\bsnm{{Elphic}}, \binits{R.C.}},
\bauthor{\bsnm{{Ernst}}, \binits{C.M.}},
\bauthor{\bsnm{{Fix}}, \binits{S.}},
\bauthor{\bsnm{{Goldsten}}, \binits{J.O.}},
\bauthor{\bsnm{{Hoffer}}, \binits{E.M.}},
\bauthor{\bsnm{{Kusano}}, \binits{H.}},
\bauthor{\bsnm{{Murchie}}, \binits{S.L.}},
\bauthor{\bsnm{{Schratz}}, \binits{B.C.}},
\bauthor{\bsnm{{Usui}}, \binits{T.}},
\bauthor{\bsnm{{Yokley}}, \binits{Z.W.}}:
\batitle{{Measuring the Elemental Composition of Phobos: The Mars-moon
  Exploration with GAmma rays and NEutrons (MEGANE) Investigation for the
  Martian Moons eXploration (MMX) Mission}}.
\bjtitle{Earth and Space Science}
\bvolume{6}(\bissue{12}),
\bfpage{2605}--\blpage{2623}
(\byear{2019})
\doiurl{10.1029/2019EA000811}
\end{barticle}
\endbibitem

%%% 395
\bibitem[\protect\citeauthoryear{{Hannah} et~al.}{2007}]{Hannah2007}
\begin{barticle}
\bauthor{\bsnm{{Hannah}}, \binits{I.G.}},
\bauthor{\bsnm{{Hurford}}, \binits{G.J.}},
\bauthor{\bsnm{{Hudson}}, \binits{H.S.}},
\bauthor{\bsnm{{Lin}}, \binits{R.P.}},
\bauthor{\bsnm{{van Bibber}}, \binits{K.}}:
\batitle{{First Limits on the 3-200 keV X-Ray Spectrum of the Quiet Sun Using
  RHESSI}}.
\bjtitle{\apjl}
\bvolume{659}(\bissue{1}),
\bfpage{77}--\blpage{80}
(\byear{2007})
\doiurl{10.1086/516750}
{\href{https://arxiv.org/abs/astro-ph/0702726}{{arXiv:astro-ph/0702726}}}
{[astro-ph]}
\end{barticle}
\endbibitem

%%% 396
\bibitem[\protect\citeauthoryear{{Peplowski}}{2016}]{Peplowski2016}
\begin{barticle}
\bauthor{\bsnm{{Peplowski}}, \binits{P.N.}}:
\batitle{{The global elemental composition of 433 Eros: First results from the
  NEAR gamma-ray spectrometer orbital dataset}}.
\bjtitle{\planss}
\bvolume{134},
\bfpage{36}--\blpage{51}
(\byear{2016})
\doiurl{10.1016/j.pss.2016.10.006}
\end{barticle}
\endbibitem

%%% 397
\bibitem[\protect\citeauthoryear{{Dohnanyi}}{1969}]{Dohnanyi1969}
\begin{barticle}
\bauthor{\bsnm{{Dohnanyi}}, \binits{J.S.}}:
\batitle{{Collisional Model of Asteroids and Their Debris}}.
\bjtitle{\jgr}
\bvolume{74},
\bfpage{2531}--\blpage{2554}
(\byear{1969})
\doiurl{10.1029/JB074i010p02531}
\end{barticle}
\endbibitem

%%% 398
\bibitem[\protect\citeauthoryear{{Kippen} et~al.}{2007}]{Kippen2007}
\begin{bchapter}
\bauthor{\bsnm{{Kippen}}, \binits{R.M.}},
\bauthor{\bsnm{{Hoover}}, \binits{A.S.}},
\bauthor{\bsnm{{Wallace}}, \binits{M.S.}},
\bauthor{\bsnm{{Pendleton}}, \binits{G.N.}},
\bauthor{\bsnm{{Meegan}}, \binits{C.A.}},
\bauthor{\bsnm{{Fishman}}, \binits{G.J.}},
\bauthor{\bsnm{{Wilson-Hodge}}, \binits{C.A.}},
\bauthor{\bsnm{{Kouveliotou}}, \binits{C.}},
\bauthor{\bsnm{{Lichti}}, \binits{G.G.}},
\bauthor{\bsnm{{von Kienlin}}, \binits{A.}},
\bauthor{\bsnm{{Steinle}}, \binits{H.}},
\bauthor{\bsnm{{Diehl}}, \binits{R.}},
\bauthor{\bsnm{{Greiner}}, \binits{J.}},
\bauthor{\bsnm{{Preece}}, \binits{R.D.}},
\bauthor{\bsnm{{Connaughton}}, \binits{V.}},
\bauthor{\bsnm{{Briggs}}, \binits{M.S.}},
\bauthor{\bsnm{{Paciesas}}, \binits{W.S.}},
\bauthor{\bsnm{{Bhat}}, \binits{P.N.}}:
\bctitle{{Instrument Response Modeling and Simulation for the GLAST Burst
  Monitor}}.
In: \beditor{\bsnm{{Ritz}}, \binits{S.}},
\beditor{\bsnm{{Michelson}}, \binits{P.}},
\beditor{\bsnm{{Meegan}}, \binits{C.A.}} (eds.)
\bbtitle{The First GLAST Symposium}.
\bsertitle{American Institute of Physics Conference Series},
vol. \bseriesno{921},
pp. \bfpage{590}--\blpage{591}.
\bpublisher{AIP}, \blocation{???}
(\byear{2007}).
\doiurl{10.1063/1.2757466}
\end{bchapter}
\endbibitem

%%% 399
\bibitem[\protect\citeauthoryear{{Share} and {Murphy}}{2001}]{Share2001}
\begin{barticle}
\bauthor{\bsnm{{Share}}, \binits{G.H.}},
\bauthor{\bsnm{{Murphy}}, \binits{R.J.}}:
\batitle{{Atmospheric gamma rays from solar energetic particles and cosmic rays
  penetrating the magnetosphere}}.
\bjtitle{\jgr}
\bvolume{106}(\bissue{A1}),
\bfpage{77}--\blpage{92}
(\byear{2001})
\doiurl{10.1029/2000JA002012}
\end{barticle}
\endbibitem

%%% 400
\bibitem[\protect\citeauthoryear{{Churazov} et~al.}{2008}]{Churazov2008}
\begin{barticle}
\bauthor{\bsnm{{Churazov}}, \binits{E.}},
\bauthor{\bsnm{{Sazonov}}, \binits{S.}},
\bauthor{\bsnm{{Sunyaev}}, \binits{R.}},
\bauthor{\bsnm{{Revnivtsev}}, \binits{M.}}:
\batitle{{Earth X-ray albedo for cosmic X-ray background radiation in the
  1-1000 keV band}}.
\bjtitle{\mnras}
\bvolume{385}(\bissue{2}),
\bfpage{719}--\blpage{727}
(\byear{2008})
\doiurl{10.1111/j.1365-2966.2008.12918.x}
\end{barticle}
\endbibitem

%%% 401
\bibitem[\protect\citeauthoryear{{Johnson} and {Haymes}}{1973}]{Johnson1973}
\begin{barticle}
\bauthor{\bsnm{{Johnson}}, \binits{W.N.} \bsuffix{III}},
\bauthor{\bsnm{{Haymes}}, \binits{R.C.}}:
\batitle{{Detection of a Gamma-Ray Spectral Line from the Galactic-Center
  Region}}.
\bjtitle{\apj}
\bvolume{184},
\bfpage{103}--\blpage{126}
(\byear{1973})
\doiurl{10.1086/152309}
\end{barticle}
\endbibitem

%%% 402
\bibitem[\protect\citeauthoryear{{Haymes} et~al.}{1969}]{Haymes1969}
\begin{barticle}
\bauthor{\bsnm{{Haymes}}, \binits{R.C.}},
\bauthor{\bsnm{{Ellis}}, \binits{D.V.}},
\bauthor{\bsnm{{Fishman}}, \binits{G.J.}},
\bauthor{\bsnm{{Glenn}}, \binits{S.W.}},
\bauthor{\bsnm{{Kurfess}}, \binits{J.D.}}:
\batitle{{Observation of Hard Radiation from the Region of the Galactic
  Center}}.
\bjtitle{\apj}
\bvolume{157},
\bfpage{1455}
(\byear{1969})
\doiurl{10.1086/150164}
\end{barticle}
\endbibitem

%%% 403
\bibitem[\protect\citeauthoryear{{Riegler} et~al.}{1981}]{Riegler1981}
\begin{barticle}
\bauthor{\bsnm{{Riegler}}, \binits{G.R.}},
\bauthor{\bsnm{{Ling}}, \binits{J.C.}},
\bauthor{\bsnm{{Mahoney}}, \binits{W.A.}},
\bauthor{\bsnm{{Wheaton}}, \binits{W.A.}},
\bauthor{\bsnm{{Willett}}, \binits{J.B.}},
\bauthor{\bsnm{{Jacobson}}, \binits{A.S.}},
\bauthor{\bsnm{{Prince}}, \binits{T.A.}}:
\batitle{{Variable positron annihilation radiation from the galactic center
  region}}.
\bjtitle{\apjl}
\bvolume{248},
\bfpage{13}--\blpage{16}
(\byear{1981})
\doiurl{10.1086/183613}
\end{barticle}
\endbibitem

%%% 404
\bibitem[\protect\citeauthoryear{{Share} et~al.}{1988}]{Share1988}
\begin{barticle}
\bauthor{\bsnm{{Share}}, \binits{G.H.}},
\bauthor{\bsnm{{Kinzer}}, \binits{R.L.}},
\bauthor{\bsnm{{Kurfess}}, \binits{J.D.}},
\bauthor{\bsnm{{Messina}}, \binits{D.C.}},
\bauthor{\bsnm{{Purcell}}, \binits{W.R.}},
\bauthor{\bsnm{{Chupp}}, \binits{E.L.}},
\bauthor{\bsnm{{Forrest}}, \binits{D.J.}},
\bauthor{\bsnm{{Reppin}}, \binits{C.}}:
\batitle{{SMM Detection of Diffuse Galactic 511 keV Annihilation Radiation}}.
\bjtitle{\apj}
\bvolume{326},
\bfpage{717}
(\byear{1988})
\doiurl{10.1086/166130}
\end{barticle}
\endbibitem

%%% 405
\bibitem[\protect\citeauthoryear{{Lingenfelter} and
  {Ramaty}}{1989}]{Lingenfelter1989}
\begin{barticle}
\bauthor{\bsnm{{Lingenfelter}}, \binits{R.E.}},
\bauthor{\bsnm{{Ramaty}}, \binits{R.}}:
\batitle{{The Nature of the Annihilation Radiation and Gamma-Ray Continuum from
  the Galactic Center Region}}.
\bjtitle{\apj}
\bvolume{343},
\bfpage{686}
(\byear{1989})
\doiurl{10.1086/167740}
\end{barticle}
\endbibitem

%%% 406
\bibitem[\protect\citeauthoryear{{Kn{\"o}dlseder}
  et~al.}{2005}]{Knoedlseder2005}
\begin{barticle}
\bauthor{\bsnm{{Kn{\"o}dlseder}}, \binits{J.}},
\bauthor{\bsnm{{Jean}}, \binits{P.}},
\bauthor{\bsnm{{Lonjou}}, \binits{V.}},
\bauthor{\bsnm{{Weidenspointner}}, \binits{G.}},
\bauthor{\bsnm{{Guessoum}}, \binits{N.}},
\bauthor{\bsnm{{Gillard}}, \binits{W.}},
\bauthor{\bsnm{{Skinner}}, \binits{G.}},
\bauthor{\bsnm{{von Ballmoos}}, \binits{P.}},
\bauthor{\bsnm{{Vedrenne}}, \binits{G.}},
\bauthor{\bsnm{{Roques}}, \binits{J.-P.}},
\bauthor{\bsnm{{Schanne}}, \binits{S.}},
\bauthor{\bsnm{{Teegarden}}, \binits{B.}},
\bauthor{\bsnm{{Sch{\"o}nfelder}}, \binits{V.}},
\bauthor{\bsnm{{Winkler}}, \binits{C.}}:
\batitle{{The all-sky distribution of 511 keV electron-positron annihilation
  emission}}.
\bjtitle{\aap}
\bvolume{441}(\bissue{2}),
\bfpage{513}--\blpage{532}
(\byear{2005})
\doiurl{10.1051/0004-6361:20042063}
{\href{https://arxiv.org/abs/astro-ph/0506026}{{arXiv:astro-ph/0506026}}}
{[astro-ph]}
\end{barticle}
\endbibitem

%%% 407
\bibitem[\protect\citeauthoryear{{Weidenspointner}
  et~al.}{2008}]{Weidenspointner2008}
\begin{barticle}
\bauthor{\bsnm{{Weidenspointner}}, \binits{G.}},
\bauthor{\bsnm{{Skinner}}, \binits{G.}},
\bauthor{\bsnm{{Jean}}, \binits{P.}},
\bauthor{\bsnm{{Kn{\"o}dlseder}}, \binits{J.}},
\bauthor{\bsnm{{von Ballmoos}}, \binits{P.}},
\bauthor{\bsnm{{Bignami}}, \binits{G.}},
\bauthor{\bsnm{{Diehl}}, \binits{R.}},
\bauthor{\bsnm{{Strong}}, \binits{A.W.}},
\bauthor{\bsnm{{Cordier}}, \binits{B.}},
\bauthor{\bsnm{{Schanne}}, \binits{S.}},
\bauthor{\bsnm{{Winkler}}, \binits{C.}}:
\batitle{{An asymmetric distribution of positrons in the Galactic disk revealed
  by {\ensuremath{\gamma}}-rays}}.
\bjtitle{\nat}
\bvolume{451}(\bissue{7175}),
\bfpage{159}--\blpage{162}
(\byear{2008})
\doiurl{10.1038/nature06490}
\end{barticle}
\endbibitem

%%% 408
\bibitem[\protect\citeauthoryear{{Bouchet} et~al.}{2010}]{Bouchet2010}
\begin{barticle}
\bauthor{\bsnm{{Bouchet}}, \binits{L.}},
\bauthor{\bsnm{{Roques}}, \binits{J.P.}},
\bauthor{\bsnm{{Jourdain}}, \binits{E.}}:
\batitle{{On the Morphology of the Electron-Positron Annihilation Emission as
  Seen by Spi/integral}}.
\bjtitle{\apj}
\bvolume{720}(\bissue{2}),
\bfpage{1772}--\blpage{1780}
(\byear{2010})
\doiurl{10.1088/0004-637X/720/2/1772}
{\href{https://arxiv.org/abs/1007.4753}{{arXiv:1007.4753}}}
{[astro-ph.HE]}
\end{barticle}
\endbibitem

%%% 409
\bibitem[\protect\citeauthoryear{Skinner et~al.}{2014}]{Skinner2014}
\begin{bchapter}
\bauthor{\bsnm{Skinner}, \binits{G.}},
\bauthor{\bsnm{Diehl}, \binits{R.}},
\bauthor{\bsnm{Zhang}, \binits{X.}},
\bauthor{\bsnm{Bouchet}, \binits{L.}},
\bauthor{\bsnm{Jean}, \binits{P.}}:
\bctitle{{The Galactic distribution of the 511 keV e+/e- annihilation
  radiation}}.
In: \bbtitle{Proceedings of the 10th INTEGRAL Workshop: ''A Synergistic View of
  the High-Energy Sky'' (INTEGRAL 2014). 15-19 September 2014. Annapolis, MD,
  USA. Published Online at
  Http://pos.sissa.it/cgi-bin/reader/conf.cgi?confid=228, Id.054},
p. \bfpage{054}
(\byear{2014})
\end{bchapter}
\endbibitem

%%% 410
\bibitem[\protect\citeauthoryear{{Siegert} et~al.}{2016}]{Siegert2016a}
\begin{barticle}
\bauthor{\bsnm{{Siegert}}, \binits{T.}},
\bauthor{\bsnm{{Diehl}}, \binits{R.}},
\bauthor{\bsnm{{Khachatryan}}, \binits{G.}},
\bauthor{\bsnm{{Krause}}, \binits{M.G.H.}},
\bauthor{\bsnm{{Guglielmetti}}, \binits{F.}},
\bauthor{\bsnm{{Greiner}}, \binits{J.}},
\bauthor{\bsnm{{Strong}}, \binits{A.W.}},
\bauthor{\bsnm{{Zhang}}, \binits{X.}}:
\batitle{{Gamma-ray spectroscopy of positron annihilation in the Milky Way}}.
\bjtitle{\aap}
\bvolume{586},
\bfpage{84}
(\byear{2016})
\doiurl{10.1051/0004-6361/201527510}
{\href{https://arxiv.org/abs/1512.00325}{{arXiv:1512.00325}}}
{[astro-ph.HE]}
\end{barticle}
\endbibitem

%%% 411
\bibitem[\protect\citeauthoryear{{Siegert} et~al.}{2019}]{Siegert2019b}
\begin{barticle}
\bauthor{\bsnm{{Siegert}}, \binits{T.}},
\bauthor{\bsnm{{Crocker}}, \binits{R.M.}},
\bauthor{\bsnm{{Diehl}}, \binits{R.}},
\bauthor{\bsnm{{Krause}}, \binits{M.G.H.}},
\bauthor{\bsnm{{Panther}}, \binits{F.H.}},
\bauthor{\bsnm{{Pleintinger}}, \binits{M.M.M.}},
\bauthor{\bsnm{{Weinberger}}, \binits{C.}}:
\batitle{{Constraints on positron annihilation kinematics in the inner
  Galaxy}}.
\bjtitle{\aap}
\bvolume{627},
\bfpage{126}
(\byear{2019})
\doiurl{10.1051/0004-6361/201833856}
{\href{https://arxiv.org/abs/1906.00498}{{arXiv:1906.00498}}}
{[astro-ph.HE]}
\end{barticle}
\endbibitem

%%% 412
\bibitem[\protect\citeauthoryear{{Siegert} et~al.}{2022}]{Siegert2022a}
\begin{barticle}
\bauthor{\bsnm{{Siegert}}, \binits{T.}},
\bauthor{\bsnm{{Crocker}}, \binits{R.M.}},
\bauthor{\bsnm{{Macias}}, \binits{O.}},
\bauthor{\bsnm{{Panther}}, \binits{F.H.}},
\bauthor{\bsnm{{Calore}}, \binits{F.}},
\bauthor{\bsnm{{Song}}, \binits{D.}},
\bauthor{\bsnm{{Horiuchi}}, \binits{S.}}:
\batitle{{Measuring the smearing of the Galactic 511-keV signal: positron
  propagation or supernova kicks?}}
\bjtitle{\mnras}
\bvolume{509}(\bissue{1}),
\bfpage{11}--\blpage{16}
(\byear{2022})
\doiurl{10.1093/mnrasl/slab113}
{\href{https://arxiv.org/abs/2109.03691}{{arXiv:2109.03691}}}
{[astro-ph.HE]}
\end{barticle}
\endbibitem

%%% 413
\bibitem[\protect\citeauthoryear{{Yoneda} et~al.}{2025}]{Yoneda2025_511}
\begin{barticle}
\bauthor{\bsnm{{Yoneda}}, \binits{H.}},
\bauthor{\bsnm{{Siegert}}, \binits{T.}},
\bauthor{\bsnm{{Mittal}}, \binits{S.}}:
\batitle{{Imaging the positron annihilation line with 20 years of INTEGRAL/SPI
  observations}}.
\bjtitle{\aap}
\bvolume{702},
\bfpage{220}
(\byear{2025})
\doiurl{10.1051/0004-6361/202555895}
{\href{https://arxiv.org/abs/2509.01066}{{arXiv:2509.01066}}}
{[astro-ph.HE]}
\end{barticle}
\endbibitem

%%% 414
\bibitem[\protect\citeauthoryear{{Churazov} et~al.}{2011}]{Churazov2011}
\begin{barticle}
\bauthor{\bsnm{{Churazov}}, \binits{E.}},
\bauthor{\bsnm{{Sazonov}}, \binits{S.}},
\bauthor{\bsnm{{Tsygankov}}, \binits{S.}},
\bauthor{\bsnm{{Sunyaev}}, \binits{R.}},
\bauthor{\bsnm{{Varshalovich}}, \binits{D.}}:
\batitle{{Positron annihilation spectrum from the Galactic Centre region
  observed by SPI/INTEGRAL revisited: annihilation in a cooling ISM?}}
\bjtitle{\mnras}
\bvolume{411}(\bissue{3}),
\bfpage{1727}--\blpage{1743}
(\byear{2011})
\doiurl{10.1111/j.1365-2966.2010.17804.x}
{\href{https://arxiv.org/abs/1010.0864}{{arXiv:1010.0864}}}
{[astro-ph.HE]}
\end{barticle}
\endbibitem

%%% 415
\bibitem[\protect\citeauthoryear{{Kinzer} et~al.}{2001}]{Kinzer2001}
\begin{barticle}
\bauthor{\bsnm{{Kinzer}}, \binits{R.L.}},
\bauthor{\bsnm{{Milne}}, \binits{P.A.}},
\bauthor{\bsnm{{Kurfess}}, \binits{J.D.}},
\bauthor{\bsnm{{Strickman}}, \binits{M.S.}},
\bauthor{\bsnm{{Johnson}}, \binits{W.N.}},
\bauthor{\bsnm{{Purcell}}, \binits{W.R.}}:
\batitle{{Positron Annihilation Radiation from the Inner Galaxy}}.
\bjtitle{\apj}
\bvolume{559}(\bissue{1}),
\bfpage{282}--\blpage{295}
(\byear{2001})
\doiurl{10.1086/322371}
\end{barticle}
\endbibitem

%%% 416
\bibitem[\protect\citeauthoryear{{Siegert} et~al.}{2022}]{Siegert2022d}
\begin{bchapter}
\bauthor{\bsnm{{Siegert}}, \binits{T.}},
\bauthor{\bsnm{{Horan}}, \binits{D.}},
\bauthor{\bsnm{{Kanbach}}, \binits{G.}}:
\bctitle{{Telescope Concepts in Gamma-Ray Astronomy}}.
In: \beditor{\bsnm{{Bambi}}, \binits{C.}},
\beditor{\bsnm{{Sangangelo}}, \binits{A.}} (eds.)
\bbtitle{Handbook of X-ray and Gamma-ray Astrophysics},
p. \bfpage{80}
(\byear{2022}).
\doiurl{10.1007/978-981-16-4544-0_43-1}
\end{bchapter}
\endbibitem

%%% 417
\bibitem[\protect\citeauthoryear{{Siegert} et~al.}{2016}]{Siegert2016c}
\begin{barticle}
\bauthor{\bsnm{{Siegert}}, \binits{T.}},
\bauthor{\bsnm{{Diehl}}, \binits{R.}},
\bauthor{\bsnm{{Vincent}}, \binits{A.C.}},
\bauthor{\bsnm{{Guglielmetti}}, \binits{F.}},
\bauthor{\bsnm{{Krause}}, \binits{M.G.H.}},
\bauthor{\bsnm{{Boehm}}, \binits{C.}}:
\batitle{{Search for 511 keV emission in satellite galaxies of the Milky Way
  with INTEGRAL/SPI}}.
\bjtitle{\aap}
\bvolume{595},
\bfpage{25}
(\byear{2016})
\doiurl{10.1051/0004-6361/201629136}
{\href{https://arxiv.org/abs/1608.00393}{{arXiv:1608.00393}}}
{[astro-ph.HE]}
\end{barticle}
\endbibitem

%%% 418
\bibitem[\protect\citeauthoryear{{Siegert} et~al.}{2022}]{Siegert2022b}
\begin{barticle}
\bauthor{\bsnm{{Siegert}}, \binits{T.}},
\bauthor{\bsnm{{Boehm}}, \binits{C.}},
\bauthor{\bsnm{{Calore}}, \binits{F.}},
\bauthor{\bsnm{{Diehl}}, \binits{R.}},
\bauthor{\bsnm{{Krause}}, \binits{M.G.H.}},
\bauthor{\bsnm{{Serpico}}, \binits{P.D.}},
\bauthor{\bsnm{{Vincent}}, \binits{A.C.}}:
\batitle{{An INTEGRAL/SPI view of reticulum II: particle dark matter and
  primordial black holes limits in the MeV range}}.
\bjtitle{\mnras}
\bvolume{511}(\bissue{1}),
\bfpage{914}--\blpage{924}
(\byear{2022})
\doiurl{10.1093/mnras/stac008}
{\href{https://arxiv.org/abs/2109.03791}{{arXiv:2109.03791}}}
{[astro-ph.HE]}
\end{barticle}
\endbibitem

%%% 419
\bibitem[\protect\citeauthoryear{{Milne} et~al.}{2001}]{Milne2001}
\begin{barticle}
\bauthor{\bsnm{{Milne}}, \binits{P.A.}},
\bauthor{\bsnm{{The}}, \binits{L.-S.}},
\bauthor{\bsnm{{Leising}}, \binits{M.D.}}:
\batitle{{Late Light Curves of Type Ia Supernovae}}.
\bjtitle{\apj}
\bvolume{559}(\bissue{2}),
\bfpage{1019}--\blpage{1031}
(\byear{2001})
\doiurl{10.1086/322352}
{\href{https://arxiv.org/abs/astro-ph/0104185}{{arXiv:astro-ph/0104185}}}
{[astro-ph]}
\end{barticle}
\endbibitem

%%% 420
\bibitem[\protect\citeauthoryear{{Sizun} et~al.}{2006}]{Sizun2006}
\begin{barticle}
\bauthor{\bsnm{{Sizun}}, \binits{P.}},
\bauthor{\bsnm{{Cass{\'e}}}, \binits{M.}},
\bauthor{\bsnm{{Schanne}}, \binits{S.}}:
\batitle{{Continuum {\ensuremath{\gamma}}-ray emission from light dark matter
  positrons and electrons}}.
\bjtitle{\prd}
\bvolume{74}(\bissue{6}),
\bfpage{063514}
(\byear{2006})
\doiurl{10.1103/PhysRevD.74.063514}
{\href{https://arxiv.org/abs/astro-ph/0607374}{{arXiv:astro-ph/0607374}}}
{[astro-ph]}
\end{barticle}
\endbibitem

%%% 421
\bibitem[\protect\citeauthoryear{{Beacom} and {Y{\"u}ksel}}{2006}]{Beacom2006}
\begin{barticle}
\bauthor{\bsnm{{Beacom}}, \binits{J.F.}},
\bauthor{\bsnm{{Y{\"u}ksel}}, \binits{H.}}:
\batitle{{Stringent Constraint on Galactic Positron Production}}.
\bjtitle{\prl}
\bvolume{97}(\bissue{7}),
\bfpage{071102}
(\byear{2006})
\doiurl{10.1103/PhysRevLett.97.071102}
{\href{https://arxiv.org/abs/astro-ph/0512411}{{arXiv:astro-ph/0512411}}}
{[astro-ph]}
\end{barticle}
\endbibitem

%%% 422
\bibitem[\protect\citeauthoryear{{Siegert} et~al.}{2022}]{Siegert2022c}
\begin{barticle}
\bauthor{\bsnm{{Siegert}}, \binits{T.}},
\bauthor{\bsnm{{Berteaud}}, \binits{J.}},
\bauthor{\bsnm{{Calore}}, \binits{F.}},
\bauthor{\bsnm{{Serpico}}, \binits{P.D.}},
\bauthor{\bsnm{{Weinberger}}, \binits{C.}}:
\batitle{{Diffuse Galactic emission spectrum between 0.5 and 8.0 MeV}}.
\bjtitle{\aap}
\bvolume{660},
\bfpage{130}
(\byear{2022})
\doiurl{10.1051/0004-6361/202142639}
{\href{https://arxiv.org/abs/2202.04574}{{arXiv:2202.04574}}}
{[astro-ph.HE]}
\end{barticle}
\endbibitem

%%% 423
\bibitem[\protect\citeauthoryear{{Aharonian} and
  {Atoyan}}{2000}]{Aharonian2000}
\begin{barticle}
\bauthor{\bsnm{{Aharonian}}, \binits{F.A.}},
\bauthor{\bsnm{{Atoyan}}, \binits{A.M.}}:
\batitle{{Broad-band diffuse gamma ray emission of the galactic disk}}.
\bjtitle{\aap}
\bvolume{362},
\bfpage{937}--\blpage{952}
(\byear{2000})
\doiurl{10.48550/arXiv.astro-ph/0009009}
{\href{https://arxiv.org/abs/astro-ph/0009009}{{arXiv:astro-ph/0009009}}}
{[astro-ph]}
\end{barticle}
\endbibitem

%%% 424
\bibitem[\protect\citeauthoryear{{Porter} et~al.}{2008}]{Porter2008}
\begin{barticle}
\bauthor{\bsnm{{Porter}}, \binits{T.A.}},
\bauthor{\bsnm{{Moskalenko}}, \binits{I.V.}},
\bauthor{\bsnm{{Strong}}, \binits{A.W.}},
\bauthor{\bsnm{{Orlando}}, \binits{E.}},
\bauthor{\bsnm{{Bouchet}}, \binits{L.}}:
\batitle{{Inverse Compton Origin of the Hard X-Ray and Soft Gamma-Ray Emission
  from the Galactic Ridge}}.
\bjtitle{\apj}
\bvolume{682}(\bissue{1}),
\bfpage{400}--\blpage{407}
(\byear{2008})
\doiurl{10.1086/589615}
{\href{https://arxiv.org/abs/0804.1774}{{arXiv:0804.1774}}}
{[astro-ph]}
\end{barticle}
\endbibitem

%%% 425
\bibitem[\protect\citeauthoryear{{Bouchet} et~al.}{1991}]{Bouchet1991}
\begin{barticle}
\bauthor{\bsnm{{Bouchet}}, \binits{L.}},
\bauthor{\bsnm{{Mandrou}}, \binits{P.}},
\bauthor{\bsnm{{Roques}}, \binits{J.P.}},
\bauthor{\bsnm{{Vedrenne}}, \binits{G.}},
\bauthor{\bsnm{{Cordier}}, \binits{B.}},
\bauthor{\bsnm{{Goldwurm}}, \binits{A.}},
\bauthor{\bsnm{{Lebrun}}, \binits{F.}},
\bauthor{\bsnm{{Paul}}, \binits{J.}},
\bauthor{\bsnm{{Sunyaev}}, \binits{R.}},
\bauthor{\bsnm{{Churazov}}, \binits{E.}},
\bauthor{\bsnm{{Gilfanov}}, \binits{M.}},
\bauthor{\bsnm{{Pavlinsky}}, \binits{M.}},
\bauthor{\bsnm{{Grebenev}}, \binits{S.}},
\bauthor{\bsnm{{Babalyan}}, \binits{G.}},
\bauthor{\bsnm{{Dekhanov}}, \binits{I.}},
\bauthor{\bsnm{{Khavenson}}, \binits{N.}}:
\batitle{{SIGMA Discovery of Variable E +- E - Annihilation Radiation from the
  Near Galactic Center Variable Compact Source 1E 1740.7-2942}}.
\bjtitle{\apjl}
\bvolume{383},
\bfpage{45}
(\byear{1991})
\doiurl{10.1086/186237}
\end{barticle}
\endbibitem

%%% 426
\bibitem[\protect\citeauthoryear{{Sunyaev} et~al.}{1991}]{Sunyaev1991}
\begin{barticle}
\bauthor{\bsnm{{Sunyaev}}, \binits{R.}},
\bauthor{\bsnm{{Churazov}}, \binits{E.}},
\bauthor{\bsnm{{Gilfanov}}, \binits{M.}},
\bauthor{\bsnm{{Pavlinsky}}, \binits{M.}},
\bauthor{\bsnm{{Grebenev}}, \binits{S.}},
\bauthor{\bsnm{{Babalyan}}, \binits{G.}},
\bauthor{\bsnm{{Dekhanov}}, \binits{I.}},
\bauthor{\bsnm{{Khavenson}}, \binits{N.}},
\bauthor{\bsnm{{Bouchet}}, \binits{L.}},
\bauthor{\bsnm{{Mandrou}}, \binits{P.}},
\bauthor{\bsnm{{Roques}}, \binits{J.P.}},
\bauthor{\bsnm{{Vedrenne}}, \binits{G.}},
\bauthor{\bsnm{{Cordier}}, \binits{B.}},
\bauthor{\bsnm{{Goldwurm}}, \binits{A.}},
\bauthor{\bsnm{{Lebrun}}, \binits{F.}},
\bauthor{\bsnm{{Paul}}, \binits{J.}}:
\batitle{{Three Spectral States of 1E 1740.7-2942: From Standard Cygnus X-1
  Type Spectrum to the Evidence of Electron-Positron Annihilation Feature}}.
\bjtitle{\apjl}
\bvolume{383},
\bfpage{49}
(\byear{1991})
\doiurl{10.1086/186238}
\end{barticle}
\endbibitem

%%% 427
\bibitem[\protect\citeauthoryear{{Sunyaev} et~al.}{1992}]{Sunyaev1992}
\begin{barticle}
\bauthor{\bsnm{{Sunyaev}}, \binits{R.}},
\bauthor{\bsnm{{Churazov}}, \binits{E.}},
\bauthor{\bsnm{{Gilfanov}}, \binits{M.}},
\bauthor{\bsnm{{Dyachkov}}, \binits{A.}},
\bauthor{\bsnm{{Khavenson}}, \binits{N.}},
\bauthor{\bsnm{{Grebenev}}, \binits{S.}},
\bauthor{\bsnm{{Kremnev}}, \binits{R.}},
\bauthor{\bsnm{{Sukhanov}}, \binits{K.}},
\bauthor{\bsnm{{Goldwurm}}, \binits{A.}},
\bauthor{\bsnm{{Ballet}}, \binits{J.}},
\bauthor{\bsnm{{Cordier}}, \binits{B.}},
\bauthor{\bsnm{{Paul}}, \binits{J.}},
\bauthor{\bsnm{{Denis}}, \binits{M.}},
\bauthor{\bsnm{{Vedrenne}}, \binits{G.}},
\bauthor{\bsnm{{Niel}}, \binits{M.}},
\bauthor{\bsnm{{Jourdain}}, \binits{E.}}:
\batitle{{X-Ray Nova in Musca (GRS 1124-68): Hard X-Ray Source with Narrow
  Annihilation Line}}.
\bjtitle{\apjl}
\bvolume{389},
\bfpage{75}
(\byear{1992})
\doiurl{10.1086/186352}
\end{barticle}
\endbibitem

%%% 428
\bibitem[\protect\citeauthoryear{{Siegert} et~al.}{2016}]{Siegert2016b}
\begin{barticle}
\bauthor{\bsnm{{Siegert}}, \binits{T.}},
\bauthor{\bsnm{{Diehl}}, \binits{R.}},
\bauthor{\bsnm{{Greiner}}, \binits{J.}},
\bauthor{\bsnm{{Krause}}, \binits{M.G.H.}},
\bauthor{\bsnm{{Beloborodov}}, \binits{A.M.}},
\bauthor{\bsnm{{Bel}}, \binits{M.C.}},
\bauthor{\bsnm{{Guglielmetti}}, \binits{F.}},
\bauthor{\bsnm{{Rodriguez}}, \binits{J.}},
\bauthor{\bsnm{{Strong}}, \binits{A.W.}},
\bauthor{\bsnm{{Zhang}}, \binits{X.}}:
\batitle{{Positron annihilation signatures associated with the outburst of the
  microquasar V404 Cygni}}.
\bjtitle{\nat}
\bvolume{531}(\bissue{7594}),
\bfpage{341}--\blpage{343}
(\byear{2016})
\doiurl{10.1038/nature16978}
{\href{https://arxiv.org/abs/1603.01169}{{arXiv:1603.01169}}}
{[astro-ph.HE]}
\end{barticle}
\endbibitem

%%% 429
\bibitem[\protect\citeauthoryear{{Beloborodov}}{1999}]{Beloborodov1999}
\begin{barticle}
\bauthor{\bsnm{{Beloborodov}}, \binits{A.M.}}:
\batitle{{Electron-positron outflows from gamma-ray emitting accretion discs}}.
\bjtitle{\mnras}
\bvolume{305}(\bissue{1}),
\bfpage{181}--\blpage{189}
(\byear{1999})
\doiurl{10.1046/j.1365-8711.1999.02384.x}
{\href{https://arxiv.org/abs/astro-ph/9901107}{{arXiv:astro-ph/9901107}}}
{[astro-ph]}
\end{barticle}
\endbibitem

%%% 430
\bibitem[\protect\citeauthoryear{{Bartels} et~al.}{2018}]{Bartels2018}
\begin{barticle}
\bauthor{\bsnm{{Bartels}}, \binits{R.}},
\bauthor{\bsnm{{Calore}}, \binits{F.}},
\bauthor{\bsnm{{Storm}}, \binits{E.}},
\bauthor{\bsnm{{Weniger}}, \binits{C.}}:
\batitle{{Galactic binaries can explain the Fermi Galactic centre excess and
  511 keV emission}}.
\bjtitle{\mnras}
\bvolume{480}(\bissue{3}),
\bfpage{3826}--\blpage{3841}
(\byear{2018})
\doiurl{10.1093/mnras/sty2135}
{\href{https://arxiv.org/abs/1803.04370}{{arXiv:1803.04370}}}
{[astro-ph.HE]}
\end{barticle}
\endbibitem

%%% 431
\bibitem[\protect\citeauthoryear{{Murphy} et~al.}{2005}]{Murphy2005}
\begin{barticle}
\bauthor{\bsnm{{Murphy}}, \binits{R.J.}},
\bauthor{\bsnm{{Share}}, \binits{G.H.}},
\bauthor{\bsnm{{Skibo}}, \binits{J.G.}},
\bauthor{\bsnm{{Kozlovsky}}, \binits{B.}}:
\batitle{{The Physics of Positron Annihilation in the Solar Atmosphere}}.
\bjtitle{\apjs}
\bvolume{161}(\bissue{2}),
\bfpage{495}--\blpage{519}
(\byear{2005})
\doiurl{10.1086/452634}
\end{barticle}
\endbibitem

%%% 432
\bibitem[\protect\citeauthoryear{{Frost} et~al.}{1966}]{Frost1966}
\begin{barticle}
\bauthor{\bsnm{{Frost}}, \binits{K.J.}},
\bauthor{\bsnm{{Rothe}}, \binits{E.D.}},
\bauthor{\bsnm{{Peterson}}, \binits{L.E.}}:
\batitle{{A Search for the Quiet-Time Solar Gamma Rays from Ballon Altitudes}}.
\bjtitle{\jgr}
\bvolume{71},
\bfpage{4079}
(\byear{1966})
\doiurl{10.1029/JZ071i017p04079}
\end{barticle}
\endbibitem

%%% 433
\bibitem[\protect\citeauthoryear{{Mazziotta} et~al.}{2020}]{Mazziotta2020}
\begin{barticle}
\bauthor{\bsnm{{Mazziotta}}, \binits{M.N.}},
\bauthor{\bsnm{{Luque}}, \binits{P.D.L.T.}},
\bauthor{\bsnm{{Di Venere}}, \binits{L.}},
\bauthor{\bsnm{{Fass{\`o}}}, \binits{A.}},
\bauthor{\bsnm{{Ferrari}}, \binits{A.}},
\bauthor{\bsnm{{Loparco}}, \binits{F.}},
\bauthor{\bsnm{{Sala}}, \binits{P.R.}},
\bauthor{\bsnm{{Serini}}, \binits{D.}}:
\batitle{{Cosmic-ray interactions with the Sun using the FLUKA code}}.
\bjtitle{\prd}
\bvolume{101}(\bissue{8}),
\bfpage{083011}
(\byear{2020})
\doiurl{10.1103/PhysRevD.101.083011}
{\href{https://arxiv.org/abs/2001.09933}{{arXiv:2001.09933}}}
{[astro-ph.HE]}
\end{barticle}
\endbibitem

%%% 434
\bibitem[\protect\citeauthoryear{{Bisnovatyi-Kogan} and
  {Pozanenko}}{2017}]{Bisnovatyi-Kogan2017}
\begin{barticle}
\bauthor{\bsnm{{Bisnovatyi-Kogan}}, \binits{G.S.}},
\bauthor{\bsnm{{Pozanenko}}, \binits{A.S.}}:
\batitle{{Can Flare Stars Explain the Annihilation Line from the Galactic
  Bulge?}}
\bjtitle{Astrophysics}
\bvolume{60}(\bissue{2}),
\bfpage{223}--\blpage{227}
(\byear{2017})
\doiurl{10.1007/s10511-017-9477-6}
\end{barticle}
\endbibitem

%%% 435
\bibitem[\protect\citeauthoryear{{McConnell} et~al.}{1997}]{McConnell1997}
\begin{bchapter}
\bauthor{\bsnm{{McConnell}}, \binits{M.}},
\bauthor{\bsnm{{Fletcher}}, \binits{S.}},
\bauthor{\bsnm{{Bennett}}, \binits{K.}},
\bauthor{\bsnm{{Bloemen}}, \binits{H.}},
\bauthor{\bsnm{{Diehl}}, \binits{R.}},
\bauthor{\bsnm{{Hermsen}}, \binits{W.}},
\bauthor{\bsnm{{Ryan}}, \binits{J.}},
\bauthor{\bsnm{{Sch{\"o}nfelder}}, \binits{V.}},
\bauthor{\bsnm{{Strong}}, \binits{A.}},
\bauthor{\bsnm{{van Dijk}}, \binits{R.}}:
\bctitle{{COMPTEL all-sky imaging at 2.2 MeV}}.
In: \beditor{\bsnm{{Dermer}}, \binits{C.D.}},
\beditor{\bsnm{{Strickman}}, \binits{M.S.}},
\beditor{\bsnm{{Kurfess}}, \binits{J.D.}} (eds.)
\bbtitle{Proceedings of the Fourth Compton Symposium}.
\bsertitle{American Institute of Physics Conference Series},
vol. \bseriesno{410},
pp. \bfpage{1099}--\blpage{1103}.
\bpublisher{AIP}, \blocation{???}
(\byear{1997}).
\doiurl{10.1063/1.54175}
\end{bchapter}
\endbibitem

%%% 436
\bibitem[\protect\citeauthoryear{{Spanier} et~al.}{2022}]{Spanier2022}
\begin{barticle}
\bauthor{\bsnm{{Spanier}}, \binits{F.}},
\bauthor{\bsnm{{Schreiner}}, \binits{C.}},
\bauthor{\bsnm{{Schlickeiser}}, \binits{R.}}:
\batitle{{Determining Pitch-Angle Diffusion Coefficients for Electrons in
  Whistler Turbulence}}.
\bjtitle{Physics}
\bvolume{4}(\bissue{1}),
\bfpage{80}--\blpage{103}
(\byear{2022})
\doiurl{10.3390/physics4010008}
{\href{https://arxiv.org/abs/2112.03999}{{arXiv:2112.03999}}}
{[astro-ph.HE]}
\end{barticle}
\endbibitem

%%% 437
\bibitem[\protect\citeauthoryear{{Alexis} et~al.}{2014}]{Alexis2014}
\begin{barticle}
\bauthor{\bsnm{{Alexis}}, \binits{A.}},
\bauthor{\bsnm{{Jean}}, \binits{P.}},
\bauthor{\bsnm{{Martin}}, \binits{P.}},
\bauthor{\bsnm{{Ferri{\`e}re}}, \binits{K.}}:
\batitle{{Monte Carlo modelling of the propagation and annihilation of
  nucleosynthesis positrons in the Galaxy}}.
\bjtitle{\aap}
\bvolume{564},
\bfpage{108}
(\byear{2014})
\doiurl{10.1051/0004-6361/201322393}
{\href{https://arxiv.org/abs/1402.6110}{{arXiv:1402.6110}}}
{[astro-ph.HE]}
\end{barticle}
\endbibitem

%%% 438
\bibitem[\protect\citeauthoryear{{Jean} et~al.}{2009}]{Jean2009}
\begin{barticle}
\bauthor{\bsnm{{Jean}}, \binits{P.}},
\bauthor{\bsnm{{Gillard}}, \binits{W.}},
\bauthor{\bsnm{{Marcowith}}, \binits{A.}},
\bauthor{\bsnm{{Ferri{\`e}re}}, \binits{K.}}:
\batitle{{Positron transport in the interstellar medium}}.
\bjtitle{\aap}
\bvolume{508}(\bissue{3}),
\bfpage{1099}--\blpage{1116}
(\byear{2009})
\doiurl{10.1051/0004-6361/200809830}
{\href{https://arxiv.org/abs/0909.4022}{{arXiv:0909.4022}}}
{[astro-ph.HE]}
\end{barticle}
\endbibitem

%%% 439
\bibitem[\protect\citeauthoryear{{Kn{\"o}dlseder}
  et~al.}{2025}]{Knoedlseder2025_aif}
\begin{barticle}
\bauthor{\bsnm{{Kn{\"o}dlseder}}, \binits{J.}},
\bauthor{\bsnm{{Sabri}}, \binits{K.}},
\bauthor{\bsnm{{Jean}}, \binits{P.}},
\bauthor{\bsnm{{von Ballmoos}}, \binits{P.}},
\bauthor{\bsnm{{Skinner}}, \binits{G.}},
\bauthor{\bsnm{{Collmar}}, \binits{W.}}:
\batitle{{Detection of positron in-flight annihilation from the Galaxy}}.
\bjtitle{\aap}
\bvolume{700},
\bfpage{257}
(\byear{2025})
\doiurl{10.1051/0004-6361/202556046}
{\href{https://arxiv.org/abs/2506.17427}{{arXiv:2506.17427}}}
{[astro-ph.HE]}
\end{barticle}
\endbibitem

%%% 440
\bibitem[\protect\citeauthoryear{{Das} et~al.}{2025}]{Das2025_aif}
\begin{botherref}
\oauthor{\bsnm{{Das}}, \binits{S.}},
\oauthor{\bsnm{{Krumholz}}, \binits{M.R.}},
\oauthor{\bsnm{{Crocker}}, \binits{R.M.}},
\oauthor{\bsnm{{Siegert}}, \binits{T.}},
\oauthor{\bsnm{{Eisenberger}}, \binits{L.}}:
{Relaxation of Energy Constraints for Positrons Generating the Galactic
  Annihilation Signal}.
arXiv e-prints,
2506--00847
(2025)
\doiurl{10.48550/arXiv.2506.00847}
{\href{https://arxiv.org/abs/2506.00847}{{arXiv:2506.00847}}}
{[astro-ph.HE]}
\end{botherref}
\endbibitem

%%% 441
\bibitem[\protect\citeauthoryear{Berteaud et~al.}{2022}]{Berteaud:2022tws}
\begin{barticle}
\bauthor{\bsnm{Berteaud}, \binits{J.}},
\bauthor{\bsnm{Calore}, \binits{F.}},
\bauthor{\bsnm{Iguaz}, \binits{J.}},
\bauthor{\bsnm{Serpico}, \binits{P.D.}},
\bauthor{\bsnm{Siegert}, \binits{T.}}:
\batitle{{Strong constraints on primordial black hole dark matter from 16~years
  of INTEGRAL/SPI observations}}.
\bjtitle{Phys. Rev. D}
\bvolume{106}(\bissue{2}),
\bfpage{023030}
(\byear{2022})
\doiurl{10.1103/PhysRevD.106.023030}
{\href{https://arxiv.org/abs/2202.07483}{{arXiv:2202.07483}}}
{[astro-ph.HE]}
\end{barticle}
\endbibitem

%%% 442
\bibitem[\protect\citeauthoryear{De~la Torre~Luque
  et~al.}{2024a}]{DelaTorreLuque:2023cef}
\begin{barticle}
\bauthor{\bsnm{Torre~Luque}, \binits{P.}},
\bauthor{\bsnm{Balaji}, \binits{S.}},
\bauthor{\bsnm{Silk}, \binits{J.}}:
\batitle{{New 511 keV Line Data Provide Strongest sub-GeV Dark Matter
  Constraints}}.
\bjtitle{Astrophys. J. Lett.}
\bvolume{973}(\bissue{1}),
\bfpage{6}
(\byear{2024})
\doiurl{10.3847/2041-8213/ad72f4}
{\href{https://arxiv.org/abs/2312.04907}{{arXiv:2312.04907}}}
{[hep-ph]}
\end{barticle}
\endbibitem

%%% 443
\bibitem[\protect\citeauthoryear{De~la Torre~Luque
  et~al.}{2024b}]{DelaTorreLuque:2023huu}
\begin{barticle}
\bauthor{\bsnm{Torre~Luque}, \binits{P.}},
\bauthor{\bsnm{Balaji}, \binits{S.}},
\bauthor{\bsnm{Carenza}, \binits{P.}}:
\batitle{{Multimessenger search for electrophilic feebly interacting particles
  from supernovae}}.
\bjtitle{Phys. Rev. D}
\bvolume{109}(\bissue{10}),
\bfpage{103028}
(\byear{2024})
\doiurl{10.1103/PhysRevD.109.103028}
{\href{https://arxiv.org/abs/2307.13731}{{arXiv:2307.13731}}}
{[hep-ph]}
\end{barticle}
\endbibitem

%%% 444
\bibitem[\protect\citeauthoryear{De~la Torre~Luque
  et~al.}{2025}]{DelaTorreLuque:2024zsr}
\begin{barticle}
\bauthor{\bsnm{Torre~Luque}, \binits{P.}},
\bauthor{\bsnm{Balaji}, \binits{S.}},
\bauthor{\bsnm{Carenza}, \binits{P.}},
\bauthor{\bsnm{Mastrototaro}, \binits{L.}}:
\batitle{{\ensuremath{\gamma} rays from in-flight positron annihilation as a
  probe of new physics}}.
\bjtitle{Phys. Rev. D}
\bvolume{111}(\bissue{6}),
\bfpage{061303}
(\byear{2025})
\doiurl{10.1103/PhysRevD.111.L061303}
{\href{https://arxiv.org/abs/2405.08482}{{arXiv:2405.08482}}}
{[hep-ph]}
\end{barticle}
\endbibitem

%%% 445
\bibitem[\protect\citeauthoryear{Slatyer and Wu}{2017}]{Slatyer:2016qyl}
\begin{barticle}
\bauthor{\bsnm{Slatyer}, \binits{T.R.}},
\bauthor{\bsnm{Wu}, \binits{C.-L.}}:
\batitle{{General Constraints on Dark Matter Decay from the Cosmic Microwave
  Background}}.
\bjtitle{Phys. Rev. D}
\bvolume{95}(\bissue{2}),
\bfpage{023010}
(\byear{2017})
\doiurl{10.1103/PhysRevD.95.023010}
{\href{https://arxiv.org/abs/1610.06933}{{arXiv:1610.06933}}}
{[astro-ph.CO]}
\end{barticle}
\endbibitem

%%% 446
\bibitem[\protect\citeauthoryear{{Mirabel} et~al.}{1992}]{Mirabel1992}
\begin{barticle}
\bauthor{\bsnm{{Mirabel}}, \binits{I.F.}},
\bauthor{\bsnm{{Rodriguez}}, \binits{L.F.}},
\bauthor{\bsnm{{Cordier}}, \binits{B.}},
\bauthor{\bsnm{{Paul}}, \binits{J.}},
\bauthor{\bsnm{{Lebrun}}, \binits{F.}}:
\batitle{{A double-sided radio jet from the compact Galactic Centre annihilator
  1E1740.7-2942}}.
\bjtitle{\nat}
\bvolume{358}(\bissue{6383}),
\bfpage{215}--\blpage{217}
(\byear{1992})
\doiurl{10.1038/358215a0}
\end{barticle}
\endbibitem

%%% 447
\bibitem[\protect\citeauthoryear{{Iliadis}}{2015}]{Iliadis2015}
\begin{bbook}
\bauthor{\bsnm{{Iliadis}}, \binits{C.}}:
\bbtitle{{Nuclear Physics of Stars}},
(\byear{2015}).
\doiurl{10.1002/9783527692668}
\end{bbook}
\endbibitem

%%% 448
\bibitem[\protect\citeauthoryear{{D'Auria} and {DRAG.~O.~N.
  Collaboration}}{2002}]{DRAGON}
\begin{barticle}
\bauthor{\bsnm{{D'Auria}}, \binits{J.M.}},
\bauthor{\bsnm{{DRAG.~O.~N. Collaboration}}}:
\batitle{{Astrophysics with a DRAGON at ISAC}}.
\bjtitle{\nphysa}
\bvolume{701},
\bfpage{625}--\blpage{631}
(\byear{2002})
\doiurl{10.1016/S0375-9474(01)01656-6}
\end{barticle}
\endbibitem

%%% 449
\bibitem[\protect\citeauthoryear{{Couder} et~al.}{2008}]{Couder2008}
\begin{barticle}
\bauthor{\bsnm{{Couder}}, \binits{M.}},
\bauthor{\bsnm{{Berg}}, \binits{G.P.A.}},
\bauthor{\bsnm{{G{\"o}rres}}, \binits{J.}},
\bauthor{\bsnm{{LeBlanc}}, \binits{P.J.}},
\bauthor{\bsnm{{Lamm}}, \binits{L.O.}},
\bauthor{\bsnm{{Stech}}, \binits{E.}},
\bauthor{\bsnm{{Wiescher}}, \binits{M.}},
\bauthor{\bsnm{{Hinnefeld}}, \binits{J.}}:
\batitle{{Design of the recoil mass separator St. George}}.
\bjtitle{Nuclear Instruments and Methods in Physics Research A}
\bvolume{587}(\bissue{1}),
\bfpage{35}--\blpage{45}
(\byear{2008})
\doiurl{10.1016/j.nima.2007.11.069}
\end{barticle}
\endbibitem

%%% 450
\bibitem[\protect\citeauthoryear{{Berg} et~al.}{2018}]{SECAR}
\begin{barticle}
\bauthor{\bsnm{{Berg}}, \binits{G.P.A.}},
\bauthor{\bsnm{{Couder}}, \binits{M.}},
\bauthor{\bsnm{{Moran}}, \binits{M.T.}},
\bauthor{\bsnm{{Smith}}, \binits{K.}},
\bauthor{\bsnm{{Wiescher}}, \binits{M.}},
\bauthor{\bsnm{{Schatz}}, \binits{H.}},
\bauthor{\bsnm{{Hager}}, \binits{U.}},
\bauthor{\bsnm{{Wrede}}, \binits{C.}},
\bauthor{\bsnm{{Montes}}, \binits{F.}},
\bauthor{\bsnm{{Perdikakis}}, \binits{G.}},
\bauthor{\bsnm{{Wu}}, \binits{X.}},
\bauthor{\bsnm{{Zeller}}, \binits{A.}},
\bauthor{\bsnm{{Smith}}, \binits{M.S.}},
\bauthor{\bsnm{{Bardayan}}, \binits{D.W.}},
\bauthor{\bsnm{{Chipps}}, \binits{K.A.}},
\bauthor{\bsnm{{Pain}}, \binits{S.D.}},
\bauthor{\bsnm{{Blackmon}}, \binits{J.}},
\bauthor{\bsnm{{Greife}}, \binits{U.}},
\bauthor{\bsnm{{Rehm}}, \binits{K.E.}},
\bauthor{\bsnm{{Janssens}}, \binits{R.V.F.}}:
\batitle{{Design of SECAR a recoil mass separator for astrophysical capture
  reactions with radioactive beams}}.
\bjtitle{Nuclear Instruments and Methods in Physics Research A}
\bvolume{877},
\bfpage{87}--\blpage{103}
(\byear{2018})
\doiurl{10.1016/j.nima.2017.08.048}
\end{barticle}
\endbibitem

%%% 451
\bibitem[\protect\citeauthoryear{{Heine} et~al.}{2018}]{STELLA}
\begin{barticle}
\bauthor{\bsnm{{Heine}}, \binits{M.}},
\bauthor{\bsnm{{Courtin}}, \binits{S.}},
\bauthor{\bsnm{{Fruet}}, \binits{G.}},
\bauthor{\bsnm{{Jenkins}}, \binits{D.G.}},
\bauthor{\bsnm{{Morris}}, \binits{L.}},
\bauthor{\bsnm{{Montanari}}, \binits{D.}},
\bauthor{\bsnm{{Rudigier}}, \binits{M.}},
\bauthor{\bsnm{{Adsley}}, \binits{P.}},
\bauthor{\bsnm{{Curien}}, \binits{D.}},
\bauthor{\bsnm{{Della Negra}}, \binits{S.}},
\bauthor{\bsnm{{Lesrel}}, \binits{J.}},
\bauthor{\bsnm{{Beck}}, \binits{C.}},
\bauthor{\bsnm{{Charles}}, \binits{L.}},
\bauthor{\bsnm{{Den{\'e}}}, \binits{P.}},
\bauthor{\bsnm{{Haas}}, \binits{F.}},
\bauthor{\bsnm{{Hammache}}, \binits{F.}},
\bauthor{\bsnm{{Heitz}}, \binits{G.}},
\bauthor{\bsnm{{Krauth}}, \binits{M.}},
\bauthor{\bsnm{{Meyer}}, \binits{A.}},
\bauthor{\bsnm{{Podoly{\'a}k}}, \binits{Z.}},
\bauthor{\bsnm{{Regan}}, \binits{P.H.}},
\bauthor{\bsnm{{Richer}}, \binits{M.}},
\bauthor{\bsnm{{de S{\'e}r{\'e}ville}}, \binits{N.}},
\bauthor{\bsnm{{Stodel}}, \binits{C.}}:
\batitle{{The STELLA apparatus for particle-Gamma coincidence fusion
  measurements with nanosecond timing}}.
\bjtitle{Nuclear Instruments and Methods in Physics Research A}
\bvolume{903},
\bfpage{1}--\blpage{7}
(\byear{2018})
\doiurl{10.1016/j.nima.2018.06.058}
{\href{https://arxiv.org/abs/1802.07679}{{arXiv:1802.07679}}}
{[physics.ins-det]}
\end{barticle}
\endbibitem

%%% 452
\bibitem[\protect\citeauthoryear{{Greife} et~al.}{1994}]{LUNA}
\begin{barticle}
\bauthor{\bsnm{{Greife}}, \binits{U.}},
\bauthor{\bsnm{{Arpesella}}, \binits{C.}},
\bauthor{\bsnm{{Barnes}}, \binits{C.A.}},
\bauthor{\bsnm{{Bartolucci}}, \binits{F.}},
\bauthor{\bsnm{{Bellotti}}, \binits{E.}},
\bauthor{\bsnm{{Broggini}}, \binits{C.}},
\bauthor{\bsnm{{Corvisiero}}, \binits{P.}},
\bauthor{\bsnm{{Fiorentini}}, \binits{G.}},
\bauthor{\bsnm{{Fubini}}, \binits{A.}},
\bauthor{\bsnm{{Gervino}}, \binits{G.}},
\bauthor{\bsnm{{Gorris}}, \binits{F.}},
\bauthor{\bsnm{{Gustavino}}, \binits{C.}},
\bauthor{\bsnm{{Junker}}, \binits{M.}},
\bauthor{\bsnm{{Kavanagh}}, \binits{R.W.}},
\bauthor{\bsnm{{Lanza}}, \binits{A.}},
\bauthor{\bsnm{{Mezzorani}}, \binits{G.}},
\bauthor{\bsnm{{Prati}}, \binits{P.}},
\bauthor{\bsnm{{Quarati}}, \binits{P.}},
\bauthor{\bsnm{{Rodney}}, \binits{W.S.}},
\bauthor{\bsnm{{Rolfs}}, \binits{C.}},
\bauthor{\bsnm{{Schulte}}, \binits{W.H.}},
\bauthor{\bsnm{{Trautvetter}}, \binits{H.P.}},
\bauthor{\bsnm{{Zahnow}}, \binits{D.}}:
\batitle{{Laboratory for Underground Nuclear Astrophysics (LUNA)}}.
\bjtitle{Nuclear Instruments and Methods in Physics Research A}
\bvolume{350}(\bissue{1-2}),
\bfpage{327}--\blpage{337}
(\byear{1994})
\doiurl{10.1016/0168-9002(94)91182-7}
\end{barticle}
\endbibitem

%%% 453
\bibitem[\protect\citeauthoryear{{Liu} et~al.}{2016}]{JUNA}
\begin{barticle}
\bauthor{\bsnm{{Liu}}, \binits{W.}},
\bauthor{\bsnm{{Li}}, \binits{Z.}},
\bauthor{\bsnm{{He}}, \binits{J.}},
\bauthor{\bsnm{{Tang}}, \binits{X.}},
\bauthor{\bsnm{{Lian}}, \binits{G.}},
\bauthor{\bsnm{{An}}, \binits{Z.}},
\bauthor{\bsnm{{Chang}}, \binits{J.}},
\bauthor{\bsnm{{Chen}}, \binits{H.}},
\bauthor{\bsnm{{Chen}}, \binits{Q.}},
\bauthor{\bsnm{{Chen}}, \binits{X.}},
\bauthor{\bsnm{{Chen}}, \binits{Z.}},
\bauthor{\bsnm{{Cui}}, \binits{B.}},
\bauthor{\bsnm{{Du}}, \binits{X.}},
\bauthor{\bsnm{{Fu}}, \binits{C.}},
\bauthor{\bsnm{{Gan}}, \binits{L.}},
\bauthor{\bsnm{{Guo}}, \binits{B.}},
\bauthor{\bsnm{{He}}, \binits{G.}},
\bauthor{\bsnm{{Heger}}, \binits{A.}},
\bauthor{\bsnm{{Hou}}, \binits{S.}},
\bauthor{\bsnm{{Huang}}, \binits{H.}},
\bauthor{\bsnm{{Huang}}, \binits{N.}},
\bauthor{\bsnm{{Jia}}, \binits{B.}},
\bauthor{\bsnm{{Jiang}}, \binits{L.}},
\bauthor{\bsnm{{Kubono}}, \binits{S.}},
\bauthor{\bsnm{{Li}}, \binits{J.}},
\bauthor{\bsnm{{Li}}, \binits{K.}},
\bauthor{\bsnm{{Li}}, \binits{T.}},
\bauthor{\bsnm{{Li}}, \binits{Y.}},
\bauthor{\bsnm{{Lugaro}}, \binits{M.}},
\bauthor{\bsnm{{Luo}}, \binits{X.}},
\bauthor{\bsnm{{Ma}}, \binits{H.}},
\bauthor{\bsnm{{Ma}}, \binits{S.}},
\bauthor{\bsnm{{Mei}}, \binits{D.}},
\bauthor{\bsnm{{Qian}}, \binits{Y.}},
\bauthor{\bsnm{{Qin}}, \binits{J.}},
\bauthor{\bsnm{{Ren}}, \binits{J.}},
\bauthor{\bsnm{{Shen}}, \binits{Y.}},
\bauthor{\bsnm{{Su}}, \binits{J.}},
\bauthor{\bsnm{{Sun}}, \binits{L.}},
\bauthor{\bsnm{{Tan}}, \binits{W.}},
\bauthor{\bsnm{{Tanihata}}, \binits{I.}},
\bauthor{\bsnm{{Wang}}, \binits{S.}},
\bauthor{\bsnm{{Wang}}, \binits{P.}},
\bauthor{\bsnm{{Wang}}, \binits{Y.}},
\bauthor{\bsnm{{Wu}}, \binits{Q.}},
\bauthor{\bsnm{{Xu}}, \binits{S.}},
\bauthor{\bsnm{{Yan}}, \binits{S.}},
\bauthor{\bsnm{{Yang}}, \binits{L.}},
\bauthor{\bsnm{{Yang}}, \binits{Y.}},
\bauthor{\bsnm{{Yu}}, \binits{X.}},
\bauthor{\bsnm{{Yue}}, \binits{Q.}},
\bauthor{\bsnm{{Zeng}}, \binits{S.}},
\bauthor{\bsnm{{Zhang}}, \binits{H.}},
\bauthor{\bsnm{{Zhang}}, \binits{H.}},
\bauthor{\bsnm{{Zhang}}, \binits{L.}},
\bauthor{\bsnm{{Zhang}}, \binits{N.}},
\bauthor{\bsnm{{Zhang}}, \binits{Q.}},
\bauthor{\bsnm{{Zhang}}, \binits{T.}},
\bauthor{\bsnm{{Zhang}}, \binits{X.}},
\bauthor{\bsnm{{Zhang}}, \binits{X.}},
\bauthor{\bsnm{{Zhang}}, \binits{Z.}},
\bauthor{\bsnm{{Zhao}}, \binits{W.}},
\bauthor{\bsnm{{Zhao}}, \binits{Z.}},
\bauthor{\bsnm{{Zhou}}, \binits{C.}},
\bauthor{\bsnm{{JUNA Collaboration}}}:
\batitle{{Progress of Jinping Underground laboratory for Nuclear Astrophysics
  (JUNA)}}.
\bjtitle{Science China Physics, Mechanics, and Astronomy}
\bvolume{59}(\bissue{4}),
\bfpage{642001}
(\byear{2016})
\doiurl{10.1007/s11433-016-5785-9}
\end{barticle}
\endbibitem

%%% 454
\bibitem[\protect\citeauthoryear{{Sz{\"u}cs} et~al.}{2019}]{Felsenkeller}
\begin{barticle}
\bauthor{\bsnm{{Sz{\"u}cs}}, \binits{T.}},
\bauthor{\bsnm{{Bemmerer}}, \binits{D.}},
\bauthor{\bsnm{{Degering}}, \binits{D.}},
\bauthor{\bsnm{{Domula}}, \binits{A.}},
\bauthor{\bsnm{{Grieger}}, \binits{M.}},
\bauthor{\bsnm{{Ludwig}}, \binits{F.}},
\bauthor{\bsnm{{Schmidt}}, \binits{K.}},
\bauthor{\bsnm{{Steckling}}, \binits{J.}},
\bauthor{\bsnm{{Turkat}}, \binits{S.}},
\bauthor{\bsnm{{Zuber}}, \binits{K.}}:
\batitle{{Background in {\ensuremath{\gamma}}-ray detectors and carbon beam
  tests in the Felsenkeller shallow-underground accelerator laboratory}}.
\bjtitle{European Physical Journal A}
\bvolume{55}(\bissue{10}),
\bfpage{174}
(\byear{2019})
\doiurl{10.1140/epja/i2019-12865-4}
{\href{https://arxiv.org/abs/1908.08945}{{arXiv:1908.08945}}}
{[nucl-ex]}
\end{barticle}
\endbibitem

%%% 455
\bibitem[\protect\citeauthoryear{{Hammache} and {de
  S{\'e}r{\'e}ville}}{2021}]{Hammache2021}
\begin{barticle}
\bauthor{\bsnm{{Hammache}}, \binits{F.}},
\bauthor{\bsnm{{de S{\'e}r{\'e}ville}}, \binits{N.}}:
\batitle{{Transfer reactions as a tool in Nuclear Astrophysics}}.
\bjtitle{Frontiers in Physics}
\bvolume{8},
\bfpage{630}
(\byear{2021})
\doiurl{10.3389/fphy.2020.602920}
{\href{https://arxiv.org/abs/2107.13228}{{arXiv:2107.13228}}}
{[nucl-ex]}
\end{barticle}
\endbibitem

%%% 456
\bibitem[\protect\citeauthoryear{{Limata} et~al.}{2010}]{Limata2010}
\begin{barticle}
\bauthor{\bsnm{{Limata}}, \binits{B.}},
\bauthor{\bsnm{{Strieder}}, \binits{F.}},
\bauthor{\bsnm{{Formicola}}, \binits{A.}},
\bauthor{\bsnm{{Imbriani}}, \binits{G.}},
\bauthor{\bsnm{{Junker}}, \binits{M.}},
\bauthor{\bsnm{{Becker}}, \binits{H.W.}},
\bauthor{\bsnm{{Bemmerer}}, \binits{D.}},
\bauthor{\bsnm{{Best}}, \binits{A.}},
\bauthor{\bsnm{{Bonetti}}, \binits{R.}},
\bauthor{\bsnm{{Broggini}}, \binits{C.}},
\bauthor{\bsnm{{Caciolli}}, \binits{A.}},
\bauthor{\bsnm{{Corvisiero}}, \binits{P.}},
\bauthor{\bsnm{{Costantini}}, \binits{H.}},
\bauthor{\bsnm{{Dileva}}, \binits{A.}},
\bauthor{\bsnm{{Elekes}}, \binits{Z.}},
\bauthor{\bsnm{{F{\"u}l{\"o}p}}, \binits{Z.}},
\bauthor{\bsnm{{Gervino}}, \binits{G.}},
\bauthor{\bsnm{{Guglielmetti}}, \binits{A.}},
\bauthor{\bsnm{{Gustavino}}, \binits{C.}},
\bauthor{\bsnm{{Gy{\"u}rky}}, \binits{G.}},
\bauthor{\bsnm{{Lemut}}, \binits{A.}},
\bauthor{\bsnm{{Marta}}, \binits{M.}},
\bauthor{\bsnm{{Mazzocchi}}, \binits{C.}},
\bauthor{\bsnm{{Menegazzo}}, \binits{R.}},
\bauthor{\bsnm{{Prati}}, \binits{P.}},
\bauthor{\bsnm{{Roca}}, \binits{V.}},
\bauthor{\bsnm{{Rolfs}}, \binits{C.}},
\bauthor{\bsnm{{Rossi Alvarez}}, \binits{C.}},
\bauthor{\bsnm{{Salvo}}, \binits{C.}},
\bauthor{\bsnm{{Somorjai}}, \binits{E.}},
\bauthor{\bsnm{{Straniero}}, \binits{O.}},
\bauthor{\bsnm{{Terrasi}}, \binits{F.}},
\bauthor{\bsnm{{Trautvetter}}, \binits{H.-P.}}:
\batitle{{New experimental study of low-energy (p,{\ensuremath{\gamma}})
  resonances in magnesium isotopes}}.
\bjtitle{\prc}
\bvolume{82}(\bissue{1}),
\bfpage{015801}
(\byear{2010})
\doiurl{10.1103/PhysRevC.82.015801}
{\href{https://arxiv.org/abs/1006.5281}{{arXiv:1006.5281}}}
{[nucl-ex]}
\end{barticle}
\endbibitem

%%% 457
\bibitem[\protect\citeauthoryear{{Strieder} et~al.}{2012}]{Strieder2012}
\begin{barticle}
\bauthor{\bsnm{{Strieder}}, \binits{F.}},
\bauthor{\bsnm{{Limata}}, \binits{B.}},
\bauthor{\bsnm{{Formicola}}, \binits{A.}},
\bauthor{\bsnm{{Imbriani}}, \binits{G.}},
\bauthor{\bsnm{{Junker}}, \binits{M.}},
\bauthor{\bsnm{{Bemmerer}}, \binits{D.}},
\bauthor{\bsnm{{Best}}, \binits{A.}},
\bauthor{\bsnm{{Broggini}}, \binits{C.}},
\bauthor{\bsnm{{Caciolli}}, \binits{A.}},
\bauthor{\bsnm{{Corvisiero}}, \binits{P.}},
\bauthor{\bsnm{{Costantini}}, \binits{H.}},
\bauthor{\bsnm{{DiLeva}}, \binits{A.}},
\bauthor{\bsnm{{Elekes}}, \binits{Z.}},
\bauthor{\bsnm{{F{\"u}l{\"o}p}}, \binits{Z.}},
\bauthor{\bsnm{{Gervino}}, \binits{G.}},
\bauthor{\bsnm{{Guglielmetti}}, \binits{A.}},
\bauthor{\bsnm{{Gustavino}}, \binits{C.}},
\bauthor{\bsnm{{Gy{\"u}rky}}, \binits{G.}},
\bauthor{\bsnm{{Lemut}}, \binits{A.}},
\bauthor{\bsnm{{Marta}}, \binits{M.}},
\bauthor{\bsnm{{Mazzocchi}}, \binits{C.}},
\bauthor{\bsnm{{Menegazzo}}, \binits{R.}},
\bauthor{\bsnm{{Prati}}, \binits{P.}},
\bauthor{\bsnm{{Roca}}, \binits{V.}},
\bauthor{\bsnm{{Rolfs}}, \binits{C.}},
\bauthor{\bsnm{{Rossi Alvarez}}, \binits{C.}},
\bauthor{\bsnm{{Somorjai}}, \binits{E.}},
\bauthor{\bsnm{{Straniero}}, \binits{O.}},
\bauthor{\bsnm{{Terrasi}}, \binits{F.}},
\bauthor{\bsnm{{Trautvetter}}, \binits{H.P.}}:
\batitle{{The $^{25}$Mg(p,{\ensuremath{\gamma}})$^{26}$Al reaction at low
  astrophysical energies}}.
\bjtitle{Physics Letters B}
\bvolume{707}(\bissue{1}),
\bfpage{60}--\blpage{65}
(\byear{2012})
\doiurl{10.1016/j.physletb.2011.12.029}
\end{barticle}
\endbibitem

%%% 458
\bibitem[\protect\citeauthoryear{{Su} et~al.}{2022}]{Su2022}
\begin{barticle}
\bauthor{\bsnm{{Su}}, \binits{J.}},
\bauthor{\bsnm{{Strieder}}, \binits{F.}},
\bauthor{\bsnm{{Formicola}}, \binits{A.}},
\bauthor{\bsnm{{Imbriani}}, \binits{G.}},
\bauthor{\bsnm{{Junker}}, \binits{M.}},
\bauthor{\bsnm{{Becker}}, \binits{H.W.}},
\bauthor{\bsnm{{Bemmerer}}, \binits{D.}},
\bauthor{\bsnm{{Best}}, \binits{A.}},
\bauthor{\bsnm{{Bonetti}}, \binits{R.}},
\bauthor{\bsnm{{Broggini}}, \binits{C.}},
\bauthor{\bsnm{{Caciolli}}, \binits{A.}},
\bauthor{\bsnm{{Corvisiero}}, \binits{P.}},
\bauthor{\bsnm{{Costantini}}, \binits{H.}},
\bauthor{\bsnm{{Dileva}}, \binits{A.}},
\bauthor{\bsnm{{Elekes}}, \binits{Z.}},
\bauthor{\bsnm{{F{\"u}l{\"o}p}}, \binits{Z.}},
\bauthor{\bsnm{{Gervino}}, \binits{G.}},
\bauthor{\bsnm{{Guglielmetti}}, \binits{A.}},
\bauthor{\bsnm{{Gustavino}}, \binits{C.}},
\bauthor{\bsnm{{Gy{\"u}rky}}, \binits{G.}},
\bauthor{\bsnm{{Lemut}}, \binits{A.}},
\bauthor{\bsnm{{Marta}}, \binits{M.}},
\bauthor{\bsnm{{Mazzocchi}}, \binits{C.}},
\bauthor{\bsnm{{Menegazzo}}, \binits{R.}},
\bauthor{\bsnm{{Prati}}, \binits{P.}},
\bauthor{\bsnm{{Roca}}, \binits{V.}},
\bauthor{\bsnm{{Rolfs}}, \binits{C.}},
\bauthor{\bsnm{{Rossi Alvarez}}, \binits{C.}},
\bauthor{\bsnm{{Salvo}}, \binits{C.}},
\bauthor{\bsnm{{Somorjai}}, \binits{E.}},
\bauthor{\bsnm{{Straniero}}, \binits{O.}},
\bauthor{\bsnm{{Terrasi}}, \binits{F.}},
\bauthor{\bsnm{{Trautvetter}}, \binits{H.-P.}}:
\batitle{{First result from the Jinping Underground Nuclear Astrophysics
  experiment JUNA: precise measurement of the 92 keV
  $^{25}$Mg(p,{\ensuremath{\gamma}})$^{26}$Al}}.
\bjtitle{Science Bulletin}
\bvolume{67},
\bfpage{125}--\blpage{132}
(\byear{2022})
\end{barticle}
\endbibitem

%%% 459
\bibitem[\protect\citeauthoryear{{Champagne} et~al.}{1989}]{Champagne1989}
\begin{barticle}
\bauthor{\bsnm{{Champagne}}, \binits{A.E.}},
\bauthor{\bsnm{{Howard}}, \binits{A.J.}},
\bauthor{\bsnm{{Smith}}, \binits{M.S.}},
\bauthor{\bsnm{{Magnus}}, \binits{P.V.}},
\bauthor{\bsnm{{Parker}}, \binits{P.D.}}:
\batitle{{The effect of weak resonances on the $^{25}$Mg(p,
  {\ensuremath{\gamma}})$^{26}$Al reaction rate}}.
\bjtitle{\nphysa}
\bvolume{505}(\bissue{2}),
\bfpage{384}--\blpage{396}
(\byear{1989})
\doiurl{10.1016/0375-9474(89)90382-5}
\end{barticle}
\endbibitem

%%% 460
\bibitem[\protect\citeauthoryear{{Rollefson} et~al.}{1990}]{Rollefson1990}
\begin{barticle}
\bauthor{\bsnm{{Rollefson}}, \binits{A.A.}},
\bauthor{\bsnm{{Wijekumar}}, \binits{V.}},
\bauthor{\bsnm{{Browne}}, \binits{C.P.}},
\bauthor{\bsnm{{Wiescher}}, \binits{M.}},
\bauthor{\bsnm{{Hausman}}, \binits{H.J.}},
\bauthor{\bsnm{{Kim}}, \binits{W.Y.}},
\bauthor{\bsnm{{Schmalbrock}}, \binits{P.}}:
\batitle{{Spectroscopic factors for proton unbound levels in $^{26}$Al and
  their influence on stellar reaction rates}}.
\bjtitle{\nphysa}
\bvolume{507}(\bissue{2}),
\bfpage{413}--\blpage{425}
(\byear{1990})
\doiurl{10.1016/0375-9474(90)90301-2}
\end{barticle}
\endbibitem

%%% 461
\bibitem[\protect\citeauthoryear{{Li} et~al.}{2020}]{Li2020}
\begin{barticle}
\bauthor{\bsnm{{Li}}, \binits{Y.J.}},
\bauthor{\bsnm{{Li}}, \binits{Z.H.}},
\bauthor{\bsnm{{Li}}, \binits{E.T.}},
\bauthor{\bsnm{{Li}}, \binits{X.Y.}},
\bauthor{\bsnm{{Ma}}, \binits{T.L.}},
\bauthor{\bsnm{{Shen}}, \binits{Y.P.}},
\bauthor{\bsnm{{Liu}}, \binits{J.C.}},
\bauthor{\bsnm{{Gan}}, \binits{L.}},
\bauthor{\bsnm{{Su}}, \binits{Y.}},
\bauthor{\bsnm{{Qiao}}, \binits{L.H.}},
\bauthor{\bsnm{{Han}}, \binits{Z.Y.}},
\bauthor{\bsnm{{Zhou}}, \binits{Y.}},
\bauthor{\bsnm{{Su}}, \binits{J.}},
\bauthor{\bsnm{{Yan}}, \binits{S.Q.}},
\bauthor{\bsnm{{Zeng}}, \binits{S.}},
\bauthor{\bsnm{{Wang}}, \binits{Y.B.}},
\bauthor{\bsnm{{Guo}}, \binits{B.}},
\bauthor{\bsnm{{Lian}}, \binits{G.}},
\bauthor{\bsnm{{Nan}}, \binits{D.}},
\bauthor{\bsnm{{Bai}}, \binits{X.X.}},
\bauthor{\bsnm{{Liu}}, \binits{W.P.}}:
\batitle{{Indirect measurement of the 57.7 keV resonance strength for the
  astrophysical {\ensuremath{\gamma}}-ray source of the
  $^{25}$Mg(p,{\ensuremath{\gamma}})$^{26}$Al reaction}}.
\bjtitle{\prc}
\bvolume{102}(\bissue{2}),
\bfpage{025804}
(\byear{2020})
\doiurl{10.1103/PhysRevC.102.025804}
\end{barticle}
\endbibitem

%%% 462
\bibitem[\protect\citeauthoryear{{Vogelaar} et~al.}{1996}]{Vogelaar1996}
\begin{barticle}
\bauthor{\bsnm{{Vogelaar}}, \binits{R.B.}},
\bauthor{\bsnm{{Mitchell}}, \binits{L.W.}},
\bauthor{\bsnm{{Kavanagh}}, \binits{R.W.}},
\bauthor{\bsnm{{Champagne}}, \binits{A.E.}},
\bauthor{\bsnm{{Magnus}}, \binits{P.V.}},
\bauthor{\bsnm{{Smith}}, \binits{M.S.}},
\bauthor{\bsnm{{Howard}}, \binits{A.J.}},
\bauthor{\bsnm{{Parker}}, \binits{P.D.}},
\bauthor{\bsnm{{O'brien}}, \binits{H.A.}}:
\batitle{{Constraining $^{26}$Al+p resonances using
  $^{26}$Al($^{3}$He,d)$^{27}$Si}}.
\bjtitle{\prc}
\bvolume{53}(\bissue{4}),
\bfpage{1945}--\blpage{1949}
(\byear{1996})
\doiurl{10.1103/PhysRevC.53.1945}
\end{barticle}
\endbibitem

%%% 463
\bibitem[\protect\citeauthoryear{{Pain} et~al.}{2015}]{Pain2015}
\begin{barticle}
\bauthor{\bsnm{{Pain}}, \binits{S.D.}},
\bauthor{\bsnm{{Bardayan}}, \binits{D.W.}},
\bauthor{\bsnm{{Blackmon}}, \binits{J.C.}},
\bauthor{\bsnm{{Brown}}, \binits{S.M.}},
\bauthor{\bsnm{{Chae}}, \binits{K.Y.}},
\bauthor{\bsnm{{Chipps}}, \binits{K.A.}},
\bauthor{\bsnm{{Cizewski}}, \binits{J.A.}},
\bauthor{\bsnm{{Jones}}, \binits{K.L.}},
\bauthor{\bsnm{{Kozub}}, \binits{R.L.}},
\bauthor{\bsnm{{Liang}}, \binits{J.F.}},
\bauthor{\bsnm{{Matei}}, \binits{C.}},
\bauthor{\bsnm{{Matos}}, \binits{M.}},
\bauthor{\bsnm{{Moazen}}, \binits{B.H.}},
\bauthor{\bsnm{{Nesaraja}}, \binits{C.D.}},
\bauthor{\bsnm{{Oko{\l}owicz}}, \binits{J.}},
\bauthor{\bsnm{{O'Malley}}, \binits{P.D.}},
\bauthor{\bsnm{{Peters}}, \binits{W.A.}},
\bauthor{\bsnm{{Pittman}}, \binits{S.T.}},
\bauthor{\bsnm{{P{\l}oszajczak}}, \binits{M.}},
\bauthor{\bsnm{{Schmitt}}, \binits{K.T.}},
\bauthor{\bsnm{{Shriner}}, \binits{J.F.}},
\bauthor{\bsnm{{Shapira}}, \binits{D.}},
\bauthor{\bsnm{{Smith}}, \binits{M.S.}},
\bauthor{\bsnm{{Stracener}}, \binits{D.W.}},
\bauthor{\bsnm{{Wilson}}, \binits{G.L.}}:
\batitle{{Constraint of the Astrophysical $^{Al}$g$_{26}$(p
  ,{\ensuremath{\gamma}} )<mml:mmultiscripts>Si 27 </mml:mmultiscripts>
  Destruction Rate at Stellar Temperatures}}.
\bjtitle{\prl}
\bvolume{114}(\bissue{21}),
\bfpage{212501}
(\byear{2015})
\doiurl{10.1103/PhysRevLett.114.212501}
\end{barticle}
\endbibitem

%%% 464
\bibitem[\protect\citeauthoryear{{Lotay} et~al.}{2020}]{Lotay2020}
\begin{barticle}
\bauthor{\bsnm{{Lotay}}, \binits{G.}},
\bauthor{\bsnm{{Woods}}, \binits{P.J.}},
\bauthor{\bsnm{{Moukaddam}}, \binits{M.}},
\bauthor{\bsnm{{Aliotta}}, \binits{M.}},
\bauthor{\bsnm{{Christian}}, \binits{G.}},
\bauthor{\bsnm{{Davids}}, \binits{B.}},
\bauthor{\bsnm{{Davinson}}, \binits{T.}},
\bauthor{\bsnm{{Doherty}}, \binits{D.T.}},
\bauthor{\bsnm{{Howell}}, \binits{D.}},
\bauthor{\bsnm{{Margerin}}, \binits{V.}},
\bauthor{\bsnm{{Ruiz}}, \binits{C.}}:
\batitle{{High-resolution radioactive beam study of the $^{26}$Al(d,p) reaction
  and measurements of single-particle spectroscopic factors}}.
\bjtitle{European Physical Journal A}
\bvolume{56}(\bissue{1}),
\bfpage{3}
(\byear{2020})
\doiurl{10.1140/epja/s10050-019-00008-8}
\end{barticle}
\endbibitem

%%% 465
\bibitem[\protect\citeauthoryear{{Ruiz} et~al.}{2006}]{Ruiz2006}
\begin{barticle}
\bauthor{\bsnm{{Ruiz}}, \binits{C.}},
\bauthor{\bsnm{{Parikh}}, \binits{A.}},
\bauthor{\bsnm{{Jos{\'e}}}, \binits{J.}},
\bauthor{\bsnm{{Buchmann}}, \binits{L.}},
\bauthor{\bsnm{{Caggiano}}, \binits{J.A.}},
\bauthor{\bsnm{{Chen}}, \binits{A.A.}},
\bauthor{\bsnm{{Clark}}, \binits{J.A.}},
\bauthor{\bsnm{{Crawford}}, \binits{H.}},
\bauthor{\bsnm{{Davids}}, \binits{B.}},
\bauthor{\bsnm{{D'Auria}}, \binits{J.M.}},
\bauthor{\bsnm{{Davis}}, \binits{C.}},
\bauthor{\bsnm{{Deibel}}, \binits{C.}},
\bauthor{\bsnm{{Erikson}}, \binits{L.}},
\bauthor{\bsnm{{Fogarty}}, \binits{L.}},
\bauthor{\bsnm{{Frekers}}, \binits{D.}},
\bauthor{\bsnm{{Greife}}, \binits{U.}},
\bauthor{\bsnm{{Hussein}}, \binits{A.}},
\bauthor{\bsnm{{Hutcheon}}, \binits{D.A.}},
\bauthor{\bsnm{{Huyse}}, \binits{M.}},
\bauthor{\bsnm{{Jewett}}, \binits{C.}},
\bauthor{\bsnm{{Laird}}, \binits{A.M.}},
\bauthor{\bsnm{{Lewis}}, \binits{R.}},
\bauthor{\bsnm{{Mumby-Croft}}, \binits{P.}},
\bauthor{\bsnm{{Olin}}, \binits{A.}},
\bauthor{\bsnm{{Ottewell}}, \binits{D.F.}},
\bauthor{\bsnm{{Ouellet}}, \binits{C.V.}},
\bauthor{\bsnm{{Parker}}, \binits{P.}},
\bauthor{\bsnm{{Pearson}}, \binits{J.}},
\bauthor{\bsnm{{Ruprecht}}, \binits{G.}},
\bauthor{\bsnm{{Trinczek}}, \binits{M.}},
\bauthor{\bsnm{{Vockenhuber}}, \binits{C.}},
\bauthor{\bsnm{{Wrede}}, \binits{C.}}:
\batitle{{Measurement of the E$_{c.m.}$=184keV Resonance Strength in the
  $^{26g}$Al(p,{\ensuremath{\gamma}})$^{27}$Si Reaction}}.
\bjtitle{\prl}
\bvolume{96}(\bissue{25}),
\bfpage{252501}
(\byear{2006})
\doiurl{10.1103/PhysRevLett.96.252501}
\end{barticle}
\endbibitem

%%% 466
\bibitem[\protect\citeauthoryear{{Lotay} et~al.}{2009}]{Lotay2009}
\begin{barticle}
\bauthor{\bsnm{{Lotay}}, \binits{G.}},
\bauthor{\bsnm{{Woods}}, \binits{P.J.}},
\bauthor{\bsnm{{Seweryniak}}, \binits{D.}},
\bauthor{\bsnm{{Carpenter}}, \binits{M.P.}},
\bauthor{\bsnm{{Janssens}}, \binits{R.V.F.}},
\bauthor{\bsnm{{Zhu}}, \binits{S.}}:
\batitle{{Identification of Key Astrophysical Resonances Relevant for the
  $^{26g}$Al(p,{\ensuremath{\gamma}})$^{27}$Si Reaction in Wolf-Rayet Stars,
  AGB stars, and Classical Novae}}.
\bjtitle{\prl}
\bvolume{102}(\bissue{16}),
\bfpage{162502}
(\byear{2009})
\doiurl{10.1103/PhysRevLett.102.162502}
\end{barticle}
\endbibitem

%%% 467
\bibitem[\protect\citeauthoryear{{Lederer-Woods} et~al.}{2021a}]{Lederer2021a}
\begin{barticle}
\bauthor{\bsnm{{Lederer-Woods}}, \binits{C.}},
\bauthor{\bsnm{{Woods}}, \binits{P.J.}},
\bauthor{\bsnm{{Davinson}}, \binits{T.}},
\bauthor{\bsnm{{Kahl}}, \binits{D.}},
\bauthor{\bsnm{{Lonsdale}}, \binits{S.J.}},
\bauthor{\bsnm{{Aberle}}, \binits{O.}},
\bauthor{\bsnm{{Amaducci}}, \binits{S.}},
\bauthor{\bsnm{{Andrzejewski}}, \binits{J.}},
\bauthor{\bsnm{{Audouin}}, \binits{L.}},
\bauthor{\bsnm{{Bacak}}, \binits{M.}},
\bauthor{\bsnm{{Balibrea}}, \binits{J.}},
\bauthor{\bsnm{{Barbagallo}}, \binits{M.}},
\bauthor{\bsnm{{Be{\v{c}}v{\'a}{\v{r}}}}, \binits{F.}},
\bauthor{\bsnm{{Berthoumieux}}, \binits{E.}},
\bauthor{\bsnm{{Billowes}}, \binits{J.}},
\bauthor{\bsnm{{Bosnar}}, \binits{D.}},
\bauthor{\bsnm{{Brown}}, \binits{A.}},
\bauthor{\bsnm{{Caama{\~n}o}}, \binits{M.}},
\bauthor{\bsnm{{Calvi{\~n}o}}, \binits{F.}},
\bauthor{\bsnm{{Calviani}}, \binits{M.}},
\bauthor{\bsnm{{Cano-Ott}}, \binits{D.}},
\bauthor{\bsnm{{Cardella}}, \binits{R.}},
\bauthor{\bsnm{{Casanovas}}, \binits{A.}},
\bauthor{\bsnm{{Cerutti}}, \binits{F.}},
\bauthor{\bsnm{{Chen}}, \binits{Y.H.}},
\bauthor{\bsnm{{Chiaveri}}, \binits{E.}},
\bauthor{\bsnm{{Colonna}}, \binits{N.}},
\bauthor{\bsnm{{Cort{\'e}s}}, \binits{G.}},
\bauthor{\bsnm{{Cort{\'e}s-Giraldo}}, \binits{M.A.}},
\bauthor{\bsnm{{Cosentino}}, \binits{L.}},
\bauthor{\bsnm{{Cristallo}}, \binits{S.}},
\bauthor{\bsnm{{Damone}}, \binits{L.A.}},
\bauthor{\bsnm{{Diakaki}}, \binits{M.}},
\bauthor{\bsnm{{Domingo-Pardo}}, \binits{C.}},
\bauthor{\bsnm{{Dressler}}, \binits{R.}},
\bauthor{\bsnm{{Dupont}}, \binits{E.}},
\bauthor{\bsnm{{Dur{\'a}n}}, \binits{I.}},
\bauthor{\bsnm{{Fern{\'a}ndez-Dom{\'\i}nguez}}, \binits{B.}},
\bauthor{\bsnm{{Ferrari}}, \binits{A.}},
\bauthor{\bsnm{{Ferreira}}, \binits{P.}},
\bauthor{\bsnm{{Ferrer}}, \binits{F.J.}},
\bauthor{\bsnm{{Finocchiaro}}, \binits{P.}},
\bauthor{\bsnm{{Furman}}, \binits{V.}},
\bauthor{\bsnm{{G{\"o}bel}}, \binits{K.}},
\bauthor{\bsnm{{Garc{\'\i}a}}, \binits{A.R.}},
\bauthor{\bsnm{{Gawlik}}, \binits{A.}},
\bauthor{\bsnm{{Gilardoni}}, \binits{S.}},
\bauthor{\bsnm{{Glodariu}}, \binits{T.}},
\bauthor{\bsnm{{Gon{\c{c}}alves}}, \binits{I.F.}},
\bauthor{\bsnm{{Gonz{\'a}lez-Romero}}, \binits{E.}},
\bauthor{\bsnm{{Griesmayer}}, \binits{E.}},
\bauthor{\bsnm{{Guerrero}}, \binits{C.}},
\bauthor{\bsnm{{Gunsing}}, \binits{F.}},
\bauthor{\bsnm{{Harada}}, \binits{H.}},
\bauthor{\bsnm{{Heinitz}}, \binits{S.}},
\bauthor{\bsnm{{Heyse}}, \binits{J.}},
\bauthor{\bsnm{{Jenkins}}, \binits{D.G.}},
\bauthor{\bsnm{{Jericha}}, \binits{E.}},
\bauthor{\bsnm{{K{\"a}ppeler}}, \binits{F.}},
\bauthor{\bsnm{{Kadi}}, \binits{Y.}},
\bauthor{\bsnm{{Kalamara}}, \binits{A.}},
\bauthor{\bsnm{{Kavrigin}}, \binits{P.}},
\bauthor{\bsnm{{Kimura}}, \binits{A.}},
\bauthor{\bsnm{{Kivel}}, \binits{N.}},
\bauthor{\bsnm{{Kokkoris}}, \binits{M.}},
\bauthor{\bsnm{{Krti{\v{c}}ka}}, \binits{M.}},
\bauthor{\bsnm{{Kurtulgil}}, \binits{D.}},
\bauthor{\bsnm{{Leal-Cidoncha}}, \binits{E.}},
\bauthor{\bsnm{{Leeb}}, \binits{H.}},
\bauthor{\bsnm{{Lerendegui-Marco}}, \binits{J.}},
\bauthor{\bsnm{{Lo Meo}}, \binits{S.}},
\bauthor{\bsnm{{Macina}}, \binits{D.}},
\bauthor{\bsnm{{Manna}}, \binits{A.}},
\bauthor{\bsnm{{Marganiec}}, \binits{J.}},
\bauthor{\bsnm{{Mart{\'\i}nez}}, \binits{T.}},
\bauthor{\bsnm{{Masi}}, \binits{A.}},
\bauthor{\bsnm{{Massimi}}, \binits{C.}},
\bauthor{\bsnm{{Mastinu}}, \binits{P.}},
\bauthor{\bsnm{{Mastromarco}}, \binits{M.}},
\bauthor{\bsnm{{Maugeri}}, \binits{E.A.}},
\bauthor{\bsnm{{Mazzone}}, \binits{A.}},
\bauthor{\bsnm{{Mendoza}}, \binits{E.}},
\bauthor{\bsnm{{Mengoni}}, \binits{A.}},
\bauthor{\bsnm{{Milazzo}}, \binits{P.M.}},
\bauthor{\bsnm{{Mingrone}}, \binits{F.}},
\bauthor{\bsnm{{Musumarra}}, \binits{A.}},
\bauthor{\bsnm{{Negret}}, \binits{A.}},
\bauthor{\bsnm{{Nolte}}, \binits{R.}},
\bauthor{\bsnm{{Oprea}}, \binits{A.}},
\bauthor{\bsnm{{Patronis}}, \binits{N.}},
\bauthor{\bsnm{{Pavlik}}, \binits{A.}},
\bauthor{\bsnm{{Perkowski}}, \binits{J.}},
\bauthor{\bsnm{{Porras}}, \binits{I.}},
\bauthor{\bsnm{{Praena}}, \binits{J.}},
\bauthor{\bsnm{{Quesada}}, \binits{J.M.}},
\bauthor{\bsnm{{Radeck}}, \binits{D.}},
\bauthor{\bsnm{{Rauscher}}, \binits{T.}},
\bauthor{\bsnm{{Reifarth}}, \binits{R.}},
\bauthor{\bsnm{{Rubbia}}, \binits{C.}},
\bauthor{\bsnm{{Ryan}}, \binits{J.A.}},
\bauthor{\bsnm{{Sabat{\'e}-Gilarte}}, \binits{M.}},
\bauthor{\bsnm{{Saxena}}, \binits{A.}},
\bauthor{\bsnm{{Schillebeeckx}}, \binits{P.}},
\bauthor{\bsnm{{Schumann}}, \binits{D.}},
\bauthor{\bsnm{{Sedyshev}}, \binits{P.}},
\bauthor{\bsnm{{Smith}}, \binits{A.G.}},
\bauthor{\bsnm{{Sosnin}}, \binits{N.V.}},
\bauthor{\bsnm{{Stamatopoulos}}, \binits{A.}},
\bauthor{\bsnm{{Tagliente}}, \binits{G.}},
\bauthor{\bsnm{{Tain}}, \binits{J.L.}},
\bauthor{\bsnm{{Tarife{\~n}o-Saldivia}}, \binits{A.}},
\bauthor{\bsnm{{Tassan-Got}}, \binits{L.}},
\bauthor{\bsnm{{Valenta}}, \binits{S.}},
\bauthor{\bsnm{{Vannini}}, \binits{G.}},
\bauthor{\bsnm{{Variale}}, \binits{V.}},
\bauthor{\bsnm{{Vaz}}, \binits{P.}},
\bauthor{\bsnm{{Ventura}}, \binits{A.}},
\bauthor{\bsnm{{Vlachoudis}}, \binits{V.}},
\bauthor{\bsnm{{Vlastou}}, \binits{R.}},
\bauthor{\bsnm{{Wallner}}, \binits{A.}},
\bauthor{\bsnm{{Warren}}, \binits{S.}},
\bauthor{\bsnm{{Weiss}}, \binits{C.}},
\bauthor{\bsnm{{Wright}}, \binits{T.}},
\bauthor{\bsnm{{{\v{Z}}ugec}}, \binits{P.}},
\bauthor{\bsnm{{n TOF Collaboration}}}:
\batitle{{Destruction of the cosmic {\ensuremath{\gamma}}-ray emitter $^{26}$Al
  in massive stars: Study of the key $^{26}$Al(n,p) reaction}}.
\bjtitle{\prc}
\bvolume{104}(\bissue{2}),
\bfpage{022803}
(\byear{2021})
\doiurl{10.1103/PhysRevC.104.L022803}
\end{barticle}
\endbibitem

%%% 468
\bibitem[\protect\citeauthoryear{{Lederer-Woods} et~al.}{2021b}]{Lederer2021b}
\begin{barticle}
\bauthor{\bsnm{{Lederer-Woods}}, \binits{C.}},
\bauthor{\bsnm{{Woods}}, \binits{P.J.}},
\bauthor{\bsnm{{Davinson}}, \binits{T.}},
\bauthor{\bsnm{{Estrade}}, \binits{A.}},
\bauthor{\bsnm{{Heyse}}, \binits{J.}},
\bauthor{\bsnm{{Kahl}}, \binits{D.}},
\bauthor{\bsnm{{Lonsdale}}, \binits{S.J.}},
\bauthor{\bsnm{{Paradela}}, \binits{C.}},
\bauthor{\bsnm{{Schillebeeckx}}, \binits{P.}},
\bauthor{\bsnm{{Aberle}}, \binits{O.}},
\bauthor{\bsnm{{Amaducci}}, \binits{S.}},
\bauthor{\bsnm{{Andrzejewski}}, \binits{J.}},
\bauthor{\bsnm{{Audouin}}, \binits{L.}},
\bauthor{\bsnm{{Bacak}}, \binits{M.}},
\bauthor{\bsnm{{Balibrea}}, \binits{J.}},
\bauthor{\bsnm{{Barbagallo}}, \binits{M.}},
\bauthor{\bsnm{{Be{\v{c}}v{\'a}{\v{r}}}}, \binits{F.}},
\bauthor{\bsnm{{Berthoumieux}}, \binits{E.}},
\bauthor{\bsnm{{Billowes}}, \binits{J.}},
\bauthor{\bsnm{{Bosnar}}, \binits{D.}},
\bauthor{\bsnm{{Brown}}, \binits{A.}},
\bauthor{\bsnm{{Caama{\~n}o}}, \binits{M.}},
\bauthor{\bsnm{{Calvi{\~n}o}}, \binits{F.}},
\bauthor{\bsnm{{Calviani}}, \binits{M.}},
\bauthor{\bsnm{{Cano-Ott}}, \binits{D.}},
\bauthor{\bsnm{{Cardella}}, \binits{R.}},
\bauthor{\bsnm{{Casanovas}}, \binits{A.}},
\bauthor{\bsnm{{Cerutti}}, \binits{F.}},
\bauthor{\bsnm{{Chen}}, \binits{Y.H.}},
\bauthor{\bsnm{{Chiaveri}}, \binits{E.}},
\bauthor{\bsnm{{Colonna}}, \binits{N.}},
\bauthor{\bsnm{{Cort{\'e}s}}, \binits{G.}},
\bauthor{\bsnm{{Cort{\'e}s-Giraldo}}, \binits{M.A.}},
\bauthor{\bsnm{{Cosentino}}, \binits{L.}},
\bauthor{\bsnm{{Cristallo}}, \binits{S.}},
\bauthor{\bsnm{{Damone}}, \binits{L.A.}},
\bauthor{\bsnm{{Diakaki}}, \binits{M.}},
\bauthor{\bsnm{{Domingo-Pardo}}, \binits{C.}},
\bauthor{\bsnm{{Dressler}}, \binits{R.}},
\bauthor{\bsnm{{Dupont}}, \binits{E.}},
\bauthor{\bsnm{{Dur{\'a}n}}, \binits{I.}},
\bauthor{\bsnm{{Fern{\'a}ndez-Dom{\'\i}nguez}}, \binits{B.}},
\bauthor{\bsnm{{Ferrari}}, \binits{A.}},
\bauthor{\bsnm{{Ferreira}}, \binits{P.}},
\bauthor{\bsnm{{Ferrer}}, \binits{F.J.}},
\bauthor{\bsnm{{Finocchiaro}}, \binits{P.}},
\bauthor{\bsnm{{Furman}}, \binits{V.}},
\bauthor{\bsnm{{G{\"o}bel}}, \binits{K.}},
\bauthor{\bsnm{{Garc{\'\i}a}}, \binits{A.R.}},
\bauthor{\bsnm{{Gawlik}}, \binits{A.}},
\bauthor{\bsnm{{Gilardoni}}, \binits{S.}},
\bauthor{\bsnm{{Glodariu}}, \binits{T.}},
\bauthor{\bsnm{{Gon{\c{c}}alves}}, \binits{I.F.}},
\bauthor{\bsnm{{Gonz{\'a}lez-Romero}}, \binits{E.}},
\bauthor{\bsnm{{Griesmayer}}, \binits{E.}},
\bauthor{\bsnm{{Guerrero}}, \binits{C.}},
\bauthor{\bsnm{{Gunsing}}, \binits{F.}},
\bauthor{\bsnm{{Harada}}, \binits{H.}},
\bauthor{\bsnm{{Heinitz}}, \binits{S.}},
\bauthor{\bsnm{{Jenkins}}, \binits{D.G.}},
\bauthor{\bsnm{{Jericha}}, \binits{E.}},
\bauthor{\bsnm{{K{\"a}ppeler}}, \binits{F.}},
\bauthor{\bsnm{{Kadi}}, \binits{Y.}},
\bauthor{\bsnm{{Kalamara}}, \binits{A.}},
\bauthor{\bsnm{{Kavrigin}}, \binits{P.}},
\bauthor{\bsnm{{Kimura}}, \binits{A.}},
\bauthor{\bsnm{{Kivel}}, \binits{N.}},
\bauthor{\bsnm{{Kokkoris}}, \binits{M.}},
\bauthor{\bsnm{{Krti{\v{c}}ka}}, \binits{M.}},
\bauthor{\bsnm{{Kurtulgil}}, \binits{D.}},
\bauthor{\bsnm{{Leal-Cidoncha}}, \binits{E.}},
\bauthor{\bsnm{{Leeb}}, \binits{H.}},
\bauthor{\bsnm{{Lerendegui-Marco}}, \binits{J.}},
\bauthor{\bsnm{{Lo Meo}}, \binits{S.}},
\bauthor{\bsnm{{Macina}}, \binits{D.}},
\bauthor{\bsnm{{Manna}}, \binits{A.}},
\bauthor{\bsnm{{Marganiec}}, \binits{J.}},
\bauthor{\bsnm{{Mart{\'\i}nez}}, \binits{T.}},
\bauthor{\bsnm{{Masi}}, \binits{A.}},
\bauthor{\bsnm{{Massimi}}, \binits{C.}},
\bauthor{\bsnm{{Mastinu}}, \binits{P.}},
\bauthor{\bsnm{{Mastromarco}}, \binits{M.}},
\bauthor{\bsnm{{Maugeri}}, \binits{E.A.}},
\bauthor{\bsnm{{Mazzone}}, \binits{A.}},
\bauthor{\bsnm{{Mendoza}}, \binits{E.}},
\bauthor{\bsnm{{Mengoni}}, \binits{A.}},
\bauthor{\bsnm{{Milazzo}}, \binits{P.M.}},
\bauthor{\bsnm{{Mingrone}}, \binits{F.}},
\bauthor{\bsnm{{Musumarra}}, \binits{A.}},
\bauthor{\bsnm{{Negret}}, \binits{A.}},
\bauthor{\bsnm{{Nolte}}, \binits{R.}},
\bauthor{\bsnm{{Oprea}}, \binits{A.}},
\bauthor{\bsnm{{Patronis}}, \binits{N.}},
\bauthor{\bsnm{{Pavlik}}, \binits{A.}},
\bauthor{\bsnm{{Perkowski}}, \binits{J.}},
\bauthor{\bsnm{{Porras}}, \binits{I.}},
\bauthor{\bsnm{{Praena}}, \binits{J.}},
\bauthor{\bsnm{{Quesada}}, \binits{J.M.}},
\bauthor{\bsnm{{Radeck}}, \binits{D.}},
\bauthor{\bsnm{{Rauscher}}, \binits{T.}},
\bauthor{\bsnm{{Reifarth}}, \binits{R.}},
\bauthor{\bsnm{{Rubbia}}, \binits{C.}},
\bauthor{\bsnm{{Ryan}}, \binits{J.A.}},
\bauthor{\bsnm{{Sabat{\'e}-Gilarte}}, \binits{M.}},
\bauthor{\bsnm{{Saxena}}, \binits{A.}},
\bauthor{\bsnm{{Schumann}}, \binits{D.}},
\bauthor{\bsnm{{Sedyshev}}, \binits{P.}},
\bauthor{\bsnm{{Smith}}, \binits{A.G.}},
\bauthor{\bsnm{{Sosnin}}, \binits{N.V.}},
\bauthor{\bsnm{{Stamatopoulos}}, \binits{A.}},
\bauthor{\bsnm{{Tagliente}}, \binits{G.}},
\bauthor{\bsnm{{Tain}}, \binits{J.L.}},
\bauthor{\bsnm{{Tarife{\~n}o-Saldivia}}, \binits{A.}},
\bauthor{\bsnm{{Tassan-Got}}, \binits{L.}},
\bauthor{\bsnm{{Valenta}}, \binits{S.}},
\bauthor{\bsnm{{Vannini}}, \binits{G.}},
\bauthor{\bsnm{{Variale}}, \binits{V.}},
\bauthor{\bsnm{{Vaz}}, \binits{P.}},
\bauthor{\bsnm{{Ventura}}, \binits{A.}},
\bauthor{\bsnm{{Vlachoudis}}, \binits{V.}},
\bauthor{\bsnm{{Vlastou}}, \binits{R.}},
\bauthor{\bsnm{{Wallner}}, \binits{A.}},
\bauthor{\bsnm{{Warren}}, \binits{S.}},
\bauthor{\bsnm{{Weiss}}, \binits{C.}},
\bauthor{\bsnm{{Wright}}, \binits{T.}},
\bauthor{\bsnm{{{\v{Z}}ugec}}, \binits{P.}},
\bauthor{\bsnm{{n TOF Collaboration}}}:
\batitle{{Destruction of the cosmic {\ensuremath{\gamma}}-ray emitter $^{26}$Al
  in massive stars: Study of the key $^{26}$Al(n,{\ensuremath{\alpha}})
  reaction}}.
\bjtitle{\prc}
\bvolume{104}(\bissue{3}),
\bfpage{032803}
(\byear{2021})
\doiurl{10.1103/PhysRevC.104.L032803}
\end{barticle}
\endbibitem

%%% 469
\bibitem[\protect\citeauthoryear{{Benamara} et~al.}{2014}]{Benamara2014}
\begin{barticle}
\bauthor{\bsnm{{Benamara}}, \binits{S.}},
\bauthor{\bsnm{{de S{\'e}r{\'e}ville}}, \binits{N.}},
\bauthor{\bsnm{{Laird}}, \binits{A.M.}},
\bauthor{\bsnm{{Hammache}}, \binits{F.}},
\bauthor{\bsnm{{Stefan}}, \binits{I.}},
\bauthor{\bsnm{{Roussel}}, \binits{P.}},
\bauthor{\bsnm{{Ancelin}}, \binits{S.}},
\bauthor{\bsnm{{Assi{\'e}}}, \binits{M.}},
\bauthor{\bsnm{{Coc}}, \binits{A.}},
\bauthor{\bsnm{{Deloncle}}, \binits{I.}},
\bauthor{\bsnm{{Fox}}, \binits{S.P.}},
\bauthor{\bsnm{{Kiener}}, \binits{J.}},
\bauthor{\bsnm{{Lefebvre}}, \binits{L.}},
\bauthor{\bsnm{{Lefebvre-Schuhl}}, \binits{A.}},
\bauthor{\bsnm{{Mavilla}}, \binits{G.}},
\bauthor{\bsnm{{Morfouace}}, \binits{P.}},
\bauthor{\bsnm{{S{\'a}nchez-Ben{\'\i}tez}}, \binits{{\'A}.M.}},
\bauthor{\bsnm{{Perrot}}, \binits{L.}},
\bauthor{\bsnm{{Sinha}}, \binits{M.}},
\bauthor{\bsnm{{Tatischeff}}, \binits{V.}},
\bauthor{\bsnm{{Vandebrouck}}, \binits{M.}}:
\batitle{{Nucleosynthesis of Al26 in massive stars: New Al27 states above
  {\ensuremath{\alpha}} and neutron emission thresholds}}.
\bjtitle{\prc}
\bvolume{89}(\bissue{6}),
\bfpage{065805}
(\byear{2014})
\doiurl{10.1103/PhysRevC.89.065805}
\end{barticle}
\endbibitem

%%% 470
\bibitem[\protect\citeauthoryear{{Diehl} et~al.}{2021}]{Diehl2021}
\begin{barticle}
\bauthor{\bsnm{{Diehl}}, \binits{R.}},
\bauthor{\bsnm{{Lugaro}}, \binits{M.}},
\bauthor{\bsnm{{Heger}}, \binits{A.}},
\bauthor{\bsnm{{Sieverding}}, \binits{A.}},
\bauthor{\bsnm{{Tang}}, \binits{X.}},
\bauthor{\bsnm{{Li}}, \binits{K.A.}},
\bauthor{\bsnm{{Li}}, \binits{E.T.}},
\bauthor{\bsnm{{Doherty}}, \binits{C.L.}},
\bauthor{\bsnm{{Krause}}, \binits{M.G.H.}},
\bauthor{\bsnm{{Wallner}}, \binits{A.}},
\bauthor{\bsnm{{Prantzos}}, \binits{N.}},
\bauthor{\bsnm{{Brinkman}}, \binits{H.E.}},
\bauthor{\bsnm{{den Hartogh}}, \binits{J.W.}},
\bauthor{\bsnm{{Wehmeyer}}, \binits{B.}},
\bauthor{\bsnm{{Yag{\"u}e L{\'o}pez}}, \binits{A.}},
\bauthor{\bsnm{{Pleintinger}}, \binits{M.M.M.}},
\bauthor{\bsnm{{Banerjee}}, \binits{P.}},
\bauthor{\bsnm{{Wang}}, \binits{W.}}:
\batitle{{The radioactive nuclei $^{26}$Al and $^{60}$Fe in the Cosmos and in
  the solar system}}.
\bjtitle{\pasa}
\bvolume{38},
\bfpage{062}
(\byear{2021})
\doiurl{10.1017/pasa.2021.48}
{\href{https://arxiv.org/abs/2109.08558}{{arXiv:2109.08558}}}
{[astro-ph.HE]}
\end{barticle}
\endbibitem

%%% 471
\bibitem[\protect\citeauthoryear{{Laird} et~al.}{2023}]{Laird2023}
\begin{barticle}
\bauthor{\bsnm{{Laird}}, \binits{A.M.}},
\bauthor{\bsnm{{Lugaro}}, \binits{M.}},
\bauthor{\bsnm{{Kankainen}}, \binits{A.}},
\bauthor{\bsnm{{Adsley}}, \binits{P.}},
\bauthor{\bsnm{{Bardayan}}, \binits{D.W.}},
\bauthor{\bsnm{{Brinkman}}, \binits{H.E.}},
\bauthor{\bsnm{{C{\^o}t{\'e}}}, \binits{B.}},
\bauthor{\bsnm{{Deibel}}, \binits{C.M.}},
\bauthor{\bsnm{{Diehl}}, \binits{R.}},
\bauthor{\bsnm{{Hammache}}, \binits{F.}},
\bauthor{\bsnm{{den Hartogh}}, \binits{J.W.}},
\bauthor{\bsnm{{Jos{\'e}}}, \binits{J.}},
\bauthor{\bsnm{{Kurtulgil}}, \binits{D.}},
\bauthor{\bsnm{{Lederer-Woods}}, \binits{C.}},
\bauthor{\bsnm{{Lotay}}, \binits{G.}},
\bauthor{\bsnm{{Meynet}}, \binits{G.}},
\bauthor{\bsnm{{Palmerini}}, \binits{S.}},
\bauthor{\bsnm{{Pignatari}}, \binits{M.}},
\bauthor{\bsnm{{Reifarth}}, \binits{R.}},
\bauthor{\bsnm{{de S{\'e}r{\'e}ville}}, \binits{N.}},
\bauthor{\bsnm{{Sieverding}}, \binits{A.}},
\bauthor{\bsnm{{Stancliffe}}, \binits{R.J.}},
\bauthor{\bsnm{{Trueman}}, \binits{T.C.L.}},
\bauthor{\bsnm{{Lawson}}, \binits{T.}},
\bauthor{\bsnm{{Vink}}, \binits{J.S.}},
\bauthor{\bsnm{{Massimi}}, \binits{C.}},
\bauthor{\bsnm{{Mengoni}}, \binits{A.}}:
\batitle{{Progress on nuclear reaction rates affecting the stellar production
  of $^{26}$Al}}.
\bjtitle{Journal of Physics G Nuclear Physics}
\bvolume{50}(\bissue{3}),
\bfpage{033002}
(\byear{2023})
\doiurl{10.1088/1361-6471/ac9cf8}
\end{barticle}
\endbibitem

%%% 472
\bibitem[\protect\citeauthoryear{{Uberseder} et~al.}{2009}]{Uberseder2009}
\begin{barticle}
\bauthor{\bsnm{{Uberseder}}, \binits{E.}},
\bauthor{\bsnm{{Reifarth}}, \binits{R.}},
\bauthor{\bsnm{{Schumann}}, \binits{D.}},
\bauthor{\bsnm{{Dillmann}}, \binits{I.}},
\bauthor{\bsnm{{Pardo}}, \binits{C.D.}},
\bauthor{\bsnm{{G{\"o}rres}}, \binits{J.}},
\bauthor{\bsnm{{Heil}}, \binits{M.}},
\bauthor{\bsnm{{K{\"a}ppeler}}, \binits{F.}},
\bauthor{\bsnm{{Marganiec}}, \binits{J.}},
\bauthor{\bsnm{{Neuhausen}}, \binits{J.}},
\bauthor{\bsnm{{Pignatari}}, \binits{M.}},
\bauthor{\bsnm{{Voss}}, \binits{F.}},
\bauthor{\bsnm{{Walter}}, \binits{S.}},
\bauthor{\bsnm{{Wiescher}}, \binits{M.}}:
\batitle{{Measurement of the $^{60}$Fe(n,{\ensuremath{\gamma}})$^{61}$Fe Cross
  Section at Stellar Temperatures}}.
\bjtitle{\prl}
\bvolume{102}(\bissue{15}),
\bfpage{151101}
(\byear{2009})
\doiurl{10.1103/PhysRevLett.102.151101}
\end{barticle}
\endbibitem

%%% 473
\bibitem[\protect\citeauthoryear{{Giron} et~al.}{2017}]{Giron2017}
\begin{barticle}
\bauthor{\bsnm{{Giron}}, \binits{S.}},
\bauthor{\bsnm{{Hammache}}, \binits{F.}},
\bauthor{\bsnm{{de S{\'e}r{\'e}ville}}, \binits{N.}},
\bauthor{\bsnm{{Roussel}}, \binits{P.}},
\bauthor{\bsnm{{Burgunder}}, \binits{J.}},
\bauthor{\bsnm{{Moukaddam}}, \binits{M.}},
\bauthor{\bsnm{{Beaumel}}, \binits{D.}},
\bauthor{\bsnm{{Caceres}}, \binits{L.}},
\bauthor{\bsnm{{Duch{\^e}ne}}, \binits{G.}},
\bauthor{\bsnm{{Cl{\'e}ment}}, \binits{E.}},
\bauthor{\bsnm{{Fernandez-Dominguez}}, \binits{B.}},
\bauthor{\bsnm{{Flavigny}}, \binits{F.}},
\bauthor{\bsnm{{de France}}, \binits{G.}},
\bauthor{\bsnm{{Franchoo}}, \binits{S.}},
\bauthor{\bsnm{{Galaviz-Redondo}}, \binits{D.}},
\bauthor{\bsnm{{Gasques}}, \binits{L.}},
\bauthor{\bsnm{{Gibelin}}, \binits{J.}},
\bauthor{\bsnm{{Gillibert}}, \binits{A.}},
\bauthor{\bsnm{{Grevy}}, \binits{S.}},
\bauthor{\bsnm{{Guillot}}, \binits{J.}},
\bauthor{\bsnm{{Heil}}, \binits{M.}},
\bauthor{\bsnm{{Kiener}}, \binits{J.}},
\bauthor{\bsnm{{Lapoux}}, \binits{V.}},
\bauthor{\bsnm{{Mar{\'e}chal}}, \binits{F.}},
\bauthor{\bsnm{{Matta}}, \binits{A.}},
\bauthor{\bsnm{{Matea}}, \binits{I.}},
\bauthor{\bsnm{{Nalpas}}, \binits{L.}},
\bauthor{\bsnm{{Pancin}}, \binits{J.}},
\bauthor{\bsnm{{Perrot}}, \binits{L.}},
\bauthor{\bsnm{{Obertelli}}, \binits{A.}},
\bauthor{\bsnm{{Raabe}}, \binits{R.}},
\bauthor{\bsnm{{Scarpaci}}, \binits{J.A.}},
\bauthor{\bsnm{{Sieja}}, \binits{K.}},
\bauthor{\bsnm{{Sorlin}}, \binits{O.}},
\bauthor{\bsnm{{Stefan}}, \binits{I.}},
\bauthor{\bsnm{{Stodel}}, \binits{C.}},
\bauthor{\bsnm{{Takechi}}, \binits{M.}},
\bauthor{\bsnm{{Thomas}}, \binits{J.C.}},
\bauthor{\bsnm{{Togano}}, \binits{Y.}}:
\batitle{{Spectroscopy of $^{61}$Fe via the neutron transfer reaction
  $^{2}$H($^{60}$Fe,p)$^{61}$Fe}}.
\bjtitle{\prc}
\bvolume{95}(\bissue{3}),
\bfpage{035806}
(\byear{2017})
\doiurl{10.1103/PhysRevC.95.035806}
\end{barticle}
\endbibitem

%%% 474
\bibitem[\protect\citeauthoryear{{Gao} et~al.}{2021}]{Gao2021}
\begin{barticle}
\bauthor{\bsnm{{Gao}}, \binits{B.}},
\bauthor{\bsnm{{Giraud}}, \binits{S.}},
\bauthor{\bsnm{{Li}}, \binits{K.A.}},
\bauthor{\bsnm{{Sieverding}}, \binits{A.}},
\bauthor{\bsnm{{Zegers}}, \binits{R.G.T.}},
\bauthor{\bsnm{{Tang}}, \binits{X.}},
\bauthor{\bsnm{{Ash}}, \binits{J.}},
\bauthor{\bsnm{{Ayyad-Limonge}}, \binits{Y.}},
\bauthor{\bsnm{{Bazin}}, \binits{D.}},
\bauthor{\bsnm{{Biswas}}, \binits{S.}},
\bauthor{\bsnm{{Brown}}, \binits{B.A.}},
\bauthor{\bsnm{{Chen}}, \binits{J.}},
\bauthor{\bsnm{{DeNudt}}, \binits{M.}},
\bauthor{\bsnm{{Farris}}, \binits{P.}},
\bauthor{\bsnm{{Gabler}}, \binits{J.M.}},
\bauthor{\bsnm{{Gade}}, \binits{A.}},
\bauthor{\bsnm{{Ginter}}, \binits{T.}},
\bauthor{\bsnm{{Grinder}}, \binits{M.}},
\bauthor{\bsnm{{Heger}}, \binits{A.}},
\bauthor{\bsnm{{Hultquist}}, \binits{C.}},
\bauthor{\bsnm{{Hill}}, \binits{A.M.}},
\bauthor{\bsnm{{Iwasaki}}, \binits{H.}},
\bauthor{\bsnm{{Kwan}}, \binits{E.}},
\bauthor{\bsnm{{Li}}, \binits{J.}},
\bauthor{\bsnm{{Longfellow}}, \binits{B.}},
\bauthor{\bsnm{{Maher}}, \binits{C.}},
\bauthor{\bsnm{{Ndayisabye}}, \binits{F.}},
\bauthor{\bsnm{{Noji}}, \binits{S.}},
\bauthor{\bsnm{{Pereira}}, \binits{J.}},
\bauthor{\bsnm{{Qi}}, \binits{C.}},
\bauthor{\bsnm{{Rebenstock}}, \binits{J.}},
\bauthor{\bsnm{{Revel}}, \binits{A.}},
\bauthor{\bsnm{{Rhodes}}, \binits{D.}},
\bauthor{\bsnm{{Sanchez}}, \binits{A.}},
\bauthor{\bsnm{{Schmitt}}, \binits{J.}},
\bauthor{\bsnm{{Sumithrarachchi}}, \binits{C.}},
\bauthor{\bsnm{{Sun}}, \binits{B.H.}},
\bauthor{\bsnm{{Weisshaar}}, \binits{D.}}:
\batitle{{New $^{59}$Fe Stellar Decay Rate with Implications for the $^{60}$Fe
  Radioactivity in Massive Stars}}.
\bjtitle{\prl}
\bvolume{126}(\bissue{15}),
\bfpage{152701}
(\byear{2021})
\doiurl{10.1103/PhysRevLett.126.152701}
\end{barticle}
\endbibitem

%%% 475
\bibitem[\protect\citeauthoryear{{Li} et~al.}{2016}]{Li2016}
\begin{barticle}
\bauthor{\bsnm{{Li}}, \binits{K.A.}},
\bauthor{\bsnm{{Lam}}, \binits{Y.H.}},
\bauthor{\bsnm{{Qi}}, \binits{C.}},
\bauthor{\bsnm{{Tang}}, \binits{X.D.}},
\bauthor{\bsnm{{Zhang}}, \binits{N.T.}}:
\batitle{{{\ensuremath{\beta}} -decay rate of $^{59}$Fe in shell burning
  environment and its influence on the production of $^{60}$Fe in a massive
  star}}.
\bjtitle{\prc}
\bvolume{94}(\bissue{6}),
\bfpage{065807}
(\byear{2016})
\doiurl{10.1103/PhysRevC.94.065807}
\end{barticle}
\endbibitem

%%% 476
\bibitem[\protect\citeauthoryear{{Langanke} and
  {Mart{\'\i}nez-Pinedo}}{2001}]{Langanke2001}
\begin{barticle}
\bauthor{\bsnm{{Langanke}}, \binits{K.}},
\bauthor{\bsnm{{Mart{\'\i}nez-Pinedo}}, \binits{G.}}:
\batitle{{Rate Tables for the Weak Processes of pf-SHELL Nuclei in Stellar
  Environments}}.
\bjtitle{Atomic Data and Nuclear Data Tables}
\bvolume{79}(\bissue{1}),
\bfpage{1}--\blpage{46}
(\byear{2001})
\doiurl{10.1006/adnd.2001.0865}
\end{barticle}
\endbibitem

%%% 477
\bibitem[\protect\citeauthoryear{{Uberseder} et~al.}{2014}]{Uberseder2014}
\begin{barticle}
\bauthor{\bsnm{{Uberseder}}, \binits{E.}},
\bauthor{\bsnm{{Adachi}}, \binits{T.}},
\bauthor{\bsnm{{Aumann}}, \binits{T.}},
\bauthor{\bsnm{{Beceiro-Novo}}, \binits{S.}},
\bauthor{\bsnm{{Boretzky}}, \binits{K.}},
\bauthor{\bsnm{{Caesar}}, \binits{C.}},
\bauthor{\bsnm{{Dillmann}}, \binits{I.}},
\bauthor{\bsnm{{Ershova}}, \binits{O.}},
\bauthor{\bsnm{{Estrade}}, \binits{A.}},
\bauthor{\bsnm{{Farinon}}, \binits{F.}},
\bauthor{\bsnm{{Hagdahl}}, \binits{J.}},
\bauthor{\bsnm{{Heftrich}}, \binits{T.}},
\bauthor{\bsnm{{Heil}}, \binits{M.}},
\bauthor{\bsnm{{Heine}}, \binits{M.}},
\bauthor{\bsnm{{Holl}}, \binits{M.}},
\bauthor{\bsnm{{Ignatov}}, \binits{A.}},
\bauthor{\bsnm{{Johansson}}, \binits{H.T.}},
\bauthor{\bsnm{{Kalantar}}, \binits{N.}},
\bauthor{\bsnm{{Langer}}, \binits{C.}},
\bauthor{\bsnm{{Le Bleis}}, \binits{T.}},
\bauthor{\bsnm{{Litvinov}}, \binits{Y.A.}},
\bauthor{\bsnm{{Marganiec}}, \binits{J.}},
\bauthor{\bsnm{{Movsesyan}}, \binits{A.}},
\bauthor{\bsnm{{Najafi}}, \binits{M.A.}},
\bauthor{\bsnm{{Nilsson}}, \binits{T.}},
\bauthor{\bsnm{{Nociforo}}, \binits{C.}},
\bauthor{\bsnm{{Panin}}, \binits{V.}},
\bauthor{\bsnm{{Pietri}}, \binits{S.}},
\bauthor{\bsnm{{Plag}}, \binits{R.}},
\bauthor{\bsnm{{Prochazka}}, \binits{A.}},
\bauthor{\bsnm{{Rastrepina}}, \binits{G.}},
\bauthor{\bsnm{{Reifarth}}, \binits{R.}},
\bauthor{\bsnm{{Ricciardi}}, \binits{V.}},
\bauthor{\bsnm{{Rigollet}}, \binits{C.}},
\bauthor{\bsnm{{Rossi}}, \binits{D.M.}},
\bauthor{\bsnm{{Savran}}, \binits{D.}},
\bauthor{\bsnm{{Simon}}, \binits{H.}},
\bauthor{\bsnm{{Sonnabend}}, \binits{K.}},
\bauthor{\bsnm{{Streicher}}, \binits{B.}},
\bauthor{\bsnm{{Terashima}}, \binits{S.}},
\bauthor{\bsnm{{Thies}}, \binits{R.}},
\bauthor{\bsnm{{Togano}}, \binits{Y.}},
\bauthor{\bsnm{{Volkov}}, \binits{V.}},
\bauthor{\bsnm{{Wamers}}, \binits{F.}},
\bauthor{\bsnm{{Weick}}, \binits{H.}},
\bauthor{\bsnm{{Weigand}}, \binits{M.}},
\bauthor{\bsnm{{Wiescher}}, \binits{M.}},
\bauthor{\bsnm{{Wimmer}}, \binits{C.}},
\bauthor{\bsnm{{Winckler}}, \binits{N.}},
\bauthor{\bsnm{{Woods}}, \binits{P.J.}}:
\batitle{{First Experimental Constraint on the
  $^{59}$Fe(n,{\ensuremath{\gamma}})$^{60}$Fe Reaction Cross Section at
  Astrophysical Energies via the Coulomb Dissociation of Fe60}}.
\bjtitle{\prl}
\bvolume{112}(\bissue{21}),
\bfpage{211101}
(\byear{2014})
\doiurl{10.1103/PhysRevLett.112.211101}
\end{barticle}
\endbibitem

%%% 478
\bibitem[\protect\citeauthoryear{{Yan} et~al.}{2021}]{Yan2021}
\begin{barticle}
\bauthor{\bsnm{{Yan}}, \binits{S.Q.}},
\bauthor{\bsnm{{Li}}, \binits{X.Y.}},
\bauthor{\bsnm{{Nishio}}, \binits{K.}},
\bauthor{\bsnm{{Lugaro}}, \binits{M.}},
\bauthor{\bsnm{{Li}}, \binits{Z.H.}},
\bauthor{\bsnm{{Makii}}, \binits{H.}},
\bauthor{\bsnm{{Pignatari}}, \binits{M.}},
\bauthor{\bsnm{{Wang}}, \binits{Y.B.}},
\bauthor{\bsnm{{Orlandi}}, \binits{R.}},
\bauthor{\bsnm{{Hirose}}, \binits{K.}},
\bauthor{\bsnm{{Tsukada}}, \binits{K.}},
\bauthor{\bsnm{{Mohr}}, \binits{P.}},
\bauthor{\bsnm{{Li}}, \binits{G.S.}},
\bauthor{\bsnm{{Wang}}, \binits{J.G.}},
\bauthor{\bsnm{{Gao}}, \binits{B.S.}},
\bauthor{\bsnm{{Han}}, \binits{Y.L.}},
\bauthor{\bsnm{{Guo}}, \binits{B.}},
\bauthor{\bsnm{{Li}}, \binits{Y.J.}},
\bauthor{\bsnm{{Shen}}, \binits{Y.P.}},
\bauthor{\bsnm{{Sato}}, \binits{T.K.}},
\bauthor{\bsnm{{Ito}}, \binits{Y.}},
\bauthor{\bsnm{{Suzaki}}, \binits{F.}},
\bauthor{\bsnm{{Su}}, \binits{J.}},
\bauthor{\bsnm{{Yang}}, \binits{Y.Y.}},
\bauthor{\bsnm{{Wang}}, \binits{J.S.}},
\bauthor{\bsnm{{Ma}}, \binits{J.B.}},
\bauthor{\bsnm{{Ma}}, \binits{P.}},
\bauthor{\bsnm{{Bai}}, \binits{Z.}},
\bauthor{\bsnm{{Xu}}, \binits{S.W.}},
\bauthor{\bsnm{{Ren}}, \binits{J.}},
\bauthor{\bsnm{{Fan}}, \binits{Q.W.}},
\bauthor{\bsnm{{Zeng}}, \binits{S.}},
\bauthor{\bsnm{{Han}}, \binits{Z.Y.}},
\bauthor{\bsnm{{Nan}}, \binits{W.}},
\bauthor{\bsnm{{Nan}}, \binits{W.K.}},
\bauthor{\bsnm{{Chen}}, \binits{C.}},
\bauthor{\bsnm{{Lian}}, \binits{G.}},
\bauthor{\bsnm{{Hu}}, \binits{Q.}},
\bauthor{\bsnm{{Duan}}, \binits{F.F.}},
\bauthor{\bsnm{{Jin}}, \binits{S.Y.}},
\bauthor{\bsnm{{Tang}}, \binits{X.D.}},
\bauthor{\bsnm{{Liu}}, \binits{W.P.}}:
\batitle{{The $^{59}$Fe(n,{\ensuremath{\gamma}})$^{60}$Fe Cross Section from
  the Surrogate Ratio Method and Its Effect on the $^{60}$Fe Nucleosynthesis}}.
\bjtitle{\apj}
\bvolume{919}(\bissue{2}),
\bfpage{84}
(\byear{2021})
\doiurl{10.3847/1538-4357/ac12ce}
{\href{https://arxiv.org/abs/2109.12654}{{arXiv:2109.12654}}}
{[astro-ph.GA]}
\end{barticle}
\endbibitem

%%% 479
\bibitem[\protect\citeauthoryear{{Spyrou} et~al.}{2024}]{Spyrou2024}
\begin{barticle}
\bauthor{\bsnm{{Spyrou}}, \binits{A.}},
\bauthor{\bsnm{{Richman}}, \binits{D.}},
\bauthor{\bsnm{{Couture}}, \binits{A.}},
\bauthor{\bsnm{{Fields}}, \binits{C.E.}},
\bauthor{\bsnm{{Liddick}}, \binits{S.N.}},
\bauthor{\bsnm{{Childers}}, \binits{K.}},
\bauthor{\bsnm{{Crider}}, \binits{B.P.}},
\bauthor{\bsnm{{DeYoung}}, \binits{P.A.}},
\bauthor{\bsnm{{Dombos}}, \binits{A.C.}},
\bauthor{\bsnm{{Gastis}}, \binits{P.}},
\bauthor{\bsnm{{Guttormsen}}, \binits{M.}},
\bauthor{\bsnm{{Hermansen}}, \binits{K.}},
\bauthor{\bsnm{{Larsen}}, \binits{A.C.}},
\bauthor{\bsnm{{Lewis}}, \binits{R.}},
\bauthor{\bsnm{{Lyons}}, \binits{S.}},
\bauthor{\bsnm{{Midtb{\o}}}, \binits{J.E.}},
\bauthor{\bsnm{{Mosby}}, \binits{S.}},
\bauthor{\bsnm{{Muecher}}, \binits{D.}},
\bauthor{\bsnm{{Naqvi}}, \binits{F.}},
\bauthor{\bsnm{{Palmisano-Kyle}}, \binits{A.}},
\bauthor{\bsnm{{Perdikakis}}, \binits{G.}},
\bauthor{\bsnm{{Prokop}}, \binits{C.}},
\bauthor{\bsnm{{Schatz}}, \binits{H.}},
\bauthor{\bsnm{{Smith}}, \binits{M.K.}},
\bauthor{\bsnm{{Sumithrarachchi}}, \binits{C.}},
\bauthor{\bsnm{{Sweet}}, \binits{A.}}:
\batitle{{Enhanced production of $^{60}$Fe in massive stars}}.
\bjtitle{Nature Communications}
\bvolume{15}(\bissue{1}),
\bfpage{9608}
(\byear{2024})
\doiurl{10.1038/s41467-024-54040-4}
{\href{https://arxiv.org/abs/2412.01723}{{arXiv:2412.01723}}}
{[nucl-ex]}
\end{barticle}
\endbibitem

%%% 480
\bibitem[\protect\citeauthoryear{{Grefenstette}
  et~al.}{2014}]{Grenfenstette2014}
\begin{barticle}
\bauthor{\bsnm{{Grefenstette}}, \binits{B.W.}},
\bauthor{\bsnm{{Harrison}}, \binits{F.A.}},
\bauthor{\bsnm{{Boggs}}, \binits{S.E.}},
\bauthor{\bsnm{{Reynolds}}, \binits{S.P.}},
\bauthor{\bsnm{{Fryer}}, \binits{C.L.}},
\bauthor{\bsnm{{Madsen}}, \binits{K.K.}},
\bauthor{\bsnm{{Wik}}, \binits{D.R.}},
\bauthor{\bsnm{{Zoglauer}}, \binits{A.}},
\bauthor{\bsnm{{Ellinger}}, \binits{C.I.}},
\bauthor{\bsnm{{Alexander}}, \binits{D.M.}},
\bauthor{\bsnm{{An}}, \binits{H.}},
\bauthor{\bsnm{{Barret}}, \binits{D.}},
\bauthor{\bsnm{{Christensen}}, \binits{F.E.}},
\bauthor{\bsnm{{Craig}}, \binits{W.W.}},
\bauthor{\bsnm{{Forster}}, \binits{K.}},
\bauthor{\bsnm{{Giommi}}, \binits{P.}},
\bauthor{\bsnm{{Hailey}}, \binits{C.J.}},
\bauthor{\bsnm{{Hornstrup}}, \binits{A.}},
\bauthor{\bsnm{{Kaspi}}, \binits{V.M.}},
\bauthor{\bsnm{{Kitaguchi}}, \binits{T.}},
\bauthor{\bsnm{{Koglin}}, \binits{J.E.}},
\bauthor{\bsnm{{Mao}}, \binits{P.H.}},
\bauthor{\bsnm{{Miyasaka}}, \binits{H.}},
\bauthor{\bsnm{{Mori}}, \binits{K.}},
\bauthor{\bsnm{{Perri}}, \binits{M.}},
\bauthor{\bsnm{{Pivovaroff}}, \binits{M.J.}},
\bauthor{\bsnm{{Puccetti}}, \binits{S.}},
\bauthor{\bsnm{{Rana}}, \binits{V.}},
\bauthor{\bsnm{{Stern}}, \binits{D.}},
\bauthor{\bsnm{{Westergaard}}, \binits{N.J.}},
\bauthor{\bsnm{{Zhang}}, \binits{W.W.}}:
\batitle{{Asymmetries in core-collapse supernovae from maps of radioactive
  $^{44}$Ti in CassiopeiaA}}.
\bjtitle{\nat}
\bvolume{506}(\bissue{7488}),
\bfpage{339}--\blpage{342}
(\byear{2014})
\doiurl{10.1038/nature12997}
{\href{https://arxiv.org/abs/1403.4978}{{arXiv:1403.4978}}}
{[astro-ph.HE]}
\end{barticle}
\endbibitem

%%% 481
\bibitem[\protect\citeauthoryear{{Subedi} et~al.}{2020}]{Subedi2020}
\begin{barticle}
\bauthor{\bsnm{{Subedi}}, \binits{S.K.}},
\bauthor{\bsnm{{Meisel}}, \binits{Z.}},
\bauthor{\bsnm{{Merz}}, \binits{G.}}:
\batitle{{Sensitivity of $^{44}$Ti and $^{56}$Ni Production in Core-collapse
  Supernova Shock-driven Nucleosynthesis to Nuclear Reaction Rate Variations}}.
\bjtitle{\apj}
\bvolume{898}(\bissue{1}),
\bfpage{5}
(\byear{2020})
\doiurl{10.3847/1538-4357/ab9745}
{\href{https://arxiv.org/abs/2005.14702}{{arXiv:2005.14702}}}
{[astro-ph.HE]}
\end{barticle}
\endbibitem

%%% 482
\bibitem[\protect\citeauthoryear{{Hermansen} et~al.}{2020}]{Hermansen2020}
\begin{barticle}
\bauthor{\bsnm{{Hermansen}}, \binits{K.}},
\bauthor{\bsnm{{Couch}}, \binits{S.M.}},
\bauthor{\bsnm{{Roberts}}, \binits{L.F.}},
\bauthor{\bsnm{{Schatz}}, \binits{H.}},
\bauthor{\bsnm{{Warren}}, \binits{M.L.}}:
\batitle{{Reaction Rate Sensitivity of the Production of
  {\ensuremath{\gamma}}-Ray Emitting Isotopes in Core-collapse Supernovae}}.
\bjtitle{\apj}
\bvolume{901}(\bissue{1}),
\bfpage{77}
(\byear{2020})
\doiurl{10.3847/1538-4357/abafb5}
{\href{https://arxiv.org/abs/2006.16181}{{arXiv:2006.16181}}}
{[astro-ph.SR]}
\end{barticle}
\endbibitem

%%% 483
\bibitem[\protect\citeauthoryear{{Foug{\`e}res} et~al.}{}]{Fougeres2023b}
\begin{botherref}
\oauthor{\bsnm{{Foug{\`e}res}}, \binits{C.}},
\oauthor{\bsnm{{de Oliveira Santos}}, \binits{F.}},
\oauthor{\bsnm{{de S{\'e}r{\'e}ville}}, \binits{N.}},
\oauthor{\bsnm{{Hammache}}, \binits{F.}}:
Experiment e13037 at FRIB
\end{botherref}
\endbibitem

%%% 484
\bibitem[\protect\citeauthoryear{{Seuthe} et~al.}{1990}]{Seuthe1990}
\begin{barticle}
\bauthor{\bsnm{{Seuthe}}, \binits{S.}},
\bauthor{\bsnm{{Rolfs}}, \binits{C.}},
\bauthor{\bsnm{{Schr{\"o}der}}, \binits{U.}},
\bauthor{\bsnm{{Schulte}}, \binits{W.H.}},
\bauthor{\bsnm{{Somorjai}}, \binits{E.}},
\bauthor{\bsnm{{Trautvetter}}, \binits{H.P.}},
\bauthor{\bsnm{{Waanders}}, \binits{F.B.}},
\bauthor{\bsnm{{Kavanagh}}, \binits{R.W.}},
\bauthor{\bsnm{{Ravn}}, \binits{H.}},
\bauthor{\bsnm{{Arnould}}, \binits{M.}},
\bauthor{\bsnm{{Paulus}}, \binits{G.}}:
\batitle{{Resonances in the $^{22}$Na(p, {\ensuremath{\gamma}})$^{23}$Mg
  reaction}}.
\bjtitle{\nphysa}
\bvolume{514}(\bissue{3}),
\bfpage{471}--\blpage{502}
(\byear{1990})
\doiurl{10.1016/0375-9474(90)90153-D}
\end{barticle}
\endbibitem

%%% 485
\bibitem[\protect\citeauthoryear{{Stegm{\"u}ller}
  et~al.}{1996}]{Stegmuller1996}
\begin{barticle}
\bauthor{\bsnm{{Stegm{\"u}ller}}, \binits{F.}},
\bauthor{\bsnm{{Rolfs}}, \binits{C.}},
\bauthor{\bsnm{{Schmidt}}, \binits{S.}},
\bauthor{\bsnm{{Schulte}}, \binits{W.H.}},
\bauthor{\bsnm{{Trautvetter}}, \binits{H.P.}},
\bauthor{\bsnm{{Kavanagh}}, \binits{R.W.}}:
\batitle{{$^{22}$Na(p,{\ensuremath{\gamma}})$^{23}$Mg resonant reaction at low
  energies}}.
\bjtitle{\nphysa}
\bvolume{601}(\bissue{2}),
\bfpage{168}--\blpage{180}
(\byear{1996})
\doiurl{10.1016/0375-9474(96)00084-X}
\end{barticle}
\endbibitem

%%% 486
\bibitem[\protect\citeauthoryear{{Sallaska} et~al.}{2010}]{Sallaska2010}
\begin{barticle}
\bauthor{\bsnm{{Sallaska}}, \binits{A.L.}},
\bauthor{\bsnm{{Wrede}}, \binits{C.}},
\bauthor{\bsnm{{Garc{\'\i}a}}, \binits{A.}},
\bauthor{\bsnm{{Storm}}, \binits{D.W.}},
\bauthor{\bsnm{{Brown}}, \binits{T.A.D.}},
\bauthor{\bsnm{{Ruiz}}, \binits{C.}},
\bauthor{\bsnm{{Snover}}, \binits{K.A.}},
\bauthor{\bsnm{{Ottewell}}, \binits{D.F.}},
\bauthor{\bsnm{{Buchmann}}, \binits{L.}},
\bauthor{\bsnm{{Vockenhuber}}, \binits{C.}},
\bauthor{\bsnm{{Hutcheon}}, \binits{D.A.}},
\bauthor{\bsnm{{Caggiano}}, \binits{J.A.}}:
\batitle{{Direct Measurements of $^{22}$Na(p,{\ensuremath{\gamma}})$^{23}$Mg
  Resonances and Consequences for $^{22}$Na Production in Classical Novae}}.
\bjtitle{\prl}
\bvolume{105}(\bissue{15}),
\bfpage{152501}
(\byear{2010})
\doiurl{10.1103/PhysRevLett.105.152501}
\end{barticle}
\endbibitem

%%% 487
\bibitem[\protect\citeauthoryear{{de Oliveira Santos}}{2025}]{deOliveira2025}
\begin{barticle}
\bauthor{\bsnm{{de Oliveira Santos}}, \binits{F.}}:
\batitle{{The $^{18}$F(p, {\ensuremath{\alpha}})$^{15}$O reaction: A textbook
  case in nuclear astrophysics}}.
\bjtitle{Progress in Particle and Nuclear Physics}
\bvolume{142},
\bfpage{104154}
(\byear{2025})
\doiurl{10.1016/j.ppnp.2025.104154}
\end{barticle}
\endbibitem

%%% 488
\bibitem[\protect\citeauthoryear{{de S{\'e}r{\'e}ville}
  et~al.}{2009}]{deSereville2009}
\begin{barticle}
\bauthor{\bsnm{{de S{\'e}r{\'e}ville}}, \binits{N.}},
\bauthor{\bsnm{{Angulo}}, \binits{C.}},
\bauthor{\bsnm{{Coc}}, \binits{A.}},
\bauthor{\bsnm{{Achouri}}, \binits{N.L.}},
\bauthor{\bsnm{{Casarejos}}, \binits{E.}},
\bauthor{\bsnm{{Davinson}}, \binits{T.}},
\bauthor{\bsnm{{Descouvemont}}, \binits{P.}},
\bauthor{\bsnm{{Figuera}}, \binits{P.}},
\bauthor{\bsnm{{Fox}}, \binits{S.}},
\bauthor{\bsnm{{Hammache}}, \binits{F.}},
\bauthor{\bsnm{{Kiener}}, \binits{J.}},
\bauthor{\bsnm{{Laird}}, \binits{A.}},
\bauthor{\bsnm{{Lefebvre-Schuhl}}, \binits{A.}},
\bauthor{\bsnm{{Leleux}}, \binits{P.}},
\bauthor{\bsnm{{Mumby-Croft}}, \binits{P.}},
\bauthor{\bsnm{{Orr}}, \binits{N.A.}},
\bauthor{\bsnm{{Stefan}}, \binits{I.}},
\bauthor{\bsnm{{Vaughan}}, \binits{K.}},
\bauthor{\bsnm{{Tatischeff}}, \binits{V.}}:
\batitle{{Low-energy $^{18}$F(p, {\ensuremath{\alpha}})$^{15}$O cross section
  measurements relevant to nova {\ensuremath{\gamma}}-ray emission}}.
\bjtitle{\prc}
\bvolume{79}(\bissue{1}),
\bfpage{015801}
(\byear{2009})
\doiurl{10.1103/PhysRevC.79.015801}
\end{barticle}
\endbibitem

%%% 489
\bibitem[\protect\citeauthoryear{{Beer} et~al.}{2011}]{Beer2011}
\begin{barticle}
\bauthor{\bsnm{{Beer}}, \binits{C.E.}},
\bauthor{\bsnm{{Laird}}, \binits{A.M.}},
\bauthor{\bsnm{{Murphy}}, \binits{A.S.J.}},
\bauthor{\bsnm{{Bentley}}, \binits{M.A.}},
\bauthor{\bsnm{{Buchman}}, \binits{L.}},
\bauthor{\bsnm{{Davids}}, \binits{B.}},
\bauthor{\bsnm{{Davinson}}, \binits{T.}},
\bauthor{\bsnm{{Diget}}, \binits{C.A.}},
\bauthor{\bsnm{{Fox}}, \binits{S.P.}},
\bauthor{\bsnm{{Fulton}}, \binits{B.R.}},
\bauthor{\bsnm{{Hager}}, \binits{U.}},
\bauthor{\bsnm{{Howell}}, \binits{D.}},
\bauthor{\bsnm{{Martin}}, \binits{L.}},
\bauthor{\bsnm{{Ruiz}}, \binits{C.}},
\bauthor{\bsnm{{Ruprecht}}, \binits{G.}},
\bauthor{\bsnm{{Salter}}, \binits{P.}},
\bauthor{\bsnm{{Vockenhuber}}, \binits{C.}},
\bauthor{\bsnm{{Walden}}, \binits{P.}}:
\batitle{{Direct measurement of the $^{18}$F(p, {\ensuremath{\alpha}})$^{15}$O
  reaction at nova temperatures}}.
\bjtitle{\prc}
\bvolume{83}(\bissue{4}),
\bfpage{042801}
(\byear{2011})
\doiurl{10.1103/PhysRevC.83.042801}
\end{barticle}
\endbibitem

%%% 490
\bibitem[\protect\citeauthoryear{{Kahl} et~al.}{2021}]{Kahl2021}
\begin{barticle}
\bauthor{\bsnm{{Kahl}}, \binits{D.}},
\bauthor{\bsnm{{Jos{\'e}}}, \binits{J.}},
\bauthor{\bsnm{{Woods}}, \binits{P.J.}}:
\batitle{{Uncertainties in the $^{18}$F(p, {\ensuremath{\alpha}})$^{15}$O
  reaction rate in classical novae}}.
\bjtitle{\aap}
\bvolume{653},
\bfpage{64}
(\byear{2021})
\doiurl{10.1051/0004-6361/202140339}
{\href{https://arxiv.org/abs/2106.02606}{{arXiv:2106.02606}}}
{[nucl-th]}
\end{barticle}
\endbibitem

%%% 491
\bibitem[\protect\citeauthoryear{{Timmes} et~al.}{2019}]{Timmes2019}
\begin{barticle}
\bauthor{\bsnm{{Timmes}}, \binits{F.}},
\bauthor{\bsnm{{Fryer}}, \binits{C.}},
\bauthor{\bsnm{{Timmes}}, \binits{F.}},
\bauthor{\bsnm{{Hungerford}}, \binits{A.L.}},
\bauthor{\bsnm{{Couture}}, \binits{A.}},
\bauthor{\bsnm{{Adams}}, \binits{F.}},
\bauthor{\bsnm{{Aoki}}, \binits{W.}},
\bauthor{\bsnm{{Arcones}}, \binits{A.}},
\bauthor{\bsnm{{Arnett}}, \binits{D.}},
\bauthor{\bsnm{{Auchettl}}, \binits{K.}},
\bauthor{\bsnm{{Avila}}, \binits{M.}},
\bauthor{\bsnm{{Badenes}}, \binits{C.}},
\bauthor{\bsnm{{Baron}}, \binits{E.}},
\bauthor{\bsnm{{Bauswein}}, \binits{A.}},
\bauthor{\bsnm{{Beacom}}, \binits{J.}},
\bauthor{\bsnm{{Blackmon}}, \binits{J.}},
\bauthor{\bsnm{{Blondin}}, \binits{S.}},
\bauthor{\bsnm{{Bloser}}, \binits{P.}},
\bauthor{\bsnm{{Boggs}}, \binits{S.}},
\bauthor{\bsnm{{Boss}}, \binits{A.}},
\bauthor{\bsnm{{Brandt}}, \binits{T.}},
\bauthor{\bsnm{{Bravo}}, \binits{E.}},
\bauthor{\bsnm{{Brown}}, \binits{E.}},
\bauthor{\bsnm{{Brown}}, \binits{P.}},
\bauthor{\bsnm{{Bruenn}}, \binits{S.}},
\bauthor{\bsnm{{Budtz-J{\o}rgensen}}, \binits{C.}},
\bauthor{\bsnm{{Burns}}, \binits{E.}},
\bauthor{\bsnm{{Calder}}, \binits{A.}},
\bauthor{\bsnm{{Caputo}}, \binits{R.}},
\bauthor{\bsnm{{Champagne}}, \binits{A.}},
\bauthor{\bsnm{{Chevalier}}, \binits{R.}},
\bauthor{\bsnm{{Chieffi}}, \binits{A.}},
\bauthor{\bsnm{{Chipps}}, \binits{K.}},
\bauthor{\bsnm{{Cinabro}}, \binits{D.}},
\bauthor{\bsnm{{Clarkson}}, \binits{O.}},
\bauthor{\bsnm{{Clayton}}, \binits{D.}},
\bauthor{\bsnm{{Coc}}, \binits{A.}},
\bauthor{\bsnm{{Connolly}}, \binits{D.}},
\bauthor{\bsnm{{Conroy}}, \binits{C.}},
\bauthor{\bsnm{{C{\^o}t{\'e}}}, \binits{B.}},
\bauthor{\bsnm{{Couch}}, \binits{S.}},
\bauthor{\bsnm{{Dauphas}}, \binits{N.}},
\bauthor{\bsnm{{deBoer}}, \binits{R.J.}},
\bauthor{\bsnm{{Deibel}}, \binits{C.}},
\bauthor{\bsnm{{Denisenkov}}, \binits{P.}},
\bauthor{\bsnm{{Desch}}, \binits{S.}},
\bauthor{\bsnm{{Dessart}}, \binits{L.}},
\bauthor{\bsnm{{Diehl}}, \binits{R.}},
\bauthor{\bsnm{{Doherty}}, \binits{C.}},
\bauthor{\bsnm{{Dom{\'\i}nguez}}, \binits{I.}},
\bauthor{\bsnm{{Dong}}, \binits{S.}},
\bauthor{\bsnm{{Dwarkadas}}, \binits{V.}},
\bauthor{\bsnm{{Fan}}, \binits{D.}},
\bauthor{\bsnm{{Fields}}, \binits{B.}},
\bauthor{\bsnm{{Fields}}, \binits{C.}},
\bauthor{\bsnm{{Filippenko}}, \binits{A.}},
\bauthor{\bsnm{{Fisher}}, \binits{R.}},
\bauthor{\bsnm{{Foucart}}, \binits{F.}},
\bauthor{\bsnm{{Fransson}}, \binits{C.}},
\bauthor{\bsnm{{Fr{\"o}hlich}}, \binits{C.}},
\bauthor{\bsnm{{Fuller}}, \binits{G.}},
\bauthor{\bsnm{{Gibson}}, \binits{B.}},
\bauthor{\bsnm{{Giryanskaya}}, \binits{V.}},
\bauthor{\bsnm{{G{\"o}rres}}, \binits{J.}},
\bauthor{\bsnm{{Goriely}}, \binits{S.}},
\bauthor{\bsnm{{Grebenev}}, \binits{S.}},
\bauthor{\bsnm{{Grefenstette}}, \binits{B.}},
\bauthor{\bsnm{{Grohs}}, \binits{E.}},
\bauthor{\bsnm{{Guillochon}}, \binits{J.}},
\bauthor{\bsnm{{Harpole}}, \binits{A.}},
\bauthor{\bsnm{{Harris}}, \binits{C.}},
\bauthor{\bsnm{{Harris}}, \binits{J.A.}},
\bauthor{\bsnm{{Harrison}}, \binits{F.}},
\bauthor{\bsnm{{Hartmann}}, \binits{D.}},
\bauthor{\bsnm{{Hashimoto}}, \binits{M.-a.}},
\bauthor{\bsnm{{Heger}}, \binits{A.}},
\bauthor{\bsnm{{Hernanz}}, \binits{M.}},
\bauthor{\bsnm{{Herwig}}, \binits{F.}},
\bauthor{\bsnm{{Hirschi}}, \binits{R.}},
\bauthor{\bsnm{{Hix}}, \binits{W.R.}},
\bauthor{\bsnm{{H{\"o}flich}}, \binits{P.}},
\bauthor{\bsnm{{Hoffman}}, \binits{R.}},
\bauthor{\bsnm{{Holcomb}}, \binits{C.}},
\bauthor{\bsnm{{Hsiao}}, \binits{E.}},
\bauthor{\bsnm{{Iliadis}}, \binits{C.}},
\bauthor{\bsnm{{Janiuk}}, \binits{A.}},
\bauthor{\bsnm{{Janka}}, \binits{T.}},
\bauthor{\bsnm{{Jerkstrand}}, \binits{A.}},
\bauthor{\bsnm{{Johns}}, \binits{L.}},
\bauthor{\bsnm{{Jones}}, \binits{S.}},
\bauthor{\bsnm{{Jos{\'e}}}, \binits{J.}},
\bauthor{\bsnm{{Kajino}}, \binits{T.}},
\bauthor{\bsnm{{Karakas}}, \binits{A.}},
\bauthor{\bsnm{{Karpov}}, \binits{P.}},
\bauthor{\bsnm{{Kasen}}, \binits{D.}},
\bauthor{\bsnm{{Kierans}}, \binits{C.}},
\bauthor{\bsnm{{Kippen}}, \binits{M.}},
\bauthor{\bsnm{{Korobkin}}, \binits{O.}},
\bauthor{\bsnm{{Kobayashi}}, \binits{C.}},
\bauthor{\bsnm{{Kozma}}, \binits{C.}},
\bauthor{\bsnm{{Krot}}, \binits{S.}},
\bauthor{\bsnm{{Kumar}}, \binits{P.}},
\bauthor{\bsnm{{Kuvvetli}}, \binits{I.}},
\bauthor{\bsnm{{Laird}}, \binits{A.}},
\bauthor{\bsnm{{Laming}}, \binits{J.M.}},
\bauthor{\bsnm{{Larsson}}, \binits{J.}},
\bauthor{\bsnm{{Lattanzio}}, \binits{J.}},
\bauthor{\bsnm{{Lattimer}}, \binits{J.}},
\bauthor{\bsnm{{Leising}}, \binits{M.}},
\bauthor{\bsnm{{Lennarz}}, \binits{A.}},
\bauthor{\bsnm{{Lentz}}, \binits{E.}},
\bauthor{\bsnm{{Limongi}}, \binits{M.}},
\bauthor{\bsnm{{Lippuner}}, \binits{J.}},
\bauthor{\bsnm{{Livne}}, \binits{E.}},
\bauthor{\bsnm{{Lloyd-Ronning}}, \binits{N.}},
\bauthor{\bsnm{{Longland}}, \binits{R.}},
\bauthor{\bsnm{{Lopez}}, \binits{L.A.}},
\bauthor{\bsnm{{Lugaro}}, \binits{M.}},
\bauthor{\bsnm{{Lutovinov}}, \binits{A.}},
\bauthor{\bsnm{{Madsen}}, \binits{K.}},
\bauthor{\bsnm{{Malone}}, \binits{C.}},
\bauthor{\bsnm{{Matteucci}}, \binits{F.}},
\bauthor{\bsnm{{McEnery}}, \binits{J.}},
\bauthor{\bsnm{{Meisel}}, \binits{Z.}},
\bauthor{\bsnm{{Messer}}, \binits{B.}},
\bauthor{\bsnm{{Metzger}}, \binits{B.}},
\bauthor{\bsnm{{Meyer}}, \binits{B.}},
\bauthor{\bsnm{{Meynet}}, \binits{G.}},
\bauthor{\bsnm{{Mezzacappa}}, \binits{A.}},
\bauthor{\bsnm{{Miller}}, \binits{J.}},
\bauthor{\bsnm{{Miller}}, \binits{R.}},
\bauthor{\bsnm{{Milne}}, \binits{P.}},
\bauthor{\bsnm{{Misch}}, \binits{W.}},
\bauthor{\bsnm{{Mitchell}}, \binits{L.}},
\bauthor{\bsnm{{M{\"o}sta}}, \binits{P.}},
\bauthor{\bsnm{{Motizuki}}, \binits{Y.}},
\bauthor{\bsnm{{M{\"u}ller}}, \binits{B.}},
\bauthor{\bsnm{{Mumpower}}, \binits{M.}},
\bauthor{\bsnm{{Murphy}}, \binits{J.}},
\bauthor{\bsnm{{Nagataki}}, \binits{S.}},
\bauthor{\bsnm{{Nakar}}, \binits{E.}},
\bauthor{\bsnm{{Nomoto}}, \binits{K.}},
\bauthor{\bsnm{{Nugent}}, \binits{P.}},
\bauthor{\bsnm{{Nunes}}, \binits{F.}},
\bauthor{\bsnm{{O'Shea}}, \binits{B.}},
\bauthor{\bsnm{{Oberlack}}, \binits{U.}},
\bauthor{\bsnm{{Pain}}, \binits{S.}},
\bauthor{\bsnm{{Parker}}, \binits{L.}},
\bauthor{\bsnm{{Perego}}, \binits{A.}},
\bauthor{\bsnm{{Pignatari}}, \binits{M.}},
\bauthor{\bsnm{{Pinedo}}, \binits{G.M.}},
\bauthor{\bsnm{{Plewa}}, \binits{T.}},
\bauthor{\bsnm{{Poznanski}}, \binits{D.}},
\bauthor{\bsnm{{Priedhorsky}}, \binits{W.}},
\bauthor{\bsnm{{Pritychenko}}, \binits{B.}},
\bauthor{\bsnm{{Radice}}, \binits{D.}},
\bauthor{\bsnm{{Ramirez-Ruiz}}, \binits{E.}},
\bauthor{\bsnm{{Rauscher}}, \binits{T.}},
\bauthor{\bsnm{{Reddy}}, \binits{S.}},
\bauthor{\bsnm{{Rehm}}, \binits{E.}},
\bauthor{\bsnm{{Reifarth}}, \binits{R.}},
\bauthor{\bsnm{{Richman}}, \binits{D.}},
\bauthor{\bsnm{{Ricker}}, \binits{P.}},
\bauthor{\bsnm{{Rijal}}, \binits{N.}},
\bauthor{\bsnm{{Roberts}}, \binits{L.}},
\bauthor{\bsnm{{R{\"o}pke}}, \binits{F.}},
\bauthor{\bsnm{{Rosswog}}, \binits{S.}},
\bauthor{\bsnm{{Ruiter}}, \binits{A.J.}},
\bauthor{\bsnm{{Ruiz}}, \binits{C.}},
\bauthor{\bsnm{{Savin}}, \binits{D.W.}},
\bauthor{\bsnm{{Schatz}}, \binits{H.}},
\bauthor{\bsnm{{Schneider}}, \binits{D.}},
\bauthor{\bsnm{{Schwab}}, \binits{J.}},
\bauthor{\bsnm{{Seitenzahl}}, \binits{I.}},
\bauthor{\bsnm{{Shen}}, \binits{K.}},
\bauthor{\bsnm{{Siegert}}, \binits{T.}},
\bauthor{\bsnm{{Sim}}, \binits{S.}},
\bauthor{\bsnm{{Smith}}, \binits{D.}},
\bauthor{\bsnm{{Smith}}, \binits{K.}},
\bauthor{\bsnm{{Smith}}, \binits{M.}},
\bauthor{\bsnm{{Sollerman}}, \binits{J.}},
\bauthor{\bsnm{{Sprouse}}, \binits{T.}},
\bauthor{\bsnm{{Spyrou}}, \binits{A.}},
\bauthor{\bsnm{{Starrfield}}, \binits{S.}},
\bauthor{\bsnm{{Steiner}}, \binits{A.}},
\bauthor{\bsnm{{Strong}}, \binits{A.W.}},
\bauthor{\bsnm{{Sukhbold}}, \binits{T.}},
\bauthor{\bsnm{{Suntzeff}}, \binits{N.}},
\bauthor{\bsnm{{Surman}}, \binits{R.}},
\bauthor{\bsnm{{Tanimori}}, \binits{T.}},
\bauthor{\bsnm{{The}}, \binits{L.-S.}},
\bauthor{\bsnm{{Thielemann}}, \binits{F.-K.}},
\bauthor{\bsnm{{Tolstov}}, \binits{A.}},
\bauthor{\bsnm{{Tominaga}}, \binits{N.}},
\bauthor{\bsnm{{Tomsick}}, \binits{J.}},
\bauthor{\bsnm{{Townsley}}, \binits{D.}},
\bauthor{\bsnm{{Tsintari}}, \binits{P.}},
\bauthor{\bsnm{{Tsygankov}}, \binits{S.}},
\bauthor{\bsnm{{Vartanyan}}, \binits{D.}},
\bauthor{\bsnm{{Venters}}, \binits{T.}}:
\batitle{{Catching Element Formation In The Act ; The Case for a New MeV
  Gamma-Ray Mission: Radionuclide Astronomy in the 2020s}}.
\bjtitle{\baas}
\bvolume{51}(\bissue{3}),
\bfpage{2}
(\byear{2019})
\doiurl{10.48550/arXiv.1902.02915}
{\href{https://arxiv.org/abs/1902.02915}{{arXiv:1902.02915}}}
{[astro-ph.HE]}
\end{barticle}
\endbibitem

%%% 492
\bibitem[\protect\citeauthoryear{FRIB}{}]{FRIB}
\begin{botherref}
\oauthor{\bsnm{FRIB}}:
https://frib.msu.edu/.
\url{https://asd.gsfc.nasa.gov/amego/}
\end{botherref}
\endbibitem

%%% 493
\bibitem[\protect\citeauthoryear{{Kim}}{2020}]{RAON}
\begin{barticle}
\bauthor{\bsnm{{Kim}}, \binits{Y.J.}}:
\batitle{{Current status of experimental facilities at RAON}}.
\bjtitle{Nuclear Instruments and Methods in Physics Research B}
\bvolume{463},
\bfpage{408}--\blpage{414}
(\byear{2020})
\doiurl{10.1016/j.nimb.2019.04.041}
\end{barticle}
\endbibitem

%%% 494
\bibitem[\protect\citeauthoryear{{Kierans} et~al.}{2017}]{Kierans2017}
\begin{botherref}
\oauthor{\bsnm{{Kierans}}, \binits{C.A.}},
\oauthor{\bsnm{{Boggs}}, \binits{S.E.}},
\oauthor{\bsnm{{Chiu}}, \binits{J.-L.}},
\oauthor{\bsnm{{Lowell}}, \binits{A.}},
\oauthor{\bsnm{{Sleator}}, \binits{C.}},
\oauthor{\bsnm{{Tomsick}}, \binits{J.A.}},
\oauthor{\bsnm{{Zoglauer}}, \binits{A.}},
\oauthor{\bsnm{{Amman}}, \binits{M.}},
\oauthor{\bsnm{{Chang}}, \binits{H.-K.}},
\oauthor{\bsnm{{Tseng}}, \binits{C.-H.}},
\oauthor{\bsnm{{Yang}}, \binits{C.-Y.}},
\oauthor{\bsnm{{Lin}}, \binits{C.-H.}},
\oauthor{\bsnm{{Jean}}, \binits{P.}},
\oauthor{\bsnm{{von Ballmoos}}, \binits{P.}}:
{The 2016 Super Pressure Balloon flight of the Compton Spectrometer and
  Imager}.
arXiv e-prints,
1701--05558
(2017)
\doiurl{10.48550/arXiv.1701.05558}
{\href{https://arxiv.org/abs/1701.05558}{{arXiv:1701.05558}}}
{[astro-ph.IM]}
\end{botherref}
\endbibitem

%%% 495
\bibitem[\protect\citeauthoryear{AMEGO}{}]{AMEGO}
\begin{botherref}
\oauthor{\bsnm{AMEGO}}:
https://asd.gsfc.nasa.gov/amego/.
\url{https://asd.gsfc.nasa.gov/amego/}
\end{botherref}
\endbibitem

%%% 496
\bibitem[\protect\citeauthoryear{ASTROGAM}{}]{ASTROGAM}
\begin{botherref}
\oauthor{\bsnm{ASTROGAM}}:
http://new-astrogam.eu/.
\url{https://asd.gsfc.nasa.gov/amego/}
\end{botherref}
\endbibitem

%%% 497
\bibitem[\protect\citeauthoryear{{Hotokezaka} et~al.}{2016}]{Hotokezaka2016}
\begin{barticle}
\bauthor{\bsnm{{Hotokezaka}}, \binits{K.}},
\bauthor{\bsnm{{Wanajo}}, \binits{S.}},
\bauthor{\bsnm{{Tanaka}}, \binits{M.}},
\bauthor{\bsnm{{Bamba}}, \binits{A.}},
\bauthor{\bsnm{{Terada}}, \binits{Y.}},
\bauthor{\bsnm{{Piran}}, \binits{T.}}:
\batitle{{Radioactive decay products in neutron star merger ejecta: heating
  efficiency and {\ensuremath{\gamma}}-ray emission}}.
\bjtitle{\mnras}
\bvolume{459}(\bissue{1}),
\bfpage{35}--\blpage{43}
(\byear{2016})
\doiurl{10.1093/mnras/stw404}
{\href{https://arxiv.org/abs/1511.05580}{{arXiv:1511.05580}}}
{[astro-ph.HE]}
\end{barticle}
\endbibitem

%%% 498
\bibitem[\protect\citeauthoryear{{Vestrand} et~al.}{1999}]{Vestrand1999}
\begin{barticle}
\bauthor{\bsnm{{Vestrand}}, \binits{W.T.}},
\bauthor{\bsnm{{Share}}, \binits{G.H.}},
\bauthor{\bsnm{{J. Murphy}}, \binits{R.}},
\bauthor{\bsnm{{Forrest}}, \binits{D.J.}},
\bauthor{\bsnm{{Rieger}}, \binits{E.}},
\bauthor{\bsnm{{Chupp}}, \binits{E.L.}},
\bauthor{\bsnm{{Kanbach}}, \binits{G.}}:
\batitle{{The Solar Maximum Mission Atlas of Gamma-Ray Flares}}.
\bjtitle{\apjs}
\bvolume{120}(\bissue{2}),
\bfpage{409}--\blpage{467}
(\byear{1999})
\doiurl{10.1086/313180}
\end{barticle}
\endbibitem

%%% 499
\bibitem[\protect\citeauthoryear{{Leventhal}}{1973}]{Leventhal1973}
\begin{barticle}
\bauthor{\bsnm{{Leventhal}}, \binits{M.}}:
\batitle{{Deuterium Formation Hypothesis for the Diffuse Gamma-ray Excess at 1
  MeV}}.
\bjtitle{\nat}
\bvolume{246}(\bissue{5429}),
\bfpage{136}--\blpage{138}
(\byear{1973})
\doiurl{10.1038/246136a0}
\end{barticle}
\endbibitem

%%% 500
\bibitem[\protect\citeauthoryear{{Guessoum} and {Dermer}}{1988}]{Guessoum1988}
\begin{bchapter}
\bauthor{\bsnm{{Guessoum}}, \binits{N.}},
\bauthor{\bsnm{{Dermer}}, \binits{C.D.}}:
\bctitle{{Properties of hydrogen/helium accretion plasmas}}.
In: \beditor{\bsnm{{Gehrels}}, \binits{N.}},
\beditor{\bsnm{{Share}}, \binits{G.H.}} (eds.)
\bbtitle{Nuclear Spectroscopy of Astrophysical Sources}.
\bsertitle{American Institute of Physics Conference Series},
vol. \bseriesno{170},
pp. \bfpage{332}--\blpage{337}.
\bpublisher{AIP}, \blocation{???}
(\byear{1988}).
\doiurl{10.1063/1.37227}
\end{bchapter}
\endbibitem

%%% 501
\bibitem[\protect\citeauthoryear{{Aharonian} and
  {Sunyaev}}{1984}]{Aharonian1984}
\begin{barticle}
\bauthor{\bsnm{{Aharonian}}, \binits{F.A.}},
\bauthor{\bsnm{{Sunyaev}}, \binits{R.A.}}:
\batitle{{Gamma-ray line emission, nuclear destruction and neutron production
  in hot astrophysical plasmas.The deuterium boiler as a gamma-ray source.}}
\bjtitle{\mnras}
\bvolume{210},
\bfpage{257}--\blpage{277}
(\byear{1984})
\doiurl{10.1093/mnras/210.2.257}
\end{barticle}
\endbibitem

%%% 502
\bibitem[\protect\citeauthoryear{{Shvartsman}}{1970}]{Shvartsman1972}
\begin{barticle}
\bauthor{\bsnm{{Shvartsman}}, \binits{V.F.}}:
\batitle{{On the generation of relativistic particles by neutron stars in the
  state of accretion.}}
\bjtitle{Astrofizika}
\bvolume{6},
\bfpage{309}--\blpage{317}
(\byear{1970})
\end{barticle}
\endbibitem

%%% 503
\bibitem[\protect\citeauthoryear{{Bildsten} et~al.}{1992}]{Bildsten1992}
\begin{barticle}
\bauthor{\bsnm{{Bildsten}}, \binits{L.}},
\bauthor{\bsnm{{Salpeter}}, \binits{E.E.}},
\bauthor{\bsnm{{Wasserman}}, \binits{I.}}:
\batitle{{The Fate of Accreted CNO Elements in Neutron Star Atmospheres: X-Ray
  Bursts and Gamma-Ray Lines}}.
\bjtitle{\apj}
\bvolume{384},
\bfpage{143}
(\byear{1992})
\doiurl{10.1086/170860}
\end{barticle}
\endbibitem

%%% 504
\bibitem[\protect\citeauthoryear{{Jean} and
  {Guessoum}}{2001}]{2001A&A...378..509J}
\begin{barticle}
\bauthor{\bsnm{{Jean}}, \binits{P.}},
\bauthor{\bsnm{{Guessoum}}, \binits{N.}}:
\batitle{{Neutron-capture and 2.22 MeV emission in the atmosphere of the
  secondary of an X-ray binary}}.
\bjtitle{\aap}
\bvolume{378},
\bfpage{509}--\blpage{521}
(\byear{2001})
\doiurl{10.1051/0004-6361:20011201}
{\href{https://arxiv.org/abs/astro-ph/0109185}{{arXiv:astro-ph/0109185}}}
{[astro-ph]}
\end{barticle}
\endbibitem

%%% 505
\bibitem[\protect\citeauthoryear{{Guessoum} and
  {Jean}}{2002}]{2002A&A...396..157G}
\begin{barticle}
\bauthor{\bsnm{{Guessoum}}, \binits{N.}},
\bauthor{\bsnm{{Jean}}, \binits{P.}}:
\batitle{{Detectability and characteristics of the 2.223 MeV line emission from
  nearby X-ray binaries}}.
\bjtitle{\aap}
\bvolume{396},
\bfpage{157}--\blpage{169}
(\byear{2002})
\doiurl{10.1051/0004-6361:20021376}
\end{barticle}
\endbibitem

%%% 506
\bibitem[\protect\citeauthoryear{{Vestrand}}{1990}]{Vestrand1990}
\begin{barticle}
\bauthor{\bsnm{{Vestrand}}, \binits{W.T.}}:
\batitle{{A New Gamma-Ray Diagnostic for Energetic Ion Distributions: The
  Compton Tail on the Neutron Capture Line}}.
\bjtitle{\apj}
\bvolume{352},
\bfpage{353}
(\byear{1990})
\doiurl{10.1086/168542}
\end{barticle}
\endbibitem

%%% 507
\bibitem[\protect\citeauthoryear{Sabti et~al.}{2020}]{Sabti:2019mhn}
\begin{barticle}
\bauthor{\bsnm{Sabti}, \binits{N.}},
\bauthor{\bsnm{Alvey}, \binits{J.}},
\bauthor{\bsnm{Escudero}, \binits{M.}},
\bauthor{\bsnm{Fairbairn}, \binits{M.}},
\bauthor{\bsnm{Blas}, \binits{D.}}:
\batitle{{Refined Bounds on MeV-scale Thermal Dark Sectors from BBN and the
  CMB}}.
\bjtitle{JCAP}
\bvolume{01},
\bfpage{004}
(\byear{2020})
\doiurl{10.1088/1475-7516/2020/01/004}
{\href{https://arxiv.org/abs/1910.01649}{{arXiv:1910.01649}}}
{[hep-ph]}
\end{barticle}
\endbibitem

%%% 508
\bibitem[\protect\citeauthoryear{Slatyer}{2016}]{Slatyer:2015jla}
\begin{barticle}
\bauthor{\bsnm{Slatyer}, \binits{T.R.}}:
\batitle{{Indirect dark matter signatures in the cosmic dark ages. I.
  Generalizing the bound on s-wave dark matter annihilation from Planck
  results}}.
\bjtitle{Phys. Rev. D}
\bvolume{93}(\bissue{2}),
\bfpage{023527}
(\byear{2016})
\doiurl{10.1103/PhysRevD.93.023527}
{\href{https://arxiv.org/abs/1506.03811}{{arXiv:1506.03811}}}
{[hep-ph]}
\end{barticle}
\endbibitem

%%% 509
\bibitem[\protect\citeauthoryear{Antel et~al.}{2023}]{Antel:2023hkf}
\begin{barticle}
\bauthor{\bsnm{Antel}, \binits{C.}}, \betal:
\batitle{{Feebly-interacting particles: FIPs 2022 Workshop Report}}.
\bjtitle{Eur. Phys. J. C}
\bvolume{83}(\bissue{12}),
\bfpage{1122}
(\byear{2023})
\doiurl{10.1140/epjc/s10052-023-12168-5}
{\href{https://arxiv.org/abs/2305.01715}{{arXiv:2305.01715}}}
{[hep-ph]}
\end{barticle}
\endbibitem

%%% 510
\bibitem[\protect\citeauthoryear{Krnjaic and McDermott}{2020}]{Krnjaic:2019dzc}
\begin{barticle}
\bauthor{\bsnm{Krnjaic}, \binits{G.}},
\bauthor{\bsnm{McDermott}, \binits{S.D.}}:
\batitle{{Implications of BBN Bounds for Cosmic Ray Upscattered Dark Matter}}.
\bjtitle{Phys. Rev. D}
\bvolume{101}(\bissue{12}),
\bfpage{123022}
(\byear{2020})
\doiurl{10.1103/PhysRevD.101.123022}
{\href{https://arxiv.org/abs/1908.00007}{{arXiv:1908.00007}}}
{[hep-ph]}
\end{barticle}
\endbibitem

%%% 511
\bibitem[\protect\citeauthoryear{Boyarsky et~al.}{2019}]{Boyarsky:2018tvu}
\begin{barticle}
\bauthor{\bsnm{Boyarsky}, \binits{A.}},
\bauthor{\bsnm{Drewes}, \binits{M.}},
\bauthor{\bsnm{Lasserre}, \binits{T.}},
\bauthor{\bsnm{Mertens}, \binits{S.}},
\bauthor{\bsnm{Ruchayskiy}, \binits{O.}}:
\batitle{{Sterile neutrino Dark Matter}}.
\bjtitle{Prog. Part. Nucl. Phys.}
\bvolume{104},
\bfpage{1}--\blpage{45}
(\byear{2019})
\doiurl{10.1016/j.ppnp.2018.07.004}
{\href{https://arxiv.org/abs/1807.07938}{{arXiv:1807.07938}}}
{[hep-ph]}
\end{barticle}
\endbibitem

%%% 512
\bibitem[\protect\citeauthoryear{{Siegert} et~al.}{2022}]{2022A&A...660A.130S}
\begin{barticle}
\bauthor{\bsnm{{Siegert}}, \binits{T.}},
\bauthor{\bsnm{{Berteaud}}, \binits{J.}},
\bauthor{\bsnm{{Calore}}, \binits{F.}},
\bauthor{\bsnm{{Serpico}}, \binits{P.D.}},
\bauthor{\bsnm{{Weinberger}}, \binits{C.}}:
\batitle{{Diffuse Galactic emission spectrum between 0.5 and 8.0 MeV}}.
\bjtitle{\aap}
\bvolume{660},
\bfpage{130}
(\byear{2022})
\doiurl{10.1051/0004-6361/202142639}
{\href{https://arxiv.org/abs/2202.04574}{{arXiv:2202.04574}}}
{[astro-ph.HE]}
\end{barticle}
\endbibitem

%%% 513
\bibitem[\protect\citeauthoryear{Siegert et~al.}{2022}]{Siegert:2021upf}
\begin{barticle}
\bauthor{\bsnm{Siegert}, \binits{T.}},
\bauthor{\bsnm{Boehm}, \binits{C.}},
\bauthor{\bsnm{Calore}, \binits{F.}},
\bauthor{\bsnm{Diehl}, \binits{R.}},
\bauthor{\bsnm{Krause}, \binits{M.G.H.}},
\bauthor{\bsnm{Serpico}, \binits{P.D.}},
\bauthor{\bsnm{Vincent}, \binits{A.C.}}:
\batitle{{An INTEGRAL/SPI view of reticulum II: particle dark matter and
  primordial black holes limits in the MeV range}}.
\bjtitle{Mon. Not. Roy. Astron. Soc.}
\bvolume{511}(\bissue{1}),
\bfpage{914}--\blpage{924}
(\byear{2022})
\doiurl{10.1093/mnras/stac008}
{\href{https://arxiv.org/abs/2109.03791}{{arXiv:2109.03791}}}
{[astro-ph.HE]}
\end{barticle}
\endbibitem

%%% 514
\bibitem[\protect\citeauthoryear{Siegert et~al.}{2016}]{Siegert:2015knp}
\begin{barticle}
\bauthor{\bsnm{Siegert}, \binits{T.}},
\bauthor{\bsnm{Diehl}, \binits{R.}},
\bauthor{\bsnm{Khachatryan}, \binits{G.}},
\bauthor{\bsnm{Krause}, \binits{M.G.H.}},
\bauthor{\bsnm{Guglielmetti}, \binits{F.}},
\bauthor{\bsnm{Greiner}, \binits{J.}},
\bauthor{\bsnm{Strong}, \binits{A.W.}},
\bauthor{\bsnm{Zhang}, \binits{X.}}:
\batitle{{Gamma-ray spectroscopy of Positron Annihilation in the Milky Way}}.
\bjtitle{Astron. Astrophys.}
\bvolume{586},
\bfpage{84}
(\byear{2016})
\doiurl{10.1051/0004-6361/201527510}
{\href{https://arxiv.org/abs/1512.00325}{{arXiv:1512.00325}}}
{[astro-ph.HE]}
\end{barticle}
\endbibitem

%%% 515
\bibitem[\protect\citeauthoryear{Siegert et~al.}{2021}]{Siegert:2021trw}
\begin{barticle}
\bauthor{\bsnm{Siegert}, \binits{T.}},
\bauthor{\bsnm{Crocker}, \binits{R.M.}},
\bauthor{\bsnm{Macias}, \binits{O.}},
\bauthor{\bsnm{Panther}, \binits{F.H.}},
\bauthor{\bsnm{Calore}, \binits{F.}},
\bauthor{\bsnm{Song}, \binits{D.}},
\bauthor{\bsnm{Horiuchi}, \binits{S.}}:
\batitle{{Measuring the smearing of the Galactic 511-keV signal: positron
  propagation or supernova kicks?}}
\bjtitle{Mon. Not. Roy. Astron. Soc.}
\bvolume{509}(\bissue{1}),
\bfpage{11}--\blpage{16}
(\byear{2021})
\doiurl{10.1093/mnrasl/slab113}
{\href{https://arxiv.org/abs/2109.03691}{{arXiv:2109.03691}}}
{[astro-ph.HE]}
\end{barticle}
\endbibitem

%%% 516
\bibitem[\protect\citeauthoryear{Prantzos et~al.}{2011}]{Prantzos:2010wi}
\begin{barticle}
\bauthor{\bsnm{Prantzos}, \binits{N.}}, \betal:
\batitle{{The 511 keV emission from positron annihilation in the Galaxy}}.
\bjtitle{Rev. Mod. Phys.}
\bvolume{83},
\bfpage{1001}--\blpage{1056}
(\byear{2011})
\doiurl{10.1103/RevModPhys.83.1001}
{\href{https://arxiv.org/abs/1009.4620}{{arXiv:1009.4620}}}
{[astro-ph.HE]}
\end{barticle}
\endbibitem

%%% 517
\bibitem[\protect\citeauthoryear{Aghaie et~al.}{2025}]{Aghaie:2025dgl}
\begin{botherref}
\oauthor{\bsnm{Aghaie}, \binits{M.}},
\oauthor{\bsnm{Torre~Luque}, \binits{P.}},
\oauthor{\bsnm{Dondarini}, \binits{A.}},
\oauthor{\bsnm{Gaggero}, \binits{D.}},
\oauthor{\bsnm{Marino}, \binits{G.}},
\oauthor{\bsnm{Panci}, \binits{P.}}:
{(H)ALPing the 511 keV line: A thermal DM interpretation of the 511 keV
  emission}
(2025)
{\href{https://arxiv.org/abs/2501.10504}{{arXiv:2501.10504}}}
{[hep-ph]}
\end{botherref}
\endbibitem

%%% 518
\bibitem[\protect\citeauthoryear{Vincent et~al.}{2012}]{Vincent:2012an}
\begin{barticle}
\bauthor{\bsnm{Vincent}, \binits{A.C.}},
\bauthor{\bsnm{Martin}, \binits{P.}},
\bauthor{\bsnm{Cline}, \binits{J.M.}}:
\batitle{{Interacting dark matter contribution to the Galactic 511 keV gamma
  ray emission: constraining the morphology with INTEGRAL/SPI observations}}.
\bjtitle{JCAP}
\bvolume{04},
\bfpage{022}
(\byear{2012})
\doiurl{10.1088/1475-7516/2012/04/022}
{\href{https://arxiv.org/abs/1201.0997}{{arXiv:1201.0997}}}
{[hep-ph]}
\end{barticle}
\endbibitem

%%% 519
\bibitem[\protect\citeauthoryear{Liu et~al.}{2016}]{Liu:2016cnk}
\begin{barticle}
\bauthor{\bsnm{Liu}, \binits{H.}},
\bauthor{\bsnm{Slatyer}, \binits{T.R.}},
\bauthor{\bsnm{Zavala}, \binits{J.}}:
\batitle{{Contributions to cosmic reionization from dark matter annihilation
  and decay}}.
\bjtitle{Phys. Rev. D}
\bvolume{94}(\bissue{6}),
\bfpage{063507}
(\byear{2016})
\doiurl{10.1103/PhysRevD.94.063507}
{\href{https://arxiv.org/abs/1604.02457}{{arXiv:1604.02457}}}
{[astro-ph.CO]}
\end{barticle}
\endbibitem

%%% 520
\bibitem[\protect\citeauthoryear{Diamanti et~al.}{2014}]{Diamanti:2013bia}
\begin{barticle}
\bauthor{\bsnm{Diamanti}, \binits{R.}},
\bauthor{\bsnm{Lopez-Honorez}, \binits{L.}},
\bauthor{\bsnm{Mena}, \binits{O.}},
\bauthor{\bsnm{Palomares-Ruiz}, \binits{S.}},
\bauthor{\bsnm{Vincent}, \binits{A.C.}}:
\batitle{{Constraining Dark Matter Late-Time Energy Injection: Decays and
  P-Wave Annihilations}}.
\bjtitle{JCAP}
\bvolume{02},
\bfpage{017}
(\byear{2014})
\doiurl{10.1088/1475-7516/2014/02/017}
{\href{https://arxiv.org/abs/1308.2578}{{arXiv:1308.2578}}}
{[astro-ph.CO]}
\end{barticle}
\endbibitem

%%% 521
\bibitem[\protect\citeauthoryear{Bartels et~al.}{2017}]{Bartels:2017dpb}
\begin{barticle}
\bauthor{\bsnm{Bartels}, \binits{R.}},
\bauthor{\bsnm{Gaggero}, \binits{D.}},
\bauthor{\bsnm{Weniger}, \binits{C.}}:
\batitle{{Prospects for indirect dark matter searches with MeV photons}}.
\bjtitle{JCAP}
\bvolume{05},
\bfpage{001}
(\byear{2017})
\doiurl{10.1088/1475-7516/2017/05/001}
{\href{https://arxiv.org/abs/1703.02546}{{arXiv:1703.02546}}}
{[astro-ph.HE]}
\end{barticle}
\endbibitem

%%% 522
\bibitem[\protect\citeauthoryear{Carenza et~al.}{2025}]{Carenza:2024ehj}
\begin{barticle}
\bauthor{\bsnm{Carenza}, \binits{P.}},
\bauthor{\bsnm{Giannotti}, \binits{M.}},
\bauthor{\bsnm{Isern}, \binits{J.}},
\bauthor{\bsnm{Mirizzi}, \binits{A.}},
\bauthor{\bsnm{Straniero}, \binits{O.}}:
\batitle{{Axion astrophysics}}.
\bjtitle{Phys. Rept.}
\bvolume{1117},
\bfpage{1}--\blpage{102}
(\byear{2025})
\doiurl{10.1016/j.physrep.2025.02.002}
{\href{https://arxiv.org/abs/2411.02492}{{arXiv:2411.02492}}}
{[hep-ph]}
\end{barticle}
\endbibitem

%%% 523
\bibitem[\protect\citeauthoryear{Calore et~al.}{2021}]{Calore:2021klc}
\begin{barticle}
\bauthor{\bsnm{Calore}, \binits{F.}},
\bauthor{\bsnm{Carenza}, \binits{P.}},
\bauthor{\bsnm{Giannotti}, \binits{M.}},
\bauthor{\bsnm{J{\"a}ckel}, \binits{J.}},
\bauthor{\bsnm{Lucente}, \binits{G.}},
\bauthor{\bsnm{Mirizzi}, \binits{A.}}:
\batitle{{Supernova bounds on axionlike particles coupled with nucleons and
  electrons}}.
\bjtitle{Phys. Rev. D}
\bvolume{104}(\bissue{4}),
\bfpage{043016}
(\byear{2021})
\doiurl{10.1103/PhysRevD.104.043016}
{\href{https://arxiv.org/abs/2107.02186}{{arXiv:2107.02186}}}
{[hep-ph]}
\end{barticle}
\endbibitem

%%% 524
\bibitem[\protect\citeauthoryear{Aramaki et~al.}{2022}]{Aramaki:2022zpw}
\begin{botherref}
\oauthor{\bsnm{Aramaki}, \binits{T.}}, et al.:
{Snowmass2021 Cosmic Frontier: The landscape of cosmic-ray and high-energy
  photon probes of particle dark matter}
(2022)
{\href{https://arxiv.org/abs/2203.06894}{{arXiv:2203.06894}}}
{[hep-ex]}
\end{botherref}
\endbibitem

%%% 525
\bibitem[\protect\citeauthoryear{Caputo et~al.}{2022}]{Caputo:2022dkz}
\begin{botherref}
\oauthor{\bsnm{Caputo}, \binits{A.}},
\oauthor{\bsnm{Negro}, \binits{M.}},
\oauthor{\bsnm{Regis}, \binits{M.}},
\oauthor{\bsnm{Taoso}, \binits{M.}}:
{Dark Matter prospects with COSI: ALPs, PBHs and sub-GeV Dark Matter}
(2022)
{\href{https://arxiv.org/abs/2210.09310}{{arXiv:2210.09310}}}
{[hep-ph]}
\end{botherref}
\endbibitem

%%% 526
\bibitem[\protect\citeauthoryear{Coogan et~al.}{2021}]{Coogan:2021sjs}
\begin{barticle}
\bauthor{\bsnm{Coogan}, \binits{A.}},
\bauthor{\bsnm{Morrison}, \binits{L.}},
\bauthor{\bsnm{Profumo}, \binits{S.}}:
\batitle{{Precision gamma-ray constraints for sub-GeV dark matter models}}.
\bjtitle{JCAP}
\bvolume{08},
\bfpage{044}
(\byear{2021})
\doiurl{10.1088/1475-7516/2021/08/044}
{\href{https://arxiv.org/abs/2104.06168}{{arXiv:2104.06168}}}
{[hep-ph]}
\end{barticle}
\endbibitem

\end{thebibliography}


\begin{thebibliography}{}

\bibitem[Canete et al.(2023)]{Can23} Canete, L., Doherty, D.~T., Lotay, G., et al.\ 2023, \prc, 108, 035807. doi:10.1103/PhysRevC.108.035807

\bibitem[Churazov et al.(2020)]{Chu20} Churazov, E., Bouchet, L., Jean, P., et al.\ 2020, \nar, 90, 101548. doi:10.1016/j.newar.2020.101548

\bibitem[Clayton \& Hoyle(1974)]{CH74} Clayton, D.~D. \& Hoyle, F.\ 1974, \apjl, 187, L101. doi:10.1086/181406

\bibitem[Clayton(1981)]{Cla81} Clayton, D.~D.\ 1981, \apjl, 244, L97. doi:10.1086/183488

\bibitem[Diehl et al.(2006)]{Die06} Diehl, R., Halloin, H., Kretschmer, K., et al.\ 2006, \nat, 439, 45. doi:10.1038/nature04364

\bibitem[Foug{\`e}res et al.(2023)]{Fou23} Foug{\`e}res, C., de Oliveira Santos, F., Jos{\'e}, J., et al.\ 2023, Nature Communications, 14, 4536. doi:10.1038/s41467-023-40121-3

\bibitem[Gomez-Gomar et al.(1998)]{Gom98} Gomez-Gomar, J., Hernanz, M., Jose, J., et al.\ 1998, \mnras, 296, 913. doi:10.1046/j.1365-8711.1998.01421.x

\bibitem[Harris et al.(1991)]{Har91} Harris, M.~J., Leising, M.~D., \& Share, G.~H.\ 1991, \apj, 375, 216. doi:10.1086/170183

\bibitem[Harris et al.(1999)]{Har99} Harris, M.~J., Naya, J.~E., Teegarden, B.~J., et al.\ 1999, \apj, 522, 424. doi:10.1086/307625

\bibitem[Harris et al.(2000)]{Har00} Harris, M.~J., Teegarden, B.~J., Cline, T.~L., et al.\ 2000, \apj, 542, 1057. doi:10.1086/317022

\bibitem[Hernanz et al.(1999)]{Her99a} Hernanz, M., Jos{\'e}, J., Coc, A., et al.\ 1999, \apjl, 526, L97. doi:10.1086/312372

\bibitem[Hernanz et al.(1999)]{Her99b} Hernanz, M., G{\'o}mez-Gomar, J., Jos{\'e}, J., et al.\ 1999, Astrophysical Letters and Communications, 38, 407. doi:10.48550/arXiv.astro-ph/9812100

\bibitem[Hernanz et al.(2000)]{Her00} Hernanz, M., Smith, D.~M., Fishman, J., et al.\ 2000, The Fifth Compton Symposium, 510, 82. doi:10.1063/1.1303179

\bibitem[Hernanz \& Jos{\'e}(2006)]{HJ06} Hernanz, M. \& Jos{\'e}, J.\ 2006, \nar, 50, 504. doi:10.1016/j.newar.2006.06.012

\bibitem[Hernanz (2014)]{Her14} Hernanz, M.\ 2014, Stellar Novae: Past and Future Decades, 490, 319. doi:10.48550/arXiv.1305.0769

\bibitem[Higdon \& Fowler(1987)]{HF87} Higdon, J.~C. \& Fowler, W.~A.\ 1987, \apj, 317, 710. doi:10.1086/165317

\bibitem[Isern et al. (2021)]{Ise21} Isern, J., Hernanz, M., Bravo, E., et al.\ 2021, \nar, 92, 101606. doi:10.1016/j.newar.2020.101606

\bibitem[Jean et al.(1999)]{Jea99} Jean, P., G{\'o}mez-Gomar, J., Hernanz, M., et al.\ 1999, Astrophysical Letters and Communications, 38, 421. doi:10.48550/arXiv.astro-ph/9903015

\bibitem[Jean et al.(2000)]{Jea00} Jean, P., Hernanz, M., G{\'o}mez-Gomar, J., et al.\ 2000, \mnras, 319, 350. doi:10.1046/j.1365-8711.2000.03587.x

\bibitem[Jean et al. (2001)]{Jea01} Jean, P., Kn{\"o}dlseder, J., von Ballmoos, P., et al.\ 2001, Exploring the Gamma-Ray Universe, 459, 73. doi:10.48550/arXiv.astro-ph/0106340

\bibitem[Jos{\'e} et al. (1997)]{Jos97} Jos{\'e}, J., Hernanz, M., \& Coc, A.\ 1997, \apjl, 479, L55. doi:10.1086/310575

\bibitem[Jos{\'e} \& Hernanz(1998)]{Jos98} Jos{\'e}, J. \& Hernanz, M.\ 1998, \apj, 494, 680. doi:10.1086/305244

\bibitem[Iyudin et al.(1995)]{Iyu95} Iyudin, A.~F., Bennett, K., Bloemen, H., et al.\ 1995, \aap, 300, 422

\bibitem[Izzo et al.(2025)]{Izz25} Izzo, L., Siegert, T., Jean, P., et al.\ 2025, \aap, submitted

\bibitem[Kn{\"o}dlseder et al.(1999)]{Kno99} Kn{\"o}dlseder, J., Bennett, K., Bloemen, H., et al.\ 1999, \aap, 344, 68

\bibitem[Leising et al.(1988)]{Lei88} Leising, M.~D., Share, G.~H., Chupp, E.~L., et al.\ 1988, \apj, 328, 755. doi:10.1086/166334

\bibitem[Leising \& Clayton(1987)]{Lei87} Leising, M.~D. \& Clayton, D.~D.\ 1987, \apj, 323, 159. doi:10.1086/165816

\bibitem[Leung \& Siegert(2022)]{LS22} Leung, S.-C. \& Siegert, T. \ 2022, \mnras, 516, 1008. doi:10.1093/mnras/stac1672

\bibitem[Mahoney et al.(1982)]{Mah82} Mahoney, W.~A., Ling, J.~C., Jacobson, A.~S., et al.\ 1982, \apj, 262, 742. doi:10.1086/160469

\bibitem[Prantzos et al.(2011)]{Pra11} Prantzos, N., Boehm, C., Bykov, A.~M., et al.\ 2011, Reviews of Modern Physics, 83, 1001. doi:10.1103/RevModPhys.83.1001

\bibitem[Senziani et al.(2008)]{Sen08} Senziani, F., Skinner, G.~K., Jean, P., et al.\ 2008, \aap, 485, 223. doi:10.1051/0004-6361:200809863

\bibitem[Siegert et al.(2018)]{Sie18} Siegert, T., Coc, A., Delgado, L., et al.\ 2018, \aap, 615, A107. doi:10.1051/0004-6361/201732514

\bibitem[Siegert et al.(2021)]{Sie21} Siegert, T., Ghosh, S., Mathur, K., et al.\ 2021, \aap, 650, A187. doi:10.1051/0004-6361/202140300

\bibitem[Tomsick et al.(2024)]{Tom24} Tomsick, J., Boggs, S., Zoglauer, A., et al.\ 2024, 38th International Cosmic Ray Conference, 745. doi:10.48550/arXiv.2308.12362

\end{thebibliography}

\end{document}